\documentclass[acmsmall]{acmart}

\setcopyright{cc}
\setcctype{by}
\acmDOI{10.1145/3871760}
\acmYear{2027}
\acmJournal{PACMPL}
\acmVolume{11}
\acmNumber{POPL}
\acmArticle{58}
\acmMonth{1}
\acmSubmissionID{popl27main-p524-p}
\received{2026-07-09}
\received[accepted]{2026-10-05}

\usepackage[utf8]{inputenc}
\usepackage{lipsum}

\usepackage{hyperref}
\usepackage{tabularx}

\usepackage{booktabs}
\usepackage{subcaption}
\usepackage{amsthm}
\usepackage{mathrsfs}

\usepackage{etoolbox}%

\usepackage[dvipsnames]{xcolor}%

\usepackage{xr}%
\usepackage{amsmath}%
\usepackage{wrapfig}%

\usepackage[nomargin,inline,index,%
  status=draft
]{fixme}
\fxusetheme{colorsig}
\FXRegisterAuthor{fxJM}{anfxJM}{JM}%
\FXRegisterAuthor{fxAB}{anfxAB}{AB}%
\FXRegisterAuthor{fxNY}{anfxNY}{NY}%
\FXRegisterAuthor{fxFZ}{anfxFZ}{FZ}
\FXRegisterAuthor{fxPH}{anfxPH}{PH}

\makeatletter
\DeclareFontFamily{OMX}{MnSymbolE}{}
\DeclareSymbolFont{MnLargeSymbols}{OMX}{MnSymbolE}{m}{n}
\SetSymbolFont{MnLargeSymbols}{bold}{OMX}{MnSymbolE}{b}{n}
\DeclareFontShape{OMX}{MnSymbolE}{m}{n}{
    <-6>  MnSymbolE5
   <6-7>  MnSymbolE6
   <7-8>  MnSymbolE7
   <8-9>  MnSymbolE8
   <9-10> MnSymbolE9
  <10-12> MnSymbolE10
  <12->   MnSymbolE12
}{}
\DeclareFontShape{OMX}{MnSymbolE}{b}{n}{
    <-6>  MnSymbolE-Bold5
   <6-7>  MnSymbolE-Bold6
   <7-8>  MnSymbolE-Bold7
   <8-9>  MnSymbolE-Bold8
   <9-10> MnSymbolE-Bold9
  <10-12> MnSymbolE-Bold10
  <12->   MnSymbolE-Bold12
}{}

\let\llangle\@undefined
\let\rrangle\@undefined
\DeclareMathDelimiter{\llangle}{\mathopen}%
                     {MnLargeSymbols}{'164}{MnLargeSymbols}{'164}
\DeclareMathDelimiter{\rrangle}{\mathclose}%
                     {MnLargeSymbols}{'171}{MnLargeSymbols}{'171}
\makeatother

\usepackage[inline]{enumitem}%

\usepackage{nicefrac}%

\usepackage{proof}%

\usepackage[most]{tcolorbox}%
\newtcolorbox{myframe}[1][]{
  enhanced,
  arc=0pt,
  outer arc=0pt,
  colback=white,
  boxrule=0.5pt,
  boxsep=0mm,
  left=1mm,
  right=1mm,
  top=0.5mm,
  bottom=0.5mm,
  #1
}

\lstdefinelanguage{Scribble}{%
  basicstyle=\footnotesize\ttfamily,
  stringstyle=\color{Blue},
  showstringspaces=false,
  keywords={nested,new,calls,and,as,at,by,catches,choice,continue,do,from,global,import,instantiates,interruptible,local,module,or,par,protocol,rec,role,sig,throws,to,type,with,int,aux,reliable,crash},
  morestring=[b]",
  morestring=[b]',
  morecomment=[l][\color{greencomments}]{//},
}

\lstdefinelanguage{nuScr}{%
  basicstyle=\footnotesize\ttfamily,
  stringstyle=\color{Blue},
  showstringspaces=false,
  keywords={
    nested,new,calls,and,as,at,by,catches,choice,continue,do,from,global,import,instantiates,interruptible,local,module,or,par,protocol,rec,role,sig,throws,to,type,with,int,aux,
    safe
  },
  morestring=[b]",
  morestring=[b]',
  morecomment=[l][\color{greencomments}]{//},
  morecomment=[s][\color{magenta}]{(*}{*)},
}

\lstdefinelanguage{effpi}{
  keywords=[1]{
    case,class,sealed,abstract,object,extends,type,def,val,if,else,new,var,match
  },
  keywords=[2]{
    InChan,OutChan,RecVar,Rec,Out,In,InErr,Loop,
  },
  keywords=[3]{
    rec,send,receive,receiveErr,eval,par,Channel
  },
  keywordstyle=[1]{\color{blue}},
  keywordstyle=[2]{\color{ImperialIris}},
  keywordstyle=[3]{\color{OliveGreen}},
  otherkeywords={=>,.type,<:,>>:},
  morecomment=[l][\color{darkgray}]{//},
}

\usepackage{stmaryrd}
\usepackage{thmtools}%
\usepackage{arydshln}

\usepackage{xifthen}%
\usepackage{xspace}%

\usepackage{mathtools}%

\usepackage{balance}%

\usepackage{hyperref}
\hypersetup{hidelinks}
\usepackage[capitalise]{cleveref}

\definecolor{ImperialBlue}{HTML}{003E74}
\definecolor{ImperialDarkGreen}{HTML}{02893B}
\definecolor{ImperialTangerine}{HTML}{EC7300}
\definecolor{ImperialIris}{HTML}{751E66}

\definecolor{RYB1}{RGB}{141, 211, 199}
\definecolor{RYB2}{RGB}{255, 255, 179}
\definecolor{RYB3}{RGB}{190, 186, 218}
\definecolor{RYB4}{RGB}{251, 128, 114}
\definecolor{RYB5}{RGB}{128, 177, 211}
\definecolor{RYB6}{RGB}{253, 180, 98}
\definecolor{RYB7}{RGB}{179, 222, 105}
\usepackage{pgfplots}
\usetikzlibrary{shapes, shapes.geometric, arrows, arrows.meta, calc, decorations, decorations.pathreplacing, calligraphy, positioning, patterns, backgrounds,fit,snakes, automata,intersections, pgfplots.fillbetween}
\tikzset{
  >=stealth,
  node distance=2cm,
  every state/.style={thick, fill=gray!10},
  initial text=$ $,
}
\tikzstyle{rect} = [rectangle, rounded corners, minimum width=3cm, minimum height=1cm,text centered, draw=black, fill=black!10]
\tikzstyle{medrect} = [rectangle, rounded corners, minimum height=.7cm,text centered, draw=black, fill=black!10]
\tikzstyle{smallrect} = [rectangle, rounded corners,text centered, draw=black, fill=black!10]
\tikzstyle{circ} = [circle, minimum height=0.55cm,text centered, draw=black, fill=black!10]
\tikzstyle{smallcirc} = [circle,text centered, draw=black, fill=black!10]
\tikzstyle{arrow} = [thick,-{Latex[length=2mm, width=1.5mm]},>=stealth]
\tikzstyle{arrow2} = [thick,{Latex[length=2mm, width=1.5mm]}-{Latex[length=2mm, width=1.5mm]},>=stealth]
\tikzstyle{garrow} = [thick,-{Latex[length=2mm, width=1.5mm]},>=stealth,draw=black!40]
\tikzstyle{garrow2} = [thick,{Latex[length=2mm, width=1.5mm]}-{Latex[length=2mm, width=1.5mm]},>=stealth,draw=black!40]
\tikzstyle{venn} = [preaction={fill, #1},opacity=0.6,anchor=south,rounded corners=2pt,draw=black]
\tikzstyle{exnode} = [circle,draw=black,inner sep=1pt,fill=gray!20]
\pgfplotscreateplotcyclelist{stackexchange}{
  {RYB1!50!black,fill=RYB1},
  {RYB4!50!black,fill=RYB4},
  {RYB3!50!black,fill=RYB3},
  {RYB2!50!black,fill=RYB2},
  {RYB5!50!black,fill=RYB5},
  {RYB6!50!black,fill=RYB6},
  {RYB7!50!black,fill=RYB7},
}
\pgfplotscreateplotcyclelist{stackexchangePlusOne}{
  {RYB4!50!black,fill=RYB4},
  {RYB3!50!black,fill=RYB3},
  {RYB2!50!black,fill=RYB2},
  {RYB5!50!black,fill=RYB5},
  {RYB6!50!black,fill=RYB6},
  {RYB7!50!black,fill=RYB7},
}
\pgfplotsset{
  compat=1.8,
  /pgfplots/bar cycle list/.style={/pgfplots/cycle list={%
    {brown!60!black,fill=brown!30!white,mark=none},
    {red,fill=red!30!white,mark=none},
    {blue,fill=blue!30!white,mark=none},
    {black,fill=gray,mark=none},
    }
  },
}
\newcolumntype{L}{>{$}l<{$}}
\newcolumntype{C}{>{$}c<{$}}
\newcolumntype{P}[1]{>{\centering\arraybackslash$}p{#1}<{$}}
\usepackage{multirow}

\usepackage[export]{adjustbox}

\Crefname{section}{\S\!}{\S\!}%
\Crefname{subsection}{\S\!}{\S\!}%
\Crefname{subsubsection}{\S\!}{\S\!}%
\Crefname{appendix}{Appendix \S\!}{Appendix \S\!}
\Crefname{definition}{Def.\@}{Defs.\@}%
\Crefname{figure}{Fig.\@}{Figs.\@}%
\Crefname{example}{Ex.\@}{Exs.\@}%
\Crefname{corollary}{Cor.\@}{Cors.\@}%
\Crefname{theorem}{Thm.\@}{Thms.\@}%
\Crefname{proposition}{Prop.\@}{Props.\@}%
\Crefname{lemma}{Lem.\@}{Lems.\@}
\Crefname{equation}{Eq.\@}{Eqs.\@}

\crefname{section}{\S\!}{\S\!}%
\crefname{subsection}{\S\!}{\S\!}%
\crefname{subsubsection}{\S\!}{\S\!}%
\crefname{appendix}{Appendix \S\!}{Appendix \S\!}
\crefname{definition}{Def.\@}{Defs.\@}%
\crefname{figure}{Fig.\@}{Figs.\@}%
\crefname{example}{Ex.\@}{Exs.\@}%
\crefname{corollary}{Cor.\@}{Cors.\@}%
\crefname{theorem}{Thm.\@}{Thms.\@}%
\crefname{proposition}{Prop.\@}{Props.\@}%
\crefname{lemma}{Lem.\@}{Lems.\@}
\crefname{equation}{Eq.\@}{Eqs.\@}

\usepackage{etoc}

\usepackage{magicvariables}

\renewcommand{\ell}{\labFmt{m}}

\newif\ifdraft%
  \draftfalse%
  \drafttrue%
  
\newcommand{\appref}[1]{\cite[{#1}]{fullversion}}

\newcommand{\ifempty}[3]{%
  \ifthenelse{\isempty{#1}}{#2}{#3}%
}%

\newtoggle{techreport}%

\newtoggle{cruft}%

\newcommand{\dom}[1]{{\color{black}\operatorname{dom}}\!\left({#1}\right)}%
\newcommand{\ran}[1]{{\color{black}\operatorname{ran}}\!\left({#1}\right)}%
\newcommand{\suchthat}{\colon}%
\newcommand{\gtFv}[1]{
\ifempty{#1}
{
\gtFmt{\operatorname{fv}}
}
{
\gtFmt{\operatorname{fv}}\!\left({#1}\right)
}
}%
\newcommand{\fpv}[1]{\operatorname{fpv}\!\left({#1}\right)}%
\newcommand{\unfoldOne}[1]{%
  {\color{black}\operatorname{unf}}\!\left({#1}\right)}%
\newcommand{\notImpliedBy}{\mathrel{{\kern .5em}{\not{\kern -1.2em}\impliedby}}}%

\newcommand{\coloncolonequals}{\Coloneqq}%
\newcommand{\bnfdef}{\coloncolonequals}%
\newcommand{\bnfsep}{\mathbin{\;\big|\;}}%

\newcommand{\eval}[2]{#1 \downarrow #2}

\def\etc{\emph{etc}\@\xspace}%
\def\eg{e.g.\@\xspace}%
\def\ie{i.e.\@\xspace}%

\definecolor{ruleColor}{rgb}{0.1, 0.3, 0.1}%
\newcommand{\inferrule}[1]{{\color{ruleColor}\textsc{\scriptsize [#1]}}}%
\newcommand{\inference}[3][]{\infer[\ifempty{#1}{}{\inferrule{#1}}]{#3}{#2}}%
\newcommand{\cinference}[3][]{\infer=[\ifempty{#1}{}{\inferrule{#1}}]{#3}{#2}}%

\newcommand{\setenum}[1]{\mathord{{\color{black}\left\{#1\right\}}}}%

\newcommand{\predP}[1][]{\ifempty{#1}{\varphi}{\varphi_{#1}}}%
\newcommand{\predPApp}[2][]{\ifempty{#1}{\predP}{\predP[{#1}]}\!\left({#2}\right)}%

\newcommand{\bisim}{\sim}%

\newcommand{\bind}[2]{{#2}/{#1}}%
\newcommand{\substenum}[1]{\mathord{\left[{#1}\right]}}%
\newcommand{\subst}[2]{\substenum{\bind{#1}{#2}}}%

\definecolor{hlColor}{rgb}{0.65, 1.0, 0.65}%

\newcommand{\lbbar}{\{\kern-0.2em|}
\newcommand{\rbbar}{|\kern-0.2em\}}

\definecolor{tyColorCustom}{rgb}{0.0, 0.0, 0.85}%
\newcommand{\tyCol}[1]{{\color{tyColorCustom}{#1}}}%
\newcommand{\tyFont}[1]{{#1}}%
\newcommand{\tyFmt}[1]{\tyCol{\tyFont{#1}}}%
\newcommand{\tyFontC}[1]{\operatorname{#1}}%
\newcommand{\tyFmtC}[1]{\tyCol{\tyFontC{#1}}}%

\newcommand{\tyGroundSet}{\stFmt{\mathcal{B}}}  
\newcommand{\tyGround}[1][]{\tyFmt{\ifempty{#1}{B}{B_{#1}}}}%
\newcommand{\tyGroundi}[1][]{\tyFmt{\ifempty{#1}{B'}{B'_{#1}}}}%
\newcommand{\tyGroundii}[1][]{\tyFmt{\ifempty{#1}{B''}{B''_{#1}}}}%
\newcommand{\tyBool}{\tyFmtC{bool}}%
\newcommand{\tyInt}{\tyFmtC{int}}%
\newcommand{\tyReal}{\tyFmtC{real}}%
\newcommand{\tyString}{\tyFmtC{str}}%

\newcommand{\tyTi}[1][]{\tyCol{\ifempty{#1}{\tyFont{T}'}{\tyFont{T}'_{#1}}}}%

\newcommand{\tySub}{\mathrel{\tyCol{\leqslant}}}%
\newcommand{\tyNotSub}{\mathrel{\tyCol{\not\leqslant}}}%
\newcommand{\tySSub}{\mathrel{\tyCol{\leqslant_{\&}}}}%
\newcommand{\tyJudge}[3]{%
  {#1} \mathrel{{\vdash}} {#2} \mathrel{\tyCol{:}} {#3}%
}%
\newcommand{\tyJudgei}[3]{%
  {#1} \mathrel{{\vdash}'} {#2} \mathrel{\tyCol{:}} {#3}%
}%

\newcommand{\muCol}[1]{{\color{red}#1}}%
\newcommand{\muWordEmpty}[1][]{\muCol{\epsilon}}%
{\centerline{\bf --- Begin Copied From Previous Paper ---} \hrule}%
{\hrule \centerline{\bf --- End Copied From Previous Paper ---}}%

{\centerline{\bf --- Begin Discussion\ifempty{#1}{}{: {#1}} ---}
  \hrule\vspace{1mm}}%
{\hrule\vspace{1mm}\centerline{\bf --- End Discussion ---}}%

\newtcolorbox{cross}{blank,breakable,parbox=false,
  overlay={\draw[red,line width=5pt] (interior.south west)--(interior.north east);
    \draw[red,line width=5pt] (interior.north west)--(interior.south east);}}

\definecolor{roleColor}{rgb}{0.5, 0.0, 0.0}%
\newcommand{\roleCol}[1]{{\color{roleColor}#1}}%
\newcommand{\roleSet}{\roleCol{\mathcal{R}}}%
\newcommand{\roleFmt}[1]{\ensuremath{{\boldsymbol{\roleCol{\mathtt{#1}}}}}\xspace}%

\newcommand{\roleP}[1][]{%
  \ifempty{#1}{{\color{roleColor}\roleFmt{p}}}{{\color{roleColor}\roleFmt{p}_{#1}}}%
}%
\newcommand{\rolePi}[1][]{%
  \ifempty{#1}{{\color{roleColor}\roleFmt{p}'}}{{\color{roleColor}\roleFmt{p}'_{#1}}}%
}%
\newcommand{\roleQ}[1][]{%
  \ifempty{#1}{{\color{roleColor}\roleFmt{q}}}{{\color{roleColor}\roleFmt{q}_{#1}}}%
}%
\newcommand{\roleQi}[1][]{%
  \ifempty{#1}{{\color{roleColor}\roleFmt{q}'}}{{\color{roleColor}\roleFmt{q}'_{#1}}}%
}%
\newcommand{\roleR}[1][]{%
  \ifempty{#1}{{\color{roleColor}\roleFmt{r}}}{{\color{roleColor}\roleFmt{r}_{\!#1}}}%
}%
\newcommand{\roleRi}[1][]{%
  \ifempty{#1}{{\color{roleColor}\roleFmt{r}'}}{{\color{roleColor}\roleFmt{r}'_{\!#1}}}%
}%
\newcommand{\roleS}[1][]{%
  \ifempty{#1}{{\color{roleColor}\roleFmt{s}}}{{\color{roleColor}\roleFmt{s}_{\!#1}}}%
}%
\newcommand{\roleSi}[1][]{%
  \ifempty{#1}{{\color{roleColor}\roleFmt{s}'}}{{\color{roleColor}\roleFmt{s}'_{\!#1}}}%
}%
\newcommand{\roleT}[1][]{%
  \ifempty{#1}{{\color{roleColor}\roleFmt{t}}}{{\color{roleColor}\roleFmt{t}_{\!#1}}}%
}%

\definecolor{gtColor}{rgb}{0.43, 0.21, 0.1}%
\newcommand{\gtFmt}[1]{\ensuremath{{\color{gtColor}#1}}\xspace}%
\newcommand{\gtMsgFmt}[1]{\gtFmt{\labFmt{#1}}}%
\newcommand{\gtLab}[1][]{%
  \ifempty{#1}{\gtMsgFmt{m}}{{\color{gtColor}\gtMsgFmt{m}_{#1}}}%
}%
\newcommand{\gtLabi}[1][]{%
  \ifempty{#1}{\gtMsgFmt{m}'}{{\color{gtColor}\gtMsgFmt{m}'_{#1}}}%
}%

\newcommand{\gtG}[1][]{\gtFmt{\ifempty{#1}{G}{G_{#1}}}}%
\newcommand{\gtGi}[1][]{\gtFmt{\ifempty{#1}{G'}{G'_{#1}}}}%
\newcommand{\gtGii}[1][]{\gtFmt{\ifempty{#1}{G''}{G''_{#1}}}}%
\newcommand{\gtGiii}[1][]{\gtFmt{\ifempty{#1}{G'''}{G'''_{#1}}}}%
\newcommand{\gtGorigin}[1][]{\gtFmt{\ifempty{#1}{G^{\text{\tiny origin}}}{G^{\text{\tiny origin}}_{#1}}}}%
\newcommand{\gtGsbal}[1][]{\gtFmt{\ifempty{#1}{G^{\text{\tiny sbal}}}{G^{\text{\tiny sbal}}_{#1}}}}%
\newcommand{\gtGcounter}[1][]{\gtFmt{\ifempty{#1}{G^{\text{\tiny coex}}}{G^{\text{\tiny coex}}_{#1}}}}%

\newcommand{\gtSeq}{\mathbin{\gtFmt{.}}}%

\newcommand{\gtPar}[2]{#1 \, \gtFmt{\parallel} \, #2}

\newcommand{\gtCommRaw}[3]{%
  \gtFmt{%
    {#1} {\to} {#2}{:}%
    \left\{%
      {#3}%
    \right\}%
  }%
}%
\newcommand{\gtComm}[6]{%
  \gtFmt{%
    \gtCommRaw{#1}{#2}{%
      \gtCommChoice{#4}{#5}{#6}%
    }_{#3}%
  }%
}%
\newcommand{\gtCommSDRaw}[2]{%
  \gtFmt{\Sigma
\left\{
    {#1} {\to}%
      {#2}%
    \right\}%
  }%
}%
\newcommand{\gtCommSD}[6]{%
  \gtFmt{%
    \gtCommSDRaw{#1}{%
      {#2}:\gtCommChoice{#4}{#5}{#6}%
    }_{#3}%
  }%
}%
\newcommand{\gtCommSmall}[6]{%
  \gtFmt{%
    \gtCommRaw{#1}{#2}{%
      \gtCommChoiceSmall{#4}{#5}{#6}%
    }_{#3}%
  }%
}%
\newcommand{\gtCommSingle}[5]{%
  \gtFmt{%
    {#1} {\to} {#2}{:}%
    \gtCommChoice{#3}{#4}{#5}%
  }%
}%

\newcommand{\gtCommRawSquig}[4]{%
  \gtFmt{%
    {#1} {\overset{#3}{\rightsquigarrow}} {#2}{:}%
    \left\{%
      {#4}%
    \right\}%
  }%
}

\newcommand{\gtCommSquig}[7]{%
  \gtFmt{%
    \gtCommRawSquig{#1}{#2}{#3}{%
      \gtCommChoice{#5}{#6}{#7}%
    }_{#4}%
  }%
}

\newcommand{\gtCommSquigSmall}[7]{%
  \gtFmt{%
    \gtCommRawSquig{#1}{#2}{#3}{%
      \gtCommChoiceSmall{#5}{#6}{#7}%
    }_{#4}%
  }%
}

\newcommand{\gtCommSquigSingle}[6]{%
  \gtFmt{%
    {#1} {\overset{#3}{\rightsquigarrow}} {#2}{:}%
    \gtCommChoice{#4}{#5}{#6}%
  }%
}

\newcommand{\gtCommChoice}[3]{%
  \gtFmt{%
    \gtMsgFmt{#1}\ifempty{#2}{}{({#2})}%
    \ifempty{#3}{}{\vphantom{x} \!\gtSeq\! {#3}}%
  }%
}%
\newcommand{\gtCommChoiceSmall}[3]{%
  \gtFmt{%
    \gtMsgFmt{#1}\ifempty{#2}{}{({#2})}%
    \ifempty{#3}{}{\vphantom{x} \!\gtSeq\! {#3}}%
  }%
}%

\newcommand{\gtEnd}{\gtFmt{\mathbf{end}}}%

\newcommand{\gtRec}[2]{\gtFmt{\mu{#1}.{#2}}}%
\newcommand{\gtRecVarBase}{\gtFmt{\mathbf{t}}}%
\newcommand{\gtRecVar}[1][]{\gtFmt{\ifempty{#1}{\gtRecVarBase}{\gtRecVarBase_{#1}}}}%
\newcommand{\gtRecVari}[1][]{\gtFmt{\ifempty{#1}{\gtRecVarBase'}{\gtRecVarBase'_{#1}}}}%
\newcommand{\gtRecVarii}[1][]{\gtFmt{\ifempty{#1}{\gtRecVarBase''}{\gtRecVarBase''_{#1}}}}

\newcommand{\gtRoles}[1]{{\color{roleColor}\operatorname{pt}}\ifempty{#1}{}{({#1})}}%
\newcommand{\gtBRoles}[1]{{\color{roleColor} \operatorname{balpt}}({#1})}%
\newcommand{\gtSendRoles}[1]{{\color{roleColor}\operatorname{spt}}
\ifempty{#1}{}
{({#1})}
}%
\newcommand{\gtRecvRoles}[1]{{\color{roleColor}\operatorname{rpt}}
\ifempty{#1}{}
{({#1})}
}%
\newcommand{\gtDepth}[2]{{\color{roleColor}\operatorname{dpt}}_{#2}
\ifempty{#1}{}
{({#1})}
}%
\newcommand{\gtMDepth}[3]{{\color{roleColor}\operatorname{mdpt}}_{{#2},{#3}}
\ifempty{#1}{}
{({#1})}
}%
\newcommand{\gtMRoles}[1]{{\color{roleColor} \operatorname{mpt}}(#1)}%

\newcommand{\gtSch}[3]{{\color{roleColor} \operatorname{inter}}\left({#1},{#2},{#3}\right)}%
\newcommand{\gtSerial}[1]{{\color{roleColor} \operatorname{serial}}
\ifempty{#1}{}{
\left({#1}\right)}
}%
\newcommand{\gtTrace}[1]{{\color{roleColor} \operatorname{Tr}}({#1})}%
\newcommand{\gtTraceN}[2]{{\color{roleColor} \operatorname{Tr}}_{#2}({#1})}%
\newcommand{\gtCount}[3]{{\color{roleColor} \operatorname{count}}_{#2,#3}({#1})}%
\newcommand{\gtARoles}[1]{{\color{roleColor} \operatorname{active}}
\ifempty{#1}{}{({#1})}
}%
\newcommand{\gtQRoles}[1]{{\color{roleColor} \operatorname{qpt}}
\ifempty{#1}{}{({#1})}
}%

\newcommand{\gtGuarded}[1]{{\color{roleColor} \operatorname{guarded}}({#1})}%

\newcommand{\mpGuarded}[2]{{\color{roleColor}\operatorname{guarded}}({#1},{#2})}%

\newcommand{\gtQu}[2]{
        {{#1}\mathbin{\mathbf{\upharpoonright}}{#2}}
}

\newcommand{\gtProj}[3]{
         {\boldsymbol{\small\color{projColor}\operatorname{proj}}}
	\ifempty{#1}{
		\ifempty{#2}{
			\ifempty{#3}{}{\langle{#3}\rangle}
		}{
			\langle{#2}
			\ifempty{#3}{}{,{#3}}
			\rangle
		}
	}{
		\langle{#1}
		\ifempty{#2}{}{,{#2}}
		\ifempty{#3}{}{,{#3}}
		\rangle
	}
}
\newcommand{\gtProji}[3]{
         {\boldsymbol{\small\color{projColor}\operatorname{proj}'}}
	\ifempty{#1}{
		\ifempty{#2}{
			\ifempty{#3}{}{\langle{#3}\rangle}
		}{
			\langle{#2}
			\ifempty{#3}{}{,{#3}}
			\rangle
		}
	}{
		\langle{#1}
		\ifempty{#2}{}{,{#2}}
		\ifempty{#3}{}{,{#3}}
		\rangle
	}
}
\newcommand{\gtProjAnnot}[4]{
         {\boldsymbol{\small\color{projColor}\operatorname{proj}_{#4}}}
	\ifempty{#1}{
		\ifempty{#2}{
			\ifempty{#3}{}{\langle{#3}\rangle}
		}{
			\langle{#2}
			\ifempty{#3}{}{,{#3}}
			\rangle
		}
	}{
		\langle{#1}
		\ifempty{#2}{}{,{#2}}
		\ifempty{#3}{}{,{#3}}
		\rangle
	}
}

\newcommand{\gtMove}[1][\phantom{\stEnvAnnotGenericSym}]{\gtFmt{\xrightarrow{#1}}}
\newcommand{\gtMoveCo}[1][\phantom{\stEnvAnnotGenericSym}]{\gtFmt{\xrightarrow{#1}_{\text{\tiny co}}}}
\newcommand{\gtMoveStar}[1][]{\ifempty{#1}{\gtMove[]{}^{\!\gtFmt{*}}}{\gtMove[]{#1}^{\!\gtFmt{*}}}}
\newcommand{\gtNotMove}[1]{\gtFmt{#1}\!\,\,\not\!\!\!\!\gtMove}%

\newcommand{\gtMoveA}[1][\phantom{\stEnvAnnotGenericSym}]{\gtFmt{\xrightarrow{#1}_{\mathbf{a}}}}

\newcommand{\iruleGtMove}[1]{GR-{#1}}

\newcommand{\iruleGtMoveComm}[0]{\iruleGtMove{$\oplus\&$}}
\newcommand{\iruleGtMoveBra}[0]{\iruleGtMove{$\&$}}
\newcommand{\iruleGtMoveSel}[0]{\iruleGtMove{$\oplus$}}
\newcommand{\iruleGtMoveRec}[0]{\iruleGtMove{$\mu$}}

\newcommand{\iruleGtMoveCtx}[0]{\iruleGtMove{Ctx}}
\newcommand{\iruleGtMoveCtxI}[0]{\iruleGtMove{Ctx-I}}
\newcommand{\iruleGtMoveCtxII}[0]{\iruleGtMove{Ctx-II}}
\newcommand{\iruleGtMovePar}[0]{\iruleGtMove{$\parallel$}}

\newcommand{\labFmt}[2][]{\ensuremath{\ifempty{#1}{\mathtt{#2}}{\mathtt{#2}\textsubscript{#1}}}\xspace}%

\definecolor{mergeColor}{rgb}{0.6, 0.0, 0.0}%
\definecolor{exprColor}{rgb}{0.6, 0.0, 0.0}%
\definecolor{projColor}{rgb}{0.54, 0.15, 0.9}%
\definecolor{stColor}{rgb}{0, 0, 0.9}%
\newcommand{\stFmt}[1]{\ensuremath{{\color{stColor}#1}}\xspace}%
\newcommand{\stFmtC}[1]{\stFmt{\operatorname{#1}}}
\definecolor{compColor}{rgb}{0, 0, 0.9}%

\newcommand{\stIn}[4]{\ifempty{#1}{}{\roleFmt{#1}}\stFmt{\&\left\{{#2}\ifempty{#3}{}{({#3})} \ifempty{#4}{}{\stSeq #4}\right\}}}%
\newcommand{\stOut}[3]{\ifempty{#1}{}{\roleFmt{#1}}\stFmt{\oplus{#2}\ifempty{#3}{}{({#3})}}}%

\newcommand{\stChoice}[2]{\stLabFmt{#1}\ifempty{#2}{}{({#2})}}%

\newcommand{\stSeq}{\mathbin{\!\stFmt{.}\!}}%
\newcommand{\stIntC}{\mathbin{\stFmt{\oplus}}}%
\newcommand{\stSum}[4]{\stFmt{\Sigma\!\left\{{#1}{#4}{#3}\right\}_{#2}}}%
\newcommand{\stIntSum}[3]{{#1}\stFmt{\oplus\!\left\{#3\right\}_{#2}}}%
\newcommand{\stExtC}{\mathbin{\stFmt{\&}}}%
\newcommand{\stExtSum}[3]{{#1}\stFmt{\&\!\left\{#3\right\}_{#2}}}%
\newcommand{\stRec}[2]{\stFmt{\mu{#1}.{#2}}}%
\newcommand{\stEnd}{\stFmt{\mathbf{end}}}%

\newcommand{\stLabFmt}[1]{\stFmt{\labFmt{#1}}}%
\newcommand{\stLab}[1][]{%
  \ifempty{#1}{\stLabFmt{m}}{\stLabFmt{m}_{{\color{stColor}#1}}}
}%
\newcommand{\stLabi}[1][]{%
  \ifempty{#1}{\stLabFmt{m'}}{\stLabFmt{m'}_{{\color{stColor}#1}}}
}%
\newcommand{\stLabii}[1][]{%
  \ifempty{#1}{\stLabFmt{m''}}{\stLabFmt{m''}_{{\color{stColor}#1}}}
}%
\newcommand{\stLabiii}[1][]{%
  \ifempty{#1}{\stLabFmt{m'''}}{\stLabFmt{m'''}_{{\color{stColor}#1}}}
}%

\newcommand{\stS}[1][]{\stFmt{\ifempty{#1}{S}{S_{#1}}}}%

\newcommand{\stT}[1][]{\stFmt{\ifempty{#1}{T}{T_{#1}}}}%
\newcommand{\stTi}[1][]{\stFmt{\ifempty{#1}{T'}{T'_{#1}}}}%
\newcommand{\stTii}[1][]{\stFmt{\ifempty{#1}{T''}{T''_{#1}}}}%
\newcommand{\stTiii}[1][]{\stFmt{\ifempty{#1}{T'''}{T'''_{#1}}}}%

\newcommand{\stW}[1][]{\stFmt{\ifempty{#1}{\mathbb{W}}{\mathbb{W}_{#1}}}}%
\newcommand{\stWi}[1][]{\stFmt{\ifempty{#1}{\mathbb{W}'}{\mathbb{W}'_{#1}}}}%

\newcommand{\stU}[1][]{\stFmt{\ifempty{#1}{U}{U_{#1}}}}%
\newcommand{\stUi}[1][]{\stFmt{\ifempty{#1}{U'}{U'_{#1}}}}%
\newcommand{\stUii}[1][]{\stFmt{\ifempty{#1}{U''}{U''_{#1}}}}%

\newcommand{\stV}[1][]{\stFmt{\ifempty{#1}{\mathbb{V}}{\mathbb{V}_{#1}}}}%
\newcommand{\stVi}[1][]{\stFmt{\ifempty{#1}{\mathbb{V}'}{\mathbb{V}'_{#1}}}}%

\newcommand{\trT}[1][]{\stFmt{\ifempty{#1}{\mathbb{T}}{\mathbb{T}_{#1}}}}%
\newcommand{\trTi}[1][]{\stFmt{\ifempty{#1}{\mathbb{T}'}{\mathbb{T}'_{#1}}}}%
\newcommand{\trTii}[1][]{\stFmt{\ifempty{#1}{\mathbb{T}''}{\mathbb{T}''_{#1}}}}%

\newcommand{\stRecv}[4]{\roleFmt{#1}\mathord{?}\labFmt{#2}\stFmt{({#3})}; {#4}}
\newcommand{\stRecvOne}[3]{\roleFmt{#1}\mathord{?}\labFmt{#2}\stFmt{({#3})}}
\newcommand{\stSend}[4]{\roleFmt{#1}\mathord{!}\labFmt{#2}\stFmt{({#3})}; {#4}}
\newcommand{\stSendOne}[3]{\roleFmt{#1}\mathord{!}\labFmt{#2}\stFmt{({#3})}}

\newcommand{\stRecVarBase}{\stFmt{\mathbf{t}}}%
\newcommand{\stRecVar}[1][]{\stFmt{\ifempty{#1}{\stRecVarBase}{\stRecVarBase_{#1}}}}%
\newcommand{\stRecVari}[1][]{\stFmt{\ifempty{#1}{\stRecVar'}{\stRecVar'_{#1}}}}%
\newcommand{\stRecVarii}[1][]{\stFmt{\ifempty{#1}{\stRecVar''}{\stRecVar''_{#1}}}}%

\newcommand{\stMerge}[2]{\stFmt{\bigsqcap}_{#1}{#2}}%
\newcommand{\stMergei}[2]{\stFmt{\bigsqcap^\star}_{#1}{#2}}%
\newcommand{\stBinMerge}{\mathbin{\stFmt{\sqcap}}}%
\newcommand{\stBinMergeStar}{\mathbin{\setbox0\hbox{\stFmt{\sqcap}}\rlap{\hbox to \wd0{\hss $\star$\hss}}\box0}}%
\newcommand{\relMerge}[2]{\ifempty{#1}{\color{mergeColor}{\operatorname{merge}}}{{\color{mergeColor}\operatorname{merge}}\langle{#1},{#2}\rangle}}
\newcommand{\funcMerge}[1]{\ifempty{#1}{\color{mergeColor}{\operatorname{merge}}}{{\color{mergeColor}\operatorname{merge}}({#1})}}

\newcommand{\ltsSel}[3]{{#1}\mathord{:}{#2}\mathord{\stFmt{\oplus}}{#3}}
\newcommand{\ltsBra}[3]{{#1}\mathord{:}{#2}\mathord{\stFmt{\&}{#3}}}

\newcommand{\ltsSendRecv}[3]{{#1}{#2}\mathord{:}#3}

\newcommand{\ltsSubject}[1]{{\color{roleColor} \operatorname{sbj}({#1})}}%

\definecolor{mpColor}{rgb}{0, 0, 0}%
\newcommand{\mpFmt}[1]{{\color{mpColor}#1}}%

\newcommand{\mpTrue}{\mpFmt{\text{\texttt{true}}}}%
\newcommand{\mpFalse}{\mpFmt{\text{\texttt{false}}}}%
\newcommand{\mpSucc}[1]{\text{\texttt{succ}}({#1})}%

\newcommand{\mpNil}{\mpFmt{\mathbf{0}}}%
\newcommand{\mpSeq}{\mathbin{\mpFmt{\!.\!}}}%
\newcommand{\mpIf}[3]{%
  \mpFmt{\mathsf{if}\,{#1}\,\mathsf{then}}\,{#2}\,\mathsf{else}\,{#3}%
}%

\newcommand{\mpPar}{\mathbin{\mpFmt{\mid}}}%
\newcommand{\mpRes}[2]{\mpFmt{\left(\mathbf{\nu}{#1}\right){#2}}}%
\newcommand{\mpRec}[2]{\mpFmt{\mu{#1}.{#2}}}%

\newcommand{\mpErr}{\mpFmt{\boldsymbol{\mathtt{err}}}}%

\makeVars{mpE}{mpFmt}{e}

\newcommand{\mpS}[1][]{\mpFmt{\ifempty{#1}{s}{s_{#1}}}}%

\newcommand{\mpX}[1][]{\mpFmt{\ifempty{#1}{X}{X_{#1}}}}%
\newcommand{\mpXi}[1][]{\mpFmt{\ifempty{#1}{X'}{X'_{#1}}}}%

\newcommand{\mpP}[1][]{\mpFmt{\ifempty{#1}{P}{P_{#1}}}}%
\newcommand{\mpPi}[1][]{\mpFmt{\ifempty{#1}{P'}{P'_{#1}}}}%
\newcommand{\mpPii}[1][]{\mpFmt{\ifempty{#1}{P''}{P''_{#1}}}}%
\newcommand{\mpQ}[1][]{\mpFmt{\ifempty{#1}{Q}{Q_{#1}}}}%
\newcommand{\mpQi}[1][]{\mpFmt{\ifempty{#1}{Q'}{Q'_{#1}}}}%
\newcommand{\mpR}[1][]{\mpFmt{\ifempty{#1}{R}{R_{#1}}}}%
\newcommand{\mpRi}[1][]{\mpFmt{\ifempty{#1}{R'}{R'_{#1}}}}%

\newcommand{\mpMove}{\to}%
\newcommand{\mpMoveTau}{\xrightarrow{\tau}}
\newcommand{\mpMoveErr}{\xrightarrow{\mpErr}}
\newcommand{\mpMoveTauStar}{\mathrel{\mpMoveTau{}^{\!\!\!*}}}%
\newcommand{\mpMoveAnnot}[4]{\xrightarrow{\ltsSendRecv{{#1}}{{#2}}{{#3}\ifempty{#4}{}{\left({#4}\right)}}}}%
\newcommand{\mpMoveGen}{\xrightarrow{\alpha}}%
\newcommand{\mpMoveStar}{\mathrel{\mpMove{}^{\!\!\!*}}}%
\newcommand{\mpMovePlus}{\mathrel{\mpMove{}^{\!\!\!+}}}%
\newcommand{\iruleSafeComm}{S-${\stIntC}{\stExtC}$}%
\newcommand{\iruleMove}{S-$\stEnvMove$}%
\newcommand{\iruleMoveSession}{S-$\stEnvMoveWithSession[\mpS]$}%

\newcommand{\iruleStSubEnd}{Sub-$\stEnd$}

\newcommand{\iruleStSubRecL}{Sub-$\stFmt{\mu}$L}
\newcommand{\iruleStSubRecR}{Sub-$\stFmt{\mu}$R}
\newcommand{\iruleStSubOut}{Sub-$\stFmt{\oplus}$}

\newcommand{\iruleStSubIn}{Sub-$\stFmt{\&}$}

\newcommand{\iruleStSubSum}{Sub-$\Sigma$}

\newcommand{\iruleLtRec}{LR-$\mu$}%
\newcommand{\iruleLtOut}{LR-$\stFmt{\oplus}$}%
\newcommand{\iruleLtIn}{LR-$\stFmt{\&}$}%
\newcommand{\iruleLtSum}{LR-$\stFmt{\Sigma}$}%

\newcommand{\stEnv}[1][]{\stFmt{\ifempty{#1}{\Delta}{\Delta_{#1}}}}%
\newcommand{\stEnvi}[1][]{\stFmt{\ifempty{#1}{\Delta'}{\Delta'_{#1}}}}%
\newcommand{\stEnvii}[1][]{\stFmt{\ifempty{#1}{\Delta''}{\Delta''_{#1}}}}%
\newcommand{\stEnviii}[1][]{\stFmt{\ifempty{#1}{\Delta'''}{\Delta'''_{#1}}}}%
\newcommand{\stEnvEmpty}{\stFmt{\emptyset}}%
\newcommand{\stEnvMap}[2]{\stFmt{\mpFmt{#1}\mathbin{:}{#2}}}%
\newcommand{\stEnvComp}{\mathpunct{\stFmt{,}}}%
\newcommand{\stEnvApp}[2]{\stFmt{#1\!\left(\mpFmt{#2}\right)}}%

\newcommand{\stEnvNew}[1][]{\stFmt{\ifempty{#1}{\Delta}{\Delta_{#1}}}}%
\newcommand{\stEnvNewi}[1][]{\stFmt{\ifempty{#1}{\Delta'}{\Delta'_{#1}}}}%
\newcommand{\stEnvNewii}[1][]{\stFmt{\ifempty{#1}{\Delta''}{\Delta''_{#1}}}}%

\newcommand{\assocs}[1]{\stFmt{\sqsubseteq_s}}
\newcommand{\assoca}[1]{\stFmt{\sqsubseteq_a}}

\newcommand{\stEnvAssoc}[3]{\stFmt{{#2} \mathrel{\stFmt{\sqsubseteq}_{#3}} {#1}}}

\newcommand{\stEnvMove}{\mathrel{\stFmt{\to}}}%

\newcommand{\stEnvMoveWithSession}[1][]{
\ifempty{#1}{}{
    \mathrel{\stFmt{\to}_{\!#1}}}}

\newcommand{\stEnvAnnotOutSym}{\stFmt{\oplus}}%
\newcommand{\stEnvAnnotQSym}{\stFmt{\dagger}}%
\newcommand{\stEnvAnnotInSym}{\stFmt{\&}}%
\newcommand{\stEnvAnnotGenericSym}[1][]{\stFmt{\ifempty{#1}{\alpha}{\alpha_{#1}}}}%
\newcommand{\stEnvAnnotGenericSymi}[1][]{\stFmt{\ifempty{#1}{\alpha'}{\alpha'_{#1}}}}%
\newcommand{\stEnvAnnotGenericSymii}[1][]{\stFmt{\ifempty{#1}{\alpha''}{\alpha''_{#1}}}}%
\newcommand{\stEnvAnnotGenericSymiii}[1][]{\stFmt{\ifempty{#1}{\alpha'''}{\alpha'''_{#1}}}}%
\newcommand{\stEnvMoveAnnot}[1]{\mathrel{\stFmt{\xrightarrow{#1}}}}
\newcommand{\stEnvMoveGenAnnot}{\stEnvMoveAnnot{\stEnvAnnotGenericSym}}%
\newcommand{\stEnvMoveCommAnnot}[3]{%
  \stEnvMoveAnnot{\ltsSendRecv{#1}{#2}{#3}}%
}%

\newcommand{\stEnvDagAnnot}[3]{{#1}:{#2}{\stEnvAnnotQSym}{#3}}%
\newcommand{\stEnvInAnnot}[3]{{#1}:{#2}{\stEnvAnnotInSym}{#3}}%
\newcommand{\stEnvOutAnnot}[3]{{#1}:{#2}{\stEnvAnnotOutSym}{#3}}%
\newcommand{\stEnvCommAnnotSmall}[3]{\ltsSendRecv{#1}{#2}{#3}}%
\newcommand{\stEnvNotMoveP}[1]{{#1}\!\!\not\stEnvMove}%
\newcommand{\stEnvMoveStar}{\mathrel{\stFmt{\stEnvMove{}^{\!\!\!*}}}}%

\newcommand{\stEnvMoveAnnotP}[2]{{#1}\!\!\stEnvMoveAnnot{#2}}%
\newcommand{\stEnvMoveGenAnnotP}[1]{{#1}\!\!\stEnvMoveGenAnnot}%

\newcommand{\stEnvEndPred}{\operatorname{end}}%
\newcommand{\stEnvEndP}[1]{\stEnvEndPred(\stFmt{#1})}%

\newcommand{\stEnvDFPred}{\operatorname{df}}%
\newcommand{\stEnvDFP}[1]{\stEnvDFPred(\stFmt{#1})}%

\newcommand{\stEnvLivePred}{\operatorname{live}}%
\newcommand{\stEnvLiveP}[1]{\stEnvLivePred(\stFmt{#1})}%

\newcommand{\stEnvSendLivePred}{\operatorname{slive}}%

\newcommand{\stEnvSendLiveRolePred}[1]{\operatorname{slive}_{{#1}}}%
\newcommand{\stEnvSendLiveRoleP}[2]{\stEnvSendLiveRolePred{#1}(\stFmt{#2})}%

\newcommand{\stEnvRecvLivePred}{\operatorname{rlive}}%

\newcommand{\stEnvRecvLiveRolePred}[1]{\operatorname{rlive}_{{#1}}}%
\newcommand{\stEnvRecvLiveRoleP}[2]{\stEnvRecvLiveRolePred{#1}(\stFmt{#2})}%

\newcommand{\stEnvSafePred}{\operatorname{safe}}%
\newcommand{\stEnvSafeP}[1]{\stEnvSafePred(\stFmt{#1})}%

\newcommand{\mpEnv}[1][]{\stFmt{\ifempty{#1}{\Gamma}{\Gamma_{#1}}}}%
\newcommand{\mpEnvEmpty}{\stFmt{\emptyset}}%
\newcommand{\mpEnvMap}[2]{\stFmt{\mpFmt{#1}\mathbin{:}\stFmt{#2}}}%
\newcommand{\mpEnvComp}{\mathpunct{\stFmt{,}}}%

\newcommand{\stJudge}[3]{%
  \stFmt{{#1}%
  \mathrel{\mpFmt{\vdash}} \mpFmt{#2} \; \mpFmt{\triangleright}\; #3}%
}%

\newcommand{\gtExChoice}{\gtG[\text{auth}]}%
\newcommand{\roleAuth}{{\color{roleColor}\roleFmt{a}}}%
\newcommand{\roleClient}{{\color{roleColor}\roleFmt{c}}}%
\newcommand{\roleServer}{{\color{roleColor}\roleFmt{s}}}%

\newcommand{\projRuleEnd}{\text{p-}\stEnd}
\newcommand{\projRuleIn}{\text{p-}\&}
\newcommand{\projRuleOut}{\text{p-}\oplus}
\newcommand{\projRuleComm}{\text{p-}{\oplus}{\&}}
\newcommand{\projRuleCtx}{\text{p-ctx}}
\newcommand{\projRulePar}{\text{p-}\parallel}
\newcommand{\projRuleMu}{\text{p-}\mu}
\newcommand{\projRuleVar}{\text{p-}\stRecVar}

\newcommand{\mergeRuleEnd}{\sqcap\text{-}\stEnd}
\newcommand{\mergeRuleOut}{\sqcap\text{-}\oplus}
\newcommand{\mergeRuleIn}{\sqcap\text{-}\&}
\newcommand{\mergeRuleComm}{\sqcap\text{-}{\oplus}{\&}}

\newcommand{\projGoodEnd}{G\text{-}\stEnd}

\newcommand{\projGoodOut}{G\text{-}\oplus}
\newcommand{\projGoodIn}{G\text{-}\&}
\newcommand{\projGoodAlt}{G\text{-}{\tySub}}

\newcommand{\projGoodType}{G\text{-}{\oplus}{\&}}

\newcommand{\projGoodOutS}{G\text{-}\oplus^\star}
\newcommand{\projGoodInS}{G\text{-}\&^\star}

\newcommand{\mergeGoodAssoc}{G-Assoc}
\newcommand{\mergeGoodRedI}{G-$\to$}
\newcommand{\mergeGoodRedII}{G-$\to'$}
\newcommand{\mergeGoodSubset}{G-$\subseteq$}
\newcommand{\mergeGoodSubI}{G-$\tySub$}
\newcommand{\mergeGoodSubII}{G-$\tySub$'}
\newcommand{\mergeGoodSubst}{G-$\stRecVar$}
\newcommand{\mergeGoodCtx}{G-Ctx}

\newcommand{\quH}[1][]{\tyCol{\ifempty{#1}{\tyFont{\sigma}}{\tyFont{\sigma}_{#1}}}}%
\newcommand{\quHi}[1][]{\tyCol{\ifempty{#1}{\tyFont{\sigma'}}{\tyFont{\sigma'}_{#1}}}}%
\newcommand{\quEmpty}[1][]{\tyCol{\tyFont{\varnothing}}}%
\newcommand{\quMsg}[3]{\tyCol{\left({#1},\tyFont{{#2}\ifempty{#3}{}{({#3})}}\right)}}%
\newcommand{\quCons}[2]{\tyCol{\tyFont{{#1}\mathbin{\!\cdot\!}{#2}}}}%

\newcommand{\pquH}[1][]{{\ifempty{#1}{\tyFont{h}}{\tyFont{h}_{#1}}}}%
\newcommand{\pquHi}[1][]{{\ifempty{#1}{\tyFont{h'}}{\tyFont{h'}_{#1}}}}%
\newcommand{\pquEmpty}[1][]{{\tyFont{\varnothing}}}%
\newcommand{\pquMsg}[3]{{\left({#1},\tyFont{{#2}\ifempty{#3}{}{({#3})}}\right)}}%
\newcommand{\pquCons}[2]{{\tyFont{{#1}\mathbin{\!\cdot\!}{#2}}}}%

\newcommand{\asubt}{\mathrel{\stFmt{\leqslant_{\text{a}}}}}
\newcommand{\psubt}[1]{\mathrel{\stFmt{\leqslant_{\text{#1}}}}}

\newcommand{\set}[1]{\{#1\}}

\newcommand{\M}{M}
\newcommand{\N}{M}

\newcommand{\subttt}{\mathrel{\stFmt{\lesssim}}}
\newcommand{\ttreeSym}{\mathfrak{T}}%
\newcommand{\ttree}[1]{\operatorname{\ttreeSym}\!\left({#1}\right)}%

\newcommand{\distinctInp}[1]{\mathcal{A}^{(#1)}}
\newcommand{\inpDistinctOut}[1]{\mathcal{B}^{(#1)}}

\newcommand{\Ap}{\distinctInp{\roleP}}
\newcommand{\Bp}{\inpDistinctOut{\roleP}}

\newcommand{\act}[1]{\texttt{act}(#1)}

\newcommand{\singleIn}[1]{{\llbracket #1 \rrbracket}_{\text{SI}}}
\newcommand{\singleOut}[1]{{\llbracket #1 \rrbracket}_{\text{SO}}}

\newcommand{\trCtxSelp}{\mathbb{L}^{(\roleP)}_{\oplus}}
\newcommand{\trCtxBrap}{\mathbb{L}^{(\roleP)}_{\&}}
\newcommand{\trCtxSelq}{\mathbb{L}^{(\roleQ)}_{\oplus}}
\newcommand{\trCtxBraq}{\mathbb{L}^{(\roleQ)}_{\&}}

\newcommand{\forgetHole}[2]{\langle#1\rangle_{#2}}

\newcommand{\myparagraph}[1]{\noindent\textbf{#1}}

\newcommand{\playsrole}[2]{{#1}\triangleleft{#2}} 

\newcommand{\projGlobal}[1]{{\color{compColor}\operatorname{gbl}}
\ifempty{#1}{}{\langle{#1}\rangle}}

\newcommand{\projLocal}[1]{{\color{compColor}\operatorname{lcl}}
\ifempty{#1}{}{\langle{#1}\rangle}}

\newcommand{\weakComm}[3]{{\color{roleColor}\operatorname{weak}}\!\left({#1},{#2},{#3}\right)}

\newcommand{\prestruct}{\Rrightarrow}

\newcommand{\proofapp}[1]{\ifextended\textcolor{blue!60!black}{\textup{[Proof in #1]}}\fi}

\newcommand{\ENCan}[1]{\langle{#1}\rangle}%

\newcommand{\labelpath}[2]{{\color{roleColor}\operatorname{paths}}\!\left({#1},{#2}\right)}

\newif\ifextended
\extendedtrue

\begin{document}

\theoremstyle{acmdefinition}
\newtheorem{remark}{Remark}[section]

\title{What Is a Good Projection?} 
\subtitle{Correctness and Separation Results of End-Point Projections}

\author{Jake Masters}
\orcid{0000-0002-3783-6149}
\affiliation{%
  \institution{University of Oxford}
  \department{Computer Science}
  \city{Oxford}
  \country{United Kingdom}
}
\email{jam275@cam.ac.uk}

\author{Nobuko Yoshida}
\correspondingauthor
\orcid{0000-0002-3925-8557}
\affiliation{%
  \institution{University of Oxford}
  \department{Computer Science}
  \city{Oxford}
  \country{United Kingdom}
}
\email{nobuko.yoshida@cs.ox.ac.uk}

\begin{abstract}
\emph{Multiparty session types} (MPST)~\cite{HYC08,HYC16} is a type
discipline for ensuring distributed programs are implemented correctly
\emph{by construction}. MPST starts by specifying a choreographic
protocol as a \emph{global type} and \emph{projects} it onto each
participant, generating \emph{local types} with a guarantee of type
soundness and deadlock-freedom.
A central issue of MPST is that its type
soundness is fragile, as it must be proven each time with respect to
a chosen projection algorithm under 
specific semantics with a specific program correctness goal. 
To overcome this problem, we propose general criteria for
a \emph{good projection} -- a set of requirements for projection
algorithms by which type soundness, deadlock-freedom and liveness of
distributed programs are \emph{automatically} derived.  Using the
criteria, we prove that 8 distinct projections in the literature are
good under both synchronous and asynchronous semantics, and that 4 are
incorrect, failing type soundness.
We extend the good projection
criteria to mixed choice global types and asynchronous communication
optimisations, proposing new ``good projection'' algorithms.  We
apply properties of good projections to prove \emph{separation} and
\emph{expressiveness} results over the projection algorithms and
MPST typing systems.
Having a generalised typing system
that is robust with respect to
the good projection criteria,
we provide an implementation
with mixed choice global types that is \emph{modular} with respect to a
chosen projection algorithm and desired program properties.
Our separation results help programmers
make informed trade-offs
between complexity and expressiveness
when choosing projection algorithms,
and our theory ensures
their chosen algorithms are
correctly applied.
\end{abstract}




\begin{abstract}
\emph{Multiparty session types} (MPST)~\cite{HYC08,HYC16} is a type
discipline for ensuring distributed programs are implemented correctly
\emph{by construction}. MPST starts by specifying a choreographic
protocol as a \emph{global type} and \emph{projects} it onto each
participant, generating \emph{local types} with a guarantee of type
soundness and deadlock-freedom.
A central issue of MPST is that its type
soundness is fragile, as it must be proven each time with respect to
a chosen projection algorithm under 
specific semantics with a specific program correctness goal. 
To overcome this problem, we propose general criteria for
a \emph{good projection} -- a set of requirements for projection
algorithms by which type soundness, deadlock-freedom and liveness of
distributed programs are \emph{automatically} derived.  Using the
criteria, we prove that 8 distinct projections in the literature are
good under both synchronous and asynchronous semantics, and that 4 are
incorrect, failing type soundness.
We extend the good projection
criteria to mixed choice global types and asynchronous communication
optimisations, proposing new ``good projection'' algorithms.  We
apply properties of good projections to prove \emph{separation} and
\emph{expressiveness} results over the projection algorithms and
MPST typing systems.
Having a generalised typing system
that is robust with respect to
the good projection criteria,
we provide an implementation
with mixed choice global types that is \emph{modular} with respect to a
chosen projection algorithm and desired program properties.
Our separation results help programmers
make informed trade-offs
between complexity and expressiveness
when choosing projection algorithms,
and our theory ensures
their chosen algorithms are
correctly applied.
\end{abstract}

\maketitle

\section{Introduction}
\label{sec:intro}
\begin{wrapfigure}{r}{0.35\textwidth}
\centering
\begin{minipage}{0.35\textwidth}
  {
  \begin{tikzpicture}
 \node (Gtext) {\small \boxed{$\gtG$}};
  \node[below= 0.5mm of Gtext, xshift=-19mm, align=center, pos=0.5] (proj) {
    \footnotesize \textbf{Step 1} \small $(\gtProj{}{}{})$};
  \node[below=6mm of Gtext, xshift=-12mm] (LA) {
  \small \boxed{\stT_{\roleP}}};
  \node[below=6mm of Gtext,xshift=-4mm] (LB) {\small \boxed{\stT_{\roleQ}}};
\node[below=6mm of Gtext,xshift=4mm] (Ldot) {};
\node[below=2.45mm of Ldot,xshift=0mm] (Ldot') {\dots};
  \node[below=6mm of Gtext, xshift=12mm] (LC) {
   \small  \boxed{\stT_{\roleR}}};
  \draw[->] (Gtext) -- (LA);
  \draw[->] (Gtext) -- (LB);
  \draw[->] (Gtext) -- (LC);
\draw[dashed,->] (Gtext) -- (Ldot);
   \node[below= 1mm of LA, xshift=-8.4mm, align=center] (sub) {
    \footnotesize \textbf{Step 2} \small $(\tySub)$};
  \node[below=6mm of LA, xshift=0mm] (L'A) {
  \small \boxed{\stTi_{\roleP}}};
  \node[below=6mm of LB] (L'B) {\small \boxed{\stTi_{\roleQ}}};
\node[below=6mm of Ldot',xshift=0mm] (L'dot) {};
\node[below=2.75mm of L'dot,xshift=0mm] (L'dot') {\dots};
  \node[below=6.4mm of LC, xshift=0mm] (L'C) {
   \small  \boxed{\stTi_{\roleR}}};
  \draw[->] (LA) -- (L'A);
  \draw[->] (LB) -- (L'B);
  \draw[->] (LC) -- (L'C);
\draw[dashed,->] (Ldot') -- (L'dot);
   \node[below= 1mm of L'A, xshift=-8.4mm, align=center] (typ) {
    \footnotesize \textbf{Step 3} \small $(\vdash)$};
   \node[below=6mm of L'A, xshift=0mm] (PA) {\small
    \boxed{P_{\roleP}}};
   \node[below=6mm of L'B, xshift=0mm] (PB) {\small
    \boxed{P_{\roleQ}}};
\node[below=6mm of L'dot',xshift=0mm] (Pdot) {};
\node[below=2.45mm of Pdot,xshift=0mm] (Pdot') {\dots};
     \node[below=6.2mm of L'C, xshift=0mm] (PC) {\small
    \boxed{P_{\roleR}}};
   \draw[->] (L'A) -- (PA);
   \draw[->] (L'B) -- (PB);
   \draw[->] (L'C) -- (PC);
\draw[dashed,->] (L'dot') -- (Pdot);
\end{tikzpicture}
}
\end{minipage}
\caption{Top-down methodology.}
\label{fig:overview-of-topdown}
\end{wrapfigure}
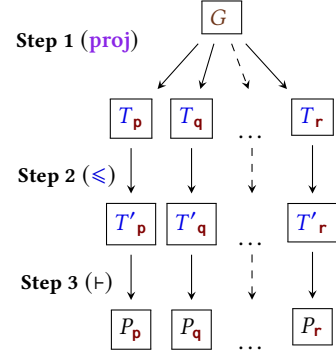

\noindent\emph{\textbf{Multiparty session types}}
(MPST)~\cite{HYC08,HYC16} is a type 
discipline for verifying distributed components ensuring
\emph{correctness by construction}.
In MPST, a communication protocol is defined
by
a \emph{global type} $\gtG$, which is \emph{projected} onto
each participant $\roleP$ (Step 1),
generating the \emph{local type} $\stT[\roleP]$ (Step 2), then 
each participant program $P_{\roleP}$ is typed by $\stT[\roleP]$ (Step
3) (Figure~\ref{fig:overview-of-topdown}).
By ensuring each participant's program conforms to its local type generated
by projection,  
MPST enables static guarantees of \emph{communication safety}
(no
communication
errors), \emph{deadlock-freedom}
and \emph{liveness} (all sent messages are eventually received
and all waiting processes are eventually sent a message).

MPST is language-agnostic;
it has been integrated in over 25 programming 
languages \cite{Yoshida24} and several systems 
are certified in theorem provers \cite{DBLP:conf/itp/EkiciY24,DBLP:journals/pacmpl/HinrichsenJK24,DBLP:conf/pldi/Castro-Perez0GY21,DBLP:journals/jar/TiroreBC25,EKY2025,Li2025,Tirore2025b,KYG2026}.

\smallskip 

\myparagraph{Issues with MPST.\ }
Over fifteen years, 
the field of MPST has grown into a jungle of separated but
interrelated concepts and methodologies.
Correctness (such as type soundness and deadlock-freedom of programs)
and complexity \cite{DBLP:journals/pacmpl/UdomsrirungruangY25}
in
an MPST system
largely depend on the chosen projection algorithm. Each time a new
projection is used, one
needs to prove type soundness again, 
taking the communication model into consideration
\ie~either
synchronous (handshake) or asynchronous (p2p) communications.
This makes the MPST framework ad hoc and fragile. 
Scalas and Yoshida discovered that proofs of type soundness in several
papers from the MPST literature,
using projections that
are common in both theory and implementations,
are \emph{flawed} \cite{POPL19LessIsMore}. 
To rescue MPST from this issue, 
researchers propose various different ``solutions'' --
a \emph{bottom-up approach} \cite{POPL19LessIsMore}
which infers
a
local type from each end-point program,
and checks
if the
collection of local types satisfies a safety property;  
a \emph{synthesis} approach \cite{PY2026,ICALP13CFSM,DBLP:journals/lmcs/DagninoGD23} which reconstructs a global type from a set of local types; 
and a \emph{direct} approach
which types a set of programs directly against a global type \cite{10.1145/3678232.3678245,10.1145/3776692},
without running a projection algorithm (Step 2).
Another direct and automata-theoretic approach is based on 
\text{deadlock-free realisability} of global protocols, 
introduced in the framework of 
message sequence charts (MSC) \cite{ALUR2005,ALUR2000}.   
This realisability problem poses
whether there is a distributed implementation that generates
(deadlock-free) communicating finite state machines (CFSMs)
\cite{Brand:1983:CFSM}, which are in an MSC language.
Those methods
are proposed to type programs which conform
to ``un-projectable''
global types \cite{Barbanera2026}. But we have now lost our way
in the jungle: 
what does ``un-projectable'' mean?
Is it because such a
projection is not expressive enough? 
An un-projectable global type might become
projectable if it is translated to another global type
with the \emph{same meaning} \ie
the underlying protocol is expressible.
A program conforming to an ``un-projectable'' global type
may still conform to a projectable one, \ie
the program is expressible.
To which projection does ``un-projectable'' refer?
How does expressibility separate
projection algorithms?

In the standard top-down MPST,
a program is \emph{typable}
if, from some global type,
the projection generates
a set of local types,
each of which successfully types each program
behaviour (\textbf{program realisability});
and the protocol specified by global type $\gtG$
is realisable if the ``correct'' projection
can project \emph{some} global type $\gtGi$ which has the same 
behaviour as $\gtG$ (\textbf{protocol realisability}).

The question is, can the MPST jungle be turned into
a carefully organised garden
without
weeds
(\ie ``bad'' projections) and
where the
projections
are
well arranged 
with respect to program and protocol realisabilities,
program properties, and different communication models?
Our starting point is to
identify a set of general requirements
on projections
for 
the typing system to \emph{automatically} guarantee type soundness and
correctness properties of typed processes such as safety,
deadlock-freedom and liveness (\textbf{the projection problem}).

\smallskip

\myparagraph{Good Projection Criteria. \ }
To ensure safety of typed processes,
we prove type soundness via a \emph{subject reduction theorem}.   
MPST is a \emph{behavioural type system} where types change 
during communications between participants.  
Let us write $\playsrole{\roleP}{\mpP}$ to mean that 
process $\mpP$ plays role $\roleP$ and
let $\gtG$ be a \emph{global type},
which specifies a protocol between the roles in $\roleSet$.
The \emph{multiparty session} 
$\M= \Pi_{\roleP\in \roleSet} \ \playsrole{\roleP}{P_{\roleP}}$
is a parallel composition of participants in a set of roles
$\roleSet$. Assume $\M$ has type $\gtG$ (denoted by
$\tyJudge{}{\M}{\gtG}$) and $\M$ reduces to $\M'$ (denoted by
$\M\rightarrow\M'$). Then:\\[1mm]
\hspace*{0.5cm}\textbf{Subject Reduction:} \quad  
If $\tyJudge{}{\M}{\gtG}$ and $\M\rightarrow\M'$, then
$\exists \gtGi.
\, \tyJudge{}{\M'}{\gtGi}$ and $\gtG\,\gtMoveStar\,\gtGi$\\[1mm]
where $\gtG\,\gtMove\,\gtGi$ denotes that $\gtG$ reduces to
$\gtGi$, $\gtMoveStar$ is 
its reflexive and transitive closure, and
$\gtG\,\gtMove$ if there is $\gtGi$ such that $\gtG\,\gtMove\,\gtGi$.

Session types are equipped with internal and external choices, which 
are adjusted by the \emph{subsumption rule} (\inferrule{T-Sub}),   
enlarging the typability of processes.
Then the main parallel composition rule 
for participants is defined by \inferrule{Sess}.\\[1mm]
\centerline{
$\inference[T-Sub]{\stJudge{\mpEnv}{\mpP}{\stT} \quad \stT \tySub \stTi}{\stJudge{\mpEnv}{\mpP}{\stTi}}
\quad\quad\quad
\inference[Sess]{
\gtProj{}{\gtG}{\{\stEnvMap{\roleP}{\stT[\roleP]}\}_{\roleP\in \roleSet}}
&
\forall \roleP\in \roleSet&
\stJudge{}{\mpP[\roleP]}{\stT[\roleP]}
}
{
\tyJudge{}{\Pi_{\roleP\in\gtRoles{\gtG}}\ \playsrole{\roleP}{\mpP[\roleP]}}{\gtG}
}$}\\[1mm]
where $\stJudge{\mpEnv}{\mpP}{\stT}$ is a typing judgement, assigning 
local type $\stT$ to process $P$,  
$\tySub$ denotes a \emph{subtyping relation}, 
$\playsrole{\roleP}{\mpP}$ denotes that
$\mpP$ plays role $\roleP$ and
$\gtRoles{\gtG}$ is a set of participants present in $\gtG$.
The relation $\gtProj{\roleP}{\gtG}{\stT}$
is a \textbf{projection relation}~(\cref{def:good-proj}), 
defining that $\stT$ is a projection of $\gtG$ onto $\roleP$.
The projection relation extends to
a collection of local types 
$\stEnv=\{\stEnvMap{\roleP}{\stT[\roleP]}\}_{\roleP\in \roleSet}$,
where $\gtRoles{\gtG}\subseteq\roleSet$,
(called \emph{typing context})
by $\gtProj{}{\gtG}{\stEnv}$
iff for each $\roleP\in\roleSet$,
$\gtProj{\roleP}{\gtG}{\stT[\roleP]}$,
and subtyping extends pointwise to typing contexts.

To ensure subject reduction, we
require that the projection is satisfied at each step of the
reductions up to $\tySub$ (by \inferrule{T-Sub}).
Hence, a global type and a typing context
must satisfy the 
\emph{association relation}:\\[1mm]
\hspace*{0.5cm}
{\bf Association}:\ 
$\stEnvAssoc{\gtG}{\stEnv}{}$
iff
there is $\stEnvi$ such that
$\gtProj{}{\gtG}{\stEnvi}$ and $\stEnv\tySub\stEnvi$
\\[1mm]
\ie
the type of participant $\roleP$ is a subtype of $\roleP$'s projection of
$\gtG$.

To prove subject reduction, 
we must prove~(\cref{thm:completeness_association}):\\[1mm]
\hspace*{0.5cm}
{\bf Completeness} (Figure~\ref{fig:proj-completeness}): \    
if $\stEnvAssoc{\gtG}{\stEnv}{}$ and $\stEnv
\stEnvMoveGenAnnot \stEnvi$, then
$\exists\gtGi.\;
\gtG\,\gtMove[\stEnvAnnotGenericSym]\,\gtGi$ 
and 
$\stEnvAssoc{\gtGi}{\stEnvi}{}$\\[1mm] 
where $\gtG\,\gtMove[\stEnvAnnotGenericSym]\,\gtGi$~(\cref{def:gtype:lts-gt}) denotes that $\gtG$
reduces to $\gtGi$ after one communication step with \emph{action} 
$\stEnvAnnotGenericSym=\ltsSendRecv{\roleP}{\roleQ}{\stChoice{\stLab}{\tyGround}}$
(role $\roleP$ sends a message with label
$\stLab$ and payload of type $\tyGround$ to role $\roleQ$, cf.~\cref{def:mpst-label-subject}).  
Matching the action $\alpha$ 
is required to precisely link the \emph{behaviours} of 
global and local types.
\begin{figure}
\centering
\begin{minipage}{0.45\textwidth}
\begin{subfigure}[t]{0.48\textwidth}
\centering
{
  \begin{tikzpicture}
	\node(G1){$\gtG[0]$};
	\node[above=4mm of G1](C1){$\stEnv[0]$};
	\draw[->] (G1) -- (C1);
	\node[above=4mm of C1](C'1){$\stEnvi[0]$};
	\path (C1) -- (C'1) node[midway](sub) {\rotatebox[origin=c]{-90}{$\tySub$}};
	\node[right=4mm of G1](G2){$\gtG[1]$};
	\draw[->](G1)--(G2) node [midway, above, draw=none] {$\stEnvAnnotGenericSym$};
	\node[right=4mm of C1](C2){$\stEnv[1]$};
	\node[right=4mm of C'1](C'2){$\stEnvi[1]$};
	\draw[->](C'1)--(C'2) node [midway, above, draw=none] {$\stEnvAnnotGenericSym$};
	\draw[->] (G2) -- (C2);
	\path (C2) -- (C'2) node[midway](sub') {\rotatebox[origin=c]{-90}{$\tySub$}};
\end{tikzpicture}
}
\caption{Completeness. 
}
\label{fig:proj-completeness}
\end{subfigure}
\begin{subfigure}[t]{0.48\textwidth}
\centering
{
  \begin{tikzpicture}
	\node(G1){$\gtG[0]$};
	\node[below=4mm of G1](C1){$\stEnv[0]$};
	\draw[->] (G1) -- (C1);
	\node[below=4mm of C1](C'1){$\stEnvi[0]$};
	\path (C1) -- (C'1) node[midway](sub) {\rotatebox[origin=c]{90}{$\tySub$}};
	\node[right=4mm of G1](G2){
	$\gtG[1]$
	};
	\draw[->](G1)--(G2) node [midway, above, draw=none] {$\stEnvAnnotGenericSym$};
	\node[right=4mm of C1](C2){
	$\stEnv[1]$
	};
	\node[right=4mm of C'1](C'2){
	$\stEnvi[1]$
	};
	\draw[->](C'1)--(C'2) node [midway, above, draw=none] {$\stEnvAnnotGenericSym$};
	\draw[->] (G2) -- (C2);
	\path (C2) -- (C'2) node[midway](sub') {\rotatebox[origin=c]{90}{$\tySub$}};
\end{tikzpicture}
}
\caption{Soundness. 
}
\label{fig:proj-soundness}
\end{subfigure}
\vspace{-3mm}
\caption{Association}
\vspace{-3mm}
\label{fig:association}
\end{minipage}
~
\begin{minipage}{0.45\textwidth}
\vspace{6mm}
\centering
    {
\begin{subfigure}[t]{0.48\textwidth}
\centering{
 \begin{tikzpicture}
	\node(G1){$\gtG$};
	\node[below=6mm of G1](C1){$\stEnv$};
	\node[right=6mm of G1](G2){$\gtGi$};
	\node[below=6mm of G2](C2){$\stEnvi$};
	\path (G1) -- (G2) node[midway](M) {};
	\node[above=4mm of M](G3){$\gtGii$};
	\draw[decorate,decoration={snake,segment length=3mm}](G1)--(G3);
	\draw[decorate,decoration={snake,segment length=2.9mm}](G2)--(G3);
	\draw[->] (G1) -- (C1) node[midway, xshift=5mm](projPoint) {\small$\gtProjAnnot{}{}{}{A}$};
	\draw[->] (G2) -- (C2) node[midway, xshift=5mm](projPoint') {\small$\gtProjAnnot{}{}{}{B}$};
\end{tikzpicture}}
\caption{Global Behaviour.}
\label{fig:expressiveness-global}
\end{subfigure}
\begin{subfigure}[t]{0.48\textwidth}
\centering{
 \begin{tikzpicture}
	\node(G1){$\gtG$};
	\node[below=6mm of G1](C1){$\stEnvNewi$};
	\node[right=6mm of G1](G2){$\gtGi$};
	\node[below=6mm of G2](C2){$\stEnvNewii$};
	\path (C1) -- (C2) node[midway](M) {};
	\node[below=4mm of M](C3){$\stEnvNew$};
	\draw[->] (G1) -- (C1) node[midway,xshift=5mm](projPoint) {\small$\gtProjAnnot{}{}{}{A}$};
	\path (C1) -- (C3) node[midway](subPoint) {\rotatebox[origin=c]{120}{$\tySub$}};
	\path (C2) -- (C3) node[midway](subPoint) {\rotatebox[origin=c]{60}{$\tySub$}};
	\draw[->] (G2) -- (C2) node[midway,xshift=5mm](projPoint') {\small$\gtProjAnnot{}{}{}{B}$};
\end{tikzpicture}}
\caption{Local Behaviour.}
\label{fig:expressiveness-local}
\end{subfigure}
}
\vspace{-3mm}
\caption{Expressiveness.}
\label{fig:expressiveness}
\end{minipage}
\end{figure}

Beyond type soundness, it is essential 
to guarantee that session $\M$ 
can simulate the behaviours specified by the global type.
This is stated as:\\[1mm]
 \hspace*{0.5cm}
\textbf{Session Fidelity}: \quad  
if $\tyJudge{}{\M}{\gtG}$ and $\gtG\gtMove$, then
$\exists\M',\gtGi.\;\M\mpMovePlus\M'$, 
$\tyJudge{}{\M'}{\gtGi}$ and $\gtG\,\gtMove\,\gtGi$.\\[1mm]
For this, we ensure the same for typing contexts \ie
if $\gtG$ moves, then $\stEnv$ \emph{surely} moves 
(\cref{thm:soundness_association}):\\[1mm]
\hspace*{0.5cm}
{\bf Soundness} (Figure~\ref{fig:proj-soundness}): \    
if $\stEnvAssoc{\gtG}{\stEnv}{}$ and
$\gtG\,\gtMove$, then
$\exists\stEnvi,\gtGi,\stEnvAnnotGenericSym.\;
\stEnv \stEnvMoveGenAnnot \stEnvi$,
$\gtG\,\gtMove[\stEnvAnnotGenericSym]\,\gtGi$
and $\stEnvAssoc{\gtGi}{\stEnvi}{}$.\\[1mm]  
The ``good projection'' criteria~(\cref{def:good-proj}) \emph{automatically} guarantee 
\emph{soundness} and \emph{completeness} of
\emph{association}
as well as safety and deadlock-freedom,
guiding the design of ``correct'' typing systems. 

\smallskip

\myparagraph{The Problems} we tackle in this paper are:
\begin{enumerate}[leftmargin=*]
\item The
\textbf{deadlock-free realisability} (\emph{implementability},
\emph{projectability}) problem \cite{Giusto2025,Li2025c} 
asks if a global type $\gtG$
has a safe and deadlock-free
typing context $\stEnv$
with the same trace as $\gtG$ (denoted by $\gtTrace{\gtG}$). 
Does there exist $\stEnv$ such that
$\gtTrace{\gtG}=\gtTrace{\stEnv}$,
$\gtRoles{\gtG}=\dom{\stEnv}$ and
${\stEnv}$ is safe and deadlock-free?
\item The \textbf{projection problem} \cite{DBLP:journals/pacmpl/UdomsrirungruangY25}
asks for \emph{correct} algorithms for producing
typing contexts given global types.
Does there exist \emph{correct}
$\gtProj{}{}{}$ \ie
if $\gtProj{}{\gtG}{\stEnv}$, then is $\gtG$ 
deadlock-free realisable with $\stEnv$?
\item The \textbf{protocol realisability}
problem asks whether given $\gtG$,
a correct projection algorithm $\gtProj{}{}{}$
can project global type $\gtGi$ which is bisimilar ($\sim$) to $\gtG$.
Given $\gtG$ and correct $\gtProj{}{}{}$,
does there exist
$\gtGi$ such that $\gtG\sim \gtGi$ and
$\gtProj{}{\gtGi}{\stEnv}$ for some $\stEnv$?
\item The \textbf{program realisability}
problem asks whether a
typing system defined with 
correct $\gtProj{}{}{}$
can type program $\M$.
Given $\M$ and typing system $\vdash$
defined with correct $\gtProj{}{}{}$ in \inferrule{Sess},
does there exist $\gtG$ such that $\tyJudge{}{\M}{\gtG}$?
\end{enumerate}
(2) asks for \emph{algorithms} solving (1).
If (3) is solved, (1) is solved indirectly.
(4) differs from
(1), (2) and
(3), asking about \emph{programs}
instead of global types.
Both (3) and (4) raise \textbf{separation problems}:
can we separate the realisable protocols
and programs of correct projections?

\smallskip
\myparagraph{This paper} answers the following questions:
\begin{itemize}[leftmargin=*]
\setlength\itemsep{0em}
\item \textbf{Q1:}
  What are the ``good projection''
  criteria that allow us to
  answer the projection problem?

\item \textbf{Q2:}
How do we guarantee program properties
by choosing global types?

\item \textbf{Q3:}
Are the good projection criteria
general enough to characterise 
all ``correct'' projections in the literature (sound), 
while rejecting incorrect ones (complete)? 

\item \textbf{Q4:}
  How can we define \emph{separation} and \emph{expressiveness} of
projections based on 
protocol realisability and
program realisability?
Which projections are separated by
their realisable protocols 
and/or their realisable programs?

\item \textbf{Q5:}
Can we extend good projection criteria
to other communication semantics and syntax, \eg asynchronous
and mixed choice types?
\end{itemize}
$\textbf{Q1}$ is answered in the previous paragraph. 
Below, we explain the rest of the questions
and summarise our contributions.

\smallskip

\myparagraph{Q2: Global type properties.}
The desired properties of typed programs
depend on their intended use.
Some use cases only require error-freedom,
others liveness \eg authentication protocols \cite{OAUTH}
and others liveness only for specific participants \eg
client-server protocols
\cite[Example 2.1]{Casetta_2026}.
The bottom-up approach \cite{POPL19LessIsMore}
uses a parameterised typing system to control
these properties.
Top-down typing systems always guarantee
safety and deadlock-freedom (\cref{cor:typingsessions}).
We propose using structural properties of
global types to \emph{efficiently} (\cref{thm:dec-bal})
guarantee
various generalisations of liveness.

\myparagraph{Q3: Generality of \emph{good projection} criteria.} 
If global type $\gtG$ is projectable,
we use it to type programs.  
If the projection algorithm is too strict, it rejects many realisable (implementable) 
global types, making 
many safe and live processes untypable. 
We demonstrate generality of our criteria,
identifying that 8 known (correct) 
projections in the literature are good~(\cref{thm:good-inductive-proj}), 
starting from the most ``limited'' inductive algorithm 
\cite{HYC08,HYC16} 
to the most ``expressive'' 
coinductive algorithm~\cite[Definition~4.23]{DBLP:journals/pacmpl/UdomsrirungruangY25}.
We then prove by general construction
(\cref{thm:incorrect-merge}) that 4 existing algorithms failing
our criteria,
the ``1-level'' \cite{GJPSY2018,10.1145/3798256} and
``disjoint'' \cite{Barbanera2021-mg,doi:10.3233/FI-2019-1863}
projections,
induce \emph{incorrect} typing systems
where subject reduction fails.
Table~\ref{table:related} in \S~\ref{sec:conclusion}
categorises the existing ``good'' projection algorithms in the
literature. 

\myparagraph{Q4: Measuring expressiveness and typability.} 
``Expressiveness'' 
of projection algorithms
can be measured~(\cref{def:expressiveness}) either by:
the set of global protocols that can be projected~(\cref{fig:expressiveness-global});
or by the set of typable sessions~(\cref{fig:expressiveness-local}).
The \emph{protocol separation problem}
for the projections
$\gtProjAnnot{}{}{}{A}$ and $\gtProjAnnot{}{}{}{B}$
asks whether there is a global type, $\gtG$, that
$\gtProjAnnot{}{}{}{A}$ can project
but $\gtProjAnnot{}{}{}{B}$ cannot up-to bisimilarity.
Asking whether $\gtG$ can be projected is less interesting
since we may simply run the algorithms and
protocol designers may
rewrite global types up-to bisimilarity.
Thus, it is more relevant to ask if the protocol
can be handled at all.
The \emph{typability separation problem}
for the projections
$\gtProjAnnot{}{}{}{A}$ and $\gtProjAnnot{}{}{}{B}$
asks whether there is a session $\M$
that $\gtProjAnnot{}{}{}{A}$ can type and
$\gtProjAnnot{}{}{}{B}$ cannot.
We prove that separating or equating the
sets of associated contexts
solves the typability separation problem.
In both cases,
we prove for the first time that
parallel composition in global types
allows strictly more expressivity for inductive projections,
no more expressivity for coinductive projections and
that coinductive projections are
strictly more expressive than inductive projections.

\myparagraph{Q5: Asynchronous communications and mixed choice. }
In MPST literature, two transport semantics are commonly used. One is 
\emph{synchronous} semantics where a sender and a receiver immediately 
interact by handshake, and
the other is \emph{asynchronous} semantics~(\cref{sec:async}) where 
a sender first enqueues a message into a corresponding queue and 
a receiver dequeues a message from a corresponding queue, hence 
sending is non-blocking but the order of messages is preserved.
Our criteria
and
separation results
work in both communication models
(\cref{def:as-good-proj,thm:async-separate}).
Additionally, extending our criteria~(\cref{def:a-good-proj}) to accommodate
decidable and sound approximations of the 
\emph{asynchronous precise subtyping}~\cite{GPPSY2023}, which 
enables the optimisation of communications of message-passing programs,
is a necessary step towards its implementation.
Adding mixed choice syntax to global types
and processes
preserves our correctness, separation and
global type property results.
We propose new projections~(\cref{def:global-proj-mixed})
for synchronous directed mixed choice global types
that are simple extensions of the standard projections,
maintaining their complexities.

\smallskip
\myparagraph{Outline.} \S~\ref{sec:sessiontypes} introduces
processes, sessions,
global and local type syntax, semantics and
characterises global trace languages.
\S~\ref{sec:projections} proposes the ``good projection'' criteria
and global type properties,
and studies their consequences. 
\S~\ref{sec:example} defines 8 good projection algorithms,
proves separation and expressiveness results amongst them,
and proves the incorrectness of 4 projections.
\S~\ref{sec:extended} extends global and local type and process syntax to
mixed choice.
\S~\ref{sec:async} extends the criteria to asynchronous
semantics with synchronous and asynchronous
subtyping relations.
\S~\ref{sec:related} includes a discussion of related works.
\S~\ref{sec:conclusion} concludes the paper with
future work.
\textbf{The full version}
contains the full proofs, 
further examples and definitions \cite{fullversion}.
\textbf{The artifact} is available from \cite{anonymous_2026_21242753}.

\section{Multiparty Session Types}
\label{sec:sessiontypes}
This section introduces syntax, semantics and properties of
the synchronous multiparty session calculus and
synchronous global and local types 
(the standard from \citep{POPL19LessIsMore,Gheri2023,DBLP:series/lncs/YoshidaH24,HYK2026}).  

\subsection{Synchronous Multiparty Session Calculus}
\label{sec:calculus}
\begin{definition}[Processes and Sessions~{\cite{GJPSY2018}}]
  \label{def:sync-processes}
\emph{Expressions} ($e,e',\ldots$) 
are variables ($x,y,z,\ldots$),
value $v$ (such as a natural $n$ or a boolean $\mpTrue/\mpFalse$)
or built by operators such as $+$, $-$ or $\neg$.
We define 
\emph{processes} ($\mpP, \mpQ, \mpP_i, \ldots$) and
\emph{sessions} ($\N,\N_i,...$) of the synchronous
multiparty session calculus by:\\[1mm]
\centerline{
$\mpP \bnfdef
\roleQ\mathord{!}\ell\langle e \rangle \mpSeq \mpP 
\bnfsep \textstyle\sum_{i\in I} \roleQ\mathord{?}\ell_i(x_i)\mpSeq \mpP[i] 
\bnfsep \mpIf{e}{\mpP}{\mpQ} 
\bnfsep \mpNil 
\bnfsep \mpRec{\mpX}{\mpP} 
\bnfsep \mpX
\bnfsep \mpErr$
\quad
$\N\bnfdef \Pi_{\roleR\in\roleSet}\:\playsrole{\roleR}{\mpP[\roleR]}$
}\\[1mm]
where
$I$ is a non-empty indexing set;
$\roleQ\mathord{!}\ell\langle e \rangle \mpSeq \mpP$ is a \emph{send} (selection) to role $\roleQ$ with label $\ell$, payload expression $e$ and continuation $\mpP$; 
$\sum_{i\in I} \roleQ\mathord{?}\ell_i(x_i)\mpSeq \mpP_i$
where $I\neq \emptyset$ and $(\ell_i=\ell_j\!\implies\!i=j)$
is a \emph{receive} (branching) from role $\roleQ$, 
where a message with label $\ell_k$ binds variable $x_k$ and continues as $\mpP[k]$;
$\mpIf{e}{\mpP}{\mpQ}$ is a conditional;
$\mpNil$ is an inactive process;
$\mpRec{\mpX}{\mpP}$ is a recursion where $\mpP$
must be \emph{guarded}, \eg $\mpRec{\mpX}{\mpX}$ is not allowed.
A multiparty session is
formally a function from
the set of participants, $\roleSet$,
to processes;
informally, it is a parallel
composition of process $\mpP[\roleR]$ which plays role $\roleR$
(denoted by $\playsrole{\roleR}{\mpP[\roleR]}$).
\end{definition}

\begin{definition}[Semantics]
\label{def:session-semantics}
Let $\eval{e}{v}$ denote $e$ is evaluated to
$v$ \cite[Table 1]{GJPSY2018}.
We define \emph{structural precongruence} on
processes and sessions to be the minimum reflexive and transitive relation, $\prestruct$,
such that $\mpRec{\mpX}{\mpP}\prestruct\mpP\subst{\mpX}{\mpRec{\mpX}{\mpP}}$
and
$\mpP\prestruct\mpPi\implies\playsrole{\roleP}{\mpP}\mpPar\M\prestruct\playsrole{\roleP}{\mpPi}\mpPar\M$.
Define reduction, $\mpMove$, inductively by the rules below.
$\mpMoveStar$/$\mpMovePlus$ is the transitive/positive closure of
$\mpMove\cup\prestruct$/$\mpMove$:
\\[2mm]
\centerline{\(
	\begin{array}{ll}
		\inferrule{MCond}&\eval{e}{v}\;\wedge\;v\in\{\mpTrue,\mpFalse\}
		\implies
		\playsrole{\roleP}{\mpIf{e}{\mpP[\mpTrue]}{\mpP[\mpFalse]}}
		\mpPar\M
		\mpMove
		\playsrole{\roleP}{\mpP[v]}
		\mpPar\M
		\\
		\inferrule{MComm}&\eval{e}{v}\;\wedge\;
		j\in I
		\implies
		\playsrole{\roleP}{
			\roleQ\mathord{!}\ell_j\langle e \rangle \mpSeq \mpP
		}
		\mpPar
		\playsrole{\roleQ}{
			\textstyle\sum_{i\in I} \roleP\mathord{?}\ell_i(x_i)\mpSeq \mpQ[i] 
		}
		\mpPar
		\M
		\mpMove
		\playsrole{\roleP}{
			\mpP
		}
		\mpPar
		\playsrole{\roleQ}{
			\mpQ[j]\subst{x_j}{v} 
		}
		\mpPar
		\M
		\\
		\inferrule{MStruct}&\M\prestruct\M_1\;\wedge \;
                \M_1\mpMove\M_2 \;\wedge \; \M_2 \prestruct\M'\implies
		\M\mpMove\M'
		\\
		\inferrule{MErrCond}&\eval{e}{v}\;\wedge\;v\notin\{\mpTrue,\mpFalse\}
		\implies
		\playsrole{\roleP}{\mpIf{e}{\mpP[\mpTrue]}{\mpP[\mpFalse]}}
		\mpPar\M
		\mpMove
		\playsrole{\roleP}{\mpErr}
		\mpPar\M
		\\
		\inferrule{MErrMism}&
		\ell\notin\{\ell_i\}_{i\in I}
		\implies
		\playsrole{\roleP}{
			\roleQ\mathord{!}\ell\langle e \rangle \mpSeq \mpP
		}
		\mpPar
		\playsrole{\roleQ}{
			\textstyle\sum_{i\in I} \roleP\mathord{?}\ell_i(x_i)\mpSeq \mpQ[i] 
		}
		\mpPar
		\M
		\mpMove
		\playsrole{\roleP}{
			\mpP
		}
		\mpPar
		\playsrole{\roleQ}{
			\mpErr
		}
		\mpPar
		\M
	\end{array}
\)}
\end{definition}

\inferrule{MCond} evaluates conditionals;
\inferrule{MComm} facilitates communication;
\inferrule{MStruct} unfolds recursions;
\inferrule{MErrCond} says that conditioning on non-booleans
causes an error; and
\inferrule{MErrMism} that sending an unexpected label
causes an error.
We want our sessions to be \emph{error-free}.

\begin{definition}[Session Properties]
\label{def:sess-prop}
\begin{enumerate}
	\item $\M$ is \emph{safe} iff
	for all $\M\mpMoveStar\M'$,
	$\M'$ has no errors.
	\item
	$\M$ is \emph{deadlock-free} iff
	$\forall\M\mpMoveStar\M'.\;
	\M'\prestruct\mpFmt{\Pi_{\roleP\in\roleSet}\playsrole{\roleP}{\mpNil}}
	\;\vee\;
	\M'\prestruct\M''\mpMove$.
	\item $\M$ is \emph{$\roleP$-send live} iff
	$\forall\M\mpMoveStar\M'.$
	if $\M'=\playsrole{\roleP}{
	\roleQ\mathord{!}\ell\langle e \rangle \mpSeq \mpP
	}\mpPar\M''$
	then $\M''\mpMoveStar\playsrole{\roleQ}{
	\textstyle\sum_{i\in I} \roleP\mathord{?}\ell_i(x_i)\mpSeq \mpP_i 
	}\mpPar\M'''$
	where $\ell\in\{\ell_i\}_{i\in I}$;
	\item $\M$ is \emph{$\roleP$-receive live} iff
	$\forall\M\mpMoveStar\M'.$
	if $\M'=\playsrole{\roleP}{
	\textstyle\sum_{i\in I} \roleQ\mathord{?}\ell_i(x_i)\mpSeq \mpP_i
	}\mpPar\M''$
	then $\M''\mpMoveStar\playsrole{\roleQ}{
	\roleP\mathord{!}\ell_k\langle e \rangle \mpSeq \mpP
	}\mpPar\M'''$
	where $k\in I$; and
	\item $\M$ is \emph{live}
	iff for all $\roleP$,
	$\M$ is $\roleP$-send live
	and $\roleP$-receive live.
\end{enumerate}
\end{definition}
\noindent Safe sessions are always error-free.
Deadlock-free sessions
can always keep reducing unless every process has terminated.
A session is $\roleP$-send/receive live iff
whenever $\roleP$ is a sending/receiving process,
a future session can be reached where $\roleP$
can send/receive.

\begin{example}
	\label{ex:calc-df-not-live}
	$
	\playsrole{\roleP}{\mpRec{\mpX}{\roleQ\mathord{!}\ell \mpSeq \mpX }}
	\mpPar
	\playsrole{\roleQ}{\mpRec{\mpX}{\roleP\mathord{?}\ell\mpSeq \mpX }}
	\mpPar
	\playsrole{\roleR}{\roleP\mathord{?}\ell\mpSeq\mpNil}
	$
	is deadlock-free
	but not
	$\roleR$-receive live.
\end{example}

\subsection{Global and Local Type Syntax}
\label{sec:syntax}
\label{sec:gtype:syntax}
\emph{Global types} specify
communication protocols,  
involving \emph{participants}
($\roleP, \roleQ, \roleS, \roleT, \ldots \in \roleSet$),
while \emph{local types} describe 
end-point specifications.
They are defined as:\\[1mm]
\centerline{\(
     \begin{array}{r@{}c@{}l}
       \gtG & \bnfdef &
         \gtComm{\roleP}{\roleQ}{i \in I}{\gtLab[i]}{\tyGround[i]}{\gtG[i]}
         \bnfsep \gtRec{\gtRecVar}{\gtG} \; \bnfsep \; \gtRecVar 
         \bnfsep \gtEnd
 	\bnfsep \gtPar{\gtG}{\gtGi}\\[1mm]
       \stT
        & \bnfdef & \stExtSum{\roleP}{i \in I}{\stChoice{\stLab[i]}{\tyGround[i]} \stSeq \stT[i]}
         \bnfsep \stIntSum{\roleP}{i \in I}{\stChoice{\stLab[i]}{\tyGround[i]} \stSeq \stT[i]}
         \bnfsep \stRec{\stRecVar}{\stT}  \bnfsep  \stRecVar 
         \bnfsep \stEnd
     \end{array}
\)}\\[1mm]
\textbf{Basic types} ($\tyGround$, $\tyGroundi$, $\ldots \in \tyGroundSet$) 
are types of values \eg $\tyInt$, $\tyReal$, \etc.
In {\bf global types} ($\gtG$, $\gtGi$, $\ldots$), 
\emph{transmission} $\gtComm{\roleP}{\roleQ}{i \in
I}{\gtLab[i]}{\tyGround[i]}{\gtG[i]}$ specifies 
a message/label $\gtLab[i]$ from $\roleP$ to $\roleQ$, 
with a payload of type $\tyGround[i]$
and a continuation $\gtG[i]$ where $I\not=\emptyset$,
$\gtLab[i]\not=\gtLab[j]$ and $\roleP \neq \roleQ$. 
\emph{Termination} $\gtEnd$, 
\emph{recursive type}
$\gtRec{\gtRecVar}{G}$ and 
\emph{type variable} $\gtRecVar$ are standard.
A \emph{parallel} composition of $\gtG$ and $\gtGi$
is written as $\gtPar{\gtG}{\gtGi}$. 
We show that the presence of the $\gtPar{\gtG}{\gtGi}$ constructor affects
the expressiveness of some of the good projections in~\cref{sec:comparison}.
$\gtRoles{\gtG}$/$\gtFv{\gtG}$ denote the set of participants/free
type variables in $\gtG$
\ifextended (\cref{def:roleset}/\cref{def:freevar}).
\else \appref{Def. A.6/Def. A.7}.
\fi
We assume 
$\gtRoles{\gtG}\cap\gtRoles{\gtGi}=\emptyset$ and 
$\gtFv{\gtG}=\gtFv{\gtGi}=\emptyset$ in
$\gtPar{\gtG}{\gtGi}$ \citep{POPL19LessIsMore,Gheri2023}.  

In {\bf local (session) types} 
($\stT, \stU, \ldots$),
an \emph{internal choice} (\emph{selection})
$\stIntSum{\roleP}{i \in
I}{\stChoice{\stLab[i]}{\tyGround[i]} \stSeq \stT[i]}$ 
specifies sending the message/label $\stLab[i]$ to $\roleP$ and
an \emph{external choice} (\emph{branching}) 
$\stExtSum{\roleP}{i \in
I}{\stChoice{\stLab[i]}{\tyGround[i]} \stSeq \stT[i]}$ 
specifies receiving 
$\stLab[i]$ from $\roleP$. 
Other type constructors are standard. 
Recursive types 
are \emph{guarded}
\ifextended (\cref{def:guarded})
\else \appref{Def. A.5}
\fi
\citep[\S 21.8]{PierceTAPL}
(\eg $\gtRec{\gtRecVar}{\gtRecVar}$ is not allowed) and
terminations are often omitted.
We define \emph{unfoldings} of global types recursively by:
$\unfoldOne{\gtRec{\gtRecVar}{\gtG}}=\unfoldOne{\gtG{}[\gtRec{\gtRecVar}{\gtG}/\gtRecVar]}$
and otherwise $\unfoldOne{\gtG}=\gtG$. Similarly for $\unfoldOne{\stT}$.

\begin{definition}[Typing Contexts]%
  \label{def:mpst-env}%
  \label{def:mpst-env-closed}%
  \label{def:mpst-env-comp}%
  \label{def:mpst-env-subtype}%
  \label{def:typing_context}
$\stEnvNew$ denotes a partial mapping from participants to session types. 
Their syntax is defined by:
  $
\stEnvNew
  \,\coloncolonequals\,
  \stEnvEmpty
  \bnfsep
   \stEnvNew \stEnvComp \stEnvMap{\roleP}{\stT}
$.
Notation 
$\stEnvNew[1] \stEnvComp \stEnvNew[2]$ 
  is defined iff $\dom{\stEnvNew[1]} \cap \dom{\stEnvNew[2]} = \emptyset$.
\end{definition}

\begin{example}[{Odd-Even Protocol~\citep[Example~2.1]{Li2023-sx}}]
\label{ex:running}
We introduce our running example.
\begin{equation*}
\begin{array}{c}
	\stT[\roleP] = \stIntSum{\roleQ}{}{
		\stLabFmt{o},\,
		\stLabFmt{m}
	}
	\;
	\stT[\roleQ] = \stExtSum{\roleP}{}{
		\stLabFmt{o}\stSeq\roleR{\oplus}\stLabFmt{o}\stSeq\stRec{\stRecVar}{
			\stIntSum{\roleR}{}{
				\stLabFmt{o}\stSeq\roleR{\oplus}\stLabFmt{o}\stSeq\stRecVar,\,
				\stLabFmt{b}\stSeq\roleR{\&}\stLabFmt{o}
			}
		},\,
		\stLabFmt{m}\stSeq\stRec{\stRecVar}{
			\stIntSum{\roleR}{}{
				\stLabFmt{o}\stSeq\roleR{\oplus}\stLabFmt{o}\stSeq\stRecVar,\,
				\stLabFmt{b}\stSeq\roleR{\&}\stLabFmt{m}
			}
		}
	}\\
	\stT[\roleR]=\stRec{\stRecVar}{
		\stExtSum{\roleQ}{}{
			\stLab\stSeq\stExtSum{\roleQ}{}{
				\stLabFmt{o}\stSeq\stRecVar,\,
				\stLabFmt{b}\stSeq\roleQ{\oplus}\stLabFmt{o}
			},\,
			\stLabFmt{b}\stSeq\roleQ{\oplus}\stLabFmt{m}}
	}\;
	\stEnv[\text{\tiny o/e}] = \stEnvMap{\roleP}{\stT[\roleP]}\stEnvComp
	\stEnvMap{\roleQ}{\stT[\roleQ]}\stEnvComp
	\stEnvMap{\roleR}{\stT[\roleR]}
	\\
	\gtG[\text{\tiny o/e}] =
	\gtCommRaw{\roleP}{\roleQ}{
		\gtMsgFmt{o}{\gtSeq}
			\gtRec{\gtRecVar}{
				\gtCommSingle{\roleQ}{\roleR}{\gtMsgFmt{o}}{}{
				\gtCommRaw{\roleQ}{\roleR}{
					\gtMsgFmt{o}{\gtSeq}\gtRecVar,\;
					\gtMsgFmt{b}{\gtSeq}\gtCommSingle{\roleR}{\roleQ}{\gtMsgFmt{o}}{}{}
				}
			}},\,
		\gtMsgFmt{m}{\gtSeq}
			\gtRec{\gtRecVar}{
				\gtCommRaw{\roleQ}{\roleR}{
					\gtMsgFmt{o}{\gtSeq}\gtCommSingle{\roleQ}{\roleR}{\gtMsgFmt{o}}{}{\gtRecVar},\,
					\gtMsgFmt{b}{\gtSeq}\gtCommSingle{\roleR}{\roleQ}{\gtMsgFmt{m}}{}{}
				}
			}
	}
	\end{array}
\end{equation*}
The protocol describes a game where $\roleP$
tells $\roleQ$ whether to send either an odd ($\gtMsgFmt{o}$)
or even ($\gtMsgFmt{m}$) number of messages ($\gtMsgFmt{o}$)
to $\roleR$,
after which $\roleQ$ will send $\gtMsgFmt{b}$ to $\roleR$,
which will respond by inferring $\roleP$'s choice.

The original type in \citep[Example~2.1]{Li2023-sx}
is structured differently.
\begin{equation*}
	\gtGorigin[\text{\tiny o/e}]=
	\gtCommRaw{\roleP}{\roleQ}{
		\gtMsgFmt{o}{\gtSeq}\gtCommSingle{\roleQ}{\roleR}{\gtMsgFmt{o}}{}{
			\gtRec{\gtRecVar}{
				\gtCommRaw{\roleQ}{\roleR}{
					\gtMsgFmt{o}{\gtSeq}\gtCommSingle{\roleQ}{\roleR}{\gtMsgFmt{o}}{}{\gtRecVar},\,
					\gtMsgFmt{b}{\gtSeq}\gtCommSingle{\roleR}{\roleQ}{\gtMsgFmt{o}}{}{}
				}
			}
		},\,
		\gtMsgFmt{m}{\gtSeq}
			\gtRec{\gtRecVar}{
				\gtCommRaw{\roleQ}{\roleR}{
					\gtMsgFmt{o}{\gtSeq}\gtCommSingle{\roleQ}{\roleR}{\gtMsgFmt{o}}{}{\gtRecVar},\,
					\gtMsgFmt{b}{\gtSeq}\gtCommSingle{\roleR}{\roleQ}{\gtMsgFmt{m}}{}{}
				}
			}
	}
\end{equation*}
The recursive binders are in different places,
so $\gtGorigin[\text{\tiny o/e}]\neq\gtG[\text{\tiny o/e}]$
but are equal up-to unfolding.
A projection algorithm may not project $\gtGorigin[\text{\tiny o/e}]$,
but can still handle the protocol if it can project $\gtG[\text{\tiny o/e}]$.
\end{example}

\subsection{Global and Local Semantics}
\label{sec:semantics}
We present Labelled Transition System (LTS)
semantics for global types, local types and typing contexts.
We introduce transition labels in~\Cref{def:mpst-env-reduction-label} below.

\begin{definition}[Transition Labels]
  \label{def:mpst-env-reduction-label}%
  \label{def:mpst-label-subject}%
We define transition labels ($\stEnvAnnotGenericSym$,
  $\stEnvAnnotGenericSymi$, $\stEnvAnnotGenericSym[i]$, \ldots) by: 
\\[1mm]\centerline{\(
  \stEnvAnnotGenericSym \bnfdef\ltsSendRecv{\roleP}{\roleQ}{\stChoice{\stLab}{\tyGround}}
\bnfsep\roleP\stEnvAnnotOutSym\stChoice{\stLab}{\tyGround}
\bnfsep\roleP\stEnvAnnotInSym\stChoice{\stLab}{\tyGround}
\bnfsep\stEnvInAnnot{\roleP}{\roleQ}{\stChoice{\stLab}{\tyGround}}
\bnfsep\stEnvOutAnnot{\roleP}{\roleQ}{\stChoice{\stLab}{\tyGround}}\)}\\[1mm]
The subject(s) of a transition label, written $\ltsSubject{\stEnvAnnotGenericSym}$, are defined as:
$\ltsSubject{\roleP\stEnvAnnotQSym\stChoice{\stLab}{\tyGround}}
=
\ltsSubject{\stEnvDagAnnot{\roleP}{\roleQ}{\stChoice{\stLab}{\tyGround}}}
=
\setenum{\roleP}$; and
$\ltsSubject{\ltsSendRecv{\roleP}{\roleQ}{\stChoice{\stLab}{\tyGround}}}
=
\setenum{\roleP, \roleQ}$
where $\stEnvAnnotQSym\in\{\stEnvAnnotInSym,\stEnvAnnotOutSym\}$.
\end{definition}
\noindent
Label $\ltsSendRecv{\roleP}{\roleQ}{\stChoice{\stLab}{\tyGround}}$ denotes a
communication between $\roleP$ and $\roleQ$ via a message 
$\stChoice{\stLab}{\tyGround}$;
$\roleQ\stEnvAnnotOutSym\stChoice{\stLab}{\tyGround}$
denotes sending a message to $\roleQ$;
$\stEnvOutAnnot{\roleP}{\roleQ}{\stChoice{\stLab}{\tyGround}}$
denotes that $\roleP$ is sending a message to $\roleQ$; and
$\{\roleP\stEnvAnnotInSym\stChoice{\stLab}{\tyGround},
\stEnvInAnnot{\roleP}{\roleQ}{\stChoice{\stLab}{\tyGround}}\}$
dually denote receiving.

\begin{definition}[Global Type Transitions~{\citep[Definition 7]{DBLP:conf/icdcit/YoshidaG20}}]
  \label{def:gtype:lts-gt}
  The global type transition
  $\gtMove[\stEnvAnnotGenericSym]$ is inductively
  defined by the rules below:\\[1mm]
\centerline{\small\(
\begin{array}{ll}
\inference[\iruleGtMovePar]
	{\gtG[i]\,\gtMove[\stEnvAnnotGenericSym]\,\gtGi[i]&\gtG[{1-i}]=\gtGi[{1-i}]}
	{\gtPar{\gtG[0]}{\gtG[1]}\,\gtMove[\stEnvAnnotGenericSym]\,\gtPar{\gtGi[0]}{\gtGi[1]}}
	&
  \inference[\iruleGtMoveComm]{j \in I}{
    \gtCommSmall{\roleP}{\roleQ}{i \in I}{\gtLab[i]}{\tyGround[i]}{\gtG[i]}
    \,\gtMove[\ltsSendRecv{\roleP}{\roleQ}{\stChoice{\stLab[j]}{\tyGround[j]}}]\, 
    \gtG[j]
  }
\\[2ex]
  \inference[\iruleGtMoveRec]{
    \gtG{}[\gtRec{\gtRecVar}{\gtG}/\gtRecVar] 
    \,\gtMove[\stEnvAnnotGenericSym]\, 
    \gtGi}
  {
    \gtRec{\gtRecVar}{\gtG} 
    \,\gtMove[\stEnvAnnotGenericSym]\, 
    \gtGi
  }
  &
  \inference[\iruleGtMoveCtx]{
    \forall i \in I. 
    \gtG[i] 
    \,\gtMove[\stEnvAnnotGenericSym]\, 
    \gtGi[i] 
    &
    \ltsSubject{\stEnvAnnotGenericSym} \cap \setenum{\roleP, \roleQ} =
    \emptyset
  }{
      \gtCommSmall{\roleP}{\roleQ}{i \in I}{\gtLab[i]}{\tyGround[i]}{\gtG[i]}
   \,\gtMove[\stEnvAnnotGenericSym]\, 
      \gtCommSmall{\roleP}{\roleQ}{i \in I}{\gtLab[i]}{\tyGround[i]}{\gtGi[i]}   
  }
\end{array}
\)}\\[1mm]
  $\gtG \,\gtMove\, 
  \gtGi$
  if there
  exists $\stEnvAnnotGenericSym$ such that
  $\gtG
  \,\gtMove[\stEnvAnnotGenericSym]\, 
  \gtGi$;
  $\gtG \,\gtMove$
  if there
  exists $\gtGi$ such that
  $\gtG
  \,\gtMove\,  
  \gtGi$,
  and $\gtNotMove{\gtG}$ for its negation;
 $\gtMoveStar$ denotes the transitive and reflexive closure of
  $\gtMove$; and
  $\gtG\,\gtMove[\vec{\stEnvAnnotGenericSym}]\,\gtGi$
  if $\vec{\stEnvAnnotGenericSym}=\stEnvAnnotGenericSym[1]\dots\stEnvAnnotGenericSym[n]$
  and
  $\gtG\,\gtMove[{\stEnvAnnotGenericSym[1]}]
  \cdot\gtMove[{\stEnvAnnotGenericSym[2]}]
  \dots\gtMove[{\stEnvAnnotGenericSym[n-1]}]\cdot
  \gtMove[{\stEnvAnnotGenericSym[n]}]\,\gtGi$.
\end{definition}
\noindent Rule $\inferrule{\iruleGtMoveComm}$ defines communication between two participants;
rule $\inferrule{\iruleGtMoveCtx}$ permits the transition of a global type if
the head's participants are disjoint from the subject of the transition, provided that all continuations 
perform the specified transition;
rule $\inferrule{\iruleGtMovePar}$ allows for transitions of
parallel composition; and 
rule $\inferrule{\iruleGtMoveRec}$ is a standard recursive rule. 

Using these semantics, we define two equivalent
notions of semantic equivalence for global types.
These are important, as we only care about global types up-to
the protocols that they represent.

\begin{definition}[Global Bisimilarity]
    \label{def:bisim}
    We define \emph{bisimilarity} $\bisim$ as the maximal relation on global types
    such that if $\gtG[L]\bisim\gtG[R]$ then:
    \begin{itemize}[leftmargin=0.2in,labelindent=-\leftmargin]
        \item If $\gtG[L]\,\gtMove[\stEnvAnnotGenericSym]\,\gtGi[L]$,
        then there exists $\gtGi[R]$ such that
        $\gtG[R]\,\gtMove[\stEnvAnnotGenericSym]\,\gtGi[R]$ and
        $\gtGi[L]\bisim\gtGi[R]$
        \item If $\gtG[R]\,\gtMove[\stEnvAnnotGenericSym]\,\gtGi[R]$,
        then there exists $\gtGi[L]$ such that
        $\gtG[L]\,\gtMove[\stEnvAnnotGenericSym]\,\gtGi[L]$ and
        $\gtGi[L]\bisim\gtGi[R]$
    \end{itemize}
\end{definition}

\begin{definition}[Trace]
	We define the \emph{trace} of a global type to be
	$
		\gtTrace{\gtG} = \{ \vec{\stEnvAnnotGenericSym}\suchthat
		\gtG\,\gtMove[\vec{\stEnvAnnotGenericSym}]
\}
$.
	We say that $\gtG$ and $\gtGi$ are \emph{trace equivalent} iff $\gtTrace{\gtG}=\gtTrace{\gtGi}$.
\end{definition}
Our semantics are deterministic i.e.
$\gtG\,\gtMove[\stEnvAnnotGenericSym]\,\gtGi\;\wedge\;
\gtG\,\gtMove[\stEnvAnnotGenericSym]\,\gtGii
\implies
\gtGi=\gtGii$.
Hence, we have:
\begin{restatable}[Bisimilarity is Trace Equivalence]{proposition}{thmBisimTrace}
\label{prop:bisimeq-traceeq}
$\gtTrace{\gtG}=\gtTrace{\gtGi}\iff\gtG\bisim\gtGi$
\end{restatable}

\begin{example}
\label{ex:running-bisim}
Recall $\gtG[\text{\tiny o/e}]$ and $\gtGorigin[\text{\tiny o/e}]$ from~\cref{ex:running}.
We have $\gtG[\text{\tiny o/e}]\bisim\gtGorigin[\text{\tiny o/e}]$.
\end{example}

We now provide a Labelled Transition System (LTS) semantics for local
types in~\Cref{def:ltype:lts-lt}.
\begin{definition}[Local Type Transitions]
\label{def:ltype:lts-lt}
The local type transition
$\gtMove[\stEnvAnnotGenericSym]$ is inductively
defined by the rules below:
\\[1ex]\centerline{\small\(
\begin{array}{c}
\inference[\iruleLtOut]
{j\in I}
{\stIntSum{\roleP}{i \in I}{\stChoice{\stLab[i]}{\tyGround[i]} \stSeq \stT[i]}\,\gtMove[{\roleP\stEnvAnnotOutSym\stChoice{\stLab[j]}{\tyGround[j]}}]\,\stT[j]}
\quad 
\inference[\iruleLtIn]
{j\in I}
{\stExtSum{\roleP}{i \in
I}{\stChoice{\stLab[i]}{\tyGround[i]} \stSeq \stT[i]}\,\gtMove[{\roleP\stEnvAnnotInSym\stChoice{\stLab[j]}{\tyGround[j]}}]\,\stT[j]}
\quad 
\inference[\iruleLtRec]
{\stT{}[\stRec{\stRecVar}{\stT}/\stRecVar]\,\gtMove[\stEnvAnnotGenericSym]\,\stTi}
{\stRec{\stRecVar}{\stT}\,\gtMove[\stEnvAnnotGenericSym]\,\stTi}
\end{array}
\)}
\\[1mm]
We define the same additional relations as in~\cref{def:gtype:lts-gt} analogously.
\end{definition}
\noindent Rule $\inferrule{\iruleLtOut}$ is for sending a message; rule
$\inferrule{\iruleLtIn}$ is for receiving a message; and 
rule $\inferrule{\iruleLtRec}$ deals with recursion.  
As with global types, we define \emph{local type bisimilarity}
analogously to~\cref{def:bisim}.

We present an LTS semantics for \emph{typing contexts} in \Cref{def:mpst-env-reduction}.
\begin{definition}[Typing Context Transition~{\citep[Definition~2.8]{POPL19LessIsMore}}]%
  \label{def:mpst-env-reduction}%
  \label{def:typing_context_reduction}
  Define \emph{typing context transition $\stEnvMoveGenAnnot$} by
  $\stT[\roleP]\,\gtMove[{\roleQ\stEnvAnnotOutSym\stChoice{\stLab}{\tyGround}}]\,\stTi[\roleP]
  \wedge
  \stT[\roleQ]\,\gtMove[{\roleP\stEnvAnnotInSym\stChoice{\stLab}{\tyGround}}]\,\stTi[\roleQ]
  \iff
  \stEnvMap{\roleP}{\stT[\roleP]}\stEnvComp\stEnvMap{\roleQ}{\stT[\roleQ]},\stEnv
  \stEnvMoveCommAnnot{\roleP}{\roleQ}{\stChoice{\stLab}{\tyGround}}
  \stEnvMap{\roleP}{\stTi[\roleP]}\stEnvComp\stEnvMap{\roleQ}{\stTi[\roleQ]},\stEnv$
  and
  $\gtTrace{\stEnv}=\{\vec{\stEnvAnnotGenericSym}\suchthat\stEnv\stEnvMoveAnnot{\vec{\stEnvAnnotGenericSym}}\}$.
\end{definition}
This semantics uses a synchronous communication model;
this means that sending and receiving seem
to have equal status.
Context safety (\cref{def:mpst-env-safe})
indicates that senders `control'
communications \ie
label mismatches occur when the sent label cannot be received.

We highlight three key properties of
typing contexts:
\emph{communication safety}, \emph{deadlock-freedom}
and \emph{liveness}.
These are crucial in ensuring that typed
processes are \emph{correct-by-construction}. 
\begin{definition}[Safety~{\cite[Definition~4.1]{POPL19LessIsMore}}]
\label{def:mpst-env-safe}%
\label{def:safety_typing_context}
  We say that
  $\predP$ is a \emph{safety property} of typing contexts
  iff, whenever $\predPApp{\stEnvNew}$, we have:
\\
  \begin{tabular}{@{\;\;}r@{\hskip 2mm}l@{\qquad}r@{\hskip 2mm}l}
    \inferrule{\iruleSafeComm}%
    &if
    $\stEnvApp{\stEnvNew}{\roleP}\,\gtMove[{\roleQ\stEnvAnnotOutSym\stChoice{\stLab}{\tyGround}}]$
    and
    $\stEnvApp{\stEnvNew}{\roleQ}\,\gtMove[{\roleP\stEnvAnnotInSym\stChoice{\stLabi}{\tyGroundi}}]$,
    then
    $\stEnvMoveAnnotP{\stEnvNew}{\stEnvCommAnnotSmall{\roleP}{\roleQ}{\stChoice{\stLab}{\tyGround}}}$;
    &
    \inferrule{\iruleMove}%
    &if
    $\stEnvNew \stEnvMove \stEnvNewi$,
    then
    $\predPApp{\stEnvNewi}$.
  \end{tabular}
\\[1mm]
  \upshape
  We say \emph{$\stEnvNew$ is safe},
  written $\stEnvSafeP{\stEnvNew}$,
  if $\predPApp{\stEnvNew}$  holds
  for some safety property $\predP$.
\end{definition}

\begin{definition}[Deadlock-Freedom~{\cite[\S~5.3]{POPL19LessIsMore}}]
  \label{def:stenv-deadlock-free}
A context $\stEnvNew$ is \emph{deadlock-free}, 
written $\stEnvDFP{\stEnvNew}$,  
iff
$\stEnvNew \!\stEnvMoveStar\! \stEnvNotMoveP{\stEnvNewi}$ 
implies $\forall \roleP\!\in\! \dom{\stEnvNewi}.\,
\unfoldOne{\stEnvApp{\stEnvNewi}{\roleP}}  =
\stEnd$. 
\end{definition}
\begin{definition}[Liveness~{\cite[\S~5.3]{POPL19LessIsMore}}]%
  \label{def:stenv-fairness}
  \label{def:stenv-live}
  \label{def:typing-ctx-live}
  A \emph{path} is a
  possibly infinite sequence of typing contexts
  $(\stEnvNew[n])_{n \in N}$,
  where $N = \{0,1,2,\ldots\}$ is a (finite or infinite) set of consecutive natural numbers,
  and, $\forall n, n+1 \!\in\! N$, $\stEnvNew[n] \stEnvMove \stEnvNew[n+1]$.
  We say that a path $(\stEnvNew[n])_{n \in N}$ is \emph{fair} iff,
  $\forall n \!\in\! N$:
    $\stEnvMoveAnnotP{\stEnvNew[n]}{\ltsSendRecv{\roleP}{\roleQ}{\stChoice{\stLab}{\tyGround}}}$
    implies $\exists k, \stLabi, \tyGroundi$
    such that $N \!\ni\! k \ge n$,
    and $\stEnvNew[k] \stEnvMoveCommAnnot{\roleP}{\roleQ}{\stChoice{\stLabi}{\tyGroundi}} \stEnvNew[k+1]$.
  We say that a path $(\stEnvNew[n])_{n \in N}$ is:
 \begin{enumerate}[leftmargin=*, nosep]
  \item
  \emph{$\roleP$-send live} iff $\forall n \!\in\! N.\left(
  \stEnvApp{\stEnvNew[n]}{\roleP}\gtMove[{\roleQ\stEnvAnnotOutSym\stChoice{\stLab}{\tyGround}}]
  \implies
  \exists k, \stLabi,\tyGroundi.
    \;N \ni k \ge n\;\wedge\;
    \stEnvNew[k] \stEnvMoveCommAnnot{\roleP}{\roleQ}{\stChoice{\stLabi}{\tyGroundi}}
      \stEnvNew[k+1]\right)$;
  \item
  \emph{$\roleP$-receive live} iff $\forall n \!\in\! N.\left(
  \stEnvApp{\stEnvNew[n]}{\roleP}\gtMove[{\roleQ\stEnvAnnotInSym\stChoice{\stLab}{\tyGround}}]
  \implies
    \exists k, \stLabi, \tyGroundi.\;
    N \ni k \ge n\;\wedge\;
    \stEnvNew[k] \!\stEnvMoveCommAnnot{\roleQ}{\roleP}{\stChoice{\stLabi}{\tyGroundi}}\! \stEnvNew[k+1]\right)$; and
    \item \emph{live} iff it is $\roleP$-send live and $\roleP$-receive live for all $\roleP$.
  \end{enumerate}
\noindent A typing context $\stEnv$ is
\emph{live},  
written
$\stEnvLiveP{\stEnvNew}$, iff
all fair paths starting with $\stEnv$ are also
live,
similarly for $\roleP$-send live and $\roleP$-receive live
written
$\stEnvSendLiveRoleP{\roleP}{\stEnvNew}$
and
$\stEnvRecvLiveRoleP{\roleP}{\stEnvNew}$
respectively.
\end{definition}
\noindent \emph{Safety} ensures that
only compatible messages are exchanged
\ie label mismatch errors do not occur;
\emph{deadlock-free} typing contexts
only terminate when each participant has completed;
\emph{$\roleP$-send/receive liveness} ensures that if $\roleP$
can send/receive, then it will send/receive; and
a context is \emph{live} if every participant performs their next action
eventually, assuming fair scheduling.

We introduce $\roleP$-send/receive liveness
for more granular analysis.
We may not require full liveness \eg
a client-server protocol requires
$\roleFmt{client}$-send and receive
liveness but not $\roleFmt{server}$ liveness.
Conditions (1) and (2)
allow the future message
to be $\stChoice{\stLabi}{\tyGroundi}$
instead of $\stChoice{\stLab}{\tyGround}$:
$\roleP$ may be able to send/receive
multiple messages
$\stChoice{\stLab}{\tyGround}$
but only one, $\stChoice{\stLabi}{\tyGroundi}$,
will be sent/received.

\begin{example}[Semantics of {$\gtG[\text{\tiny o/e}]$}]
\label{ex:semantics-running}
Recall $\gtG[\text{\tiny o/e}]$ and $\stEnv[\text{\tiny o/e}]$ from~\cref{ex:running}.
$\gtTrace{\gtG[\text{\tiny o/e}]}=\gtTrace{\stEnv[\text{\tiny o/e}]}$
and are given by the prefixes of the regular expression:
$
\ltsSendRecv{\roleP}{\roleQ}{\stLabFmt{o}}
\ltsSendRecv{\roleQ}{\roleR}{\stLabFmt{o}}
{(\ltsSendRecv{\roleQ}{\roleR}{\stLabFmt{o}}
\ltsSendRecv{\roleQ}{\roleR}{\stLabFmt{o}})}^*
\ltsSendRecv{\roleQ}{\roleR}{\stLabFmt{b}}
\ltsSendRecv{\roleR}{\roleQ}{\stLabFmt{o}}+
\ltsSendRecv{\roleP}{\roleQ}{\stLabFmt{m}}
{(\ltsSendRecv{\roleQ}{\roleR}{\stLabFmt{o}}
\ltsSendRecv{\roleQ}{\roleR}{\stLabFmt{o}})}^*
\ltsSendRecv{\roleQ}{\roleR}{\stLabFmt{b}}
\ltsSendRecv{\roleR}{\roleQ}{\stLabFmt{m}}
$.
We observe that $\stEnv[\text{\tiny o/e}]$ is safe, deadlock-free and live.
\end{example}

It is well-known that not all global types have
corresponding typing contexts,
but we show more.
Typing contexts have regular semantics,
while global type semantics need not be context-free.
\begin{restatable}[Global Type Semantics]{theorem}{thmGlobalSem}
\label{thm:global-sem}
\begin{enumerate*}
\item There is a global type $\gtG$ with $\gtTrace{\gtG}$ not context-free.
\item For all global types $\gtG$,
$\gtTrace{\gtG}\in LINSPACE$,
in particular
$\gtTrace{\gtG}$ is context-sensitive.
\end{enumerate*}
\end{restatable}

\begin{proof}
	The trace of $\gtG$ below,
	$\gtTrace{\gtG}$,
	is not context-free
	by the pumping lemma on the word
	$
	\left(\ltsSendRecv{\roleP}{\roleQ}{\stLab[1]}\right)^p
	\left(\ltsSendRecv{\rolePi}{\roleQi}{\stLab[1]}\right)^p
	\left(\ltsSendRecv{\roleR}{\roleS}{\stLab[3]}\right)^p
	\left(\ltsSendRecv{\roleRi}{\roleSi}{\stLab[3]}\right)^p
	$.
	\\[1mm]
	\centerline{\(
	\gtG=\gtPar{\gtRec{\gtRecVar}{
	\gtCommRaw{\roleP}{\roleQ}{
	\begin{array}{l}
		\gtLab[0]{\gtSeq}
		\gtCommRaw{\roleR}{\roleS}{
			\gtLab[2]{\gtSeq}\gtRecVar
		}
		\\
		\gtLab[1]{\gtSeq}
		\gtCommRaw{\roleR}{\roleS}{
			\gtLab[2]{\gtSeq}\gtRecVar,\,
			\gtLab[3]{\gtSeq}\gtRecVar
		}
	\end{array}
	}}}
	{
	\gtRec{\gtRecVar}{
	\gtCommRaw{\rolePi}{\roleQi}{
	\begin{array}{l}
		\gtLab[0]{\gtSeq}
		\gtCommRaw{\roleRi}{\roleSi}{
			\gtLab[2]{\gtSeq}\gtRecVar
		}
		\\
		\gtLab[1]{\gtSeq}
		\gtCommRaw{\roleRi}{\roleSi}{
			\gtLab[2]{\gtSeq}\gtRecVar,\,
			\gtLab[3]{\gtSeq}\gtRecVar
		}
	\end{array}
	}}
	}
	\)}\\[1mm]
	Let $\gtGi$ be a global type.
	$\gtTrace{\gtGi}\in LINSPACE$,
	see the artifact \cite{anonymous_2026_21242753}.
\end{proof}

\subsection{Subtyping}
\label{sec:subtyping}
\begin{definition}[Subtyping]
\label{def:subtyping}
The \emph{subtyping relation $\tySub$} on local types is coinductively defined by 
the rules:\\[1mm]
\centerline{\(
\begin{array}{c}
\cinference[\iruleStSubIn]{
  \forall i \in I
  &
  \stT[i] \tySub \tyTi[i]
}{
  \stExtSum{\roleP}{j \in I\cup J}{\stChoice{\stLab[j]}{\tyGround[j]} \stSeq \stT[j]}%
  \tySub
  \stExtSum{\roleP}{i \in I}{\stChoice{\stLab[i]}{\tyGround[i]} \stSeq \stTi[i]}%
}
\quad
\cinference[\iruleStSubOut]{
  \forall i \in I
  &
  \stT[i] \tySub \stTi[i]
}{
  \stIntSum{\roleP}{i \in I}{\stChoice{\stLab[i]}{\tyGround[i]} \stSeq \stT[i]}
  \tySub
  \stIntSum{\roleP}{j \in I\cup J}{\stChoice{\stLab[j]}{\tyGround[j]} \stSeq \stTi[j]}
}
\\[2ex]
\cinference[\iruleStSubEnd]{}{
  \stEnd \tySub \stEnd
}
\quad
\cinference[\iruleStSubRecL]{
  \stT{}[\stRec{\stRecVar}{\stT}/\stRecVar] \tySub \stTi
}{
  \stRec{\stRecVar}{\stT} \tySub \stTi
}
\quad
\cinference[\iruleStSubRecR]{
  \stT \tySub \stTi{}[\stRec{\stRecVar}{\stTi}/\stRecVar]
}{
  \stT \tySub \stRec{\stRecVar}{\stTi}
}
\end{array}
\)}\\[1mm]
Write
  $\stEnvNew \!\tySub\! \stEnvNewi$
  iff
  $\dom{\stEnvNew} \!=\! \dom{\stEnvNewi}$
  and
  $\forall \roleP \!\in\! \dom{\stEnvNew}.\,
  \stEnvApp{\stEnvNew}{\roleP} \!\tySub\! \stEnvApp{\stEnvNewi}{\roleP}$.
\end{definition}
For simplicity, we consider neither subtyping basic types~\cite[Definition~3.15]{GJPSY2018}
nor local types as payload types~\cite[Definition~2.5]{POPL19LessIsMore};
these are easy extensions. See \cref{sec:delegate}.
\begin{example}[Odd-Even]
Recall $\stT[\roleR]$ and $\stT[\roleP]$ (\cref{ex:running}):
$
\stRec{\stRecVar}{
		\stExtSum{\roleQ}{}{
			\stLabFmt{o}\stSeq\stExtSum{\roleQ}{}{
				\stLabFmt{o}\stSeq\stRecVar,\,
				\stLabFmt{b}\stSeq\roleQ{\oplus}\stLabFmt{o}
			},\,
			\stLabFmt{b}\stSeq\roleQ{\oplus}\stLabFmt{m},\,
			\stLabFmt{c}
		}
	}\tySub\stT[\roleR]$
and
	$\roleQ\stFmt{\oplus}\stLabFmt{o}\tySub\stT[\roleP]$
by increasing the external choices and
decreasing the internal choices.
\end{example}

\section{Good Projections and Association}
\label{sec:projections}
This section introduces projection relations between global types, participants and local types, alongside the
notions of association and good projections.

We consider projections to be relations between participants, global types and local types,
written $\gtProj{\roleP}{\gtG}{\stT}$.
This extends the standard approach~\citep{DBLP:conf/icdcit/YoshidaG20}
where projections are functions from participants and global types to local types.
We write $\gtProj{}{\gtG}{\stEnvNew}$ when $\dom{\stEnvNew}\supseteq\gtRoles{\gtG}$ and
for all $\roleP\in\dom{\stEnvNew}$, $\gtProj{\roleP}{\gtG}{\stEnvApp{\stEnvNew}{\roleP}}$.
We say that $\gtG$ is \emph{projectable} if there is $\stEnvNew$ such that
$\gtProj{}{\gtG}{\stEnvNew}$.

\subsection{Good Projections}
\label{sec:good-proj}
We provide the `good projection' conditions
that ensure an operational correspondence between
global types and projected contexts.
These allow more flexibility than simply equating
projected traces with global traces
and are more robust as they prevent
deadlocked participants from occurring.
The notion of a correct projection is
crucial to be able to
compare projections
and analyse
the impacts of global type properties.
\begin{definition}[Good Projection]
\label{def:good-proj}
	Fix a projection relation.
	We say that this is a \emph{good projection} if
	whenever $\gtProj{}{\gtG}{\stEnv}$
	with $\stEnvApp{\stEnv}{\roleP}=\stT$:
    \begin{itemize}[leftmargin=0.4in,labelindent=-\leftmargin]
        \item[\inferrule{$\projGoodEnd$}]
            If $\roleP\notin \gtRoles{\gtG}$, then $\unfoldOne{\stT}=\stEnd$
        \item[\inferrule{$\projGoodOut$}]
            If $\gtG\,\gtMove[ \ltsSendRecv{\roleP}{\roleQ}{\stChoice{\stLab}{\tyGround}} ]
            \,\gtGi$ then there is $\stTi$ such that
            $\gtProj{\roleP}{\gtGi}{\stTi}$ and
            $\stT\,\gtMove[{\roleQ\stEnvAnnotOutSym\stChoice{\stLab}{\tyGround}}]
            \,\cdot\tySub\stTi$
        \item[\inferrule{$\projGoodIn$}]
            If $\gtG\,\gtMove[ \ltsSendRecv{\roleQ}{\roleP}{\stChoice{\stLab}{\tyGround}}]
            \,\gtGi$
            then there is $\stTi$ such that
            $\gtProj{\roleP}{\gtGi}{\stTi}$ and
            $\stT\,\gtMove[{\roleQ\stEnvAnnotInSym\stChoice{\stLab}{\tyGround}}]
            \,\cdot\tySub\stTi$
        \item[\inferrule{$\projGoodAlt$}]
            If $\gtG\,\gtMove[ \ltsSendRecv{\rolePi}{\roleQi}{\stChoice{\stLab}{\tyGround}} ]
            \,\gtGi$
            with $\roleP\notin\{\rolePi,\roleQi\}$,
            then there is $\stTi$ such that
            $\gtProj{\roleP}{\gtGi}{\stTi}$ and
            $\stT \tySub \stTi$
        \item[\inferrule{$\projGoodType$}]
            If
$\stT\,\gtMove[{\roleQ\stEnvAnnotOutSym\stChoice{\stLab}{\tyGround}}]$ and
$\stEnvApp{\stEnv}{\roleQ}\,\gtMove[{\roleP\stEnvAnnotInSym\stChoice{\stLabi}{\tyGroundi}}]$,
            then $\gtG\,\gtMove[ \ltsSendRecv{\roleP}{\roleQ}{\stChoice{\stLab}{\tyGround}} ]$
	\end{itemize}
\end{definition}
\inferrule{$\projGoodEnd$} says that
participants with no role in a global type
have no behaviour locally;
\inferrule{$\projGoodOut$}, \inferrule{$\projGoodIn$} and
\inferrule{$\projGoodAlt$} that projectability is preserved
by reductions;
\inferrule{$\projGoodOut$} and \inferrule{$\projGoodIn$} that
global type reductions are mirrored locally;
\inferrule{$\projGoodAlt$} that after a transition,
the projection may be less specialised; and
\inferrule{$\projGoodType$} that
\emph{attempted} context transitions
are mirrored globally.
\begin{example}[Projecting {$\gtG[\text{\tiny o/e}]$}]
	\label{ex:proj-running}
	Recall the types from \cref{ex:running} and \cref{ex:semantics-running}.
	Define $\gtProj{}{}{}$ by:\\[1mm]
\centerline{
	\small
		$\begin{array}{r@{\,}c@{\,}l}
			\gtProj{\roleP}{}{}&=&\{(\gtG[\text{\tiny o/e}],\stT[\roleP]),
			(\gtG[\text{\tiny o}],\stEnd),
			(\gtGi[\text{\tiny o}],\stEnd),
			(\gtG[\text{\tiny o,end}],\stEnd),
			(\gtG[\text{\tiny e}],\stEnd),
			(\gtGi[\text{\tiny e}],\stEnd),
			(\gtG[\text{\tiny e,end}],\stEnd),
			(\gtEnd,\stEnd)\}\\[1ex]
			\gtProj{\roleQ}{}{}&=&\{(\gtG[\text{\tiny o/e}],\stT[\roleQ]),
			(\gtG[\text{\tiny o}],\stT[\text{o},\roleQ]),
			(\gtGi[\text{\tiny o}],\stTi[\text{o},\roleQ]),
			(\gtG[\text{\tiny o,end}],\stT[\text{o,end},\roleQ]),
			(\gtG[\text{\tiny e}],\stT[\text{e},\roleQ]),
			(\gtGi[\text{\tiny e}],\stTi[\text{e},\roleQ]),
			(\gtG[\text{\tiny e,end}],\stT[\text{e,end},\roleQ]),
			(\gtEnd,\stEnd)\}\\[1ex]
			\gtProj{\roleR}{}{}&=&\{(\gtG[\text{\tiny o/e}],\stT[\roleR]),
			(\gtG[\text{\tiny o}],\stT[\roleR]),
			(\gtGi[\text{\tiny o}],\stT[\text{o},\roleR]),
			(\gtG[\text{\tiny o,end}],\stT[\text{o,end},\roleR]),
			(\gtG[\text{\tiny e}],\stT[\roleR]),
			(\gtGi[\text{\tiny e}],\stT[\text{o},\roleR]),
			(\gtG[\text{\tiny e,end}],\stT[\text{e,end},\roleR]),
			(\gtEnd,\stEnd)\}
		\end{array}
$}\\[1mm]
	where
	\(
		\begin{array}{r@{\,}c@{\,}lr@{\,}c@{\,}lr@{\,}c@{\,}l}
	\gtG[\text{\tiny o}] & = &
			\gtRec{\gtRecVar}{
				\gtCommSingle{\roleQ}{\roleR}{\gtMsgFmt{o}}{}{
				\gtCommRaw{\roleQ}{\roleR}{
					\gtMsgFmt{o}{\gtSeq}\gtRecVar,\,
					\gtMsgFmt{b}{\gtSeq}\gtCommSingle{\roleR}{\roleQ}{\gtMsgFmt{o}}{}{}
				}
			}}&
	\gtGi[\text{\tiny o}] & = &
				\gtCommRaw{\roleQ}{\roleR}{
					\gtMsgFmt{o}{\gtSeq}\gtG[\text{\tiny o}],\,
					\gtMsgFmt{b}{\gtSeq}\gtCommSingle{\roleR}{\roleQ}{\gtMsgFmt{o}}{}{}
				}
				&
	\gtG[\text{\tiny o,end}] & = &\gtCommSingle{\roleR}{\roleQ}{\gtMsgFmt{o}}{}{}\\
	\gtG[\text{\tiny e}] & = &
			\gtRec{\gtRecVar}{
				\gtCommRaw{\roleQ}{\roleR}{
					\gtMsgFmt{o}{\gtSeq}\gtCommSingle{\roleQ}{\roleR}{\gtMsgFmt{o}}{}{\gtRecVar},\,
					\gtMsgFmt{b}{\gtSeq}\gtCommSingle{\roleR}{\roleQ}{\gtMsgFmt{m}}{}{}
				}
			}
			&
	\gtGi[\text{\tiny e}] & = &
					\gtCommSingle{\roleQ}{\roleR}{\gtMsgFmt{o}}{}{\gtG[\text{\tiny e}]}
					&
	\gtG[\text{\tiny e,end}] & = &
					\gtCommSingle{\roleR}{\roleQ}{\gtMsgFmt{m}}{}{}
	\end{array}
	\)
	\\[1mm]
	\centerline{\(
    \begin{array}{c}
    \stT[\roleR]=\stRec{\stRecVar}{
		\stExtSum{\roleQ}{}{
			\stLabFmt{o}\stSeq\stExtSum{\roleQ}{}{
				\stLabFmt{o}\stSeq\stRecVar,\,
				\stLabFmt{b}\stSeq\roleQ{\oplus}\stLabFmt{o}
			},\,
			\stLabFmt{b}\stSeq\roleQ{\oplus}\stLabFmt{m}
		}
	}
	\;\;
	\stT[\text{o},\roleR]=
	\stExtSum{\roleQ}{}{
				\stLabFmt{o}\stSeq\stT[\roleR],\,
				\stLabFmt{b}\stSeq\roleQ{\oplus}\stLabFmt{o}
			}\;\;
	\stT[\text{e,end},\roleR]=\roleQ\stFmt{\oplus}\stLabFmt{m}\;\;
	\stT[\text{o,end},\roleR]=\roleQ\stFmt{\oplus}\stLabFmt{o}
    \\
	\stT[\roleP] = \stIntSum{\roleQ}{}{
		\stLabFmt{o},\,
		\stLabFmt{m}
	}\quad
	\stT[\roleQ] = \stExtSum{\roleP}{}{
	\begin{array}{l}
		\stLabFmt{o}\stSeq\roleR{\oplus}\stLabFmt{o}\stSeq\stRec{\stRecVar}{
			\stIntSum{\roleR}{}{
				\stLabFmt{o}\stSeq\roleR{\oplus}\stLabFmt{o}\stSeq\stRecVar,\,
				\stLabFmt{b}\stSeq\roleR{\&}\stLabFmt{o}
			}
		},\,
		\stLabFmt{m}\stSeq\stRec{\stRecVar}{
			\stIntSum{\roleR}{}{
				\stLabFmt{o}\stSeq\roleR{\oplus}\stLabFmt{o}\stSeq\stRecVar,\,
				\stLabFmt{b}\stSeq\roleR{\&}\stLabFmt{m}
			}
		}
	\end{array}
	}
	\\
	\begin{array}{r@{\,}c@{\,}lr@{\,}c@{\,}lr@{\,}c@{\,}l}
	\stT[\text{o},\roleQ]&=& 
		\roleR\stFmt{\oplus}\stLabFmt{o}\stSeq\stRec{\stRecVar}{
			\stIntSum{\roleR}{}{
				\stLabFmt{o}\stSeq\roleR{\oplus}\stLabFmt{o}\stSeq\stRecVar,\,
				\stLabFmt{b}\stSeq\roleR{\&}\stLabFmt{o}
			}
		}&
	\stTi[\text{o},\roleQ]&=& 
		\stRec{\stRecVar}{
			\stIntSum{\roleR}{}{
				\stLabFmt{o}\stSeq\roleR{\oplus}\stLabFmt{o}\stSeq\stRecVar,\,
				\stLabFmt{b}\stSeq\roleR{\&}\stLabFmt{o}
			}
		}&
		\stT[\text{o,end},\roleQ]&=& 
		\roleR\stFmt{\&}\stLabFmt{o}
		\\
	\stT[\text{e},\roleQ]&=& 
		\stRec{\stRecVar}{
			\stIntSum{\roleR}{}{
				\stLabFmt{o}\stSeq\roleR{\oplus}\stLabFmt{o}\stSeq\stRecVar,\,
				\stLabFmt{b}\stSeq\roleR{\&}\stLabFmt{m}
			}
		}&
	\stTi[\text{e},\roleQ]&=& 
		\roleR\stFmt{\oplus}\stLabFmt{o}\stSeq\stRec{\stRecVar}{
			\stIntSum{\roleR}{}{
				\stLabFmt{o}\stSeq\roleR{\oplus}\stLabFmt{o}\stSeq\stRecVar,\,
				\stLabFmt{b}\stSeq\roleR{\&}\stLabFmt{m}
			}
		}&
	\stT[\text{e,end},\roleQ] &=& 
		\roleR\stFmt{\&}\stLabFmt{m}
		\end{array}
	\end{array}
\)}\\[1mm]
	$\gtProj{}{}{}$ is a good projection
	such that
	$\gtProj{}{\gtG[\text{\tiny o/e}]}{\stEnv[\text{\tiny o/e}]}$.
\end{example}
Later when we measure expressiveness and separation
in \cref{sec:example},
we consider restrictions on global types \eg
balancedness or
no parallel compositions.
Good projections are robust with respect to such restrictions.
\begin{restatable}[Robustness of Good Projections]{proposition}{thmRestrictGlobal}
\label{prop:good-robust}
Suppose that $\predPApp{\gtG}\;\wedge\;\gtG\,\gtMove\,\gtGi
\implies\predPApp{\gtGi}$.
If $\gtProj{}{}{}$ is a good projection, then so is
$\gtProji{}{}{}=\{(\roleP,\gtG,\stT)\in\gtProj{}{}{}\suchthat \predPApp{\gtG}\}$.
\end{restatable}

\subsection{Association and Typing for Sessions}
\label{sec:association}
Fix a good projection relation.
We use this projection alongside our subtyping~(\cref{sec:subtyping}),
to relate global types to typing contexts.
\begin{definition}[Association~{\cite[Definition~3.7]{HYK2026}}]
\label{def:assoc}
\label{def:association}
We write $\stEnvAssoc{\gtG}{\stEnvNew}{}$
iff there exists $\stEnvNewi$ such that $\stEnvNew\tySub\stEnvNewi$
and $\gtProj{}{\gtG}{\stEnvNewi}$,
in which case we say that
$\stEnvNew$ is \emph{associated with $\gtG$}
or simply that $\stEnvNew$ is associated.
\end{definition}

\begin{example}[Projection is Associated]
$\forall\roleP \in \gtRoles{\gtG}.\;
\gtProj{\roleP}{\gtG}{\stT[\roleP]}$ implies
$
  \stEnvAssoc{\gtG}{\setenum{\stEnvMap{\roleP}{\stT[\roleP]}}_{%
    \roleP \in \gtRoles{\gtG}}}{}$. 
\end{example}

The association relation provides an operational
correspondence between global types and typing contexts
via~\cref{thm:soundness_association} and~\cref{thm:completeness_association}.
Suppose that $\gtProj{}{}{}$ is good:

\begin{restatable}[Soundness]{theorem}{thmProjSoundness}%
  \label{thm:gtype:proj-sound}
  \label{thm:soundness_association}
	$\stEnvAssoc{\gtG}{\stEnv}{}
	\;\wedge\;
	\gtG \,\gtMove[\ltsSendRecv{\roleP}{\roleQ}{\stChoice{\stLab}{\tyGround}}]
	\implies
	\exists \stEnvi,\stLabi,\tyGroundi.\;
	\stEnv \gtMove[\ltsSendRecv{\roleP}{\roleQ}{\stChoice{\stLabi}{\tyGroundi}}]\stEnvi
	$
\end{restatable}
\begin{restatable}[Completeness]{theorem}{thmProjCompleteness}%
  \label{thm:gtype:proj-comp}
  \label{thm:completeness_association}
  $\stEnvAssoc{\gtG}{\stEnv}{}
  \;\wedge\;
  \stEnv \stEnvMoveGenAnnot \stEnvi
  \implies
  \exists\gtGi.\;
  \stEnvAssoc{\gtGi}{\stEnvi}{}
  \;\wedge\;
  \gtG  \,\gtMove[\stEnvAnnotGenericSym]\, \gtGi$
\end{restatable}

We explain how rules \inferrule{$\projGoodAlt$}
and \inferrule{$\projGoodType$}
are used to handle participants that are
not involved in the transition.
If $\gtProj{}{\gtG}{\stEnv}$,
$\stEnvi\tySub\stEnv$,
$\stEnvi\stEnvMoveAnnot{\ltsSendRecv{\roleP}{\roleQ}{\stChoice{\stLab}{\tyGround}}}\stEnvii$
and $\roleR\in\dom{\stEnvi}\setminus\{\roleP,\roleQ\}$,
then by \inferrule{$\projGoodType$}
there is $\gtG \,\gtMove[\ltsSendRecv{\roleP}{\roleQ}{\stChoice{\stLab}{\tyGround}}]\,\gtGi$
and by \inferrule{$\projGoodAlt$}
there is $\stT$ such that
$\gtProj{\roleR}{\gtGi}{\stT}$
and
$\stEnvApp{\stEnvii}{\roleR}
=
\stEnvApp{\stEnvi}{\roleR}
\tySub
\stEnvApp{\stEnv}{\roleR}
\tySub
\stT
$.

For the typing system, it is crucial that being associated is a safety property:
\begin{restatable}[Associated Contexts]{theorem}{thmAssocCorrect}
\label{thm:good-correct}
Let
$\predP=\{\stEnv\suchthat\exists\gtG.\;\stEnvAssoc{\gtG}{\stEnv}{}\}$.
$\predP\subseteq\stEnvDFPred$ and is a safety property.
\end{restatable}

Rule \inferrule{$\projGoodType$} guarantees that mismatches
cannot occur, guaranteeing safety.
Rule \inferrule{$\projGoodEnd$} and \cref{thm:soundness_association}
guarantee deadlock-freedom:
if $\stEnvAssoc{\gtG}{\stEnv}{}$
and $\stEnvNotMoveP{\stEnv}$,
then $\gtNotMove{\gtG}$
so $\unfoldOne{\gtG}=\gtEnd$
so for all $\stEnvMap{\roleP}{\stT}\in\stEnv$,
$\unfoldOne{\stT}=\stEnd$.

For a projection to be meaningful,
certainly
\cref{thm:soundness_association,thm:completeness_association,thm:good-correct}
must hold.
Indeed, if we add a few other conditions that are often desired of projections,
then \cref{def:good-proj} is automatically satisfied.

\begin{theorem}[Meaningful Projections are Good]
\label{thm:meaningful-implies-good}
If $\gtProj{}{}{}$ is a projection such that
\cref{thm:completeness_association} holds and
whenever $\gtProj{\gtG}{}{\stEnv}$,
$\gtTrace{\gtG}=\gtTrace{\stEnv}$ and
$\stEnvSafeP{\stEnv}$
and
$\stEnvLiveP{\stEnv}$, then
$\gtProj{}{}{}$ is good.
\end{theorem}

\begin{proof}
Suppose that $\gtProj{}{\gtG}{\stEnv}$
with $\stEnvApp{\stEnv}{\roleP}=\stT$.

\noindent
\inferrule{$\projGoodEnd$}.
$\unfoldOne{\stT}\neq\stEnd$ and liveness implies that there is
$\vec{\stEnvAnnotGenericSym}\in\gtTrace{\stEnv}=\gtTrace{\gtG}$ such that
$\roleP\in\ltsSubject{\vec{\stEnvAnnotGenericSym}}$.
Thus, $\roleP\in\gtRoles{\gtG}$.

\noindent
\inferrule{$\projGoodOut$}, \inferrule{$\projGoodIn$} and \inferrule{$\projGoodAlt$}
follow immediately from
\cref{thm:completeness_association} as $\gtTrace{\gtG}=\gtTrace{\stEnv}$.

\noindent
\inferrule{$\projGoodType$} follows from \cref{thm:completeness_association}
as $\stEnvSafeP{\stEnv}$.

If assuming $\stEnvLiveP{\stEnv}$ is too strong,
then we could assume \inferrule{$\projGoodEnd$} instead.
\end{proof}

The assumptions in \cref{thm:meaningful-implies-good}
may be too strong in some settings;
this is the purpose of \cref{def:good-proj}.
The good projection conditions weaken these conditions
while guaranteeing correctness.

Theorems 
\ref{thm:gtype:proj-sound}, \ref{thm:gtype:proj-comp} and \ref{thm:good-correct}
are sufficient to prove type correctness.

\begin{definition}[Typing Rule~\cite{GJPSY2018}]
\label{def:type-rules}
We denote typing environments by
$
	\mpEnv\bnfdef\mpEnvEmpty
	\bnfsep \mpEnv\mpEnvComp \mpX:\stT
	\bnfsep \mpEnv\mpEnvComp x:\tyGround
$
and
we assume the expression typing judgement of the form $\stJudge{\mpEnv}{e}{\tyGround}$
from~\cite[Table~4]{GJPSY2018}.
\\[2mm]
\centerline{
\(
\small
\begin{array}{c}
\inference[T-$\mpNil$]{}{\stJudge{\mpEnv}{\mpNil}{\stEnd}}
\qquad
\inference[T-Var]{\mpX : \stT \in \mpEnv}{\stJudge{\mpEnv}{\mpX}{\stT}}
\qquad
\inference[T-Rec]{
  \stJudge{\mpEnv, \mpX : \stT}{\mpP}{\stT}
}{
  \stJudge{\mpEnv}{\mpRec{\mpX}{\mpP}}{{\stT}}
}
\qquad 
\inference[T-Sub]{\stJudge{\mpEnv}{\mpP}{\stT} \quad \stT \tySub \stTi}{\stJudge{\mpEnv}{\mpP}{\stTi}}
\\[1ex]
\inference[T-Out]{
  \stJudge{\mpEnv}{e}{\tyGround} \quad 
  \stJudge{\mpEnv}{\mpP}{\stT}
}{
  \stJudge{\mpEnv}{\roleQ\mathord{!}\ell\langle e \rangle \mpSeq \mpP}{
    \stFmt{\roleQ{\oplus}\stChoice{\ell}{\tyGround} \stSeq \stT}
  }
}
\qquad
\inference[T-In]{
J\subseteq I &
  \forall i \in J \quad \stJudge{\mpEnv, x_i : \tyGround[i]}{\mpP_i}{\stT[i]}
}{
  \stJudge{\mpEnv}{\textstyle\sum_{i\in I} \roleQ\mathord{?}\ell_i(x_i)\mpSeq \mpP_i}{
    \stExtSum{\roleQ}{i \in J}{\stChoice{\ell_i}{\tyGround[i]} \stSeq \stT[i]}
  }}
\\[1.5ex]
\inference[T-Cond]{
  \stJudge{\mpEnv}{e}{\tyBool} \quad 
  \stJudge{\mpEnv}{\mpP_1}{\stT} \quad
  \stJudge{\mpEnv}{\mpP_2}{\stT}
}{
  \stJudge{\mpEnv}{\mpIf{e}{\mpP_1}{\mpP_2}}{\stT}
}
\qquad
\inference[Sess]{
\gtProj{}{\gtG}{\stEnv} & 
\forall \roleP\in \dom{\stEnv}&
\stJudge{}{\mpP[\roleP]}{\stEnvApp{\stEnv}{\roleP}}
}
{
\tyJudge{}{\Pi_{\roleP\in\dom{\stEnv}}{\ \playsrole{\roleP}{\mpP[\roleP]}}}{\gtG}
}
\end{array}
\)
}
\end{definition}
\noindent\inferrule{T-$\mpNil$} types terminated processes with terminated types;
\inferrule{T-Rec} and \inferrule{T-Var} handle recursion;
\inferrule{T-Sub} subtyping;
\inferrule{T-Out} and \inferrule{T-In}
output and input respectively; and
\inferrule{T-Cond} conditionals.
\inferrule{Sess} says that
if the session $\M$
is composed of processes
following specifications given by a
projection of $\gtG$,
then $\M$ follows the protocol
specified by $\gtG$.
Note that $\gtProj{}{\gtG}{\stEnv}$
could be replaced by $\stEnvAssoc{\gtG}{\stEnv}{}$
by \inferrule{T-Sub}
\ifextended (\cref{lem:bottom-top}).
\else \appref{Lem. B.9}.
\fi
By \cref{thm:soundness_association,thm:completeness_association,thm:good-correct},
we derive type correctness.

\begin{restatable}[Properties of Multiparty Sessions \proofapp{\cref{app:association}}]{theorem}{corTypingSessions}%
  \label{cor:typingsessions}
Let $\gtProj{}{}{}$ in $\inferrule{Sess}$ in \cref{def:type-rules}
be a good projection and
$\tyJudge{}{\M}{\gtG}$.
\begin{enumerate*}
	\item (Subject Reduction)
	$
	\M\mpMoveStar\M'
	\implies
	\exists \gtG\,\gtMoveStar\,\gtGi.\;
	\tyJudge{}{\M'}{\gtGi}
	$;
	\item (Session Fidelity)
	$
	\gtG\,\gtMove
	\implies
	\exists
	\gtG\,\gtMove\,\gtGi,
	\M\mpMovePlus\M'.\;
	\tyJudge{}{\M'}{\gtGi}
	$; and
	\item $\M$ is safe and deadlock-free.
\end{enumerate*}
\end{restatable}
\begin{example}[Associating Odd-Even]
\label{ex:assoc-running}
	Using $\gtProj{}{}{}$ from~\cref{ex:proj-running},
	$\stEnvAssoc{\gtG[\text{\tiny o/e}]}{\stEnv[\text{\tiny o/e}]}{}$.
\end{example}

\begin{remark}[Correctness of Typing]
\label{rem:correctness-of-typing}
Unlike global types,
multiparty sessions use unlabelled reductions.
Thus, modifying labels during projection
maintains correctness in the sense of \cref{cor:typingsessions}
while the projection fails to be good.
Extend the transposition $(\roleP\ \roleQ)$
to global types and,
using $\gtProj{}{}{}$ from \cref{ex:proj-running},
define $\gtProji{\roleR}{\gtG}{\stT}$
iff $\gtProj{\roleR}{(\roleP\ \roleQ)\gtG}{\stT}$.
$\gtProji{}{}{}$ is not good since it swaps
participants during projection,
but its typing system satisfies the conclusions of \cref{cor:typingsessions}.
The purpose of \cref{def:good-proj}
is to define meaningful projections
not just to ensure correct typing systems.
\end{remark}

\subsection{Delegations and Session Interleaving}
\label{sec:delegate}
\cref{cor:typingsessions} extends
to the full session $\pi$-calculi with delegations
and interleaving
\cite{HYK2026,POPL19LessIsMore}.
The only modifications would be to
allow session valued payloads,
$\stS\bnfdef\tyGround\bnfsep\stT$,
and use the subtyping of session valued payload
in the subtyping rules, and
changing typing contexts to have domains of
\emph{channels} \cite[Def. 2.1]{HYK2026} 
instead of participants.
Details on these can be found in \cite{HYK2026}.

Let $\gtProj{}{}{}$ be a good projection.
Using the multiparty session $\pi$-calculus in \cite{HYK2026},
the typing system in \cite[Figure 8]{HYK2026}
with \inferrule{T-$\gtG$-$\mpFmt{\nu}$}
satisfies subject reduction and type safety.\\[1mm]
\centerline{\(
	\left(
	\gtProj{}{\gtG}{\stEnvi}
	\;\wedge\;
	\mpS\notin\stEnv
	\;\wedge\;
	\stJudge{\mpEnv}{\mpP}{\stEnv\stEnvComp\stEnvi}
	\right)
	\implies
	\stJudge{\mpEnv}{\mpRes{\mpS}{\mpP}}{\stEnv}
	\qquad
	\inferrule{T-$\gtG$-$\mpFmt{\nu}$}
\)}\\[1mm]
Under the single session assumption in \cite[Thm. 4.4]{HYK2026},
typed processes enjoy
session fidelity; deadlock-freedom; and,
under the conditions discussed in \cref{sec:balanced},
liveness.
There are alternative typing systems to ensure those properties:
\cite{Padovani2014-zb} uses a \emph{strict partial order on events};
\cite{CDYP2015} proposes an \emph{interaction type system}; and
\cite{Van_den_Heuvel2022-ex} takes a \emph{decentralised} approach.

\subsection{Global Type Properties}
\label{sec:balanced}
With the notion of a `good projection',
we now analyse how
the structure of global types
impacts the semantics of sessions and
associated contexts.
No liveness property is guaranteed by general global types,
but the standard balanced property
\citep{GJPSY2018,DBLP:journals/pacmpl/UdomsrirungruangY25}
guarantees liveness.
\begin{definition}[Balanced Global Types]
\label{def:balanced}
	For each participant, $\roleR$, we define the partial function $\gtDepth{}{\roleR}$ on global types recursively by
	$\gtDepth{\gtG}{\roleR}=1$ if $\roleR\notin\gtRoles{\gtG}$
	and $\gtFv{\gtG}=\emptyset$,
	and otherwise:
	\\[1mm]
        \centerline{\(
	\begin{array}{c}
\gtDepth{\gtComm{\roleP}{\roleQ}{i \in I}{\gtLab[i]}{\tyGround[i]}{\gtG[i]}}{\roleR}=
			\left\{\!\!\!\begin{array}{ll}
				1+max\{\gtDepth{\gtG[i]}{\roleR}\}_{i \in I} & \!\!\roleR\notin\{\roleP,\roleQ\}\\
				1 & \!\!\roleR\in\{\roleP,\roleQ\}
			\end{array}\right.
			\\
		\gtDepth{\gtPar{\gtG[0]}{\gtG[1]}}{\roleR}=
		\gtDepth{\gtG[0]}{\roleR}\;\text{if}\;\roleR\in\gtRoles{\gtG[0]}
		\;\text{and}\;
                \gtDepth{\gtG[1]}{\roleR}\;\text{if}\;\roleR\in\gtRoles{\gtG[1]}
		\quad
		\gtDepth{\gtRec{\gtRecVar}{\gtG}}{\roleR} = \gtDepth{\gtG}{\roleR}
	\end{array}
	\)}\\[1mm]
A global type, $\gtG$, is \emph{balanced} iff for all $\gtG\,\gtMoveStar\,\gtGi$ and $\roleR\in\gtRoles{\gtGi}$,
$\gtDepth{\gtGi}{\roleR}$ exists.
\end{definition}
\begin{example}[Balanced]
	\label{ex:non-bal}
	$\gtG[\text{\tiny o/e}]$ from~\cref{ex:running}
	is balanced:
	for all $\roleP$ and $\gtG$ in \cref{ex:proj-running},
	$\gtDepth{\gtG}{\roleP}$ exists.
	
	Consider:
        \(
		\gtG_{\text{\tiny unbal}} =
			\gtRec{\gtRecVar}{
				\gtCommRaw{\roleFmt{a}}{\roleFmt{b}}{
						\gtCommChoice{go}{}{\gtRecVar},
						\gtCommChoice{cancel}{}{
							\gtCommSingle{\roleFmt{a}}{\roleFmt{c}}{cancel}{}{\gtEnd}
						}
				}	
			}
		\).
		$\gtG_{\text{\tiny unbal}}$
		is not balanced as
		$\gtDepth{\gtG_{\text{\tiny unbal}}}{\roleFmt{c}}$
		does not exist,
		since $\gtDepth{\gtRecVar}{\roleFmt{c}}$
		is not defined.
\end{example}

This condition is restrictive,
being the strongest one that we propose.
We propose the weakest possible condition,
weak balanced global types.

\begin{definition}[Weak Balanced Global Types]
\label{def:weak-bal}
We define the predicate $\weakComm{\gtG}{\roleP}{\roleQ}$
inductively by the rules:
\\\centerline{\(\begin{array}{c}
	\inference[WComm]{}{\weakComm{\gtComm{\roleP}{\roleQ}{i\in I}{\stLab[i]}{\tyGround[i]}{\gtG[i]}}{\roleP}{\roleQ}}
	\quad
	\inference[WPar]{
	i\in\{0,1\}
	&
	\weakComm{\gtG[i]}{\roleP}{\roleQ}
	}{\weakComm{\gtPar{\gtG[0]}{\gtG[1]}}{\roleP}{\roleQ}}
	\\[1ex]
	\inference[W$\mu$]{\weakComm{\gtG}{\roleP}{\roleQ}}{\weakComm{\gtRec{\gtRecVar}{\gtG}}{\roleP}{\roleQ}}
	\quad
	\inference[WCtx]{
	\exists j\in I&\weakComm{\gtG[j]}{\roleP}{\roleQ}
	&\{\roleP,\roleQ\}\cap\{\rolePi,\roleQi\}=\emptyset
	}{\weakComm{\gtComm{\rolePi}{\roleQi}{i\in I}{\stLab[i]}{\tyGround[i]}{\gtG[i]}}{\roleP}{\roleQ}}
\end{array}\)}\\[1mm]
$\gtG$ is \emph{weak balanced} iff
$\forall\gtG\,\gtMoveStar\,\gtGi.\;
\weakComm{\gtGi}{\roleP}{\roleQ}
\implies$
$\gtDepth{\gtGi}{\roleP}$ and $\gtDepth{\gtGi}{\roleQ}$ exist.
\end{definition}

Intuitively,
$\weakComm{\gtG}{\roleP}{\roleQ}$
means that, from the perspective of $\roleP$ and $\roleQ$,
$\roleP$ can send to $\roleQ$.
\inferrule{WComm} says that
$\roleP$ and $\roleQ$ think that $\roleP$ can send to $\roleQ$
in $\gtComm{\roleP}{\roleQ}{i\in I}{\stLab[i]}{\tyGround[i]}{\gtG[i]}$.
\inferrule{WCtx} says that
if
$\roleP$ and $\roleQ$ are not involved with the head,
then they think that communications,
that may occur in the continuations,
may occur.
\inferrule{WPar} handles parallel composition and \inferrule{W$\mu$} handles recursion.

\begin{example}[Weak Balanced]
\label{ex:weakbalanced}
	$\gtG[\text{\tiny unbal}]$
	from \cref{ex:non-bal}
	is weak balanced.
	$\gtDepth{\gtG[\text{\tiny unbal}]}{\roleFmt{c}}$
	does not need to exist
	as only $\weakComm{\gtG[\text{\tiny unbal}]}{\roleFmt{a}}{\roleFmt{b}}$
	holds.
\end{example}

The standard inductive global type semantics, $\gtMove[\stEnvAnnotGenericSym]$ (\cref{def:gtype:lts-gt}),
and coinductive global type semantics, $\gtMoveCo[\stEnvAnnotGenericSym]$
\ifextended (\cref{def:gtype-coind}),
\else \appref{Def. B.15},
\fi
differ,
causing critical differences when using coinductive techniques.
Earlier work used balancedness to remedy this,
but weak balancedness is sufficient.

\begin{restatable}[Semantics]{proposition}{propWeakBal}
	\label{prop:weak-bal}
	If $\gtG$ is weak balanced, then
	$
	\gtG\,\gtMove[\stEnvAnnotGenericSym]\,\gtGi
	\iff
	\gtG\,\gtMoveCo[\stEnvAnnotGenericSym]\,\gtGi
	$.
\end{restatable}

This fact is crucial for projections:
all projectable global types are weak balanced.

\begin{restatable}[Weak Balanced]{proposition}{propProjWeakBal}
\label{prop:weak-bal-nec}
	If $\gtProj{}{}{}$ is good and
	$\gtProj{}{\gtG}{\stEnv}$, then
	$\gtG$ is weak balanced.
\end{restatable}

\begin{example}[Not Weak Balanced]
\label{ex:not-weak-bal}
The global type
$
\gtG =\gtRec{\gtRecVar}{
	\gtCommRaw{\roleP}{\roleQ}{
	\gtLab\gtSeq\gtRecVar,
	\;
	\gtLabi\gtSeq\gtCommSingle{\roleR}{\roleS}{\gtLab}{}{}
	}
	}
$
is not weak balanced,
since $\weakComm{\gtG}{\roleR}{\roleS}$
and $\gtDepth{\gtG}{\roleR}$
does not exist.
Under inductive semantics,
any projection
of $\gtG$ onto $\roleR$ and $\roleS$
would have a communication,
which cannot occur in the global type.
\end{example}

We define global analogues to
$\stEnvSendLiveRolePred{\roleP}$
and
$\stEnvRecvLiveRolePred{\roleP}$.

\begin{definition}[Send/Receive Balanced Global Types]
\label{def:send-recv-bal}
	Define $\gtSendRoles{}$ and $\gtRecvRoles{}$ similarly to $\gtRoles{}$ except
	on communication constructors:
	$
	\gtSendRoles{\gtComm{\roleP}{\roleQ}{i\in I}{\stLab[i]}{\tyGround[i]}{\gtG[i]}}
	=
	\{\roleP\}\cup\bigcup_{i\in I}\gtSendRoles{\gtG[i]}\setminus\{\roleQ\}
	$ and
	$
	\gtRecvRoles{\gtComm{\roleP}{\roleQ}{i\in I}{\stLab[i]}{\tyGround[i]}{\gtG[i]}}
	=
	\{\roleQ\}\cup\bigcup_{i\in I}\gtRecvRoles{\gtG[i]}\setminus\{\roleP\}
	$.
	
	$\gtG$ is:
	\begin{enumerate*}
	\item \emph{$\roleP$-send balanced} iff
	for all $\gtG\,\gtMoveStar\,\gtGi$,
	if $\roleP\in\gtSendRoles{\gtGi}$ then
	$\gtDepth{\gtGi}{\roleP}$ exists;
	\item \emph{$\roleP$-receive balanced} iff
	for all $\gtG\,\gtMoveStar\,\gtGi$,
	if $\roleP\in\gtRecvRoles{\gtGi}$ then
	$\gtDepth{\gtGi}{\roleP}$ exists; and
	\emph{send/receive balanced} iff
	for all $\roleP$,
	$\gtG$ is $\roleP$-send/receive balanced.
	\end{enumerate*}
\end{definition}

\begin{example}[Send Balanced]
	Consider a modification of the Odd-Even protocol (\cref{ex:running})
	where the guess is returned to $\roleP$:
	$\gtGsbal =
	\gtCommRaw{\roleP}{\roleQ}{
		\gtMsgFmt{o}{\gtSeq}\gtGsbal[\text{\tiny o}],\,
		\gtMsgFmt{m}{\gtSeq}\gtGsbal[\text{\tiny e}]
	}$ where\\
	\centerline{\(
	\gtGsbal[\text{\tiny o}] = \gtRec{\gtRecVar}{
				\gtCommSingle{\roleQ}{\roleR}{\gtMsgFmt{o}}{}{
				\gtCommRaw{\roleQ}{\roleR}{
					\gtMsgFmt{o}{\gtSeq}\gtRecVar,\;
					\gtMsgFmt{b}{\gtSeq}\gtCommSingle{\roleR}{\roleP}{\gtMsgFmt{o}}{}{}
				}
			}}
	\qquad
	\gtGsbal[\text{\tiny e}] = \gtRec{\gtRecVar}{
				\gtCommRaw{\roleQ}{\roleR}{
					\gtMsgFmt{o}{\gtSeq}\gtCommSingle{\roleQ}{\roleR}{\gtMsgFmt{o}}{}{\gtRecVar},\,
					\gtMsgFmt{b}{\gtSeq}\gtCommSingle{\roleR}{\roleP}{\gtMsgFmt{m}}{}{}
				}
			}
	\)}
	$\gtGsbal$ is send balanced
	but not $\roleP$-receive balanced.
	In $\gtGsbal[\text{\tiny o}]$ and $\gtGsbal[\text{\tiny e}]$,
	the message $\roleR$ to $\roleP$
	can be indefinitely delayed
	\ie
	$\roleP\in\gtRecvRoles{\gtGsbal[\text{\tiny o}]}$ and
	$\roleP\in\gtRecvRoles{\gtGsbal[\text{\tiny e}]}$ but
	neither $\gtDepth{\gtGsbal[\text{\tiny o}]}{\roleP}$
	nor $\gtDepth{\gtGsbal[\text{\tiny e}]}{\roleP}$
	exist.
\end{example}

\begin{restatable}[Send/Receive Balanced]{proposition}{propSendRecvBalImplyWeak}
	Send or receive balanced implies
	weak balanced.
\end{restatable}

The \emph{size} of a type, written $|\gtG|$,
is the number of constructors that it contains.

\begin{restatable}[Deciding Balanced]{theorem}{thmDecideBal}
\label{thm:dec-bal}
	We can decide if $\gtG$ is:
	\begin{enumerate*}
		\item balanced in time $\mathcal{O}\!\left(|\gtG|\cdot|\gtRoles{\gtG}|\right)$;
		\item weak balanced in time $\mathcal{O}\!\left(|\gtG|\cdot|\gtRoles{\gtG}|^2\right)$; and
		\item $\roleP$-send/receive balanced in time
		$\mathcal{O}\!\left(|\gtG|\right)$
	\end{enumerate*}
\end{restatable}

\begin{restatable}[Projecting Balanced]{theorem}{thmProjBal}
	\label{thm:proj-bal}
	Suppose that $\gtProj{}{}{}$ is good
	and $\stEnvAssoc{\gtG}{\stEnv}{}$.
	If $\gtG$ is:
	\begin{enumerate*}
		\item \label{bal-env1} balanced, then $\stEnvLiveP{\stEnv}$;
		\item \label{bal-env2} $\roleP$-send balanced, then $\stEnvSendLiveRoleP{\roleP}{\stEnv}$; and
		\item \label{bal-env3} $\roleP$-receive balanced, then $\stEnvRecvLiveRoleP{\roleP}{\stEnv}$.
	\end{enumerate*}
\end{restatable}

\begin{restatable}[Liveness]{theorem}{thmTypeBal}
	\label{thm:type-bal}
	Suppose $\gtProj{}{}{}$ in $\inferrule{Sess}$ in \cref{def:type-rules}
	is a good projection and
	$\tyJudge{}{\M}{\gtG}$.
	If $\gtG$ is:
	\begin{enumerate*}
		\item balanced, then $\M$ is live; and
		\item $\roleP$-send/receive balanced, then $\M$ is $\roleP$-send/receive live.
	\end{enumerate*}
\end{restatable}

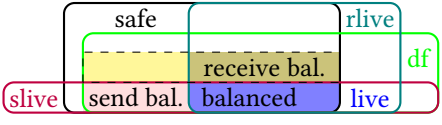
\begin{figure}
\centering{
    \begin{tikzpicture}[baseline={(0,1.8cm)}]

\fill[pink!50] (0.6cm, 1.2cm) rectangle (3cm, 1.6cm);
\fill[white, rounded corners] (2cm,1.2cm) rectangle (4cm,2.65cm);

\fill[blue!50, rounded corners] (2cm,1.2cm) rectangle (4cm,2.65cm);
\fill[white!50, rounded corners] (2cm,1.6cm) rectangle (4cm,2.65cm);

\fill[olive!50] (1.7, 1.6cm) rectangle (4cm, 2cm);

\fill[yellow!50] (0.6cm,1.6cm) rectangle (2cm,2cm);

\draw[rounded corners, name path = safe, thick] (0.35, 1.2cm) rectangle (4cm, 2.65cm);

\draw[rounded corners, green,name path = df, thick] (0.6, 1.2cm) rectangle (5.3cm,2.25cm);

\draw[rounded corners, purple, name path = slive, thick] (-0.45, 1.2cm) rectangle (5.3cm, 1.6cm);

\draw[rounded corners,teal,name path = rlive, thick] (4.8cm, 2.65cm) rectangle (2cm, 1.2cm);

\draw[dashed, name path = wbal] (0.6,1.6cm) rectangle (4cm,2cm);

\node at (1.3cm,2.45cm) {$\stEnvSafePred$};
\node at (5.05cm,1.925cm) {\color{green}$\stEnvDFPred$};
\node at (-0.05cm,1.4cm) {\color{purple}$\stEnvSendLivePred$};
\node at (4.4cm,2.45cm) {\color{teal}$\stEnvRecvLivePred$};
\node at (4.4cm,1.4cm) {\color{blue}$\stEnvLivePred$};
\node at (2.8cm,1.4cm) {balanced};

\node at (3cm,1.8cm) {receive bal.};

\node at (1.3cm,1.4cm) {send bal.};

\end{tikzpicture}
\hfill%
 \begin{minipage}{\dimexpr 0.55\linewidth-\columnsep}%
    \captionsetup{singlelinecheck=off, skip=0pt, justification=raggedright}%
        \caption{Typing context 
and global type properties.
Boxes correspond to typing context properties 
({\color{blue} live}$=${\color{purple} slive}$\cap${\color{teal} rlive}). 
The entire shaded area corresponds to weak balanced. 
Examples for each region are given in
\ifextended \cref{ex:liveness-fig}.
\else \appref{Ex. B.19}.
\fi
}
\label{fig:liveness}
    \end{minipage}}
\end{figure}

\section{Expressiveness and Separation}
\label{sec:example}
This section: introduces merge operators,
which are used to construct inductive and coinductive projections;
shows that inductive and coinductive projections are good;
and proves separation results between
various inductive and coinductive projections.

\subsection{Inductive and Coinductive Projections}
\label{sec:indcoind}
\emph{Merge operations} are functions from finite and non-empty sets of local types to
sets of local types, denoted by $\stMerge{}{\mathcal{T}}$.
Having fixed a merge operation,
we say that a set $\mathcal{T}$ is mergeable if $\stMerge{}{\mathcal{T}}$ is non-empty.
In practice,
merges are
\emph{greatest lower bound} operators
that satisfy some
\emph{algebraic and operational properties}.
We show that, assuming a list of merge properties (\cref{def:good-merge}),
projections constructed using merges are good.
\begin{definition}[Good Merge]
\label{def:good-merge}
	Let $\stMerge{}{}$ be a merge operation.
    We say that $\stMerge{}{}$ is a
    \emph{good merge} if the following conditions hold:
	\begin{enumerate}[leftmargin=0.5in,labelindent=-\leftmargin]
		\item[\inferrule{\mergeGoodSubI}]
        $\stMerge{}{\{\stT[i]\}_{i\in I}}\ni \stT
        \implies
        \forall j\in I.\; \stT \tySub \stT[j]$
		\item[\inferrule{\mergeGoodSubII}]
        $\left(\stMerge{}{\{\stT[i]\}_{i\in I}}\ni \stT
        \;\wedge\;
        \forall i\in I.\;\stTi\tySub \stT[i]\right)
        \implies
        \stTi \tySub \stT$
		\item[\inferrule{\mergeGoodAssoc}]
        $\forall j\in J.\;\stMerge{}{\{\stT[i,j]\}_{i\in I}}\ni\stT[j]
        \implies
        \stMerge{}{\{\stT[j]\}_{j\in J}}=\stMerge{}{\{\stT[i,j]\}_{i\in I, j\in J}}$
        \item[\inferrule{\mergeGoodSubset}]
        $\{\stT[i]\}_{i\in I}\text{ is mergeable}
        \implies
        \forall \emptyset\neq J\subseteq I.\;
        \{\stT[i]\}_{i\in J}\text{ is mergeable}$
        \item[\inferrule{\mergeGoodSubst}]
        $%
        \stMerge{}{\{\stT[i]\}_{i\in I}}\ni \stT
        \implies
        \stMerge{}{\{\stT[i][\stTi/\stRecVar]\}_{i\in I}}\ni \stT{}[\stTi/\stRecVar]$
        \item[\inferrule{\mergeGoodCtx}]
        $\forall i\in I.\;\stMerge{}{\{\stT[j,i]\}_{j\in J}}\ni\stTi[i]
        \implies
        \stMerge{}{\{
        \stFmt{
        \roleP{\dagger}\!\left\{\stChoice{\stLab[i]}{\tyGround[i]}\stSeq\stT[j,i]\right\}_{i\in I}
        }
        \}_{j\in J}}
        \ni
        \stFmt{
        \roleP{\dagger}\!\left\{\stChoice{\stLab[i]}{\tyGround[i]}\stSeq\stTi[i]\right\}_{i\in I}
        }
        $
		\item[\inferrule{\mergeGoodRedI}]
        $\left(\forall i\in I.\;\stT[i]\,\gtMove[\stEnvAnnotGenericSym]\,\stTi[i]
        \;\wedge\;
        \stMerge{}{\{\stT[i]\}_{i\in I}}\ni\stT\right)
        \implies
        \exists \stTi.\;
        \stT\,\gtMove[\stEnvAnnotGenericSym]\,\stTi
        \;\wedge\;
        \stMerge{}{\{\stTi[i]\}_{i\in I}}\ni \stTi$
		\item[\inferrule{\mergeGoodRedII}]
        $\left(\stMerge{}{\{\stT[i]\}_{i\in I}}\ni \stT
        \;\wedge\;
        \stT\,\gtMove[\stEnvAnnotGenericSym]\,\stTi\right)
        \implies
        \exists \emptyset\neq J\subseteq I.
        \;\forall i\in J.\;\stT[i]\,\gtMove[\stEnvAnnotGenericSym]\,\stTi[i]
        \;\wedge\;
        \stMerge{}{\{\stTi[i]\}_{i\in J}}\ni\stTi$
	\end{enumerate}
\end{definition}
\inferrule{\mergeGoodSubI} and \inferrule{\mergeGoodSubII}
say that merges are greatest lower bounds.
\inferrule{\mergeGoodAssoc}, \inferrule{\mergeGoodSubset},
\inferrule{\mergeGoodSubst}, and \inferrule{\mergeGoodCtx}
are the algebraic properties, saying that:
associativity holds;
mergeability is closed under subsets;
merges are closed under substitution; and
merges are closed under adding type constructors.
\inferrule{\mergeGoodRedI} and \inferrule{\mergeGoodRedII}
are the operational properties, saying that
there is an operational correspondence between merges and their components.

\noindent
The plain/full inductive/coinductive merges are provided below:

\begin{definition}[Plain Inductive Merge {\cite[Def. 4.1]{HYC08}}]
\label{def:plainindmerge}
We define $\stMerge{}{\{\stT\}}=\{\stT\}$.
\end{definition}

\begin{definition}[Full Inductive Merge {\cite[Def. 3.1]{ParameterisedYDBH12}}]
\label{def:fullindmerge}
We define $\funcMerge{}$
to be a partial function from non-empty and finite sets of
local types to a local type recursively by:\\[1mm]
$\funcMerge{\left\{
		\stEnd
	\right\}}=\stEnd$;
$\funcMerge{\left\{
		\stRecVar
	\right\}}=\stRecVar$;
$\funcMerge{\left\{
		\stRec{\stRecVar}{\stT[i]}\suchthat i\in I
	\right\}}=\stRec{\stRecVar}{\funcMerge{\left\{\stT[i]\suchthat i\in I\right\}}}$;\\
$\funcMerge{\left\{
		\stExtSum{\roleP}{j \in J_i}{\stChoice{\stLab[j]}{\tyGround[j]} \stSeq \stT[i,j]} \suchthat i\in I
	\right\}} =\stExtSum{\roleP}{j \in \bigcup_{i\in I}J_i}{\stChoice{\stLab[j]}{\tyGround[j]} \stSeq
	\funcMerge{\left\{ \stT[i,j] \suchthat j\in J_i \right\} } }$; and\\
	$\funcMerge{\left\{
		\stIntSum{\roleP}{j \in J}{\stChoice{\stLab[j]}{\tyGround[j]} \stSeq \stT[i,j]} \suchthat i \in I 
	\right\}} =\stIntSum{\roleP}{j \in J}{\stChoice{\stLab[j]}{\tyGround[j]} \stSeq 
	\funcMerge{ \left\{ \stT[i,j] \suchthat i\in I \right\} } } $.\\[1mm]
We define \emph{full inductive merge}
by $\stMerge{}{\mathcal{T}}=\{\stT\suchthat \stT=\funcMerge{\mathcal{T}}\}$.
\end{definition}
\begin{definition}[Plain Coinductive Merge {\cite[Fig. 1]{DBLP:journals/jar/TiroreBC25}}]
\label{def:plain-merge}
We define
$\stMerge{}{\{\stT[i]\}_{i\in I}}=\{\stT\suchthat \forall i\in I,\stT\bisim\stT[i]\}$.
\end{definition}
\begin{definition}[Full Coinductive Merge {\cite[Def. 4.23]{DBLP:journals/pacmpl/UdomsrirungruangY25}}]
\label{def:local-type-merge}
\label{def:full-merge}
Define $\relMerge{}{}$ to be the largest relation from non-empty finite sets of local types
to closed local types such that $\relMerge{\{\stT[i]\}_{i\in I}}{\stT}$ implies:
\begin{itemize}[leftmargin=0.4in,labelindent=-\leftmargin]
    \item[\inferrule{$\mergeRuleEnd$}]
        $\unfoldOne{\stT}=\stEnd
        \implies
        \forall i\in I.\;\unfoldOne{\stT[i]}=\stEnd$
    \item[\inferrule{$\mergeRuleOut$}]
        $\unfoldOne{\stT}=\stIntSum{\roleQ}{j \in J}{\stChoice{\stLab[j]}{\tyGround[j]} \stSeq \stTi[j]}
        \implies
        \begin{array}{l}
        \exists \{\stT[i,j]\}_{i\in I,j\in J}
        \forall i\in I.\;
        \unfoldOne{\stT[i]}=\stIntSum{\roleQ}{j \in J}{\stChoice{\stLab[j]}{\tyGround[j]} \stSeq \stT[i,j]}\;\wedge\\
        \forall j\in J.\;
        \relMerge{\left\{\stT[i,j]\right\}_{i\in I}}{\stTi[j]}
        \end{array}$
    \item[\inferrule{$\mergeRuleIn$}]
        $\unfoldOne{\stT} = \stExtSum{\roleP}{j \in J}{\stChoice{\stLab[j]}{\tyGround[j]}\stSeq \stTi[j]}
        \implies
        \begin{array}{l}
        \exists \{\stT[i,j]\}_{i\in I,j\in J_i}
        \forall i\in I.\;
        \unfoldOne{\stT[i]}=\stExtSum{\roleP}{j \in J_{i}}{\stChoice{\stLab[j]}{\tyGround[j]} \stSeq \stT[i,j]}\;
        \wedge\\
        J=\bigcup_{i\in I}J_{i}
        \wedge
        \forall j\in J.\;
        \relMerge{\left\{\stT[i,j]\suchthat j\in J_{i} \right\}}{\stTi[j]}
        \end{array}$
\end{itemize}
We define \emph{full coinductive merge} by
$\stMerge{}{\mathcal{T}}=\left\{\stT\suchthat \relMerge{\mathcal{T}}{\stT}\right\}$.
\end{definition}
\inferrule{$\mergeRuleEnd$} ensures
that $\stEnd$-types are only compatible with $\stEnd$-types;
\inferrule{$\mergeRuleOut$} send types are compatible with send types
with the same choices and compatible continuations; and
\inferrule{$\mergeRuleIn$} receive types are compatible with receive types
where continuations with the same labels are compatible.

\begin{restatable}[Standard Merges]{theorem}{thmMergeGood}
	\label{thm:merge-good}
	The full/plain inductive/coinductive
	merges are good.
\end{restatable}

Different merges construct different projections.
They are compared in \cref{sec:comparison}.

The inductive \citep[Definition~5]{DBLP:conf/icdcit/YoshidaG20}
and coinductive \citep{DBLP:journals/jar/TiroreBC25}
constructions are defined
in~\cref{def:global-proj}: 

\begin{definition}[Inductive and Coinductive Projections]
\label{def:global-proj}
\label{def:global-ind-proj}
Let $\stMerge{}{}$ be a good merge.

The \emph{inductive projection} is the smallest relation between
	participants, global types and local types such that $\gtProj{\roleR}{\gtG}{\stT}$ is implied by:
\begin{itemize}[leftmargin=0.4in,labelindent=-\leftmargin]
    \item[\inferrule{$\projRuleEnd$}] 
        $\roleR\notin\gtRoles{\gtG}\;\wedge\;\gtFv{\gtG}=\emptyset\;\wedge\;\stT=\stEnd$ \qquad \quad \inferrule{$\projRuleVar$} \quad 
        	$\gtG=\gtRecVar\;\wedge\;\stT=\stRecVar$
	\item[\inferrule{$\projRuleMu$}] 
		$\gtG=\gtRec{\gtRecVar}{\gtGi}\;\wedge\;\stT=\stRec{\stRecVar}{\stTi}
		\;\wedge\;\left(\roleR\in\gtRoles{\gtGi}\vee\gtFv{\gtG}\neq\emptyset\right)
		\;\wedge\;\gtProj{\roleR}{\gtGi\subst{\gtRecVar}{\gtG}}{\stTi\subst{\stRecVar}{\stT}}$
    \item[\inferrule{$\projRuleIn$}] 
        $\gtG=\gtComm{\roleP}{\roleR}{i \in I}{\gtLab[i]}{\tyGround[i]}{\gtG[i]}\;\wedge\;\stT=\stExtSum{\roleP}{i \in I}{\stChoice{\stLab[i]}{\tyGround[i]} \stSeq \stT[i]}\;\wedge\;\forall i\in I.\;\gtProj{\roleR}{\gtG[i]}{\stT[i]}$
    \item[\inferrule{$\projRuleOut$}] 
        $\gtG=\gtComm{\roleR}{\roleQ}{i \in I}{\gtLab[i]}{\tyGround[i]}{\gtG[i]}\;\wedge\;\stT=\stIntSum{\roleQ}{i \in I}{\stChoice{\stLab[i]}{\tyGround[i]} \stSeq \stT[i]}\;\wedge\;\forall i\in I.\;\gtProj{\roleR}{\gtG[i]}{\stT[i]}$
	\item[\inferrule{$\projRulePar$}]  $\gtG=\gtPar{\gtG[0]}{\gtG[1]}\;\wedge\;\exists i\in\{0,1\}.\;\roleR\in\gtRoles{\gtG[i]}\wedge\gtProj{\roleR}{\gtG[i]}{\stT}$
    \item[\inferrule{$\projRuleCtx$}] 
    $
    \gtG=
    \gtComm{\roleP}{\roleQ}{i \in I}{\gtLab[i]}{\tyGround[i]}{\gtG[i]}
    \;\wedge\;
        \exists \{\stT[i]\}_{i\in I}.\;\gtProj{\roleR}{\gtG[i]}{\stT[i]}\;\wedge\;\stT\in\stMerge{}{\{\stT[i]\}_{i\in I}}
    $
\end{itemize}
The \emph{coinductive projection} is the largest relation between
	participants, closed global types and closed local types such that $\gtProj{\roleR}{\gtG}{\stT}$ implies:
\begin{itemize}[leftmargin=0.4in,labelindent=-\leftmargin]
    \item[\inferrule{$\projRuleEnd$}] 
        $\roleR\notin\gtRoles{\gtG}\implies\unfoldOne{\stT}=\stEnd$
    \item[\inferrule{$\projRuleIn$}]
        $\unfoldOne{\gtG}=\gtComm{\roleP}{\roleR}{i \in I}{\gtLab[i]}{\tyGround[i]}{\gtG[i]}\implies\unfoldOne{\stT}=\stExtSum{\roleP}{i \in I}{\stChoice{\stLab[i]}{\tyGround[i]} \stSeq \stT[i]}\;\wedge\;\forall i\in I.\;\gtProj{\roleR}{\gtG[i]}{\stT[i]}$
    \item[\inferrule{$\projRuleOut$}] 
        $\unfoldOne{\gtG}=\gtComm{\roleR}{\roleQ}{i \in I}{\gtLab[i]}{\tyGround[i]}{\gtG[i]}\implies\unfoldOne{\stT}=\stIntSum{\roleQ}{i \in I}{\stChoice{\stLab[i]}{\tyGround[i]} \stSeq \stT[i]}\;\wedge\;\forall i\in I.\;\gtProj{\roleR}{\gtG[i]}{\stT[i]}$
        
	\item[\inferrule{$\projRulePar$}] $\unfoldOne{\gtG}=\gtPar{\gtG[0]}{\gtG[1]}\implies\exists i\in\{0,1\}\;\gtProj{\roleR}{\gtG[i]}{\stT}\;\wedge\;\roleR\notin\gtRoles{\gtG[i-1]}$
    \item[\inferrule{$\projRuleCtx$}] 
    $
\unfoldOne{\gtG}=
\gtComm{\roleP}{\roleQ}{i\in I}{\gtLab[i]}{\tyGround[i]}{\gtG[i]}\wedge
\roleR\!\in\!\gtRoles{\gtG}\setminus\{\roleP,\roleQ\}
    \Longrightarrow 
\exists \{\stT[i]\}_{i\in I}.\;\gtProj{\roleR}{\gtG[i]}{\stT[i]}\wedge
\stT\!\in\!\stMerge{}{\{\stT[i]\}_{i\in I}}
    $
\end{itemize}
\end{definition}
\noindent
\inferrule{$\projRuleEnd$} ensures that a non-present participant
cannot have non-trivial behaviour;
\inferrule{$\projRuleIn$} deals with receiving;
\inferrule{$\projRuleOut$} sending;
\inferrule{$\projRulePar$} parallel composition; and
\inferrule{$\projRuleCtx$} allows projection
for participants independent of the head communication,
given that the continuation projections are compatible via the merge.
\inferrule{$\projRuleVar$} and \inferrule{$\projRuleMu$}
are specific to the inductive projection and deal
with recursive types,
which the coinductive projection handles by unfoldings.

\begin{example}[Full Coinductive Projection]\ \\
\label{ex:full-co-proj}
\centerline{\(%
		\gtG=\gtRec{\gtRecVari}{
			\gtCommRaw{\roleP}{\roleQ}{
					\gtCommChoice{\gtLab}{}{ \gtGi },\,
					\gtCommChoice{\gtLabi}{}{\gtRecVari}
			}
		}\ \,
		\gtGi=\gtRec{\gtRecVar}{
				\gtCommRaw{\roleP}{\roleQ}{
						\gtCommChoice{\gtLab}{}{\gtRecVar},\,
						\gtCommChoice{\gtLabi}{}{\gtRecVar}
				}
			}
\quad
		\stT[\roleP]=\stRec{\stRecVar}{
						\stIntSum{\roleQ}{}{
								\stChoice{\stLab}{}\stSeq\stRecVar,\,
								\stChoice{\stLabi}{}\stSeq\stRecVar
						}
					}\ \,
		\stT[\roleQ]=\stRec{\stRecVar}{
						\stExtSum{\roleP}{}{
								\stChoice{\stLab}{}\stSeq\stRecVar,\,
								\stChoice{\stLabi}{}\stSeq\stRecVar
						}
					}
\)}\\[1mm]
We derive $\stMerge{}{\{\stT[\roleP]\}}\ni\stT[\roleP]$ and
$\stMerge{}{\{\stT[\roleQ]\}}\ni\stT[\roleQ]$
as the relation given by $\mathcal{R}\langle\{\stT\},\stT\rangle$
satisfies the rules for $\relMerge{}{}$. 
We derive $\gtProj{\roleP}{\gtG}{\stT[\roleP]}$ and $\gtProj{\roleQ}{\gtG}{\stT[\roleQ]}$
as the relations: 
$\{(\gtGi,\stT[\roleP]),(\gtG,\stT[\roleP])\}$ and\\
$\{(\gtGi,\stT[\roleQ]),(\gtG,\stT[\roleQ])\}$
satisfy $\gtProj{\roleP}{}{}$
and $\gtProj{\roleQ}{}{}$, respectively.
	
Recall $\gtG[\text{\tiny o/e}]$,
$\gtGorigin[\text{\tiny o/e}]$,
and $\stEnv[\text{\tiny o/e}]$
from~\cref{ex:running}.
$\gtG[\text{\tiny o/e}]$ projects onto $\stEnv[\text{\tiny o/e}]$
by both the full inductive and full coinductive projections.
$\gtGorigin[\text{\tiny o/e}]$ is not projectable using
the full inductive projection, but
$\gtGorigin[\text{\tiny o/e}]$ satisfies 
$\gtProj{}{\gtGorigin[\text{\tiny o/e}]}{\stEnv[\text{\tiny o/e}]}$.
However, the projection from~\cref{ex:proj-running}
is not a subrelation of the full coinductive projection,
as some local types have been given additional branches.
\end{example}

\begin{restatable}[Balanced]{proposition}{propProjBalIndCo}
\label{prop:proj-bal-ind-co}
	\begin{enumerate}
	\item
	If $\gtProj{}{\gtG}{\stEnv}$ by inductive $\gtProj{}{}{}$, then $\gtG$ is balanced.
	\item
	Let $\gtProj{}{}{}$ be the plain coinductive projection.
	There is projectable $\gtG$
	that is not weak balanced.
	\end{enumerate}
\end{restatable}

In light of \cref{prop:proj-bal-ind-co} and \cref{prop:weak-bal-nec},
we formulate \cref{thm:good-inductive-proj}.

\begin{restatable}[Inductive and Coinductive Projections]{theorem}{thmProjGood}
\label{thm:good-inductive-proj}
\label{thm:good-coinductive-proj}
Inductive projections are good and
coinductive projections, restricted to weak balanced global types, are good.
\end{restatable}

By \cref{prop:good-robust,thm:good-inductive-proj,thm:merge-good},
we conclude that the eight standard projections are good.

\begin{restatable}[Standard Projections]{theorem}{thmProjStandard}
\label{thm:good-proj-standard}
Using plain or full merge
and global types with or without parallel composition,
the inductive projection and coinductive projection,
restricted to weak balanced global types,
are good.
\end{restatable}

\begin{remark}[Odd-Even]
\label{rem:sendbalancedOddEven}
The original Odd-Even protocol, $\gtGorigin[\text{\tiny{o/e}}]$,
(\Cref{ex:running})
is not projectable by the inductive projection but 
is projectable by the full coinductive projection.  
\end{remark}

\subsection{Constructing Counterexamples}
\label{sec:counter}
The good merge conditions (\cref{def:good-merge})
are key elements in existing correctness proofs.
Failures of \cref{def:good-merge}
inform the construction of counterexamples
to type correctness.
We have identified four distinct works,
\cite{GJPSY2018,Barbanera2021-mg,doi:10.3233/FI-2019-1863,10.1145/3798256},
and two types of merge,
\emph{1-level}
and
\emph{disjoint},
where
\inferrule{\mergeGoodCtx}
fails.
Using the proof of \cref{thm:good-coinductive-proj}
as a guide,
we construct a global type,
typing context
and session,
where
completeness of association
(\cref{thm:completeness_association})
and
subject reduction fail.

\begin{definition}[1-Level Merge ({\cite[Def.~3.3]{10.1145/3798256}\cite[Def.~3.6]{GJPSY2018}})]
\label{def:1-level-merge}
	\emph{Coinductive 1-level merge} is defined by:
	\\[1mm]
	\centerline{\(\small
	\stMerge{}{\{\stT[i]\}_{i\in I}}=\left\{
		\begin{array}{ll}
			\{\stT\suchthat
			\unfoldOne{\stT}
			=
			\stExtSum{\roleP}{j\in \bigcup_{i\in I} J_i}{\stChoice{\stLab[j]}{\tyGround[j]}\stSeq\stTii[j]}
			\}
			&
			\forall i\in I.\;
			\unfoldOne{\stT[i]}=\stExtSum{\roleP}{j\in J_i}{\stChoice{\stLab[j]}{\tyGround[j]}\stSeq\stTi[i,j]}
			\;\wedge\;
			\stTi[i,j]\bisim\stTii[j]
			\\
			\{\stT\suchthat\forall i\in I.\;\stT\bisim\stT[i]\}&
			\text{otherwise}
		\end{array}
	\right.
	\)}\\[1mm]
	\emph{Inductive 1-level merge} is defined by:
	\\[1mm]
	\centerline{\(\small
	\stMerge{}{\{\stT[i]\}_{i\in I}}=\left\{
		\begin{array}{ll}
			\{\stT\suchthat
			\stT
			=
			\stRec{\stRecVar[1]}{\dots\stRec{\stRecVar[n]}{\stExtSum{\roleP}{j\in \bigcup_{i\in I} J_i}{\stChoice{\stLab[j]}{\tyGround[j]}\stSeq\stTi[j]}}}
			\}
			&
			\forall i\in I.\;
			\stT[i]=\stRec{\stRecVar[1]}{\dots\stRec{\stRecVar[n]}{\stExtSum{\roleP}{j\in J_i}{\stChoice{\stLab[j]}{\tyGround[j]}\stSeq\stTi[j]}}}
			\\
			\{\stT\suchthat\forall i\in I.\;\stT=\stT[i]\}&
			\text{otherwise}
		\end{array}
	\right.
	\)}
\end{definition}
Combining external choices can only be done at the first `level'
they occur.
The coinductive merge considers type up-to their unfoldings and
the inductive merge does not.

\begin{definition}[Disjoint Merge ({\cite[Def.~3.4]{Barbanera2021-mg}\cite[Appendix~E.2]{doi:10.3233/FI-2019-1863}})]
\label{def:disjoint-merge}
	\emph{Coinductive disjoint merge} is defined by:
	\\[1mm]
	\centerline{\(\small
	\stMerge{}{\{\stT[i]\}_{i\in I}}=\left\{
		\begin{array}{ll}
			\{\stT\suchthat
			\unfoldOne{\stT}
			=
			\stExtSum{\roleP}{j\in \bigcup_{i\in I} J_i}{\stChoice{\stLab[j]}{\tyGround[j]}\stSeq\stTii[j]}
			\}
			&
			\begin{array}{l}
			\forall i\in I.\;
			\unfoldOne{\stT[i]}=\stExtSum{\roleP}{j\in J_i}{\stChoice{\stLab[j]}{\tyGround[j]}\stSeq\stTi[j]}
			\\
			\forall i\neq i' \in I.\;
			J_i\cap J_{i'} =\emptyset
			\;\wedge\;
			\stTi[j]\bisim\stTii[j]
			\end{array}
			\\
			\{\stT\suchthat\forall i\in I.\;\stT\bisim\stT[i]\}&
			\text{otherwise}
		\end{array}
	\right.
	\)}\\[1mm]
	\emph{Inductive disjoint merge} is defined by:
	\\[1mm]
	\centerline{\(\small
	\stMerge{}{\{\stT[i]\}_{i\in I}}=\left\{
		\begin{array}{ll}
			\{\stT\suchthat
			\stT
			=
			\stRec{\stRecVar[1]}{\dots\stRec{\stRecVar[n]}{\stExtSum{\roleP}{j\in \bigcup_{i\in I} J_i}{\stChoice{\stLab[j]}{\tyGround[j]}\stSeq\stTi[j]}}}
			\}
			&
			\begin{array}{l}
			\forall i\in I.\;
			\stT[i]=\stRec{\stRecVar[1]}{\dots\stRec{\stRecVar[n]}{\stExtSum{\roleP}{j\in J_i}{\stChoice{\stLab[j]}{\tyGround[j]}\stSeq\stTi[j]}}}
\\
			\forall i\neq i' \in I.\;
			J_i\cap J_{i'}=\emptyset
			\end{array}
			\\
			\{\stT\suchthat\forall i\in I.\;\stT=\stT[i]\}&
			\text{otherwise}
		\end{array}
	\right.
	\)}
\end{definition}
Distinct sets of choices can only be combined if their labels
are disjoint. Since merged types must be well-formed,
the compatibility of continuations in the coinductive cases
are implied.

\begin{restatable}[Bad Merges]{proposition}{propMergesNotGood}
	The merge operations in
	\cref{def:1-level-merge,def:disjoint-merge}
	do not satisfy \inferrule{\mergeGoodCtx}.
\end{restatable}

\begin{proof}
	Let $\stMerge{}{}$ be a merge in
	\cref{def:1-level-merge} or \cref{def:disjoint-merge}.
	Let:
		$\stT[0] = \stFmt{
			\roleFmt{\roleP[0]}{\&}\stLab[0]
		}$;
		$\stT[1] = \stFmt{
			\roleP[0]{\&}\stLab[1]
		}$;
		$\stT = \stFmt{\roleP[0]{\&}\!\left\{\stLab[i]\right\}_{i\in \{0,1\}}}$; and
		$\stTi[i] = \stFmt{\roleP[0]{\&}\stLab\stSeq\stT[i]}$.
		Thus, $\stMerge{}{\{\stT[0],\stT[1]\}}\ni\stT$
	and $\stMerge{}{\{\stTi[0],\stTi[1]\}}=\emptyset$.
\end{proof}

This allows us to construct a global type violating
subject reduction.
Let:\\[1mm]
	\centerline{\(\gtGcounter =
	\gtCommRaw{\roleP[0]}{\roleQ[0]}{
		\gtLab[i]{\gtSeq}
		\gtCommSingle{\roleP[0]}{\roleQ[1]}{\gtLab}{}{
			\gtCommRaw{\roleP[2]}{\roleQ[2]}{
					\gtLab[j]
					{\gtSeq}
					\gtCommSingle{\roleP[2]}{\roleP[0]}{\gtLab[j]}{}{
					\gtCommSingle{\roleP[0]}{\roleQ[1]}{\gtLab[i\,xor\,j]}{}{}
					}
			}_{j\in \{0,1\}}
		}
	}_{i\in \{0,1\}}\)}\\[1mm]
$\gtGcounter$ is projectable by
the inductive projection with the inductive disjoint merge
(hence all other projections using these merges).
This construction
generalises
\ifextended (\cref{thm:failure-type}).
\else \appref{Thm. C.23}.
\fi

\begin{restatable}[Incorrect Merges]{theorem}{thmIncorrectMerges}
\label{thm:incorrect-merge}
Let $\gtProj{}{}{}$ use a merge in
\cref{def:1-level-merge}/\cref{def:disjoint-merge}.
\begin{enumerate*}
	\item $\exists\gtGcounter\,\gtMove\,\gtGcounter[1].$
	$\gtGcounter[1]$ is not projectable;
	\item association is neither sound nor complete; and
	\item there is $\M$ such that
	$\tyJudge{}{\M}{\gtGcounter}$ and
	$\M\mpMove\M'$ but
	$\forall\gtGcounter\,\gtMoveStar\,\gtGcounter[1].
	\neg(\tyJudge{}{\M'}{\gtGcounter[1]})$
	\ie subject reduction fails.
\end{enumerate*}
\end{restatable}

\begin{proof}
	$\gtGcounter
	\,\gtMove[\ltsSendRecv{\roleP[2]}{\roleQ[2]}{\stLab[0]}]\,
	\gtGcounter[1]=
	\gtCommRaw{\roleP[0]}{\roleQ[0]}{
		\gtLab[i]{\gtSeq}
		\gtCommSingle{\roleP[0]}{\roleQ[1]}{\gtLab}{}{
					\gtCommSingle{\roleP[2]}{\roleP[0]}{\gtLab[0]}{}{
					\gtCommSingle{\roleP[0]}{\roleQ[1]}{\gtLab[i]}{}{}
					}
		}
	}_{i\in \{0,1\}}$\\
	$\gtGcounter[1]$ does not project onto $\roleQ[1]$,
	since
	$\{\stTi[0],\stTi[1]\}$
	is not mergeable.
	(2) and (3) follow.
\end{proof}

Although the presentations in \cref{def:1-level-merge,def:disjoint-merge}
differ slightly from \cite{GJPSY2018,Barbanera2021-mg,doi:10.3233/FI-2019-1863,10.1145/3798256},
they are the same when restricted to our common syntax,
and our synchronous transitions are performed in two asynchronous transitions in
\cite{10.1145/3798256}.
Thus, \cref{thm:incorrect-merge} holds regardless
of the variations between these papers.
Importantly, these papers are easily corrected by using full merging.

\subsection{Expressiveness and Separation Results}
\label{sec:comparison}
The purpose of types is to type programs,
guaranteeing correctness properties for as many programs as possible.
With a notion of correctness for projections,
we
ask about
\emph{program separation}
and
\emph{protocol separation}.
To this end,
we introduce two candidate notions for expressiveness: local and global.
These correspond to sessions and global protocols respectively.
\begin{definition}[Expressiveness]
\label{def:expressiveness}
	We define \emph{global and local expressiveness maps},
	$\projGlobal{}$ and $\projLocal{}$,
	by:
	\\[1mm]\centerline{\(
		\projGlobal{\gtProj{}{}{}}=
		\left\{
		\gtG\suchthat
		\exists\gtGi,\stEnv\text{ s.t. }
		\gtG\bisim\gtGi\wedge\gtProj{}{\gtGi}{\stEnv}
		\right\}\;\;
		\projLocal{\gtProj{}{}{}}=
		\left\{
		\stEnv\suchthat
		\exists\gtG,\stEnvi.\;
		\gtProj{}{\gtG}{\stEnvi}
		\wedge
		\stEnv\tySub\stEnvi
		\right\}
	\)}\\[1mm]
	We say that:
\begin{itemize}        
\item	$\gtProj{}{}{}$ can \emph{express} $\stEnv$ iff
	  $\stEnvNew\in\projLocal{\gtProj{}{}{}}$;
	$\gtProj{}{}{}$ can \emph{express} $\gtG$ iff
	  $\gtG\in\projGlobal{\gtProj{}{}{}}$;
\item	$\gtProj{}{}{}$ and $\gtProji{}{}{}$ are \emph{equally expressive} iff
	$\projGlobal{\gtProj{}{}{}}=\projGlobal{\gtProji{}{}{}}$ and
	  $\projLocal{\gtProj{}{}{}}=\projLocal{\gtProji{}{}{}}$;
\item	  $\gtG$ \emph{separates} $\gtProj{}{}{}$ from $\gtProji{}{}{}$ iff
  $\gtG\in\projGlobal{\gtProj{}{}{}}\setminus\projGlobal{\gtProji{}{}{}}$; and
\item  $\stEnvNew$ \emph{separates} $\gtProj{}{}{}$ from $\gtProji{}{}{}$ iff
  $\stEnvNew\in\projLocal{\gtProj{}{}{}}\setminus\projLocal{\gtProji{}{}{}}$.
\end{itemize}  
\end{definition}
$\projGlobal{\gtProj{}{}{}}$
is the set of global types
that are \emph{projectable up-to bisimilarity}.
Separation and expressivity are less interesting
if they are based on sets of projectable global types,
as we can simply check whether a projection algorithm
succeeds or fails.
$\projLocal{\gtProj{}{}{}}$ is the set of typing contexts
that $\gtProj{}{}{}$ associates,
corresponding
to typable sessions
when using $\gtProj{}{}{}$ in \inferrule{Sess}
in \cref{def:type-rules}.
We want our typing systems to be as powerful as possible;
comparing local expressivenesses facilitates this.

\begin{restatable}[Separation]{theorem}{thmSessionSeparation}
	\label{thm:app:strong-process-sep}
	Let $\tyJudge{}{\N}{\gtG}$
	use $\gtProj{}{}{}$
	and
	$\tyJudgei{}{\N}{\gtG}$
	use
	$\gtProji{}{}{}$
	in \inferrule{Sess} in \cref{def:type-rules}.
	\begin{enumerate}
	\item
	If $\stEnv$ separates $\gtProj{}{}{}$
	from $\gtProji{}{}{}$,
	then
	there is $\N$ such that
	$\exists\gtG.\;\tyJudge{}{\N}{\gtG}$ and
	$\not\exists\gtG.\;\tyJudgei{}{\N}{\gtG}$.
	\item
	If $\projLocal{\gtProj{}{}{}}=\projLocal{\gtProji{}{}{}}$,
	then for all $\N$,
	$
	\exists\gtG.\;\tyJudge{}{\N}{\gtG}
	\iff
	\exists\gtG.\;\tyJudgei{}{\N}{\gtG}
	$.
	\end{enumerate}
\end{restatable}

\begin{example}[Expressing Odd-Even]
Recall $\gtG[\text{\tiny o/e}]$
and $\gtGorigin[\text{\tiny o/e}]$ from~\cref{ex:running}.
$\gtGorigin[\text{\tiny o/e}]$ is not projectable by the full inductive projection,
but it is expressible as $\gtG[\text{\tiny o/e}]\bisim\gtGorigin[\text{\tiny o/e}]$
and $\gtG[\text{\tiny o/e}]$ is projectable by the full inductive
projection. Hence, projectability is stricter than
expressibility.
\end{example}
\cref{fig:proj_strength}
characterises the separation and expressiveness results
between
plain/full
inductive/coinductive
with/without parallel compositions; and shows 
separations between 
inductive projections without parallel composition,
inductive projections with parallel composition
and 
the
coinductive projections.
The middle column in \cref{fig:global-express}
gives examples of global types that separate areas
of the diagram identified by the left column,
proved below in \cref{thm:separating-projections},
and the right column gives the time complexities
of the projection algorithms.
Unless parallel composition is specifically included,
it is excluded from the domain of projection.
We restrict the coinductive projections to balanced global types.

\newcounter{ExampleCounter}
\newcommand{\exlab}[1]{\refstepcounter{ExampleCounter}\arabic{ExampleCounter}\label{#1}}
\newcommand*\circled[1]{\tikz[baseline=(char.base)]{
    \node[shape=circle,draw,inner sep=1pt] (char) {#1};}}

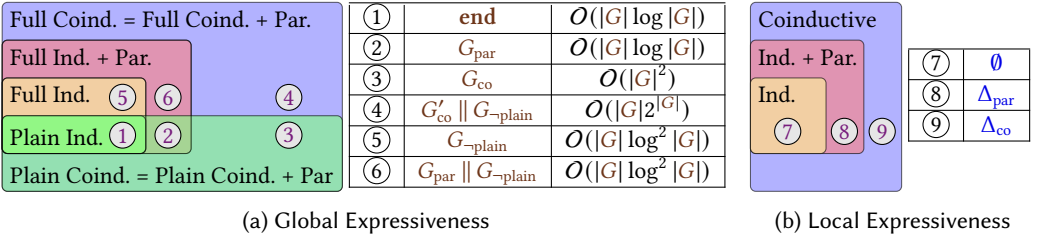
\begin{figure}
	\small
\begin{subfigure}{0.7\textwidth}
\centering{
    \begin{tikzpicture}[baseline={(0,0.7cm)}]
	\newcommand{\vennnode}[6]{\node (#1) at (#2, #4) [minimum width=(#3)-(#2), minimum height=(#5)-(#4), venn=#6, anchor=south west] {}}
	
	\vennnode{fullcoinductive}{0cm}{4.5cm}{-0.5cm}{2cm}{blue!50};
	\node[below right] at (fullcoinductive.north west) {Full Coind. = Full Coind. + Par.};
	\node[exnode] at (3.8cm, 0.75cm) {\ref{venn:fullcoind}};

	\vennnode{fullindpar}{0cm}{2.5cm}{0cm}{1.5cm}{red!50};
	\node[below right] at (fullindpar.north west) {Full Ind. + Par.};
	\node[exnode] at (2.2cm, 0.75cm) {\ref{venn:fullindpar}};

	\vennnode{fullind}{0cm}{1.9cm}{0cm}{1cm}{yellow!50};
	\node[below right] at (fullind.north west) {Full Ind.};
	\node[exnode] at (1.6cm, 0.75cm) {\ref{venn:fullind}};

	\vennnode{plaincoind}{0cm}{4.5cm}{-0.5cm}{0.5cm}{green!50};
	\node[above right] at (plaincoind.south west) {Plain Coind. = Plain Coind. + Par};

	\vennnode{plainind}{0cm}{1.9cm}{0cm}{0.5cm}{green!50};
	\node[above right] at (plainind.south west) {Plain Ind.};
	\node[exnode] at (1.6cm, 0.25cm) {\ref{venn:plainind}};
	\node[exnode] at (2.2cm, 0.25cm) {\ref{venn:plainindpar}};
	\node[exnode] at (3.8cm, 0.25cm) {\ref{venn:plaincoind}};
	
	\vennnode{}{0cm}{2.5cm}{0cm}{1.5cm}{transparent};
	\vennnode{}{0cm}{1.9cm}{0cm}{1cm}{transparent};
\end{tikzpicture}
$
\begin{array}{|c|c|c|}
\hline
\circled{\exlab{venn:plainind}} & \gtEnd & \mathcal{O}(|\gtG|\log{|\gtG|})\\
\hline
\circled{\exlab{venn:plainindpar}} & \gtG[\text{\tiny par}]&\mathcal{O}(|\gtG|\log{|\gtG|})\\
\hline
\circled{\exlab{venn:plaincoind}} & \gtG[\text{\tiny co}]&\mathcal{O}({|\gtG|}^2)\\
\hline
\circled{\exlab{venn:fullcoind}} & \gtPar{\gtGi[\text{\tiny co}]}{\gtG[{\neg\text{\tiny plain}}]}&\mathcal{O}(|\gtG|2^{|\gtG|})\\
\hline
\circled{\exlab{venn:fullind}} & \gtG[\neg\text{\tiny plain}]&\mathcal{O}(|\gtG|\log^2{|\gtG|})\\
\hline
\circled{\exlab{venn:fullindpar}} & \gtPar{\gtG[\text{\tiny par}]}{\gtG[\neg\text{\tiny plain}]}&\mathcal{O}(|\gtG|\log^2{|\gtG|})\\
\hline
\end{array}$
}
\caption{Global Expressiveness}
\label{fig:global-express}
\end{subfigure}
~
\begin{subfigure}{0.29\textwidth}
\centering{
    \begin{tikzpicture}[baseline={(0,0.7cm)}]
	\newcommand{\vennnode}[6]{\node (#1) at (#2, #4) [minimum width=(#3)-(#2), minimum height=(#5)-(#4), venn=#6, anchor=south west] {}}

	\vennnode{coinductive}{0cm}{2cm}{-0.5cm}{2cm}{blue!50};
	\node[below right] at (coinductive.north west) {Coinductive};
	\node[exnode] at (1.75cm, 0.3cm) {\ref{venn:coinductive}};

	\vennnode{indpar}{0cm}{1.5cm}{0cm}{1.5cm}{red!50};
	\node[below right] at (indpar.north west) {Ind. + Par.};
	\node[exnode] at (1.25cm, 0.3cm) {\ref{venn:indpar}};

	\vennnode{ind}{0cm}{1cm}{0cm}{1cm}{yellow!50};
	\node[below right] at (fullind.north west) {Ind.};
	\node[exnode] at (0.5cm, 0.3cm) {\ref{venn:ind}};
\end{tikzpicture}
{$
\begin{array}{|c|c|}
\hline
\circled{\exlab{venn:ind}} & \stEnvEmpty\\
\hline
\circled{\exlab{venn:indpar}} & \stEnv[\text{\tiny par}]\\
\hline
\circled{\exlab{venn:coinductive}} & \stEnv[\text{\tiny co}]\\
\hline
\end{array}
$}
}
\caption{Local Expressiveness}
\end{subfigure}
\caption{Relative Expressiveness of Projections and their
Time Complexities.
Coinductive projections are restricted to balanced global types.
Projections exclude parallel composition
except where stated.
}\label{fig:proj_strength}
\end{figure}

\begin{restatable}[Separating Projections]{theorem}{thmSeparatingProjections}
\label{thm:separating-projections}
	The separations in \cref{fig:proj_strength} hold.
\end{restatable}

\begin{proof}
	$
		\gtG[\neg\text{\tiny plain}] = \gtCommRaw{\roleP}{\roleQ}{
			\gtCommChoice{m}{}{ 
				\gtCommSingle{\roleQ}{\roleR}{\gtLab}{}{}
			 },\,
			\gtCommChoice{m'}{}{ 
				\gtCommSingle{\roleQ}{\roleR}{\gtLabi}{}{}
			 }
	}
	\quad
	\gtG[\text{\tiny par}] = \gtPar
		{
			\gtRec{\gtRecVar}
			{
				\gtCommRaw{\roleFmt{a}}{\roleFmt{b}}
				{
						\gtCommChoice{m}{}{\gtRecVar},\,
						\gtCommChoice{m'}{}{}
				}
			}
		}
		{
			\gtRec{\gtRecVar}
			{
				\gtCommRaw{\roleFmt{c}}{\roleFmt{d}}
				{
						\gtCommChoice{m}{}{\gtRecVar},\,
						\gtCommChoice{m'}{}{}
				}
			}
		}
		$
		\\
		\centerline{\(\gtG[\text{\tiny co}]=
\gtRec{\gtRecVar}
{
    \gtCommRaw{\roleFmt{a}}{\roleFmt{b}}
    {
    \begin{array}{l}
        \gtCommChoice{m}{}
        {
            \gtCommRaw{\roleFmt{c}}{\roleFmt{d}}
            {
                \gtCommChoice{m}{}
                {
                    \gtCommSingle{\roleFmt{a}}{\roleFmt{c}}{m}{}
                    {
                        \gtRecVar
                    }
                },\,
                \gtCommChoice{m'}{}
                {
                    \gtRec{\gtRecVar}
                    {
                        \gtCommSingle{\roleFmt{a}}{\roleFmt{c}}{m}{}
                        {
                            \gtCommRaw{\roleFmt{a}}{\roleFmt{b}}{
                            \gtCommChoice{m}{}
                                {
                                    \gtRecVar
                                },\,
                                \gtCommChoice{m'}{}
                                {
                                    \gtRec{\gtRecVar}
                                    {
                                        \gtCommSingle{\roleFmt{a}}{\roleFmt{c}}{m}{}
                                        {
                                            \gtRecVar
                                        }
                                    }
                                }
                            }
                        }
                    }
                }
            }
        }
        \\
        \gtCommChoice{m'}{}
        {
            \gtRec{\gtRecVar}
            {
                \gtCommRaw{\roleFmt{c}}{\roleFmt{d}}
                {
                    \gtCommChoice{m}{}
                    {
                        \gtCommSingle{\roleFmt{a}}{\roleFmt{c}}{m}{}
                        {
                            \gtRecVar
                        }
                    },\,
                    \gtCommChoice{m'}{}
                    {
                        \gtRec{\gtRecVar}
                        {
                            \gtCommSingle{\roleFmt{a}}{\roleFmt{c}}{m}{}
                            {
                                \gtRecVar
                            }
                        }
                    }
                }
            }
        }
    \end{array}
    }
}
	\)}
	Plain projections impose a restriction
	on the trace of projectable global types:
	$\gtG[\neg\text{\tiny plain}]\,
	\gtMove[\ltsSendRecv{\roleP}{\roleQ}{\stLab}\,\ltsSendRecv{\roleQ}{\roleR}{\stLab}]
	$ and
	$\gtG[\neg\text{\tiny plain}]\,
	\gtMove[\ltsSendRecv{\roleP}{\roleQ}{\stLabi}\,\ltsSendRecv{\roleQ}{\roleR}{\stLabi}]
	$
	and
	$\gtG[\neg\text{\tiny plain}]\,
	\,\,\,\,\,\,\,\,\not\!\!\!\!\!\!\!\!
	\gtMove[\ltsSendRecv{\roleP}{\roleQ}{\stLab}\,\ltsSendRecv{\roleQ}{\roleR}{\stLabi}]
	$,
	which cannot hold for any projectable global type.

	Suppose that $\gtG[\text{\tiny par}]$ has a projectable bisimilar type without parallel composition.
	When projecting onto $\roleFmt{c}$ we must merge
	a branch, where $\roleFmt{a}$ has finished,
	with a branch, where $\roleFmt{a}$ will continue after
	unfolding a recursive variable.
	This recursive variable cannot appear in both branches
	so they cannot be inductively mergeable.

To obtain $\gtG[\text{\tiny co}]$,
interlace the parallel composition in $\gtG[\text{\tiny par}]$
and add synchronisation between $\roleFmt{a}$ and
$\roleFmt{c}$ so that no parallel composition
can occur in a bisimilar type (excluding composition with $\gtEnd$-types).
If $\gtG[\text{\tiny co}]$ is expressible by the full inductive projection, then
we can remove the synchronisation both locally and globally
to observe that $\gtG[\text{\tiny par}]$ can be expressed by
the full inductive projection without parallel composition.
Let $\gtGi[\text{\tiny co}] = \gtCommRaw{\roleFmt{a}}{\roleFmt{b}}{
		\gtCommChoice{m}{}{\gtEnd},\,
		\gtCommChoice{m'}{}{\gtG[\text{\tiny co}]}
	}$
so if $\gtG\bisim\gtPar{\gtGi[\text{\tiny co}]}{\gtG[\neg\text{\tiny plain}]}$
then $\gtG\,\gtMoveStar\,\gtGi,\gtGii$
where $\gtGi\bisim\gtG[\text{\tiny co}]$
and $\gtGii\bisim\gtG[\neg\text{\tiny plain}]$.\\
\centerline{\(
	\begin{array}{lcl}
	\stEnv[\text{\tiny par}] &=&\set{
	\stEnvMap{\roleFmt{a}}{\stRec{\stRecVar}{\stIntSum{\roleFmt{b}}{}{
		\stLab\stSeq\stRecVar,\, \stLabi
	}}}
	\stEnvComp \stEnvMap{\roleFmt{b}}{\stRec{\stRecVar}{\stExtSum{\roleFmt{a}}{}{
		\stLab\stSeq\stRecVar,\, \stLabi
	}}}
	\stEnvComp \stEnvMap{\roleFmt{c}}{\stRec{\stRecVar}{\stIntSum{\roleFmt{d}}{}{
		\stLab\stSeq\stRecVar,\, \stLabi
	}}}
	\stEnvComp \stEnvMap{\roleFmt{d}}{\stRec{\stRecVar}{\stExtSum{\roleFmt{c}}{}{
		\stLab\stSeq\stRecVar,\, \stLabi
	}}}}
	\\
	\stEnv[\text{\tiny co}]&=&\set{
	\begin{array}{l}
	\stEnvMap{\roleFmt{a}}{\stRec{\stRecVar}{\stIntSum{\roleFmt{b}}{}{
		\stLab\stSeq \roleFmt{c}{\oplus}\stLab\stSeq\stRecVar,\,
		\stLabi\stSeq\stRec{\stRecVar}{\roleFmt{c}{\oplus}\stLab\stSeq\stRecVar}
	}}}
	\stEnvComp \stEnvMap{\roleFmt{b}}{\stRec{\stRecVar}{\stExtSum{\roleFmt{a}}{}{
		\stLab\stSeq\stRecVar,\, \stLabi
	}}}
	\stEnvComp
	\\
	\stEnvMap{\roleFmt{c}}{\stRec{\stRecVar}{\stIntSum{\roleFmt{d}}{}{
		\stLab\stSeq\roleFmt{a}{\&}\stLab\stSeq\stRecVar,\, \stLabi\stSeq\stRec{\stRecVar}{\roleFmt{a}{\&}\stLab\stSeq\stRecVar}
	}}}
	\stEnvComp \stEnvMap{\roleFmt{d}}{\stRec{\stRecVar}{\stExtSum{\roleFmt{c}}{}{
		\stLab\stSeq\stRecVar,\, \stLabi
	}}}
	\end{array}}
	\end{array}
\)}
We have that $\gtProj{}{\gtG[\text{\tiny par}]}{\stEnv[\text{\tiny par}]}$
by the plain inductive projection with parallel composition
and $\gtProj{}{\gtG[\text{\tiny co}]}{\stEnv[\text{\tiny co}]}$
by the plain coinductive projection without parallel composition.
All of these projections guarantee that if $\gtProj{}{\gtG}{\stEnv}$,
then $\gtTrace{\gtG}=\gtTrace{\stEnv}$.
The plain projections have the property that if $\gtProj{}{\gtG}{\stEnv}$,
$\stEnv\tySub\stEnvi$ and
$\stEnvSafeP{\stEnvi}$, then
$\stEnvi\tySub\stEnv$.
Therefore, $\stEnv[\text{\tiny par}]$
separates the plain inductive projection with parallel composition
from the plain inductive projection without parallel composition, and
$\stEnv[\text{\tiny co}]$
separates the plain coinductive projection without parallel composition
from the plain inductive projection with parallel composition.
\end{proof}

Full/plain coinductive projections
	with and without parallel
	are equally expressive.
        \begin{restatable}[Removing Parallels]{theorem}{thmRemovePar}
	\label{thm:remPar}
	Let $\gtProj{}{}{}$ be the full or plain coinductive projection.
	There is a computable function, $\gtSerial{}$,
	from (balanced) global types to global types without parallel composition
	such that
	$\gtProj{}{\gtG}{\stEnv}\implies\gtProj{}{\gtSerial{\gtG}}{\stEnv}$.
	The worst case output size
	is bounded by $\mathcal{O}(^{|\gtG|}b)$,
	where $b\geq 3$ is the maximum number of choices
	appearing in a transmission in $\gtG$.
\end{restatable}

        \begin{example}[Serialisation]\label{ex:serialisation}
\ \\[1mm]          
\centerline{$
		\gtSerial{\gtG[\text{\tiny par}]}=
		\gtRec{\gtRecVar}
{
    \gtCommRaw{\roleFmt{a}}{\roleFmt{b}}
    {
        \gtCommChoice{m}{}
        {
            \gtCommRaw{\roleFmt{c}}{\roleFmt{d}}
            {
                \gtCommChoice{m}{}{\gtRecVar},\,
                \gtCommChoice{m'}{}
                {
                    \gtRec{\gtRecVari}
                    {
                            \gtCommRaw{\roleFmt{a}}{\roleFmt{b}}{
                                \gtCommChoice{m}{}
                                {
                                    \gtRecVari
                                },\,
                                \gtCommChoice{m'}{}{}
                            }
                    }
                }
            }
        },\,
        \gtCommChoice{m'}{}
        {
            \gtRec{\gtRecVarii}
            {
	            \gtCommRaw{\roleFmt{c}}{\roleFmt{d}}{\gtCommChoice{m}{}{\gtRecVarii},\,
	            \gtCommChoice{m'}{}{}}
			}
		}
}}$\\[1mm]}
The communications of the components are alternated between,
	looping as soon as possible.
\end{example}

\begin{remark}[Time Complexity \cite{DBLP:journals/pacmpl/UdomsrirungruangY25}]
  \label{rem:projection-time-complexity}
  	As shown in \cref{fig:proj_strength}, 
	the more expressive projections
	have higher time complexities.
	Full coinductive, while the most expressive,
	is intractable.
	Plain coinductive is tractable and,
	using subtyping,
	can type the same sessions
	as full coinductive.
	Removing parallel composition from global types 
	leads to exponential blow-up.
	Thus, using the plain coinductive projection with parallel 
	is both tractable and expressive w.r.t. typability.
\end{remark}

\section{Mixed Choice Global Types}
\label{sec:extended}
This section proposes 
good projections on 
directed mixed choice.
\subsection{Mixed Choice Types and Processes}
Mixed choice session types have general
choice syntax, allowing
multiple actions
within the same choice constructor.
We replace the global transmission syntax and local choice syntax (\cref{sec:syntax}) with:
\\\centerline{\(
	\gtCommSD{\roleP[i]}{\roleQ[i]}{i\in I}{\gtLab[i]}{\tyGround[i]}{\gtG[i]}
\quad\text{and}\quad
	\stSum{\roleP[i]}{i\in I}{\stChoice{\stLab[i]}{\tyGround[i]}\stSeq\stT[i]}{\dagger_i}
\)}
Globally, for
exactly one $i\in I$,
$\roleP[i]$ sends message $\stChoice{\stLab[i]}{\tyGround[i]}$
to $\roleQ[i]$.
Locally, we allow
sends and receives to/from different participants.
We use the subtyping in \cite[Def. 3.2]{PY2024}
\ifextended (\cref{def:subtyping-mixed}).
\else \appref{Def. D.6}.
\fi
\begin{definition}[Mixed Choice Type Semantics]
\label{def:extended-types}
	We define the semantics of mixed choice types inductively by
	extending the rules in \cref{def:gtype:lts-gt,def:ltype:lts-lt} with:
	\\\centerline{\(\begin{array}{c}
	\inference[\iruleGtMoveComm']{j \in I}{
    \gtCommSD{\roleP[i]}{\roleQ[i]}{i \in I}{\gtLab[i]}{\tyGround[i]}{\gtG[i]}
    \,\gtMove[\ltsSendRecv{\roleP[j]}{\roleQ[j]}{\stChoice{\stLab[j]}{\tyGround[j]}}]\, 
    \gtG[j]
  }
  \quad
  \inference[\iruleLtSum]{j\in I}{
 \stSum{\roleP[i]}{i\in I}{\stChoice{\stLab[i]}{\tyGround[i]}\stSeq\stT[i]}{\dagger_i}
 \gtMove[\stFmt{\roleP[j]{\dagger}\stChoice{\stLab[j]}{\tyGround[j]}}]
 \stT[j]}
  \\
  \inference[\iruleGtMoveCtx']{
    \forall i \in I.\;
    \gtG[i] 
    \,\gtMove[\stEnvAnnotGenericSym]\, 
    \gtGi[i] 
    &
    \ltsSubject{\stEnvAnnotGenericSym} \cap \set{\roleP[i], \roleQ[i]} =
    \emptyset
  }{
      \gtCommSD{\roleP[i]}{\roleQ[i]}{i \in I}{\gtLab[i]}{\tyGround[i]}{\gtG[i]}
   \,\gtMove[\stEnvAnnotGenericSym]\, 
      \gtCommSD{\roleP[i]}{\roleQ[i]}{i \in I}{\gtLab[i]}{\tyGround[i]}{\gtGi[i]}   
  }
 \end{array}
 \)}
\end{definition}
\begin{definition}[Mixed Choice Processes \cite{PY2024}]
\label{def:mixed-processes}
Define prefixes by: $\pi\bnfdef \roleQ\mathord{!}\ell\langle e \rangle \bnfsep
\roleQ\mathord{?}\ell( x )$.
Extend processes with mixed choice syntax:
$\textstyle\sum_{i\in I} \pi_i \mpSeq \mpP_i$.
Extend $\stJudge{}{\mpP}{\stT}$ by the
rule \inferrule{T$\textstyle\sum$} in \cite[Fig. 4]{PY2024}.
\end{definition}

\begin{restatable}[Properties of Multiparty Sessions \proofapp{\cref{app:extended}}]{theorem}{corTypingSessionsMixed}%
  \label{cor:typingsessionsmixed}
Let $\gtProj{}{}{}$ in $\inferrule{Sess}$ in \cref{def:type-rules}
be a good projection defined in \cref{def:good-proj}.
Suppose $\tyJudge{}{\M}{\gtG}$,
then $\M$ satisfies the subject reduction theorem, 
safety, session fidelity and deadlock-freedom.
If $\gtG$ is balanced/$\roleP$-send balanced/$\roleP$-receive balanced,
then $\M$ is live/$\roleP$-send live/$\roleP$-receive live.
\end{restatable}
\subsection{Directed Mixed Choice}
\label{sec:directed-choice}
We extend our projections to
\emph{directed mixed choice global types}
\ie global types
where each transmission contains exactly two participants:
$\gtCommSD{\roleP[i]}{\roleQ[i]}{i\in I}{\gtLab[i]}{\tyGround[i]}{\gtG[i]}$
where $|\bigcup_{i\in I}\{\roleP[i],\roleQ[i]\}|=2$.

\begin{definition}[Full Coinductive Merge]
\label{def:full-merge-directed}
We define $\relMerge{}{}$ to be the largest relation between non-empty finite sets of local types
and closed local types such that if $\relMerge{\{\stT[i]\}_{i\in I}}{\stT}$, then
the rules in \cref{def:full-merge} and \inferrule{$\mergeRuleComm$'} hold:
$\unfoldOne{\stT}=
		\stSum{\roleQ}{j\in J}{\stChoice{\stLab[j]}{\tyGround[j]}\stSeq\stT[j]}{\oplus}
		+
		\stSum{\roleQ}{j\in J'}{\stChoice{\stLabi[j]}{\tyGroundi[j]}\stSeq\stTi[j]}{\&}
		\implies$
		\\
		\centerline{\(\begin{array}{c}
\forall i \in I.\;
		\unfoldOne{\stT[i]}=
		\stSum{\roleQ}{j\in J}{\stChoice{\stLab[j]}{\tyGround[j]}\stSeq\stT[i,j]}{\oplus}
		+
		\stSum{\roleQ}{j\in J'_i}{\stChoice{\stLabi[j]}{\tyGroundi[j]}\stSeq\stTi[i,j]}{\&}
		\;\wedge\;
		J'=\bigcup_{i\in I}J'_i
		\;\wedge\\
		\forall j \in J.\;
		\relMerge{\{\stT[i,j]\}_{i\in I}}{\stT[j]}
		\;\wedge\;
		\forall j \in J'.\;
        \relMerge{\left\{\stTi[i,j]\suchthat j\in J'_{i} \right\}}{\stTi[j]}
        \end{array}\)}
We define \emph{full coinductive merge} by
$\stMerge{}{\mathcal{T}}=\left\{\stT\suchthat \relMerge{\mathcal{T}}{\stT}\right\}$.
\end{definition}

\begin{definition}[Inductive and Coinductive Projections]
\label{def:global-proj-mixed}
\label{def:global-ind-proj-mixed}
Let $\stMerge{}{}$ be a good merge.
The \emph{inductive projection} is the smallest relation between
	participants, global types, and local types such that $\gtProj{\roleR}{\gtG}{\stT}$ is implied by the rules in \cref{def:global-proj}
	and:\\[2mm]
	\centerline{\(\begin{array}{l@{\,}l}
		\inferrule{$\projRuleComm$'}&\ \gtG=
		\gtCommSD{\roleR}{\roleQ}{i\in I}{\gtLab[i]}{\tyGround[i]}{\gtG[i]}
		+
		\gtCommSD{\roleQ}{\roleR}{i\in I'}{\gtLabi[i]}{\tyGroundi[i]}{\gtGi[i]}
		\;\wedge\;
		\forall i\in I.\,
		\gtProj{\roleR}{\gtG[i]}{\stT[i]}
		\;\wedge\\&
		\forall i\in I'.\,
		\gtProj{\roleR}{\gtGi[i]}{\stTi[i]}
		\;\wedge\;
		\stT=
		\stSum{\roleQ}{i\in I}{\stChoice{\stLab[i]}{\tyGround[i]}\stSeq\stT[i]}{\oplus}
		+
		\stSum{\roleQ}{i\in I'}{\stChoice{\stLabi[i]}{\tyGroundi[i]}\stSeq\stTi[i]}{\&}
		\\
		\inferrule{$\projRuleCtx$'}&\ 
		\gtG=\gtCommSD{\roleP[i]}{\roleQ[i]}{i\in I}{\gtLab[i]}{\tyGround[i]}{\gtG[i]}
		\,\wedge\,
		\roleR\notin\bigcup_{i\in I}\{\roleP[i],\roleQ[i]\}
		\,\wedge\,
		\forall i\in I.\,
		\gtProj{\roleR}{\gtG[i]}{\stT[i]}
		\,\wedge\,
		\stT\in\stMerge{}{\{\stT[i]\}_{i\in I}}
		\end{array}
	\)}
	The \emph{coinductive projection} is the largest relation between
	participants, closed global types, and local types such that if $\gtProj{\roleR}{\gtG}{\stT}$ then the rules below and in \cref{def:global-proj} hold:
	\\[1mm]
	\centerline{\(
        \small
        \begin{array}{l@{\ }l}
	\inferrule{$\projRuleComm$'} & 
		\unfoldOne{\gtG}=
		\gtCommSD{\roleR}{\roleQ}{i\in I}{\gtLab[i]}{\tyGround[i]}{\gtG[i]}
		+
		\gtCommSD{\roleQ}{\roleR}{i\in I'}{\gtLabi[i]}{\tyGroundi[i]}{\gtGi[i]}
		\;\implies\\
		& \unfoldOne{\stT}=
		\stSum{\roleQ}{i\in I}{\stChoice{\stLab[i]}{\tyGround[i]}\stSeq\stT[i]}{\oplus}
		+
		\stSum{\roleQ}{i\in I'}{\stChoice{\stLabi[i]}{\tyGroundi[i]}\stSeq\stTi[i]}{\&}
		\wedge
		\forall i\in I.\,
		\gtProj{\roleR}{\gtG[i]}{\stT[i]}
		\wedge
		\forall i\in I'.\,
		\gtProj{\roleR}{\gtGi[i]}{\stTi[i]}
		\\
		\inferrule{$\projRuleCtx$'} & 
		\unfoldOne{\gtG}=\gtCommSD{\roleP[i]}{\roleQ[i]}{i\in I}{\gtLab[i]}{\tyGround[i]}{\gtG[i]}
		\,\wedge\,
		\roleR\notin\bigcup_{i\in I}\{\roleP[i],\roleQ[i]\}	
		\implies
		 \forall i\in I.\,
		\gtProj{\roleR}{\gtG[i]}{\stT[i]}
		\,\wedge\,
		\stT\in\stMerge{}{\{\stT[i]\}_{i\in I}}
		\end{array}
\)}
\end{definition}
\begin{restatable}[Inductive and Coinductive Projections]{theorem}{thmProjGoodMixed}
\label{thm:good-inductive-proj-mixed}
\label{thm:good-coinductive-proj-mixed}
Inductive projections are good and
coinductive projections, restricted to weak balanced global types, are good.
\end{restatable}

\begin{example}[Two Clients]
\label{ex:directed-mixed-two-clients}
Using directed mixed choice,
we express a protocol
where two participants
act as clients and servers
of equal priority
for each other:\\[1mm]
\centerline{
\(
\begin{array}{c}
	\stT[i] =
	\stFmt{
	\stRec{\stRecVar}{
	\Sigma\!\left\{
		\roleP[1-i]{\oplus}\stChoice{\stLabFmt{query}}{\tyString}\stSeq
		\roleP[1-i]{\&}\stChoice{\stLabFmt{res}}{\tyString}\stSeq
		\stRecVar,\,
		\roleP[1-i]{\&}\stChoice{\stLabFmt{query}}{\tyString}\stSeq
		\roleP[1-i]{\oplus}\stChoice{\stLabFmt{res}}{\tyString}\stSeq
		\stRecVar
	\right\}
	}}
	\\
	\gtG[\text{\tiny C}]=\gtRec{\gtRecVar}{
		\gtCommSD{\roleP[i]}{\roleP[1-i]}{i\in\{0,1\}}{\gtMsgFmt{query}}{\tyString}{
		\gtCommSingle{\roleP[1-i]}{\roleP[i]}{\gtMsgFmt{res}}{\tyString}{\gtRecVar}
		}
	}
	\quad
	\stEnv[\text{\tiny C}]=
	\set{
	\stEnvMap{\roleP[0]}{\stT[0]}
	\stEnvComp
	\stEnvMap{\roleP[1]}{\stT[1]}
	}
\end{array}
\)
}\\[1mm]
	Using the full coinductive projection,
	we derive $\gtProj{}{\gtG[\text{\tiny C}]}{\stEnv[\text{\tiny C}]}$.
\end{example}

\begin{example}[Timeout]
\label{ex:directed-mixed-timeout}
	Consider the adapted timeout protocol
	\cite{10.1145/3798256}
	where
	$\roleFmt{A}$ starts some
	communication protocol,
	unless $\roleFmt{B}$ times out.
	Directed mixed choice is necessary to express this.\\[1mm]
	\centerline{\(\begin{array}{c}
		\gtG[\text{\tiny TO}] =
		\gtFmt{\Sigma
		\!\left\{
			\roleFmt{A}\to\roleFmt{B}:\gtMsgFmt{a_1}{\gtSeq}
			\roleFmt{A}\to\roleFmt{C}:\gtMsgFmt{a_2}{\gtSeq}
			\roleFmt{B}\to\roleFmt{C}:\gtMsgFmt{a_3}{\gtSeq}
			\roleFmt{B}\to\roleFmt{A}:\gtMsgFmt{a_4}{\gtSeq}
			\roleFmt{C}\to\roleFmt{A}:\gtMsgFmt{a_5}
			,\,
			\roleFmt{B}\to\roleFmt{A}:\gtMsgFmt{TOa}{\gtSeq}
			\roleFmt{A}\to\roleFmt{C}:\gtMsgFmt{TOc}
		\right\}
		}
	\\
	\begin{array}{r@{\,}c@{\,}lr@{\,}c@{\,}l}
		\stT[\roleFmt{A}]&=&
		\stFmt{
		\Sigma\!\left\{
			\roleFmt{B}{\oplus}\stLabFmt{a_1}\stSeq
			\roleFmt{C}{\oplus}\stLabFmt{a_2}\stSeq
			\roleFmt{B}{\&}\stLabFmt{a_4}\stSeq
			\roleFmt{C}{\&}\stLabFmt{a_5}
			,\,
			\roleFmt{B}{\&}\stLabFmt{TOa}\stSeq
			\roleFmt{C}{\oplus}\stLabFmt{TOc}
		\right\}
		}
		&
		\stT[\roleFmt{B}]&=&
		\stFmt{
		\Sigma\!\left\{
			\roleFmt{A}{\&}\stLabFmt{a_1}\stSeq
			\roleFmt{C}{\oplus}\stLabFmt{a_3}\stSeq
			\roleFmt{A}{\oplus}\stLabFmt{a_4}
			,\,
			\roleFmt{A}{\oplus}\stLabFmt{TOa}
		\right\}
		}
		\\
		\stT[\roleFmt{C}]&=&
		\stFmt{
		\Sigma\!\left\{
			\roleFmt{A}{\&}\stLabFmt{a_2}\stSeq
			\roleFmt{B}{\&}\stLabFmt{a_3}\stSeq
			\roleFmt{A}{\oplus}\stLabFmt{a_5}
			,\,
			\roleFmt{A}{\&}\stLabFmt{TOc}
		\right\}
		}
		&
		\stEnv[\text{\tiny TO}]&=&
		\set{
		\stEnvMap{\roleFmt{A}}{\stT[\roleFmt{A}]}
		\stEnvComp
		\stEnvMap{\roleFmt{B}}{\stT[\roleFmt{B}]}
		\stEnvComp
		\stEnvMap{\roleFmt{C}}{\stT[\roleFmt{C}]}
		}
	\end{array}
	\end{array}\)}\\[1mm]
	Using the full coinductive projection,
	we derive $\gtProj{}{\gtG[\text{\tiny TO}]}{\stEnv[\text{\tiny TO}]}$.
\end{example}

\begin{remark}[Complexity and Separation]
\label{rem:directed-complexity-sep}
The separation results in \cref{sec:comparison}
remain true on directed mixed choice global types;
if $\gtProj{}{\gtG}{\stEnv}$ then
$\stEnv$ uses the extended syntax iff
$\gtG$ uses the extended syntax for
the projections in \cref{def:global-proj-mixed}.
The algorithms to compute the projections in
\cref{def:global-proj-mixed}
require slight modifications to the algorithms
for \cref{def:global-proj}.
Thus the time complexities stay the same.
\end{remark}

\section{Good Projection in Asynchronous Semantics}
\label{sec:async}
This section studies ``good projections'' in asynchronous FIFO semantics,
where messages are non-blocking but ordered. 
We introduce asynchronous semantics of global and local types
and study:
(1) the relative strengths of projection algorithms
and
(2) approximations of the precise asynchronous subtyping.

\subsection{Global and Local Asynchronous Semantics}
\label{sec:asyntax}\
We extend the global type syntax with
\emph{en-route transmissions}~\cite[Figure~4]{pischke2026asynchronousglobalprotocolsprecisely},
denoting
that a message $\gtLab$ with payload of type $\tyGround$
has been sent to $\roleQ$ but has not been received:
$\gtCommSquig{\roleP}{\roleQ}{\gtLab}{}{\gtLab}{\tyGround}{\gtG}$.

We define queues to be maps from 
participants to finite lists of messages;
we write queues as single lists up-to commutativity of participants~\cite{GPPSY2023}.
$\gtARoles{\gtG}$/$\gtQRoles{\gtG}$
\ifextended (\cref{def:aroles})
\else \appref{Def. E.1}
\fi
is 
the set of roles with further actions/roles that have sent en-route transmissions.
They are defined by:
\\[1mm]
\centerline{\(
\begin{array}{r@{\,}c@{\,}l@{\;\;}r@{\,}c@{\,}l}
\gtARoles{\gtComm{\roleP}{\roleQ}{i \in
I}{\gtLab[i]}{\tyGround[i]}{\gtG[i]}} & = & \{\roleP,\roleQ\}\cup\bigcup_{i\in I}\gtARoles{\gtG[i]}&
\gtQRoles{\gtComm{\roleP}{\roleQ}{i \in
I}{\gtLab[i]}{\tyGround[i]}{\gtG[i]}}& =& \bigcup_{i\in I}\gtQRoles{\gtG[i]}
\\
\gtARoles{\gtCommSquig{\roleP}{\roleQ}{\gtLab}{}{\gtLab}{\tyGround}{\gtG}}
& = & \{\roleQ\}\cup\gtARoles{\gtG}&
\gtQRoles{\gtCommSquig{\roleP}{\roleQ}{\gtLab}{}{\gtLab}{\tyGround}{\gtG}}&
= & \{\roleP\}\cup\gtQRoles{\gtG}
\end{array}
\)}\\[1mm]
and similar to the definition of $\gtRoles{}$ elsewhere.

\begin{definition}[Global Type Transitions]
  \label{def:a-gtype:lts-gt}
The asynchronous global type transition
  $\gtMove[\stEnvAnnotGenericSym]$ is inductively
  defined by the rules below
  and \inferrule{\iruleGtMoveRec} and \inferrule{\iruleGtMovePar}
  from \cref{def:gtype:lts-gt}:
\\\centerline{\small\(
\begin{array}{c}
  \inference[\iruleGtMoveBra]{}{
    \gtCommSquigSmall{\roleP}{\roleQ}{\gtLab}{}{\gtLab}{\tyGround}{\gtG}
    \,\gtMove[\ltsBra{\roleQ}{\roleP}{\stChoice{\stLab}{\tyGround}}]\, 
    \gtG
  }\quad
  \inference[\iruleGtMoveSel]{
    j \in I
  }{
    \gtCommSmall{\roleP}{\roleQ}{i \in I}{\gtLab[i]}{\tyGround[i]}{\gtG[i]}
    \,\gtMove[\ltsSel{\roleP}{\roleQ}{\stChoice{\stLab[j]}{\tyGround[j]}}]\, 
    \gtCommSquigSmall{\roleP}{\roleQ}{\gtLab[j]}{}{\gtLab[j]}{\tyGround[j]}{\gtG[j]}
  }
  \\[1ex]
  \inference[\iruleGtMoveCtxI]{
    \forall i \in I.\;
    \gtG[i] 
    \,\gtMove[\stEnvAnnotGenericSym]\, 
    \gtGi[i] 
    &
    \roleP,\roleQ
    \notin
	\ltsSubject{\stEnvAnnotGenericSym}
  }{
      \gtCommSmall{\roleP}{\roleQ}{i \in I}{\gtLab[i]}{\tyGround[i]}{\gtG[i]}
   \,\gtMove[\stEnvAnnotGenericSym]\, 
      \gtCommSmall{\roleP}{\roleQ}{i \in I}{\gtLab[i]}{\tyGround[i]}{\gtGi[i]}   
  }
\;
  \inference[\iruleGtMoveCtxII]{
    \gtG
    \,\gtMove[\stEnvAnnotGenericSym]\, 
    \gtGi
    &
    \roleQ\notin\ltsSubject{
    \stEnvAnnotGenericSym}
  }{
      \gtCommSquigSmall{\roleP}{\roleQ}{\gtLab}{}{\gtLab}{\tyGround}{\gtG}
   \,\gtMove[\stEnvAnnotGenericSym]\, 
      \gtCommSquigSmall{\roleP}{\roleQ}{\gtLab}{}{\gtLab}{\tyGround}{\gtGi}   
  }
\end{array}
\)}
\end{definition}
The transition rules
are designed to mimic asynchronous process reduction by
simulating queues with en-route transmissions.
\inferrule{\iruleGtMoveSel} allows sending asynchronous messages, producing an en-route transmission;
\inferrule{\iruleGtMoveBra} receiving asynchronous messages, allowing
	en-route messages to be received;
\inferrule{\iruleGtMoveCtxI} allows transitions inside communication types so long as the participant
	undertaking the transition does not have specified behaviour at the head; and
  \inferrule{\iruleGtMoveCtxII} allows transmissions under an en-route transmission
  assuming
  that they are independent.

Typing contexts are partial maps from participants to
pairs of queues and session types~\cite{GPPSY2023}.
The subtyping relation extends to typing contexts by
asserting equality of queues.
The queue syntax is defined by:
$
  \quH \bnfdef\quEmpty \! \bnfsep \! \quMsg{\roleQ}{\stLab}{\tyGround} \! \bnfsep \! \quCons{\quH}{\quHi}$
  where $\quEmpty$ denotes the empty queue;
$\quMsg{\roleQ}{\stLab}{\tyGround}$ the queue
containing a single element, a message to $\roleQ$
with label $\stLab$, and payload $\tyGround$;
and the concatenation of
$\quH$ with $\quHi$ is denoted by $\quCons{\quH}{\quHi}$. 
The typing context syntax is defined by: 
$\stEnvNew\bnfdef
  \stEnvEmpty
  \bnfsep
   \stEnvNew \stEnvComp \stEnvMap{\roleP}{(\quH,\stT)}
$. 

\begin{definition}[Queue Projection]
\label{def:queue-projection}
Define the partial function $\gtQu{\gtG}{\roleR}$
recursively:
$\gtQu{\gtEnd}{\roleR}=\quEmpty$;
$\gtQu{\gtRec{\gtRecVar}{\gtG}}{\roleR}=
\gtQu{\gtG}{\roleR}
$ if $\roleR\in\gtQRoles{\gtG}$ else $\quEmpty$;
$\gtQu{\gtComm{\roleP}{\roleQ}{i\in I}{\gtLab[i]}{\tyGround[i]}{\gtG[i]}}{\roleR}=\quH$
if $\forall i\in I.\;\gtQu{\gtG[i]}{\roleR}=\quH$ and
($\roleR=\roleP\implies$ no message in $\quH$ is to $\roleQ$);
$\gtQu{\gtCommSquig{\roleP}{\roleQ}{\gtLab}{}{\gtLab}{\tyGround}{\gtG}}{\roleR}=
\gtQu{\gtG}{\roleR}
$ if $\roleR\neq\roleP$ else
$\quCons{\quMsg{\roleQ}{\stLab}{\tyGround}}{\gtQu{\gtG}{\roleR}}$.
\end{definition}

Queues are extracted from en-route transmissions and
this is preserved under reductions.

\begin{restatable}[Reductions]{proposition}{propQueueReductions}
\label{prop:queue-reductions}
	$\gtQu{\gtG}{\roleR}$ exists and $\gtG\,\gtMove[\stEnvAnnotGenericSym]\,\gtGi$
	imply:
	\begin{enumerate*}
	\item $\stEnvAnnotGenericSym
	=
	\ltsBra{\roleP}{\roleR}{\stChoice{\stLab}{\tyGround}}
	\implies
	\gtQu{\gtG}{\roleR}
	=$\\$
	\quCons{\quMsg{\roleP}{\stLab}{\tyGround}}{\gtQu{\gtGi}{\roleR}}$;
	\item $\stEnvAnnotGenericSym=\ltsSel{\roleR}{\roleP}{\stChoice{\stLab}{\tyGround}}\implies\gtQu{\gtGi}{\roleR}=\quCons{\quMsg{\roleP}{\stLab}{\tyGround}}{\gtQu{\gtG}{\roleR}}$; otherwise
	\item $\gtQu{\gtG}{\roleR}=\gtQu{\gtGi}{\roleR}$
	\end{enumerate*}
\end{restatable}

\subsection{Association and Good Projections}
\label{sec:agood}
We introduce the `asynchronously-good projection' conditions (`a-good'),
analogously to~\cref{def:good-proj},
concluding that the separation results in \cref{sec:comparison}
still hold.
We write
$\gtProj{\roleP}{\gtG}{(\gtQu{\gtG}{\roleP},\stT)}$
if $\gtProj{\roleP}{\gtG}{\stT}$,
and $\gtProj{}{\gtG}{\stEnv}$
if $\dom{\stEnv}\supseteq\gtRoles{\gtG}$
and $\forall \roleP\in\dom{\stEnv}.\;\gtProj{\roleP}{\gtG}{\stEnvApp{\stEnv}{\roleP}}$.

\begin{definition}[Asynchronously Good Projection]
\label{def:as-good-proj}
	Fix a projection relation.
	We say that this is an \emph{asynchronously-good projection} if
	whenever $\gtProj{}{\gtG}{\stEnv}$
	with $\stEnvApp{\stEnv}{\roleP}=(\quH[\roleP],\stT)$:
    \begin{itemize}[leftmargin=0.4in,labelindent=-\leftmargin]
\item[\inferrule{$\projGoodEnd$}]
            $\roleP\notin \gtARoles{\gtG}\implies\unfoldOne{\stT}=\stEnd$
        \item[\inferrule{$\projGoodOut$}]
            $\gtG\,\gtMove[ \ltsSel{\roleP}{\roleQ}{\stChoice{\stLab}{\tyGround}} ]
            \,\gtGi\implies\exists\stTi.\;
            \gtProj{\roleP}{\gtGi}{\stTi}\;\wedge\;
            \stT\,\gtMove[{\roleQ\stEnvAnnotOutSym\stChoice{\stLab}{\tyGround}}]
            \,\cdot\tySub\stTi$
        \item[\inferrule{$\projGoodIn$}]
	$\gtG\,\gtMove[ \ltsBra{\roleP}{\roleQ}{\stChoice{\stLab}{\tyGround}} ]
            \,\gtGi
            \implies
            \exists \stTi.\;
			\stT\,\gtMove[{\roleQ\stEnvAnnotInSym\stChoice{\stLab}{\tyGround}}]
            \,\cdot\tySub\stTi\;\wedge\;
			\gtProj{\roleP}{\gtGi}{\stTi}$
        \item[\inferrule{$\projGoodAlt$}]
            $\gtG\,\gtMove[ \stEnvAnnotGenericSym ]
            \,\gtGi\;\wedge\;\roleP\not\in\ltsSubject{\stEnvAnnotGenericSym}
            \implies\begin{array}{l}
            \exists \stTi.\;
		\gtProj{\roleP}{\gtGi}{\stTi}\;\wedge\;
	\stT \tySub \stTi
            \end{array}$
        \item[\inferrule{$\projGoodType$}]
            $\stT\,\gtMove[{\roleQ\stEnvAnnotOutSym\stChoice{\stLab}{\tyGround}}]
            \implies
            \gtG\,\gtMove[ \ltsSel{\roleP}{\roleQ}{\stChoice{\stLab}{\tyGround}}]$ and
            $\stT\,\gtMove[{\roleQ\stEnvAnnotInSym\stChoice{\stLabi}{\tyGroundi}}]
            \;\wedge\;
            \gtQu{\gtG}{\roleQ}=\quCons{\quMsg{\roleP}{\stLab}{\tyGround}}{\quH}
            \implies
            \gtG\,\gtMove[ \ltsBra{\roleP}{\roleQ}{\stChoice{\stLab}{\tyGround}} ]$
	\end{itemize}
	\inferrule{$\projGoodEnd$} ensures that
	participants with no global actions have no local actions;
	\inferrule{$\projGoodOut$}/\inferrule{$\projGoodIn$}
	reflect global message sending/receiving locally;
	\inferrule{$\projGoodType$} reflects local sending and receiving
	globally; and
	\inferrule{$\projGoodAlt$} says that after a transition,
the projection may be less specialised.
\end{definition}
Given an a-good projection,
define the association as in~\cref{def:assoc}
and obtain:
\begin{restatable}[Soundness of Association]{theorem}{thmASProjSoundness}%
  \label{thm:as_soundness_association}
Suppose that $\stEnvAssoc{\gtG}{\stEnv}{}$.\\
\begin{tabular}{clcl}
(1)&$\gtG\,\gtMove[\ltsSel{\roleP}{\roleQ}{\stChoice{\stLab}{\tyGround}}]
$&$\implies$&$
\exists \stLabi, \tyGroundi, \gtGi, \stEnvi.\;
\stEnvAssoc{\gtGi}{\stEnvi}{}
\;\wedge\;
\stEnv \stEnvMoveAnnot{\ltsSel{\roleP}{\roleQ}{\stChoice{\stLabi}{\tyGroundi}}}\stEnvi
\;\wedge\;
\gtG\,\gtMove[\ltsSel{\roleP}{\roleQ}{\stChoice{\stLabi}{\tyGroundi}}]\,\gtGi$\\
(2)&$\gtG\,\gtMove[\ltsBra{\roleP}{\roleQ}{\stChoice{\stLab}{\tyGround}}]\,\gtGi
$&$\implies$&$
\exists\stEnvi.\;
\stEnvAssoc{\gtGi}{\stEnvi}{}
\;\wedge\;
\stEnv \stEnvMoveAnnot{\ltsBra{\roleP}{\roleQ}{\stChoice{\stLab}{\tyGround}}}\stEnvi$
\end{tabular}
\end{restatable}
\begin{restatable}[Completeness]{theorem}{thmASProjCompleteness}%
  \label{thm:as_completeness_association}
  $\stEnvAssoc{\gtG}{\stEnvNew}{}
  \;\wedge\;
  \stEnvNew \stEnvMoveGenAnnot \stEnvNewi
  \implies
  \exists\gtGi.\;
  \stEnvAssoc{\gtGi}{\stEnvNewi}{}
  \;\wedge\;
  \gtG  \,\gtMove[\stEnvAnnotGenericSym]\, \gtGi$
\end{restatable}
\begin{restatable}[Associated Contexts]{theorem}{thmAAssocCorrect}
\label{thm:a-good-correct}
Let $\predP=\{\stEnv\suchthat\exists\gtG\text{ with }\stEnvAssoc{\gtG}{\stEnv}{}\}$.
(1) $\predP\subseteq\stEnvDFPred$ and $\predP$ is a safety property.
	(2) If $\stEnvAssoc{\gtG}{\stEnv}{}$ then
	$\stEnvDFP{\stEnv}$ and
	if $\gtG$ is balanced then
	$\stEnvLiveP{\stEnv}$.
\end{restatable}

The inductive projections and coinductive projections
extend to asynchronous global types.

\begin{definition}[Asynchronous Projections]
	\label{def:async-proj}
	Let $\stMerge{}{}$ be a good merge.
	
	\noindent
	Inductive projection is minimal such that $\gtProj{\roleR}{\gtG}{\stT}$
	is implied by the conditions in \cref{def:global-proj} and:
	\\[1mm]\centerline{\(
		\begin{array}{ll}
			\inferrule{P-$\stBinMerge$-II}&
			\roleR\neq\roleQ\;\wedge\;	
			{\gtG}=\gtCommSquigSmall{\roleP}{\roleQ}{\gtLab}{}{\gtLab}{\tyGround}{\gtGi}\;\wedge\;
			\gtProj{\roleR}{\gtGi}{\stT}
			\\
			\inferrule{P-$\&$-II} &
			{\gtG}=\gtCommSquigSmall{\roleP}{\roleR}{\gtLab}{}{\gtLab}{\tyGround}{\gtGi}\;\wedge\;
			\stT=\stExtSum{\roleP}{}{\stChoice{\stLab}{\tyGround} \stSeq \stTi}
			\;\wedge\;
			\gtProj{\roleR}{\gtGi}{\stTi}
		\end{array}
	\)}\\[1mm]
	Coinductive projection is maximal such that
	$\gtProj{\roleR}{\gtG}{\stT}$ implies the rules below and in \cref{def:global-proj}:\\[1mm]
        \centerline{\(
		\begin{array}{ll}
			\inferrule{P-$\stBinMerge$-II}&
			\roleR\neq\roleQ\;\wedge\;
			\unfoldOne{\gtG}=\gtCommSquigSmall{\roleP}{\roleQ}{\gtLab}{}{\gtLab}{\tyGround}{\gtGi}
			\implies
			\gtProj{\roleR}{\gtGi}{\stT}
			\\
			\inferrule{P-$\&$-II}&
			\unfoldOne{\gtG}=\gtCommSquigSmall{\roleP}{\roleR}{\gtLab}{}{\gtLab}{\tyGround}{\gtGi}
			\implies
			\unfoldOne{\stT}=\stExtSum{\roleP}{}{\stChoice{\stLab}{\tyGround} \stSeq \stTi}
			\;\wedge\;
			\gtProj{\roleR}{\gtGi}{\stTi}
		\end{array}
	\)}
\end{definition}
We obtain an analogue to \cref{thm:good-inductive-proj}
\ifextended (\cref{thm:a-good-inductive-proj}).
\else \appref{Thm. E.18}.
\fi
\begin{definition}[Asynchronous Sessions \cite{PPGSY2025}]
\label{def:async-session}
Define queues by
$\pquH\bnfdef\pquEmpty\bnfsep\pquMsg{\roleP}{\ell}{v}\bnfsep\pquCons{\pquH}{\pquHi}$ up-to commutativity of elements with distinct roles.
Extend sessions with $\playsrole{\roleP}{\pquH}$
and type queues by
$\stJudge{}{\pquCons{\pquMsg{\roleP[1]}{\stLab[1]}{v_1}}{\dots\pquMsg{\roleP[n]}{\stLab[n]}{v_n}}}{\quCons{\quMsg{\roleP[1]}{\stLab[1]}{\tyGround[1]}}{\dots\quMsg{\roleP[n]}{\stLab[n]}{\tyGround[n]}}}$
if $\forall 1\leq i \leq n.\; \stJudge{}{v_i}{\tyGround[i]}$.
Type asynchronous sessions by:
\\[1mm]
\centerline{\(
	\inference[SessAsync]{\forall \roleP\in \dom{\stEnv}&
\stJudge{}{\mpP[\roleP]}{\stT[\roleP]}&
\stJudge{}{\pquH[\roleP]}{\quH[\roleP]}&
\stEnvApp{\stEnv}{\roleP}=(\quH[\roleP],\stT[\roleP])&
\gtProj{}{\gtG}{\stEnv}}
{
\tyJudge{}{\Pi_{\roleP\in\dom{\stEnv}}{(\playsrole{\roleP}{\mpP[\roleP]}\mpPar\playsrole{\roleP}{\pquH[\roleP]})}}{\gtG}
}
\)}
\end{definition}
\begin{restatable}[Asynchronous Multiparty Sessions]{theorem}{thmAsyncTypingSessions}%
  \label{thm:asynctypingsessions}
Let $\gtProj{}{}{}$ in $\inferrule{SessAsync}$ in \cref{def:async-session}
be an a-good projection.
Suppose $\tyJudge{}{\M}{\gtG}$,
then $\M$ satisfies the subject reduction theorem, 
session fidelity, safety, deadlock-freedom
\ifextended (\cref{def:async-session-df})
\fi
and,
if $\gtG$ is balanced,
asynchronous liveness\ifextended\ (\cref{def:async-session-live})\fi.
\end{restatable}

\begin{remark}[Asynchronous liveness]
\label{rem:async-live}
Note that unlike \cref{def:stenv-live},
FIFO always guarantees send liveness, and
en-route transmissions appear at bounded depths globally
so are
always dequeued in fair paths, 
thus receive liveness implies asynchronous liveness for typed $\N$.
\end{remark}

\noindent Interestingly, the separations
and expressivenesses
as in \cref{sec:comparison} 
and summarised in Figure~\ref{fig:proj_strength}
still hold. 
\begin{restatable}[Separating Projections]{theorem}{thmSeparatingProjectionsAsync}
\label{thm:async-separate}
There are sessions $\M$ and $\M'$ such that
\begin{enumerate*}
	\item $\M$ is typed by the plain inductive projection with parallel
	and not by the full inductive projection and
	\item $\M'$ is typed by the plain coinductive projection
	and not by the full inductive projection with parallel.
\end{enumerate*}
\end{restatable}

\begin{proof}
	$\tyJudge{}{
		\Pi_{\roleR\in\roleSet}(
			\playsrole{\roleR}{\mpP[\roleR]}\mpPar
			\playsrole{\roleR}{\pquEmpty}
		)
	}{\gtG}
	\iff
	\tyJudge{}{
		\Pi_{\roleR\in\roleSet}
			\playsrole{\roleR}{\mpP[\roleR]}
	}{\gtG}
	$,
	so apply
	\cref{thm:separating-projections}.
\end{proof}

\subsection{Precise Asynchronous Subtyping}
\label{sec:aprecise}
So far, we have analysed how our choice of projection
changes program realisability.
In asynchronous semantics,
the \emph{precise asynchronous subtyping}, $\asubt$,
enlarges the set of typable sessions
by allowing \inferrule{T-Sub}
to admit safe and live asynchronous message reorderings.
For example,
$
\stFmt{\roleP\oplus\stLab\stSeq\roleQ\&\stLabi}
\asubt
\stFmt{\roleQ\&\stLabi\stSeq\roleP\oplus\stLab}
$:
messages can be sent preemptively.
The full definition is in
\ifextended \cref{def:precise_subtyping} and
\fi
\cite[Definition~3.7]{GPPSY2023}.
We use $\stEnvAssoc{}{}{a}$
to denote association using $\asubt$. 
$\asubt$ does
not provide a direct operational correspondence
as $\tySub$ does~\cite[Example~12]{pischke2026asynchronousglobalprotocolsprecisely},
so context traces are not monotone with respect to $\asubt$.
Thus,
a-good (\cref{def:as-good-proj})
is now insufficient. 
Our contributions are: a proposal of `$\text{a-good}^+$',
which captures the inductive/coinductive, plain/full merge projections
\eg the projection
from~\cite{pischke2026asynchronousglobalprotocolsprecisely}; 
and an application of `$\text{a-good}^+$' 
to design decidable sound type-checking systems with message reordering. 
\begin{definition}[Precise Global Type Transitions~{\cite[Definition~7]{pischke2026asynchronousglobalprotocolsprecisely}}]
  \label{def:precise-gtype:lts-gt}
 The precise asynchronous global type transition
  $\gtMoveA[\stEnvAnnotGenericSym]$ is inductively
  defined by the rules from~\cref{def:a-gtype:lts-gt},
  replacing \inferrule{\iruleGtMoveCtxI}
  and \inferrule{\iruleGtMoveCtxII}
  by the below rules:
	\\[1mm]\centerline{\(\small
		\begin{array}{c}
  \inference[\iruleGtMoveCtxI]{
  \ltsSubject{\stEnvAnnotGenericSym}\not\ni \roleP
  &
    \forall i \in I.\,
    \gtGi[i] 
    \,\gtMoveA[\stEnvAnnotGenericSym]\, 
    \gtGii[i] 
  }{
      \gtCommSmall{\roleP}{\roleQ}{i \in I}{\gtLab[i]}{\tyGround[i]}{\gtGi[i]}
   \,\gtMoveA[\stEnvAnnotGenericSym]\, 
      \gtCommSmall{\roleP}{\roleQ}{i \in I}{\gtLab[i]}{\tyGround[i]}{\gtGii[i]}   
  }
  \quad
\inference[\iruleGtMoveCtxII]{
  \stEnvAnnotGenericSym \ne \ltsBra{\roleQ}{\roleP}{\stChoice{\stLabi}{\tyGroundi}}
  &
    \gtGi
    \,\gtMove[\stEnvAnnotGenericSym]\, 
    \gtGii
  }{
      \gtCommSquigSmall{\roleP}{\roleQ}{\gtLab}{}{\gtLab}{\tyGround}{\gtGi}
   \,\gtMove[\stEnvAnnotGenericSym]\, 
      \gtCommSquigSmall{\roleP}{\roleQ}{\gtLab}{}{\gtLab}{\tyGround}{\gtGii}   
  }  
\\[0.5ex]
  \inference[\iruleGtMoveCtxI']{
  \roleR\ne\roleQ
  &
    \forall i \in J\subseteq I.\,
    \gtGi[i] 
    \,\gtMoveA[\ltsSel{\roleP}{\roleR}{\stChoice{\stLabi}{\tyGroundi}}]\, 
    \gtGii[i] 
    &
    }{
      \gtCommSmall{\roleP}{\roleQ}{i \in I}{\gtLab[i]}{\tyGround[i]}{\gtGi[i]}
   \,\gtMoveA[\ltsSel{\roleP}{\roleR}{\stChoice{\stLabi}{\tyGroundi}}]\, 
      \gtCommSmall{\roleP}{\roleQ}{i \in J}{\gtLab[i]}{\tyGround[i]}{\gtGii[i]}   }
\end{array}
	\)}
\end{definition}
\noindent\inferrule{\iruleGtMoveCtxI} is similar to 
\inferrule{\iruleGtMoveCtxI} in \cref{def:a-gtype:lts-gt} but also
permits $\roleQ$ to perform deeper actions;
\inferrule{\iruleGtMoveCtxII} permits $\roleQ$
to perform deeper actions as long as
it is not receiving from $\roleP$; and
\inferrule{\iruleGtMoveCtxI'} permits
$\roleP$ to make deeper selections,
possibly limiting its earlier choices.
These reflect local semantics up-to $\asubt$
and are \emph{non-deterministic} due to the choice of $J\subseteq I$.

To relate precise subtypes with projections,
we introduce two forms of local context depending
on the head of the subtype:
$\trCtxSelp$ for sending; and
$\trCtxBrap$ for receiving.
\begin{definition}[Local Type Context~{\cite[Definition~21]{pischke2026asynchronousglobalprotocolsprecisely}}]
    \label{def:local-ctx}
    Define local type contexts as follows:
    \\\centerline{\(
   \begin{array}{r@{\;}c@{\;}l} 
        \trCtxSelp& \bnfdef& [\cdot]_i
	\bnfsep \stExtSum{\roleR}{i \in I}{\stChoice{\stLab[i]}{\tyGround[i]} \stSeq \trCtxSelp}
	 \bnfsep \stIntSum{\roleQ}{i \in I_L}{\stChoice{\stLab[i]}{\tyGround[i]} \stSeq \forgetHole{\cdot}{l_i}} + 
	\stIntSum{\roleQ}{i \in I_R}{\stChoice{\stLab[i]}{\tyGround[i]} \stSeq \trCtxSelp}
	\\
        \trCtxBrap&\bnfdef& [\cdot]_i \bnfsep \stExtSum{\roleQ}{i \in I}{\stChoice{\stLab[i]}{\tyGround[i]} \stSeq \trCtxBrap}
        \qquad\qquad\qquad\qquad(\roleQ \ne \roleP\;\wedge\;I_R, I\ne\emptyset)
\end{array}
 \)}
$[\cdot]_i$ is a hole labelled with $i$
and $\forgetHole{\cdot}{i}$ is a \emph{forgetful hole}
labelled with $i$,
indicating that the corresponding choice is
forgotten by the subtyping instance.
We write $\overline{\mathbb{L}}$ for
the context given by removing the
choices corresponding to forgetful holes.
\end{definition}
\begin{example}[Precise Asynchronous Odd-Even]
\label{ex:precise:oddeven}
	Recall $\gtG[\text{\tiny o/e}]$ from \cref{ex:running} and
	$\gtGi[\text{\tiny o}]$, $\gtG[\text{\tiny e}]$ from~\cref{ex:proj-running}.
	$\roleQ$ is able to send messages
	to $\roleR$ before receiving
	a message from $\roleP$
	by \inferrule{\iruleGtMoveCtxI},
	but it cannot send $\gtMsgFmt{b}$
	before receiving from $\roleP$ \ie
	$
		\gtG[\text{\tiny o/e}]\,
		\gtMoveA[\ltsSel{\roleQ}{\roleR}{\stLabFmt{o}}]\,
	\gtCommRaw{\roleP}{\roleQ}{
		\gtMsgFmt{o}{\gtSeq}\gtCommSquigSingle{\roleQ}{\roleR}{\gtMsgFmt{o}}{\gtMsgFmt{o}}{}{
			\gtGi[\text{\tiny o}]
		},\,
		\gtMsgFmt{m}{\gtSeq}
			\gtCommSquigSingle{\roleQ}{\roleR}{\gtMsgFmt{o}}{\gtMsgFmt{o}}{}{
				\gtCommSingle{\roleQ}{\roleR}{\gtMsgFmt{o}}{}{
					\gtG[\text{\tiny o}]
				}
			}
	}
		\quad\not\!\!\!\!\!\!\gtMoveA[\ltsSel{\roleQ}{\roleR}{\gtMsgFmt{b}}]
	$.
	This action corresponds to the subtyping:
	\\[1mm]\centerline{\(
		\stFmt{\roleR{\oplus}}\stLabFmt{o}\stSeq\stExtSum{\roleP}{}{
		\stLabFmt{o}\stSeq\stRec{\stRecVar}{
			\stIntSum{\roleR}{}{
				\stLabFmt{o}\stSeq\roleR{\oplus}\stLabFmt{o}\stSeq\stRecVar,\,
				\stLabFmt{b}\stSeq\roleR{\&}\stLabFmt{o}
			}
		},\,
		\stLabFmt{m}\stSeq\roleR{\oplus}\stLabFmt{o}\stSeq\stRec{\stRecVar}{
			\stIntSum{\roleR}{}{
				\stLabFmt{o}\stSeq\roleR{\oplus}\stLabFmt{o}\stSeq\stRecVar,\,
				\stLabFmt{b}\stSeq\roleR{\&}\stLabFmt{m}
			}
		}
	}
		\asubt\stT[\roleQ]
	\)}\\[1mm]
	Observe that with
		$\trCtxSelp = \stExtSum{\roleP}{}{
		\stLabFmt{o}\stSeq [\cdot]_0,
		\stLabFmt{m}\stSeq
			\stIntSum{\roleR}{}{
				\stLabFmt{o}\stSeq\roleR{\oplus}\stLabFmt{o}\stSeq 
				\stRec{\stRecVar}{
			\stIntSum{\roleR}{}{
				\stLabFmt{o}\stSeq[\cdot]_1
			}
		}
				,\,
				\stLabFmt{b}\stSeq\forgetHole{\cdot}{2}
			}
	}$, we have:
	\\[1mm]\centerline{\(\stT[\roleQ]\bisim\trCtxSelp[\stFmt{
	\roleR{\oplus}\stLabFmt{o}\stSeq\stRec{\stRecVar}{
			\stIntSum{\roleR}{}{
				\stLabFmt{o}\stSeq\roleR{\oplus}\stLabFmt{o}\stSeq\stRecVar,\,
				\stLabFmt{b}\stSeq\roleR{\&}\stLabFmt{o}
			}
		}}
	]_0
	[\stFmt{
		\roleR{\oplus}\stLabFmt{o}\stSeq\stRecVar,
				\stLabFmt{b}\stSeq\roleR{\&}\stLabFmt{m}}
	]_1
	\forgetHole{
	 \stFmt{\roleR{\&}\stLabFmt{m}}
	}{2}\)}
\end{example}

This generalises:

\begin{restatable}[Inversion of subtyping {\cite[Lemma~16]{pischke2026asynchronousglobalprotocolsprecisely}}]{lemma}{lemInvSub}
    \label{lem:inv-subtyping}
    \ \\
    \begin{itemize}
    \item[(1)] If $\stIntSum{\roleP}{i \in I}{\stChoice{\stLab[i]}{\tyGround[i]} \stSeq \stT[i]} \asubt \stT$,
    then
    $\stT\bisim\trCtxSelp[\stIntSum{\roleP}{i \in I_j}{\stChoice{\stLab[i]}{\tyGround[i]} \stSeq \trTi[i,j]}]_{j\in J}\forgetHole{\trTii[j]}{j\in J'}$
    with $I \subseteq I_j$ for all $j\in J$ and for all $i \in I$ and $j\in J$,
    we have
    that if $\stTi[i]\bisim\trCtxSelp[\trTi[i,j]]_{j\in J}\forgetHole{\trTii[j]}{j\in J'}$
    then $\stT[i] \asubt \stTi[i] $.
    \item[(2)] If $\stExtSum{\roleP}{i \in I}{\stChoice{\stLab[i]}{\tyGround[i]} \stSeq \stT[i]} \asubt \stT$,
    then
    $\stT\bisim \trCtxBrap[\stExtSum{\roleP}{i \in I_j}{\stChoice{\stLab[i]}{\tyGround[i,j]} \stSeq \trTi[i]}]_{j\in J}$
    with $I_j \subseteq I$ for all $j\in J$,
    $I=\bigcup_{j\in J}I_j$,
    for all $j\in J$ and $i\in I$
    if $\stTi[i]\bisim\trCtxBrap[\trTi[i,j]]_{j\in J}$ for some $i\in\bigcap_{j\in J}I_j$ then
    $\stT[i] \asubt \stTi[i]$.
    \end{itemize}
\end{restatable}
\begin{definition}[$\text{A-Good}^+$ Projection]
\label{def:a-good-proj}
	Fix a projection relation.
	We say that this is an \emph{$\text{a-good}^+$ projection},
	if it is a-good and
	whenever
	$\gtProj{}{\gtG}{\stEnv}$
	with $\stEnvApp{\stEnv}{\roleP}=(\quH[\roleP],\stT)$,
	$\gtG$ is balanced and:
    \begin{itemize}[leftmargin=0.4in,labelindent=-\leftmargin]
\item[\inferrule{$\projGoodAlt$}]
            If
            $\gtG\,\gtMoveA[ \stEnvAnnotGenericSym ]
            \,\gtGi
            \;\wedge\;
            \roleP\not\in\ltsSubject{\stEnvAnnotGenericSym}
            $
            then there is
            $\stTi$
            such that
            $\gtProj{\roleP}{\gtGi}{\stTi}$
            and
            $\stT \tySub \stTi$
     \item[\inferrule{$\projGoodOutS$}]
     	If
     	$\stT\bisim\trCtxSelq[\stIntSum{\roleQ}{i \in I_j}{\stChoice{\stLab[i]}{\tyGround[i]} \stSeq \stT[i,j]}]_{j\in J}\forgetHole{\stTi[j]}{j\in J'}$
     	then
     	for $k\in\bigcap_{j\in J}I_j$ there exist $\gtGi$ and $\stTi$ such that
     	$\gtG\,\gtMoveA[\ltsSel{\roleP}{\roleQ}{\stChoice{\stLab[k]}{\tyGround[k]}}]\,\gtGi$,
     	$\gtProj{\roleP}{\gtGi}{\stTi}$
     	and $\overline{\trCtxSelq}[\stT[k,j]]_{j\in J}\tySub\stTi$
	\item[\inferrule{$\projGoodInS$}]
If $\stT\bisim\trCtxBraq[\stExtSum{\roleQ}{i\in I_j}{\stChoice{\stLab[i]}{\tyGround[i]}\stSeq\stT[i,j]}]_{j\in J}$ and
	$
	\stEnvApp{\stEnv}{\roleQ}=(\quCons{\quMsg{\roleP}{\stLab}{\tyGround}}{\quH[\roleQ]},\cdot)$,
	then for $j\in J$ there is $k_j\in I_j$ such that $\stLab[k_j]=\stLab$ and
	$\tyGround[k_j]=\tyGround$, and there is $\gtGi$ and $\stTi$ such that $\gtG\,\gtMoveA[\ltsBra{\roleP}{\roleQ}{\stChoice{\stLab}{\tyGround}}]\,\gtGi$,
     	$\gtProj{\roleP}{\gtGi}{\stTi}$
     	and $\trCtxBraq[\stT[k_j,j]]_{j\in J}\tySub\stTi$
	\end{itemize}
\end{definition}
The idea is that subtyped reducts
are reflected as reducts in a context for the projection,
which can be reflected globally.
$\text{A-good}^+$ projections provide the following:
\begin{restatable}[Soundness]{theorem}{thmAAProjSoundness}%
  \label{thm:aa_soundness_association}
  $\stEnvAssoc{\gtG}{\stEnvNew}{a}\;\wedge\;    
	\gtG\,\gtMoveA[]\implies
	\exists\stEnvAnnotGenericSym, \stEnvNewi, \gtGi.\;
	\stEnvAssoc{\gtGi}{\stEnvNewi}{a} \;\wedge\;
	\gtG\,\gtMoveA[\stEnvAnnotGenericSym]\,\gtGi\;\wedge\;
	\stEnvNew \stEnvMoveGenAnnot\stEnvNewi$
\end{restatable}
\begin{restatable}[Completeness]{theorem}{thmAAProjCompleteness}%
  \label{thm:aa_completeness_association}
    $\stEnvAssoc{\gtG}{\stEnvNew}{a}\;\wedge\;
  \stEnvNew \stEnvMoveGenAnnot \stEnvNewi\implies
  \exists\gtGi.\;
  \stEnvAssoc{\gtGi}{\stEnvNewi}{a}\;\wedge\;\gtG  \,\gtMoveA[\stEnvAnnotGenericSym]\, \gtGi$
\end{restatable}
\begin{restatable}[Precisely Associated Contexts]{theorem}{thmPreciseAssocCorrect}
\label{thm:aa-good-correct}
	Let $\predP=\{\stEnv\suchthat\exists\gtG\text{ with }\stEnvAssoc{\gtG}{\stEnv}{a}\}$.
	$\predP\subseteq\stEnvLivePred$ and $\predP$ is a safety property.
\end{restatable}
Unlike the standard subtyping, $\tySub$,
the precise asynchronous subtyping
has been proven undecidable \cite{LY2017}.
Thus instead of separating typing systems by projections,
we fix an $\text{a-good}^+$ projection,
and use decidable approximations $\psubt{p}$ of $\asubt$ to obtain
a decidable sound type-checking system. 
Candidates for $\psubt{p}$ are
the sound decidable subtyping relations
proposed in \cite{Bocchi2024-hs,bocchi_et_al:LIPIcs.CONCUR.2025.10,CYV2022}. 

\begin{restatable}[Asynchronous Multiparty Sessions with
Message Reordering]{theorem}{thmAsyncTypingSessionsSubset}%
  \label{thm:asyncprecisetyping}
Let $\gtProj{}{}{}$ in $\inferrule{SessAsync}$ in \cref{def:async-session}
be an a-good$^{+}$ projection.
Suppose a preorder $\psubt{p}$ such that 
$\tySub\subseteq\psubt{p}\subseteq\asubt$
and~\cref{lem:inv-subtyping} holds with $\psubt{p}$.
Suppose $\tyJudge{}{\M}{\gtG}$ using $\psubt{p}$ instead of $\asubt$, 
then $\M$ satisfies the subject reduction theorem, 
session fidelity, safety, deadlock-freedom and
asynchronous liveness.
\end{restatable}
\begin{proof}
Thms.~\ref{thm:aa_soundness_association},~\ref{thm:aa_completeness_association},
and~\ref{thm:aa-good-correct}
	only require that:
	$\asubt$ is a preorder;
	$\tySub\,\subseteq\,\asubt$;
	$\stEnvSafePred\cap\stEnvDFPred$ and $\stEnvLivePred$
	are downward closed under $\asubt$; 
	and~\cref{lem:inv-subtyping} (inversion of subtyping) holds.
	Therefore, if we have a preorder $\psubt{p}$ such that 
	$\tySub\subseteq\psubt{p}\subseteq\asubt$
	and~\cref{lem:inv-subtyping} holds, then
	Thms.
	\ref{thm:aa_soundness_association},
	\ref{thm:aa_completeness_association}, 
	and \ref{thm:aa-good-correct} hold
	for association ${\stEnvAssoc{}{}{\text{p}}}$, defined by 
	$\psubt{p}$.
\end{proof}

\section{Related Work}
\label{sec:related}
\myparagraph{Top-Down Typing Systems.}
Multiparty session type systems~\cite{HYC08,HYC16}
were
introduced for asynchronous p2p communication 
as in~\cref{sec:async},
and simplified in~\cite{CDYP2015}.
They use \emph{plain inductive projection} and
exclude subtyping,
ensuring a strong operational
correspondence between global and local types.
The use of full inductive projection
with synchronous subtyping
has become the most common approach.
A more expressive class of
\emph{coinductive} projections
are defined over
coinductive types
with plain \cite{EKY2025,DBLP:journals/jar/TiroreBC25}
and full \cite{DBLP:journals/pacmpl/UdomsrirungruangY25,pischke2026asynchronousglobalprotocolsprecisely}
merging;
we have the first presentation of coinductive projections on
\emph{inductive types} necessitating
viewing projections as relations.
We identify the projections 
in 
\cite{GJPSY2018,Barbanera2021-mg,doi:10.3233/FI-2019-1863,10.1145/3798256}
as \emph{not} good,
disproving the subject reduction theorem (\cref{sec:counter}).

Barbanera et al.~\cite{10.1145/3678232.3678245}
present \emph{full corecursive projection},
defined \emph{corecursively} rather than coinductively, 
ensuring that projectable global types are balanced.
The full coinductive projection
is equal to the full corecursive projection 
on balanced global types,
and
projects their `un-projectable' global types, 
including the examples
in \cite[Fig. 4]{POPL19LessIsMore} (see
\ifextended \cref{prop:full-coind-is-corec}).
\else \appref{Prop. G.2}.
\fi

Castro-Perez et al.~\cite{10.1145/3776692}
propose a \emph{synthetic approach} similar to 
\cite{10.1145/3678232.3678245}
where processes are typed directly against synchronous 
global LTSs. Our analysis has identified several 
issues with the claims of the paper;
these are due to a gap between
the definitions of the paper and
the Agda formalisation. 
The global type:\\[1mm]
\centerline{$\gtG[0]=\gtRec{\gtRecVar}{
\gtCommRaw{\roleP}{\roleQ}{
\gtLab[0]\gtSeq\gtCommSingle{\roleR}{\roleS}{\gtLab[2]}{}{\gtRecVar},\,
\gtLab[1]\gtSeq\gtCommRaw{\roleR}{\roleS}{
\gtLab[2]\gtSeq\gtRecVar,\,\gtLab[3]\gtSeq\gtRecVar
}
}
}$}\\[1mm]
does not satisfy their well-behavedness conditions
and
types the session:\\[1mm]
\centerline{
$
\M_0 =
\playsrole{\roleP}{
		\mpRec{\mpX}{
		\roleQ\mathord{!}\stLab[0]\mpSeq\mpX
		}
		}
		\mpPar
		\playsrole{\roleQ}{\mpRec{\mpX}{
		\textstyle\sum_{i\in \{0,1\}} \roleP\mathord{?}\stLab[i]\mpSeq \mpX
		}}
		\mpPar
		\playsrole{\roleR}{
		\mpRec{\mpX}{
		\roleS\mathord{!}\stLab[2]\mpSeq\mpX
		}
		}
		\mpPar
		\playsrole{\roleS}{\mpRec{\mpX}{
		\textstyle\sum_{i\in \{2,3\}} \roleR\mathord{?}\stLab[i]\mpSeq \mpX
		}}
$,}
\\[1mm]
violating \cite[Thm.~5.9]{10.1145/3776692}.
In their framework, decidability of type-checking only holds when global types have finite LTSs,
which is not guaranteed (\Cref{thm:global-sem}). 
For example, $\gtG[0]$ above has a non-regular trace language,
hence it does \emph{not} have a finite LTS
while also typing $\M_0$.
See
\ifextended \cref{sec:app:synthetic}
\else \appref{\S G.2}
\fi
for details.

Castagna et al. \cite{Castagna2012GlobalTypes} proposed
semantic conditions for projecting directly onto typing contexts.
They remark that projection algorithms are necessary
since programs can be infinite state.
This is also true for global types (\cref{thm:global-sem})
with projection algorithms rejecting `flawed' protocols.

\smallskip

\myparagraph{Global Types and Flexible Choice.}
Following the introduction of standard global types
\cite{HYC08,HYC16,DBLP:conf/icdcit/YoshidaG20,CDPY2015,HYK2026},
research has focussed on extending global types with
full mixed choice \cite{DBLP:journals/corr/abs-2508-13616,Barbanera2026-ar}
or
restrictions \eg
sender-driven choice
(transmissions may have multiple destinations) \cite{majumdar_et_al:LIPIcs.CONCUR.2021.35,Stutz2024,Li2026,Giusto2025}
and
directed choice \cite{10.1145/3798256}.
Deadlock-free realisability of general mixed choice
has been found undecidable \cite[Chapter~9]{Stutz2024}. Hence
to define sound 
decidable typing systems,
we require
restrictions/conditions on global types/projections. 
Barbanera and Dezani-Ciancaglini
\cite{DBLP:journals/corr/abs-2508-13616,Barbanera2026-ar}
introduce the notion of \emph{coherence}
for mixed choice contexts in
synchronous and asynchronous semantics, and 
define projections directly from global types 
to typing contexts.
These rely on choosing partitions of the participants
and finding witnesses for coherence;
it is unclear how these witnesses are found
and partitions must be chosen by the programmer.
Bocchi et al. \cite{10.1145/3798256} propose a projection
for asynchronous directed mixed choice.
They define a \emph{transient inconsistency}
which designates one side of the mixed choice
as a converging side that
all participants are made aware of
if activated. 
Consequently, the example in \cref{ex:directed-mixed-two-clients}
is not projectable by the
projection in
\cite{10.1145/3798256}.
Some works
\cite{Castagna2012GlobalTypes,jongmans_et_al:LIPIcs.ECOOP.2023.42,JY2020}
define global types as
regular expressions with non-deterministic choice,
sequencing and Kleene star,
to incorporate all possible choice.
All three approach typing differently:
\cite{Castagna2012GlobalTypes} is described above,
\cite{jongmans_et_al:LIPIcs.ECOOP.2023.42} uses synthetic typing
and \cite{JY2020} uses projections with bisimilarity checking.

\smallskip

\myparagraph{Expressiveness and Separation.}
The parallel composition constructor introduced in \cite{HYC08,HYC16} 
is absent in many MPST systems, but 
is occasionally used for conciseness \cite{Castagna2012GlobalTypes,Gheri2023}.  
The necessity of the parallel constructor has been questioned;
\cite{VanGlabbeekLICS2021} 
claims that:
\begin{enumerate*}
\item most systems without parallel 
are incomplete, using \cite[Figure 4 (2)]{POPL19LessIsMore} 
as an example; and 
\item the coinductive projection in~\cite{Barbanera2021-mg} 
is expressive enough to simulate parallel global behaviours.
\end{enumerate*}
\cref{sec:comparison} justifies (1,2).%

\smallskip

\myparagraph{Automata-based Approaches.}
The realisability problem
for asynchronous sender-driven global protocols
onto communicating finite state
machines (CFSMs) \cite{Brand:1983:CFSM} 
has been actively studied \cite{majumdar_et_al:LIPIcs.CONCUR.2021.35,Stutz2024,Li2026,Giusto2025},
with the development of a decidable sound and complete algorithm
\cite{Li2023-sx}.
Unfortunately, the projection algorithm in \cite{Li2023-sx}
is \textbf{(1) incomplete in synchronous semantics} and
\textbf{(2) incomplete with respect to program realisability} in
asynchronous semantics. 
To show \textbf{(1)}, consider:\\
\centerline{
$\gtG[1] = \gtFmt{
		(\roleP[0]\to\roleQ[1]:\gtLab{\gtSeq}
		\roleP[0]\to\roleP[1]:\gtLab{\gtSeq}
		\roleP[1]\to\roleQ[0]:\gtLab) \ + \
                (\roleP[0]\to\roleQ[0]:\gtLab{\gtSeq}
		\roleP[0]\to\roleP[1]:\gtLab{\gtSeq}
		\roleP[1]\to\roleQ[1]:\gtLab)}$}\\[1mm] 
There exists good projection $\gtProj{\gtG[1]}{\stEnv}{}$ 
such that $\gtTrace{\stEnv}=\gtTrace{\gtG[1]}$
and $\stEnv$ is safe and live, 
but the projection algorithm, $\gtProji{}{}{}$, in \cite{Li2023-sx}
fails to project (realise) $\gtG[1]$ (see
\ifextended \cref{thm:proj-incomplete}).
\else \appref{Thm. G.15}
\fi
For \textbf{(2)},
there exists a sound and complete (computable)
projection algorithm $\gtProj{}{}{}$
(in the sense of \cite{Li2023-sx})   
from asynchronous sender-driven global types, 
that types a program which $\gtProji{}{}{}$ cannot
\ifextended (\cref{prop:complete-proj-sep}),
\else \appref{Prop. G.5},
\fi
\ie 
assume $\gtG=
\gtCommSingle{\roleP}{\roleQ}{\gtLab}{}{}$ and 
let:\\[1mm]
\centerline{
$
\M =
		\playsrole{\roleP}{
			\roleQ\mathord{!}\stLab 
		}
		\mpPar
		\playsrole{\roleQ}{
			\roleP\mathord{?}\stLab\mpSeq \mpNil
			+
			\roleR\mathord{?}\stLab\mpSeq \mpNil
		}
		\mpPar
		\playsrole{\roleR}{\mpNil}
$,
and
$\stEnv=
\set{
			\stEnvMap{\roleP}{(\quEmpty,\stFmt{\roleQ{\oplus}\stLab})}
			\stEnvComp
			\stEnvMap{\roleQ}{(\quEmpty,\stFmt{\Sigma\!\left\{\roleP{\&}\stLab,\,\roleR{\&}\stLab\right\}})}
			\stEnvComp
			\stEnvMap{\roleR}{(\quEmpty,\stEnd)}
			}
$}\\[1mm]
The synchronisation between $\roleQ$ and $\roleR$
will never be used, so we modify $\gtProji{}{}{}$
to $\gtProj{}{}{}$ by setting
$\gtProj{}{\gtG}{\stEnv}$,
thus
$\tyJudge{}{\M}{\gtG}$ by $\gtProj{}{}{}$
but $\forall \gtGi.(\neg \tyJudge{}{\M}{\gtGi})$ by $\gtProji{}{}{}$.

In contrast to \textbf{(1,2)}, 
the work \cite{PY2026} has proven
that in synchronous semantics,
a top-down typing system with the coinductive full merge projection
gives \emph{sound and complete program realisability} with respect to liveness: if $\M$ is typed as live by the bottom-up system  
\cite{POPL19LessIsMore}, 
then $\M$ is typable with some balanced global type
by the top-down system, and vice versa. 

Similarly to~\cite{Denielou2012}, 
Stutz and D'Osualdo~\cite{10.1007/978-3-031-91121-7_13}
develop an automata-based typing system, 
replacing global types with \emph{tame} protocol state machines
(PSMs). 
While \cite{10.1007/978-3-031-91121-7_13} notes that standard
asynchronous multiparty session types are half-duplex \cite{ICALP13CFSM}, 
asynchronous global semantics with message reordering 
(\Cref{sec:aprecise}) is \emph{not} half-duplex, 
handling \cref{ex:precise:oddeven}, which 
the subtyping relation in \cite{Li2024-mu} fails to justify. 
The good projection criteria address the limitations of the top-down
MPST framework stated in 
\cite[\S~1]{10.1007/978-3-031-91121-7_13}:
projection algorithms can be \emph{decoupled} from type systems, 
handling extended global types and 
asynchronous semantics with flexible message reordering 
(\S\ref{sec:extended},~\ref{sec:async} and~\ref{sec:implementation}). 

\smallskip

\myparagraph{Bottom-Up Systems and Associations.}
Scalas and Yoshida~\cite{POPL19LessIsMore} proposed
a general MPST system
(\emph{bottom-up approach}),
using verification of safety properties
of typing contexts instead of global types.
They discovered that proofs for subject reduction of top-down systems with
the full inductive merge in the literature \cite{ParameterisedYDBH12,CHEN2015708} are flawed. 
Hou et al.~\cite{HYK2026} corrected these proofs
with the \emph{association relation}. 
This proof technique has been applied 
to p2p systems~\cite{lmcs2025,pischke2026asynchronousglobalprotocolsprecisely}.
Deciding safety, deadlock-freedom and liveness 
of synchronous typing contexts
is PSPACE-complete \cite{DBLP:journals/pacmpl/UdomsrirungruangY25}.
The top-down approach is often
more efficient,
especially with the inductive projections
(\cref{rem:projection-time-complexity}).
In asynchronous p2p, 
bottom-up systems and asynchronous subtyping are undecidable.
However, \cite{bocchi_et_al:LIPIcs.CONCUR.2025.10,CYV2022}
propose sound and decidable approximations of subtyping.
Larsen et al. \cite{larsen_et_al:LIPIcs.ECOOP.2026.17}
recently proposed `output-liveness' to prevent
orphan-messages in network protocols;
this is similar to our `send liveness',
which is always guaranteed by a-good projections.

\smallskip 
\myparagraph{Scalas-Yoshida Benchmarks~\cite{POPL19LessIsMore} and
  Projections. }
Scalas and Yoshida \cite[Fig. 4]{POPL19LessIsMore}
present four examples for evaluating MPST systems.
These have been effective:
Hou et al.~\cite{HYK2026,DBLP:series/lncs/YoshidaH24}
show that the full inductive projection handles the first example;
Pischke and Yoshida \cite{PY2026} that the full coinductive projection
handles all four examples;
Castro-Perez et al.~\cite{10.1145/3776692}
and
Barbanera et al.~\cite{Barbanera2026,10.1145/3678232.3678245}
that their typing systems 
handle all four examples.
However,
the processes of Barbanera et al.
\cite{Barbanera2026,10.1145/3678232.3678245}
correspond to our local types
instead of MPST calculi
\cite{HYK2026,DBLP:series/lncs/YoshidaH24,10.1145/3776692,PY2026,DBLP:journals/pacmpl/UdomsrirungruangY25}. 
Junge et al.~\cite{Junge2026} have proven
that safety checking in 
our calculus 
(\cref{def:sync-processes})
is \emph{undecidable}.
This result establishes a strict separation in computability between
MPST calculi
and processes in
\cite{Barbanera2026,10.1145/3678232.3678245},
where safety checking is decidable.

\section{Conclusion and Future Work}
\label{sec:conclusion}
We have now reached the
point to return to the question from the introduction:
are we
turning
the MPST jungle into a garden?
To the best of our knowledge, this work is the first to provide criteria
for projections on syntactic session types to be
\emph{correct}
(safe, deadlock-free, live, \etc)
for either synchronous or asynchronous semantics.
In this paper, we focus on
the \emph{program realisability} problem,
whether a program can be typed in a given top-down typing system,
which required us to consider
the \emph{projection} problem \cite{DBLP:journals/pacmpl/UdomsrirungruangY25}:
does a given projection induce a correct typing system?
This is one step away from the
deadlock-free realisability problem \cite{ALUR2005}:
if the projection problem is solved, then 
the deadlock-free realisability problem is solved, but not vice versa.

\smallskip 

\myparagraph{Q1.}
In \cref{sec:projections},
we introduced the \emph{good projection criteria}
to define correctness.
Typing systems defined by good projections
always satisfy subject reduction, session fidelity,
safety and deadlock-freedom.

\smallskip 

\myparagraph{Q2.}
In \cref{sec:balanced},
we introduced
the global type properties
weak balanced,
$\roleP$-send/receive balanced and
balanced,
found efficient decision procedures (\cref{thm:dec-bal}) and
determined their relationships with
liveness properties (\cref{fig:liveness}).

\smallskip 

\myparagraph{Q3.}
In \cref{sec:indcoind}, we showed that
the 8 standard projection algorithms
(inductive/coinductive projections with
plain/full merging on global types
with/without parallel)
are good,
unifying their correctness proofs.
We followed this
in \cref{sec:counter},
by showing that
the standard projection algorithms that are not good
are \emph{incorrect} with
counterexamples derived from failure of our correctness proofs.

\smallskip 

\myparagraph{Q4.}
We introduce semantic-based/typing-based criteria for
global/local expressiveness and separation of projections (\Cref{def:expressiveness}),
and apply them to
session typing systems (\Cref{thm:app:strong-process-sep}) and,
for the first time,
categorise the 8 projection algorithms (\Cref{fig:proj_strength}).
Hence, we solve the \emph{typability and protocol separation problems}.
These results
have clear implications for
implementing top-down session types.
Using coinductive projections
allows more expressive programs to be typed compared
to using inductive projections,
and likewise for inductive projections with parallel composition
over inductive projections without parallel composition.
This justifies the use of coinductive projections
over the simpler inductive projections.
Our separation results identify parallel compositions 
as an essential construct of tractable and expressive global types; 
it adds no further complexity to projections
and, for inductive projections,
strictly increases the set of typable programs.

\smallskip 

\myparagraph{Q5.}
	We extended our results to asynchronous semantics (\cref{sec:async})
	and to global types with mixed choice constructors (\cref{sec:extended}).
	We proposed new ``good projection'' algorithms
  for directed mixed choice global types.
  We have shown $\text{a-good}^+$ ensures the correctness of  
  systems with
precise asynchronous subtyping~\cite{GPPSY2023},
enabling us to define general properties
for correctness of decidable process typing systems
using
approximations of $\asubt$~\cite{Bocchi2024-hs,bocchi_et_al:LIPIcs.CONCUR.2025.10,CYV2022}
(\cref{thm:asyncprecisetyping}).

\newcounter{CounterRelatedTaxonomy}%
\renewcommand{\theCounterRelatedTaxonomy}{\alph{CounterRelatedTaxonomy}}%
\newcommand{\RelTax}[1]{%
  \refstepcounter{CounterRelatedTaxonomy}\label{#1}%
  \textbf{(\theCounterRelatedTaxonomy)}%
}%
\newcommand{\refRelTax}[1]{\textbf{(\ref{#1})}}%
\begin{table}[t]
\caption{Projections. 
    ``Restricted inductive'' means 
more restrictive than the plain inductive.
``Direct'' means relations directly defined
between global types and contexts that cannot be written  
as participant-wise. 
}
\begin{center}
  \small%
  \setlength\tabcolsep{1mm}%
  \begin{tabular}{@{\hskip 0mm}l|l@{\hskip 0mm}}
  	\textbf{good?} & \text{projection}\\[1mm]\hline
  	good&
  	full coinductive \cite{DBLP:journals/pacmpl/UdomsrirungruangY25}
  	and inductive \cite{CHEN2015708,ParameterisedYDBH12,TY2017,SDHY2017,Lanese2024}
  	and corecursive \cite{10.1145/3678232.3678245},
  	\\&
  	plain coinductive \cite{DBLP:journals/jar/TiroreBC25,DBLP:journals/pacmpl/UdomsrirungruangY25}
  	and inductive \cite{DBLP:conf/icdcit/YoshidaG20},
  	restricted inductive \cite{CarboneMSY15,CarboneLMSW16}
  	\\\hline
  	a-good&
  	asynchronous full inductive \cite{BSYZ2022,HLY2024},\\&
  	asynchronous plain coinductive \cite{Tirore2025b,DBLP:conf/pldi/Castro-Perez0GY21}
  	and inductive \cite{HYC08,HYC16,CDPY2015,CDYP2015}
  	\\\hline
  	$\text{a-good}^+$&
  	asynchronous full coinductive \cite{pischke2026asynchronousglobalprotocolsprecisely}
  	\\\hline
  	incorrect&
  	disjoint coinductive \cite{Barbanera2021-mg}
  	and inductive \cite{doi:10.3233/FI-2019-1863},
  	1-level coinductive \cite{GJPSY2018}
  	and inductive \cite{10.1145/3798256}
  	\\\hline
  	direct&
  	mixed choice coherence \cite{DBLP:journals/corr/abs-2508-13616},
  	asynchronous mixed choice coherence \cite{Barbanera2026-ar},\\&
  	complete asynchronous sender-driven \cite{Li2023-sx},
  	empty path elimination \cite{majumdar_et_al:LIPIcs.CONCUR.2021.35},\\&
  	continuation semantic and algorithmic \cite{Castagna2012GlobalTypes} 
  	\\\hline
  \end{tabular}
\end{center}
\label{table:related}
\end{table}
\label{sec:summary}

\smallskip

\noindent\textbf{\cref{table:related}}
categorises the existing projections and 
states which criteria they satisfy.
The projections listed in  ``a-good'' are introduced using $\psubt{}$
in the literature, but
likely
satisfy ``$\text{a-good}^+$''.

\smallskip

\myparagraph{Future Work.} \ The MPST framework has been implemented
in more than 25 programming languages~\cite{Yoshida24}, but most
toolkits (e.g., Scribble~\cite{scribble,scribble10,YHNN2013,nuScrPaper}) support only
full inductive merging and do not support parallel composition.
Extending existing tools to incorporate full coinductive merging and
parallel composition would be effective in increasing
their expressiveness.
In MPST certification work
\cite{DBLP:journals/pacmpl/HinrichsenJK24,DBLP:conf/pldi/Castro-Perez0GY21,DBLP:journals/jar/TiroreBC25,EKY2025,Tirore2025b,KYG2026},
subject reduction has been proved from scratch for each
projection and typing system. Correctness of an
existing typing system with a new projection can be proven 
by verifying that the projection satisfies the good projection
criteria, significantly reducing proof effort.
The precise asynchronous subtyping ($\asubt$) has not been investigated for
systems with mixed choice. Extending $\asubt$ 
and $\text{a-good}^+$ to mixed choice is another  
future direction.

\section*{Data Availability Statement}
\label{sec:dataavailability}
\label{sec:implementation}
We have implemented the following algorithms (1-5) and judgements
in this paper. A prototype implementation 
together with instructions 
to test the claims of this paper 
is available from \cite{anonymous_2026_21242753}. 
\begin{enumerate}
	\item The $LINSPACE$ decision algorithm for $\gtTrace{\gtG}$ (\cref{thm:global-sem}).
	\item The algorithms for checking balanced, weak balanced and $\roleP$-send/receive balanced (\cref{thm:dec-bal}).
	\item The inductive, coinductive, full and plain projection algorithms on directed mixed choice global
types (\cref{def:global-proj-mixed}).
	\item The type judgement on sessions (\cref{def:mixed-processes})
	parameterised by a projection algorithm.
	\item The serialisation function (\cref{thm:remPar}).
\end{enumerate}
(5) is implemented in GHC 9.0.1,
and the rest in Scala 3.8.2.
Notably, the parametrisation of type judgement in (4)
by projection algorithms in (3)
and global type property checking in (2)
provides an implementation of a
`decoupled' typing system.
This modularity over good projections
and global type properties enables both
our session liveness results (\cref{thm:type-bal}) and
the trade-offs between
expressivity and complexity in \cref{sec:comparison}
to be safely and correctly applied
within a unified system.
Deciding a type judgement in (4) is mostly syntax driven,
excepting recursive processes which require subtyping.
We check recursive processes via a
computable inversion lemma
\ifextended (\cref{lem:decide-inv-rec}).
\else \appref{Lem. D.8}.
\fi

\textbf{The full version}
of this paper, which contains the full proofs, 
further examples and definitions, is available from \cite{fullversion}.

\subsubsection*{Acknowledgements.}
We thank the POPL reviewers for their detailed comments
and helpful suggestions.
We thank Mariangiola Dezani-Ciancaglini,
Laura Bocchi, Franco Barbanera and Ugo de'Liguoro for discussions on
their projections and typing systems.
We thank David Castro-Perez,
Francisco Ferreira and
Sung-Shik Jongmans for discussions on \cite{10.1145/3776692}
and its mechanisation. 
We thank Ping Hou and Jessica Richards 
for testing the artifact.

The introduction was inspired by 
Joachim Parrow's talk at the LIX Colloquium (2006) and Ichiro
Hasuo's invited talk at FM~2026.

This work is supported by
an EPSRC-VCA PhD Studentship;
EPSRC
EP/T006544/2,
EP/T014709/2,
EP/Z533749/1 and 
EP/Y005244/1;
ARIA MSAI-PR01-P020 and MSAI-PR01-P029 (PR-SGAI-00110); 
Horizon EU TaRDIS 101093006 (UKRI number 10066667);
and a grant from the Simons Foundation.

\bibliography{main}

@misc{junge2026,
      title={Model Checking Probabilistic Multiparty Session Types}, 
      author={Aleksander Junge and Leo Seojun Lee and Joseph Paulus and Nobuko Yoshida and David Parker},
      year={2026},
      eprint={2611.14795},
      archivePrefix={arXiv},
      primaryClass={cs.PL},
      url={https://arxiv.org/abs/2611.14795},
      note={To appear in POPL'27},
}

@inproceedings{Lanese2024,
author = {Lanese, Ivan and Dal Lago, Ugo and Choudhury, Vikraman},
title = {Towards Quantum Multiparty Session Types},
year = {2024},
isbn = {978-3-031-77381-5},
publisher = {Springer-Verlag},
address = {Berlin, Heidelberg},
url = {https://doi.org/10.1007/978-3-031-77382-2_22},
doi = {10.1007/978-3-031-77382-2_22},
booktitle = {Software Engineering and Formal Methods: 22nd International Conference, SEFM 2024, Aveiro, Portugal, November 6-8, 2024, Proceedings},
pages = {385–403},
numpages = {19},
location = {Aveiro, Portugal}
}

@article{DBLP:journals/lmcs/DagninoGD23,
  author       = {Francesco Dagnino and
                  Paola Giannini and
                  Mariangiola Dezani{-}Ciancaglini},
  title        = {Deconfined Global Types for Asynchronous Sessions},
  journal      = {Log. Methods Comput. Sci.},
  volume       = {19},
  number       = {1},
  year         = {2023},
  url          = {https://doi.org/10.46298/lmcs-19(1:3)2023},
  doi          = {10.46298/LMCS-19(1:3)2023},
bibsource    = {dblp computer science bibliography, https://dblp.org}
}

@article{PY2026,
author = {Pischke, Kai and Yoshida, Nobuko},
title = {Top-Down = Bottom-Up: Sound and Complete Characterisations of Liveness by Multiparty Global Protocols},
year = {2026},
issue_date = {October 2026},
publisher = {Association for Computing Machinery},
address = {New York, NY, USA},
volume = {10},
number = {OOPSLA2},
url = {https://doi.org/10.1145/3839538},
doi = {10.1145/3839538},
journal = {Proc. ACM Program. Lang.},
month = oct,
articleno = {406},
numpages = {32}
}

@phdthesis{Stutz2024,
  author      = {Felix Stutz},
  title       = {Implementability of Asynchronous Communication Protocols - The Power of Choice},
  type        = {doctoral thesis},
  pages       = {XVII, 257},
  school      = {Rheinland-Pf{\"a}lzische Technische Universit{\"a}t Kaiserslautern-Landau},
  doi       = {10.26204/KLUEDO/8077},
  url  = {https://nbn-resolving.de/urn:nbn:de:hbz:386-kluedo-80778},
  year        = {2024},
}

@article{Li2026,
author = {Elaine Li},
title = {{Decision Problems for Global Protocol Specifications}},
Number={149},
year = {2026},
month={June},
journal = {EATCS Bulletin},
publisher = {EATCS},
url = {https://bulletin.eatcs.org/index.php/beatcs/article/view/891},
}

@article{Barbanera2026,
author = {Barbanera, Franco and Dezani-Ciancaglini, Mariangiola and de'Liguoro, Ugo},
title = {{Typing Scalas-Yoshida Benchmarks}},
Number={149},
year = {2026},
month={June},
journal = {EATCS Bulletin},
publisher = {EATCS},
url = {https://bulletin.eatcs.org/index.php/beatcs/article/view/896},
}

@inproceedings{CDPY2015,
  author = {Mario Coppo and Mariangiola Dezani-Ciancaglini and Luca Padovani and Nobuko Yoshida},
  title = {{A Gentle Introduction to Multiparty Asynchronous Session Types}},
  booktitle = {15th International School on Formal Methods for the Design of Computer, Communication and Software Systems: Multicore Programming},
  series = {LNCS},
  volume = {9104},
  pages = {146--178},
  publisher = {Springer},
  year = 2015,
  doi="10.1007/978-3-319-18941-3\_4"
}

@article{Gheri2023,
author = {Gheri, Lorenzo and Yoshida, Nobuko},
title = {Hybrid Multiparty Session Types: Compositionality for Protocol Specification through Endpoint Projection},
year = {2023},
issue_date = {April 2023},
publisher = {Association for Computing Machinery},
address = {New York, NY, USA},
volume = {7},
number = {OOPSLA1},
url = {https://doi.org/10.1145/3586031},
doi = {10.1145/3586031},
journal = {Proc. ACM Program. Lang.},
month = apr,
articleno = {79},
numpages = {31}
}

@inproceedings{DBLP:conf/itp/EkiciY24,
  author       = {Burak Ekici and
                  Nobuko Yoshida},
  editor       = {Yves Bertot and
                  Temur Kutsia and
                  Michael Norrish},
  title        = {Completeness of Asynchronous Session Tree Subtyping in Coq},
  booktitle    = {15th International Conference on Interactive Theorem Proving, {ITP}
                  2024, September 9-14, 2024, Tbilisi, Georgia},
  series       = {LIPIcs},
  volume       = {309},
  pages        = {13:1--13:20},
  publisher    = {Schloss Dagstuhl - Leibniz-Zentrum f{\"{u}}r Informatik},
  year         = {2024},
  url          = {https://doi.org/10.4230/LIPIcs.ITP.2024.13},
  doi          = {10.4230/LIPICS.ITP.2024.13},
  bibsource    = {dblp computer science bibliography, https://dblp.org}
}

@article{DBLP:journals/pacmpl/HinrichsenJK24,
  author       = {Jonas Kastberg Hinrichsen and
                  Jules Jacobs and
                  Robbert Krebbers},
  title        = {Multris: Functional Verification of Multiparty Message Passing in
                  Separation Logic},
  journal      = {Proc. {ACM} Program. Lang.},
  volume       = {8},
  number       = {{OOPSLA2}},
  pages        = {1446--1474},
  year         = {2024},
  url          = {https://doi.org/10.1145/3689762},
  doi          = {10.1145/3689762},
  bibsource    = {dblp computer science bibliography, https://dblp.org}
}

@inproceedings{DBLP:conf/pldi/Castro-Perez0GY21,
  author       = {David Castro{-}Perez and
                  Francisco Ferreira and
                  Lorenzo Gheri and
                  Nobuko Yoshida},
  editor       = {Stephen N. Freund and
                  Eran Yahav},
  title        = {Zooid: a {DSL} for certified multiparty computation: from mechanised
                  metatheory to certified multiparty processes},
  booktitle    = {{PLDI} '21: 42nd {ACM} {SIGPLAN} International Conference on Programming
                  Language Design and Implementation, Virtual Event, Canada, June 20-25,
                  2021},
  pages        = {237--251},
  publisher    = {{ACM}},
  year         = {2021},
  url          = {https://doi.org/10.1145/3453483.3454041},
  doi          = {10.1145/3453483.3454041},
  bibsource    = {dblp computer science bibliography, https://dblp.org}
}

@InProceedings{Tirore2025b,
  author =	{Tirore, Dawit and Bengtson, Jesper and Carbone, Marco},
  title =	{{Multiparty Asynchronous Session Types: A Mechanised Proof of Subject Reduction}},
  booktitle =	{39th European Conference on Object-Oriented Programming (ECOOP 2025)},
  pages =	{31:1--31:30},
  series =	{Leibniz International Proceedings in Informatics (LIPIcs)},
  ISBN =	{978-3-95977-373-7},
  ISSN =	{1868-8969},
  year =	{2025},
  volume =	{333},
  editor =	{Aldrich, Jonathan and Silva, Alexandra},
  publisher =	{Schloss Dagstuhl -- Leibniz-Zentrum f{\"u}r Informatik},
  address =	{Dagstuhl, Germany},
  URL =		{https://drops.dagstuhl.de/entities/document/10.4230/LIPIcs.ECOOP.2025.31},
  URN =		{urn:nbn:de:0030-drops-233238},
  doi =		{10.4230/LIPIcs.ECOOP.2025.31}
}

@article{DBLP:journals/jar/TiroreBC25,
  author       = {Dawit Legesse Tirore and
                  Jesper Bengtson and
                  Marco Carbone},
  title        = {A Sound and Complete Projection for Global Types},
  journal      = {J. Autom. Reason.},
  volume       = {69},
  number       = {2},
  pages        = {14},
  year         = {2025},
  url          = {https://doi.org/10.1007/s10817-025-09726-9},
  doi          = {10.1007/S10817-025-09726-9},
  bibsource    = {dblp computer science bibliography, https://dblp.org}
}

@inproceedings{EKY2025,
  author = {Burak Ekici and Tadayoshi Kamegai and Nobuko Yoshida},
  title = {{Formalising Subject Reduction and Progress for Multiparty Session Processes}},
  booktitle = {16th International Conference on Interactive Theorem Proving, 2025, Reykjavik, Iceland},
  series = {LIPIcs},
  volume = {352},
  pages = {19:1--19:23},
  publisher = {Schloss Dagstuhl - Leibniz-Zentrum f{\"u}r Informatik},
  doi = "https://doi.org/10.4230/LIPIcs.ITP.2025.19",
  year = 2025
}

@InProceedings{Li2025,
  author =	{Li, Elaine and Wies, Thomas},
  title =	{{Certified Implementability of Global Multiparty Protocols}},
  booktitle =	{16th International Conference on Interactive Theorem Proving (ITP 2025)},
  pages =	{15:1--15:20},
  series =	{Leibniz International Proceedings in Informatics (LIPIcs)},
  ISBN =	{978-3-95977-396-6},
  ISSN =	{1868-8969},
  year =	{2025},
  volume =	{352},
  editor =	{Forster, Yannick and Keller, Chantal},
  publisher =	{Schloss Dagstuhl -- Leibniz-Zentrum f{\"u}r Informatik},
  address =	{Dagstuhl, Germany},
  URL =		{https://drops.dagstuhl.de/entities/document/10.4230/LIPIcs.ITP.2025.15},
  URN =		{urn:nbn:de:0030-drops-246139},
  doi =		{10.4230/LIPIcs.ITP.2025.15}
}

@INPROCEEDINGS{ALUR2000,
  author={Alur, R. and Etessami, K. and Yannakakis, M.},
  booktitle={Proceedings of the 2000 International Conference on Software Engineering. ICSE 2000 the New Millennium}, 
  title={Inference of message sequence charts}, 
  year={2000},
  volume={},
  number={},
  pages={304-313},
  doi={10.1145/337180.337215}
 }

@article{ALUR2005,
title = {Realizability and verification of MSC graphs},
journal = {Theoretical Computer Science},
volume = {331},
number = {1},
pages = {97-114},
year = {2005},
issn = {0304-3975},
doi = {https://doi.org/10.1016/j.tcs.2004.09.034},
url = {https://www.sciencedirect.com/science/article/pii/S030439750400670X},
author = {Rajeev Alur and Kousha Etessami and Mihalis Yannakakis}
}

@inproceedings{Giusto2025,
author = {Di Giusto, Cinzia and Lozes, Etienne and Urso, Pascal},
title = {Realisability and Complementability of Multiparty Session Types},
year = {2025},
isbn = {9798400720857},
publisher = {Association for Computing Machinery},
address = {New York, NY, USA},
url = {https://doi.org/10.1145/3756907.3756918},
doi = {10.1145/3756907.3756918},
booktitle = {Proceedings of the 27th International Symposium on Principles and Practice of Declarative Programming},
articleno = {11},
numpages = {12},
location = {
},
series = {PPDP '25}
}

@article{Li2025c,
author = {Li, Elaine and Stutz, Felix and Wies, Thomas and Zufferey, Damien},
title = {Characterizing Implementability of Global Protocols with Infinite States and Data},
year = {2025},
issue_date = {April 2025},
publisher = {Association for Computing Machinery},
address = {New York, NY, USA},
volume = {9},
number = {OOPSLA1},
url = {https://doi.org/10.1145/3720493},
doi = {10.1145/3720493},
journal = {Proc. ACM Program. Lang.},
month = apr,
articleno = {131},
numpages = {30}
}

@book{PierceTAPL,
  author = {Pierce, Benjamin C.},
  title = {Types and Programming Languages},
  year = {2002},
  isbn = {0262162091, 9780262162098},
  edition = {1st},
  publisher = {The MIT Press},
  doi = "10.5555/509043"
}

@article{POPL19LessIsMore,
 author = {Scalas, Alceste and Yoshida, Nobuko},
 title = {{Less is More: Multiparty Session Types Revisited}},
 journal = {Proc. ACM Program. Lang.},
 issue_date = {January 2019},
 volume = {3},
 number = {POPL},
 month = jan,
 year = {2019},
 issn = {2475-1421},
 pages = {30:1--30:29},
 articleno = {30},
 numpages = {29},
 _url = {http://doi.acm.org/10.1145/3290343},
 doi = {10.1145/3290343},
 acmid = {3290343},
 publisher = {ACM},
 address = {New York, NY, USA},
 keywords = {deadlock-freedom, duality, liveness, session types},
}

@INPROCEEDINGS{VanGlabbeekLICS2021,
  author={Glabbeek, Rob van and Höfner, Peter and Horne, Ross},
  booktitle={2021 36th Annual ACM/IEEE Symposium on Logic in Computer Science (LICS)},
  title={Assuming Just Enough Fairness to make Session Types Complete for Lock-freedom},
  year={2021},
  volume={},
  number={},
  pages={1-13},
  doi={10.1109/LICS52264.2021.9470531}}

@InProceedings{nuScrPaper,
author="Yoshida, Nobuko
and Zhou, Fangyi
and Ferreira, Francisco",
editor="Bampis, Evripidis
and Pagourtzis, Aris",
title="{Communicating Finite State Machines and an Extensible Toolchain
          for Multiparty Session Types}",
booktitle="Fundamentals of Computation Theory",
year="2021",
publisher="Springer International Publishing",
address="Cham",
pages="18--35",
isbn="978-3-030-86593-1",
doi = "10.1007/978-3-030-86593-1_2",
}

@inproceedings{DBLP:conf/icdcit/YoshidaG20,
  author    = {Nobuko Yoshida and
               Lorenzo Gheri},
  editor    = {Dang Van Hung and
               Meenakshi D'Souza},
  title     = {A Very Gentle Introduction to Multiparty Session Types},
  booktitle = {Distributed Computing and Internet Technology - 16th International
               Conference, {ICDCIT} 2020, Bhubaneswar, India, January 9-12, 2020,
               Proceedings},
  series    = {Lecture Notes in Computer Science},
  volume    = {11969},
  pages     = {73--93},
  publisher = {Springer},
  year      = {2020},
  _url       = {https://doi.org/10.1007/978-3-030-36987-3\_5},
  doi       = {10.1007/978-3-030-36987-3\_5},
}

@inproceedings{MFYZ2021,
  author = {Anson Miu and Francisco Ferreira and Nobuko Yoshida and Fangyi Zhou},
  title = {{Communication-Safe Web Programming in TypeScript with Routed Multiparty Session Types}},
  booktitle = {International Conference on Compiler Construction},
  series = {CC},
  pages = {94--106},
  doi = "10.1145/3446804.3446854",
  year = 2021
}

@inproceedings{BSYZ2022,
  author = {Adam Barwell and Alceste Scalas and Nobuko Yoshida and Fangyi Zhou},
  title = {{Generalised Multiparty Session Types with Crash-Stop Failures}},
  booktitle = {33rd International Conference on Concurrency Theory},
  series = {LIPIcs},
  volume = {243},
  pages = {35:1--35:25},
  publisher = {Dagstuhl},
  doi = "10.4230/LIPIcs.CONCUR.2022.35",
  year = 2022
}

@article{GPPSY2023,
  author = {Silvia Ghilezan and Jovanka Pantović and Ivan Prokić and Alceste Scalas and Nobuko Yoshida},
  title = {{Precise Subtyping for Asynchronous Multiparty Sessions}},
  journal = {ACM Transactions on Computational Logic},
  volume = {Volume 24, Issue 2}, 
  number    = {14},
  pages = {1--73},
  publisher = {Association for Computing Machinery},
  doi = "10.1145/3568422",
  year = 2023
}

@inproceedings{ICALP13CFSM,
  author = {Pierre-Malo Deniélou and Nobuko Yoshida},
  title = {{Multiparty Compatibility in Communicating Automata: Characterisation and Synthesis of Global Session Types}},
  booktitle = {40th International Colloquium on Automata, Languages and Programming},
  series = {LNCS},
  volume = {7966},
  pages = {174--186},
  publisher = {Springer},
  address = {Berlin, Heidelberg},
  doi = {10.1007/978-3-642-39212-2\_18},
  year = 2013
}

@incollection{Yoshida24,
  author       = {Nobuko Yoshida},
  editor       = {Frank S. de Boer and
                  Ferruccio Damiani and
                  Reiner H{\"{a}}hnle and
                  Einar Broch Johnsen and
                  Eduard Kamburjan},
  title        = {Programming Language Implementations with Multiparty Session Types},
  booktitle    = {Active Object Languages: Current Research Trends},
  series       = {Lecture Notes in Computer Science},
  volume       = {14360},
  pages        = {147--165},
  publisher    = {Springer},
  year         = {2024},
  url          = {https://doi.org/10.1007/978-3-031-51060-1\_6},
  doi          = {10.1007/978-3-031-51060-1\_6},
  bibsource    = {dblp computer science bibliography, https://dblp.org}
}

@inproceedings{CYV2022,
author = {Cutner, Zak and Yoshida, Nobuko and Vassor, Martin},
title = {Deadlock-free asynchronous message reordering in rust with multiparty session types},
year = {2022},
isbn = {9781450392044},
publisher = {Association for Computing Machinery},
address = {New York, NY, USA},
url = {https://doi.org/10.1145/3503221.3508404},
doi = {10.1145/3503221.3508404},
booktitle = {Proceedings of the 27th ACM SIGPLAN Symposium on Principles and Practice of Parallel Programming},
pages = {246–261},
numpages = {16},
location = {Seoul, Republic of Korea},
series = {PPoPP '22}
}

@article{lmcs2025,
  author = {Adam D. Barwell and Ping Hou and Nobuko Yoshida and Fangyi Zhou},
   title = {{Crash-Stop Failures in Asynchronous Multiparty Session Types}}, 
  journal = {Logical Methods in Computer Science},
  volume = {21},
  issue = {2},
  doi = "10.46298/lmcs-21(2:5)2025",
  year = 2025
}

@article{HYK2026,
title = {Less is more revisited: Association with global protocols and multiparty sessions},
journal = {Theoretical Computer Science},
volume = {1076},
pages = {115873},
year = {2026},
issn = {0304-3975},
doi = {https://doi.org/10.1016/j.tcs.2026.115873},
url = {https://www.sciencedirect.com/science/article/pii/S0304397526001325},
author = {Ping Hou and Nobuko Yoshida and Iona Kuhn}
}

@incollection{DBLP:series/lncs/YoshidaH24,
  author       = {Nobuko Yoshida and
                  Ping Hou},
  editor       = {Ana Cavalcanti and
                  James Baxter},
  title        = {Less is More Revisited: Association with Global Multiparty Session
                  Types},
  booktitle    = {The Practice of Formal Methods: Essays in Honour of Cliff Jones, Part
                  {II}},
  series       = {Lecture Notes in Computer Science},
  volume       = {14781},
  pages        = {268--291},
  publisher    = {Springer},
  year         = {2024},
  url          = {https://doi.org/10.1007/978-3-031-66673-5\_14},
  doi          = {10.1007/978-3-031-66673-5\_14},
  bibsource    = {dblp computer science bibliography, https://dblp.org}
}

@ARTICLE{Barbanera2021-mg,
  title     = "Composition and decomposition of multiparty sessions",
  author    = "Barbanera, Franco and Dezani-Ciancaglini, Mariangiola and
               Lanese, Ivan and Tuosto, Emilio",
  journal   = "J. Log. Algebr. Methods Program.",
  publisher = "Elsevier BV",
  volume    =  119,
  number    =  100620,
  pages     = "100620",
  month     =  feb,
  year      =  2021,
  copyright = "http://www.elsevier.com/open-access/userlicense/1.0/",
  language  = "en",
  doi = "10.1016/j.jlamp.2020.100620"
}

@InProceedings{majumdar_et_al:LIPIcs.CONCUR.2021.35,
  author =	{Majumdar, Rupak and Mukund, Madhavan and Stutz, Felix and Zufferey, Damien},
  title =	{{Generalising Projection in Asynchronous Multiparty Session Types}},
  booktitle =	{32nd International Conference on Concurrency Theory (CONCUR 2021)},
  pages =	{35:1--35:24},
  series =	{Leibniz International Proceedings in Informatics (LIPIcs)},
  ISBN =	{978-3-95977-203-7},
  ISSN =	{1868-8969},
  year =	{2021},
  volume =	{203},
  editor =	{Haddad, Serge and Varacca, Daniele},
  publisher =	{Schloss Dagstuhl -- Leibniz-Zentrum f{\"u}r Informatik},
  address =	{Dagstuhl, Germany},
  URL =		{https://drops.dagstuhl.de/entities/document/10.4230/LIPIcs.CONCUR.2021.35},
  URN =		{urn:nbn:de:0030-drops-144125},
  doi =		{10.4230/LIPIcs.CONCUR.2021.35}
}

@inproceedings{10.1145/3678232.3678245,
author = {Barbanera, Franco and Dezani-Ciancaglini, Mariangiola and de'Liguoro, Ugo},
title = {Un-projectable Global Types for Multiparty Sessions},
year = {2024},
isbn = {9798400709692},
publisher = {Association for Computing Machinery},
address = {New York, NY, USA},
url = {https://doi.org/10.1145/3678232.3678245},
doi = {10.1145/3678232.3678245},
booktitle = {Proceedings of the 26th International Symposium on Principles and Practice of Declarative Programming},
articleno = {15},
numpages = {13},
location = {Milano, Italy},
series = {PPDP '24}
}

@inproceedings{PY2024,
author = {Peters, Kirstin and Yoshida, Nobuko},
title = {Separation and Encodability in Mixed Choice Multiparty Sessions},
year = {2024},
isbn = {9798400706608},
publisher = {Association for Computing Machinery},
address = {New York, NY, USA},
url = {https://doi.org/10.1145/3661814.3662085},
doi = {10.1145/3661814.3662085},
booktitle = {Proceedings of the 39th Annual ACM/IEEE Symposium on Logic in Computer Science},
articleno = {62},
numpages = {15},
location = {Tallinn, Estonia},
series = {LICS '24}
}

@ARTICLE{pischke2026asynchronousglobalprotocolsprecisely,
  title     = "Asynchronous global protocols, precisely",
  author    = "Pischke, Kai and Masters, Jake and Yoshida, Nobuko",
  journal   = "Inf. Comput.",
  publisher = "Elsevier BV",
  volume    =  312,
  number    =  105501,
  pages     = "105501",
  month     =  sep,
  year      =  2026,
  copyright = "http://creativecommons.org/licenses/by/4.0/",
  language  = "en",
  doi = "10.1016/j.ic.2026.105501"
}

@inproceedings{LY2017,
  author = {Julien Lange and Nobuko Yoshida},
  title = {{On the Undecidability of Asynchronous Session Subtyping}},
  booktitle = {20th International Conference on Foundations of Software Science and Computation Structures},
  series = {LNCS},
  volume = {10203},
  pages = {441--457},
  publisher = {Springer},
  doi = "10.1007/978-3-662-54458-7_26",
  year = 2017
}

@InProceedings{bocchi_et_al:LIPIcs.CONCUR.2025.10,
  author =	{Bocchi, Laura and King, Andy and Murgia, Maurizio and Thompson, Simon},
  title =	{{Abstract Subtyping for Asynchronous Multiparty Sessions}},
  booktitle =	{36th International Conference on Concurrency Theory (CONCUR 2025)},
  pages =	{10:1--10:19},
  series =	{Leibniz International Proceedings in Informatics (LIPIcs)},
  ISBN =	{978-3-95977-389-8},
  ISSN =	{1868-8969},
  year =	{2025},
  volume =	{348},
  editor =	{Bouyer, Patricia and van de Pol, Jaco},
  publisher =	{Schloss Dagstuhl -- Leibniz-Zentrum f{\"u}r Informatik},
  address =	{Dagstuhl, Germany},
  URL =		{https://drops.dagstuhl.de/entities/document/10.4230/LIPIcs.CONCUR.2025.10},
  URN =		{urn:nbn:de:0030-drops-239605},
  doi =		{10.4230/LIPIcs.CONCUR.2025.10}
}

@InProceedings{Bocchi2024-hs,
author="Bocchi, Laura
and King, Andy
and Murgia, Maurizio",
editor="Finkbeiner, Bernd
and Kov{\'a}cs, Laura",
title="Asynchronous Subtyping by Trace Relaxation",
booktitle="Tools and Algorithms for the Construction and Analysis of Systems",
year="2024",
publisher="Springer Nature Switzerland",
address="Cham",
pages="207--226",
isbn="978-3-031-57246-3",
doi = {10.1007/978-3-031-57246-3_12}
}

@InProceedings{Li2024-mu,
author="Li, Elaine
and Stutz, Felix
and Wies, Thomas",
editor="Weirich, Stephanie",
title="Deciding Subtyping for Asynchronous Multiparty Sessions",
booktitle="Programming Languages and Systems",
year="2024",
publisher="Springer Nature Switzerland",
address="Cham",
pages="176--205",
isbn="978-3-031-57262-3",
doi = {10.1007/978-3-031-57262-3_8}
}

@inproceedings{10.1007/978-3-031-91121-7_13,
author = {Stutz, Felix and D’Osualdo, Emanuele},
title = {An Automata-theoretic Basis for Specification and Type Checking of Multiparty Protocols},
year = {2025},
isbn = {978-3-031-91120-0},
publisher = {Springer-Verlag},
address = {Berlin, Heidelberg},
url = {https://doi.org/10.1007/978-3-031-91121-7_13},
doi = {10.1007/978-3-031-91121-7_13},
booktitle = {Programming Languages and Systems: 34th European Symposium on Programming, ESOP 2025, Held as Part of the International Joint Conferences on Theory and Practice of Software, ETAPS 2025, Hamilton, ON, Canada, May 3–8, 2025, Proceedings, Part II},
pages = {314–346},
numpages = {33},
location = {Hamilton, ON, Canada}
}

@article{10.1145/3776692,
author = {Castro-Perez, David and Ferreira, Francisco and Jongmans, Sung-Shik},
title = {A Synthetic Reconstruction of Multiparty Session Types},
year = {2026},
issue_date = {January 2026},
publisher = {Association for Computing Machinery},
address = {New York, NY, USA},
volume = {10},
number = {POPL},
url = {https://doi.org/10.1145/3776692},
doi = {10.1145/3776692},
journal = {Proc. ACM Program. Lang.},
month = jan,
articleno = {50},
numpages = {29}
}

@InProceedings{HLY2024,
  author =	{Hou, Ping and Lagaillardie, Nicolas and Yoshida, Nobuko},
  title =	{{Fearless Asynchronous Communications with Timed Multiparty Session Protocols}},
  booktitle =	{38th European Conference on Object-Oriented Programming (ECOOP 2024)},
  pages =	{19:1--19:30},
  series =	{Leibniz International Proceedings in Informatics (LIPIcs)},
  ISBN =	{978-3-95977-341-6},
  ISSN =	{1868-8969},
  year =	{2024},
  volume =	{313},
  editor =	{Aldrich, Jonathan and Salvaneschi, Guido},
  publisher =	{Schloss Dagstuhl -- Leibniz-Zentrum f{\"u}r Informatik},
  address =	{Dagstuhl, Germany},
  URL =		{https://drops.dagstuhl.de/entities/document/10.4230/LIPIcs.ECOOP.2024.19},
  URN =		{urn:nbn:de:0030-drops-208681},
  doi =		{10.4230/LIPIcs.ECOOP.2024.19}
}

@ARTICLE{doi:10.3233/FI-2019-1863,
  title     = "Observational Equivalence for Multiparty Sessions",
  author    = "Severi, Paula and Dezani-Ciancaglini, Mariangiola",
  journal   = "Fundam. Inform.",
  publisher = "SAGE Publications",
  volume    =  170,
  number    = "1-3",
  pages     = "267--305",
  month     =  oct,
  year      =  2019,
  doi = {10.3233/FI-2019-1863},
  URL = {https://journals.sagepub.com/doi/abs/10.3233/FI-2019-1863}
}

@article{10.1145/3798256,
author = {Bocchi, Laura and Hu, Raymond and Voinea, Adriana Laura and Thompson, Simon},
title = {Mixed Choice in Asynchronous Multiparty Session Types},
year = {2026},
issue_date = {April 2026},
publisher = {Association for Computing Machinery},
address = {New York, NY, USA},
volume = {10},
number = {OOPSLA1},
url = {https://doi.org/10.1145/3798256},
doi = {10.1145/3798256},
journal = {Proc. ACM Program. Lang.},
month = apr,
articleno = {148},
numpages = {28},
note = {Full version available at \url{https://arxiv.org/abs/2602.23927}}
}

@inproceedings{Li2023-sx,
author = {Li, Elaine and Stutz, Felix and Wies, Thomas and Zufferey, Damien},
title = {Complete Multiparty Session Type Projection with Automata},
year = {2023},
isbn = {978-3-031-37708-2},
publisher = {Springer-Verlag},
address = {Berlin, Heidelberg},
url = {https://doi.org/10.1007/978-3-031-37709-9_17},
doi = {10.1007/978-3-031-37709-9_17},
booktitle = {Computer Aided Verification: 35th International Conference, CAV 2023, Paris, France, July 17–22, 2023, Proceedings, Part III},
pages = {350–373},
numpages = {24},
location = {Paris, France}
}

@InProceedings{Padovani2014-zb,
author="Padovani, Luca
and Vasconcelos, Vasco Thudichum
and Vieira, Hugo Torres",
editor="K{\"u}hn, Eva
and Pugliese, Rosario",
title="Typing Liveness in Multiparty Communicating Systems",
booktitle="Coordination Models and Languages",
year="2014",
publisher="Springer Berlin Heidelberg",
address="Berlin, Heidelberg",
pages="147--162",
isbn="978-3-662-43376-8",
doi = "10.1007/978-3-662-43376-8_10"
}

@ARTICLE{Van_den_Heuvel2022-ex,
  title     = "A decentralized analysis of multiparty protocols",
  author    = "van den Heuvel, Bas and P{\'e}rez, Jorge A",
  journal   = "Sci. Comput. Program.",
  publisher = "Elsevier BV",
  volume    =  222,
  number    =  102840,
  pages     = "102840",
  month     =  oct,
  year      =  2022,
  copyright = "http://creativecommons.org/licenses/by/4.0/",
  language  = "en",
  doi = "10.1016/j.scico.2022.102840"
}

@inproceedings{PPGSY2025,
  author = {Ivan Prokić and Simona Prokić and Silvia Ghilezan and Alceste Scalas and Nobuko Yoshida},
  title = {{On Asynchronous Multiparty Session Types for Federated Learning}},
  booktitle = {22nd International Colloquium on Theoretical Aspects of Computing, 2025, Marrakech, Morocco},
  series = {Lecture Notes in Computer Science},
  volume = {16237},
  pages = {164--182},
  publisher = {Springer Nature},
  doi = "https://doi.org/10.1007/978-3-032-11176-0_11",
  year = 2025
}

@inproceedings{DBLP:journals/corr/abs-2508-13616,
  author       = {Franco Barbanera and
                  Mariangiola Dezani{-}Ciancaglini},
  editor       = {Cl{\'{e}}ment Aubert and
                  Cinzia Di Giusto and
                  Simon Fowler and
                  Violet Ka I Pun},
  title        = {Modular Multiparty Sessions with Mixed Choice},
  booktitle    = {Proceedings 18th Interaction and Concurrency Experience, {ICE} 2025,
                  Lille, France, 20th June 2025},
  series       = {{EPTCS}},
  volume       = {425},
  pages        = {2--20},
  year         = {2025},
  month        = aug,
  url          = {https://doi.org/10.4204/EPTCS.425.2},
  doi          = "10.4204/EPTCS.425.2",
  bibsource    = {dblp computer science bibliography, https://dblp.org}
}

@article{Brand:1983:CFSM,
 author = {Brand, Daniel and Zafiropulo, Pitro},
 title = {On Communicating Finite-State Machines},
 journal = {JACM},
 optissue_date = {April 1983},
 volume = {30},
 number = {2},
 optmonth = apr,
 year = {1983},
 optissn = {0004-5411},
 optpages = {323--342},
 optnumpages = {20},
 opturl = {http://doi.acm.org/10.1145/322374.322380},
 doi = {10.1145/322374.322380},
 optacmid = {322380},
 optpublisher = {ACM},
 optaddress = {New York, NY, USA},
}

@article{GJPSY2018,
title = {Precise subtyping for synchronous multiparty sessions},
journal = {Journal of Logical and Algebraic Methods in Programming},
volume = {104},
pages = {127-173},
year = {2019},
issn = {2352-2208},
doi = {10.1016/j.jlamp.2018.12.002},
url = {https://www.sciencedirect.com/science/article/pii/S2352220817302237},
author = {Silvia Ghilezan and Svetlana Jakšić and Jovanka Pantović and Alceste Scalas and Nobuko Yoshida}
}

@inproceedings{Denielou2012,
author="Deni{\'e}lou, Pierre{-}Malo
and Yoshida, Nobuko",
opteditor="Seidl, Helmut",
title="{Multiparty Session Types Meet Communicating Automata}",
bookTitle={{ESOP}},
year="2012",
optpublisher="Springer Berlin Heidelberg",
optaddress="Berlin, Heidelberg",
optpages="194--213",
optisbn="978-3-642-28869-2",
doi="10.1007/978-3-642-28869-2\_10",
}

@inproceedings{HYC08,
  author    = {Kohei Honda and
               Nobuko Yoshida and
               Marco Carbone},
  title     = {Multiparty asynchronous session types},
  booktitle = {{POPL}},
  year      = {2008},
  optpages     = {273-284},
  optvolume    = {43(1)},
  optpublisher = {ACM},
  doi       = {10.1145/1328438.1328472},
  note = {{J}ournal version: \cite{HYC16}},
  OPTee        = {http://doi.acm.org/10.1145/1328438.1328472},
  OPTbibsource = {DBLP, http://dblp.uni-trier.de}
}

@article{HYC16,
 author = {Honda, Kohei and Yoshida, Nobuko and Carbone, Marco},
 title = {{Multiparty Asynchronous Session Types}},
 journal = {J.~ACM},
 optissue_date = {March 2016},
 volume = {63},
 number = {1},
 optmonth = mar,
 year = {2016},
 optissn = {0004-5411},
 optpages = {9:1--9:67},
 articleno = {9},
 optnumpages = {67},
 opturl = {http://doi.acm.org/10.1145/2827695},
 doi = {10.1145/2827695},
 optacmid = {2827695},
 optpublisher = {ACM},
 optaddress = {New York, NY, USA},
}

@article{ParameterisedYDBH12,
  author    = {Pierre{-}Malo Deni{\'{e}}lou and
               Nobuko Yoshida and
               Andi Bejleri and
               Raymond Hu},
  title     = {{Parameterised Multiparty Session Types}},
  journal   = {{LMCS}},
  volume    = {8},
  number    = {4},
  year      = {2012},
  opturl       = {http://dx.doi.org/10.2168/LMCS-8(4:6)2012},
  doi       = {10.2168/LMCS-8(4:6)2012},
  bibsource = {dblp computer science bibliography, http://dblp.org}
}

@misc{scribble,
  author={Scribble},
  year={2017},
  title = {project home page},
  howpublished = {\url{http://www.scribble.org}}
}

@inproceedings{SDHY2017,
  author = {Alceste Scalas and Ornela Dardha and Raymond Hu and Nobuko Yoshida},
  title = {{A Linear Decomposition of Multiparty Sessions for Safe Distributed Programming}},
  booktitle = {{ECOOP}},
  optseries = {LIPIcs},
  optvolume = {74},
  optpages = {24:1--24:31},
  optpublisher = {Schloss Dagstuhl},
  doi = "10.4230/LIPIcs.ECOOP.2017.24",
  year = 2017
}

@inproceedings{YHNN2013,
  author = "Nobuko Yoshida and Raymond Hu and Rumyana Neykova and Nicholas Ng",
  title = {{The Scribble Protocol Language}},
  booktitle = "Trustworthy Global Computing",
  series = "LNCS",
  volume = "8358",
  publisher = "Springer",
  pages = "22--41",
  year = "2013",
  address = {Buenos Aires, Argentina},
  doi = "10.1007/978-3-319-05119-2_3"
}

@inproceedings{scribble10,
	Author = {Kohei Honda and Aybek Mukhamedov and Gary Brown and Tzu-Chun Chen and Nobuko Yoshida},
	Booktitle = {ICDCIT},
	Publisher = {Springer},
	Series = {LNCS},
	Title = {Scribbling Interactions with a Formal Foundation},
	Volume = {6536},
	Year = {2011},
	Pages= {55-75},
    doi = "10.1007/978-3-642-19056-8_4"
	}

@article{TY2017,
  author = {Bernardo Toninho and Nobuko Yoshida},
  title = {{Certifying data in multiparty session types}},
  journal = {JLAMP},
  volume = {90},
  optpublisher = {Elsevier},
  doi = {10.1016/j.jlamp.2016.11.005},
  year = 2017
}

@inproceedings{CarboneLMSW16,
  author    = {Marco Carbone and
               Sam Lindley and
               Fabrizio Montesi and
               Carsten Sch{\"{u}}rmann and
               Philip Wadler},
  title     = {Coherence Generalises Duality: {A} Logical Explanation of Multiparty
               Session Types},
  booktitle = {{CONCUR}},
  optpages     = {33:1--33:15},
  year      = {2016},
  optcrossref  = {DBLP:conf/concur/2016},
  opturl       = {https://doi.org/10.4230/LIPIcs.CONCUR.2016.33},
  doi       = {10.4230/LIPIcs.CONCUR.2016.33},
  bibsource = {dblp computer science bibliography, http://dblp.org}
}

@inproceedings{CarboneMSY15,
  author    = {Marco Carbone and
               Fabrizio Montesi and
               Carsten Sch{\"{u}}rmann and
               Nobuko Yoshida},
  title     = {Multiparty Session Types as Coherence Proofs},
  booktitle = {{CONCUR}},
  optpages     = {412--426},
  year      = {2015},
  optcrossref  = {DBLP:conf/concur/2015},
  opturl       = {https://doi.org/10.4230/LIPIcs.CONCUR.2015.412},
  doi       = {10.4230/LIPIcs.CONCUR.2015.412},
  bibsource = {dblp computer science bibliography, http://dblp.org}
}

@article{CHEN2015708,
title = "Lightening global types",
optjournal = "Journal of Logical and Algebraic Methods in Programming",
journal = "JLAMP",
volume = "84",
number = "5",
optpages = "708 - 729",
year = "2015",
optissn = "2352-2208",
doi = "https://doi.org/10.1016/j.jlamp.2015.06.003",
opturl = "http://www.sciencedirect.com/science/article/pii/S235222081500053X",
author = "Tzu-Chun Chen",
optkeywords = "Communication centred programming",
optkeywords = "Session types",
optkeywords = "Global types"
}

@misc{OAUTH,
  author = {{OAuth Working Group}},
  title = {{RFC} 6749: {OAuth} 2.0 Framework},
  howpublished = {\url{http://tools.ietf.org/html/rfc6749}},
  year = 2012
}

@article{Castagna2012GlobalTypes,
  author    = {Giuseppe Castagna and
               Mariangiola Dezani{-}Ciancaglini and
               Luca Padovani},
  title     = {On Global Types and Multi-Party Session},
  journal   = {Logical Methods in Computer Science},
  volume    = {8},
  number    = {1},
  year      = {2012},
  opturl       = {https://doi.org/10.2168/LMCS-8(1:24)2012},
  doi       = {10.2168/LMCS-8(1:24)2012},
  bibsource = {dblp computer science bibliography, https://dblp.org}
}

@article{DBLP:journals/pacmpl/UdomsrirungruangY25,
  author       = {Thien Udomsrirungruang and
                  Nobuko Yoshida},
  title        = {{Top-Down or Bottom-Up? Complexity Analyses of Synchronous Multiparty
                  Session Types}},
  journal      = {Proc. {ACM} Program. Lang.},
  volume       = {9},
  number       = {{POPL}},
  pages        = {1040--1071},
  year         = {2025},
  url          = {https://doi.org/10.1145/3704872},
  doi          = {10.1145/3704872},
  bibsource    = {dblp computer science bibliography, https://dblp.org}
}

@ARTICLE{Barbanera2026-ar,
  title     = "Asynchronous multiparty sessions with mixed choice",
  author    = "Barbanera, Franco and Dezani-Ciancaglini, Mariangiola",
  journal   = "Electron. Proc. Theor. Comput. Sci.",
  publisher = "Open Publishing Association",
  volume    =  444,
  pages     = "11--22",
  month     =  apr,
  year      =  2026,
  language  = "en",
  doi = "10.4204/EPTCS.444.2"
}

@inproceedings{KYG2026,
  author = {Omer Keskin and Nobuko Yoshida and Rob van Glabbeek},
  title = {{ Formally Verified Liveness with Multiparty Session Types in Rocq}},
  booktitle = {17th International Conference on Interactive Theorem Proving, 2026, Lisbon, Portugal},
  series = {LIPIcs},
  volume = {382},
  pages = {4:1--4:21},
  publisher = {Schloss Dagstuhl - Leibniz-Zentrum f{\"u}r Informatik},
  doi = "https://doi.org/10.4230/LIPIcs.ITP.2026.4",
  year = 2026
}

@software{anonymous_2026_21242753,
  author       = {Masters, Jake and
                  Yoshida, Nobuko},
  title        = {Top-Down Multiparty Session Types: Serialisation;
                   Balanced Checking; Global Type Languages; and
                   Projections
                  },
  month        = oct,
  year         = 2026,
  publisher    = {Zenodo},
  doi          = {10.5281/zenodo.23232673},
  url          = {https://doi.org/10.5281/zenodo.23232673},
}

@article{Casetta_2026,
   title={Towards Multiparty Session Types for Highly-Concurrent and Fault-Tolerant Web Applications},
   volume={444},
   ISSN={2075-2180},
   url={http://dx.doi.org/10.4204/EPTCS.444.7},
   DOI={10.4204/eptcs.444.7},
   journal={Electronic Proceedings in Theoretical Computer Science},
   publisher={Open Publishing Association},
   author={Casetta, Richard and Gesbert, Nils and Genevès, Pierre},
   year={2026},
   month=Apr, pages={68–78} }

@InProceedings{jongmans_et_al:LIPIcs.ECOOP.2023.42,
  author =	{Jongmans, Sung-Shik and Ferreira, Francisco},
  title =	{{Synthetic Behavioural Typing: Sound, Regular Multiparty Sessions via Implicit Local Types}},
  booktitle =	{37th European Conference on Object-Oriented Programming (ECOOP 2023)},
  pages =	{42:1--42:30},
  series =	{Leibniz International Proceedings in Informatics (LIPIcs)},
  ISBN =	{978-3-95977-281-5},
  ISSN =	{1868-8969},
  year =	{2023},
  volume =	{263},
  editor =	{Ali, Karim and Salvaneschi, Guido},
  publisher =	{Schloss Dagstuhl -- Leibniz-Zentrum f{\"u}r Informatik},
  address =	{Dagstuhl, Germany},
  URL =		{https://drops.dagstuhl.de/entities/document/10.4230/LIPIcs.ECOOP.2023.42},
  URN =		{urn:nbn:de:0030-drops-182358},
  doi =		{10.4230/LIPIcs.ECOOP.2023.42}
}

@inproceedings{JY2020,
  author = {Sung-Shik Jongmans and Nobuko Yoshida},
  title = {{Exploring Type-Level Bisimilarity towards More Expressive Multiparty Session Types}},
  booktitle = {29th European Symposium on Programming},
  series = {LNCS},
  volume = {12075},
  pages = {251--279},
  publisher = {Springer},
  doi = "10.1007/978-3-030-44914-8_10",
  year = 2020
}

@InProceedings{larsen_et_al:LIPIcs.ECOOP.2026.17,
  author =	{Larsen, Jens Kanstrup and Scalas, Alceste and Amir, Guy and Jacobs, Jules and Wagemaker, Jana and Foster, Nate},
  title =	{{NEST: Network Enforced Session Types}},
  booktitle =	{40th European Conference on Object-Oriented Programming (ECOOP 2026)},
  pages =	{17:1--17:31},
  series =	{Leibniz International Proceedings in Informatics (LIPIcs)},
  ISBN =	{978-3-95977-423-9},
  ISSN =	{1868-8969},
  year =	{2026},
  volume =	{372},
  editor =	{Krebbers, Robbert and Silva, Alexandra},
  publisher =	{Schloss Dagstuhl -- Leibniz-Zentrum f{\"u}r Informatik},
  address =	{Dagstuhl, Germany},
  URL =		{https://drops.dagstuhl.de/entities/document/10.4230/LIPIcs.ECOOP.2026.17},
  URN =		{urn:nbn:de:0030-drops-261137},
  doi =		{10.4230/LIPIcs.ECOOP.2026.17}
}

@article{CDYP2015,
  author = {Mario Coppo and Mariangiola Dezani-Ciancaglini and Nobuko Yoshida and Luca Padovani},
  title = {{Global Progress for Dynamically Interleaved Multiparty Sessions}},
  journal = {Mathematical Structures in Computer Science},
  volume = {26},
  issue = {2},
  pages = {238--302},
  publisher = {Cambridge University Press},
  doi = "10.1017/S0960129514000188",
  year = 2015
}

@misc{fullversion,
      title={What is a Good Projection?: Correctness and Separation Results of End-Point Projections}, 
      author={Jake Masters and Nobuko Yoshida},
      year={2026},
      eprint={2610.28525},
      archivePrefix={arXiv},
      primaryClass={cs.PL},
      url={https://arxiv.org/abs/2610.28525}, 
}

\ifextended

\newpage

\appendix

\part[Appendices]{}

\etocsettocstyle{\vskip0.3\baselineskip}{\noindent\rule{\linewidth}{0.5pt}\vskip0.5\baselineskip}

\localtableofcontents

\newpage

\renewcommand{\thesection}{\Alph{section}}

\section{Appendix for \cref{sec:sessiontypes} (Multiparty Session Types)}
\label{app:sessiontypes-calc}
\label{app:sessiontypes}

\subsection{Appendix for \cref{sec:calculus} (Synchronous Multiparty Session Calculus)}

\begin{definition}[Expressions]
\label{def:expressions}
Variables are denoted by $(x,y,z,\dots)$
and values by $v$.

Expressions are defined by:
\(
e\bnfdef
x\bnfsep
v\bnfsep
e + e'\bnfsep
e - e'\bnfsep
\neg e \bnfsep
e < e'\bnfsep
\mpSucc{e}
\)
\end{definition}

\begin{definition}[Guarded Processes]
\label{def:guard-proc}
We define the property $\mpGuarded{\mpP}{S}$
on processes, $\mpP$, and
sets of process variables, $S$,
by the inductive rules:
\[
\begin{array}{c}
	\inference{}{\mpGuarded{\mpNil}{S}}
	\quad
	\inference{\mpX\notin S}{\mpGuarded{\mpX}{S}}
	\quad
	\inference{\mpGuarded{{\mpP}}{S\cup\{\mpX\}}}{\mpGuarded{\mpRec{\mpX}{\mpP}}{S}}
	\\[2ex]
	\inference{\mpGuarded{\mpP}{S}&\mpGuarded{\mpPi}{S}}{
	\mpGuarded{\mpIf{e}{\mpP}{\mpPi}}{S}
	}
	\quad
	\inference{\mpGuarded{\mpP}{\emptyset}}{\mpGuarded{
		\roleQ\mathord{!}\ell\langle e \rangle \mpSeq \mpP 
	}{S}}
	\\[2ex]
	\inference{\forall i\in I.\;\mpGuarded{\mpP[i]}{\emptyset}}{\mpGuarded{
		\textstyle\sum_{i\in I} \roleQ\mathord{?}\ell_i(x_i)\mpSeq \mpP[i] 
	}{S}}
\end{array}
\]
We say that $\mpP$ is guarded if $\mpGuarded{\mpP}{\emptyset}$.
\end{definition}

\begin{definition}[Precongruence]
\label{def:precongruence}
Sessions are considered equal up-to reordering of parallel compositions.

We define the \emph{structural precongruence relation},
$\prestruct$, on processes and sessions by the inductive rules:
\[
\begin{array}{c}
	\inference{}{\mpP\prestruct\mpP}
	\quad
	\inference{\mpP\prestruct\mpPi&\mpPi\prestruct\mpPii}{\mpP\prestruct\mpPii}
	\quad
	\inference{}{\mpRec{\mpX}{\mpP}\prestruct\mpP\subst{\mpX}{\mpRec{\mpX}{\mpP}}}
	\quad
	\inference{\forall \roleR\in\roleSet.&\mpP[\roleR]\prestruct\mpPi[\roleR]}{
	\Pi_{\roleR\in\roleSet}\playsrole{\roleR}{\mpP[\roleR]}\prestruct\Pi_{\roleR\in\roleSet}\playsrole{\roleR}{\mpPi[\roleR]}
	}
\end{array}
\]
\end{definition}

\begin{definition}[Session Semantics]
	We define the transition
	relations
	$\mpMoveTau$,
	$\mpMoveAnnot{\roleP}{\roleQ}{\ell}{v}$, and
	$\mpMoveErr$
	by the rules:
	\[
	\begin{array}{c}
		\inference{\eval{e}{v}&v\in\{\mpTrue,\mpFalse\}}{\playsrole{\roleP}{\mpIf{e}{\mpP[\mpTrue]}{\mpP[\mpFalse]}\mpPar\M}\mpMoveTau\playsrole{\roleP}{\mpP[v]}\mpPar\M}
		\\[1ex]
		\inference{\eval{e}{v}&k\in I}{
		\playsrole{\roleP}{
			\textstyle\sum_{i\in I} \roleQ\mathord{?}\ell_i(x_i)\mpSeq \mpP[i]
		}
		\mpPar
		\playsrole{\roleQ}{
			\roleP\mathord{!}\ell_k\langle e \rangle \mpSeq \mpP 
		}
		\mpPar\M
		\mpMoveAnnot{\roleQ}{\roleP}{\ell}{v}
		\playsrole{\roleP}{
			\mpP[k]\subst{x_k}{v}
		}
		\mpPar
		\playsrole{\roleQ}{
			\mpP 
		}
		\mpPar\M
		}
		\\[1ex]
		\inference{\eval{e}{v}&v\notin\{\mpTrue,\mpFalse\}}{\playsrole{\roleP}{\mpIf{e}{\mpP[\mpTrue]}{\mpP[\mpFalse]}\mpPar\M}\mpMoveErr\playsrole{\roleP}{\mpErr}\mpPar\M}
		\\[1ex]
		\inference{\ell\notin\{\ell_i\}_{i\in I}}{
		\playsrole{\roleP}{
			\textstyle\sum_{i\in I} \roleQ\mathord{?}\ell_i(x_i)\mpSeq \mpP[i]
		}
		\mpPar
		\playsrole{\roleQ}{
			\roleP\mathord{!}\ell\langle e \rangle \mpSeq \mpP 
		}
		\mpPar\M
		\mpMoveErr
		\playsrole{\roleP}{
			\mpErr
		}
		\mpPar
		\playsrole{\roleQ}{
			\mpP 
		}
		\mpPar\M
		}
	\end{array}
	\]
	Define the reduction relation, $\mpMove$, by:
	\[
		\inference
		{\mpP\prestruct\mpPi\mpMoveGen\mpQi\prestruct\mpQ}
		{\mpP\mpMove\mpQ}
	\]
	We take $\mpMoveTauStar$ and $\mpMoveStar$
	to be the reflexive and transitive closures of $\mpMoveTau\cup\prestruct$
	and $\mpMove\cup\prestruct$,
	respectively.
\end{definition}

\subsection{Appendix for~\cref{sec:syntax} (Global and Local Type Syntax)}
\label{app:syntax}
\begin{definition}[{Guarded Types~\cite{DBLP:conf/pldi/Castro-Perez0GY21}}]
	\label{def:guarded}
	We define the global type property $\gtGuarded{\gtG}$ by the following inductive definition:
	\[
		\begin{array}{c}
			\inference
			{}
			{\gtGuarded{\gtEnd}}\qquad
			\inference
			{}
			{\gtGuarded{\gtRecVar}}\qquad
			\inference
			{\gtGuarded{\gtG} & \gtGuarded{\gtGi}}
			{\gtGuarded{\gtPar{\gtG}{\gtGi}}}\\[4ex]
			\inference
			{\forall i\in I & \gtGuarded{\gtG[i]}}
			{\gtGuarded{\gtComm{\roleP}{\roleQ}{i \in I}{\gtLab[i]}{\tyGround[i]}{\gtG[i]}}}\qquad
			\inference
			{\gtGuarded{\gtG} & \gtG\neq \gtRec{\gtRecVar[1]}{ \dots \gtRec{\gtRecVar[n]}{\gtRecVar} }  \text{ for any }n}
			{\gtGuarded{\gtRec{\gtRecVar}{\gtG}}}
		\end{array}
	\]
\end{definition}

\begin{definition}[Roles]
\label{def:roleset}
	We define
	the function $\gtRoles{}$,
	returning the set of participants in a global type,
	by the following recursive definition: 
	$\gtRoles{\gtEnd}=\emptyset$;
	$\gtRoles{\gtRecVar}=\emptyset$;
	$\gtRoles{\gtRec{\gtRecVar}{\gtG}}=\gtRoles{\gtG}$;
	$\gtRoles{\gtPar{\gtG[L]}{\gtG[R]}}=\gtRoles{\gtG[L]}\cup\gtRoles{\gtG[R]}$; and
	$\gtRoles{\gtComm{\roleP}{\roleQ}{i \in I}{\gtLab[i]}{\tyGround[i]}{\gtG[i]}}=\{\roleP,\roleQ\}\cup\bigcup_{i\in I}\gtRoles{\gtG[i]}$.
\end{definition}

\begin{definition}[Free Variables]
\label{def:freevar}
The set of free variables in $\gtG$ is returned by the function $\gtFv{\gtG}$, defined recursively by:
	$\gtFv{\gtEnd}=\emptyset$;
	$\gtFv{\gtRecVar}=\{\gtRecVar\}$;
	$\gtFv{\gtRec{\gtRecVar}{\gtG}}=\gtFv{\gtG}\setminus\{\gtRecVar\}$;
$\gtFv{\gtPar{\gtG[L]}{\gtG[R]}}=\gtFv{\gtG[L]}\cup\gtFv{\gtG[R]}$; and
	$\gtFv{\gtComm{\roleP}{\roleQ}{i \in I}{\gtLab[i]}{\tyGround[i]}{\gtG[i]}}=\bigcup_{i\in I}\gtFv{\gtG[i]}$.
\end{definition}

\begin{definition}[Size of Types]
	\label{def:typesize}
	We define the map $|\cdot|$ from global and local types to integers by:
	$|\gtEnd|=1$;
	$|\gtRecVar|=1$;
	$|\gtRec{\gtRecVar}{\gtG}|=|\gtG|+1$;
	$|\gtPar{\gtG[L]}{\gtG[R]}|=|\gtG[L]|+|\gtG[R]|+1$;
	$|\gtComm{\roleP}{\roleQ}{i \in I}{\gtLab[i]}{\tyGround[i]}{\gtG[i]}|=
	1+\sum_{i\in I}|\gtG[i]|$;
	$|\stEnd|=1$;
	$|\stRecVar|=1$;
	$|\stRec{\stRecVar}{\stT}|=|\stT|+1$;
	$|\stExtSum{\roleP}{i\in I}{\stT[i]}|=
	1+\sum_{i\in I}|\stT[i]|$; and
	$|\stIntSum{\roleP}{i\in I}{\stT[i]}|=
	1+\sum_{i\in I}|\stT[i]|$.
\end{definition}

\subsection{Appendix for~\cref{sec:semantics} (Global and Local Semantics)}
\label{app:semantic}
\thmBisimTrace*
\begin{proof}
By induction on $n$, we show that if $\gtG[L]\bisim\gtG[R]$, then
$\gtTraceN{\gtG[L]}{n}=\gtTraceN{\gtG[R]}{n}$.
$\gtTraceN{\gtG[L]}{0}=\gtTraceN{\gtG[R]}{0}=\left\{()\right\}$.
Suppose that $n>0$, let $\stEnvAnnotGenericSymi : \vec{\stEnvAnnotGenericSym} \in \gtTraceN{\gtG[L]}{n}$.
Let $\gtG[L]\,\gtMove[\stEnvAnnotGenericSymi]\,\gtGi[L]$.
By bisimilarity, there is $\gtGi[R]$ such that
$\gtG[R]\,\gtMove[\stEnvAnnotGenericSymi]\,\gtGi[R]$ and
$\gtGi[L]\bisim\gtGi[R]$.
By the I.H., $\vec{\stEnvAnnotGenericSym} \in \gtTraceN{\gtGi[L]}{n-1}=\gtTraceN{\gtGi[R]}{n-1}$.
So, $\stEnvAnnotGenericSymi : \vec{\stEnvAnnotGenericSym} \in \gtTraceN{\gtG[R]}{n}$.
Therefore, $\gtTraceN{\gtG[L]}{n}\subseteq \gtTraceN{\gtG[R]}{n}$
and vice versa as bisimilarity is symmetric.
Suppose that $\gtG[L]\bisim\gtG[R]$, then
$\gtTrace{\gtG[L]}=\bigcup_{n\geq 0}\gtTraceN{\gtG[L]}{n}
=\bigcup_{n\geq 0}\gtTraceN{\gtG[R]}{n}
=\gtTrace{\gtG[R]}$.
\newline
Consider the relation $\cong = \left\{\left(\gtG[L],\gtG[R]\right)\suchthat \gtTrace{\gtG[L]}=\gtTrace{\gtG[R]}\right\}$.
Suppose that $\gtG[L] \cong \gtG[R]$.
If $\gtG[L]\,\gtMove[\stEnvAnnotGenericSym]\,\gtGi[L]$,
then $(\stEnvAnnotGenericSym)\in\gtTrace{\gtG[L]}=\gtTrace{\gtG[R]}$,
so there is $\gtGi[R]$ such that
$\gtG[R]\,\gtMove[\stEnvAnnotGenericSym]\,\gtGi[R]$.
As our semantics are deterministic,
$\gtTrace{\gtGi[L]}=
\left\{\vec{\stEnvAnnotGenericSymi}\suchthat
\stEnvAnnotGenericSym:\vec{\stEnvAnnotGenericSymi}\in\gtTrace{\gtG[L]}\right\}=
\left\{\vec{\stEnvAnnotGenericSymi}\suchthat
\stEnvAnnotGenericSym:\vec{\stEnvAnnotGenericSymi}\in\gtTrace{\gtG[R]}\right\}=
\gtTrace{\gtGi[R]}$.
So, as $\cong$ is symmetric,
$\cong$ satisfies the coinductive rules for $\bisim$.
Therefore, $\cong\subseteq \bisim$.
\end{proof}

\begin{example}[Details for~\cref{ex:semantics-running}]
\label{ex:detail-semantics-running}
Recall $\gtG[\text{\tiny o/e}]$ and $\stEnv[\text{\tiny o/e}]$ from~\cref{ex:running}.
$\gtTrace{\gtG[\text{\tiny o/e}]}=\gtTrace{\stEnv[\text{\tiny o/e}]}$
and are given by the prefixes of the regular expression:
$
\ltsSendRecv{\roleP}{\roleQ}{\stLabFmt{o}}
\ltsSendRecv{\roleQ}{\roleR}{\stLabFmt{o}}
{(\ltsSendRecv{\roleQ}{\roleR}{\stLabFmt{o}}
\ltsSendRecv{\roleQ}{\roleR}{\stLabFmt{o}})}^*
\ltsSendRecv{\roleQ}{\roleR}{\stLabFmt{b}}
\ltsSendRecv{\roleR}{\roleQ}{\stLabFmt{o}}+
\ltsSendRecv{\roleP}{\roleQ}{\stLabFmt{m}}
{(\ltsSendRecv{\roleQ}{\roleR}{\stLabFmt{o}}
\ltsSendRecv{\roleQ}{\roleR}{\stLabFmt{o}})}^*
\ltsSendRecv{\roleQ}{\roleR}{\stLabFmt{b}}
\ltsSendRecv{\roleR}{\roleQ}{\stLabFmt{m}}
$.
We observe that $\stEnv[\text{\tiny o/e}]$ is safe, deadlock-free and live.
Following the trace of $\stEnv[\text{\tiny o/e}]$:
\\\centerline{
\begin{tikzpicture}
	\node(E0){$\stEnv[\text{\tiny o/e}]$};
	\node[above=1mm of E0, xshift=15mm](Eo){$\stEnv[\text{\tiny o}]$};
	\draw[->] (E0) -- (Eo) node[midway, xshift=-4mm,yshift=2mm]{$\ltsSendRecv{\roleP}{\roleQ}{\stLabFmt{o}}$};
	\node[right=15mm of Eo](Eo'){$\stEnvi[\text{\tiny o}]$};
	\draw[->] (Eo) -- (Eo');
	\draw[->] (Eo') to [out=150,in=30] (Eo);
	\path (Eo) -- (Eo') node[midway, yshift=2mm]{$\ltsSendRecv{\roleQ}{\roleR}{\stLabFmt{o}}$};
	\node[right=15mm of Eo'](EO){$\stEnv[\text{\tiny o,end}]$};
	\draw[->] (Eo') -- (EO) node[midway,yshift=2mm]{$\ltsSendRecv{\roleQ}{\roleR}{\stLabFmt{b}}$};
	\node[below=1mm of E0, xshift=15mm](Ee){$\stEnv[\text{\tiny e}]$};
	\draw[->] (E0) -- (Ee) node[midway,, xshift=-4mm,yshift=-2mm]{$\ltsSendRecv{\roleP}{\roleQ}{\stLabFmt{m}}$};
	\node[right=15mm of Ee](Ee'){$\stEnvi[\text{\tiny e}]$};
	\draw[->] (Ee) -- (Ee');
	\draw[->] (Ee') to [out=150,in=30] (Ee);
	\path (Ee) -- (Ee') node[midway, yshift=2mm]{$\ltsSendRecv{\roleQ}{\roleR}{\stLabFmt{o}}$};
	\node[right=15mm of Ee'](EE){$\stEnv[\text{\tiny e,end}]$};
	\draw[->] (Ee) to [out=40,in=150] (EE);
	\path (Ee') -- (EE) node[midway,xshift=-3mm,yshift=3mm]{$\ltsSendRecv{\roleQ}{\roleR}{\stLabFmt{b}}$};
	\path (EO) -- (EE) node[midway,xshift=15mm](EEnd){$\stEnv[{\stEnd}]$};
	\draw[->] (EO) -- (EEnd) node[midway, xshift=4mm,yshift=2mm]{$\ltsSendRecv{\roleR}{\roleQ}{\stLabFmt{o}}$};
	\draw[->] (EE) -- (EEnd) node[midway,, xshift=4mm,yshift=-2mm]{$\ltsSendRecv{\roleR}{\roleQ}{\stLabFmt{m}}$};
\end{tikzpicture}
}
where
\begin{equation*}
    \begin{array}{c}
    \stT[\roleR]=\stRec{\stRecVar}{
		\stExtSum{\roleQ}{}{
			\stLabFmt{o}\stSeq\stExtSum{\roleQ}{}{
				\stLabFmt{o}\stSeq\stRecVar,\,
				\stLabFmt{b}\stSeq\roleQ{\oplus}\stLabFmt{o}
			},\,
			\stLabFmt{b}\stSeq\roleQ{\oplus}\stLabFmt{m}
		}
	}
	\;\;
	\stT[\text{o},\roleR]=
	\stExtSum{\roleQ}{}{
				\stLabFmt{o}\stSeq\stT[\roleR],\,
				\stLabFmt{b}\stSeq\roleQ{\oplus}\stLabFmt{o}
			}\;\;
	\stT[\text{e,end},\roleR]=\roleQ\stFmt{\oplus}\stLabFmt{m}\;\;
	\stT[\text{o,end},\roleR]=\roleQ\stFmt{\oplus}\stLabFmt{o}
    \\[1ex]
	\stT[\roleP] = \stIntSum{\roleQ}{}{
		\stLabFmt{o},\,
		\stLabFmt{m}
	}\quad
	\stT[\roleQ] = \stExtSum{\roleP}{}{
	\begin{array}{l}
		\stLabFmt{o}\stSeq\roleR{\oplus}\stLabFmt{o}\stSeq\stRec{\stRecVar}{
			\stIntSum{\roleR}{}{
				\stLabFmt{o}\stSeq\roleR{\oplus}\stLabFmt{o}\stSeq\stRecVar,\,
				\stLabFmt{b}\stSeq\roleR{\&}\stLabFmt{o}
			}
		},\,
		\stLabFmt{m}\stSeq\stRec{\stRecVar}{
			\stIntSum{\roleR}{}{
				\stLabFmt{o}\stSeq\roleR{\oplus}\stLabFmt{o}\stSeq\stRecVar,\,
				\stLabFmt{b}\stSeq\roleR{\&}\stLabFmt{m}
			}
		}
	\end{array}
	}
	\\
	\begin{array}{r@{\,}c@{\,}lr@{\,}c@{\,}lr@{\,}c@{\,}l}
	\stT[\text{o},\roleQ]&=& 
		\roleR\stFmt{\oplus}\stLabFmt{o}\stSeq\stRec{\stRecVar}{
			\stIntSum{\roleR}{}{
				\stLabFmt{o}\stSeq\roleR{\oplus}\stLabFmt{o}\stSeq\stRecVar,\,
				\stLabFmt{b}\stSeq\roleR{\&}\stLabFmt{o}
			}
		}&
	\stTi[\text{o},\roleQ]&=& 
		\stRec{\stRecVar}{
			\stIntSum{\roleR}{}{
				\stLabFmt{o}\stSeq\roleR{\oplus}\stLabFmt{o}\stSeq\stRecVar,\,
				\stLabFmt{b}\stSeq\roleR{\&}\stLabFmt{o}
			}
		}&
		\stT[\text{o,end},\roleQ]&=& 
		\roleR\stFmt{\&}\stLabFmt{o}
		\\
	\stT[\text{e},\roleQ]&=& 
		\stRec{\stRecVar}{
			\stIntSum{\roleR}{}{
				\stLabFmt{o}\stSeq\roleR{\oplus}\stLabFmt{o}\stSeq\stRecVar,\,
				\stLabFmt{b}\stSeq\roleR{\&}\stLabFmt{m}
			}
		}&
	\stTi[\text{e},\roleQ]&=& 
		\roleR\stFmt{\oplus}\stLabFmt{o}\stSeq\stRec{\stRecVar}{
			\stIntSum{\roleR}{}{
				\stLabFmt{o}\stSeq\roleR{\oplus}\stLabFmt{o}\stSeq\stRecVar,\,
				\stLabFmt{b}\stSeq\roleR{\&}\stLabFmt{m}
			}
		}&
	\stT[\text{e,end},\roleQ] &=& 
		\roleR\stFmt{\&}\stLabFmt{m}
		\end{array}\\[1ex]
		\stEnv[\text{\tiny o/e}]= \stEnvMap{\roleP}{\stT[\roleP]}\stEnvComp
	\stEnvMap{\roleQ}{\stT[\roleQ]}\stEnvComp
	\stEnvMap{\roleR}{\stT[\roleR]}\qquad
	\stEnv[\stEnd] = \stEnvMap{\roleP}{\stEnd}\stEnvComp
	\stEnvMap{\roleQ}{\stEnd}\stEnvComp
	\stEnvMap{\roleR}{\stEnd}
	\\
		\begin{array}{r@{\,}c@{\,}lr@{\,}c@{\,}lr@{\,}c@{\,}l}
	\stEnv[\text{\tiny o}] &=& \stEnvMap{\roleP}{\stEnd}\stEnvComp
	\stEnvMap{\roleQ}{\stT[\text{o},\roleQ]}\stEnvComp
	\stEnvMap{\roleR}{\stT[\roleR]}&
	\stEnvi[\text{\tiny o}] &=& \stEnvMap{\roleP}{\stEnd}\stEnvComp
	\stEnvMap{\roleQ}{\stTi[\text{o},\roleQ]}\stEnvComp
	\stEnvMap{\roleR}{\stT[\text{o},\roleR]}&
	\stEnv[\text{\tiny o,end}] &=& \stEnvMap{\roleP}{\stEnd}\stEnvComp
	\stEnvMap{\roleQ}{\stT[\text{o,end},\roleQ]}\stEnvComp
	\stEnvMap{\roleR}{\stT[\text{o,end},\roleR]}\\
	\stEnv[\text{\tiny e}] &=& \stEnvMap{\roleP}{\stEnd}\stEnvComp
	\stEnvMap{\roleQ}{\stT[\text{e},\roleQ]}\stEnvComp
	\stEnvMap{\roleR}{\stT[\roleR]}&
	\stEnvi[\text{\tiny e}] &=& \stEnvMap{\roleP}{\stEnd}\stEnvComp
	\stEnvMap{\roleQ}{\stTi[\text{e},\roleQ]}\stEnvComp
	\stEnvMap{\roleR}{\stT[\text{o},\roleR]}&
	\stEnv[\text{\tiny e,end}] &=& \stEnvMap{\roleP}{\stEnd}\stEnvComp
	\stEnvMap{\roleQ}{\stT[\text{e,end},\roleQ]}\stEnvComp
	\stEnvMap{\roleR}{\stT[\text{e,end},\roleR]}
	\end{array}
	\end{array}
\end{equation*}

The conditions for safety and deadlock freedom clearly apply.
At any reachable context, every non-$\stEnd$ participant
must act within the next two transitions.
Therefore, $\stEnv[\text{\tiny o/e}]$ is live.
\end{example}

\thmGlobalSem*

\begin{proof}[1]
	The trace of $\gtG$ below,
	$\gtTrace{\gtG}$,
	is not context-free.
	\[
	\gtG=\gtPar{\gtRec{\gtRecVar}{
	\gtCommRaw{\roleP}{\roleQ}{
	\begin{array}{l}
		\gtLab[0]\gtSeq
		\gtCommRaw{\roleR}{\roleS}{
			\gtLab[2]\gtSeq\gtRecVar
		}
		\\
		\gtLab[1]\gtSeq
		\gtCommRaw{\roleR}{\roleS}{
		\begin{array}{l}
			\gtLab[2]\gtSeq\gtRecVar
			\\
			\gtLab[3]\gtSeq\gtRecVar
		\end{array}
		}
	\end{array}
	}}}
	{
	\gtRec{\gtRecVar}{
	\gtCommRaw{\rolePi}{\roleQi}{
	\begin{array}{l}
		\gtLab[0]\gtSeq
		\gtCommRaw{\roleRi}{\roleSi}{
			\gtLab[2]\gtSeq\gtRecVar
		}
		\\
		\gtLab[1]\gtSeq
		\gtCommRaw{\roleRi}{\roleSi}{
		\begin{array}{l}
			\gtLab[2]\gtSeq\gtRecVar
			\\
			\gtLab[3]\gtSeq\gtRecVar
		\end{array}
		}
	\end{array}
	}}
	}
	\]
	Suppose that $\gtTrace{\gtG}$
	is context-free.
	By the pumping lemma,
	take $p\geq 1$ to be the pumping length.
	Consider the word
	$s=\left(\ltsSendRecv{\roleP}{\roleQ}{\stLab[1]}\right)^p
	\left(\ltsSendRecv{\rolePi}{\roleQi}{\stLab[1]}\right)^p
	\left(\ltsSendRecv{\roleR}{\roleS}{\stLab[3]}\right)^p
	\left(\ltsSendRecv{\roleRi}{\roleSi}{\stLab[3]}\right)^p
	\in
	\gtTrace{\gtG}$.
	Find substrings
	$u$, $v$, $w$, $x$, and $y$
	such that
	$s=uvwxy$,
	$|vx|\geq 1$,
	$|vwx|\leq p$, and
	$\forall n\geq 0.\;uv^nwx^ny\in\gtTrace{\gtG}$.
	We consider all cases for the placement of $vwx$.
	\begin{enumerate}
		\item
		If
		$v=\left(\ltsSendRecv{\roleP}{\roleQ}{\stLab[1]}\right)^i$,
		$w=\left(\ltsSendRecv{\roleP}{\roleQ}{\stLab[1]}\right)^j$, and
		$x=\left(\ltsSendRecv{\roleP}{\roleQ}{\stLab[1]}\right)^k$, then
		$i+k\geq 1$ so
		\[uwy=\left(\ltsSendRecv{\roleP}{\roleQ}{\stLab[1]}\right)^{p-i-k}
	\left(\ltsSendRecv{\rolePi}{\roleQi}{\stLab[1]}\right)^p
	\left(\ltsSendRecv{\roleR}{\roleS}{\stLab[3]}\right)^p
	\left(\ltsSendRecv{\roleRi}{\roleSi}{\stLab[3]}\right)^p
	\notin
	\gtTrace{\gtG}\]
		\item
		If
		$v=\left(\ltsSendRecv{\roleP}{\roleQ}{\stLab[1]}\right)^i$,
		$w=\left(\ltsSendRecv{\roleP}{\roleQ}{\stLab[1]}\right)^j$, and
		$x=\left(\ltsSendRecv{\roleP}{\roleQ}{\stLab[1]}\right)^k
		\left(\ltsSendRecv{\rolePi}{\roleQi}{\stLab[1]}\right)^{k'}
		$, then
		$i+k+k'\geq 1$ so
		\[uwy=\left(\ltsSendRecv{\roleP}{\roleQ}{\stLab[1]}\right)^{p-i-k}
	\left(\ltsSendRecv{\rolePi}{\roleQi}{\stLab[1]}\right)^{p-k'}
	\left(\ltsSendRecv{\roleR}{\roleS}{\stLab[3]}\right)^p
	\left(\ltsSendRecv{\roleRi}{\roleSi}{\stLab[3]}\right)^p
	\notin
	\gtTrace{\gtG}\]
	\item
		If
		$v=\left(\ltsSendRecv{\roleP}{\roleQ}{\stLab[1]}\right)^i$,
		$w=\left(\ltsSendRecv{\roleP}{\roleQ}{\stLab[1]}\right)^j
		\left(\ltsSendRecv{\rolePi}{\roleQi}{\stLab[1]}\right)^{j'}
		$, and
		$x=
		\left(\ltsSendRecv{\rolePi}{\roleQi}{\stLab[1]}\right)^{k}
		$, then
		$i+k\geq 1$ so
		\[uwy=\left(\ltsSendRecv{\roleP}{\roleQ}{\stLab[1]}\right)^{p-i}
	\left(\ltsSendRecv{\rolePi}{\roleQi}{\stLab[1]}\right)^{p-k}
	\left(\ltsSendRecv{\roleR}{\roleS}{\stLab[3]}\right)^p
	\left(\ltsSendRecv{\roleRi}{\roleSi}{\stLab[3]}\right)^p
	\notin
	\gtTrace{\gtG}\]
	\item
		If
		$v=\left(\ltsSendRecv{\roleP}{\roleQ}{\stLab[1]}\right)^i
		\left(\ltsSendRecv{\rolePi}{\roleQi}{\stLab[1]}\right)^{i'}
		$,
		$w=
		\left(\ltsSendRecv{\rolePi}{\roleQi}{\stLab[1]}\right)^{j}
		$, and
		$x=
		\left(\ltsSendRecv{\rolePi}{\roleQi}{\stLab[1]}\right)^{k}
		$, then
		$i+i'+k\geq 1$ so
		\[uwy=\left(\ltsSendRecv{\roleP}{\roleQ}{\stLab[1]}\right)^{p-i}
	\left(\ltsSendRecv{\rolePi}{\roleQi}{\stLab[1]}\right)^{p-i'-k}
	\left(\ltsSendRecv{\roleR}{\roleS}{\stLab[3]}\right)^p
	\left(\ltsSendRecv{\roleRi}{\roleSi}{\stLab[3]}\right)^p
	\notin
	\gtTrace{\gtG}\]
	\item
		If
		$v=
		\left(\ltsSendRecv{\rolePi}{\roleQi}{\stLab[1]}\right)^{i}
		$,
		$w=
		\left(\ltsSendRecv{\rolePi}{\roleQi}{\stLab[1]}\right)^{j}
		$, and
		$x=
		\left(\ltsSendRecv{\rolePi}{\roleQi}{\stLab[1]}\right)^{k}
		$, then
		$i+k\geq 1$ so
		\[uwy=\left(\ltsSendRecv{\roleP}{\roleQ}{\stLab[1]}\right)^{p}
	\left(\ltsSendRecv{\rolePi}{\roleQi}{\stLab[1]}\right)^{p-i-k}
	\left(\ltsSendRecv{\roleR}{\roleS}{\stLab[3]}\right)^p
	\left(\ltsSendRecv{\roleRi}{\roleSi}{\stLab[3]}\right)^p
	\notin
	\gtTrace{\gtG}\]
	\item
		If
		$v=
		\left(\ltsSendRecv{\rolePi}{\roleQi}{\stLab[1]}\right)^{i}
		$,
		$w=
		\left(\ltsSendRecv{\rolePi}{\roleQi}{\stLab[1]}\right)^{j}
		$, and
		$x=
		\left(\ltsSendRecv{\rolePi}{\roleQi}{\stLab[1]}\right)^{k}
		\left(\ltsSendRecv{\roleR}{\roleS}{\stLab[3]}\right)^{k'}
		$, then
		$i+k+k'\geq 1$.
		
		If $k'\geq 1$, then
		\[uv^2wx^2y=\left(\ltsSendRecv{\roleP}{\roleQ}{\stLab[1]}\right)^{p}
	\left(\ltsSendRecv{\rolePi}{\roleQi}{\stLab[1]}\right)^{p+i+k}
	\left(\ltsSendRecv{\roleR}{\roleS}{\stLab[3]}\right)^{p+k'}
	\left(\ltsSendRecv{\roleRi}{\roleSi}{\stLab[3]}\right)^p
	\notin
	\gtTrace{\gtG}\]
		If $k'=0$, then the previous case applies.
		
		\item
		If
		$v=
		\left(\ltsSendRecv{\rolePi}{\roleQi}{\stLab[1]}\right)^{i}
		$,
		$w=
		\left(\ltsSendRecv{\rolePi}{\roleQi}{\stLab[1]}\right)^{j}
		\left(\ltsSendRecv{\roleR}{\roleS}{\stLab[3]}\right)^{j'}
		$, and
		$x=
		\left(\ltsSendRecv{\roleR}{\roleS}{\stLab[3]}\right)^{k}
		$, then
		$i+k\geq 1$.
		
		If $k\geq 1$, then
		\[uv^2wx^2y=\left(\ltsSendRecv{\roleP}{\roleQ}{\stLab[1]}\right)^{p}
	\left(\ltsSendRecv{\rolePi}{\roleQi}{\stLab[1]}\right)^{p+i}
	\left(\ltsSendRecv{\roleR}{\roleS}{\stLab[3]}\right)^{p+k}
	\left(\ltsSendRecv{\roleRi}{\roleSi}{\stLab[3]}\right)^p
	\notin
	\gtTrace{\gtG}\]
		If $k=0$, then the previous case applies.
		
	\item
		If
		$v=
		\left(\ltsSendRecv{\rolePi}{\roleQi}{\stLab[1]}\right)^{i}
		\left(\ltsSendRecv{\roleR}{\roleS}{\stLab[3]}\right)^{i'}
		$,
		$w=
		\left(\ltsSendRecv{\roleR}{\roleS}{\stLab[3]}\right)^{j}
		$, and
		$x=
		\left(\ltsSendRecv{\roleR}{\roleS}{\stLab[3]}\right)^{k}
		$, then
		$i+i'+k\geq 1$.
		
		If $i'+k\geq 1$, then
		\[uv^2wx^2y=\left(\ltsSendRecv{\roleP}{\roleQ}{\stLab[1]}\right)^{p}
	\left(\ltsSendRecv{\rolePi}{\roleQi}{\stLab[1]}\right)^{p+i}
	\left(\ltsSendRecv{\roleR}{\roleS}{\stLab[3]}\right)^{p+i'+k}
	\left(\ltsSendRecv{\roleRi}{\roleSi}{\stLab[3]}\right)^p
	\notin
	\gtTrace{\gtG}\]
		If $i'+k=0$, then the previous case applies.
		
	\item
		If
		$v=
		\left(\ltsSendRecv{\roleR}{\roleS}{\stLab[3]}\right)^{i}
		$,
		$w=
		\left(\ltsSendRecv{\roleR}{\roleS}{\stLab[3]}\right)^{j}
		$, and
		$x=
		\left(\ltsSendRecv{\roleR}{\roleS}{\stLab[3]}\right)^{k}
		$, then
		$i+k\geq 1$ so
		\[uv^2wx^2y=\left(\ltsSendRecv{\roleP}{\roleQ}{\stLab[1]}\right)^{p}
	\left(\ltsSendRecv{\rolePi}{\roleQi}{\stLab[1]}\right)^{p}
	\left(\ltsSendRecv{\roleR}{\roleS}{\stLab[3]}\right)^{p+i+k}
	\left(\ltsSendRecv{\roleRi}{\roleSi}{\stLab[3]}\right)^p
	\notin
	\gtTrace{\gtG}\]
		
	\item
		If
		$v=
		\left(\ltsSendRecv{\roleR}{\roleS}{\stLab[3]}\right)^{i}
		$,
		$w=
		\left(\ltsSendRecv{\roleR}{\roleS}{\stLab[3]}\right)^{j}
		$, and
		$x=
		\left(\ltsSendRecv{\roleR}{\roleS}{\stLab[3]}\right)^{k}
		\left(\ltsSendRecv{\roleRi}{\roleSi}{\stLab[3]}\right)^{k'}
		$, then
		$i+k+k'\geq 1$ so
		\[uv^2wx^2y=\left(\ltsSendRecv{\roleP}{\roleQ}{\stLab[1]}\right)^{p}
	\left(\ltsSendRecv{\rolePi}{\roleQi}{\stLab[1]}\right)^{p}
	\left(\ltsSendRecv{\roleR}{\roleS}{\stLab[3]}\right)^{p+i+k}
	\left(\ltsSendRecv{\roleRi}{\roleSi}{\stLab[3]}\right)^{p+k'}
	\notin
	\gtTrace{\gtG}\]
	
	\item
		If
		$v=
		\left(\ltsSendRecv{\roleR}{\roleS}{\stLab[3]}\right)^{i}
		$,
		$w=
		\left(\ltsSendRecv{\roleR}{\roleS}{\stLab[3]}\right)^{j}
		\left(\ltsSendRecv{\roleRi}{\roleSi}{\stLab[3]}\right)^{j'}
		$, and
		$x=
		\left(\ltsSendRecv{\roleRi}{\roleSi}{\stLab[3]}\right)^{k}
		$, then
		$i+k\geq 1$ so
		\[uv^2wx^2y=\left(\ltsSendRecv{\roleP}{\roleQ}{\stLab[1]}\right)^{p}
	\left(\ltsSendRecv{\rolePi}{\roleQi}{\stLab[1]}\right)^{p}
	\left(\ltsSendRecv{\roleR}{\roleS}{\stLab[3]}\right)^{p+i}
	\left(\ltsSendRecv{\roleRi}{\roleSi}{\stLab[3]}\right)^{p+k}
	\notin
	\gtTrace{\gtG}\]
	
	\item
		If
		$v=
		\left(\ltsSendRecv{\roleR}{\roleS}{\stLab[3]}\right)^{i}
		\left(\ltsSendRecv{\roleRi}{\roleSi}{\stLab[3]}\right)^{i'}
		$,
		$w=
		\left(\ltsSendRecv{\roleRi}{\roleSi}{\stLab[3]}\right)^{j}
		$, and
		$x=
		\left(\ltsSendRecv{\roleRi}{\roleSi}{\stLab[3]}\right)^{k}
		$, then
		$i+i'+k\geq 1$ so
		\[uv^2wx^2y=\left(\ltsSendRecv{\roleP}{\roleQ}{\stLab[1]}\right)^{p}
	\left(\ltsSendRecv{\rolePi}{\roleQi}{\stLab[1]}\right)^{p}
	\left(\ltsSendRecv{\roleR}{\roleS}{\stLab[3]}\right)^{p+i}
	\left(\ltsSendRecv{\roleRi}{\roleSi}{\stLab[3]}\right)^{p+i'+k}
	\notin
	\gtTrace{\gtG}\]
	
	\item
		If
		$v=
		\left(\ltsSendRecv{\roleRi}{\roleSi}{\stLab[3]}\right)^{i}
		$,
		$w=
		\left(\ltsSendRecv{\roleRi}{\roleSi}{\stLab[3]}\right)^{j}
		$, and
		$x=
		\left(\ltsSendRecv{\roleRi}{\roleSi}{\stLab[3]}\right)^{k}
		$, then
		$i+k\geq 1$ so
		\[uv^2wx^2y=\left(\ltsSendRecv{\roleP}{\roleQ}{\stLab[1]}\right)^{p}
	\left(\ltsSendRecv{\rolePi}{\roleQi}{\stLab[1]}\right)^{p}
	\left(\ltsSendRecv{\roleR}{\roleS}{\stLab[3]}\right)^{p}
	\left(\ltsSendRecv{\roleRi}{\roleSi}{\stLab[3]}\right)^{p+i+k}
	\notin
	\gtTrace{\gtG}\]
	\end{enumerate}
	This contradicts the pumping lemma,
	so $\gtTrace{\gtG}$
	is not context-free.
	
	Later, we will see that
	there is a good projection making $\gtG$ projectable (\cref{ex:non-reg-global-proj}).
\end{proof}

\begin{proof}[2]
	We give an algorithm
	that, given $\gtG$,
	decides if $w\in\gtTrace{\gtG}$
	in linear space with respect to $|w|$.
	$LINSPACE\subseteq NLINSPACE$,
	which is the class of context-sensitive languages.
	
	For all $\gtG$,
	$\epsilon\in\gtTrace{\gtG}$.
	
	For all $\gtG$,
	$\stEnvAnnotGenericSym
	\vec{\stEnvAnnotGenericSymi}
	\in
	\gtTrace{\gtG}$ iff
	$\gtG\,\gtMove[\stEnvAnnotGenericSym]$
	and
	if
	$\unfoldOne{\gtG}=
	\gtComm{\roleP}{\roleQ}{i\in I}{\stLab[i]}{\tyGround[i]}{\gtG[i]}
	$, then:
	\begin{enumerate*}
		\item if $\{\roleP,\roleQ\}\cap\ltsSubject{
		\stEnvAnnotGenericSym
		\vec{\stEnvAnnotGenericSymi}
		}\neq\emptyset$
		then
		the first label containing $\roleP$ or $\roleQ$ is
		$\ltsSendRecv{\roleP}{\roleQ}{\stChoice{\stLab[k]}{\tyGround[k]}}$
		for some $k\in I$,
		$\stEnvAnnotGenericSym
		\vec{\stEnvAnnotGenericSymi}=
		\vec{\stEnvAnnotGenericSym[0]}
		\ltsSendRecv{\roleP}{\roleQ}{\stChoice{\stLab[k]}{\tyGround[k]}}
		\vec{\stEnvAnnotGenericSym[1]}
		$
		and
		$\gtG[k]\,\gtMove[\vec{\stEnvAnnotGenericSym[0]}\vec{\stEnvAnnotGenericSym[1]}]$
		and
		$\forall i\in I.\;\gtG[i]\,\gtMove[\vec{\stEnvAnnotGenericSym[0]}]$;
		\item if $\{\roleP,\roleQ\}\cap\ltsSubject{
		\stEnvAnnotGenericSym
		\vec{\stEnvAnnotGenericSymi}
		}=\emptyset$
		then
		$\forall i\in I.\;\gtG[i]\,\gtMove[
		\stEnvAnnotGenericSym
		\vec{\stEnvAnnotGenericSymi}
		]$.
	\end{enumerate*}
	
	Fix a global type $\gtG$.
	Compute the set of head reductions of $\gtG$,
	$\mathcal{G}$.
	This is finite.
	The procedure above on $\gtG$
	only uses head reductions of $\gtG$
	and,
	if $\gtG\,\gtMove[\stEnvAnnotGenericSym]$
	then running the procedure to a depth of $|\gtG|$
	will decrease the size of the parameter word by at least $1$.
	Therefore, the procedure halts for $w$
	and has recursion depth at most $|w||\gtG|$.
	Each recursion frame needs constant space with respect to $w$.
	Thus, the algorithm is linear space.
\end{proof}

An implementation is provided in this paper's artifact \cite{anonymous_2026_21242753}.

\subsection{Appendix for~\cref{sec:subtyping} (Subtyping)}

We introduce a `strong subtyping'
that allows us to increase the external choices
but not decrease the internal choices.
This will be useful in analysing the full projections.

\begin{definition}[Strong Subtyping]
\label{def:ssubtyping}
The \emph{strong subtyping relation $\tySub$} on local types is coinductively defined by 
the rules:
\newline
\centerline{\(
\begin{array}{c}
\cinference[\iruleStSubIn]{
  \forall i \in I
  &
  \stT[i] \tySSub \tyTi[i]
}{
  \stExtSum{\roleP}{i \in I\cup J}{\stChoice{\stLab[i]}{\tyGround[i]} \stSeq \stT[i]}%
  \tySSub
  \stExtSum{\roleP}{i \in I}{\stChoice{\stLab[i]}{\tyGround[i]} \stSeq \stTi[i]}%
}
\;
\cinference[\iruleStSubOut]{
  \forall i \in I
  &
  \stT[i] \tySSub \stTi[i]
}{
  \stIntSum{\roleP}{i \in I}{\stChoice{\stLab[i]}{\tyGround[i]} \stSeq \stT[i]}
  \tySSub
  \stIntSum{\roleP}{i \in I}{\stChoice{\stLab[i]}{\tyGround[i]} \stSeq \stTi[i]}
}
\\[2ex]
\cinference[\iruleStSubEnd]{}{
  \stEnd \tySSub \stEnd
}
\;
\cinference[\iruleStSubRecL]{
  \stT{}[\stRec{\stRecVar}{\stT}/\stRecVar] \tySSub \stTi
}{
  \stRec{\stRecVar}{\stT} \tySSub \stTi
}
\;
\cinference[\iruleStSubRecR]{
  \stT \tySSub \stTi{}[\stRec{\stRecVar}{\stTi}/\stRecVar]
}{
  \stT \tySSub \stRec{\stRecVar}{\stTi}
}
\end{array}
\)}
  \\Write
  $\stEnvNew \!\tySSub\! \stEnvNewi$
  iff
  $\dom{\stEnvNew} \!=\! \dom{\stEnvNewi}$
  and
  $\forall \roleP \!\in\! \dom{\stEnvNew}{:}\,
  \stEnvApp{\stEnvNew}{\roleP} \!\tySSub\! \stEnvApp{\stEnvNewi}{\roleP}$.
\end{definition}

\begin{lemma}[Strong Subtyping Operational Correspondence]
\label{lem:sub-op-cor}
	Let $\stT$ and $\stTi$ be two closed types,
	then $\stTi\tySSub\stT$ iff:
	\begin{itemize}
		\item if $\stT\,\gtMove[\vec{\stEnvAnnotGenericSym}]$,
		then $\stTi\,\gtMove[\vec{\stEnvAnnotGenericSym}]$
		\item if $\stTi\,\gtMove[\vec{\stEnvAnnotGenericSym}]\,\stTi[1]$
	and $\stT\,\gtMove[\vec{\stEnvAnnotGenericSym}]\,\stT[1]$, then:
		\begin{itemize}
			\item if $\stTi[1]\,\gtMove[{\roleQ\stEnvAnnotOutSym\stChoice{\stLab}{\tyGround}}]$, then $\stT[1]\,\gtMove[{\roleQ\stEnvAnnotOutSym\stChoice{\stLab}{\tyGround}}]$
			\item if $\stTi[1]\,\gtMove[{\roleQ\stEnvAnnotInSym\stChoice{\stLab}{\tyGround}}]$, then  there exist $\stLabi$ and $\tyGroundi$ s.t.
			$\stT[1]\,\gtMove[{\roleQ\stEnvAnnotInSym\stChoice{\stLabi}{\tyGroundi}}]$
		\end{itemize}
	
	\end{itemize}
\end{lemma}

\begin{proof}
	Define $\stTi \Re \stT$ iff the above conditions hold.
	$\Re$ satisfies the coinductive rules for $\tySSub$.
	So, $\Re\subseteq\tySSub$.
	
	Suppose that $\stTi\tySSub\stT$.
	If $\stT\,\gtMove[\stEnvAnnotGenericSym]\,\stT[1]$, then
	there exists $\stTi[1]$ s.t.
	$\stTi\,\gtMove[\stEnvAnnotGenericSym]\,\stTi[1]$
	and $\stTi[1]\tySSub\stT[1]$.
	So if $\stT\,\gtMove[\vec{\stEnvAnnotGenericSym}]\stT[1]$, then
	there exists $\stTi[1]$ s.t.
	$\stTi\,\gtMove[\vec{\stEnvAnnotGenericSym}]\stTi[1]$
	by induction on the length of $\vec{\stEnvAnnotGenericSym}$.
	
	Since the transition relation is deterministic,
	if $\stTi\,\gtMove[\vec{\stEnvAnnotGenericSym}]\,\stTi[1]$
	and $\stT\,\gtMove[\vec{\stEnvAnnotGenericSym}]\,\stT[1]$, then
	$\stTi[1]\tySSub\stT[1]$.
	Therefore: if $\stTi[1]\,\gtMove[{\roleQ\stEnvAnnotOutSym\stChoice{\stLab}{\tyGround}}]$, then $\stT[1]\,\gtMove[{\roleQ\stEnvAnnotOutSym\stChoice{\stLab}{\tyGround}}]$; and if $\stTi[1]\,\gtMove[{\roleQ\stEnvAnnotInSym\stChoice{\stLab}{\tyGround}}]$, then  there exist $\stLabi$ and $\tyGroundi$ s.t.
			$\stT[1]\,\gtMove[{\roleQ\stEnvAnnotOutSym\stChoice{\stLab}{\tyGround}}]$.
	So, $\tySSub\subseteq\Re$.
\end{proof}

\begin{lemma}[Strong Subtyping Contexts]
\label{lem:subtyping-contexts}
	Suppose that $\stEnv\tySSub\stEnvi$
	and $\stEnvSafeP{\stEnvi}$.
	\begin{enumerate*}
		\item $\stEnv\stEnvMoveGenAnnot\iff\stEnvi\stEnvMoveGenAnnot$ and
		\item$\stEnv\stEnvMoveGenAnnot\stEnv[1]
		\;\wedge\;
		\stEnvi\stEnvMoveGenAnnot\stEnvi[1]
		\implies
		\stEnv[1]\tySSub\stEnvi[1]$.
	\end{enumerate*}
\end{lemma}

\begin{proof}[Proof. (1)]
	If $\stEnv\stEnvMoveAnnot{\ltsSendRecv{\roleP}{\roleQ}{\stChoice{\stLab}{\tyGround}}}$,
	then
	$\stEnvApp{\stEnv}{\roleP}\gtMove[\roleQ\stFmt{\oplus}\stChoice{\stLab}{\tyGround}]$
	and
	$\stEnvApp{\stEnv}{\roleQ}\gtMove[\roleP\stFmt{\&}\stChoice{\stLab}{\tyGround}]$.
	Thus,
	$\stEnvApp{\stEnvi}{\roleP}\gtMove[\roleQ\stFmt{\oplus}\stChoice{\stLab}{\tyGround}]$
	and
	$\stEnvApp{\stEnvi}{\roleQ}\gtMove[\roleP\stFmt{\&}\stChoice{\stLabi}{\tyGroundi}]$.
	By $\stEnvSafeP{\stEnvi}$,
	$\stEnvi\stEnvMoveAnnot{\ltsSendRecv{\roleP}{\roleQ}{\stChoice{\stLab}{\tyGround}}}$.
	
	If $\stEnvi\stEnvMoveAnnot{\ltsSendRecv{\roleP}{\roleQ}{\stChoice{\stLab}{\tyGround}}}$,
	then
	$\stEnvApp{\stEnvi}{\roleP}\gtMove[\roleQ\stFmt{\oplus}\stChoice{\stLab}{\tyGround}]$
	and
	$\stEnvApp{\stEnvi}{\roleQ}\gtMove[\roleP\stFmt{\&}\stChoice{\stLab}{\tyGround}]$.
	Thus,
	$\stEnvApp{\stEnv}{\roleP}\gtMove[\roleQ\stFmt{\oplus}\stChoice{\stLab}{\tyGround}]$
	and
	$\stEnvApp{\stEnv}{\roleQ}\gtMove[\roleP\stFmt{\&}\stChoice{\stLab}{\tyGround}]$.
	Hence,
	$\stEnv\stEnvMoveAnnot{\ltsSendRecv{\roleP}{\roleQ}{\stChoice{\stLab}{\tyGround}}}$.
\end{proof}

\begin{proof}[Proof. (2)]
	Suppose that
	$\stEnv\stEnvMoveAnnot{\ltsSendRecv{\roleP}{\roleQ}{\stChoice{\stLab}{\tyGround}}}\stEnv[1]$
	and
	$\stEnvi\stEnvMoveAnnot{\ltsSendRecv{\roleP}{\roleQ}{\stChoice{\stLab}{\tyGround}}}\stEnvi[1]$.
	For $\roleR\in\dom{\stEnv}\setminus\{\roleP,\roleQ\}$,
	$\stEnvApp{\stEnv[1]}{\roleR}
	=
	\stEnvApp{\stEnv}{\roleR}
	\tySSub
	\stEnvApp{\stEnvi}{\roleR}
	=
	\stEnvApp{\stEnvi[1]}{\roleR}
	$.
	Write
	$
	\unfoldOne{\stEnvApp{\stEnv}{\roleP}}
	=
	\stIntSum{\roleQ}{i\in I}{\stChoice{\stLab[i]}{\tyGround[i]}\stSeq\stT[i]}
	$,
	$
	\unfoldOne{\stEnvApp{\stEnvi}{\roleP}}
	=
	\stIntSum{\roleQ}{i\in I}{\stChoice{\stLab[i]}{\tyGround[i]}\stSeq\stTi[i]}
	$, and
	$\forall i\in I.\;\stT[i]\tySSub\stTi[i]$.
	There is $k\in I$ such that $\stChoice{\stLab[k]}{\tyGround[k]}=\stChoice{\stLab}{\tyGround}$.
	Thus, $\stEnvApp{\stEnv[1]}{\roleP}=\stT[k]\tySSub\stTi[k]=\stEnvApp{\stEnvi[1]}{\roleP}$.
	Write
	$
	\unfoldOne{\stEnvApp{\stEnv}{\roleP}}
	=
	\stExtSum{\roleQ}{i\in I}{\stChoice{\stLab[i]}{\tyGround[i]}\stSeq\stT[i]}
	$,
	$
	\unfoldOne{\stEnvApp{\stEnvi}{\roleP}}
	=
	\stExtSum{\roleQ}{i\in I'}{\stChoice{\stLab[i]}{\tyGround[i]}\stSeq\stTi[i]}
	$,
	$I'\subseteq I$,
	and
	$\forall i\in I'.\;\stT[i]\tySSub\stTi[i]$.
	There is $k\in I'$ such that $\stChoice{\stLab[k]}{\tyGround[k]}=\stChoice{\stLab}{\tyGround}$.
	Thus, $\stEnvApp{\stEnv[1]}{\roleQ}=\stT[k]\tySSub\stTi[k]=\stEnvApp{\stEnvi[1]}{\roleQ}$.
	
	Thus, $\stEnv[1]\tySSub\stEnvi[1]$.
\end{proof}%
\section{Appendix for~\cref{sec:projections} (Good Projections and Association)}
\label{app:projections}
\subsection{Appendix for~\cref{sec:good-proj} (Good Projections)}

\begin{lemma}
\label{lem:roles-maintained}
	Suppose that $\gtProj{}{}{}$
	is good,
	$\roleP\in\gtRoles{\gtG}$ and
	$\gtProj{}{\gtG}{\stEnv}$.
	For all $\gtG\,\gtMove[\vec{\stEnvAnnotGenericSymi}]\,\gtGi$
	with $\roleP\notin\ltsSubject{\vec{\stEnvAnnotGenericSymi}}$,
	$\roleP\in\gtRoles{\gtGi}$.
\end{lemma}

\begin{proof}
	We have $\stT$ such that
	$\gtProj{\roleP}{\gtGi}{\stT}$
	and
	$\stEnvApp{\stEnv}{\roleP}\tySub\stT$.
	
	Since $\roleP\in\gtRoles{\gtG}$,
	there is $\gtG\,\gtMoveStar\,\gtGii\,\gtMove[\stEnvAnnotGenericSym]$
	with $\roleP\in\ltsSubject{\stEnvAnnotGenericSym}$.
	Taking the path $\gtG\,\gtMoveStar\,\gtGii$
	to be minimal,
	we get
	$\stTi$ such that
	$\gtProj{\roleP}{\gtGii}{\stTi}$
	and
	$\stEnvApp{\stEnv}{\roleP}\tySub\stTi$.
	We have that $\stTi\,\gtMove$
	so $\stEnvApp{\stEnv}{\roleP}\,\gtMove$
	so $\stT\,\gtMove$.
	Thus, $\unfoldOne{\stT}\neq\stEnd$
	and $\roleP\in\gtRoles{\gtGi}$.
\end{proof}

\thmRestrictGlobal*
\begin{proof}
Conditions \inferrule{$\projGoodEnd$} and \inferrule{$\projGoodType$}
hold as $\gtProji{\roleP}{\gtG}{\stT}$ implies $\gtProj{\roleP}{\gtG}{\stT}$,
which is good.
The other conditions hold as $\gtProji{\roleP}{\gtG}{\stT}$ implies $\gtProj{\roleP}{\gtG}{\stT}$,
which is good, and if $\gtG\,\gtMove[]\,\gtGi$ then $\predPApp{\gtGi}$,
as $\predP$ is closed under transitions.
\end{proof}

\subsection{Appendix for~\cref{sec:association} (Association and Typing for Sessions)}
\label{app:association}

\thmProjSoundness*

\begin{proof}
Let $\stEnvAssoc{\gtG}{\stEnvNew}{}$ and
for $\roleR \in \gtRoles{\gtG}$,
let $\gtProj{\roleR}{\gtG}{\stT[\roleR]}$ and
$\stEnvApp{\stEnvNew}{\roleR}\tySub \stT[\roleR]$.
\\
Suppose that $\gtG \,\gtMove[\ltsSendRecv{\roleP}{\roleQ}{\stChoice{\stLab}{\tyGround}}]\, \gtGi$.
\\
There is $\stTi$ such that
$\stT[\roleP]\,\gtMove[{\roleQ\stEnvAnnotOutSym\stChoice{\stLab}{\tyGround}}]\,\stTi$ and
$\stEnvApp{\stEnvNew}{\roleP}\tySub \stT[\roleP]$
(\inferrule{$\projGoodOut$}),
so there are $\stLabi$, $\tyGroundi$, $\stTi[\roleP]$, and $\stTii[\roleP]$ such that
$\stEnvApp{\stEnvNew}{\roleP}\,\gtMove[{\roleQ\stEnvAnnotOutSym\stChoice{\stLabi}{\tyGroundi}}]\,\stTi[\roleP]$,
$\stT[\roleP]\,\gtMove[{\roleQ\stEnvAnnotOutSym\stChoice{\stLabi}{\tyGroundi}}]\,\stTii[\roleP]$ and
$\stTi[\roleP]\tySub\stTii[\roleP]$.
\\
Also $\stT[\roleQ]\,\gtMove[{\roleQ\stEnvAnnotInSym\stChoice{\stLab}{\tyGround}}]$,
so there is $\gtGii$ such that $\gtG\,\gtMove[ \ltsSendRecv{\roleP}{\roleQ}{\stChoice{\stLabi}{\tyGroundi}} ]\,\gtGii$
(\inferrule{$\projGoodType$}).
\\
Therefore, there is $\stTiii[\roleP]$ such that $\gtProj{\roleP}{\gtGii}{\stTiii[\roleP]}$ and
there are $\stTiii[\roleR]$ for all $\roleR\in\gtRoles{\gtG}\setminus\{\roleP\}$
and $\stTii[\roleQ]$ such that 
$\gtProj{\roleR}{\gtGii}{\stTiii[\roleR]}$,
$\stT[\roleQ]\,\gtMove[{\roleP\stEnvAnnotInSym\stChoice{\stLabi}{\tyGroundi}}]\,\stTii[\roleQ]\tySub\stTiii[\roleQ]$,
and $\stT[\roleR]\tySub\stTiii[\roleR]$ when
$\roleR\notin\{\roleP,\roleQ\}$
(\inferrule{$\projGoodAlt$},\inferrule{$\projGoodIn$}).
\\
$\stEnvApp{\stEnvNew}{\roleQ}\tySub \stT[\roleQ]$,
so there is $\stTi[\roleQ]$ such that
$\stEnvApp{\stEnvNew}{\roleQ}\,\gtMove[{\roleP\stEnvAnnotInSym\stChoice{\stLabi}{\tyGroundi}}]\,\stTi[\roleQ]$ and
$\stTi[\roleQ]\tySub\stTii[\roleQ]\tySub\stTiii[\roleQ]$.
\\
Now, define $\stEnvNewi$ to be $\stEnvNew$ with
$\stEnvApp{\stEnvNew}{\roleP}=\stTi[\roleP]$ and
$\stEnvApp{\stEnvNew}{\roleQ}=\stTi[\roleQ]$, so that
$\stEnvNew \stEnvMoveAnnot{ \stEnvCommAnnotSmall{\roleP}{\roleQ}{\stChoice{\stLabi}{\tyGroundi}} } \stEnvNewi$ and
$\stEnvAssoc{\gtGii}{\stEnvNewi}{}$.
\end{proof}

 \thmProjCompleteness*

\begin{proof}
Let $\stEnvAssoc{\gtG}{\stEnvNew}{}$ and for $\roleR \in \gtRoles{\gtG}$,
let $\gtProj{\roleR}{\gtG}{\stT[\roleR]}$ and
$\stEnvApp{\stEnvNew}{\roleR}\tySub \stT[\roleR]$.
\\
Suppose that $\stEnvNew \stEnvMoveAnnot{ \stEnvCommAnnotSmall{\roleP}{\roleQ}{\stChoice{\stLab}{\tyGround}} } \stEnvNewi$.
\\
$\stEnvApp{\stEnvNew}{\roleP}\,\gtMove[{\roleQ\stEnvAnnotOutSym\stChoice{\stLab}{\tyGround}}]\,\stEnvApp{\stEnvNewi}{\roleP}$ and
$\stEnvApp{\stEnvNew}{\roleP}\tySub\stT[\roleP]$,
so there is $\stTi[\roleP]$ such that
$\stT[\roleP]\,\gtMove[{\roleQ\stEnvAnnotOutSym\stChoice{\stLab}{\tyGround}}]\,\stTi[\roleP]$ and
$\stEnvApp{\stEnvNewi}{\roleP}\tySub\stTi[\roleP]$.
\\
Also,
$\stT[\roleQ]\,\gtMove[{\roleQ\stEnvAnnotInSym\stChoice{\stLab}{\tyGround}}]$ as
$\stEnvApp{\stEnvNew}{\roleQ}\tySub \stT[\roleQ]$,
so there is $\gtGi$ such that
$\gtG\,\gtMove[ \ltsSendRecv{\roleP}{\roleQ}{\stChoice{\stLab}{\tyGround}} ]\,\gtGi$
(\inferrule{$\projGoodType$}).
\\
So, there is $\stTii[\roleP]$ such that $\gtProj{\roleP}{\gtGi}{\stTi[\roleP]}$ and $\stTi[\roleP]\tySub\stTii[\roleP]$
(\inferrule{$\projGoodOut$}).
\\
Now, find $\stTi[\roleQ]$ and $\stTii[\roleQ]$ such that
$\gtProj{\roleQ}{\gtGi}{\stTii[\roleQ]}$,
$\stT[\roleQ]\,\gtMove[{\roleP\stEnvAnnotInSym\stChoice{\stLab}{\tyGround}}]\,\stTi[\roleQ]$,
and $\stTi[\roleQ]\tySub\stTii[\roleQ]$
(\inferrule{$\projGoodIn$}).
\\
$\stEnvApp{\stEnvNew}{\roleQ}\tySub\stT[\roleQ]$, so
$\stEnvApp{\stEnvNewi}{\roleQ}\tySub\stTi[\roleQ]\tySub\stTii[\roleQ]$.
\\
For all $\roleR \in \gtRoles{\gtG}$ with
$\roleR\notin\{\roleP,\roleQ\}$,
find $\stTii[\roleR]$ such that
$\gtProj{\roleR}{\gtGi}{\stTii[\roleR]}$ and
$\stT[\roleR] \tySub \stTii[\roleR]$
(\inferrule{$\projGoodAlt$}).
\\
$\stEnvApp{\stEnvNewi}{\roleR}=\stEnvApp{\stEnvNew}{\roleR}\tySub\stT[\roleR]\tySub\stTii[\roleR]$.
\\
Therefore, $\stEnvAssoc{\gtGi}{\stEnvNewi}{}$.
\end{proof}

\thmAssocCorrect*

We prove this theorem in parts.
\begin{theorem}
	If a projection is good, then $\predP=\{\stEnvNew\suchthat\exists\gtG\text{ with }\stEnvAssoc{\gtG}{\stEnvNew}{}\}$ is a safety property.
\end{theorem}
\begin{proof}
	\begin{itemize}
		\item[\inferrule{\iruleSafeComm}] Suppose that $\stEnvAssoc{\gtG}{\stEnvNew}{}$,
        $\stEnvApp{\stEnvNew}{\roleP}\gtMove[{\roleQ\stEnvAnnotOutSym\stChoice{\stLab}{\tyGround}}]$, and
        $\stEnvApp{\stEnvNew}{\roleQ}\gtMove[{\roleP\stEnvAnnotInSym\stChoice{\stLabi}{\tyGroundi}}]$.
        \\
	Let $\stT[\roleP]$ and $\stT[\roleQ]$ be such that
    $\gtProj{\roleP}{\gtG}{\stT[\roleP]}$, $\gtProj{\roleQ}{\gtG}{\stT[\roleQ]}$,
    $\stEnvApp{\stEnvNew}{\roleP}\tySub \stT[\roleP]$, and
    $\stEnvApp{\stEnvNew}{\roleQ}\tySub \stT[\roleQ]$.
    \\
	Find $\tyGroundii$ and $\stLabii$ such that
    $\stT[\roleP]\,\gtMove[{\roleQ\stEnvAnnotOutSym\stChoice{\stLab}{\tyGround}}]$ and
    $\stT[\roleQ]\,\gtMove[{\roleP\stEnvAnnotInSym\stChoice{\stLabii}{\tyGroundii}}]$.
    \\
	Now, $\gtG\,\gtMove[ \stEnvCommAnnotSmall{\roleP}{\roleQ}{\stChoice{\stLab}{\tyGround}} ]$ (\inferrule{$\projGoodType$}),
    so $\stT[\roleQ]\,\gtMove[{\roleP\stEnvAnnotInSym\stChoice{\stLab}{\tyGround}}]$
    (\inferrule{$\projGoodIn$}).
    \\
	Therefore,
    $\stEnvApp{\stEnvNew}{\roleQ}\,\gtMove[{\roleP\stEnvAnnotInSym\stChoice{\stLab}{\tyGround}}]$.
    \\
	We now have that $\stEnvMoveAnnotP{\stEnvNew}{\stEnvCommAnnotSmall{\roleP}{\roleQ}{\stChoice{\stLab}{\tyGround}}}$.
		\item[\inferrule{\iruleMoveSession}] Suppose that $\stEnvAssoc{\gtG}{\stEnvNew}{}$ and
        $\stEnv\,\gtMove[{\stEnvCommAnnotSmall{\roleP}{\roleQ}{\stChoice{\stLab}{\tyGround}}}]\stEnvi$.
        \\
        By~\Cref{thm:completeness_association},
        there is $\gtGi$ with $\stEnvAssoc{\gtGi}{\stEnvi}{}$.
	\end{itemize}
\end{proof}
\begin{theorem}
	If a projection is good, then associated contexts are deadlock-free.
\end{theorem}
\begin{proof}
	Suppose that $\stEnvAssoc{\gtG}{\stEnv}{}$ and
    $\stEnv \!\stEnvMoveStar\! \stEnvNotMoveP{\stEnvi}$.
    \\
	By completeness,
    there is $\gtGi$ with $\stEnvAssoc{\gtGi}{\stEnvi}{}$.
    \\
	By soundness,
    $\gtNotMove{\gtGi}$ so
    $\gtGi\bisim\gtEnd$.
    \\
	So for all $\roleP \!\in\! \dom{\stEnvi}$,
    $\roleP\notin\gtRoles{\gtGi}$,
    so $\unfoldOne{\stEnvApp{\stEnvNew}{\roleP}}=\stEnd$.
\end{proof}

\begin{definition}[Full Typing Rules for Processes~\cite{GJPSY2018}]
\label{def:full-type-rules}
We define typing environments to be
partial mappings from
process variables to closed session types and
expression variables to basic types.
We denote typing environments by the syntax:
\[
	\mpEnv\bnfdef\mpEnvEmpty
	\bnfsep \mpEnv\mpEnvComp \mpX:\stT
	\bnfsep \mpEnv\mpEnvComp x:\tyGround
\]
We assume the typing judgement of the form $\stJudge{\mpEnv}{e}{\tyGround}$
from~\cite[Table~4]{GJPSY2018}.
Let $\gtProj{}{}{}$ be a projection.
\[
\begin{array}{c}
\inference[T-$\mpNil$]{}{\stJudge{\mpEnv}{\mpNil}{\stEnd}}
\qquad
\inference[T-Var]{\mpX : \stT \in \mpEnv}{\stJudge{\mpEnv}{\mpX}{\stT}}
\qquad
\inference[T-Sub]{\stJudge{\mpEnv}{\mpP}{\stT} \quad \stT \tySub \stTi}{\stJudge{\mpEnv}{\mpP}{\stTi}}
\\[3ex]
\inference[T-Out]{
  \stJudge{\mpEnv}{e}{\tyGround} \quad 
  \stJudge{\mpEnv}{\mpP}{\stT}
}{
  \stJudge{\mpEnv}{\roleQ\mathord{!}\ell\langle e \rangle \mpSeq \mpP}{
    \stFmt{\roleQ{\oplus}\stChoice{\ell}{\tyGround} \stSeq \stT}
  }
}
\qquad
\inference[T-In]{
J\subseteq I &
  \forall i \in J \quad \stJudge{\mpEnv, x_i : \tyGround[i]}{\mpP_i}{\stT[i]}
}{
  \stJudge{\mpEnv}{\textstyle\sum_{i\in I} \roleQ\mathord{?}\ell_i(x_i)\mpSeq \mpP_i}{
    \stExtSum{\roleQ}{i \in J}{\stChoice{\ell_i}{\tyGround[i]} \stSeq \stT[i]}
  }
}
\\[3ex]
\inference[T-Cond]{
  \stJudge{\mpEnv}{e}{\tyBool} \quad 
  \stJudge{\mpEnv}{\mpP_1}{\stT} \quad
  \stJudge{\mpEnv}{\mpP_2}{\stT}
}{
  \stJudge{\mpEnv}{\mpIf{e}{\mpP_1}{\mpP_2}}{\stT}
}
\qquad
\inference[T-Rec]{
  \stJudge{\mpEnv, \mpX : \stT}{\mpP}{\stT}
}{
  \stJudge{\mpEnv}{\mpRec{\mpX}{\mpP}}{{\stT}}
}
\\[3ex]
\inference[Sess]{\forall \roleP\in \dom{\stEnv}&
\stJudge{}{\mpP[\roleP]}{\stT[\roleP]}&
\stEnvApp{\stEnv}{\roleP}=\stT[\roleP]&
\gtProj{}{\gtG}{\stEnv}}
{
\tyJudge{}{\Pi_{\roleP\in\dom{\stEnv}}{\ \playsrole{\roleP}{P_{\roleP}}}}{\gtG}
}
\end{array}
\]
\end{definition}
\inferrule{T-$\mpNil$} types terminated processes with terminated types;
\inferrule{T-Rec} and \inferrule{T-Var} handle recursion;
\inferrule{T-Sub} subtyping;
\inferrule{T-Out} and \inferrule{T-In}
output and input respectively;
\inferrule{T-Cond} conditionals;
and \inferrule{Sess} combines processes into sessions.
These rules are standard, see
\cite[\S~3.3]{GJPSY2018}.

\begin{lemma}[Expressions]
	\label{lem:typed-expr}
	If $\stJudge{}{e}{\tyGround}$, then
	there is $v$ such that $\eval{e}{v}$
	and
	$\forall v'.\;\eval{e}{v'}
	\implies
	\stJudge{}{v'}{\tyGround}$.
\end{lemma}

\begin{lemma}[Substitution {\cite[Lemma A.21 and Lemma A.24]{GJPSY2018}}]\ \\
\label{lem:proc-subst}
\begin{enumerate}
\item	If $\stJudge{\mpEnv\mpEnvComp\mpEnvMap{x}{\tyGround}}{\mpP}{\stT}$
and $\stJudge{\mpEnv}{v}{\tyGround}$, then
$\stJudge{\mpEnv}{\mpP\subst{x}{v}}{\stT}$.
\item If $\stJudge{\mpEnv\mpEnvComp\mpEnvMap{\mpX}{\stTi}}{\mpP}{\stT}$
and $\stJudge{\mpEnv}{\mpPi}{\stTi}$, then
$\stJudge{\mpEnv}{\mpP\subst{\mpX}{\mpPi}}{\stT}$.
\end{enumerate}
\end{lemma}

\begin{lemma}[Subject Inversion {\cite[Lemma A.23]{GJPSY2018}}]
\label{lem:subj-inv}
Suppose that $\stJudge{\mpEnv}{\mpP}{\stT}$.
\begin{enumerate}
	\item If $\mpP=\mpNil$, then $\unfoldOne{\stT}=\stEnd$.
	\item If $\mpP=\mpIf{e}{\mpPi}{\mpPii}$, then
	$\stJudge{\mpEnv}{e}{\tyBool}$,
	$\stJudge{\mpEnv}{\mpPi}{\stT}$ and
	$\stJudge{\mpEnv}{\mpPii}{\stT}$.
	\item If $\mpP=\roleQ\mathord{!}\stLab\langle e \rangle \mpSeq \mpPi $, then
	$\unfoldOne{\stT}=\stIntSum{\roleQ}{i\in I}{\stChoice{\stLab[i]}{\tyGround[i]}\stSeq\stT[i]}$ and there is $k\in I$ such that
	$\stLab=\stLab[k]$,
	$\stJudge{\mpEnv}{e}{\tyGround[k]}$ and
	$\stJudge{\mpEnv}{\mpPi}{\stT[k]}$.
	\item If $\mpP=\textstyle\sum_{i\in I} \roleQ\mathord{?}\stLab[i](x_i)\,\mpSeq\, \mpP[i] $, then
	$\unfoldOne{\stT}=\stExtSum{\roleQ}{i\in I'}{\stChoice{\stLab[i]}{\tyGround[i]}\stSeq\stT[i]}$,
	$I\supseteq I'$
	and for all $i\in I'$,
	$\stJudge{\mpEnv\mpEnvComp\mpEnvMap{x_i}{\tyGround[i]}}{\mpP[i]}{\stT[i]}$.
\end{enumerate}
\end{lemma}

We refer to \cite{EKY2025}
for the mechanisation and technical details.

\begin{definition}[Context Typing Rule]
\label{def:type-rule-bot}
	We define the judgement $\tyJudge{}{\M}{\stEnv}$
	by the rule:
	\[
		\inference[SessBottom]{\forall \roleP\in \dom{\stEnv}&
\stJudge{}{\mpP[\roleP]}{\stEnvApp{\stEnv}{\roleP}}&
\stEnvAssoc{\gtG}{\stEnv}{}}
{
\tyJudge{}{\Pi_{\roleP\in\dom{\stEnv}}{\ \playsrole{\roleP}{P_{\roleP}}}}{\stEnv}
}
	\]
\end{definition}

\begin{lemma}[Bottom-Up and Top-Down]
\label{lem:bottom-top}
$\exists\stEnv.\;
\tyJudge{}{\M}{\stEnv}\;\wedge\;
\stEnvAssoc{\gtG}{\stEnv}{}$
iff
$\tyJudge{}{\M}{\gtG}$.
\end{lemma}

\begin{proof}
	If $\tyJudge{}{\Pi_{\roleP\in\dom{\stEnv}}{\playsrole{\roleP}{P_{\roleP}}}}{\stEnv}$
	and $\stEnvAssoc{\gtG}{\stEnv}{}$,
	then
	$\forall \roleP\in \dom{\stEnv}.\;
\stJudge{}{\mpP[\roleP]}{\stEnvApp{\stEnv}{\roleP}}$.
	Let $\gtProj{}{\gtG}{\stEnvi}$
	such that $\stEnv\tySub\stEnvi$.
	By \inferrule{T-Sub},
	$\forall \roleP\in \dom{\stEnvi}.\;
\stJudge{}{\mpP[\roleP]}{\stEnvApp{\stEnvi}{\roleP}}$.
	Thus, $\tyJudge{}{\Pi_{\roleP\in\dom{\stEnv}}{\playsrole{\roleP}{P_{\roleP}}}}{\gtG}$.
	
	If $\tyJudge{}{\M}{\gtG}$,
	then
	there exists $\stEnv$ such that
	$\M=\Pi_{\roleP\in\dom{\stEnv}}{\playsrole{\roleP}{P_{\roleP}}}$
	and
	$\forall \roleP\in \dom{\stEnv}.\;
\stJudge{}{\mpP[\roleP]}{\stEnvApp{\stEnv}{\roleP}}$
and
$\gtProj{}{\gtG}{\stEnv}$.
	So, $\stEnvAssoc{\gtG}{\stEnv}{}$
	and
	$\tyJudge{}{\M}{\playsrole{\roleP}{P_{\roleP}}}{\stEnv}$
\end{proof}

\corTypingSessions*

\begin{theorem}[Subject Reduction - Bottom]
\label{thm:sr-bot}
Let $\gtProj{}{}{}$ in \inferrule{SessBottom} in \cref{def:type-rule-bot}
be a good projection.
Suppose that $\tyJudge{}{\M}{\stEnv}$ and $\stEnvAssoc{\gtG}{\stEnv}{}$.
\begin{enumerate*}
\item If $\M\prestruct\M'$, then $\tyJudge{}{\M'}{\stEnv}$;
\item if $\M\mpMoveTau\M'$, then $\tyJudge{}{\M'}{\stEnv}$;
\item if $\M\mpMoveAnnot{\roleP}{\roleQ}{\stLab}{v}\M'$,
then
$\stEnv\stEnvMoveCommAnnot{\roleP}{\roleQ}{\stChoice{\stLab}{\tyGround}}\stEnvi$,
$\tyJudge{}{v}{\tyGround}$, and
$\tyJudge{}{\M'}{\stEnvi}$; and
\item $\M\,\,\,\,\not\!\!\!\!\mpMoveErr$.
\end{enumerate*}
\end{theorem}

\begin{proof}[1]
	Induction on the derivation of $\M\prestruct\M'$.
	See \cite{EKY2025}.
\end{proof}

\begin{proof}[2]
	$\M=\playsrole{\roleP}{\mpIf{e}{\mpP[\mpTrue]}{\mpP[\mpFalse]}}\mpPar
	\Pi_{\roleR\in\dom{\stEnv}\setminus\{\roleP\}}{\playsrole{\roleR}{\mpP[\roleR]}}
	$,
	$\eval{e}{v}$ for some $v\in\{\mpTrue,\mpFalse\}$ and
	$\M'=\playsrole{\roleP}{\mpP[v]}\mpPar
	\Pi_{\roleR\in\dom{\stEnv}\setminus\{\roleP\}}{\playsrole{\roleR}{\mpP[\roleR]}}
	$.
	We have that
	$\stJudge{}{\mpIf{e}{\mpP[\mpTrue]}{\mpP[\mpFalse]}}{\stEnvApp{\stEnv}{\roleP}}$ and
	for $\roleR\in\dom{\stEnv\setminus\{\roleP\}}$,
	$\stJudge{}{\mpP[\roleR]}{\stEnvApp{\stEnv}{\roleR}}$.
	By inversion,
	$\stJudge{}{\mpP[v]}{\stEnvApp{\stEnv}{\roleP}}$.
	Thus,
	$\tyJudge{}{
	\M'
	}{\stEnv}$.
\end{proof}

\begin{proof}[3]
	$\M=
	\playsrole{\roleP}{\roleQ\mathord{!}\stLab[k]\langle e \rangle \mpSeq \mpP}
	\mpPar
	\playsrole{\roleQ}{\textstyle\sum_{i\in I} \roleP\mathord{?}\stLab[i](x_i)\mpSeq \mpQ[i]}
	\mpPar
	\Pi_{\roleR\in\dom{\stEnv}\setminus\{\roleP\}}{\playsrole{\roleR}{\mpP[\roleR]}}
	$,
	$\eval{e}{v}$,
	$k\in I$ and
	$\M'=
	\playsrole{\roleP}{\mpP}
	\mpPar
	\playsrole{\roleQ}{\mpQ[k]\subst{x_k}{v}}
	\mpPar
	\Pi_{\roleR\in\dom{\stEnv}\setminus\{\roleP,\roleQ\}}{\playsrole{\roleR}{\mpP[\roleR]}}
	$.
	We have that
	$\stJudge{}{
		\textstyle\sum_{i\in I} \roleP\mathord{?}\stLab[i](x_i)\mpSeq \mpQ[i]
	}{\stEnvApp{\stEnv}{\roleQ}}$,
	$
	\stJudge{}{\roleQ\mathord{!}\stLab[k]\langle e \rangle \mpSeq \mpP}{\stEnvApp{\stEnv}{\roleP}}
	$ and
	for $\roleR\in\dom{\stEnv\setminus\{\roleP,\roleQ\}}$,
	$\stJudge{}{\mpP[\roleR]}{\stEnvApp{\stEnv}{\roleR}}$.
	By inversion,
	$\unfoldOne{\stEnvApp{\stEnv}{\roleQ}}=
	\stExtSum{\roleP}{i\in I'}{
	\stChoice{\stLab[i]}{\tyGround[i]}\stSeq\stT[i]
	}
	$,
	$I'\subseteq I$
	and for $i\in I'$,
	$\stJudge{\mpEnvMap{x_i}{\tyGround[i]}}{\mpQ[i]}{\stT[i]}$.
	By inversion,
	$\unfoldOne{\stEnvApp{\stEnv}{\roleP}}=
	\stIntSum{\roleQ}{i\in J}{
	\stChoice{\stLabi[i]}{\tyGroundi[i]}\stSeq\stTi[i]
	}
	$
	and there is $k'\in J$ such that
	$\stLab[k]=\stLabi[k']$,
	$\stJudge{}{e}{\tyGroundi[k']}$ and
	$\stJudge{}{\mpP}{\stTi[k']}$.
	$\stEnvSafeP{\stEnv}$, so
	$k\in I'$ and $\tyGroundi[k']=\tyGround[k]$
	and $\stEnv\stEnvMoveCommAnnot{\roleP}{\roleQ}{\stChoice{\stLab[k]}{\tyGround[k]}}\stEnvi$.
	By \cref{thm:completeness_association},
	$\stEnvi$ is associated to some global type.
	For $\roleR\in\dom{\stEnvi}\setminus\{\roleP,\roleQ\}$,
	$\stEnvApp{\stEnv}{\roleR}=\stEnvApp{\stEnvi}{\roleR}$.
	$\stEnvApp{\stEnvi}{\roleQ}=\stT[k]$ and
	$\stEnvApp{\stEnvi}{\roleP}=\stTi[k']$.
	$\stJudge{}{\mpQ[k]\subst{x_k}{v}}{\stT[k]}$
	by substitution and
	$\stJudge{}{\mpP}{\stTi[k']}$.
	Thus,
	$\tyJudge{}{\M'}{\stEnvi}$.
\end{proof}

\begin{proof}[4]
	Suppose that $\M=\playsrole{\roleP}{\mpIf{e}{\mpPi}{\mpPii}}\mpPar\M_0$,
	$\eval{e}{v}$ and $\tyJudge{}{\M}{\stEnv}$.
	So, $\stJudge{}{\mpIf{e}{\mpPi}{\mpPii}}{\stEnvApp{\stEnv}{\roleP}}$.
	So by inversion, $\stJudge{}{e}{\tyBool}$.
	So, $v\in\{\mpTrue,\mpFalse\}$.
	
	Suppose that
	$\M=
	\playsrole{\roleP}{\roleQ\mathord{!}\stLab\langle e \rangle \mpSeq \mpP}
	\mpPar
	\playsrole{\roleQ}{\textstyle\sum_{i\in I} \roleP\mathord{?}\stLab[i](x_i)\mpSeq \mpQ[i]}
	\mpPar
	\M_0
	$
	and
	$\tyJudge{}{\M}{\stEnv}$.
	So,
	$\stJudge{}{
		\textstyle\sum_{i\in I} \roleP\mathord{?}\stLab[i](x_i)\mpSeq \mpQ[i]
	}{\stEnvApp{\stEnv}{\roleQ}}$ and
	$
	\stJudge{}{\roleQ\mathord{!}\stLab\langle e \rangle \mpSeq \mpP}{\stEnvApp{\stEnv}{\roleP}}
	$.
	By inversion
	$\unfoldOne{\stEnvApp{\stEnv}{\roleQ}}=
	\stExtSum{\roleP}{i\in I'}{
	\stChoice{\stLab[i]}{\tyGround[i]}\stSeq\stT[i]
	}
	$ and
	$I'\subseteq I$.
	By inversion,
	$\unfoldOne{\stEnvApp{\stEnv}{\roleP}}=
	\stIntSum{\roleQ}{i\in J}{
	\stChoice{\stLabi[i]}{\tyGroundi[i]}\stSeq\stTi[i]
	}
	$
	and there is $k'\in J$ such that
	$\stLab=\stLabi[k']$.
	$\stEnvSafeP{\stEnv}$,
	so there is $k\in I'\subseteq I$ such that
	$\stLab=\stLab[i]$.
	
	Therefore, if $\tyJudge{}{\M}{\stEnv}$ then
	$\M\,\,\,\,\not\!\!\!\!\mpMoveErr$.
\end{proof}

\begin{theorem}[Subject Reduction - Top]
\label{thm:sr-top}
Let $\gtProj{}{}{}$ in \inferrule{Sess} in \cref{def:type-rules}
be a good projection.
Suppose that $\tyJudge{}{\M}{\gtG}$.
\begin{enumerate*}
\item If $\M\prestruct\M'$, then $\tyJudge{}{\M'}{\gtG}$;
\item if $\M\mpMoveTau\M'$, then $\tyJudge{}{\M'}{\gtG}$;
\item if $\M\mpMoveAnnot{\roleP}{\roleQ}{\stLab}{v}\M'$,
then $\gtG\,\gtMove[\ltsSendRecv{\roleP}{\roleQ}{\stChoice{\stLab}{\tyGround}}]\,\gtGi$,
$\stJudge{}{v}{\tyGround}$, and
$\tyJudge{}{\M'}{\gtGi}$; and
\item $\M\,\,\,\,\not\!\!\!\!\mpMoveErr$.
\end{enumerate*}
\end{theorem}

\begin{proof}
	Suppose that $\tyJudge{}{\M}{\gtG}$.
	By \cref{lem:bottom-top},
	find $\stEnv$ such that
	$\stEnvAssoc{\gtG}{\stEnv}{}$
	and
	$\tyJudge{}{\M}{\stEnv}$.
	So, $\M\,\,\,\,\not\!\!\!\!\mpMoveErr$.
	
	If $\M\prestruct\M'$, then
	$\tyJudge{}{\M'}{\stEnv}$.
	So by \cref{lem:bottom-top},
	$\tyJudge{}{\M}{\gtG}$.
	
	If $\M\mpMoveTau\M'$, then
	$\tyJudge{}{\M'}{\stEnv}$.
	So by \cref{lem:bottom-top},
	$\tyJudge{}{\M}{\gtG}$.
	
	If $\M\mpMoveAnnot{\roleP}{\roleQ}{\stLab}{v}\M'$,
	then
	$\stEnv\stEnvMoveCommAnnot{\roleP}{\roleQ}{\stChoice{\stLab}{\tyGround}}\stEnvi$,
	$\tyJudge{}{v}{\tyGround}$ and
	$\tyJudge{}{\M'}{\stEnvi}$.
	By \cref{thm:completeness_association},
	$\gtG\,\gtMove[\ltsSendRecv{\roleP}{\roleQ}{\stChoice{\stLab}{\tyGround}}]\,\gtGi$
	and
	$\stEnvAssoc{\gtGi}{\stEnvi}{}$.
	By \cref{lem:bottom-top},
	$\tyJudge{}{\M'}{\gtGi}$.
\end{proof}

\begin{theorem}[Session Fidelity - Bottom]
\label{thm:sf-bot}
Let $\gtProj{}{}{}$ in \inferrule{SessBottom} in \cref{def:type-rule-bot}
be a good projection.
Suppose that $\tyJudge{}{\M}{\stEnv}$.
If
$\stEnv\stEnvMoveCommAnnot{\roleP}{\roleQ}{\stChoice{\stLabi}{\tyGroundi}}$,
then
$\M\mpMoveTauStar\cdot\mpMoveAnnot{\roleP}{\roleQ}{\stLab}{v}\M'$,
$\stEnv\stEnvMoveCommAnnot{\roleP}{\roleQ}{\stChoice{\stLab}{\tyGround}}\stEnvi$,
$\tyJudge{}{v}{\tyGround}$, and
$\tyJudge{}{\M'}{\stEnvi}$.
\end{theorem}

\begin{proof}
	Suppose that $\tyJudge{}{\M}{\stEnv}$
	and
	$\stEnv\stEnvMoveCommAnnot{\roleP}{\roleQ}{\stChoice{\stLabi}{\tyGroundi}}$.
	Write $\M=\playsrole{\roleP}{\mpP}\mpPar\playsrole{\roleQ}{\mpQ}\mpPar\M'$.
	$\stJudge{}{\mpP}{\stEnvApp{\stEnv}{\roleP}}$
	and
	$\stJudge{}{\mpQ}{\stEnvApp{\stEnv}{\roleQ}}$.
	
	We now proceed by induction on the derivation of
	$\mpGuarded{\mpR}{\fpv{\mpR}}$
	that if
	\[
	\stJudge{}{\mpR\subst{\mpR[1]}{\mpX[1]}\dots\subst{\mpR[n]}{\mpX[n]}}{\stT}
	\] with $\fpv{\mpR}=\{\mpX[1],\dots,\mpX[n]\}$,
	then
	\[
	\playsrole{\roleR}{\mpR\subst{\mpR[1]}{\mpX[1]}\dots\subst{\mpR[n]}{\mpX[n]}}
	\mpPar\M''
	\mpMoveTauStar
	\playsrole{\roleR}{\mpRi}
	\mpPar\M''
	\]
	where $\mpRi$ is a selection, branching, or $\mpNil$.
	
	So, $\playsrole{\roleP}{\mpP}\mpPar
	\playsrole{\roleQ}{\mpQ}\mpPar
	\M'	
	\mpMoveTauStar
	\playsrole{\roleP}{\mpPi}\mpPar
	\playsrole{\roleQ}{\mpQi}\mpPar
	\M'$.
	
	By \cref{thm:sr-bot},
	$\tyJudge{}{\playsrole{\roleP}{\mpPi}\mpPar
	\playsrole{\roleQ}{\mpQi}\mpPar
	\M'}{\stEnv}$.
	By inversion,
	$\mpPi$ is a selection towards $\roleQ$
	and
	$\mpQi$ is a branching from $\roleP$.
	
	By \cref{thm:sr-bot}, no error occurs
	so if $\mpPi=\roleQ\mathord{!}\stLab\langle e \rangle \mpSeq \mpPii$
	and $\mpQi=\textstyle\sum_{i\in I} \roleP\mathord{?}\stLab[i](x_i)\mpSeq \mpQ[i]$,
	then $\stLab=\stLab[k]$ for some $k\in I$
	and $\eval{e}{v}$ as $\stJudge{}{e}{\tyGround}$ for some $\tyGround$.
	Therefore,
	$
	\playsrole{\roleP}{\mpPi}\mpPar\playsrole{\roleQ}{\mpQi}\mpPar\M'
	\mpMoveAnnot{\roleP}{\roleQ}{\stLab}{v}
	\playsrole{\roleP}{\mpPii}\mpPar\playsrole{\roleQ}{\mpQ[k]\subst{x_k}{v}}\mpPar\M'
	$.
	By \cref{thm:sr-bot},
	$\stEnv\stEnvMoveCommAnnot{\roleP}{\roleQ}{\stChoice{\stLab}{\tyGroundii}}\stEnvi$,
	$\tyJudge{}{v}{\tyGroundii}$ and
	$\tyJudge{}{\playsrole{\roleP}{\mpPii}\mpPar\playsrole{\roleQ}{\mpQ[k]\subst{x_k}{v}}\mpPar\M'}{\stEnvi}$.
\end{proof}

\begin{theorem}[Session Fidelity - Top]
\label{thm:sf-top}
Let $\gtProj{}{}{}$ in \inferrule{Sess} in \cref{def:type-rules}
be a good projection.
Suppose that $\tyJudge{}{\M}{\gtG}$.
If $\gtG\,\gtMove[\ltsSendRecv{\roleP}{\roleQ}{\stChoice{\stLabi}{\tyGroundi}}]$,
then
$\M\mpMoveTauStar\cdot\mpMoveAnnot{\roleP}{\roleQ}{\stLab}{v}\M'$,
$\gtG\,\gtMove[\ltsSendRecv{\roleP}{\roleQ}{\stChoice{\stLab}{\tyGround}}]\,\gtGi$,
$\stJudge{}{v}{\tyGround}$, and
$\tyJudge{}{\M'}{\gtGi}$.
\end{theorem}

\begin{proof}
	Suppose that $\tyJudge{}{\M}{\gtG}$.
	By \cref{lem:bottom-top},
	find $\stEnv$ such that
	$\stEnvAssoc{\gtG}{\stEnv}{}$
	and
	$\tyJudge{}{\M}{\stEnv}$.
	Suppose that
	$\gtG\,\gtMove[\ltsSendRecv{\roleP}{\roleQ}{\stChoice{\stLabi}{\tyGroundi}}]$.
	By \cref{thm:soundness_association},
	$\stEnv\stEnvMoveCommAnnot{\roleP}{\roleQ}{\stChoice{\stLabii}{\tyGroundii}}$.
	
	By \cref{thm:sf-bot},
	$\M\mpMoveTauStar\cdot\mpMoveAnnot{\roleP}{\roleQ}{\stLab}{v}\M'$,
	$\stEnv\stEnvMoveCommAnnot{\roleP}{\roleQ}{\stChoice{\stLab}{\tyGround}}\stEnvi$,
	$\stJudge{}{v}{\tyGround}$, and
	$\tyJudge{}{\M'}{\stEnvi}$.
	
	By \cref{thm:completeness_association},
	$\gtG\,\gtMove[\ltsSendRecv{\roleP}{\roleQ}{\stChoice{\stLab}{\tyGround}}]\gtGi$
	and
	$\stEnvAssoc{\gtGi}{\stEnvi}{}$.
	
	By \cref{lem:bottom-top},
	$\tyJudge{}{\M'}{\gtGi}$.
\end{proof}

\begin{theorem}[Safety and Deadlock-freedom]
\label{thm:proc-safe-df}
\begin{enumerate}
\item
Let $\gtProj{}{}{}$ in \inferrule{SessBottom} in \cref{def:type-rule-bot}
be a good projection.
If $\tyJudge{}{\M}{\stEnv}$, then
$\M$ is safe and deadlock-free.
\item
Let $\gtProj{}{}{}$ in \inferrule{Sess} in \cref{def:type-rules}
be a good projection.
If $\tyJudge{}{\M}{\gtG}$, then
$\M$ is safe and deadlock-free.
\end{enumerate}
\end{theorem}

\begin{proof}[1]
Suppose that $\M\mpMoveStar\M'$.
By \cref{thm:sr-bot},
there is $\stEnvi$ such that
$\tyJudge{}{\M'}{\stEnvi}$.
Typed sessions do not contain $\mpErr$
as it is not typable.
Therefore $\M$ is safe.

Suppose that $\M\mpMoveStar\M'$.
By \cref{thm:sr-bot},
there is $\stEnvi$ such that
$\tyJudge{}{\M'}{\stEnvi}$.
If $\stEnvi\stEnvMove$, then by \cref{thm:sf-bot},
$\M'\prestruct\M''\mpMove$.
If $\stEnvi\not\stEnvMove$, then $\stEnvEndP{\stEnvi}$ as
$\stEnvDFP{\stEnv}$.

By induction on the derivation of
$\mpGuarded{\mpR}{\fpv{\mpR}}$,
if
\[
\stJudge{}{\mpR\subst{\mpR[1]}{\mpX[1]}\dots\subst{\mpR[n]}{\mpX[n]}}{\stEnd}
\]
with $\fpv{\mpR}=\{\mpX[1],\dots,\mpX[n]\}$ and
$\playsrole{\roleR}{\mpR}\not\mpMove$,
then
\[
\mpR\subst{\mpR[1]}{\mpX[1]}\dots\subst{\mpR[n]}{\mpX[n]}\prestruct\mpNil
\]

Therefore, $\M'\prestruct\mpFmt{\Pi_{\roleP\in\roleSet}\playsrole{\roleP}{\mpNil}}$,
where $\roleSet$ is the set of participants in $\M'$.
So, $\M$ is deadlock-free.
\end{proof}

\begin{proof}[2]
	If $\tyJudge{}{\M}{\gtG}$, then by \cref{lem:bottom-top},
	there is $\stEnv$ such that $\tyJudge{}{\M}{\stEnv}$.
	By (1),
	$\M$ is safe and deadlock-free.
\end{proof}

\subsection{Appendix for \cref{sec:balanced} (Global Type Properties)}

\thmDecideBal*

\begin{proof}
	We adapt the algorithm in \cite[Theorem 4.19]{DBLP:journals/pacmpl/UdomsrirungruangY25}.
	
	First, assume closed $\gtG$ is given as a graph
	where:
	choice constructors and $\gtEnd$
	are nodes;
	choices are edges; and
	recursive variables refer to the nodes where they were bound.
	This graph is of size at most $|\gtG|$.
	
	For each participant $\roleP$,
	we determine the nodes that can reach $\roleP$
	but do not contain $\roleP$ in time $\mathcal{O}\!\left(|\gtG|\right)$.
	We determine if this subgraph contains a cycle
	in time $\mathcal{O}\!\left(|\gtG|\right)$.
	$\gtG$ is not balanced iff one of these subgraphs has a cycle.
	
	For each participant $\roleP$,
	we determine the nodes such that
	$\weakComm{\gtGi}{\roleP}{\roleQ}$ or $\weakComm{\gtGi}{\roleQ}{\roleP}$
	for some $\roleQ$
	but do not contain $\roleP$ in time $\mathcal{O}\!\left(|\gtRoles{\gtG}|\cdot|\gtG|\right)$.
	We extend this to nodes that can reach $\roleP$ and are reachable
	without passing
	a node with $\roleP$.
	We determine if this subgraph contains a cycle
	in time $\mathcal{O}\!\left(|\gtG|\right)$.
	$\gtG$ is not weak balanced iff one of these subgraphs have a cycle.
	
	For each participant $\roleP$,
	we determine the nodes such that
	$\roleP$ is reachable as the sender
	without crossing $\roleP$ as a receiver
	but do not contain $\roleP$ in time $\mathcal{O}\!\left(|\gtRoles{\gtG}|\cdot|\gtG|\right)$.
	We extend this to nodes that can reach $\roleP$ and are reachable
	without passing
	a node with $\roleP$.
	We determine if this subgraph contains a cycle
	in time $\mathcal{O}\!\left(|\gtG|\right)$.
	$\gtG$ is not $\roleP$-send balanced iff this subgraph has a cycle.
	
	For each participant $\roleP$,
	we determine the nodes such that
	$\roleP$ is reachable as the receiver
	without crossing $\roleP$ as a sender
	but do not contain $\roleP$ in time $\mathcal{O}\!\left(|\gtRoles{\gtG}|\cdot|\gtG|\right)$.
	We extend this to nodes that can reach $\roleP$ and are reachable
	without passing
	a node with $\roleP$.
	We determine if this subgraph contains a cycle
	in time $\mathcal{O}\!\left(|\gtG|\right)$.
	$\gtG$ is not $\roleP$-receive balanced iff this subgraph has a cycle.
\end{proof}

\begin{definition}[Coinductive Global Type Transitions]
\label{def:gtype-coind}
The coinductive global type transition
  $\gtMoveCo[\stEnvAnnotGenericSym]$ is coinductively
  defined by the rules in \cref{def:gtype:lts-gt},
  replacing
  \inferrule{\iruleGtMoveCtx}
  by
  \inferrule{\iruleGtMoveCtx-co}:
  \[
  	\cinference[\iruleGtMoveCtx-co]{
    \forall i \in I. 
    \gtG[i] 
    \,\gtMoveCo[\stEnvAnnotGenericSym]\, 
    \gtGi[i] 
    &
    \ltsSubject{\stEnvAnnotGenericSym} \cap \setenum{\roleP, \roleQ} =
    \emptyset
    &
    \ltsSubject{\stEnvAnnotGenericSym}\subseteq\gtRoles{\gtGi[i]}
  }{
      \gtCommSmall{\roleP}{\roleQ}{i \in I}{\gtLab[i]}{\tyGround[i]}{\gtG[i]}
   \,\gtMoveCo[\stEnvAnnotGenericSym]\, 
      \gtCommSmall{\roleP}{\roleQ}{i \in I}{\gtLab[i]}{\tyGround[i]}{\gtGi[i]}   
  }
  \]
\end{definition}

\propWeakBal*

\begin{proof}
	If $\gtG\gtMove[\ltsSendRecv{\roleP}{\roleQ}{\stChoice{\stLab}{\tyGround}}]$,
	then
	$\{\roleP,\roleQ\}\subseteq\gtRoles{\gtG}$.
	So $\gtMove\subseteq\gtMoveCo$.
	
	Suppose that $\gtG\gtMoveCo[\ltsSendRecv{\roleP}{\roleQ}{\stChoice{\stLab}{\tyGround}}]\gtGi$
	and $\gtG$ is weak balanced.
	$\gtG$ is guarded, so
	$\inferrule{\iruleGtMovePar}$
	and
	$\inferrule{\iruleGtMoveRec}$
	may only be used finitely many times
	in the derivation between
	uses of \inferrule{\iruleGtMoveCtx-co}
	or \inferrule{\iruleGtMoveComm}.
	So, $\roleP\in\gtRoles{\gtG}$.
	By induction on the derivation of $\roleP\in\gtRoles{\gtG}$,
	$\weakComm{\gtG}{\roleP}{\roleQ}$.
	$\gtG$ is weak balanced, so
	$\gtDepth{\gtG}{\roleP}$
	and $\gtDepth{\gtG}{\roleQ}$
	exist.
	By induction on $\gtDepth{\gtG}{\roleP}$,
	if
	$\gtG\gtMoveCo[\ltsSendRecv{\roleP}{\roleQ}{\stChoice{\stLab}{\tyGround}}]\gtGi$
	then
	$\gtG\gtMove[\ltsSendRecv{\roleP}{\roleQ}{\stChoice{\stLab}{\tyGround}}]\gtGi$.
\end{proof}

\propProjWeakBal*

\begin{proof}
	Suppose that $\gtProj{}{}{}$ is a good projection.
	Assume that $\gtProj{}{}{}$ is closed under global unfolding
	\ie if $\gtProj{\roleP}{\gtRec{\gtRecVar}{\gtG}}{\stT}$ then $\gtProj{\roleP}{\gtG\subst{\gtRecVar}{\gtRec{\gtRecVar}{\gtG}}}{\stT}$.
	Adding this to the projection maintains \cref{def:good-proj}.
	
	Suppose that
	$\gtProj{}{\gtG}{\stEnv}$.
	Suppose $\gtG\gtMoveStar\gtGi$
	and
	$\weakComm{\gtGi}{\roleP}{\roleQ}$
	but either $\gtDepth{\gtGi}{\roleP}$ or $\gtDepth{\gtGi}{\roleQ}$
	does not exist.
	
	By the good projection conditions,
	there is $\stEnvi$ such that
	$\gtProj{}{\gtGi}{\stEnvi}$.
	
	We proceed by induction on the derivation of $\weakComm{\gtGi}{\roleP}{\roleQ}$,
	that if $\gtProj{}{\gtGi\subst{\gtRecVar[1]}{\gtGi[1]}\dots\subst{\gtRecVar[n]}{\gtGi[n]}}{\stEnvi}$,
	then there are $\gtLab$ and $\tyGround$
	such that $\stEnvi\stEnvMoveCommAnnot{\roleP}{\roleQ}{\stChoice{\stLab}{\tyGround}}$.
	\begin{enumerate}
	\item If
	\[\gtGi=\gtComm{\roleP}{\roleQ}{i\in I}{\gtLab[i]}{\tyGround[i]}{\gtG[i]}\]
	then
	take $i\in I$ and
	\[\gtGi\subst{\gtRecVar[1]}{\gtGi[1]}\dots\subst{\gtRecVar[n]}{\gtGi[n]}\,\gtMove[\ltsSendRecv{\roleP}{\roleQ}{\gtLab[i]}{\tyGround[i]}]\]
	By \inferrule{$\projGoodOut$} and \inferrule{$\projGoodIn$},
	\[\stEnvi\stEnvMoveCommAnnot{\roleP}{\roleQ}{\stChoice{\stLab[i]}{\tyGround[i]}}\]
	
	\item If \[\gtGi=\gtComm{\rolePi}{\roleQi}{i\in I}{\gtLab[i]}{\tyGround[i]}{\gtG[i]}\]
	with $\{\rolePi,\roleQi\}\cap\{\roleP,\roleQ\}=\emptyset$, then
	there is $j\in I$ such that
	$\weakComm{\gtG[j]}{\roleP}{\roleQ}$.
	\[\gtGi\subst{\gtRecVar[1]}{\gtGi[1]}\dots\subst{\gtRecVar[n]}{\gtGi[n]}
	\gtMove[\ltsSendRecv{\rolePi}{\roleQi}{\stChoice{\gtLab[j]}{\tyGround[j]}}]
	\gtG[j]\subst{\gtRecVar[1]}{\gtGi[1]}\dots\subst{\gtRecVar[n]}{\gtGi[n]}
	\]
	By \inferrule{$\projGoodOut$}, \inferrule{$\projGoodIn$}, and
	\inferrule{$\projGoodAlt$},
	there is $\stEnvii$ such that
	$\gtProj{}{\gtG[j]\subst{\gtRecVar[1]}{\gtGi[1]}\dots\subst{\gtRecVar[n]}{\gtGi[n]}}{\stEnvii}$,
	$\stEnvApp{\stEnvii}{\roleP}\tySub\stEnvApp{\stEnvi}{\roleP}$
	and
	$\stEnvApp{\stEnvii}{\roleQ}\tySub\stEnvApp{\stEnvi}{\roleQ}$.
	By the I.H.,
	\[\stEnvii\stEnvMoveCommAnnot{\roleP}{\roleQ}{\stChoice{\stLab}{\tyGround}}\]
	By subtyping, there are $\stLabi$ and $\tyGroundi$ such that
	\[\stEnvi\stEnvMoveCommAnnot{\roleP}{\roleQ}{\stChoice{\stLabi}{\tyGroundi}}\]
	
	\item If $\gtGi=\gtRec{\gtRecVar}{\gtGii}$, then
	$\weakComm{\gtGii}{\roleP}{\roleQ}$.
	We have
	\[\gtProj{}{
	\gtGii\subst{\gtRecVar[1]}{\gtGi[1]}\dots\subst{\gtRecVar[n]}{\gtGi[n]}\subst{\gtRecVar}{
	\gtGi\subst{\gtRecVar[1]}{\gtGi[1]}\dots\subst{\gtRecVar[n]}{\gtGi[n]}
	}
	}{\stEnvi}\]
	By the I.H.,
	\[\stEnvi\stEnvMoveCommAnnot{\roleP}{\roleQ}{\stChoice{\stLab}{\tyGround}}\]
	\end{enumerate}
	
	Now by \inferrule{$\projGoodType$},
	we deduce that
	\[\gtG\,\gtMove[\ltsSendRecv{\roleP}{\roleQ}{\stChoice{\stLab}{\tyGround}}]\]
	By induction on the derivation of this transition,
	$\gtDepth{\gtG}{\roleP}$
	and
	$\gtDepth{\gtG}{\roleQ}$
	exist.
\end{proof}

\begin{example}[Unbalanced Global Types]
\label{ex:app:balanced}
	The following types are unbalanced:
	\begin{equation*}
		\gtG_{\text{\tiny unbal}} =
			\gtRec{\gtRecVar}{
				\gtCommRaw{\roleFmt{a}}{\roleFmt{b}}{
						\gtCommChoice{go}{}{\gtRecVar},\,
						\gtCommChoice{cancel}{}{
							\gtCommSingle{\roleFmt{c}}{\roleFmt{d}}{cancel}{}{\gtEnd}
						}
				}	
			}
	\end{equation*}
	\begin{equation*}
		\gtGi_{\text{\tiny unbal}} =
			\gtRec{\gtRecVar}{
				\gtCommRaw{\roleFmt{a}}{\roleFmt{b}}{
						\gtCommChoice{go}{}{\gtRecVar},\,
						\gtCommChoice{cancel}{}{
							\gtCommSingle{\roleFmt{a}}{\roleFmt{c}}{cancel}{}{\gtEnd}
						}
				}	
			}
	\end{equation*}
	However, $\gtGi_{\text{\tiny unbal}}$ is send-balanced, unlike $\gtG_{\text{\tiny unbal}}$.
\end{example}

\begin{lemma}[Approaching Participant]
	\label{lem:glb-approach}
	Suppose that
	$\roleP\in\gtRoles{\gtG}$,
	$\gtDepth{\gtG}{\roleP}$ exists and
	for all $\gtG\,\gtMove[\vec{\stEnvAnnotGenericSymi}]\,\gtGi$
	with $\roleP\notin\ltsSubject{\vec{\stEnvAnnotGenericSymi}}$,
	$\roleP\in\gtRoles{\gtGi}$.
	
	There is $\stEnvAnnotGenericSym$ such that
	$\gtG\,\gtMove[\stEnvAnnotGenericSym]$
	and if $\gtG\,\gtMove[\vec{\stEnvAnnotGenericSymi}]\,\gtGi$
	with $\ltsSubject{\stEnvAnnotGenericSym}\subseteq\ltsSubject{\vec{\stEnvAnnotGenericSymi}}$,
	then either $\roleP\in\ltsSubject{\vec{\stEnvAnnotGenericSymi}}$
	or $\gtDepth{\gtGi}{\roleP}<\gtDepth{\gtG}{\roleP}$.
\end{lemma}

\begin{proof}
	We proceed by induction on $|\gtG|$
	that if $\unfoldOne{\gtG}\neq \gtRecVar$, then
	the hypothesis and conclusion hold for
	\[\gtG\subst{\gtRecVar[1]}{\gtG[1]}\dots\subst{\gtRecVar[n]}{\gtG[n]}\]
	where $\gtFv{\gtG}=\{\gtRecVar[1],\dots,\gtRecVar[n]\}$.
	
	If $\gtG=\gtPar{\gtG[0]}{\gtG[1]}$,
	then
	let $i\in\{0,1\}$ such that
	$\roleP\in\gtRoles{\gtG[i]}\setminus\gtRoles{\gtG[1-i]}$.
	By I.H.,
	there is $\stEnvAnnotGenericSym$ such that
	$\gtG[i]\,\gtMove[\stEnvAnnotGenericSym]$
	and if $\gtG[i]\,\gtMove[\vec{\stEnvAnnotGenericSymi}]\,\gtGi[i]$
	with $\ltsSubject{\stEnvAnnotGenericSym}\subseteq\ltsSubject{\vec{\stEnvAnnotGenericSymi}}$,
	then either $\roleP\in\ltsSubject{\vec{\stEnvAnnotGenericSymi}}$
	or $\gtDepth{\gtGi[i]}{\roleP}<\gtDepth{\gtG[i]}{\roleP}$.
	Let $\gtGi[1-i]=\gtG[1-i]$.
	Thus,
	$\gtPar{\gtG[0]}{\gtG[1]}\,\gtMove[\stEnvAnnotGenericSym]$
	and if $\gtPar{\gtG[0]}{\gtG[1]}\,\gtMove[\vec{\stEnvAnnotGenericSymi}]\,\gtPar{\gtGi[0]}{\gtGi[1]}$
	with $\ltsSubject{\stEnvAnnotGenericSym}\subseteq\ltsSubject{\vec{\stEnvAnnotGenericSymi}}$,
	then
	let $\vec{\stEnvAnnotGenericSymii}$
	be the labels in $\vec{\stEnvAnnotGenericSymi}$
	with roles in $\gtG[i]$,
	$\gtG[i]\,\gtMove[\vec{\stEnvAnnotGenericSymii}]\,\gtGi[i]$,
	so either $\roleP\in\ltsSubject{\vec{\stEnvAnnotGenericSymi}}$
	or $\gtDepth{\gtPar{\gtGi[0]}{\gtGi[1]}}{\roleP}
	=1+\gtDepth{\gtGi[i]}{\roleP}
	<1+\gtDepth{\gtG[i]}{\roleP}
	=\gtDepth{\gtPar{\gtG[0]}{\gtG[1]}}{\roleP}$.
	
	If $\gtG=\gtRec{\gtRecVar[n+1]}{\gtGi}$,
	then by the I.H.,
	there is $\stEnvAnnotGenericSym$ such that
	$\gtGi\subst{\gtRecVar[1]}{\gtG[1]}\dots\subst{\gtRecVar[n]}{\gtG[n]}\subst{\gtRecVar[n+1]}{\gtG}\,\gtMove[\stEnvAnnotGenericSym]$
	and if $\gtGi\subst{\gtRecVar[1]}{\gtG[1]}\dots\subst{\gtRecVar[n]}{\gtG[n]}\subst{\gtRecVar[n+1]}{\gtG}\,\gtMove[\vec{\stEnvAnnotGenericSymi}]\,\gtGii$
	with $\ltsSubject{\stEnvAnnotGenericSym}\subseteq\ltsSubject{\vec{\stEnvAnnotGenericSymi}}$,
	then either $\roleP\in\ltsSubject{\vec{\stEnvAnnotGenericSymi}}$
	or $\gtDepth{\gtGii}{\roleP}<\gtDepth{\gtGi\subst{\gtRecVar[1]}{\gtG[1]}\dots\subst{\gtRecVar[n]}{\gtG[n]}\subst{\gtRecVar[n+1]}{\gtG}}{\roleP}$.
	So,
	$\gtG\subst{\gtRecVar[1]}{\gtG[1]}\dots\subst{\gtRecVar[n]}{\gtG[n]}\,\gtMove[\stEnvAnnotGenericSym]$
	and if $\gtG\subst{\gtRecVar[1]}{\gtG[1]}\dots\subst{\gtRecVar[n]}{\gtG[n]}\,\gtMove[\vec{\stEnvAnnotGenericSymi}]\,\gtGii$
	with $\ltsSubject{\stEnvAnnotGenericSym}\subseteq\ltsSubject{\vec{\stEnvAnnotGenericSymi}}$,
	then
	$\gtGi\subst{\gtRecVar[1]}{\gtG[1]}\dots\subst{\gtRecVar[n]}{\gtG[n]}\subst{\gtRecVar[n+1]}{\gtG}\,\gtMove[\vec{\stEnvAnnotGenericSymi}]\,\gtGii$
	so
	either $\roleP\in\ltsSubject{\vec{\stEnvAnnotGenericSymi}}$
	or $\gtDepth{\gtGii}{\roleP}<\gtDepth{\gtGi\subst{\gtRecVar[1]}{\gtG[1]}\dots\subst{\gtRecVar[n]}{\gtG[n]}\subst{\gtRecVar[n+1]}{\gtG}}{\roleP}=\gtDepth{\gtG\subst{\gtRecVar[1]}{\gtG[1]}\dots\subst{\gtRecVar[n]}{\gtG[n]}}{\roleP}-1$.
	
	If $\gtG=\gtComm{\rolePi}{\roleQi}{i\in I}{\gtLab[i]}{\tyGround[i]}{\gtG[i]}$
	and $\roleP\notin\{\rolePi,\roleQi\}$, then
	for $i\in I$, let $\stEnvAnnotGenericSym[i]=\ltsSendRecv{\rolePi}{\roleQi}{\stChoice{\stLab[i]}{\tyGround[i]}}$.
	Pick $j\in I$ and set $\stEnvAnnotGenericSym=\stEnvAnnotGenericSym[j]$.
	If $\ltsSubject{\vec{\stEnvAnnotGenericSymi}}\supseteq\ltsSubject{\stEnvAnnotGenericSym}$ and
	$\roleP\notin\ltsSubject{\vec{\stEnvAnnotGenericSymi}}$
	and $\gtG\subst{\gtRecVar[1]}{\gtG[1]}\dots\subst{\gtRecVar[n]}{\gtG[n]}\,\gtMove[\vec{\stEnvAnnotGenericSymi}]\,\gtGi$, then
	let $\vec{\stEnvAnnotGenericSymii}\stEnvAnnotGenericSymiii$
	be the minimal prefix of $\vec{\stEnvAnnotGenericSymi}$
	such that $\ltsSubject{\vec{\stEnvAnnotGenericSymi}}\cap\ltsSubject{\stEnvAnnotGenericSymiii}\neq\emptyset$.
	So,
	$\gtG\subst{\gtRecVar[1]}{\gtG[1]}\dots\subst{\gtRecVar[n]}{\gtG[n]}\,\gtMove[\vec{\stEnvAnnotGenericSymii}]\,
	\gtComm{\rolePi}{\roleQi}{i\in I}{\gtLab[i]}{\tyGround[i]}{\gtGi[i]}
	$
	with, for $i\in I$,
	$\gtDepth{\gtGi[i]}{\roleP}\leq\gtDepth{\gtG[i]}{\roleP}$.
	So $\stEnvAnnotGenericSymiii=\stEnvAnnotGenericSym[k]$
	for some $k\in I$.
	So,
	$\gtG\subst{\gtRecVar[1]}{\gtG[1]}\dots\subst{\gtRecVar[n]}{\gtG[n]}\,\gtMove[\vec{\stEnvAnnotGenericSymii}\stEnvAnnotGenericSymiii]\,
	\gtGi[k]
	$
	and
	$\gtDepth{\gtGi[k]}{\roleP}\leq \gtDepth{\gtG\subst{\gtRecVar[1]}{\gtG[1]}\dots\subst{\gtRecVar[n]}{\gtG[n]}}{\roleP}-1$.
	$\roleP\notin\ltsSubject{\vec{\stEnvAnnotGenericSymi}}$,
	so we can complete the path
	to get that
	$\gtG\subst{\gtRecVar[1]}{\gtG[1]}\dots\subst{\gtRecVar[n]}{\gtG[n]}\,\gtMove[\vec{\stEnvAnnotGenericSymi}]\,
	\gtGi
	$
	and $\gtDepth{\gtGi}{\roleP}<\gtDepth{\gtG\subst{\gtRecVar[1]}{\gtG[1]}\dots\subst{\gtRecVar[n]}{\gtG[n]}}{\roleP}-1$.
	
	If $\gtG=\gtComm{\roleP}{\roleQ}{i\in I}{\gtLab[i]}{\tyGround[i]}{\gtG[i]}$, then
	pick $j\in I$, let $\stEnvAnnotGenericSym=\ltsSendRecv{\roleP}{\roleQ}{\stChoice{\stLab[j]}{\tyGround[j]}}$.
	
	If $\gtG=\gtComm{\roleQ}{\roleP}{i\in I}{\gtLab[i]}{\tyGround[i]}{\gtG[i]}$, then
	pick $j\in I$, let $\stEnvAnnotGenericSym=\ltsSendRecv{\roleQ}{\roleP}{\stChoice{\stLab[j]}{\tyGround[j]}}$.
\end{proof}

\thmProjBal*

\begin{proof}
	Note that
	$\stEnv$ is live iff
	$\forall\roleP.\;
	\stEnvSendLiveRoleP{\roleP}{\stEnv}
	\;\wedge\;
	\stEnvRecvLiveRoleP{\roleP}{\stEnv}$
	and
	if $\gtG$ is balanced then
	for all $\roleP$,
	$\gtG$ is $\roleP$-send balanced
	and
	$\roleP$-receive balanced.
	Thus, proving
	(\ref{bal-env2}) and
	(\ref{bal-env3})
	is sufficient.
	
	Suppose that $\stEnvAssoc{\gtG}{\stEnv}{}$ and $\gtG$ is $\roleP$-send balanced.
	\\
	Let $(\stEnv[n])_{n \in N}$ be a fair path with $\stEnv=\stEnv[0]$.
	\\
	For $n,n+1\in N$,
    let $\roleP[n]$, $\roleQ[n]$, $\stLab[n]$, and $\tyGround[n]$ be such that
	$\stEnv[n] \stEnvMoveCommAnnot{\roleP[n]}{\roleQ[n]}{\stChoice{\stLab[n]}{\tyGround[n]}} \stEnv[n]$.
	\\
	By~\Cref{thm:completeness_association},
    find $\gtG[n]$ for $n\in N$ such that
    $\stEnvAssoc{\gtG[n]}{\stEnv[n]}{}$ such that
    $\gtG[n]\,\gtMove[\ltsSendRecv{\roleP[n]}{\roleQ[n]}{\stChoice{\stLab[n]}{\tyGround[n]}}]\,\gtG[n+1]$.
    \\
	Suppose that $\stEnvApp{\stEnv[i]}{\roleP}\,\gtMove[{\roleQ\stEnvAnnotOutSym\stChoice{\stLab}{\tyGround}}]$,
    so $\unfoldOne{\stEnvApp{\stEnv[i]}{\roleP}}\neq\stEnd$
    so $\roleP\in\gtRoles{\gtG[i]}$.

    Let $\gtG[i]\,\gtMove[\vec{\stEnvAnnotGenericSym}]\,\gtGi$
    with
    $\roleP\notin\ltsSubject{\vec{\stEnvAnnotGenericSym}}$
    $\gtGi\,\gtMove[\stEnvAnnotGenericSymi]$,
    if $\roleP\in\gtSendRoles{\gtG[i]}$ then
    $\stEnvAnnotGenericSymi=\ltsSendRecv{\roleP}{\roleR}{\stChoice{\stLabi}{\tyGroundi}}$, and
    if $\roleP\in\gtRecvRoles{\gtG[i]}$ then
    $\stEnvAnnotGenericSymi=\ltsSendRecv{\roleR}{\roleP}{\stChoice{\stLabi}{\tyGroundi}}$,
    since $\roleP\in\gtRoles{\gtG[i]}$.
    Note that $\roleP\in\gtRecvRoles{\gtG[i]}\cup\gtSendRoles{\gtG[i]}=\gtRoles{\gtG[i]}$.
    By the good projection criteria,
    there is $\stEnvi$ such that
    $\gtProj{}{\gtGi}{\stEnvi}$,
    $\stEnvApp{\stEnv[i]}{\roleP}\tySub\stEnvApp{\stEnvi}{\roleP}$.
    If $\stEnvAnnotGenericSym=\ltsSendRecv{\roleR}{\roleP}{\stChoice{\stLabi}{\tyGroundi}}$, then
    $\stEnvApp{\stEnvi}{\roleP}$ is a receive from $\roleR$,
    so $\stEnvApp{\stEnvi}{\roleP}$ is a receive from $\roleR$,
    contradicting that $\stEnvApp{\stEnv[i]}{\roleP}\,\gtMove[{\roleQ\stEnvAnnotOutSym\stChoice{\stLab}{\tyGround}}]$.
    So $\roleP\in\gtSendRoles{\gtG[i]}$,
    so $\gtDepth{\gtG[i]}{\roleP}$ exists.
    
    By induction on $\gtDepth{\gtG[i]}{\roleP}$,
    there is $j\geq i$ such that
    $\stEnv[j] \stEnvMoveCommAnnot{\roleP}{\roleQ}{\stChoice{\stLabii}{\tyGroundii}}$.
    
    By \cref{lem:glb-approach},
    there is $\stEnvAnnotGenericSym$ such that
	$\gtG[i]\,\gtMove[\ltsSendRecv{\rolePi}{\roleQi}{\stChoice{\stLabi}{\tyGroundi}}]$
	and if $\gtG[i]\,\gtMove[\vec{\stEnvAnnotGenericSymi}]\,\gtGi$
	with $\{\rolePi,\roleQi\}\subseteq\ltsSubject{\stEnvAnnotGenericSymi}$,
	then either $\roleP\in\ltsSubject{\vec{\stEnvAnnotGenericSymi}}$
	or $\gtDepth{\gtGi[i]}{\roleP}<\gtDepth{\gtG[i]}{\roleP}$.
	By~\Cref{thm:soundness_association} and fairness,
    there is $k\geq i$ with $\rolePi=\roleP[k]$,
    $\roleQi=\roleQ[k]$,
    $\stLab[k]=\stLabi$,
    and $\tyGround[k]=\tyGroundi$.
    If there is $i\leq j \leq k$
    such that $\roleP\in\{\roleQ[j],\roleP[j]\}$
    then we are done;
    by picking the least $j$
    we would get $\roleQ[j]=\roleQ$,
    $\roleP[j]=\roleP$.
    Otherwise,
    $\gtDepth{\gtG[k+1]}{\roleP}<\gtDepth{\gtG[i]}{\roleP}$.
    By the I.H.,
     there is $j\geq k+1>i $ such that
    $\stEnv[j] \stEnvMoveCommAnnot{\roleP}{\roleQ}{\stChoice{\stLabii}{\tyGroundii}}$.
	
	Hence, $\stEnv$ is $\roleP$-send live.
	
	The case of $\gtG$ being $\roleP$-receive live is similar.
\end{proof}

\begin{theorem}[Liveness of Multiparty Sessions - Bottom]
\label{thm:sess-live-bot}
Let $\gtProj{}{}{}$ in \inferrule{SessBottom} in \cref{def:type-rule-bot}
be a good projection.
Suppose that $\tyJudge{}{\M}{\stEnv}$.
If $\stEnv$ is $\roleP$-send/receive live,
then so is $\M$.
If $\stEnv$ is live, then so is $\M$.
\end{theorem}

\begin{proof}
	Suppose that $\tyJudge{}{\M}{\stEnv}$,
	$\M\mpMoveStar\M'$, and write
	$\M'=\Pi_{\roleR\in\dom{\stEnv}}{\playsrole{\roleR}{\mpP[\roleR]}}$.
	By \cref{thm:sr-bot}, there is $\stEnvi$
	such that $\stEnv\stEnvMoveStar\stEnvi$
	and $\tyJudge{}{\M'}{\stEnvi}$.
	
	Let $\M_0=\M'$
	and $\stEnv[0]=\stEnvi$.
	
	Given $\M_i$ and $\stEnv[i]$ with $\tyJudge{}{\M_i}{\stEnv[i]}$ for $i\leq n$,
	if $\stEnv[n]\not\stEnvMove$
	then $N=\{0,1,\dots,n\}$.
	Otherwise,
	consider $S=\bigcup\{\ltsSubject{\stEnvAnnotGenericSym}\suchthat\stEnv[n]\stEnvMoveAnnot{\stEnvAnnotGenericSym}\}$.
	Pick $\roleP\in S$ such that
	$\max{\{i< n\suchthat \exists\stEnvAnnotGenericSym.\;\stEnv[i]\stEnvMoveAnnot{\stEnvAnnotGenericSym}\stEnv[i+1]\}\cup\{-1\}}$
	is least.
	So there are $\roleQ$, $\stLabi$, and $\tyGroundi$
	such that
	$\stEnv[n]\stEnvMoveAnnot{\ltsSendRecv{\roleP}{\roleQ}{\stChoice{\stLabi}{\tyGroundi}}}$
	or
	$\stEnv[n]\stEnvMoveAnnot{\ltsSendRecv{\roleQ}{\roleP}{\stChoice{\stLabi}{\tyGroundi}}}$.
	If the latter, swap $\roleP$ and $\roleQ$.
	By \cref{thm:sf-bot},
	we have
	$\M_n\mpMoveTauStar\M'_n\mpMoveAnnot{\roleP}{\roleQ}{\stLab}{v}\M_{n+1}$,
	$\stEnv[n]\stEnvMoveAnnot{\ltsSendRecv{\roleP}{\roleQ}{\stChoice{\stLab}{\tyGround}}}\stEnv[n+1]$, and
	$\tyJudge{}{\stEnv[n+1]}{\stEnv[n+1]}$.
	
	If $\stEnv[n]\stEnvMove$ for all $n\in \mathbb{N}$,
	then $N=\mathbb{N}$.
	
	By our choice of transitions,
	$\stEnv\stEnvMoveStar\stEnvi,\{\stEnv[i]\}_{i\in N\setminus\{0\}}$ is a fair path.
	
	Suppose that $\stEnv$ is $\roleP$-send live.
	If $\mpP[\roleP]=\roleQ\mathord{!}\stLab\langle e \rangle \mpSeq \mpP$,
	then by inversion,
	$\stEnvApp{\stEnvi}{\roleP}\gtMove[\roleQ\stFmt{\oplus}\stChoice{\stLab}{\tyGround}]$.
	By fairness of the path, there is $i\in N$
	such that $\stEnv[i]\stEnvMoveAnnot{\ltsSendRecv{\roleP}{\roleQ}{\stLabi}{\tyGroundi}}\stEnv[i+1]$.
	Let $i$ be minimal,
	so $\M_i=\playsrole{\roleP}{\mpP[\roleP]}\mpPar\Pi_{\roleR\in\dom{\stEnv}\setminus\{\roleP\}}{\playsrole{\roleR}{\mpPi[\roleR]}}$,
	$\M_i\mpMoveTauStar\M'_i=
	\playsrole{\roleP}{\mpP[\roleP]}\mpPar\Pi_{\roleR\in\dom{\stEnv}\setminus\{\roleP\}}{\playsrole{\roleR}{\mpPii[\roleR]}}
	\mpMoveAnnot{\roleP}{\roleQ}{\stLabi}{v}
	\M_{i+1}
	$.
	Thus, $\mpPii[\roleQ]=\textstyle\sum_{i\in I} \roleP\mathord{?}\stLab[i](x_i)\mpSeq \mpQ[i]$
	with $\stLab=\stLabi\in\{\stLab[i]\}_{i\in I}$.
	So $\M$ is $\roleP$-send live.
	
	Suppose that $\stEnv$ is $\roleP$-receive live.
	If $\mpP[\roleP]=\textstyle\sum_{i\in I} \roleQ\mathord{?}\stLab[i](x_i)\mpSeq \mpQ[i]$,
	then by inversion,
	$\stEnvApp{\stEnvi}{\roleP}\gtMove[\roleQ\stFmt{\&}\stChoice{\stLab[j]}{\tyGround[j]}]$ for any $j\in I$.
	By fairness of the path, there is $i\in N$
	such that $\stEnv[i]\stEnvMoveAnnot{\ltsSendRecv{\roleQ}{\roleP}{\stLab}{\tyGround}}\stEnv[i+1]$.
	Let $i$ be minimal,
	so $\M_i=\playsrole{\roleP}{\mpP[\roleP]}\mpPar\Pi_{\roleR\in\dom{\stEnv}\setminus\{\roleP\}}{\playsrole{\roleR}{\mpPi[\roleR]}}$,
	$\M_i\mpMoveTauStar\M'_i=
	\playsrole{\roleP}{\mpP[\roleP]}\mpPar\Pi_{\roleR\in\dom{\stEnv}\setminus\{\roleP\}}{\playsrole{\roleR}{\mpPii[\roleR]}}
	\mpMoveAnnot{\roleQ}{\roleP}{\stLab}{v}
	\M_{i+1}
	$.
	Thus, $\mpPii[\roleQ]=\roleP\mathord{!}\stLab\langle e \rangle \mpSeq \mpP$
	with $\stLab\in\{\stLab[i]\}_{i\in I}$.
	So $\M$ is $\roleP$-receive live.
	
	If $\stEnv$ is live, then for all $\roleP$ it is $\roleP$-receive live and $\roleP$-send live.
	Thus for all $\roleP$,
	$\M$ is $\roleP$-receive live and $\roleP$-send live.
	Thus, $\M$ is live.
\end{proof}

\thmTypeBal*

\begin{proof}
	This follows immediately from
	\cref{thm:sess-live-bot}
	by applying \cref{lem:bottom-top}
	and
	\cref{thm:proj-bal}.
\end{proof}

\begin{example}[Examples for each region of \cref{fig:liveness}]
\label{ex:liveness-fig}
Let $\gtProj{}{}{}$ be the full coinductive projection.
\\
\centerline{
\begin{tabular}{c||cccc}
$\stEnv$&$\stEnvSafePred$&$\stEnvDFPred$&$\stEnvSendLivePred$&$\stEnvRecvLivePred$
\\
\hline
$
		\set{\stEnvMap{\roleP}{\stFmt{\roleR{\oplus}\stLab}}
		\stEnvComp
		\stEnvMap{\roleQ}{\stFmt{\roleR{\&}\stLabi}}}
$
& $\times$ & $\times$ & $\times$ & $\times$
\\
$
		\set{\stEnvMap{\roleP}{\stFmt{\roleR{\oplus}\stLab}}
		\stEnvComp
		\stEnvMap{\roleQ}{\stFmt{\roleR{\&}\stLab}}}
$
& \checkmark & $\times$ & $\times$ & $\times$
\\
$
		\set{
		\stEnvMap{\roleP}{\stFmt{\roleQ{\oplus}\stLab}}
		\stEnvComp
		\stEnvMap{\roleQ}{\stFmt{\roleP{\&}\stLabi}}
		\stEnvComp
		\stEnvMap{\roleFmt{a}}{\stRec{\stRecVar}{\roleFmt{b}{\oplus}\stLab\stSeq\stRecVar}}
		\stEnvComp
		\stEnvMap{\roleFmt{b}}{\stRec{\stRecVar}{\roleFmt{a}{\&}\stLab\stSeq\stRecVar}}
		}
$
& $\times$ & \checkmark & $\times$ & $\times$
\\
$
		\set{
		\stEnvMap{\roleP}{\stIntSum{\roleQ}{}{\stLab,\,\stLabi}}
		\stEnvComp
		\stEnvMap{\roleQ}{\stFmt{\roleP{\&}\stLabi\stSeq\roleP{\&}\stLab}}
		}
$
& $\times$ & $\times$ & \checkmark & $\times$
\\
$
		\set{
		\stEnvMap{\roleP}{\stIntSum{\roleQ}{}{\stLab,\,\stLabi}}
		\stEnvComp
		\stEnvMap{\roleQ}{\stFmt{\roleP{\&}\stLabi\stSeq\roleP{\oplus}\stLab}}
		}
$
& $\times$ & $\times$ & $\times$ & \checkmark
\\
$
\set{
		\stEnvMap{\roleP}{\stFmt{\roleFmt{a}{\oplus}\stLab}}
		\stEnvComp
		\stEnvMap{\roleQ}{\stFmt{\roleFmt{a}{\&}\stLab}}
		\stEnvComp
		\stEnvMap{\roleFmt{a}}{\stRec{\stRecVar}{
		\stIntSum{\roleFmt{b}}{}{
		\stLab\stSeq\stRecVar,\,\stLabi\stSeq
		\roleP{\&}\stLab\stSeq
		\roleQ{\oplus}\stLab
		}
		}}
		\stEnvComp
		\stEnvMap{\roleFmt{b}}{\stRec{\stRecVar}{
		\stExtSum{\roleFmt{a}}{}{\stLab\stSeq\stRecVar,\,\stLabi}}}
		}
$
& \checkmark & \checkmark & $\times$ & $\times$
\\
$
\set{
		\stEnvMap{\roleP}{\stFmt{\roleQ{\&}\stLab}}
		}
$
& \checkmark & $\times$ & \checkmark & $\times$
\\
$
\set{
		\stEnvMap{\roleP}{\stFmt{\roleQ{\oplus}\stLab}}
		}
$
& \checkmark & $\times$ & $\times$ & \checkmark
\\
$
\set{
		\stEnvMap{\roleP}{\stFmt{\roleQ{\&}\stLab}}
		\stEnvComp
		\stEnvMap{\roleFmt{a}}{\stRec{\stRecVar}{
		\stIntSum{\roleFmt{b}}{}{\stLab\stSeq\stRecVar,\,\stLabi}
		}}
		\stEnvComp
		\stEnvMap{\roleFmt{b}}{\stRec{\stRecVar}{\roleFmt{a}{\&}\stLab\stSeq\stRecVar}}
		}
$
& $\times$ & \checkmark & \checkmark & $\times$
\\
$
\set{
		\stEnvMap{\roleP}{\stFmt{\roleQ{\oplus}\stLab}}
		\stEnvComp
		\stEnvMap{\roleFmt{a}}{\stRec{\stRecVar}{
		\stIntSum{\roleFmt{b}}{}{\stLab\stSeq\stRecVar,\,\stLabi}
		}}
		\stEnvComp
		\stEnvMap{\roleFmt{b}}{\stRec{\stRecVar}{\roleFmt{a}{\&}\stLab\stSeq\stRecVar}}
		}
$
& $\times$ & \checkmark & $\times$ & \checkmark
\\
N/A
& $\times$ & $\times$ & \checkmark & \checkmark
\\
$
\set{
		\stEnvMap{\roleP}{\stFmt{\roleFmt{a}{\&}\stLab}}
		\stEnvComp
		\stEnvMap{\roleFmt{a}}{\stRec{\stRecVar}{
		\stIntSum{\roleFmt{b}}{}{
		\stLab\stSeq\stRecVar,\,\stLabi\stSeq
		\roleP{\oplus}\stLab
		}
		}}
		\stEnvComp
		\stEnvMap{\roleFmt{b}}{\stRec{\stRecVar}{
		\stExtSum{\roleFmt{a}}{}{\stLab\stSeq\stRecVar,\,\stLabi}}}
		}
$
& \checkmark & \checkmark & \checkmark & $\times$
\\
$
\set{
		\stEnvMap{\roleP}{\stFmt{\roleFmt{a}{\oplus}\stLab}}
		\stEnvComp
		\stEnvMap{\roleFmt{a}}{\stRec{\stRecVar}{
		\stIntSum{\roleFmt{b}}{}{
		\stLab\stSeq\stRecVar,\,\stLabi\stSeq
		\roleP{\&}\stLab
		}
		}}
		\stEnvComp
		\stEnvMap{\roleFmt{b}}{\stRec{\stRecVar}{
		\stExtSum{\roleFmt{a}}{}{\stLab\stSeq\stRecVar,\,\stLabi}}}
		}
$
& \checkmark & \checkmark & $\times$ & \checkmark
\\
N/A
& \checkmark & $\times$ & \checkmark & \checkmark
\\
$
\set{
		\stEnvMap{\roleP}{\stIntSum{\roleQ}{}{\stLab,\,\stLabi}}
		\stEnvComp
		\stEnvMap{\roleQ}{\stFmt{\roleP{\&}\stLab}}
		}
$
& $\times$ & \checkmark & \checkmark & \checkmark
\\
$
\set{
		\stEnvMap{\roleP}{\stEnd}
		}
$
& \checkmark & \checkmark & \checkmark & \checkmark
\end{tabular}}
\begin{itemize}
	\item Weak balanced and neither send balanced nor receive balanced:
	\[
		\gtG
		=\gtRec{\gtRecVar}{
		\gtCommRaw{\roleFmt{a}}{\roleFmt{b}}{
			\gtLab{\gtSeq}\gtRecVar,\,
			\gtLabi{\gtSeq}
			\gtCommSingle{\roleP}{\roleFmt{a}}{\gtLab}{}{
			\gtCommSingle{\roleFmt{a}}{\roleQ}{\gtLab}{}{}
			}
		}
		}
	\]
	\[
		\gtProj{}{\gtG}{
		\set{
		\stEnvMap{\roleP}{\stFmt{\roleFmt{a}{\oplus}\stLab}}
		\stEnvComp
		\stEnvMap{\roleQ}{\stFmt{\roleFmt{a}{\&}\stLab}}
		\stEnvComp
		\stEnvMap{\roleFmt{a}}{\stRec{\stRecVar}{
		\stIntSum{\roleFmt{b}}{}{
		\stLab\stSeq\stRecVar,\,\stLabi\stSeq
		\roleP{\&}\stLab\stSeq
		\roleQ{\oplus}\stLab
		}
		}}
		\stEnvComp
		\stEnvMap{\roleFmt{b}}{\stRec{\stRecVar}{
		\stExtSum{\roleFmt{a}}{}{\stLab\stSeq\stRecVar,\,\stLabi}}}
		}
		}
	\]
\item Weak balanced and send balanced but not receive balanced:
	\[
		\gtG
		=\gtRec{\gtRecVar}{
		\gtCommRaw{\roleFmt{a}}{\roleFmt{b}}{
			\gtLab{\gtSeq}\gtRecVar,\,
			\gtLabi{\gtSeq}
			{
			\gtCommSingle{\roleFmt{a}}{\roleP}{\gtLab}{}{}
			}
		}
		}
	\]
	\[
		\gtProj{}{\gtG}{
		\set{
		\stEnvMap{\roleP}{\stFmt{\roleFmt{a}{\&}\stLab}}
		\stEnvComp
		\stEnvMap{\roleFmt{a}}{\stRec{\stRecVar}{
		\stIntSum{\roleFmt{b}}{}{
		\stLab\stSeq\stRecVar,\,\stLabi\stSeq
		\roleP{\oplus}\stLab
		}
		}}
		\stEnvComp
		\stEnvMap{\roleFmt{b}}{\stRec{\stRecVar}{
		\stExtSum{\roleFmt{a}}{}{\stLab\stSeq\stRecVar,\,\stLabi}}}
		}
		}
	\]
\item Weak balanced and receive balanced but not send balanced:
	\[
		\gtG
		=\gtRec{\gtRecVar}{
		\gtCommRaw{\roleFmt{a}}{\roleFmt{b}}{
			\gtLab{\gtSeq}\gtRecVar,\,
			\gtLabi{\gtSeq}
			{
			\gtCommSingle{\roleP}{\roleFmt{a}}{\gtLab}{}{}
			}
		}
		}
	\]
	\[
		\gtProj{}{\gtG}{
		\set{
		\stEnvMap{\roleP}{\stFmt{\roleFmt{a}{\oplus}\stLab}}
		\stEnvComp
		\stEnvMap{\roleFmt{a}}{\stRec{\stRecVar}{
		\stIntSum{\roleFmt{b}}{}{
		\stLab\stSeq\stRecVar,\,\stLabi\stSeq
		\roleP{\&}\stLab
		}
		}}
		\stEnvComp
		\stEnvMap{\roleFmt{b}}{\stRec{\stRecVar}{
		\stExtSum{\roleFmt{a}}{}{\stLab\stSeq\stRecVar,\,\stLabi}}}
		}
		}
	\]
\item Balanced:
\[
	\gtG = \gtEnd
	\qquad\qquad
	\gtProj{}{\gtEnd}{\set{\stEnvMap{\roleP}{\stEnd}}}
\]
Alternatively:
\[
	\gtG
		=\gtRec{\gtRecVar}{
		\gtCommSingle{\roleFmt{a}}{\roleFmt{b}}{\gtLab}{}{\gtRecVar}}
	\qquad\qquad
	\gtProj{}{\gtG}{
	\set{
		\stEnvMap{\roleFmt{a}}{\stRec{\stRecVar}{\roleFmt{b}{\oplus}\stLab\stSeq\stRecVar}}
		\stEnvComp
		\stEnvMap{\roleFmt{b}}{\stRec{\stRecVar}{\roleFmt{a}{\&}\stLab\stSeq\stRecVar}}
	}
	}
\]
\end{itemize}
\end{example}
\section{Appendix for~\cref{sec:example} (Expressiveness and Separation)}

\label{app:example}
\subsection{Appendix for~\cref{sec:indcoind} (Inductive and Coinductive Projections)}

The good criteria are used to
unify our treatment of
projections,
\emph{not} as conditions to check
(although failures to satisfy the conditions
may be used to produce counterexamples
to type correctness).
Being a non-trivial property of arbitrary algorithms,
the conditions are undecidable.

\begin{restatable}[Good Criteria are Undecidable]{theorem}{thmGoodUndec}
\label{thm:good-undec}
\
\begin{enumerate}[leftmargin=0.15in,labelindent=-\leftmargin]
\item There is a Turing reduction from
the Halting problem to the Good Projection predicate
on computable projections
and from the Halting problem to its negation.
\item There is a Turing reduction from the Halting Problem
to the negation of the Good Merge predicate
on computable merge operations.
\end{enumerate}
\end{restatable}

\begin{proof}
	Let $\langle T, x\rangle$ be
	an instance of the halting problem.
	\[
		\stT[0] = \roleP\stFmt{\&}\stLab
		\qquad \stT[n+1]=\roleP\stFmt{\&}\stLab\stSeq\stT[n]\\
	\]
	Define the projection
	$\gtProj{}{}{}$
	by $\gtProj{\roleP}{\gtEnd}{\stT[n]}$ iff
	$\langle T, x\rangle$ halts within $n$ steps.
	$\gtProj{}{}{}$ is good iff $\langle T, x\rangle$ does not halt.
	\[
		\stTi[0]=\stEnd
		\qquad
		\stTi[n+1]=\stRec{\stRecVar}{\stEnd}
	\]
	Define the projection
	$\gtProji{}{}{}$ by
	$\gtProji{\roleQ}{\gtCommSingle{\roleP}{\roleQ}{\gtLab}{}{}}{\roleP\stFmt{\&}\stLab}$,
	$\gtProji{\roleP}{\gtCommSingle{\roleP}{\roleQ}{\gtLab}{}{}}{\roleQ\stFmt{\oplus}\stLab}$,
	and for all $\roleR$,\\
	$\gtProji{\roleR}{\gtEnd}{\stTi[n]}$
	if $\langle T, x\rangle$ halts within $n$ steps.
	$\gtProji{}{}{}$ is good iff $\langle T, x\rangle$ halts.
	\\[2ex]
	Define $\stMerge{}{}$ by
	$\stMerge{}{\{\stT[n]\}}=\{\stT[n+1]\}$ if $\langle T, x\rangle$
	halts within $n$ steps.
	$\stMerge{}{}$ is good iff $\langle T, x\rangle$ does not halt. 
\end{proof}

\begin{example}[Type Equivalence with Merge and Projection]
\label{ex:equiv-types}
Let $\gtProj{}{}{}$ be the full coinductive projection.
Consider the following types:
\[
		\stT = \stRec{\stRecVar}{
			\stOut{\roleFmt{b}}{\stChoice{\stLabFmt{m}}{}\stSeq\stRecVar}{}
		}\qquad
		\stTi = \stRec{\stRecVar}{
			\stOut{\roleFmt{b}}{\stChoice{\stLabFmt{m}}{}\stSeq
			\stOut{\roleFmt{b}}{\stChoice{\stLabFmt{m}}{}\stSeq
			\stRecVar}{}
			}{}
		}
\]
We have that $\stT\bisim\stTi$, so that $\stMerge{}{\left\{\stT,\stTi\right\}}\ni \stT,\stTi$.
An even simpler example would be $\stEnd\bisim \stRec{\stRecVar}{\stEnd}$, with a similar conclusion.
As we want the coinductive merges to be defined semantically,
we must allow multiple outputs from merges.
\newline
Now consider the following global type:
\[
		\gtG = \gtRec{\gtRecVar}{
			\gtCommSingle{\roleFmt{a}}{\roleFmt{b}}{m}{}{\gtRecVar}
		}
\]
We have that $\gtProj{\roleFmt{a}}{\gtG}{\stT}$ and $\gtProj{\roleFmt{a}}{\gtG}{\stTi}$.
As the coinductive projections are defined up-to unfoldings,
we must allow global types to project onto multiple
local types.

It is relevant to note that the output of the full coinductive merge will either be the empty set or a single bisimilarity class of local types.
\end{example}

The use of merge operations, to ensure that the projections of all continuations are ``compatible''
along branches in a global type, is explored in~\Cref{ex:types_oath}, ~\Cref{ex:merge}, and~\Cref{ex:another-merge-proj}.

\begin{example}[Authentication Protocol]
\label{ex:types_oath}
Consider a service with two profiles, Alice and Bob.
The default for this service is that a client may make password attempts until
they are authenticated.
This service provides a security option to prevent brute-force attacks, 
where entering a password may only be done once per session.
Alice has left on the default setting, while Bob has enabled the security option.
The service now has the following login protocol:
\newline
The $\roleFmt{s}$ervice sends
to the $\roleFmt{c}$lient \emph{either} a request
to $\gtMsgFmt{loginAlice}$
\emph{or} $\gtMsgFmt{loginBob}$.
In the first case,
$\roleFmt{c}$ continues by sending $\gtMsgFmt{passwd}$
(carrying a $\stFmtC{str}$ing)
to the $\roleFmt{a}$uthorisation server,
which then sends $\emph{either}$ $\gtMsgFmt{pass}$
or $\gtMsgFmt{fail}$
to $\roleFmt{s}$.
If $\roleFmt{s}$ receives $\gtMsgFmt{pass}$, then
$\roleFmt{s}$ sends $\gtMsgFmt{success}$ to $\roleFmt{c}$ and the session $\gtEnd$s.
If $\roleFmt{s}$ receives $\gtMsgFmt{fail}$, then
$\roleFmt{s}$ sends $\gtMsgFmt{fail}$ to $\roleFmt{c}$,
$\roleFmt{s}$ sends $\gtMsgFmt{continue}$ to $\roleFmt{a}$, and
the protocol loops back to $\roleFmt{c}$ sending a $\gtMsgFmt{passwd}$.
In the second case,
$\roleFmt{c}$ continues by sending $\gtMsgFmt{passwd}$
(carrying a $\stFmtC{str}$ing)
to the $\roleFmt{a}$uthorisation server.
In response, $\roleFmt{a}$ sends $\emph{either}$ $\gtMsgFmt{pass}$
or $\gtMsgFmt{fail}$ to $\roleFmt{s}$.
If $\roleFmt{s}$ receives $\gtMsgFmt{pass}$, then the session $\gtEnd$s.
If $\roleFmt{s}$ receives $\gtMsgFmt{fail}$, then
$\roleFmt{s}$ sends $\gtMsgFmt{stop}$ to $\roleFmt{a}$,
and the session $\gtEnd$s.
This protocol can be represented by the global type $\gtExChoice$.
\newline
\smallskip
\centerline{\(
\gtExChoice =
	\gtCommRaw{\roleServer}{\roleClient}
	{
		\begin{array}{l}%
			\gtCommChoice{loginAlice}{}
			{
				\gtRec{\gtRecVar}
				{
					\gtCommSingle{\roleClient}{\roleAuth}{passwd}{\stFmtC{str}}
					{
						\gtCommRaw{\roleAuth}{\roleServer}
						{
							\begin{array}{l}
								\gtCommChoice{pass}{}
								{
									\gtCommSingle{\roleServer}{\roleClient}{success}{}{}
								}
								\\
								\gtCommChoice{fail}{}
								{
									\gtCommSingle{\roleServer}{\roleClient}{fail}{}{
									\gtCommSingle{\roleServer}{\roleAuth}{continue}{}{\gtRecVar}
									}
								}
							\end{array}
						}
					}
				}
			}
	        \\
			\gtCommChoice{loginBob}{}
			{
				\gtCommSingle{\roleClient}{\roleAuth}{passwd}{\stFmtC{str}}
				{
					\gtCommRaw{\roleAuth}{\roleServer}
					{
						\begin{array}{l}
							\gtCommChoice{pass}{}{}
							\\
							\gtCommChoice{fail}{}
							{
								\gtCommSingle{\roleServer}{\roleAuth}{stop}{}{}
							}
						\end{array}
					}
				}
			}
		\end{array}
	}
\)}
\smallskip
\noindent
$\gtExChoice$  is projected onto the following local types:
\newline
\smallskip
\centerline
{\(
	\begin{array}{lcl}
		\stT[\roleFmt{s}]&=& 
			\stIntSum{\roleClient}{}
			{
				\begin{array}{@{\hskip 0mm}l@{\hskip 0mm}}
					\stChoice{\stLabFmt{loginAlice}}{} \stSeq
					\stRec{\stRecVar}
					{
						\stExtSum{\roleAuth}{}
						{
							\begin{array}{@{\hskip 0mm}l@{\hskip 0mm}}
								\stChoice{\stLabFmt{pass}}{} \stSeq
								\stOut{\roleClient}{\stChoice{\stLabFmt{success}}{}}{}\\
								\stChoice{\stLabFmt{fail}}{} \stSeq
								\stOut{\roleClient}
								{
									\stChoice{\stLabFmt{fail}}{}\stSeq
									\stOut{\roleAuth}
									{
										\stChoice{\stLabFmt{continue}}{} \stSeq
										\stRecVar
									}{}
								}{}
							\end{array}
						}
					}\\
					\stChoice{\stLabFmt{loginBob}}{} \stSeq
					\stExtSum{\roleAuth}{}
					{
						\begin{array}{@{\hskip 0mm}l@{\hskip 0mm}}
							\stChoice{\stLabFmt{pass}}{}\\
							\stChoice{\stLabFmt{fail}}{} \stSeq
							\stOut{\roleAuth}{\stChoice{\stLabFmt{stop}}{}}{}
						\end{array}
					}
				\end{array}
			}
		\\[8ex]
		\stT[\roleClient] &=&
			\stExtSum{\roleServer}{}
			{
				\begin{array}{@{\hskip 0mm}l@{\hskip 0mm}}
					\stChoice{\stLabFmt{loginAlice}}{} \stSeq
					\stRec{\stRecVari}
					{
						\stOut{\roleAuth}
						{
							\stChoice{\stLabFmt{passwd}}{\stFmtC{str}}\stSeq
							\stExtSum{\roleServer}{}
							{
								\begin{array}{@{\hskip 0mm}l@{\hskip 0mm}}
									\stChoice{\stLabFmt{success}}{}\\
									\stChoice{\stLabFmt{fail}}{} \stSeq \stRecVari
								\end{array}
							}
						}{}
					}\\
					\stChoice{\stLabFmt{loginBob}}{} \stSeq
					\stOut{\roleAuth}
					{
						\stChoice{\stLabFmt{passwd}}{\stFmtC{str}}
					}{}
				\end{array}
			}
		\\[8ex]
		\stT[\roleAuth]& =& 
			\stRec{\stRecVarii}
			{
				\stIn{\roleClient}
				{
					\stChoice{\stLabFmt{passwd}}{\stFmtC{str}}\stSeq
					\stIntSum{\roleServer}{}
					{
						\begin{array}{@{\hskip 0mm}l@{\hskip 0mm}}
							\stChoice{\stLabFmt{pass}}{}\\
							\stChoice{\stLabFmt{fail}}{} \stSeq
							\stExtSum{\roleServer}{}
							{
								\begin{array}{@{\hskip 0mm}l@{\hskip 0mm}}
									\stChoice{\stLabFmt{continue}}{} \stSeq \stRecVarii\\
									\stChoice{\stLabFmt{stop}}{}
								\end{array}
							}
						\end{array}
					}
				}{}{}
			}
	\end{array}
\)}
\end{example}

\begin{example}[Merge and Projection~\citep{MFYZ2021}]
\label{ex:merge}
Let $\stMerge{}{}$ be the full coinductive merge.
\\[2ex]
Two external choice~(branching) types from the same role with disjoint labels can be 
merged into a type
carrying both labels, 
\eg 
\[
\stMerge{}{\left\{\roleFmt{a} 
 \stFmt{\&}
  \stLabFmt{greet(\stFmtC{str})} 
  , 
  \roleFmt{a} 
 \stFmt{\&}
  \stLabFmt{farewell(\stFmtC{bool})} 
\right\}}
   \ni 
    \stExtSum{\roleFmt{a}}{}{
  \begin{array}{@{}l@{}}
   \stLabFmt{greet(\stFmtC{str})},\,
   \stLabFmt{farewell(\stFmtC{bool})} 
  \end{array}
  }
 \]
 
 However, this does not apply to internal choices~(selections), \eg 
 \[\left\{\roleFmt{a} 
 \stFmt{\oplus}
  \stLabFmt{greet(\stFmtC{str})} 
  , 
  \roleFmt{a} 
 \stFmt{\oplus}
 \stLabFmt{farewell(\stFmtC{bool})}\right\}\] has no merge,
since the only rule allowing types with a send at the head to merge is \inferrule{$\mergeRuleOut$},
which requires that the messages of all the types being merged are the same. 

Two external choices from different roles cannot be merged. Same 
for internal choices;  \eg
 \[\left\{\roleFmt{a} 
 \stFmt{\oplus}
  \stLabFmt{greet(\stFmtC{str})}
  , 
  \roleFmt{b} 
 \stFmt{\oplus}
 \stLabFmt{greet(\stFmtC{str})}
   \right\}\]
   has no merge,
since the only rule allowing types with a send at the head to merge is \inferrule{$\mergeRuleOut$},
which requires that the roles are the same. 

Furthermore, two external choices from the same role with same labels but different payloads cannot 
be merged;  \eg
  \[\left\{\roleFmt{a} 
 \stFmt{\&}
  \stLabFmt{greet(\stFmtC{str})} 
  ,
  \roleFmt{a} 
 \stFmt{\&}
 \stLabFmt{greet(\stFmtC{bool})}
   \right\}\] has no merge,
since the only rule allowing types with a receive at the head to merge is \inferrule{$\mergeRuleIn$},
which requires that the external choices can be combined to form the head of the merged type,
but well-formedness demands that we cannot have the same label with distinct message types.
This also applies to internal choices.

Additionally, two local types with same prefixes 
but unmergeable continuations 
cannot be merged;  \eg
\[\left\{\roleFmt{a} 
 \stFmt{\oplus}
  \stLabFmt{greet(\stFmtC{str})} 
  , 
  \roleFmt{a} 
 \stFmt{\oplus}
 \stLabFmt{greet(\stFmtC{str})}
   \stSeq
   \roleFmt{b} 
 \stFmt{\&}
 \stLabFmt{farewell(\stFmtC{bool})}
\right\}
 \] is not mergeable as we would need an application of rule \inferrule{$\mergeRuleOut$}, which would require
 \[\left\{\stEnd
  ,
  \roleFmt{b} 
 \stFmt{\&}
 \stLabFmt{farewell(\stFmtC{bool})}\right\}
   \] to be mergeable.
However, this is not mergeable as no rule allows $\stEnd$ to merge with a type that does not unfold to $\stEnd$.
   
   Consider the global type:
\begin{equation*}
\gtG = \gtCommRaw{\roleFmt{a}}{\roleFmt{b}}{
    \begin{array}{@{}l@{}}
    \gtCommChoice{\gtMsgFmt{greet}}{\stFmtC{str}}{
    \gtCommSingle{\roleFmt{c}}{\roleFmt{a}}
    {\gtMsgFmt{farewell}}{\stFmtC{bool}}{
    }}
   \\
     \gtCommChoice{\gtMsgFmt{farewell}}{\stFmtC{bool}}{
    \gtCommSingle{\roleFmt{c}}{\roleFmt{a}}{\gtMsgFmt{greet}}{\stFmtC{str}}{
    }
    }
    \end{array}
    }
\end{equation*}
$\gtG$ cannot be projected onto role $\roleFmt{c}$
since, if $\gtProj{\roleFmt{c}}{\gtG}{\stT}$,
then $\stMerge{}{\left\{\stTi,\stTii\right\}}\ni \stT$
where
\[
\gtProj{\roleFmt{c}}{\gtCommSingle{\roleFmt{c}}{\roleFmt{a}}{\gtMsgFmt{farewell}}{\stFmtC{bool}}{}}{\stTi}
\;\text{and}\;
\gtProj{\roleFmt{c}}{\gtCommSingle{\roleFmt{c}}{\roleFmt{a}}{\gtMsgFmt{greet}}{\stFmtC{str}}{}}{\stTii}\]
So, $\unfoldOne{\stTi}=\roleFmt{a}\stFmt{\oplus}\stLabFmt{farewell(\stFmtC{bool})}$ and 
$\unfoldOne{\stTii}=\roleFmt{a}\stFmt{\oplus}\stLabFmt{greet(\stFmtC{str})}$,
so that $\left\{\stTi,\stTii\right\}$ is not mergeable.
$\gtG$ doesn't project onto $\roleFmt{c}$ because, in $\gtG$, depending on 
whether $\roleFmt{a}$ and $\roleFmt{b}$ transmit $\gtMsgFmt{greet}$ or 
$\gtMsgFmt{farewell}$, $\roleFmt{c}$ is expected to send either 
$\gtMsgFmt{farewell}$ or 
$\gtMsgFmt{greet}$ to $\roleFmt{a}$. 
However, $\roleFmt{c}$ is not involved in the interactions between $\roleFmt{a}$ and 
$\roleFmt{b}$, so $\roleFmt{c}$ is not aware of which message to send
\ie $\gtG$ does not provide a valid specification for $\roleFmt{c}$. 
 \end{example}

\begin{example}[Dinner Party Conversation]
Let $\gtProj{}{}{}$ be the full coinductive projection.

Dave has invited Carol to a dinner party.
However he is overly talkative,
so Carol has decided to ignore him and type his behaviour.
At the start of the night,
Dave will talk repeatedly about either the canap\'es or the main.
Talking about the main reminds him to serve it,
after which he continues to talk repeatedly about the
canap\'es or the main.
We can type this communication via the following global and local types:
\[
	\begin{array}{lcllcl}
	\gtG&=&\gtRec{\gtRecVari}{
		\gtCommRaw{\roleFmt{Dave}}{\roleFmt{Carol}}{
			\begin{array}{@{}l@{}}
				\gtCommChoice{\gtMsgFmt{main}}{}{ \gtGi }\\
				\gtCommChoice{\gtMsgFmt{canapes}}{}{\gtRecVari}
			\end{array}
		}
	}&
	\gtGi&=&\gtRec{\gtRecVar}{
				\gtCommRaw{\roleFmt{Dave}}{\roleFmt{Carol}}{
					\begin{array}{@{}l@{}}
						\gtCommChoice{\gtMsgFmt{main}}{}{\gtRecVar}\\
						\gtCommChoice{\gtMsgFmt{canapes}}{}{\gtRecVar}
					\end{array}
				}
			}\\[4ex]
	\stT[\roleFmt{Dave}]&=&\stRec{\stRecVar}{
						\stIntSum{\roleFmt{Carol}}{}{
							\begin{array}{@{\hskip 0mm}l@{\hskip 0mm}}
								\stChoice{\stLabFmt{main}}{}\stSeq\stRecVar\\
								\stChoice{\stLabFmt{canapes}}{}\stSeq\stRecVar
							\end{array}
						}
					}
	&
	\stT[\roleFmt{Carol}]&=&\stRec{\stRecVar}{
						\stExtSum{\roleFmt{Dave}}{}{
							\begin{array}{@{\hskip 0mm}l@{\hskip 0mm}}
								\stChoice{\stLabFmt{main}}{}\stSeq\stRecVar\\
								\stChoice{\stLabFmt{canapes}}{}\stSeq\stRecVar
							\end{array}
						}
					}
	\end{array}
\]
The relations
\[\{(\gtGi,\stT[\roleFmt{Dave}]),(\gtG,\stT[\roleFmt{Dave}])\}
\;\text{and}\;
\{(\gtGi,\stT[\roleFmt{Carol}]),(\gtG,\stT[\roleFmt{Carol}])\}\]
satisfy the coinductive rules defining $\gtProj{\roleFmt{Dave}}{}{}$ and $\gtProj{\roleFmt{Carol}}{}{}$ respectively.

Therefore, $\gtProj{\roleFmt{Dave}}{\gtG}{\stT[\roleFmt{Dave}]}$ and $\gtProj{\roleFmt{Carol}}{\gtG}{\stT[\roleFmt{Carol}]}$.

\end{example}
\begin{remark}
The global type in the above example is semantically equivalent to $\gtGi$,
which is projectable using the plain inductive projection.
\end{remark}

\begin{example}[Another Dinner Party]
\label{ex:another-merge-proj}
Dave has invited Carol to another dinner party.
This time, Carol's friend Eloise is also invited.
Eloise may hint to Dave that last time he talked too much about the canap\'es and the main.
In this case, 
Dave will misunderstand the message
and will stop talking about canap\'es once he starts talking about the main.
Eloise, being a bad friend, may instead
tell Dave that Carol loved their conversation last time.
In this case,
Dave will talk about canap\'es and the main as much as he wants.
We can describe this situation using the following global and local types:
\[
	\begin{array}{lcllcl}
		\gtG&=&\gtCommRaw{\roleFmt{Eloise}}{\roleFmt{Dave}}{
			\begin{array}{@{}l@{}}
					\gtCommChoice{\gtMsgFmt{talkLess}}{}{
						\gtGi
					}\\
					\gtCommChoice{\gtMsgFmt{talkMore}}{}{
						\gtGii
					}
				\end{array}
		}
	&
		\gtGi&=&\gtRec{\gtRecVari}{
			\gtCommRaw{\roleFmt{Dave}}{\roleFmt{Carol}}{
				\begin{array}{@{}l@{}}
					\gtCommChoice{\gtMsgFmt{main}}{}{\gtGiii}\\
					\gtCommChoice{\gtMsgFmt{canapes}}{}{\gtRecVari}
				\end{array}
			}
		}\\[2mm]
		\gtGii&=&\gtRec{\gtRecVari}{
				\gtCommRaw{\roleFmt{Dave}}{\roleFmt{Carol}}{
					\begin{array}{@{}l@{}}
						\gtCommChoice{\gtMsgFmt{main}}{}{\gtRecVari}\\
						\gtCommChoice{\gtMsgFmt{canapes}}{}{\gtRecVari}
					\end{array}
				}
			}
	&
		\gtGiii&=&\gtRec{\gtRecVar}{\gtCommSingle{\roleFmt{Dave}}{\roleFmt{Carol}}{\gtMsgFmt{main}}{}{\gtRecVar}}
\\[2ex]
		\stT[\roleFmt{Dave}]&=&\stExtSum{\roleFmt{Eloise}}{}{
				\begin{array}{@{\hskip 0mm}l@{\hskip 0mm}}
					\stChoice{\stLabFmt{talkLess}}{}\stSeq\stTi[\roleFmt{Dave}]\\
					\stChoice{\stLabFmt{talkMore}}{}\stSeq\stTii[\roleFmt{Dave}]
				\end{array}
			}
&
		\stTi[\roleFmt{Dave}]&=&\stRec{\stRecVari}{
						\stIntSum{\roleFmt{Carol}}{}{
							\begin{array}{@{\hskip 0mm}l@{\hskip 0mm}}
								\stChoice{\stLabFmt{main}}{}\stSeq\stTiii[\roleFmt{Dave}]\\
								\stChoice{\stLabFmt{canapes}}{}\stSeq\stRecVari
							\end{array}
						}
					}
\\[2ex]
		\stTii[\roleFmt{Dave}]&=&\stRec{\stRecVar}{
						\stIntSum{\roleFmt{Carol}}{}{
							\begin{array}{@{\hskip 0mm}l@{\hskip 0mm}}
								\stChoice{\stLabFmt{main}}{}\stSeq\stRecVar\\
								\stChoice{\stLabFmt{canapes}}{}\stSeq\stRecVar
							\end{array}
						}
					}
&
		\stTiii[\roleFmt{Dave}]&=&\stRec{\stRecVar}{\roleFmt{Carol}\stFmt{\oplus}\stLabFmt{main}\stSeq\stRecVar}
\\[2ex]
		\stT[\roleFmt{Carol}]&=&\stRec{\stRecVar}{
			\stExtSum{\roleFmt{Dave}}{}{
				\begin{array}{@{\hskip 0mm}l@{\hskip 0mm}}
					\stChoice{\stLabFmt{main}}{}\stSeq\stRecVar\\
					\stChoice{\stLabFmt{canapes}}{}\stSeq\stRecVar\\
				\end{array}
			}
		}
&
		\stTi[\roleFmt{Carol}]&=&\stRec{\stRecVar}{\roleFmt{Dave}\stFmt{\&}\stLabFmt{main}\stSeq\stRecVar}
\\[2ex]
		\stTii[\roleFmt{Carol}]&=&\stRec{\stRecVari}{
			\stExtSum{\roleFmt{Dave}}{}{
				\begin{array}{@{\hskip 0mm}l@{\hskip 0mm}}
					\stChoice{\stLabFmt{main}}{}\stSeq\stTi[\roleFmt{Carol}]\\
					\stChoice{\stLabFmt{canapes}}{}\stSeq\stRecVari
				\end{array}
			}
		}
&
		\stT[\roleFmt{Eloise}]&=&\stIntSum{\roleFmt{Dave}}{}{
			\begin{array}{@{\hskip 0mm}l@{\hskip 0mm}}
				\stChoice{talkLess}{}\\
				\stChoice{talkMore}{}
			\end{array}
		}
	\end{array}
\]
Define the following relation:
\begin{equation*}
	\stBinMergeStar = 
	\left\{\begin{array}{ccc}
		
		(\left\{\stT[\roleFmt{Carol}],\stTii[\roleFmt{Carol}]\right\},\stT[\roleFmt{Carol}]),
&
		(\left\{\stT[\roleFmt{Carol}],\stTi[\roleFmt{Carol}]\right\},\stT[\roleFmt{Carol}]),
&
		(\left\{\stT[\roleFmt{Carol}],\stT[\roleFmt{Carol}]\right\},\stT[\roleFmt{Carol}])
	\end{array}\right\}
\end{equation*}
Observe that $\stBinMergeStar$ satisfies the rules of the full coinductive merge~\Cref{def:full-merge},
so $\stBinMergeStar\subseteq\relMerge{}{}$.
So, the relations
\begin{equation*}
\begin{array}{c}
\{(\gtG,\stT[\roleFmt{Dave}]),(\gtGi,\stTi[\roleFmt{Dave}]),(\gtGii,\stTii[\roleFmt{Dave}]),(\gtGiii,\stTiii[\roleFmt{Dave}])\}
\\[2ex]
\{(\gtG,\stT[\roleFmt{Carol}]),(\gtGi,\stTii[\roleFmt{Carol}]),(\gtGii,\stT[\roleFmt{Carol}]),(\gtGiii,\stTi[\roleFmt{Carol}])\}
\\[2ex]
\{(\gtG,\stT[\roleFmt{Eloise}]),(\gtGi,\stEnd),(\gtGii,\stEnd),(\gtGiii,\stEnd)\}
\end{array}
\end{equation*}
satisfy the coinductive rules defining $\gtProj{\roleFmt{Dave}}{}{}$, $\gtProj{\roleFmt{Carol}}{}{}$, and $\gtProj{\roleFmt{Eloise}}{}{}$ respectively,
so $\gtProj{\roleFmt{Dave}}{\gtG}{\stT[\roleFmt{Dave}]}$, $\gtProj{\roleFmt{Carol}}{\gtG}{\stT[\roleFmt{Carol}]}$, and $\gtProj{\roleFmt{Eloise}}{\gtG}{\stT[\roleFmt{Eloise}]}$.
\end{example}

\begin{example}[Coinductively Projecting Odd-Even]
\label{ex:app:proj-odd-even}
We claim that $\gtProj{}{\gtG[\text{\tiny o/e}]}{\stEnv[\text{\tiny o/e}]}$.
\begin{equation*}
	\cinference[$\projRuleOut$]
	{
		\cinference[$\projRuleEnd$]
		{}
		{\gtProj{\roleP}{\gtG[\text{\tiny o}]}{\stEnd}}
		&
		\cinference[$\projRuleEnd$]
		{}
		{\gtProj{\roleP}{\gtG[\text{\tiny e}]}{\stEnd}}
	}
	{
		\gtProj{\roleP}{\gtG[\text{\tiny o/e}]}{\stIntSum{\roleQ}{}{
		\stLabFmt{o},
		\stLabFmt{m}
	}}
	}
\end{equation*}
\begin{equation*}
	\cinference[$\projRuleIn$]
	{
		\cinference[$\projRuleOut$]
		{
			\cinference[$\projRuleOut$]
			{\gtProj{\roleQ}{\gtG[\text{\tiny e}]}{\stT[\text{e},\roleQ]}}
			{\gtProj{\roleQ}{\gtGi[\text{\tiny e}]}{\stTi[\text{e},\roleQ]}}
			&
			\cinference[$\projRuleOut$]
			{
				\cinference[$\projRuleEnd$]
					{}
					{\gtProj{\roleQ}{\gtEnd}{\stEnd}}
			}
			{\gtProj{\roleQ}{\gtG[\text{\tiny e,end}]}{\stT[\text{e,end},\roleQ]}}
		}
		{\gtProj{\roleQ}{\gtG[\text{\tiny e}]}{\stT[\text{e},\roleQ]}}
		&
		\cinference[$\projRuleOut$]
		{
			\cinference[$\projRuleOut$]
			{
				\gtProj{\roleQ}{\gtG[\text{\tiny o}]}{\stT[\text{o},\roleQ]}
				&
				\cinference[$\projRuleIn$]
				{
					\cinference[$\projRuleEnd$]
					{}
					{\gtProj{\roleQ}{\gtEnd}{\stEnd}}
				}
				{\gtProj{\roleQ}{\gtG[\text{\tiny o,end}]}{\stTi[\text{o},\roleQ]}}
			}
			{\gtProj{\roleQ}{\gtGi[\text{\tiny o}]}{\stTi[\text{o},\roleQ]}}
		}
		{\gtProj{\roleQ}{\gtG[\text{\tiny o}]}{\stT[\text{o},\roleQ]}}
	}
	{\gtProj{\roleQ}{\gtG[\text{\tiny o/e}]}{\stT[\roleQ]}}
\end{equation*}
\begin{equation*}
\begin{array}{c}
	\cinference[$\projRuleIn$]
		{
			\cinference[$\projRuleIn$]
			{
				\gtProj{\roleR}{\gtG[\text{\tiny e}]}{\stRec{\stRecVar}{
					\stExtSum{\roleQ}{}{
						\stLab\stSeq\roleQ{\&}\stLab\stSeq\stRecVar,\,
						\stLabFmt{b}\stSeq\roleQ{\oplus}\stLabFmt{m}
					}
				}}
			}
			{
				\gtProj{\roleR}{\gtGi[\text{\tiny e}]}{\roleQ\stFmt{\&}\stLab\stSeq\stRec{\stRecVar}{
					\stExtSum{\roleQ}{}{
						\stLab\stSeq\roleQ{\&}\stLab\stSeq\stRecVar,\,
						\stLabFmt{b}\stSeq\roleQ{\oplus}\stLabFmt{m}
					}
					}
				}
			}
			&
			\cinference[$\projRuleOut$]
			{
				\cinference[$\projRuleEnd$]
					{}
					{\gtProj{\roleR}{\gtEnd}{\stEnd}}
			}
			{\gtProj{\roleR}{\gtG[\text{\tiny e,end}]}{\roleQ\stFmt{\oplus}\stLabFmt{m}}}
		}
		{
			\gtProj{\roleR}{\gtG[\text{\tiny e}]}{\stRec{\stRecVar}{
				\stExtSum{\roleQ}{}{
					\stLab\stSeq\roleQ{\&}\stLab\stSeq\stRecVar,\,
					\stLabFmt{b}\stSeq\roleQ{\oplus}\stLabFmt{m}
				}
			}}
		}
		\\[2ex]
		\cinference[$\projRuleIn$]
		{
			\cinference[$\projRuleIn$]
			{
				\gtProj{\roleR}{\gtG[\text{\tiny o}]}{
			\stRec{\stRecVar}{
				\roleQ{\&}\stLab\stSeq
					\stExtSum{\roleQ}{}{
						\stLab\stSeq\stRecVar,\,
						\stLabFmt{b}\stSeq\roleQ{\oplus}\stLabFmt{o}
				}
		}}
				&
				\cinference[$\projRuleOut$]
				{
					\cinference[$\projRuleEnd$]
					{}
					{\gtProj{\roleR}{\gtEnd}{\stEnd}}
				}
				{\gtProj{\roleR}{\gtG[\text{\tiny o,end}]}{
					\roleQ\stFmt{\oplus}\stLabFmt{o}
				}}
			}
			{
				\gtProj{\roleR}{\gtGi[\text{\tiny o}]}{
							\stExtSum{\roleQ}{}{
								\stLab\stSeq\stRec{\stRecVar}{
						\roleQ{\&}\stLab\stSeq
							\stLab\stSeq\stExtSum{\roleQ}{}{
								\stLab\stSeq\stRecVar,\,
								\stLabFmt{b}\stSeq\roleQ{\oplus}\stLabFmt{o}
						}
				},\,
								\stLabFmt{b}\stSeq\roleQ{\oplus}\stLabFmt{o}
						}
				}
			}
		}
		{\gtProj{\roleR}{\gtG[\text{\tiny o}]}{
			\stRec{\stRecVar}{
				\roleQ{\&}\stLab\stSeq
					\stExtSum{\roleQ}{}{
						\stLab\stSeq\stRecVar,\,
						\stLabFmt{b}\stSeq\roleQ{\oplus}\stLabFmt{o}
				}
		}}}
		\\[2ex]
	\cinference[$\projRuleCtx$]
	{
		\gtProj{\roleR}{\gtG[\text{\tiny e}]}{\stRec{\stRecVar}{
				\stExtSum{\roleQ}{}{
					\stLab\stSeq\roleQ{\&}\stLab\stSeq\stRecVar,\,
					\stLabFmt{b}\stSeq\roleQ{\oplus}\stLabFmt{m}
				}
			}}
		&
		\gtProj{\roleR}{\gtG[\text{\tiny o}]}{
			\stRec{\stRecVar}{
				\roleQ{\&}\stLab\stSeq
					\stExtSum{\roleQ}{}{
						\stLab\stSeq\stRecVar,\,
						\stLabFmt{b}\stSeq\roleQ{\oplus}\stLabFmt{o}
				}
		}}
	}
	{\gtProj{\roleR}{\gtG[\text{\tiny o/e}]}{\stT[\roleR]}}
	\end{array}
\end{equation*}
\begin{equation*}
\begin{array}{c}
	\cinference[$\mergeRuleIn$]{
			\cinference[$\mergeRuleOut$]
			{\cinference[$\mergeRuleEnd$]{}{\relMerge{\{\stEnd\}}{\stEnd}}}
			{\relMerge{\{\roleQ\stFmt{\oplus}\stLabFmt{o}\}}{\roleQ\stFmt{\oplus}\stLabFmt{o}}}
			&
			{\relMerge{\{
	\stRec{\stRecVar}{
				\roleQ{\&}\stLab\stSeq
					\stExtSum{\roleQ}{}{
						\stLab\stSeq\stRecVar,
						\stLabFmt{b}\stSeq\roleQ{\oplus}\stLabFmt{o}
				}
		},\,
		\stRec{\stRecVar}{
				\stExtSum{\roleQ}{}{
					\stLab\stSeq\roleQ{\&}\stLab\stSeq\stRecVar,
					\stLabFmt{b}\stSeq\roleQ{\oplus}\stLabFmt{m}
				}
			}
	\}}{\stT[\roleR]}}
		}{
			\relMerge{\{
			\stExtSum{\roleQ}{}{
								\stLab\stSeq\stRec{\stRecVar}{
						\roleQ{\&}\stLab\stSeq
							\stLab\stSeq\stExtSum{\roleQ}{}{
								\stLab\stSeq\stRecVar,
								\stLabFmt{b}\stSeq\roleQ{\oplus}\stLabFmt{o}
						}
				},\,
								\stLabFmt{b}\stSeq\roleQ{\oplus}\stLabFmt{o}
						},\,
				\roleQ\stFmt{\&}\stLab\stSeq\stRec{\stRecVar}{
					\stExtSum{\roleQ}{}{
						\stLab\stSeq\roleQ{\&}\stLab\stSeq\stRecVar,
						\stLabFmt{b}\stSeq\roleQ{\oplus}\stLabFmt{m}
					}
					}
			\}}{\stT[\text{o},\roleR]}
		}
	\\[2ex]
	\mathcal{T}=
	\{
			\stExtSum{\roleQ}{}{
								\stLab\stSeq\stRec{\stRecVar}{
						\roleQ{\&}\stLab\stSeq
							\stExtSum{\roleQ}{}{
								\stLab\stSeq\stRecVar,
								\stLabFmt{b}\stSeq\roleQ{\oplus}\stLabFmt{o}
						}
				},\,
								\stLabFmt{b}\stSeq\roleQ{\oplus}\stLabFmt{o}
						},\,
				\roleQ\stFmt{\&}\stLab\stSeq\stRec{\stRecVar}{
					\stExtSum{\roleQ}{}{
						\stLab\stSeq\roleQ{\&}\stLab\stSeq\stRecVar,
						\stLabFmt{b}\stSeq\roleQ{\oplus}\stLabFmt{m}
					}
					}
			\}
	\\[2ex]
	\cinference[$\mergeRuleIn$]
	{
		\cinference[$\mergeRuleOut$]
		{\cinference[$\mergeRuleEnd$]{}{\relMerge{\{\stEnd\}}{\stEnd}}}
		{\relMerge{\{\roleQ\stFmt{\oplus}\stLabFmt{m}\}}{\roleQ\stFmt{\oplus}\stLabFmt{m}}}
		&
		\relMerge{\mathcal{T}}{\stT[\text{o},\roleR]}
	}
	{\relMerge{\{
	\stRec{\stRecVar}{
				\roleQ{\&}\stLab\stSeq
					\stExtSum{\roleQ}{}{
						\stLab\stSeq\stRecVar,\,
						\stLabFmt{b}\stSeq\roleQ{\oplus}\stLabFmt{o}
				}
		},
		\stRec{\stRecVar}{
				\stExtSum{\roleQ}{}{
					\stLab\stSeq\roleQ{\&}\stLab\stSeq\stRecVar,\,
					\stLabFmt{b}\stSeq\roleQ{\oplus}\stLabFmt{m}
				}
			}
	\}}{\stT[\roleR]}}
\end{array}
\end{equation*}
\end{example}

The full coinductive projection is invariant up-to type semantics:

\begin{proposition}[Projection is Invariant under Local Bisimilarity]
\label{prop:proj-inv-local}
Let $\gtProj{}{}{}$ be the coinductive projection with the full (or plain) coinductive merge.
	If $\stT\bisim\stTi$, then
	$\gtProj{\roleR}{\gtG}{\stT}$ implies
	$\gtProj{\roleR}{\gtG}{\stTi}$.
\end{proposition}
\begin{proof}
	Define $\gtProji{}{}{}$ by
	$\gtProji{\roleR}{\gtG}{\stT}$
	if there exists $\stTi\bisim\stT$ s.t.
	$\gtProji{\roleR}{\gtG}{\stTi}$.
	If $\stT\bisim\stTi$ then:
	\begin{itemize}
		\item If $\unfoldOne{\stT}=\stEnd$, then $\unfoldOne{\stTi}=\stEnd$.
		\item If $\unfoldOne{\stT}=\stExtSum{\roleP}{i\in I}{\stChoice{\stLab[i]}{\tyGround}\stSeq\stT[i]}$, then $\unfoldOne{\stTi}=\stExtSum{\roleP}{i\in I}{\stChoice{\stLab[i]}{\tyGround}\stSeq\stTi[i]}$ for some $\stTi[i]\bisim\stT[i]$ for $i\in I$.
		\item If $\unfoldOne{\stT}=\stIntSum{\roleP}{i\in I}{\stChoice{\stLab[i]}{\tyGround}\stSeq\stT[i]}$, then $\unfoldOne{\stTi}=\stIntSum{\roleP}{i\in I}{\stChoice{\stLab[i]}{\tyGround}\stSeq\stTi[i]}$ for some $\stTi[i]\bisim\stT[i]$ for $i\in I$.
	\end{itemize}
	Therefore, $\gtProji{}{}{}$ satisfies the rules defining $\gtProj{}{}{}$,
	so $\gtProji{}{}{}=\gtProj{}{}{}$.
\end{proof}

For~\cref{thm:proj-bisim} and~\cref{cor:proj-bisim}
we assume~\cref{thm:co-proj-good}
and~\cref{thm:co-merge}.

\begin{restatable}[Projection is Invariant under Global Bisimilarity]{theorem}{thmProjBisim}
	\label{thm:proj-bisim}
	Let $\gtProj{}{}{}$ be the coinductive projection with the full coinductive merge.
	Suppose that $\gtG$ and $\gtGi$ are global types
	without parallel composition and
	$\gtG\bisim\gtGi$.
	$\gtProj{\roleR}{\gtG}{\stT}$ implies
	$\gtProj{\roleR}{\gtGi}{\stT}$.
\end{restatable}

\begin{lemma}
	\label{lem:technical-proj-bisim}
	Let $\gtG$ be a type without parallel composition.
	Suppose that
	$\gtG\bisim\gtComm{\roleP}{\roleQ}{i \in I}{\gtLab[i]}{\tyGround[i]}{\gtGi[i]}$,
	$\gtProj{\roleR}{\gtG}{\stT}$, and
	$\roleR\not\in\{\roleP,\roleQ\}$.
	Let $\gtG\,\gtMove[\ltsSendRecv{\roleP}{\roleQ}{\stChoice{\stLab[i]}{\tyGround[i]}}]\,\gtG[i]$ for $i\in I$.
	Then, there are $\stT[i]$ for $i\in I$ such that
	$\gtProj{\roleR}{\gtG[i]}{\stT[i]}$ and
	$\stMerge{}{\{\stT[i]\suchthat i\in I\}}\ni \stT$.
\end{lemma}

\begin{proof}
	We induct on the length of the reductions
	$\gtG\,\gtMove[\ltsSendRecv{\roleP}{\roleQ}{\stChoice{\stLab[i]}{\tyGround[i]}}]\,\gtG[i]$.
	\begin{itemize}
		\item $\gtG\bisim\gtComm{\roleP}{\roleQ}{i \in I}{\gtLab[i]}{\tyGround[i]}{\gtGi[i]}$ so
		$\gtG\neq\gtEnd$.

		\item If $\gtG=\gtRec{\gtRecVar}{\gtGii}$, then
		$\gtGii{}[\gtG/\gtRecVar]\,\gtMove[\ltsSendRecv{\roleP}{\roleQ}{\stChoice{\stLab[i]}{\tyGround[i]}}]\,\gtG[i]$ for $i\in I$,
		$\gtGii{}[\gtG/\gtRecVar]\bisim
		\gtG\bisim
		\gtComm{\roleP}{\roleQ}{i \in I}{\gtLab[i]}{\tyGround[i]}{\gtGi[i]}$, and
		$\gtProj{\roleR}{\gtGii{}[\gtG/\gtRecVar]}{\stT}$.
		By the I.H.,
		there are $\stT[i]$ for $i\in I$ such that
		$\gtProj{\roleR}{\gtG[i]}{\stT[i]}$ and
		$\stMerge{}{\{\stT[i]\suchthat i\in I\}}\ni \stT$.

		\item Suppose that $\gtG=\gtComm{\roleP}{\roleQ}{j \in J}{\gtLabi[j]}{\tyGround[j]}{\gtGii[j]}$.
		As $\gtG\bisim \gtComm{\roleP}{\roleQ}{i \in I}{\gtLab[i]}{\tyGround[i]}{\gtGi[i]}$, we can write
		$\gtG=\gtComm{\roleP}{\roleQ}{i \in I}{\gtLab[i]}{\tyGround[i]}{\gtG[i]}$.
		$\gtProj{\roleR}{\gtG}{\stT}$ so
		there are $\stT[i]$ for $i\in I$ such that $\gtProj{\roleR}{\gtG[i]}{\stT[i]}$ and
		$\stMerge{}{\{\stT[i]\suchthat i\in I\}}\ni \stT$.

		\item Suppose that $\gtG=\gtComm{\rolePi}{\roleQi}{j \in J}{\gtLabi[j]}{\tyGroundi[j]}{\gtGii[j]}$
		where $(\roleP,\roleQ)\neq(\rolePi,\roleQi)$.
		$\gtG\bisim\gtComm{\roleP}{\roleQ}{i \in I}{\gtLab[i]}{\tyGround[i]}{\gtGi[i]}$, so
		$\{\roleP,\roleQ\}\cap\{\rolePi,\roleQi\}=\emptyset$.
		Now, for $i\in I$ find $\gtGii[i,j]$ such that
		$\gtGii[j]\,\gtMove[\ltsSendRecv{\roleP}{\roleQ}{\stChoice{\stLab[i]}{\tyGround[i]}}]\,\gtGii[i,j]$ and
		$\gtG[i]=\gtComm{\rolePi}{\roleQi}{j \in J}{\gtLabi[j]}{\tyGroundi[j]}{\gtGii[i,j]}$.
		\newline
		If $\roleR=\rolePi$, then
		$\unfoldOne{\stT}=\stIntSum{\roleQi}{j \in J}{\stChoice{\stLabi[j]}{\tyGroundi[j]} \stSeq \stT[j]}$, where
		$\gtProj{\roleR}{\gtGii[j]}{\stT[j]}$.
		By the I.H., find $\stT[i,j]$ for $i\in I, j \in J$ such that $\gtProj{\roleR}{\gtGii[i,j]}{\stT[i,j]}$ and
		$\stMerge{}{\{\stT[i,j]\suchthat i\in I\}}\ni \stT[j]$.
		So $\{\stT[i,j]\}_{i\in I, j\in J}$ are compatible.
		Now, find $\stTi[i]\in \stMerge{}{\{\stT[i,j]\suchthat j\in J\}}$ for $i\in I$, so that
		$\gtProj{\roleR}{\gtG[i]}{\stTi[i]}$ and $\stMerge{}{\{\stTi[i]\suchthat i\in I\}}\ni \stT$.
	\end{itemize}
\end{proof}

\thmProjBisim*

\begin{proof}
	Define $\gtProji{\roleR}{\gtG}{\stT}$ iff there is $\gtGi\bisim\gtG$
	such that $\gtProj{\roleR}{\gtGi}{\stT}$
	and neither $\gtG$ nor $\gtGi$ contain parallel composition.
	We show that $\gtProji{\roleR}{}{}$ satisfies the rules defining $\gtProj{\roleR}{}{}$,
	hence $\gtProji{\roleR}{}{}\subseteq\gtProj{\roleR}{}{}$.
	Suppose that $\gtGi\bisim\gtG$, $\gtGi$ has no parallel composition,
	and $\gtProj{\roleR}{\gtGi}{\stT}$.
	\begin{itemize}[leftmargin=0.5in,labelindent=-\leftmargin]
	    \item[\inferrule{$\projRuleEnd$}]
	        If $\roleR\notin\gtRoles{\gtG}$, then $\roleR\notin\gtRoles{\gtGi}$, so $\unfoldOne{\stT}=\stEnd$.
	    \item[\inferrule{$\projRuleIn$}]
	        Suppose that $\unfoldOne{\gtG}=\gtComm{\roleP}{\roleR}{i \in I}{\gtLab[i]}{\tyGround[i]}{\gtG[i]}$.
		Let $\gtGi\,\gtMove[\ltsSendRecv{\roleP}{\roleR}{\stChoice{\stLab[i]}{\tyGround[i]}}]\,\gtGi[i]$ for $i\in I$.
		For $i\in I$, $\gtGi[i]\bisim\gtG[i]$ and there is $\stT[i]$ such that
		$\stT\,\gtMove[\roleP\stEnvAnnotInSym\stChoice{\stLab[i]}{\tyGround[i]}]\,\stT[i]$ and
		$\gtProj{\roleR}{\gtGi[i]}{\stT[i]}$.

	    \item[\inferrule{$\projRuleOut$}]
		Suppose that $\unfoldOne{\gtG}=\gtComm{\roleR}{\roleQ}{i \in I}{\gtLab[i]}{\tyGround[i]}{\gtG[i]}$.
		Let $\gtGi\,\gtMove[\ltsSendRecv{\roleR}{\roleQ}{\stChoice{\stLab[i]}{\tyGround[i]}}]\,\gtGi[i]$ for $i\in I$.
		For $i\in I$, $\gtGi[i]\bisim \gtG[i]$ and there is $\stT[i]$ such that
		$\stT\,\gtMove[\roleQ\stEnvAnnotOutSym\stChoice{\stLab[i]}{\tyGround[i]}]\,\stT[i]$ and
		$\gtProj{\roleR}{\gtGi[i]}{\stT[i]}$.

	    \item[\inferrule{$\projRuleCtx$}]
	        Suppose that
		$\unfoldOne{\gtG}=\gtComm{\roleP}{\roleQ}{i \in I}{\gtLab[i]}{\tyGround[i]}{\gtG[i]}$,
		$\roleR\notin \{\roleP,\roleQ\}$, and
		$\roleR\in \gtRoles{\gtG}$.\\
		Let $\gtGi\,\gtMove[\ltsSendRecv{\roleP}{\roleQ}{\stChoice{\stLab[i]}{\tyGround[i]}}]\,\gtGi[i]$ for $i\in I$, so
		$\gtGi[i]\bisim\gtG[i]$ for $i\in I$.
		Then by~\cref{lem:technical-proj-bisim},
		there are $\stT[i]$ for $i\in I$ such that
		$\gtProj{\roleR}{\gtGi[i]}{\stT[i]}$ and
		$\stMerge{}{\{\stT[i]\suchthat i\in I\}}\ni \stT$.
	\end{itemize}
\end{proof}

\begin{corollary}
\label{cor:proj-bisim}
	Suppose that $\gtG$ is weak balanced and neither $\gtG$ nor $\gtGi$ use parallel composition.
	Consider the full coinductive projection.
	If $\gtProj{}{\gtG}{\stEnvNew}$, then
	$\gtG\bisim\gtGi$ iff $\gtProj{}{\gtGi}{\stEnvNew}$
\end{corollary}

\thmMergeGood*

These are standard properties of the full inductive merge and
the plain merges are trivially good.

\begin{theorem}[Full Coinductive Merge]
\label{thm:co-merge}
	The full coinductive merge is good.
\end{theorem}

\inferrule{\mergeGoodSubst} trivially holds, since the full coinductive merge
only involves closed types.
\inferrule{\mergeGoodCtx} is immediate from
\inferrule{$\mergeRuleOut$} and \inferrule{$\mergeRuleIn$}.

\begin{lemma}[Associativity of Merge]
	\label{lem:good-merge-assoc}
	Full coinductive merge is associative \ie
	\\
	\centerline{\(
	\left(
	\forall j\in J.\;\stMerge{}{\{\stT[i]\}_{i\in I_j}}\ni \stTi[j]
	\right)
	\implies
	\left(
	\stMerge{}{\{\stT[i]\suchthat j\in J, i\in I_j\}}=\stMerge{}{\{\stTi[j]\}_{j\in J}}
	\right)
	\)}
\end{lemma}
\begin{proof}
Consider
\begin{equation*}
	\Re \ =
		\stMerge{}{}
		\cup
		\left\{
			\begin{array}{ccc}
				(\mathcal{T},\stT)
				&\suchthat&
				\begin{array}{c}
					\bigcup_{i\in I} \mathcal{T}_i=\mathcal{T}\\
					\forall i\in I, \stMerge{}{\mathcal{T}_i}\ni\stTi[i]\\
					\stMerge{}{\{\stTi[i]\suchthat i\in I\}}\ni\stT
				\end{array}
			\end{array}
		\right\}
	\cup
		\left\{
			\begin{array}{ccc}
				(\{\stTi[i]\suchthat i\in I\},\stT)
				&\suchthat&
				\begin{array}{c}
					\bigcup_{i\in I}\mathcal{T}_i=\mathcal{T}\\
					\forall i\in I, \stMerge{}{\mathcal{T}_i}\ni\stTi[i]\\
					\stMerge{}{\mathcal{T}}\ni\stT
				\end{array}
			\end{array}
		\right\}
\end{equation*}
This satisfies the coinductive rules defining full merge, and so $\Re\,
=\,\stMerge{}{}$.
Suppose that $\{\stT[i]\}_{i\in I} \Re\, \stT$.
If $\stMerge{}{\{\stT[i]\}_{i\in I}} \ni \stT$, then:
\begin{itemize}[leftmargin=0.5in,labelindent=-\leftmargin]
    \item[\inferrule{$\mergeRuleEnd$}]
        If $\unfoldOne{\stT}=\stEnd$, then
        $\unfoldOne{\stT[i]}=\stEnd$ for all $i\in I$
    \item[\inferrule{$\mergeRuleOut$}]
        If $\unfoldOne{\stT}=\stIntSum{\roleQ}{j \in J}{\stChoice{\stLab[j]}{\tyGround[j]} \stSeq \stTi[j]}$, then 
        $\unfoldOne{\stT[i]}=\stIntSum{\roleQ}{j \in J}{\stChoice{\stLab[j]}{\tyGround[j]} \stSeq \stT[i,j]}$ for all $i\in I$
        where for all $j\in J$ $\stMerge{}{\left\{\stT[i,j]\suchthat i\in I\right\}}\ni\stTi[j]$ so
	$\left\{\stT[i,j]\suchthat i\in I\right\}\Re\,\stTi[j]$
    \item[\inferrule{$\mergeRuleIn$}]
        If $\unfoldOne{\stT} = \stExtSum{\roleP}{j \in J}{\stChoice{\stLab[j]}{\tyGround[j]}\stSeq \stTi[j]}$, then
        $\unfoldOne{\stT[i]}=\stExtSum{\roleP}{j \in J_{i}}{\stChoice{\stLab[j]}{\tyGround[j]} \stSeq \stT[i,j]}$ for all $i\in I$
        where $J=\bigcup_{i\in I}J_{i}$ and for all $j\in J$
        $\stMerge{}{\left\{\stT[i,j]\suchthat j\in J_{i} \right\}}\ni \stTi[j]$ so
	$\left\{\stT[i,j]\suchthat j\in J_{i} \right\}\Re\,\stTi[j]$
\end{itemize}
If $\{I_j\}_{j\in J}$ is a cover of $I$ of non-empty sets and for $j\in J$
there is $\stTi[j]$ with $\stMerge{}{\{\stT[i]\}_{i\in I_j}}\ni\stTi[j]$
and $\stMerge{}{\{\stTi[j]\}_{j\in J}}\ni\stT$, then:
\begin{itemize}[leftmargin=0.5in,labelindent=-\leftmargin]
    \item[\inferrule{$\mergeRuleEnd$}]
        If $\unfoldOne{\stT}=\stEnd$, then
        $\unfoldOne{\stTi[j]}=\stEnd$ for all $j\in J$, then
	$\unfoldOne{\stT[i]}=\stEnd$ for all $i\in I$.
    \item[\inferrule{$\mergeRuleOut$}]
        If $\unfoldOne{\stT}=\stIntSum{\roleQ}{k \in K}{\stChoice{\stLab[k]}{\tyGround[k]} \stSeq \stTii[k]}$, then 
        $\unfoldOne{\stTi[j]}=\stIntSum{\roleQ}{k \in K}{\stChoice{\stLab[k]}{\tyGround[k]} \stSeq \stTii[{j,k}]}$ for all $j\in J$
        where for all $k\in K$ $\stMerge{}{\left\{\stTii[{j,k}]\suchthat j\in J\right\}}\ni\stTii[k]$.
	So for all $j\in J$,
	$\unfoldOne{\stT[i]}=\stIntSum{\roleQ}{k \in K}{\stChoice{\stLab[k]}{\tyGround[k]} \stSeq \stTii[{i,k}]}$ for all $i\in I_j$
	where for all $k\in K$ $\stMerge{}{\left\{\stTii[i,k]\suchthat i\in I_j\right\}}\ni\stTii[j,k]$.
	Therefore, $\left\{\stTii[i,k]\right\}_{i\in I} \Re\, \stTii[k]$.
    \item[\inferrule{$\mergeRuleIn$}]
        If $\unfoldOne{\stT} = \stExtSum{\roleP}{k \in K}{\stChoice{\stLab[k]}{\tyGround[k]}\stSeq \stTii[k]}$, then
        $\unfoldOne{\stTi[j]}=\stExtSum{\roleP}{k \in K_{j}}{\stChoice{\stLab[k]}{\tyGround[k]} \stSeq \stTii[j,k]}$ for all $j\in J$
        where $K=\bigcup_{j\in J}K_{j}$ and for all $k\in K$
        $\stMerge{}{\left\{\stTii[j,k]\suchthat k\in K_{j} \right\}}\ni \stTii[k]$.
	So for all $j\in J$,
	$\unfoldOne{\stT[i]}=\stExtSum{\roleP}{k \in K_{i}}{\stChoice{\stLab[k]}{\tyGround[k]} \stSeq \stTii[i,k]}$ for all $i\in I_j$
	where $K_j=\bigcup_{i\in I_j}K_i$ and for all $k\in K_j$,\\
	$\stMerge{}{\left\{\stTii[i,k]\suchthat k\in K_{i}, i\in I_j \right\}}\ni \stTii[j,k]$.
	Therefore, $\left\{\stTii[i,k]\suchthat k\in K_i \right\} \Re\, \stTii[k]$.
\end{itemize}
If $\stMerge{}{\{\stTi[j]\}_{j\in J_i}}\ni \stT[i]$ for $i\in I$
and $\stMerge{}{\{\stTi[j]\}_{j\in J_i, i\in I}}\ni\stT$, then:
\begin{itemize}[leftmargin=0.5in,labelindent=-\leftmargin]
    \item[\inferrule{$\mergeRuleEnd$}]
        If $\unfoldOne{\stT}=\stEnd$, then
        $\unfoldOne{\stTi[j]}=\stEnd$ for all $j\in J_i$ and $i\in I$.
	So, $\unfoldOne{\stT[i]}=\stEnd$ for all $i\in I$.
    \item[\inferrule{$\mergeRuleOut$}]
        If $\unfoldOne{\stT}=\stIntSum{\roleQ}{k \in K}{\stChoice{\stLab[k]}{\tyGround[k]} \stSeq \stTii[k]}$, then 
        $\unfoldOne{\stTi[j]}=\stIntSum{\roleQ}{k \in K}{\stChoice{\stLab[k]}{\tyGround[k]} \stSeq \stTii[j,k]}$ for all $j\in J_i$ and $i\in I$
        where for all $k\in K$
        \\$\stMerge{}{\left\{\stTii[j,k]\suchthat j\in J_i, i\in I\right\}}\ni\stTii[k]$.
	So for $i\in I$, $\unfoldOne{\stT[i]}=\stIntSum{\roleQ}{k \in K}{\stChoice{\stLab[k]}{\tyGround[k]} \stSeq \stTii[i,k]}$
	where for all $k\in K$ $\stMerge{}{\left\{\stTii[j,k]\suchthat j\in J_i\right\}}\ni\stTii[i,k]$.
	Therefore, $\left\{\stTii[j,k]\suchthat j\in J_i, i\in I\right\}\Re\,\stTii[k]$.
    \item[\inferrule{$\mergeRuleIn$}]
        If $\unfoldOne{\stT} = \stExtSum{\roleP}{k \in K}{\stChoice{\stLab[k]}{\tyGround[k]}\stSeq \stTii[k]}$, then
        $\unfoldOne{\stTi[j]}=\stExtSum{\roleP}{k \in K_{j}}{\stChoice{\stLab[k]}{\tyGround[k]} \stSeq \stTii[j,k]}$ for all $j\in J_i$ and $i\in I$
        where $K=\bigcup_{j\in J}K_{j}$ and for all $k\in K$
        $\stMerge{}{\left\{\stTii[j,k]\suchthat k\in K_{j} \right\}}\ni \stTii[k]$.
	So for $i\in I$, $\unfoldOne{\stT[i]} = \stExtSum{\roleP}{k \in K_i}{\stChoice{\stLab[k]}{\tyGround[k]}\stSeq \stTii[i,k]}$
	where $K_i = \bigcup_{j\in J_i}K_{j}$ and for all $k\in K_j$
	$\stMerge{}{\left\{\stTii[j,k]\suchthat k\in K_{j}, j\in J_i \right\}}\ni \stTii[i,k]$.
	Therefore, $\left\{\stTii[i,k]\suchthat k\in K_i\right\}\Re\,\stTii[k]$.
\end{itemize}
\end{proof}

\begin{lemma}[Merge has an Operational Correspondence]
	\label{lem:good-merge-corr}
	\begin{itemize}
		\item If $\stT[i]\,\gtMove[\stEnvAnnotGenericSym]\,\stTi[i]$
        for all $i\in I$ and $\stMerge{}{\{\stT[i]\}_{i\in I}}\ni \stT$, then
        there exists $\stTi$ such that
        $\stT\,\gtMove[\stEnvAnnotGenericSym]\,\stTi$ and
        $\stMerge{}{\{\stTi[i]\suchthat i\in I\}}\ni \stTi$.
		\item If $\stMerge{}{\{\stT[i]\suchthat i\in I\}}\ni \stT$ and
        $\stT\,\gtMove[\stEnvAnnotGenericSym]\,\stTi$, then
        there is some non-empty $J\subseteq I$ such that
        $\stT[i]\,\gtMove[\stEnvAnnotGenericSym]\,\stTi[i]$ for $i\in J$ and
        $\stMerge{}{\{\stTi[i]\suchthat i\in J\}}\ni\stTi$.
	\end{itemize}
\end{lemma}
\begin{proof}
	Suppose that $\stMerge{}{\{\stT[i]\suchthat i\in I\}}\ni\stT$.
	\begin{itemize}
		\item Suppose that $\stT[i]\,\gtMove[\roleP\stEnvAnnotOutSym\stChoice{\stLab}{\tyGround}]\,\stTi[i]$ for all $i\in I$.
So, for $i\in I$, $\unfoldOne{\stT[i]}=\stIntSum{\roleP}{j \in J_i}{\stChoice{\stLab[i,j]}{\tyGround[i,j]} \stSeq \stTi[i,j]}$ and there is $j\in J$ with $\stLab[i,j]=\stLab$, $\tyGround[i,j]=\tyGround$, and $\stTi[i,j]=\stTi[i]$.
$\stMerge{}{\{\stT[i]\suchthat i\in I\}}\ni \stT$, so we may assume that $J_i=J$, $\stLab[i,j]=\stLab[j]$, and $\tyGround[i,j]=\tyGround[j]$ for all $j\in J$.
Say $k\in J$ has $\stLab[k]=\stLab$, $\tyGround[k]=\tyGround$, and $\stTi[i,k]=\stTi[i]$.
We also have that $\unfoldOne{\stT}=\stIntSum{\roleP}{j \in J}{\stChoice{\stLab[j]}{\tyGround[j]} \stSeq \stTii[j]}$, where $\stMerge{}{\{\stTi[i,j]\suchthat i\in I\}}\ni\stTii[j]$ for $j \in J$.
So, $\stT\,\gtMove[\roleP\stEnvAnnotOutSym\stChoice{\stLab}{\tyGround}]\,\stTii[k]$ and $\stMerge{}{\{\stTi[i]\suchthat i\in I\}}\ni \stTii[k]$. 
The proof in the case that $\stT[i]\,\gtMove[\roleP\stEnvAnnotInSym\stChoice{\stLab}{\tyGround}]\,\stTi[i]$ for all $i\in I$ is similar.
		\item Suppose that $\stT\,\gtMove[\roleP\stEnvAnnotOutSym\stChoice{\stLab}{\tyGround}]\,\stTi$.
So, $\unfoldOne{\stT}=\stIntSum{\roleP}{j \in J}{\stChoice{\stLab[j]}{\tyGround[j]} \stSeq \stTi[j]}$ and there is $k\in J$ such that
$\stLab[k]=\stLab$, $\tyGround[k]=\tyGround$, and $\stTi[k]=\stTi$.
$\stMerge{}{\{\stT[i]\suchthat i\in I\}}\ni\stT$ so for $i\in I$, $\unfoldOne{\stT[i]}=\stIntSum{\roleP}{j \in J}{\stChoice{\stLab[j]}{\tyGround[j]} \stSeq \stTi[i,j]}$,
where $\stMerge{}{\{\stTi[i,j]\suchthat i\in I\}}\ni\stTi[j]$ for $j\in J$.
So, $\stT[i]\,\gtMove[\roleP\stEnvAnnotOutSym\stChoice{\stLab}{\tyGround}]\,\stTi[i,k]$ for $i\in I$,
$\stMerge{}{\{\stTi[i,k]\suchthat i\in I\}}\ni\stTi$, and
$\emptyset\not = I\subseteq I$.
		\item Suppose that $\stT\,\gtMove[\roleP\stEnvAnnotInSym\stChoice{\stLab}{\tyGround}]\,\stTi$.
So, $\unfoldOne{\stT}=\stExtSum{\roleP}{j \in J}{\stChoice{\stLab[j]}{\tyGround[j]} \stSeq \stTi[j]}$ and there is $k\in J$ such that
$\stLab[k]=\stLab$, $\tyGround[k]=\tyGround$, and $\stTi[k]=\stTi$.
$\stMerge{}{\{\stT[i]\suchthat i\in I\}}\ni\stT$ so there are $J_j\subseteq I$, non-empty, for $j\in J$ such that
for $i\in I$, $\unfoldOne{\stT[i]}=\stIntSum{\roleP}{j \suchthat i\in J_j}{\stChoice{\stLab[j]}{\tyGround[j]} \stSeq \stTi[i,j]}$,
where $\stMerge{}{\{\stTi[i,j]\suchthat i\in J_j\}}\ni \stTi[j]$ for $j\in J$.
So, $\stT[i]\,\gtMove[\roleP\stEnvAnnotOutSym\stChoice{\stLab}{\tyGround}]\,\stTi[i,k]$ for $i\in J_k$,
$\stMerge{}{\{\stTi[i,k]\suchthat i\in J_k\}}\ni \stTi$, and
$\emptyset\not = J_k\subseteq I$.
	\end{itemize}
\end{proof}

\begin{lemma}[Merge is a Subtype]
	If $\stMerge{}{\{\stT[i]\suchthat i\in I\}}\ni \stT$,
    then for all $j\in I$ $\stT \tySSub \stT[j]$.
\end{lemma}
\begin{proof}
	Consider $\Re=\{(\stT,\stTi)\suchthat \exists \mathcal{T}.\;
	\mathcal{T}\ni \stTi
	\;\text{and}\;
	\stMerge{}{\mathcal{T}}\ni \stT \}$.
	This clearly satisfies the laws for strong subtyping by~\Cref{lem:good-merge-corr}, so $\Re\subseteq\tySSub$.
\end{proof}

Note that $\tySSub\subseteq\tySub$.

\begin{lemma}[Merge preserves Strong Subtypes]
	If $\stMerge{}{\{\stT[i]\suchthat i\in I\}}\ni \stT$
    and $\stTi\tySSub \stT[i]$ for all $i\in I$, then $\stTi \tySSub \stT$.
\end{lemma}
\begin{proof}
	Consider
	\[\Re=\{(\stT,\stTi)\suchthat \exists \mathcal{T}.\;\stMerge{}{\mathcal{T}}\ni \stTi
	\;\text{and}\;
	\forall\stTii\in\mathcal{T}.\;
	\stT\tySSub\stTii\}\]
	By~\Cref{lem:good-merge-corr}, this clearly satisfies the laws for strong subtyping, so $\Re\subseteq\tySSub$.
\end{proof}

\begin{lemma}[Merge preserves Subtypes]
	If $\stMerge{}{\{\stT[i]\suchthat i\in I\}}\ni \stT$
    and $\stTi\tySub \stT[i]$ for all $i\in I$, then $\stTi \tySub \stT$.
\end{lemma}
\begin{proof}
	Consider
	\[\Re=\{(\stT,\stTi)\suchthat \exists \mathcal{T}.\;
	\stMerge{}{\mathcal{T}}\ni \stTi\;\text{and}\;
	\forall \stTii\in\mathcal{T}.\;
	\stT\tySub\stTii\}\]
	By~\Cref{lem:good-merge-corr}, this clearly satisfies the laws for subtyping, so $\Re\subseteq\tySub$.
\end{proof}

\begin{lemma}[Mergeability respects Set Inclusion]
	If $\{\stT[i]\suchthat i\in I\}$ is mergeable, then $\{\stT[i]\suchthat i\in J\}$ is mergeable for all non-empty $J\subseteq I$.
\end{lemma}
\begin{proof}
	Suppose that $\stMerge{}{\{\stT[i]\suchthat i\in I\}}\ni\stT$. 
	We construct an LTS as follows:
	\begin{enumerate}
		\item Start at node $(\stT, \stT[i] : i\in I)$
		\item At node $(\stTi, \stTi[i] :i\in I')$, if $\stTi\,\gtMove[\stEnvAnnotGenericSym]\,\stTii$,
 then let $I''=\{i\in I\suchthat \exists \stTii[i],\stTi[i]\,\gtMove[\stEnvAnnotGenericSym]\,\stTii[i]\}$.
Add an edge from $(\stTi, \stTi[i] :i\in I')$ to $(\stTii, \stTii[i] :i\in I'')$ labelled by $\stEnvAnnotGenericSym$.
	\end{enumerate}
	Now, let $J\subseteq I$ be non-empty.
	Remove all nodes, $(\stTi, \stTi[i] :i\in I')$, with $I'\cap J=\emptyset$.
	This graph is a correct type graph, so let $\stU$ be the corresponding type.
	We relabel the nodes.
	\begin{enumerate}
		\item The root node is relabelled $(\stU, \stT[i] : i\in J)$
		\item If there is an edge from $(\stUi, \stTi[i] : i\in J')$ to $(\stTii, \stTii[i] :i\in I'')$ with $\stEnvAnnotGenericSym$, 
then there is $\stUii$ such that $\stUi\,\gtMove[\stEnvAnnotGenericSym]\,\stUii$.
Relabel $(\stTii, \stTii[i] :i\in I'')$ to $(\stUii, \stTii[i] :i\in I''\cap J)$.
	\end{enumerate}
	The nodes of this graph satisfy the coinductive rules of $\relMerge{}{}$,
    so $\stMerge{}{\{\stT[i]\suchthat i\in J\}}\ni \stU$.
\end{proof}

\propProjBalIndCo*

\begin{proof}[1]
	Let $\stMerge{}{}$ be a good merge
	and $\gtProj{}{}{}$ be the corresponding inductive projection.
	
	By induction on $|\gtG|$,
	if $\gtProj{\roleP}{\gtG}{\stT}$
	and
	$\unfoldOne{\stT}$ is an internal or external choice,
	then $\gtDepth{\gtG}{\roleP}$ exists.
	\\
	If $\gtG=\gtEnd$, then $\stT=\stEnd$.
	\\
	If $\gtG=\gtRecVar$, then $\stT=\stRecVar$.
	\\
	Suppose that $\gtG=\gtComm{\rolePi}{\roleQi}{i\in I}{\gtLab[i]}{\tyGround[i]}{\gtG[i]}$.
	If $\roleP\in\{\rolePi,\roleQi\}$, then $\gtDepth{\gtG}{\roleP}=1$.
	If $\roleP\notin\{\rolePi,\roleQi\}$, then
	there is $\{\stT[i]\}_{i\in I}$
	such that for $i\in I$, $\gtProj{\roleP}{\gtG[i]}{\stT[i]}$
	and $\stMerge{}{\{\stT[i]\}_{i\in I}}\ni\stT$.
	Let $\{\stRecVar[1],\dots,\stRecVar[n]\}=\bigcup_{i\in I}\gtFv{\stT[i]}\cup\gtFv{\stT}$.
	Use \inferrule{\mergeGoodSubst} to obtain
	$\stMerge{}{\{
	\stT[i]\subst{\stRecVar[1]}{\stEnd}\dots\subst{\stRecVar[n]}{\stEnd}
	\}_{i\in I}}\ni
	\stT\subst{\stRecVar[1]}{\stEnd}\dots\subst{\stRecVar[n]}{\stEnd}
	$.
	By \inferrule{\mergeGoodSubI},
	for $i\in I$,
	$\stT[i]\subst{\stRecVar[1]}{\stEnd}\dots\subst{\stRecVar[n]}{\stEnd}$
	is an internal or external choice.
	Thus for $i\in I$,
	$\stT[i]$ is an internal or external choice.
	By the I.H., for $i\in I$, $\gtDepth{\gtG[i]}{\roleP}$
	exists.
	Hence, $\gtDepth{\gtG}{\roleP}$ exists.
	
	By induction on $|\gtG|$,
	if $\gtProj{\roleP}{\gtG}{\stT}$
	and
	$\unfoldOne{\stT}=\stEnd$
	or
	$\unfoldOne{\stT}=\stRecVar$,
	then $\roleP\notin\gtRoles{\gtG}$.
	\\
	If $\gtG=\gtEnd$, then $\roleP\notin\gtRoles{\gtG}$
	\\
	If $\gtG=\gtRecVar$, then $\roleP\notin\gtRoles{\gtG}$.
	\\
	Suppose that $\gtG=\gtComm{\rolePi}{\roleQi}{i\in I}{\gtLab[i]}{\tyGround[i]}{\gtG[i]}$.
	If $\roleP\in\{\rolePi,\roleQi\}$, then
	$\stT$ is an internal or external choice.
	If $\roleP\notin\{\rolePi,\roleQi\}$, then
	there is $\{\stT[i]\}_{i\in I}$
	such that for $i\in I$, $\gtProj{\roleP}{\gtG[i]}{\stT[i]}$
	and $\stMerge{}{\{\stT[i]\}_{i\in I}}\ni\stT$.
	Let
	\[\{\stRecVar[1],\dots,\stRecVar[n]\}=\bigcup_{i\in I}\gtFv{\stT[i]}\cup\gtFv{\stT}\]
	Use \inferrule{\mergeGoodSubst} to obtain
	\[
	\stMerge{}{\{
	\stT[i]\subst{\stRecVar[1]}{\stEnd}\dots\subst{\stRecVar[n]}{\stEnd}
	\}_{i\in I}}\ni
	\stT\subst{\stRecVar[1]}{\stEnd}\dots\subst{\stRecVar[n]}{\stEnd}
	\]
	By \inferrule{\mergeGoodSubI},
	for $i\in I$,
	$\stT[i]\subst{\stRecVar[1]}{\stEnd}\dots\subst{\stRecVar[n]}{\stEnd}$
	is an $\stEnd$-type.
	Thus for $i\in I$,
	$\stT[i]$ is an $\stEnd$-type or a recursive variable.
	By the I.H., for $i\in I$, $\roleP\notin\gtRoles{\gtG[i]}$.
	Therefore, $\roleP\notin\gtRoles{\gtG}$.
	
	Now suppose that $\gtProj{}{\gtG}{\stEnv}$.
	For $\roleP\in\gtRoles{\gtG}$, $\unfoldOne{\stEnvApp{\stEnv}{\roleP}}$
	is not a recursive variable or $\stEnd$
	so is an internal or external choice.
	Thus, $\gtDepth{\gtG}{\roleP}$ exists.
	Using \inferrule{\mergeGoodSubst},
	for $\roleP\in\dom{\stEnv}$
	we obtain
	$\gtProj{\roleP}{\unfoldOne{\gtG}}{\unfoldOne{\stEnvApp{\stEnv}{\roleP}}}$.
	If $\unfoldOne{\gtG}$ has $\gtGi$ as a continuation,
	$\gtGi$ is projectable.
	So, every head reduction of $\gtG$ is projectable
	so has depths for every participant.
	This is sufficient to guarantee balancedness.
\end{proof}

\begin{proof}[2]
\label{ex:bad-co-proj}
Let $\gtProj{}{}{}$ be the plain coinductive projection.
\begin{equation*}
	\gtG=\gtRec{\gtRecVar}{
		\gtCommRaw{\roleFmt{a}}{\roleFmt{b}}{
		\begin{array}{l}
			\gtLab\gtSeq\gtRecVar\\
			\gtLabi\gtSeq\gtCommSingle{\roleFmt{c}}{\roleFmt{d}}{\gtLab}{}{}
		\end{array}
		}
	}
\end{equation*}
\begin{equation*}
	\stT[\roleFmt{a}]=\stRec{\stRecVar}{
		\stIntSum{\roleFmt{b}}{}{
			\begin{array}{l}
				\stLab\stSeq\stRecVar\\
				\stLabi
			\end{array}
		}
	}
	\qquad
	\stT[\roleFmt{b}]=\stRec{\stRecVar}{
		\stExtSum{\roleFmt{a}}{}{
			\begin{array}{l}
				\stLab\stSeq\stRecVar\\
				\stLabi
			\end{array}
		}
	}
	\qquad
	\stT[\roleFmt{c}]=\stFmt{
		\roleFmt{d}{\oplus}\stLab
	}
	\qquad
	\stT[\roleFmt{d}]=\stFmt{
		\roleFmt{c}{\&}\stLab
	}
\end{equation*}
\begin{equation*}
	\stEnv=\stEnvMap{\roleFmt{a}}{\stT[\roleFmt{a}]}
	\stEnvComp\stEnvMap{\roleFmt{b}}{\stT[\roleFmt{b}]}
	\stEnvComp\stEnvMap{\roleFmt{c}}{\stT[\roleFmt{c}]}
	\stEnvComp\stEnvMap{\roleFmt{d}}{\stT[\roleFmt{d}]}
\end{equation*}
$\gtG$ is not weak balanced as
$\weakComm{\gtG}{\roleFmt{c}}{\roleFmt{d}}$
and neither $\gtDepth{\gtG}{\roleFmt{c}}$
nor $\gtDepth{\gtG}{\roleFmt{d}}$ exist.
We have that $\gtProj{}{\gtG}{\stEnv}$
and $\stEnv\stEnvMoveAnnot{\ltsSendRecv{\roleFmt{c}}{\roleFmt{d}}{\stLab}}$,
but $\gtG\,\,\not\!\!\!\!\gtMove[\ltsSendRecv{\roleFmt{c}}{\roleFmt{d}}{\stLab}]$.
Thus, $\gtProj{}{}{}$ is not good
as it projects a type that is not weak balanced.
\end{proof}

\thmProjGood*

\begin{theorem}[Coinductive Projection]
\label{thm:co-proj-good}
If the coinductive projection is defined using a good merge operation and we restrict the global types to weak balanced types, then it is a good projection.
\end{theorem}

\begin{proof}
\begin{itemize}
	\item We prove that if $\gtG \,\gtMove[\ltsSendRecv{\roleP}{\roleQ}{\stChoice{\stLab}{\tyGround}}]\, \gtGi$, then:
	\begin{enumerate}
	\item If $\gtProj{\roleP}{\gtG}{\stT}$, then there is $\stTi$ such that $\gtProj{\roleP}{\gtGi}{\stTi}$ and $\stT\,\gtMove[{\roleQ\stEnvAnnotOutSym\stChoice{\stLab}{\tyGround}}]\,\stTi$.
	\item If $\gtProj{\roleQ}{\gtG}{\stT}$, then there is $\stTi$ such that $\gtProj{\roleQ}{\gtGi}{\stTi}$ and $\stT\,\gtMove[{\roleP\stEnvAnnotInSym\stChoice{\stLab}{\tyGround}}]\,\stTi$.
	\item If $\gtProj{\roleR}{\gtG}{\stT}$, then there is $\stTi$ such that $\gtProj{\roleR}{\gtGi}{\stTi}$ and $\stMerge{}{\{\stT,\stTi\}} \ni \stT$.
	\end{enumerate}
	We induct on the definition of $\gtG \,\gtMove[\ltsSendRecv{\roleP}{\roleQ}{\stChoice{\stLab}{\tyGround}}]\, \gtGi$.\\
	
	If $\gtG \,\gtMove[\ltsSendRecv{\roleP}{\roleQ}{\stChoice{\stLab}{\tyGround}}]\, \gtGi$ by
	$\inferrule{\iruleGtMovePar}$ using the left component
	and $\gtProj{\roleR}{\gtG}{\stT}$, then
	write $\gtG=\gtPar{\gtG[L]}{\gtG[R]}$ and $\gtGi=\gtPar{\gtGi[L]}{\gtG[R]}$,
	where $\gtG[L] \,\gtMove[\ltsSendRecv{\roleP}{\roleQ}{\stChoice{\stLab}{\tyGround}}]\, \gtGi[L]$.\\
	If $\roleR\in\gtRoles{\gtG[L]}$, then $\gtProj{\roleR}{\gtG[L]}{\stT}$.\\
	By the I.H., there is $\stTi$ such that $\gtProj{\roleR}{\gtGi[L]}{\stTi}$, so
	$\gtProj{\roleR}{\gtGi}{\stTi}$
	and the other required conclusions hold.\\
	If $\roleR\notin\gtRoles{\gtG[L]}$, then $\gtProj{\roleR}{\gtGi[R]}{\stT}$ so,
	$\gtProj{\roleR}{\gtGi}{\stT}$,
	$\roleR\notin\{\roleP,\roleQ\}$, and $\stMerge{}{\{\stT\}}\ni\stT$.\\
	If $\gtG \,\gtMove[\ltsSendRecv{\roleP}{\roleQ}{\stChoice{\stLab}{\tyGround}}]\, \gtGi$ by
	$\inferrule{\iruleGtMovePar}$ using the right component
	and $\gtProj{\roleR}{\gtG}{\stT}$, then
	write $\gtG=\gtPar{\gtG[L]}{\gtG[R]}$ and $\gtGi=\gtPar{\gtG[L]}{\gtGi[R]}$,
	where $\gtG[R] \,\gtMove[\ltsSendRecv{\roleP}{\roleQ}{\stChoice{\stLab}{\tyGround}}]\, \gtGi[R]$.\\
	If $\roleR\in\gtRoles{\gtG[R]}$, then $\gtProj{\roleR}{\gtG[R]}{\stT}$.\\
	By the I.H., there is $\stTi$ such that $\gtProj{\roleR}{\gtGi[R]}{\stTi}$, so
	$\gtProj{\roleR}{\gtGi}{\stTi}$
	and the other required conclusions hold.\\
	If $\roleR\notin\gtRoles{\gtG[R]}$, then $\gtProj{\roleR}{\gtGi[L]}{\stT}$ so,
	$\gtProj{\roleR}{\gtGi}{\stT}$,
	$\roleR\notin\{\roleP,\roleQ\}$, and $\stMerge{}{\{\stT\}}\ni\stT$.\\
	If $\gtG \,\gtMove[\ltsSendRecv{\roleP}{\roleQ}{\stChoice{\stLab}{\tyGround}}]\, \gtGi$ by
    $\inferrule{\iruleGtMoveRec}$ and
    $\gtProj{\roleR}{\gtG}{\stT}$,
    then $\gtG=\gtRec{\gtRecVar}{\gtGii}$ and
    $\gtGii{}[\gtRec{\gtRecVar}{\gtGii}/\gtRecVar] \,\gtMove[\ltsSendRecv{\roleP}{\roleQ}{\stChoice{\stLab}{\tyGround}}]\, \gtGi$.\\
	$\unfoldOne{\gtGii{}[\gtRec{\gtRecVar}{\gtGii}/\gtRecVar]}=\unfoldOne{G}$,
    so $\gtProj{\roleR}{\gtGii{}[\gtRec{\gtRecVar}{\gtGii}/\gtRecVar]}{\stT}$.\\
	By the I.H.,
    there is $\stTi$ such that $\gtProj{\roleR}{\gtGi}{\stTi}$
    and the other required conclusions hold.\\
	If $\gtG \,\gtMove[\ltsSendRecv{\roleP}{\roleQ}{\stChoice{\stLab}{\tyGround}}]\, \gtGi$ by
    $\inferrule{\iruleGtMoveCtx}$ and
    $\gtProj{\roleR}{\gtG}{\stT}$,
	then $\gtG=\gtCommSmall{\rolePi}{\roleQi}{i \in I}{\gtLab[i]}{\tyGround[i]}{\gtGi[i]}$ and\\
    $\gtGi=\gtCommSmall{\rolePi}{\roleQi}{i \in I}{\gtLab[i]}{\tyGround[i]}{\gtGii[i]}$ and
    $\gtGi[i] \,\gtMove[\ltsSendRecv{\roleP}{\roleQ}{\stChoice{\stLab}{\tyGround}}]\, \gtGii[i]$ for $i\in I$.\\
	$\gtProj{\roleR}{\gtG}{\stT}$, so
    there are $\stT[i]$ for $i\in I$ with
    $\gtProj{\roleR}{\gtGi[i]}{\stT[i]}$ and
    $\stMerge{}{\{\stT[i]\suchthat i\in I\}}\ni\stT$.\\
	By the I.H., there are $\stTi[i]$ with
    $\gtProj{\roleR}{\gtGii[i]}{\stTi[i]}$.\\
	If $\roleR=\roleP$, then
    $\stT[i]\,\gtMove[{\roleQ\stEnvAnnotOutSym\stChoice{\stLab}{\tyGround}}]\,\stTi[i]$ so 
    there is $\stTi$ with
    $\stMerge{}{\{\stTi[i]\suchthat i\in I\}}\ni\stTi$ and
    $\stT\,\gtMove[{\roleQ\stEnvAnnotOutSym\stChoice{\stLab}{\tyGround}}]\, \stTi$.\\
	If $\roleR=\roleQ$, then
    $\stT[i]\,\gtMove[{\roleP\stEnvAnnotInSym\stChoice{\stLab}{\tyGround}}]\,\stTi[i]$ so
    there is $\stTi$ with
    $\stMerge{}{\{\stTi[i]\suchthat i\in I\}}\ni\stTi$ and
    $\stT\,\gtMove[{\roleP\stEnvAnnotInSym\stChoice{\stLab}{\tyGround}}]\,\stTi$.\\
	If $\roleR\notin\{\roleP,\rolePi,\roleQ,\roleQi\}$,
    $\stMerge{}{\{\stT[i],\stTi[i]\}}\ni\stT[i]$ for $i\in I$, so
    $\stMerge{}{\left(\{\stT[i]\suchthat i\in I\} \cup \{\stTi[i]\suchthat i\in I\}\right)}\ni \stT$, so
    let $\stMerge{}{\{\stTi[i]\suchthat i\in I\}}\ni \stTi$, so
    $\gtProj{\roleR}{\gtGi}{\stTi}$ and
    $\stMerge{}{\{\stTi,\stT\}} \ni \stT$.\\
	If $\roleR\in\{\rolePi,\roleQi\}$,
    then we remove the head of $\stT$ and proceed similar to the above case
    before adding the head onto a merge of $\{\stT[i]\suchthat i\in I\}$
    by \inferrule{\mergeGoodCtx}
    and
    reaching the same conclusion.\\
	If $\gtG \,\gtMove[\ltsSendRecv{\roleP}{\roleQ}{\stChoice{\stLab}{\tyGround}}]\, \gtGi$ by
    $\inferrule{\iruleGtMoveComm}$ and
    $\gtProj{\roleR}{\gtG}{\stT}$, then
    $\gtG=\gtCommSmall{\roleP}{\roleQ}{i \in I}{\gtLab[i]}{\tyGround[i]}{\gtGi[i]}$ and
    there is some $j\in I$ such that
    $\gtGi=\gtGi[j]$, $\stLab=\stLab[j]$, and
    $\tyGround=\tyGround[j]$.\\
	If $\roleR=\roleP$, then
    $\unfoldOne{\stT}=\stIntSum{\roleQ}{i \in I}{\stChoice{\stLab[i]}{\tyGround[i]} \stSeq
	        \stTi[i]}$ and
            $\gtProj{\roleP}{\gtGi[i]}{\stTi[i]}$ for $i\in I$, so
            $\stT\,\gtMove[{\roleQ\stEnvAnnotOutSym\stChoice{\stLab[j]}{\tyGround[j]}}]\,\stTi[j]$.\\
	If $\roleR=\roleQ$, then
    $\unfoldOne{\stT}=\stExtSum{\roleP}{i \in I}{\stChoice{\stLab[i]}{\tyGround[i]} \stSeq
	        \stTi[i]}$ and
            $\gtProj{\roleQ}{\gtGi[i]}{\stTi[i]}$ for $i\in I$, so
            $\stT\,\gtMove[{\roleP\stEnvAnnotInSym\stChoice{\stLab[j]}{\tyGround[j]}}]\,\stTi[j]$.\\
	Otherwise, there are $\stTi[i]$ for $i\in I$ with
    $\gtProj{\roleR}{\gtGi[i]}{\stTi[i]}$ and
    $\stMerge{}{\{\stTi[i]\suchthat i\in I\}}\ni \stT$, so
    $\gtProj{\roleR}{\gtGi[j]}{\stTi[j]}$ and
    $\stMerge{}{\{\stT,\stTi[j]\}}\ni\stT$.
	\item If $\gtProj{\roleP}{\gtG}{\stT}$ and
    $\roleP\notin\gtRoles{\gtG}$, we immediately get that $\unfoldOne{\stT}=\stEnd$ by~\Cref{def:global-proj}.
	\item Suppose that $\gtProj{\roleP}{\gtG}{\stT}$,
    $\gtProj{\roleQ}{\gtG}{\stTi}$,
    $\stT\,\gtMove[{\roleQ\stEnvAnnotOutSym\stChoice{\stLab}{\tyGround}}]$, and
    $\stTi\,\gtMove[{\roleP\stEnvAnnotInSym\stChoice{\stLabi}{\tyGroundi}}]$.\\
    So, $\roleP\in\gtRoles{\gtG}$ and
	$\unfoldOne{\stT}$ has a send to $\roleQ$ as its head.
	By induction on the derivation of
	$\roleP\in\gtRoles{\gtG}$,
	$\weakComm{\gtG}{\roleP}{\roleQ}$.
	Thus, $\gtDepth{\gtG}{\roleP}$ exists.\\
	We induct on $\gtDepth{\gtG}{\roleP}$.\\
	If $\gtDepth{\gtG}{\roleP}=1$, then
    $\unfoldOne{\gtG}$ is a parallel composition of 1 or more components, one of which is
of the form $\gtComm{\roleP}{\roleQi}{i \in I}{\gtLab[i]}{\tyGround[i]}{\gtG[i]}$.\\
	So, $\unfoldOne{\stT}=\stIntSum{\roleQi}{i \in I}{\stChoice{\stLab[i]}{\tyGround[i]} \stSeq \stT[i]}$ and
    $\stT\,\gtMove[{\roleQ\stEnvAnnotOutSym\stChoice{\stLab}{\tyGround}}]$,
    hence $\roleQi=\roleQ$ and there is $j\in I$ such that
    $\stLab=\stLab[i]$ and
    $\tyGround=\tyGround[i]$.\\
	Thus, $\gtG \,\gtMove[\ltsSendRecv{\roleP}{\roleQ}{\stChoice{\stLab}{\tyGround}}]$.\\
	If $\gtDepth{\gtG}{\roleP}>1$,
    then $\unfoldOne{\gtG}$ is a parallel composition of 1 or more components, one of which contains
$\roleP$ as a participant.\\
Say this component is
$\gtComm{\rolePi}{\roleQi}{i \in I}{\gtLab[i]}{\tyGround[i]}{\gtG[i]}$ where
    $\roleP\not=\rolePi$ (as $\gtDepth{\gtG}{\roleP}>1$) and
    $\roleP\not=\roleQi$ (as the head of $\stT$ is not a receive).\\
	If $\roleQ\in\{\rolePi,\roleQi\}$,
    then $\stTi$ will make a reduction involving the other participant, which is not $\roleP$,
    contradicting $\stTi\,\gtMove[{\roleP\stEnvAnnotInSym\stChoice{\stLabi}{\tyGroundi}}]$.\\
	So, there are $\stT[i]$, $\stTi[i]$ for $i\in I$ such that
    $\gtProj{\roleP}{\gtG[i]}{\stT[i]}$,
    $\gtProj{\roleQ}{\gtG[i]}{\stTi[i]}$,
    $\stMerge{}{\{\stT[i]\suchthat i\in I\}}\ni\stT$,
    and $\stMerge{}{\{\stTi[i]\suchthat i\in I\}}\ni\stTi$.\\
	Fix $i\in I$.\\
    We have $\gtDepth{\gtG[i]}{\roleP}\leq \gtDepth{\gtG}{\roleP}-1$,
    $\stT[i]\,\gtMove[{\roleQ\stEnvAnnotOutSym\stChoice{\stLab}{\tyGround}}]$, and
    $\stTi[i]\,\gtMove[{\roleP\stEnvAnnotInSym\stChoice{\stLabii}{\tyGroundii}}]$ for some $\stLabii$ and $\tyGroundii$.\\
	By the I.H., $\gtG[i] \,\gtMove[\ltsSendRecv{\roleP}{\roleQ}{\stChoice{\stLab}{\tyGround}}]$.\\
	This is true for all $i\in I$ so $\gtG \,\gtMove[\ltsSendRecv{\roleP}{\roleQ}{\stChoice{\stLab}{\tyGround}}]$.
	\end{itemize}
\end{proof}

\begin{theorem}[Inductive Projection]
Inductive projections defined using a good merge are good.
\end{theorem}
\begin{proof}

If $\gtProj{\roleP}{\gtG}{\stT}$
and $\gtProj{\roleP}{\gtGi}{\stTi}$ with $\stTi$ closed, then
$\gtProj{\roleP}{\gtG{}[\gtGi/\gtRecVar]}{\stT{}[\stTi/\stRecVar]}$
by induction on $\gtG$, using \inferrule{\mergeGoodSubst}
at merging steps.
So if $\gtProj{\roleP}{\gtRec{\gtRecVar}{\gtG}}{\stRec{\stRecVar}{\stT}}$, then
$\gtProj{\roleP}{\gtG{}[\gtRec{\gtRecVar}{\gtG}/\gtRecVar]}{\stT{}[\stRec{\stRecVar}{\stT}/\stRecVar]}$
if $\stRec{\stRecVar}{\stT}$ is closed.

The proof now proceeds similarly to the above proof, explicitly
unfolding types when necessary.
\end{proof}

\thmProjStandard*

\begin{proof}
By \cref{thm:good-inductive-proj,thm:merge-good},
the inductive projection with inductive plain merge,
the inductive projection with inductive full merge,
the coinductive projection with coinductive plain merge restricted to weak balanced global types and
the coinductive projection with coinductive full merge restricted to weak balanced global types
are all good projections.

By \cref{prop:good-robust} with
$\predP=\{\gtG\suchthat \gtG\,\text{contains no parallel composition}\}$,
the inductive projection with inductive plain merge restricted to global types without parallel composition,
the inductive projection with inductive full merge restricted to global types without parallel composition,
the coinductive projection with coinductive plain merge restricted to weak balanced global types without parallel composition and
the coinductive projection with coinductive full merge restricted to weak balanced global types without parallel composition
are all good projections.
\end{proof}

\begin{example}[Global Type with Not Context-Free Trace is Projectable]
\label{ex:non-reg-global-proj}
The global type from \cref{thm:global-sem}
has trace language that is not context-free.
\[
	\gtG[\text{\tiny irreg}]=\gtPar{\gtRec{\gtRecVar}{
	\gtCommRaw{\roleP}{\roleQ}{
	\begin{array}{l}
		\gtLab[0]\gtSeq
		\gtCommRaw{\roleR}{\roleS}{
			\gtLab[2]\gtSeq\gtRecVar
		}
		\\
		\gtLab[1]\gtSeq
		\gtCommRaw{\roleR}{\roleS}{
		\begin{array}{l}
			\gtLab[2]\gtSeq\gtRecVar
			\\
			\gtLab[3]\gtSeq\gtRecVar
		\end{array}
		}
	\end{array}
	}}}
	{
	\gtRec{\gtRecVar}{
	\gtCommRaw{\rolePi}{\roleQi}{
	\begin{array}{l}
		\gtLab[0]\gtSeq
		\gtCommRaw{\roleRi}{\roleSi}{
			\gtLab[2]\gtSeq\gtRecVar
		}
		\\
		\gtLab[1]\gtSeq
		\gtCommRaw{\roleRi}{\roleSi}{
		\begin{array}{l}
			\gtLab[2]\gtSeq\gtRecVar
			\\
			\gtLab[3]\gtSeq\gtRecVar
		\end{array}
		}
	\end{array}
	}}
	}
	\]
	
We define the $\text{full}^+$ merge operation
that can remove internal choices when merging.

We define $\stBinMerge$
	to be a partial binary operation from
	two
	local types to a local type recursively by:\\
	$\stEnd\stBinMerge\stEnd=\stEnd$;
	$
			\stRecVar\stBinMerge\stRecVar
		=\stRecVar$;
	$
	\stRec{\stRecVar}{\stT}\stBinMerge\stRec{\stRecVar}{\stTi}
	=
	\stRec{\stRecVar}{\stT\stBinMerge\stTi}
	$;\\
	$
	\stExtSum{\roleP}{i \in I}{\stChoice{\stLab[i]}{\tyGround[i]} \stSeq \stT[i]}
	\stBinMerge
	\stExtSum{\roleP}{i \in I'}{\stChoice{\stLab[i]}{\tyGround[i]} \stSeq \stTi[i]}
	=
	\stExtSum{\roleP}{i \in I\cup I'}{\stChoice{\stLab[i]}{\tyGround[i]} \stSeq \stTii[i]}
	$
	where
	$\stTii[i]=\stT[i]$ for $i\in I\setminus I'$,
	$\stTi[i]$ for $i\in I'\setminus I$ and
	$\stT[i]\stBinMerge\stTi[i]$ for $i\in I\cap I'$
	; and\\
		$
	\stIntSum{\roleP}{i \in I}{\stChoice{\stLab[i]}{\tyGround[i]} \stSeq \stT[i]}
	\stBinMerge
	\stIntSum{\roleP}{i \in I'}{\stChoice{\stLab[i]}{\tyGround[i]} \stSeq \stTi[i]}
	=
	\stIntSum{\roleP}{i \in I\cap I' \suchthat
	\stT[i]\stBinMerge\stTi[i]\;\text{exists}
	}{\stChoice{\stLab[i]}{\tyGround[i]} \stSeq \stT[i]\stBinMerge\stTi[i]}
	$.\\
	We define the partial function $\funcMerge{\mathcal{T}}$
	by:\\
	$\funcMerge{\{\stT\}}=\stT$ and
	$\funcMerge{\{\stT[1],\dots,\stT[n+2]\}}=
	\funcMerge{\{\stT[1],\dots,\stT[n+1]\}}\stBinMerge\stT[n+2]$ ($n\geq 0$).
	\\
	We define \emph{$\text{full}^+$ inductive merge}
	by $\stMerge{}{\mathcal{T}}=
		\{\stT\suchthat \stT=\funcMerge{\mathcal{T}}\}$.
		
	Similarly to the full inductive merge,
	the $\text{full}^+$ inductive merge is good
	so the $\text{full}^+$ inductive projection is good.
	
	Using the $\text{full}^+$ inductive projection,
	\[
		\gtProj{}{\gtG[\text{\tiny irreg}]}{
		\set{\begin{array}{l}
			\stEnvMap{\roleP}{
			\stRec{\stRecVar}{\stIntSum{\roleQ}{i\in\{0,1\}}{\stLab[i]\stSeq\stRecVar}}
			}
			\stEnvComp
			\stEnvMap{\roleQ}{
			\stRec{\stRecVar}{\stExtSum{\roleP}{i\in\{0,1\}}{\stLab[i]\stSeq\stRecVar}}
			}
			\stEnvComp
			\stEnvMap{\roleR}{
			\stRec{\stRecVar}{\roleS{\oplus}\stLab[2]\stSeq\stRecVar}
			}
			\stEnvComp
			\stEnvMap{\roleS}{
			\stRec{\stRecVar}{\stExtSum{\roleR}{i\in\{2,3\}}{\stLab[i]\stSeq\stRecVar}}
			}
			\stEnvComp\\
			\stEnvMap{\rolePi}{
			\stRec{\stRecVar}{\stIntSum{\roleQi}{i\in\{0,1\}}{\stLab[i]\stSeq\stRecVar}}
			}
			\stEnvComp
			\stEnvMap{\roleQi}{
			\stRec{\stRecVar}{\stExtSum{\rolePi}{i\in\{0,1\}}{\stLab[i]\stSeq\stRecVar}}
			}
			\stEnvComp
			\stEnvMap{\roleRi}{
			\stRec{\stRecVar}{\roleSi{\oplus}\stLab[2]\stSeq\stRecVar}
			}
			\stEnvComp
			\stEnvMap{\roleSi}{
			\stRec{\stRecVar}{\stExtSum{\roleRi}{i\in\{2,3\}}{\stLab[i]\stSeq\stRecVar}}
			}
		\end{array}}
		}
	\]
\end{example}

\begin{proposition}[Projected Trace]
	\label{prop:proj-trace}
	\begin{enumerate}
		\item If $\gtProj{}{}{}$
		is the plain inductive/coinductive projection,
		$\gtProj{}{\gtG}{\stEnv}$, and
		$\gtG\,\gtMove[\stEnvAnnotGenericSym]\,\gtGi$,
		then there exist $\stEnvi$ such that
		$\stEnv\stEnvMoveGenAnnot\stEnvi$ and
		$\gtProj{}{\gtGi}{\stEnvi}$.
		\item If $\gtProj{}{}{}$
		is the full inductive/coinductive projection,
		$\gtProj{}{\gtG}{\stEnv}$, and
		$\gtG\,\gtMove[\stEnvAnnotGenericSym]\,\gtGi$,
		then there exist $\stEnvi$ and $\stEnvii$ such that
		$\stEnv\stEnvMoveGenAnnot\stEnvi\tySSub\stEnvii$ and
		$\gtProj{}{\gtGi}{\stEnvii}$.
		\item Suppose that $\gtProj{}{}{}$
	is the full/plain inductive/coinductive projection.
	If $\gtProj{}{\gtG}{\stEnv}$ and $\gtG$ is weak balanced,
	then $\gtTrace{\gtG}=\gtTrace{\stEnv}$.
	\end{enumerate}
\end{proposition}

\begin{proof}[Proof. (1) and (2)]
	Immediate from the proof of \cref{thm:co-proj-good},
	since
	if $\gtProj{}{\gtG}{\stEnv}$ and
		$\gtG\,\gtMove[\stEnvAnnotGenericSym]\,\gtGi$,
		then there exist $\stEnvi$ and $\stEnvii$ such that
		\[
		\stEnv\stEnvMoveGenAnnot\stEnvi,\;
		\gtProj{}{\gtGi}{\stEnvii},\;
		\dom{\stEnvi}=\dom{\stEnvii}
		\;\text{and}\;
		\forall\roleP\in\dom{\stEnvi}.\;
		\stMerge{}{\{\stEnvApp{\stEnvi}{\roleP},\stEnvApp{\stEnvii}{\roleP}\}}
		\ni
		\stEnvApp{\stEnvi}{\roleP}
		\]
		
		For the plain inductive projection,
		$\forall\roleP\in\dom{\stEnvi}.\;
		\stEnvApp{\stEnvii}{\roleP}=
		\stEnvApp{\stEnvi}{\roleP}
		$.
		
		For the plain coinductive projection,
		$\forall\roleP\in\dom{\stEnvi}.\;
		\stEnvApp{\stEnvii}{\roleP}\bisim
		\stEnvApp{\stEnvi}{\roleP}
		$,
		so $\gtProj{}{\gtGi}{\stEnvi}$ by \cref{prop:proj-inv-local}.
		
		For the full inductive/coinductive projection,
		$\forall\roleP\in\dom{\stEnvi}.\;
		\stEnvApp{\stEnvii}{\roleP}\tySSub
		\stEnvApp{\stEnvi}{\roleP}
		$.
\end{proof}

\begin{proof}[Proof. (3)]
	By \cref{lem:subtyping-contexts}, (1) and (2),
	if $\stEnv\tySSub\stEnvi$ and $\gtProj{}{\gtG}{\stEnvi}$
	then $\stEnv\stEnvMoveGenAnnot$ iff $\gtG\,\gtMove[\stEnvAnnotGenericSym]$
	and ($\stEnv\stEnvMoveGenAnnot\stEnv[1]$ and $\gtG\,\gtMove[\stEnvAnnotGenericSym]\,\gtGi$)
	implies that there is $\stEnvi[1]$
	such that $\stEnv[1]\tySSub\stEnvi[1]$ and $\gtProj{}{\gtGi}{\stEnvi[1]}$.
	
	Therefore, $\gtTrace{\gtG}=\gtTrace{\stEnv}$.
\end{proof}

\subsection{Appendix for \cref{sec:counter} (Constructing Counterexamples)}
\label{app:counter}

\thmIncorrectMerges*

\begin{proof}[Proof. (1)]
	Suppose that
	$\gtProj{\roleQ[1]}{\gtGcounter[1]}{\stT}$.
	All the possible projections are included within
	the coinductive projection
	with coinductive 1-level merge
	so we assume that $\gtProj{}{}{}$
	is the coinductive projection
	with coinductive 1-level merge.
	
	For $i\in \{0,1\}$, there exist $\stT[i]$
	such that $\gtProj{\roleQ[1]}{
		\gtCommSingle{\roleP[0]}{\roleQ[1]}{\gtLab}{}{
					\gtCommSingle{\roleP[2]}{\roleP[0]}{\gtLab[0]}{}{
					\gtCommSingle{\roleP[0]}{\roleQ[1]}{\gtLab[i]}{}{}
					}
		}
	}{\stT[i]}$
	and
	$\stT\in\stMerge{}{\{\stT[0],\stT[1]\}}$.
	
	Thus,
	there exists $\stT[1,i]$
	such that
	$\gtProj{\roleQ[0]}{\gtCommSingle{\roleP[2]}{\roleP[0]}{\gtLab[0]}{}{
					\gtCommSingle{\roleP[0]}{\roleQ[1]}{\gtLab[i]}{}{}
					}}{\stT[1,i]}$
	and
	$\unfoldOne{\stT[i]}=
	\stFmt{\roleP[0]{\&}\stLab\stSeq\stT[1,i]}$.
	
	Thus, there exists $\stT[2,i]$ such that
	$\gtProj{\roleQ[0]}{
					\gtCommSingle{\roleP[0]}{\roleQ[1]}{\gtLab[i]}{}{}
					}{\stT[2,i]}$
	and $\stT[1,i]\in\stMerge{}{\{\stT[2,i]\}}$.
	
	Thus,
	$\unfoldOne{\stT[2,i]}=
	\stFmt{\roleP[0]{\&}\stLab[i]\stSeq\stT[3,i]}
	$
	where $\unfoldOne{\stT[3,i]}=\stEnd$.
	
	Thus,
	$\stT[1,i]\bisim
	\stFmt{
	\roleP[0]{\&}\stLab[i]}
	$.
	Therefore,
	$\stMerge{}{\{\stT[0],\stT[1]\}}=\emptyset$
	as $\stT[1,0]\not\bisim\stT[1,1]$.
	
	This contradicts the assumption that $\stT\in\stMerge{}{\{\stT[0],\stT[1]\}}$.
\end{proof}

\begin{proof}[Proof. (2)]
	\[
		\begin{array}{c}
			\stT[{\roleP[0]}]
			=
			\stFmt{
				\roleQ[0]{\oplus}\!\left\{
					\stLab[i]\stSeq
					\roleQ[1]{\oplus}\stLab\stSeq
					\roleP[2]{\&}\!\left\{
					\stLab[j]\stSeq
					\roleQ[1]{\oplus}\stLab[i\,xor\,j]
					\right\}_{j\in\{0,1\}}
				\right\}_{i\in\{0,1\}}
			}
			\quad
			\stT[{\roleQ[0]}]
			=
			\stFmt{
				\roleP[0]{\oplus}\!\left\{
				\stLab[i]
				\right\}_{i\in\{0,1\}}}
			\\[1ex]
			\stT[{\roleQ[1]}]=
			\stFmt{
					\roleP[0]{\&}\stLab\stSeq
					\roleP[0]{\&}\!\left\{
					\stLab[i]
					\right\}_{i\in\{0,1\}}
			}
			\quad
			\stT[{\roleP[2]}]=
			\stFmt{
				\roleQ[2]{\oplus}\!\left\{
					\stLab[i]\stSeq
					\roleP[0]{\oplus}\stLab[i]
				\right\}_{i\in\{0,1\}}
			}
			\quad
			\stT[{\roleQ[2]}]=
			\stFmt{
				\roleP[2]{\&}\!\left\{
					\stLab[i]
				\right\}_{i\in\{0,1\}}
			}
			\\[1ex]
			\stEnv=
			\set{
			\stEnvMap{\roleR}{\stT[\roleR]}
			\suchthat
			\roleR\in\{\roleP[0], \roleP[2], \roleQ[0],\roleQ[1],\roleQ[2]\}
			}
		\end{array}
	\]
	Let $\gtProj{}{}{}$
	be one of the projections,
	so that
	$\gtProj{}{\gtGcounter}{\stEnv}$
	and
	$\stEnvAssoc{\gtGcounter}{\stEnv}{}$.
	
	We have
	$\stEnv\stEnvMoveAnnot{\ltsSendRecv{\roleP[2]}{\roleQ[2]}{\stLab[0]}}\stEnvi$
	and
	$\gtGcounter\,\gtMove[\ltsSendRecv{\roleP[2]}{\roleQ[2]}{\stLab[0]}]\,\gtGcounter[1]$,
	but $\gtGcounter[1]$ is not projectable (and transitions are deterministic)
	so completeness of association fails.
	
	We have that
	$\gtGcounter\,\gtMove[\ltsSendRecv{\roleP[2]}{\roleQ[2]}{\stLab[0]}]$,
	$\stEnv\stEnvMoveAnnot{\ltsSendRecv{\roleP[2]}{\roleQ[2]}{\stLab[0]}}\stEnvi[0]$,
	$\stEnv\stEnvMoveAnnot{\ltsSendRecv{\roleP[2]}{\roleQ[2]}{\stLab[1]}}\stEnvi[1]$,
	and
	$\stEnv\stEnvMoveAnnot{\ltsSendRecv{\roleP[2]}{\roleQ[2]}{\stLab[i]}}$
	only if $i\in\{0,1\}$.
	$\gtGcounter\,\gtMove[\ltsSendRecv{\roleP[2]}{\roleQ[2]}{\stLab[0]}]\,\gtGcounter[1]$ is not projectable,
	and
	$\gtGcounter\,\gtMove[\ltsSendRecv{\roleP[2]}{\roleQ[2]}{\stLab[1]}]\,\gtGcounter[2]$ is not projectable by symmetry.
	Thus, soundness of association fails.
\end{proof}

\begin{proof}[Proof. (3)]
	\[
	\begin{array}{r@{\,}c@{\,}l}
	\mpP[{\roleP[0]}]
	&=&
	\mpIf{\mpFalse \oplus \mpTrue}
	{
		\roleQ[0]\mathord{!}\stLab[0] \mpSeq
		\roleQ[1]\mathord{!}\stLab \mpSeq
		\textstyle\sum_{i\in \{0,1\}} \roleP[2]\mathord{?}\stLab[i]\mpSeq
		\roleQ[0]\mathord{!}\stLab[i]
	}
	{
		\roleQ[0]\mathord{!}\stLab[1] \mpSeq
		\roleQ[1]\mathord{!}\stLab \mpSeq
		\textstyle\sum_{i\in \{0,1\}} \roleP[2]\mathord{?}\stLab[i]\mpSeq
		\roleQ[0]\mathord{!}\stLab[1-i]
	}
	\\[1ex]
	\mpP[{\roleQ[0]}]
	&=&
	\textstyle\sum_{i\in \{0,1\}} \roleP[0]\mathord{?}\stLab[i]\mpSeq\mpNil
	\\[1ex]
	\mpP[{\roleQ[1]}]
	&=&
	\roleP[0]\mathord{?}\stLab
	\mpSeq
	\textstyle\sum_{i\in \{0,1\}} \roleP[0]\mathord{?}\stLab[i]\mpSeq\mpNil
	\\[1ex]
	\mpP[{\roleP[2]}]
	&=&
	\mpIf{ \mpFalse \oplus \mpTrue }
	{
		\roleQ[2]\mathord{!}\stLab[0] \mpSeq
		\roleP[0]\mathord{!}\stLab[0]
	}
	{
		\roleQ[2]\mathord{!}\stLab[1] \mpSeq
		\roleP[0]\mathord{!}\stLab[1]
	}
	\\[1ex]
	\mpPi[{\roleP[2]}]
	&=&
		\roleQ[2]\mathord{!}\stLab[0] \mpSeq
		\roleP[0]\mathord{!}\stLab[0]
	\\[1ex]
	\mpPii[{\roleP[2]}]
	&=&
		\roleP[0]\mathord{!}\stLab[0]
	\\[1ex]
	\mpP[{\roleQ[2]}]
	&=&
	\textstyle\sum_{i\in \{0,1\}} \roleP[2]\mathord{?}\stLab[i]\mpSeq\mpNil
	\end{array}
	\]
	\[\begin{array}{rcl}
		\M
		&=&
		\playsrole{\roleP[0]}{\mpP[{\roleP[0]}]}
		\mpPar
		\playsrole{\roleQ[0]}{\mpP[{\roleQ[0]}]}
		\mpPar
		\playsrole{\roleQ[1]}{\mpP[{\roleQ[1]}]}
		\mpPar
		\playsrole{\roleP[2]}{\mpP[{\roleP[2]}]}
		\mpPar
		\playsrole{\roleQ[2]}{\mpP[{\roleQ[2]}]}
		\\[1ex]
		\M'
		&=&
		\playsrole{\roleP[0]}{\mpP[{\roleP[0]}]}
		\mpPar
		\playsrole{\roleQ[0]}{\mpP[{\roleQ[0]}]}
		\mpPar
		\playsrole{\roleQ[1]}{\mpP[{\roleQ[1]}]}
		\mpPar
		\playsrole{\roleP[2]}{\mpPi[{\roleP[2]}]}
		\mpPar
		\playsrole{\roleQ[2]}{\mpP[{\roleQ[2]}]}
		\\[1ex]
		\M''
		&=&
		\playsrole{\roleP[0]}{\mpP[{\roleP[0]}]}
		\mpPar
		\playsrole{\roleQ[0]}{\mpP[{\roleQ[0]}]}
		\mpPar
		\playsrole{\roleQ[1]}{\mpP[{\roleQ[1]}]}
		\mpPar
		\playsrole{\roleP[2]}{\mpPii[{\roleP[2]}]}
		\mpPar
		\playsrole{\roleQ[2]}{\mpNil}
		\end{array}
	\]
	\[
		\M
		\mpMoveTau
		\M'
		\mpMoveAnnot{\roleP[2]}{\roleQ[2]}{\stLab[0]}{}
		\M''
	\]
	For all $\roleR\in\{\roleP[0], \roleP[2], \roleQ[0],\roleQ[1],\roleQ[2]\}$,
	$\stJudge{}{\mpP[\roleR]}{\stT[\roleR]}$.
	Thus, $\tyJudge{}{\M}{\gtGcounter}$
	and
	$\tyJudge{}{\M'}{\gtGcounter}$.
	
	For $\gtGcounter\gtMoveStar\gtG$,
	we can compute the full coinductive projections of $\gtG$
	up-to bisimilarity.
	All the possible projections are subrelations of the full coinductive projection.
	We observe that the only $\gtG$ that can type $\M''$
	using the full coinductive projection is $\gtGcounter[1]$,
	which is not projectable by the
	coinductive projection with coinductive 1-level merge.
	Therefore, $\forall \gtGcounter\gtMoveStar\gtG.\;\neg(\tyJudge{}{\M''}{\gtG})$.
\end{proof}

We generalise the construction of $\gtGcounter$
to any failure of \inferrule{\mergeGoodCtx}.
	
For all $\stT$ and $\roleP\notin\gtRoles{\stT}$,
	let $\mathcal{G}\!\left(\stT,\roleP\right)$
	be a projectable balanced global type such that
	$\gtProj{\roleP}{\mathcal{G}\!\left(\stT,\roleP\right)}{\stT}$,
	where $\gtProj{}{}{}$
	is the inductive projection with
	the inductive disjoint merge.
	See \cite[Table 9]{GJPSY2018}
	for a full definition.

\begin{theorem}[Context Rule Failure]
\label{thm:failure-type}
	Let $\stMerge{}{}$
	be a merge operator
	containing
	the inductive disjoint merge and
	contained within
	the coinductive full merge.
	
	Suppose that
	for $i\in I$,
	$\stTi[i]\in\stMerge{}{\{\stT[i,j]\}_{j\in J}}$.
	If $\stMerge{}{
	\{
	\stTii[j]
	\}_{j\in J}
	}=\emptyset$
	for all $\stTii[j]\bisim \roleQ[0]\stFmt{{\dagger}\left\{\stChoice{\stLab[i]}{\tyGround[i]}\stSeq\stT[i,j]\right\}_{i\in I}}$,
	then the inductive and coinductive projections
	with $\stMerge{}{}$ have
	(1), (2), and (3)
	from \cref{thm:incorrect-merge}.
\end{theorem}

If we work over coinductive types, then the hypotheses
could be reformulated to
instead
require that
\inferrule{\mergeGoodCtx} fails.

\begin{proof}
	Let:
	\[
		\bigcup_{i\in I, j\in J}\gtRoles{\stT[i,j]}\setminus\{\roleQ[0]\}=\{\roleQ[1],\dots,\roleQ[n]\}
	\]
	Let $\roleP,\roleR[0],\roleR[1],\roleR[2],\roleR[3]\notin\{\roleQ[0],\dots,\roleQ[n]\}$
	be distinct.
	Let $\sigma$ be a bijection on $J$
	with order $|J|$.
	For $i\in I, j\in J$, let:
	\[
		\mathbb{G}_{i,j}
		=
		\roleR[0]\gtFmt{\to}\roleQ[0]\gtFmt{:}\gtLab[i,j]\gtSeq\gtFmt{\dots}
		\roleR[0]\gtFmt{\to}\roleQ[n]\gtFmt{:}\gtLab[i,j]\gtSeq
		\mathcal{G}\!\left(\stT[i,j],\roleP\right)
	\]
	Consider:
	\[
		\gtG[\oplus] =
			\gtComm{\roleR[0]}{\roleR[1]}{0\leq k \leq |J|-1}{\gtLab[k]}{}{
				\gtComm{\roleP}{\roleQ[0]}{i\in I}{\gtLab[i]}{\tyGround[i]}{
					\gtComm{\roleR[2]}{\roleR[3]}{j\in J}{\gtLab[j]}{}{
						\gtCommSingle{\roleR[2]}{\roleR[0]}{\gtLab[j]}{}{
						\gtCommSingle{\roleQ[0]}{\roleR[0]}{\gtLab[i]}{}{
							\mathbb{G}_{i,\sigma^k\!\left(j\right)}
						}}
					}
				}
			}
	\]
	\[
		\gtG[\&] =
			\gtComm{\roleR[0]}{\roleR[1]}{0\leq k \leq |J|-1}{\gtLab[k]}{}{
				\gtComm{\roleQ[0]}{\roleP}{i\in I}{\gtLab[i]}{\tyGround[i]}{
					\gtComm{\roleR[2]}{\roleR[3]}{j\in J}{\gtLab[j]}{}{
						\gtCommSingle{\roleR[2]}{\roleR[0]}{\gtLab[j]}{}{
						\gtCommSingle{\roleQ[0]}{\roleR[0]}{\gtLab[i]}{}{
							\mathbb{G}_{i,\sigma^k\!\left(j\right)}}
						}
					}
				}
			}
	\]
	Suppose that for all $i\in I$, $j\in J$, $0\leq k\leq n$,
	$\gtProj{\roleQ[k]}{\mathcal{G}\!\left(\stT[i,j],\roleP\right)}{\stT[{\roleQ[k]},i,j]}$.
	
	$\gtG[\dagger]$ is projectable:
	\[
	\begin{array}{rcl}
		\stT[{\roleQ[0]}]
		&=&
		\stFmt{
			\roleP{\overline{\dagger}}\!\left\{\stChoice{\stLab[i]}{\tyGround[i]}
			\stSeq
			\roleR[0]{\oplus}\stLab[i]\stSeq
			\roleR[0]{\&}\!\left\{\stLab[i,j]\stSeq \stT[{\roleQ[0]},i,j]\right\}_{j\in J}
			\right\}_{i\in I}
		}
		\\
		\stT[{\roleQ[k]}]
		&=&
		\stFmt{
			\roleR[0]{\&}\!\left\{\stLab[i,j]\stSeq \stT[{\roleQ[k]},i,j]\right\}_{i\in I,\,j\in J}
		}
		\\
		\stT[\roleP]
		&=&
		\stFmt{
			\roleQ[0]{{\dagger}}\!\left\{\stChoice{\stLab[i]}{\tyGround[i]}
			\stSeq
			\stTi[i]
			\right\}_{i\in I}
		}
		\\
		\stT[{\roleR[0]}]
		&=&
		\stFmt{
			\roleR[1]{\oplus}\!\left\{
				\stLab[k]\stSeq
				\roleR[2]{\&}\!\left\{
					\stLab[j]\stSeq
					\roleQ[0]{\&}\!\left\{
					\stLab[i]\stSeq
					
					\roleQ[0]{\oplus}\stLab[i,\sigma^k\!\left(j\right)]\stSeq\dots
					\roleQ[n]{\oplus}\stLab[i,\sigma^k\!\left(j\right)]
					
				\right\}_{i\in I}
				\right\}_{j\in J}
			\right\}_{0\leq k \leq |J|-1}
		}
		\\
		\stT[{\roleR[1]}]&=&
		\stFmt{
			\roleR[0]{\&}\!\left\{
				\stLab[k]\right\}_{0\leq k \leq |J|-1}
		}
		\\
		\stT[{\roleR[2]}]&=&\stFmt{
			\roleR[3]{\oplus}\!\left\{
				\stLab[j]\stSeq
				\roleR[0]{\oplus}\stLab[j]
				\right\}_{j\in J}}
		\\
		\stT[{\roleR[3]}]&=&\stFmt{
			\roleR[2]{\&}\!\left\{
				\stLab[j]\right\}_{j\in J}}
		\end{array}
	\]
	Since $\stMerge{}{}$ contains the
	inductive disjoint merge,
	we derive
	\[
	\forall \roleR\in\{\roleQ[0],\dots,\roleQ[n],\roleP,\roleR[0],\dots,\roleR[3]\}
	.\;
	\gtProj{\roleR}{\gtG[\dagger]}{\stT[\roleR]}
	\]
	
	\[
		\gtGi[j,\oplus] =
			\gtComm{\roleR[0]}{\roleR[1]}{0\leq k \leq |J|-1}{\gtLab[k]}{}{
				\gtComm{\roleP}{\roleQ[0]}{i\in I}{\gtLab[i]}{\tyGround[i]}{
						\gtCommSingle{\roleR[2]}{\roleR[0]}{\gtLab[j]}{}{
						\gtCommSingle{\roleQ[0]}{\roleR[0]}{\gtLab[i]}{}{
							\mathbb{G}_{i,\sigma^k\!\left(j\right)}
						}}
				}
			}
	\]
	\[
		\gtGi[j,\&] =
			\gtComm{\roleR[0]}{\roleR[1]}{0\leq k \leq |J|-1}{\gtLab[k]}{}{
				\gtComm{\roleQ[0]}{\roleP}{i\in I}{\gtLab[i]}{\tyGround[i]}{
						\gtCommSingle{\roleR[2]}{\roleR[0]}{\gtLab[j]}{}{
						\gtCommSingle{\roleQ[0]}{\roleR[0]}{\gtLab[i]}{}{
							\mathbb{G}_{i,\sigma^k\!\left(j\right)}}
						}
				}
			}
	\]
	Pick $j\in J$,
	\[
		\gtG[\dagger]\,\gtMove\,\gtGi[j,\dagger]
	\]
	$\gtGi[j,\dagger]$
	does not project onto $\roleP$.
	
	Suppose that $\dagger=\oplus$.
	If $\gtProj{\roleP}{\gtGi[j,\dagger]}{\stT}$,
	then
	$\stT\in\stMerge{}{\{\stTii[k]\}_{0\leq k\leq |J|-1}}$
	where
	\[\gtProj{\roleP}{
	\gtComm{\roleP}{\roleQ[0]}{i\in I}{\gtLab[i]}{\tyGround[i]}{
						\gtCommSingle{\roleR[2]}{\roleR[0]}{\gtLab[j]}{}{
						\gtCommSingle{\roleQ[0]}{\roleR[0]}{\gtLab[i]}{}{
							\mathbb{G}_{i,\sigma^k\!\left(j\right)}
						}}
				}
	}{\stTii[k]}\]
	Thus, this holds for the full coinductive projection,
	which is unique up-to bisimilarity on balanced types.
	Hence,
	$\stTii[k]\bisim
	\stFmt{
			\roleQ[0]{{\dagger}}\!\left\{\stChoice{\stLab[i]}{\tyGround[i]}
			\stSeq
			\stT[i,\sigma^k\!\left(j\right)]
			\right\}_{i\in I}
		}
	$.
	\[
		\{\sigma^k\!\left(j\right)\}_{0\leq k \leq |J|-1}
		=
		J
	\]
	Hence by assumption,
	$\stMerge{}{\{\stTii[k]\}_{0\leq k\leq |J|-1}}=\emptyset$,
	contradicting that $\stT\in\stMerge{}{\{\stTii[k]\}_{0\leq k\leq |J|-1}}$.
	
	The case of $\dagger = \&$ is similar.
	
	We derive (2) and (3) similarly to
	\cref{thm:incorrect-merge}.
\end{proof}

\subsection{Appendix for~\cref{sec:comparison} (Expressiveness and Separation Results)}

\begin{definition}[Characteristic Processes~\cite{GJPSY2018}]
\label{def:char-proc}
We define the characteristic map from types to processes
recursively.
We assume that recursive variables are never rebound, by $\alpha$-equivalence
(we replace local types with reused recursive variables by bisimilar
local types where we have applied $\alpha$-conversion).
For each basic type $\tyGround$,
let $\chi(\tyGround)$ be a value of type $\tyGround$, and let:
\[
\begin{array}{rcll}
\chi(\stEnd)&=&\mpNil&;\\
\chi(\stRecVar)&=&\mpX_{\stRecVar}&;\\
\chi(\stRec{\stRecVar}{\stT})&=&\mpRec{\mpX_{\stRecVar}}{\chi(\stT)}&;\\
\chi(\stExtSum{\roleP}{i\in I}{\stChoice{\stLab[i]}{\tyGround[i]}\stSeq\stT[i]})&=&
\textstyle\sum_{i\in I} \roleP\mathord{?}\ell_{i}(x_i)\mpSeq 
\mpIf{x_i=\chi(\tyGround[i])}{\chi(\stT[i])}{\chi(\stT[i])}&;\\
\chi(\stIntSum{\roleP}{i\in I}{\stChoice{\stLab[i]}{\tyGround[i]}\stSeq\stT[i]})&=&
\mpIf{\mpTrue}{
	\roleP\mathord{!}{\ell_j}\langle \chi(\tyGround[j]) \rangle \mpSeq \chi(\stT[j])
}{
	\chi(
		\stIntSum{\roleP}{i\in I\setminus\{j\}}{\stChoice{\stLab[i]}{\tyGround[i]}\stSeq\stT[i]}
	)
}
\\
&&\quad\text{if}\;j\in I\;\text{and}\;|I|>1&;\;\text{and}\\
\chi(\stIntSum{\roleP}{}{\stChoice{\stLab}{\tyGround}\stSeq\stT})&=&
\roleP\mathord{!}{\ell}\langle \chi(\tyGround) \rangle \mpSeq \chi(\stT).
\end{array}
\]
\end{definition}

\begin{lemma}[Typing Characteristic Processes]
	$\stJudge{\mpEnv\cup\{\mpX_{\stRecVar} : \stRecVar\}_{\stRecVar\in\gtFv{\stT}}}{\chi(\stT)}{\stTi}$ iff
	$\stT\tySub\stTi$
\end{lemma}
\begin{proof}
	Induction on the typing system.
\end{proof}

\thmSessionSeparation*

\begin{proof}[1]
	Let $\N=\Pi_{\roleP\in\dom{\stEnv}}{\playsrole{\roleP}{\chi(\stEnvApp{\stEnv}{\roleP})}}$,
	so that $\tyJudge{}{\N}{\stEnv}$.
	There is $\gtG$ such that $\stEnvAssoc{\gtG}{\stEnv}{}$,
	thus $\tyJudge{}{\N}{\gtG}$.
	
	If $\tyJudge{}{\N}{\gtGi}$, then
	there exists $\stEnvi\tySub\stEnvii$ such that
	$\gtProji{}{\gtGi}{\stEnvii}$ and $\stEnvi\tySub\stEnvii$ and
	$\tyJudge{}{\N}{\stEnvi}$.
	So, $\stEnv\tySub\stEnvi\tySub\stEnvii$.
	Thus, $\stEnv\in\projLocal{\gtProji{}{}{}}$.
	This contradicts our assumption.
\end{proof}

\begin{proof}[2]
	\[\begin{array}{cl}
	&\exists\gtG.\;\tyJudge{}{\Pi_{\roleP\in\roleSet}\playsrole{\roleP}{\mpP[\roleP]}}{\gtG}
	\\
	\iff&
	\exists\stEnv.\;\tyJudge{\Pi_{\roleP\in\roleSet}\playsrole{\roleP}{\mpP[\roleP]}}{\N}{\stEnv}\\
	\iff&
	\exists\stEnv.\left(
	\dom{\stEnv}=\roleSet
	\;\wedge\;
	\forall \roleP\in\roleSet.\;\stJudge{}{\mpP[\roleP]}{\stEnvApp{\stEnv}{\roleP}}
	\;\wedge\;
	\stEnv\in\projLocal{\gtProj{}{}{}}
	\right)
	\\
	\iff&
	\exists\stEnv.\left(
	\dom{\stEnv}=\roleSet
	\;\wedge\;
	\forall \roleP\in\roleSet.\;\stJudge{}{\mpP[\roleP]}{\stEnvApp{\stEnv}{\roleP}}
	\;\wedge\;
	\stEnv\in\projLocal{\gtProji{}{}{}}
	\right)
	\\
	\iff&
	\exists\stEnv.\;\tyJudgei{\Pi_{\roleP\in\roleSet}\playsrole{\roleP}{\mpP[\roleP]}}{\N}{\stEnv}
	\\
	\iff&
	\exists\gtG.\;\tyJudge{}{\Pi_{\roleP\in\roleSet}\playsrole{\roleP}{\mpP[\roleP]}}{\gtG}
	\end{array}\]
\end{proof}

\begin{lemma}[Maximal Safe Context]
	\label{lem:max-safe}
	Let $\gtProj{}{}{}$ be the plain coinductive projection restricted to balanced types.
	If $\gtProj{}{\gtG}{\stEnv}$ and $\stEnv\tySub\stEnvi$ and $\stEnvi$ is safe,
	then $\stEnvApp{\stEnv}{\roleP}\bisim\stEnvApp{\stEnvi}{\roleP}$
	for all $\roleP\in\dom{\stEnv}$.
\end{lemma}

\begin{proof}
	Let $\stTi \Re \stT$ iff
	there exists $\gtG$, $\stEnv$, $\stEnvi$, and $\roleP$ s.t.
	$\gtProj{}{\gtG}{\stEnv}$, $\stEnv\tySub\stEnvi$,
	$\stEnvApp{\stEnv}{\roleP}=\stT$, and
	$\stEnvApp{\stEnvi}{\roleP}=\stTi$.

	We show that $\Re$ satisfies the rules for $\tySub$,
	hence $\Re\subseteq\tySub$.

	Clearly if $\stTi \Re \stT$ then:
	$\unfoldOne{\stTi} = \stEnd$ iff
	$\unfoldOne{\stT} = \stEnd$;
	if $\stTi=\stRec{\stRecVar}{\stTii}$ then
	$\stTii{}[\stTi/\stRecVar] \Re \stT$ by $\tySub$; and
	if $\stT=\stRec{\stRecVar}{\stTii}$ then
	$\stTi \Re \stTii{}[\stT/\stRecVar]$ by~\cref{prop:proj-inv-local}.

	Suppose that $\gtProj{}{\gtG}{\stEnv}$ and $\stEnv\tySub\stEnvi$ and $\stEnvi$ is safe.
	
	By replacing $\tySub$ with $\bisim$ in the proof
	that the plain coinductive projection is good,
	we have that:
	if $\gtG\,\gtMove[\stEnvAnnotGenericSym]\,\gtGi$, then there exists
	$\stEnvii$ s.t. $\stEnv\stEnvMoveGenAnnot\stEnvii$ and
	$\gtProj{}{\gtGi}{\stEnvii}$; and
	 if $\stEnv\stEnvMoveGenAnnot\stEnvii$, then there exists
	$\gtGi$ s.t. $\gtG\,\gtMove[\stEnvAnnotGenericSym]\,\gtGi$ and
	$\gtProj{}{\gtGi}{\stEnvii}$.
	
	Also, if $\stEnv\stEnvMoveGenAnnot\stEnv[1]$ then, by $\tySub$ at
	$\ltsSubject{\stEnvAnnotGenericSym}$ and safety,
	$\stEnvi\stEnvMoveGenAnnot\stEnvi[1]$ with $\stEnv[1]\tySub\stEnvi[1]$.

	Let $\stT=\stEnvApp{\stEnv}{\roleP}$
	and $\stTi=\stEnvApp{\stEnvi}{\roleP}$.
	
	Suppose that $\unfoldOne{\stTi}=\stExtSum{\roleQ}{i\in I}{
		\stChoice{\stLab[i]}{\tyGround[i]}\stSeq\stTi[i]
	}$ and so
	$\unfoldOne{\stT}=\stExtSum{\roleQ}{i\in I\cup J}{
		\stChoice{\stLab[i]}{\tyGround[i]}\stSeq\stT[i]
	}$ with $\stT[i]\tySub\stTi[i]$ for $i\in I$.
	By transitioning globally towards the shallowest
	occurrence of $\roleP$ in $\gtG$,
	there are $\stEnv[1]$ and $\gtG[1]$ s.t.
	$\stEnv\stEnvMoveStar\stEnv[1]$ without involving $\roleP$
	(so $\stT=\stEnvApp{\stEnv[1]}{\roleP}$),
	$\gtProj{}{\gtG[1]}{\stEnv[1]}$, and
	$\roleP$ is at the head of $\gtG[1]$.
	So by the definition of $\gtProj{}{}{}$,
	for $i\in I\cup J$ there exists $\gtG[1,i]$ s.t.
	$\gtG[1]\,\gtMove[\stEnvCommAnnotSmall{\roleQ}{\roleP}{\stChoice{\stLab[i]}{\tyGround[i]}}]\,\gtG[1,i]$
	so there exists $\stEnv[1,i]$ s.t.
	$\stEnv[1] \stEnvMoveAnnot{ \stEnvCommAnnotSmall{\roleQ}{\roleP}{\stChoice{\stLab[i]}{\tyGround[i]}} }\stEnv[1,i]$ and
	$\gtProj{}{\gtG[1,i]}{\stEnv[1,i]}$.
	So there exist $\stEnvi[1]$ and $\stEnvi[1,i]$ for $i\in I\cup J$ s.t.
	$\stEnvi\stEnvMoveStar\stEnvi[1]$ without involving $\roleP$
	(so $\stTi=\stEnvApp{\stEnvi[1]}{\roleP}$),
	$\stEnv[1] \stEnvMoveAnnot{ \stEnvCommAnnotSmall{\roleQ}{\roleP}{\stChoice{\stLab[i]}{\tyGround[i]}} }\stEnv[1,i]$,
	$\stEnv[1]\tySub\stEnvi[1]$, and
	$\stEnv[1,i]\tySub\stEnvi[1,i]$.
	Therefore, $I=I\cup J$ and $\stTi[i]\Re\stT[i]$ for $i\in I$.
	
	Suppose that $\unfoldOne{\stT}=\stIntSum{\roleQ}{i\in I}{
		\stChoice{\stLab[i]}{\tyGround[i]}\stSeq\stT[i]
	}$ and so $\unfoldOne{\stTi}=\stIntSum{\roleQ}{i\in I\cup J}{
		\stChoice{\stLab[i]}{\tyGround[i]}\stSeq\stTi[i]
	}$ with $\stT[i]\tySub\stTi[i]$ for $i\in I$.
	By transitioning globally towards the shallowest
	occurrence of $\roleP$ in $\gtG$,
	there are $\stEnv[1]$ and $\gtG[1]$ s.t.
	$\stEnv\stEnvMoveStar\stEnv[1]$ without involving $\roleP$
	(so $\stT=\stEnvApp{\stEnv[1]}{\roleP}$),
	$\gtProj{}{\gtG[1]}{\stEnv[1]}$, and
	$\roleP$ is at the head of $\gtG[1]$.
	So by the definition of $\gtProj{}{}{}$,
	for $i\in I$ there exists $\gtG[1,i]$ s.t.
	$\gtG[1]\,\gtMove[ \stEnvCommAnnotSmall{\roleP}{\roleQ}{\stChoice{\stLab[i]}{\tyGround[i]}} ]\,\gtG[1,i]$
	so there exists $\stEnv[1,i]$ s.t.
	$\stEnv[1] \stEnvMoveAnnot{ \stEnvCommAnnotSmall{\roleP}{\roleQ}{\stChoice{\stLab[i]}{\tyGround[i]}} }\stEnv[1,i]$ and
	$\gtProj{}{\gtG[1,i]}{\stEnv[1,i]}$.
	So there exist $\stEnvi[1]$ and $\stEnvi[1,i]$ for $i\in I$ s.t.
	$\stEnvi\stEnvMoveStar\stEnvi[1]$ without involving $\roleP$
	(so $\stTi=\stEnvApp{\stEnvi[1]}{\roleP}$),
	$\stEnv[1] \stEnvMoveAnnot{ \stEnvCommAnnotSmall{\roleQ}{\roleP}{\stChoice{\stLab[i]}{\tyGround[i]}} }\stEnv[1,i]$,
	$\stEnv[1]\tySub\stEnvi[1]$, and
	$\stEnv[1,i]\tySub\stEnvi[1,i]$.
	By the definition of $\gtProj{}{}{}$,
	$\unfoldOne{\stEnvApp{\stEnv[1]}{\roleQ}}=\stExtSum{\roleQ}{i\in I}{
		\stChoice{\stLab[i]}{\tyGround[i]}\stSeq\stTii[i]
	}$ as $\roleP$ and $\roleQ$ are communicating at the head of $\gtG[1]$,
	so, by $\tySub$,
	$\unfoldOne{\stEnvApp{\stEnvi[1]}{\roleQ}}=\stExtSum{\roleQ}{i\in I'}{
		\stChoice{\stLab[i]}{\tyGround[i]}\stSeq\stTiii[i]
	}$ with $I'\subseteq I$.
	$\stEnvi[1]$ is safe, so $I'=I\cup J=I$
	and $\stTi[i]\Re\stT[i]$ for $i\in I$.
	
	We have that $\Re$ satisfies the rules for $\tySub$, so that $\Re\subseteq\tySub$.
	
	Therefore, if $\gtProj{}{\gtG}{\stEnv}$ and $\stEnv\tySub\stEnvi$ and $\stEnvi$ is safe,
	then $\stEnvApp{\stEnv}{\roleP}\tySub\stEnvApp{\stEnvi}{\roleP}$
	and $\stEnvApp{\stEnvi}{\roleP}\tySub\stEnvApp{\stEnv}{\roleP}$ so
	$\stEnvApp{\stEnv}{\roleP}\bisim\stEnvApp{\stEnvi}{\roleP}$
	for all $\roleP\in\dom{\stEnv}$.
\end{proof}

\thmSeparatingProjections*

\begin{lemma}[Global Traces and Local Traces]
	\label{lem:globallocaltrace}
	Let $\roleP\in\roleSet$.
	We define the function $f_{\roleP}$ from
	sequences of labels to sequences of labels by recursion:
	$f_{\roleP}(\langle\rangle)=\langle\rangle$;
	$f_{\roleP}(\ltsSendRecv{\roleP}{\roleQ}{\stChoice{\stLab}{\tyGround}}:\vec{\stEnvAnnotGenericSym})=
	{\roleQ\stEnvAnnotOutSym\stChoice{\stLab}{\tyGround}}:f_{\roleP}(\vec{\stEnvAnnotGenericSym})$;
	$f_{\roleP}(\ltsSendRecv{\roleQ}{\roleP}{\stChoice{\stLab}{\tyGround}}:\vec{\stEnvAnnotGenericSym})=
	{\roleQ\stEnvAnnotInSym\stChoice{\stLab}{\tyGround}}:f_{\roleP}(\vec{\stEnvAnnotGenericSym})$;
	if $\roleP\not\in\{\rolePi,\roleQi\}$ then
	$f_{\roleP}(\ltsSendRecv{\rolePi}{\roleQi}{\stChoice{\stLab}{\tyGround}}:\vec{\stEnvAnnotGenericSym})=
	f_{\roleP}(\vec{\stEnvAnnotGenericSym})$.
	
	Let $\gtProj{}{}{}$ be the full coinductive projection and
	$\gtG$ be balanced.
	If $\gtProj{\roleP}{\gtG}{\stT}$, then
	$\gtTrace{\stT}=\{f_{\roleP}(\vec{\stEnvAnnotGenericSym})\suchthat
	\vec{\stEnvAnnotGenericSym}\in\gtTrace{\gtG}\}$.
\end{lemma}
\begin{proof}
	Let $\vec{\stEnvAnnotGenericSym}\in\gtTrace{\gtG}$.
	By induction on the length of $\vec{\stEnvAnnotGenericSym}$,
	$\stT\,\gtMove[f_{\roleP}(\vec{\stEnvAnnotGenericSym})]$,
	as $\gtProj{}{}{}$ satisfies the good projection criteria
	even when the global type is not projectable.
	\\[2ex]
	If $\stT\,\gtMove[\stEnvAnnotGenericSymi]$, then,
	$\unfoldOne{\stT}\neq\stEnd$ so
	$\roleP\in\gtRoles{\gtG}$ so
	$\gtDepth{\gtG}{\roleP}$ exists.
	Induct on $\gtDepth{\gtG}{\roleP}$
	using~\cref{def:good-merge} and~\cref{def:global-proj}
	to find $\gtGi$ and $\stTi$ such that $\gtG\,\gtMoveStar\,\gtGi$ (by head reductions
	so excluding $\roleP$),
	$\gtProj{\roleP}{\gtGi}{\stTi}$,
	$\stT\tySSub\stTi$,
	$\stTi\,\gtMove[\stEnvAnnotGenericSymi]$, and
	$\gtDepth{\gtGi}{\roleP}=1$.
	Therefore, $\roleP$ appears at the head of $\gtGi$
	and, by~\cref{def:global-proj},
	it must induce an action $\stEnvAnnotGenericSymii$ s.t.
	$f(\langle\stEnvAnnotGenericSymii\rangle)=
	\langle\stEnvAnnotGenericSymi\rangle$.
	Hence we have $\vec{\stEnvAnnotGenericSym}\in\gtTrace{\gtG}$ s.t.
	$f(\vec{\stEnvAnnotGenericSym})=\langle\stEnvAnnotGenericSymi\rangle$.
	So if $\stT\,\gtMove[\vec{\stEnvAnnotGenericSymi}]$, then,
	by induction on the length of $\vec{\stEnvAnnotGenericSymi}$,
	there is $\vec{\stEnvAnnotGenericSym}\in\gtTrace{\stEnvNew}$
	such that $f_{\roleP}(\vec{\stEnvAnnotGenericSym})=\vec{\stEnvAnnotGenericSymi}$.
\end{proof}

\begin{lemma}[Context Traces and Local Traces]
	\label{lem:contextlocaltrace}
	Let $\roleP\in\roleSet$.
	We define the function $f_{\roleP}$ from
	sequences of labels to sequences of labels by recursion:
	$f_{\roleP}(\langle\rangle)=\langle\rangle$;
	$f_{\roleP}(\ltsSendRecv{\roleP}{\roleQ}{\stChoice{\stLab}{\tyGround}}:\vec{\stEnvAnnotGenericSym})=
	{\roleQ\stEnvAnnotOutSym\stChoice{\stLab}{\tyGround}}:f_{\roleP}(\vec{\stEnvAnnotGenericSym})$;
	$f_{\roleP}(\ltsSendRecv{\roleQ}{\roleP}{\stChoice{\stLab}{\tyGround}}:\vec{\stEnvAnnotGenericSym})=
	{\roleQ\stEnvAnnotInSym\stChoice{\stLab}{\tyGround}}:f_{\roleP}(\vec{\stEnvAnnotGenericSym})$;
	if $\roleP\not\in\{\rolePi,\roleQi\}$ then
	$f_{\roleP}(\ltsSendRecv{\rolePi}{\roleQi}{\stChoice{\stLab}{\tyGround}}:\vec{\stEnvAnnotGenericSym})=
	f_{\roleP}(\vec{\stEnvAnnotGenericSym})$.
	
	Let $\stEnv$ be safe, live, and $\roleP\in\dom{\stEnv}$, then
	$\stEnvApp{\stEnv}{\roleP}\tySSub\stT$ for $\stT$ s.t.
	$\gtTrace{\stT}=\{f_{\roleP}(\vec{\stEnvAnnotGenericSym})\suchthat
	\vec{\stEnvAnnotGenericSym}\in\gtTrace{\stEnv}\}$.
\end{lemma}

\begin{proof}
	Clearly, if $\stT\,\gtMove[\vec{\stEnvAnnotGenericSym}]$
	then $\stTi\,\gtMove[\vec{\stEnvAnnotGenericSym}]$.
	
	Suppose that $\stTi\,\gtMove[\vec{\stEnvAnnotGenericSym}]\,\stTi[1]$
	and $\stT\,\gtMove[\vec{\stEnvAnnotGenericSym}]\,\stT[1]$.
	By liveness $\stTi[1]\bisim\stEnd$ iff $\stT[1]\bisim\stEnd$.
	If $\stT[1]\,\gtMove[\stEnvAnnotGenericSym]$ then $\stTi[1]\,\gtMove[\stEnvAnnotGenericSym]$.
	By liveness and safety, if $\stTi[1]\,\gtMove[{\roleQ\stEnvAnnotOutSym\stChoice{\stLab}{\tyGround}}]$, then $\stT[1]\gtMove[{\roleQ\stEnvAnnotOutSym\stChoice{\stLab}{\tyGround}}]$.
	By liveness, if $\stTi[1]\,\gtMove[{\roleQ\stEnvAnnotInSym\stChoice{\stLab}{\tyGround}}]$, then
	there exist $\stLabi$ and $\tyGroundi$ s.t. $\stT[1]\,\gtMove[{\roleQ\stEnvAnnotInSym\stChoice{\stLabi}{\tyGroundi}}]$.
	
	Therefore, $\stTi\tySSub\stT$ by~\cref{lem:sub-op-cor}.
\end{proof}

\begin{restatable}[Plain Projections Trace Invariant]{theorem}{thmProjPlainInv}
	\label{thm:plain_proj_inv}
	Suppose that $\gtProj{\roleP}{\gtG}{\stExtSum{\roleQ}{i \in I}{\stChoice{\stLab[i]}{\tyGround[i]} \stSeq \stT[i]}}$
	by the plain coinductive or plain inductive projections,
	$\gtG$ is balanced,
	and
	there exist $\stEnvAnnotGenericSym[1],\dots,\stEnvAnnotGenericSym[n],\stEnvAnnotGenericSym$
	such that $\roleP\in\ltsSubject{\stEnvAnnotGenericSym}$,
	$\roleP\not\in\ltsSubject{\stEnvAnnotGenericSym[i]}$ for $1\leq i\leq n$
	and $\gtG\,\gtMove[{\stEnvAnnotGenericSym[1],\dots,\stEnvAnnotGenericSym[n],\stEnvAnnotGenericSym}]$.
	Then $\gtG\,\gtMove[{\stEnvAnnotGenericSym[1],\dots,\stEnvAnnotGenericSym[n],\stEnvAnnotGenericSymi}]$
	for all $\stEnvAnnotGenericSymi$ of the form
	$\stEnvAnnotGenericSymi=
	\stEnvCommAnnotSmall{\roleQ}{\roleP}{\stChoice{\stLab[i]}{\tyGround[i]}}$.
	\newline
	The trace of $\gtG[\neg\text{\tiny plain}]$ does not satisfy this invariant,
	so cannot be expressed by the plain coinductive projection.
\end{restatable}

\begin{proof}
	Let $\gtProj{}{}{}$ be the plain coinductive projection
	and $\gtProj{\roleP}{\gtG}{\stExtSum{\roleQ}{i \in I}{\stChoice{\stLab[i]}{\tyGround[i]} \stSeq \stT[i]}}$.

	Suppose that
	$\gtG\,\gtMove[{\stEnvAnnotGenericSym[1],\dots,\stEnvAnnotGenericSym[n]}]\,\gtGi$
	with $\roleP\not\in\ltsSubject{\stEnvAnnotGenericSym[i]}$ for $1\leq i\leq n$.

	By replacing $\tySub$ by $\bisim$ in the proof that the plain coinductive projection is good
	and applying \cref{prop:proj-inv-local},
	we get that $\gtProj{\roleP}{\gtGi}{\stExtSum{\roleQ}{i \in I}{\stChoice{\stLab[i]}{\tyGround[i]} \stSeq \stT[i]}}$ by \inferrule{$\projGoodAlt$} and induction on $n$.

	If $\gtGi\gtMove[\stEnvAnnotGenericSym]$ with $\roleP\in\stEnvAnnotGenericSym$,
	then
	we induct on $\gtMove[\stEnvAnnotGenericSym]$ to show that
	if $\gtGi\,\gtMove[\stEnvAnnotGenericSym]$
	and $\gtProj{\roleP}{\gtGi}{\stExtSum{\roleQ}{i \in I}{\stChoice{\stLab[i]}{\tyGround[i]} \stSeq \stT[i]}}$
	then $\gtGi\,\gtMove[\stEnvAnnotGenericSymi]$
	for $1\leq j\leq n+1$ and for all $\stEnvAnnotGenericSymi$ of the form
	$\stEnvAnnotGenericSymi=
	\stEnvCommAnnotSmall{\roleQ}{\roleP}{\stChoice{\stLab[i]}{\tyGround[i]}}$.

	The base case is when
	$\gtGi=\gtComm{\roleQ}{\roleP}{j\in J}{\stLabi[j]}{\tyGroundi[j]}{\gtGi[j]}$,
	from which we deduce that
	\\$\gtGi=\gtComm{\roleQ}{\roleP}{i\in I}{\stLab[i]}{\tyGround[i]}{\gtGii[i]}$ by \inferrule{$\projRuleIn$}.
\end{proof}

\begin{definition}[Parallel Context]
	We define parallel composition contexts by the following grammar:
	\\\centerline{\(
		\mathbb{C}\bnfdef [\cdot] \bnfsep \gtPar{\mathbb{C}}{\gtG}
		\bnfsep \gtPar{\gtG}{\mathbb{C}}
		\bnfsep \gtRec{\gtRecVar}{\mathbb{C}}
	\)}
	Given a global type $\gtG$,
	$\mathbb{C}[\gtG]$ is the global type formed by
	filling the hole $[\cdot]$ with $\gtG$ (if it is well-formed).
\end{definition}

\begin{lemma}[Parallel Context]
	If $\gtG\not\bisim\gtEnd$, then
	$\unfoldOne{\gtG}=\mathbb{C}[\gtGi]$
	where\\
	$\unfoldOne{\gtGi}=\gtComm{\roleP}{\roleQ}{i\in I}{\gtLab[i]}{\tyGround[i]}{\gtG[i]}$ and
	($\gtG\,\gtMove[\ltsSendRecv{\roleP}{\roleQ}{\stChoice{\stLab}{\tyGround}}]$
	iff $(\stLab,\tyGround)=(\stLab[i],\tyGround[i])$ for some $i\in I$).
\end{lemma}

\begin{proof}
	$\gtG\neq \gtEnd$ so $\gtG\,\gtMove[\stEnvAnnotGenericSym]$
	for some $\stEnvAnnotGenericSym$.
	The result is by induction on the derivation of
	$\gtG\,\gtMove[\stEnvAnnotGenericSym]$
	until reaching a communication.
\end{proof}

\begin{theorem}[Inductive Projections without Parallel Composition are Less Expressive]
	\label{thm:par-nec}
	$\gtG[\text{\tiny par}]$
	can be expressed by the plain inductive projection with parallel composition in the global types and cannot be expressed by the full inductive projection without parallel composition.
\end{theorem}
\begin{proof}
	The full inductive projection is a partial function,
	so we write $\gtProj{\roleP}{\gtG}{}$
	for the unique type $\stT$ such that $\gtProj{\roleP}{\gtG}{\stT}$
	if it exists.
	\begin{equation*}
		\gtG = \gtPar
		{
			\gtRec{\gtRecVar}
			{
				\gtCommRaw{\roleFmt{a}}{\roleFmt{b}}
				{
					\begin{array}{l}%
						\gtCommChoice{first}{}{\gtRecVar}\\
						\gtCommChoice{second}{}{}
					\end{array}
				}
			}
		}
		{
			\gtRec{\gtRecVar}
			{
				\gtCommRaw{\roleFmt{c}}{\roleFmt{d}}
				{
					\begin{array}{l}%
						\gtCommChoice{first}{}{\gtRecVar}\\
						\gtCommChoice{second}{}{}
					\end{array}
				}
			}
		}
	\end{equation*}
	$\gtG$ is projectable onto the local types:
\begin{equation*}
	\stT[\roleFmt{a}] = \stRec{\stRecVar}
	{
		\stIntSum{\roleFmt{b}}{}
		{
			\begin{array}{@{\hskip 0mm}l@{\hskip 0mm}}
				\stChoice{\stLabFmt{first}}{} \stSeq \stRecVar\\
				\stChoice{\stLabFmt{second}}{}
			\end{array}
		}
	}\quad
	\stT[\roleFmt{b}] = \stRec{\stRecVar}
	{
		\stExtSum{\roleFmt{a}}{}
		{
			\begin{array}{@{\hskip 0mm}l@{\hskip 0mm}}
				\stChoice{\stLabFmt{first}}{} \stSeq \stRecVar\\
				\stChoice{\stLabFmt{second}}{}
			\end{array}
		}
	}\quad
	\stT[\roleFmt{c}] = \stRec{\stRecVar}
	{
		\stIntSum{\roleFmt{d}}{}
		{
			\begin{array}{@{\hskip 0mm}l@{\hskip 0mm}}
				\stChoice{\stLabFmt{first}}{} \stSeq \stRecVar\\
				\stChoice{\stLabFmt{second}}{}
			\end{array}
		}
	}\quad
	\stT[\roleFmt{d}] = \stRec{\stRecVar}
	{
		\stExtSum{\roleFmt{c}}{}
		{
			\begin{array}{@{\hskip 0mm}l@{\hskip 0mm}}
				\stChoice{\stLabFmt{first}}{} \stSeq \stRecVar\\
				\stChoice{\stLabFmt{second}}{}
			\end{array}
		}
	}
\end{equation*}
Suppose that $\gtGi$ is an equivalent global type, without parallelism, that is projectable.

We induct on the complexity of $\gtGi$, that there is projectable $\gtRec{\gtRecVari}{\gtGii}\bisim\gtG$ without parallelism such that
$\gtRecVari\in\gtFv{\gtGii}$.
$\gtGi\bisim\gtG$, so $\gtGi\neq \gtEnd$.
If $\gtGi=\gtRec{\gtRecVari}{\gtGii}$ and $\gtRecVari\in\gtFv{\gtGii}$, then we are done.
If $\gtGi=\gtRec{\gtRecVari}{\gtGii}$ and $\gtRecVari\not\in\gtFv{\gtGii}$, then $\gtGi\bisim\gtGii$ so $\gtGii\bisim\gtG$
and apply the I.H.
If $\gtGi=\gtComm{\roleP}{\roleQ}{i \in I}{\gtLab[i]}{\tyGround[i]}{\gtG[i]}$, then
as $\gtGi\bisim\gtG$,
\[
\gtGi=\gtCommRaw{\roleFmt{a}}{\roleFmt{b}}
				{
					\begin{array}{l}%
						\gtCommChoice{first}{}{\gtG[A]}\\
						\gtCommChoice{second}{}{\gtG[B]}
					\end{array}
				}
\qquad\text{or}\qquad
\gtGi=\gtCommRaw{\roleFmt{c}}{\roleFmt{d}}
				{
					\begin{array}{l}%
						\gtCommChoice{first}{}{\gtG[C]}\\
						\gtCommChoice{second}{}{\gtG[D]}
					\end{array}
				}
\]
By symmetry of the situation between $(\roleFmt{a},\roleFmt{b})$ and $(\roleFmt{c},\roleFmt{d})$,
we may assume that
\[\gtGi=\gtCommRaw{\roleFmt{a}}{\roleFmt{b}}
				{
					\begin{array}{l}%
						\gtCommChoice{first}{}{\gtG[A]}\\
						\gtCommChoice{second}{}{\gtG[B]}
					\end{array}
				}\]
Now, $\gtGi\,\gtMove[\ltsSendRecv{\roleFmt{a}}{\roleFmt{b}}{\stLabFmt{first}}]\,\gtG[A]$ and
$\gtG\,\gtMove[\ltsSendRecv{\roleFmt{a}}{\roleFmt{b}}{\stLabFmt{first}}]\,\gtG$,
so $\gtG[A]\bisim\gtG$ and apply the I.H.

Therefore, we may assume that we have $\gtRec{\gtRecVari}{\gtGi}\bisim\gtG$ such that
$\gtRecVari\in\gtFv{\gtGi}$ and $\gtGi$ has no parallel composition.

If $\gtGi=\gtRec{\gtRecVarii}{\gtGii}$, then 
$\gtRec{\gtRecVari}{\unfoldOne{\gtGi}}\bisim\gtG$,
$\gtRecVari\in\gtFv{\unfoldOne{\gtGi}}$,
$\unfoldOne{\gtGi}$ has no parallel composition, and
$\unfoldOne{\gtGi}$ is projectable.
Therefore, we additionally assume that $\gtGi$ begins with a transmission.

Since $\gtRec{\gtRecVari}{\gtGi}\bisim\gtG$,
\[
\gtGi=\gtCommRaw{\roleFmt{a}}{\roleFmt{b}}
				{
					\begin{array}{l}%
						\gtCommChoice{first}{}{\gtG[A]}\\
						\gtCommChoice{second}{}{\gtG[B]}
					\end{array}
				}
\quad\text{or}\quad
\gtGi=\gtCommRaw{\roleFmt{c}}{\roleFmt{d}}
				{
					\begin{array}{l}%
						\gtCommChoice{first}{}{\gtG[C]}\\
						\gtCommChoice{second}{}{\gtG[D]}
					\end{array}
				}
\]
By symmetry of the situation between $(\roleFmt{a},\roleFmt{b})$ and $(\roleFmt{c},\roleFmt{d})$,
we may assume that
\[
\gtGi=\gtCommRaw{\roleFmt{a}}{\roleFmt{b}}
				{
					\begin{array}{l}%
						\gtCommChoice{first}{}{\gtG[A]}\\
						\gtCommChoice{second}{}{\gtG[B]}
					\end{array}
				}
\]
We know that
$\gtRec{\gtRecVari}{\gtGi}\bisim\gtG$, so
\[\gtG[B][\gtGi/\gtRecVari]\bisim \gtRec{\gtRecVar}
			{
				\gtCommRaw{\roleFmt{c}}{\roleFmt{d}}
				{
					\begin{array}{l}%
						\gtCommChoice{first}{}{\gtRecVar}\\
						\gtCommChoice{second}{}{}
					\end{array}
				}
			}\text{ with }
\roleFmt{a}\not\in\gtRoles{
\gtRec{\gtRecVar}
			{
				\gtCommRaw{\roleFmt{c}}{\roleFmt{d}}
				{
					\begin{array}{l}%
						\gtCommChoice{first}{}{\gtRecVar}\\
						\gtCommChoice{second}{}{}
					\end{array}
				}
			}
}\]
So $\roleFmt{a}\not\in\gtRoles{\gtG[B][\gtGi/\gtRecVari]}$,
so $\gtRecVari\not\in\gtFv{\gtG[B]}$ and $\gtRecVari\in\gtFv{\gtG[A]}$.
$\gtRecVari\in\gtFv{\gtG[A]}$ so $\stRecVari\in\gtFv{\gtProj{\roleFmt{c}}{\gtG[A]}{}}$.
$\gtRecVari\not\in\gtFv{\gtG[B]}$ so $\stRecVari\not\in\gtFv{\gtProj{\roleFmt{c}}{\gtG[B]}{}}$.

Now, $\roleFmt{c}$ has only sending behaviour in $\gtG$, so in $\gtGi$, so in $\gtProj{\roleFmt{c}}{\gtGi}{}$
(as inductive projection is good).
\[\gtProj{\roleFmt{c}}{\gtGi}{}=\stMerge{}{ \left\{ \gtProj{\roleFmt{c}}{\gtG[A]}{}
,\;
\gtProj{\roleFmt{c}}{\gtG[B]}{}\right\}}\]
A local type, with only sends, is only mergeable with itself and
$\gtProj{\roleFmt{c}}{\gtG[A]}{}\neq\gtProj{\roleFmt{c}}{\gtG[B]}{}$ as their free variables differ.

Therefore, $\stMerge{}{ \left\{ \gtProj{\roleFmt{c}}{\gtG[A]}{}, \gtProj{\roleFmt{c}}{\gtG[B]}{}\right\}}$ does not exist,
contradicting the existence of a projectable $\gtGi\bisim \gtG$ without parallelism.
\end{proof}

\begin{restatable}[Inductive Projection with Parallel is Incomplete]{theorem}{thmIndPlainIncomplete}
\label{thm:indplain-incomp}
$\gtG[\text{\tiny co}]$ is projectable by
the plain coinductive projection but
there is no equivalent global type that is
projectable by the full inductive projection.
\end{restatable}

\begin{proof}
Let $\gtG$ be from~\cref{thm:par-nec}.
Consider the global type:
\begin{equation*}
\gtGi=
\gtRec{\gtRecVar}
{
    \gtCommRaw{\roleFmt{a}}{\roleFmt{b}}
    {
    \begin{array}{l}
        \gtCommChoice{first}{}
        {
            \gtCommRaw{\roleFmt{c}}{\roleFmt{d}}
            {
            \begin{array}{l}
                \gtCommChoice{first}{}
                {
                    \gtCommSingle{\roleFmt{a}}{\roleFmt{c}}{m}{}
                    {
                        \gtRecVar
                    }
                }
                \\
                \gtCommChoice{second}{}
                {
                    \gtRec{\gtRecVar}
                    {
                        \gtCommSingle{\roleFmt{a}}{\roleFmt{c}}{m}{}
                        {
                            \gtCommRaw{\roleFmt{a}}{\roleFmt{b}}
                            {
                            \begin{array}{l}
                                \gtCommChoice{first}{}
                                {
                                    \gtRecVar
                                }
                                \\
                                \gtCommChoice{second}{}
                                {
                                    \gtRec{\gtRecVar}
                                    {
                                        \gtCommSingle{\roleFmt{a}}{\roleFmt{c}}{m}{}
                                        {
                                            \gtRecVar
                                        }
                                    }
                                }
                            \end{array}
                            }
                        }
                    }
                }
            \end{array}
            }
        }
        \\
        \gtCommChoice{second}{}
        {
            \gtRec{\gtRecVar}
            {
                \gtCommRaw{\roleFmt{c}}{\roleFmt{d}}
                {
                \begin{array}{l}
                    \gtCommChoice{first}{}
                    {
                        \gtCommSingle{\roleFmt{a}}{\roleFmt{c}}{m}{}
                        {
                            \gtRecVar
                        }
                    }
                    \\
                    \gtCommChoice{second}{}
                    {
                        \gtRec{\gtRecVar}
                        {
                            \gtCommSingle{\roleFmt{a}}{\roleFmt{c}}{m}{}
                            {
                                \gtRecVar
                            }
                        }
                    }
                \end{array}
                }
            }
        }
    \end{array}
    }
}
\end{equation*}
Observe that for all $\gtGi\,\gtMoveStar\,\gtGii$:
$\gtGii$ contains a communication between $\roleFmt{a}$ and $\roleFmt{b}$;
if $\roleFmt{b}\in\gtRoles{\gtGii}$ then $\gtGii$ contains a communication between $\roleFmt{a}$ and $\roleFmt{b}$; and
if $\roleFmt{d}\in\gtRoles{\gtGii}$ then $\gtGii$ contains a communication between $\roleFmt{c}$ and $\roleFmt{d}$.
Since parallel composition requires the participants of each components to be disjoint, and
as $\gtPar{\gtGiii}{\gtEnd}$, $\gtPar{\gtEnd}{\gtGiii}$, and ${\gtGiii}$ are indistinguishable
by the full inductive projection,
if $\gtGi\bisim\gtGii$, with $\gtGii$ projectable, then
we can assume that $\gtGii$ contains no parallel composition.
We can now reduce the existence of $\gtGii$ to a counterexample to~\cref{thm:par-nec},
hence no such $\gtGii$ exists.
\newline
Define a function on global types and local types by:
\begin{itemize}
    \item $F(\gtCommSingle{\roleFmt{a}}{\roleFmt{c}}{m}{}{\gtG})=F(\gtG)$
    \item $F(\gtComm{\roleP}{\roleQ}{i\in I}{\gtLab[i]}{\tyGround[i]}{\gtG[i]})
    =\gtComm{\roleP}{\roleQ}{i\in I}{\gtLab[i]}{\tyGround[i]}{F(\gtG[i])}$
    if the above case does not hold.
    \item $F(\gtEnd)=\gtEnd$
    \item $F(\gtRecVar)=\gtRecVar$
    \item $F(\gtPar{\gtG}{\gtGi})=\gtPar{F(\gtG)}{F(\gtGi)}$
    \item $F(\gtRec{\gtRecVar}{\gtG})=\gtRec{\gtRecVar}{F(\gtG)}$ if $F(\gtG)\neq\gtRecVar$ and $\gtEnd$ otherwise.
    \item $F(\roleFmt{c}\stFmt{\oplus}\stLabFmt{m}\stSeq\stT)=F(\stT)$
    \item $F(\stIntSum{\roleP}{i\in I}{\stChoice{\stLab[i]}{\tyGround[i]}\stSeq\stT[i]})=
    \stIntSum{\roleP}{i\in I}{\stChoice{\stLab[i]}{\tyGround[i]}\stSeq F(\stT[i])}$
    if the above case does not hold.
    \item $F(\roleFmt{a}\stFmt{\&}\stLabFmt{m}\stSeq\stT)=F(\stT)$
    \item $F(\stExtSum{\roleP}{i\in I}{\stChoice{\stLab[i]}{\tyGround[i]}\stSeq\stT[i]})=
    \stExtSum{\roleP}{i\in I}{\stChoice{\stLab[i]}{\tyGround[i]}\stSeq F(\stT[i])}$
    if the above case does not hold.
    \item $F(\stEnd)=\stEnd$
    \item $F(\stRecVar)=\stRecVar$
    \item $F(\stRec{\stRecVar}{\stT})=\stRec{\stRecVar}{F(\stT)}$
    if $F(\stT)\neq\stRecVar$ and $\stEnd$ otherwise.
\end{itemize}

If $\stT[0]\tySSub\stT[1]$ then $F(\stT[0])\tySSub F(\stT[1])$
as the relation given by $\Re\langle F(\stT[0]),F(\stT[1])\rangle$ if
$\stT[0]\tySSub\stT[1]$ satisfies the rules defining $\tySSub$.
By induction on $\stT[0]$,
if $\stT[0]\in\stMerge{}{\{\stT[i]\}_{i\in I}}$ then
$F(\stT[0])\in\stMerge{}{\{F(\stT[i])\}_{i\in I}}$
for the full inductive merge.
By induction on $\gtG[0]$,
if $\gtProj{\roleR}{\gtG[0]}{\stT[0]}$ then
$\gtProj{\roleR}{F(\gtG[0])}{F(\stT[0])}$.

If $\gtGii$ is projectable, then let $\gtProj{}{\gtGii}{\stEnv}$.
$\stEnv$ is safe and live so, by~\cref{lem:contextlocaltrace},
\begin{equation*}
\stEnv\tySSub
\set{
\begin{array}{l}
\stEnvMap{\roleFmt{a}}{\stRec{\stRecVar}{\stIntSum{\roleFmt{b}}{}{
		\stLab\stSeq \roleFmt{c}{\oplus}\stLab\stSeq\stRecVar,\,
		\stLabi\stSeq\stRec{\stRecVar}{\roleFmt{c}{\oplus}\stLab\stSeq\stRecVar}
	}}
}\stEnvComp
\stEnvMap{\roleFmt{b}}{\stRec{\stRecVar}{\stExtSum{\roleFmt{a}}{}{
		\stLab\stSeq \stRecVar,\,
		\stLabi
	}}
}\stEnvComp\\
\stEnvMap{\roleFmt{c}}{\stRec{\stRecVar}{\stIntSum{\roleFmt{d}}{}{
		\stLab\stSeq \roleFmt{a}{\&}\stLab\stSeq\stRecVar,\,
		\stLabi\stSeq\stRec{\stRecVar}{\roleFmt{a}{\&}\stLab\stSeq\stRecVar}
	}}
}\stEnvComp
\stEnvMap{\roleFmt{d}}{\stRec{\stRecVar}{\stExtSum{\roleFmt{c}}{}{
		\stLab\stSeq \stRecVar,\,
		\stLabi
	}}
}\end{array}}
\end{equation*}
Therefore,
$\gtProj{}{F(\gtGii)}{F(\stEnv)}$
and $F(\stEnv)\tySSub\stEnvi$
where:
\begin{equation*}
\stEnvi=
\stEnvMap{\roleFmt{a}}{\stRec{\stRecVar}{\stIntSum{\roleFmt{b}}{}{
		\stLab\stSeq\stRecVar,\,
		\stLabi\stSeq
	}}
}\stEnvComp
\stEnvMap{\roleFmt{b}}{\stRec{\stRecVar}{\stExtSum{\roleFmt{a}}{}{
		\stLab\stSeq \stRecVar,\,
		\stLabi
	}}
}\stEnvComp
\stEnvMap{\roleFmt{c}}{\stRec{\stRecVar}{\stIntSum{\roleFmt{d}}{}{
		\stLab\stSeq \stRecVar,\,
		\stLabi
	}}
}\stEnvComp
\stEnvMap{\roleFmt{d}}{\stRec{\stRecVar}{\stExtSum{\roleFmt{c}}{}{
		\stLab\stSeq \stRecVar,\,
		\stLabi
	}}
}
\end{equation*}
So, $\gtTrace{\gtG}=\gtTrace{\stEnvi}=\gtTrace{F(\stEnv)}=\gtTrace{F(\gtGii)}$.
Therefore, $F(\gtGii)\bisim\gtG$
contradicting~\cref{thm:par-nec}.
\end{proof}

\begin{theorem}[Remaining Separations]
\label{thm:remsep}
All the other separations in~\cref{fig:proj_strength}
are correct.
\end{theorem}
\begin{proof}
	$\gtG[\text{\tiny par}]\,\gtMoveStar\,\gtEnd$,
	$\gtGi[\text{\tiny co}]\,\gtMoveStar\,\gtEnd$, and
	$\gtG[\neg\text{\tiny plain}]\,\gtMoveStar\,\gtEnd$.
	Therefore, if:
	$\gtPar{\gtG[\text{\tiny par}]}{\gtG[\neg{\text{\tiny plain}}]}$
	is projectable then so are $\gtG[\text{\tiny par}]$ and $\gtG[\neg{\text{\tiny plain}}]$; and
	$\gtPar{\gtGi[\text{\tiny co}]}{\gtG[\neg{\text{\tiny plain}}]}$
	is projectable then so are $\gtGi[\text{\tiny co}]$ and $\gtG[\neg{\text{\tiny plain}}]$.
\end{proof}

\begin{proof}[Proof. (Context separations)]
Let:
\begin{equation*}
	\begin{array}{lcl}
	\stEnv[\text{\tiny par}] &=&\set{
	\stEnvMap{\roleFmt{a}}{\stRec{\stRecVar}{\stIntSum{\roleFmt{b}}{}{
		\stLab\stSeq\stRecVar,\, \stLabi
	}}}
	\stEnvComp \stEnvMap{\roleFmt{b}}{\stRec{\stRecVar}{\stExtSum{\roleFmt{a}}{}{
		\stLab\stSeq\stRecVar,\, \stLabi
	}}}
	\stEnvComp \stEnvMap{\roleFmt{c}}{\stRec{\stRecVar}{\stIntSum{\roleFmt{d}}{}{
		\stLab\stSeq\stRecVar,\, \stLabi
	}}}
	\stEnvComp \stEnvMap{\roleFmt{d}}{\stRec{\stRecVar}{\stExtSum{\roleFmt{c}}{}{
		\stLab\stSeq\stRecVar,\, \stLabi
	}}}}
	\\
	\stEnv[\text{\tiny co}]&=&
	\set{
	\begin{array}{l}
	\stEnvMap{\roleFmt{a}}{\stRec{\stRecVar}{\stIntSum{\roleFmt{b}}{}{
		\stLab\stSeq \roleFmt{c}{\oplus}\stLab\stSeq\stRecVar,\,
		\stLabi\stSeq\stRec{\stRecVar}{\roleFmt{c}{\oplus}\stLab\stSeq\stRecVar}
	}}}
	\stEnvComp \stEnvMap{\roleFmt{b}}{\stRec{\stRecVar}{\stExtSum{\roleFmt{a}}{}{
		\stLab\stSeq\stRecVar,\, \stLabi
	}}}
	\stEnvComp \\\stEnvMap{\roleFmt{c}}{\stRec{\stRecVar}{\stIntSum{\roleFmt{d}}{}{
		\stLab\stSeq\roleFmt{a}{\&}\stLab\stSeq\stRecVar,\, \stLabi\stSeq\stRec{\stRecVar}{\roleFmt{a}{\&}\stLab\stSeq\stRecVar}
	}}}
	\stEnvComp \stEnvMap{\roleFmt{d}}{\stRec{\stRecVar}{\stExtSum{\roleFmt{c}}{}{
		\stLab\stSeq\stRecVar,\, \stLabi
	}}}
	\end{array}}
	\end{array}
\end{equation*}

From~\cref{fig:proj_strength},
we have that
$\gtProjAnnot{}{\gtG[\text{\tiny par}]}{\stEnv[\text{\tiny par}]}{\text{plain ind+par}}$,
$\gtProjAnnot{}{\gtG[\text{\tiny co}]}{\stEnv[\text{\tiny co}]}{\text{plain co}}$,
$\forall\stEnv.\;\neg\gtProjAnnot{}{\gtG[\text{\tiny par}]}{\stEnv}{\text{full ind}}$, and
$\forall\stEnv.\;\neg\gtProjAnnot{}{\gtG[\text{\tiny co}]}{\stEnv}{\text{full ind+par}}$.

If $\stEnv[\text{\tiny par}]\in\projLocal{\gtProjAnnot{}{}{}{\text{full ind}}}$,
then there exists $\gtG$ and $\stEnv$ such that
$\gtProjAnnot{}{\gtG}{\stEnv}{\text{full ind}}$ and
$\stEnv[\text{\tiny par}]\tySub\stEnv$.
By \cref{lem:max-safe},
$\gtTrace{\gtG[\text{\tiny par}]}
=\gtTrace{\stEnv[\text{\tiny par}]}
=\gtTrace{\stEnv}
=\gtTrace{\gtG}$.
Thus $\gtG\bisim\gtG[\text{\tiny par}]$,
contradicting
$\gtG[\text{\tiny par}]\notin\projGlobal{\gtProjAnnot{}{}{}{\text{full ind}}}$.

If $\stEnv[\text{\tiny co}]\in\projLocal{\gtProjAnnot{}{}{}{\text{full ind+par}}}$,
then there exists $\gtG$ and $\stEnv$ such that
$\gtProjAnnot{}{\gtG}{\stEnv}{\text{full ind+par}}$ and
$\stEnv[\text{\tiny co}]\tySub\stEnv$.
By \cref{lem:max-safe},
$\gtTrace{\gtG[\text{\tiny co}]}
=\gtTrace{\stEnv[\text{\tiny co}]}
=\gtTrace{\stEnv}
=\gtTrace{\gtG}$.
Thus $\gtG\bisim\gtG[\text{\tiny co}]$,
contradicting
$\gtG[\text{\tiny co}]\notin\projGlobal{\gtProjAnnot{}{}{}{\text{full ind+par}}}$.

We now find corresponding sessions by applying the characteristic map:
\begin{equation*}
\begin{array}{cc}
	\mpP_{\text{par},\roleFmt{a}}=\mpRec{\mpX}{
		\mpIf{\mpTrue}
			{\roleFmt{b}\mathord{!}\stLab\mpSeq\mpX}
			{\roleFmt{b}\mathord{!}\stLabi\mpSeq\mpNil}
	}
	&
	\mpP_{\text{par},\roleFmt{b}}=\mpRec{\mpX}{
		\roleFmt{a}\mathord{?}\stLab\mpSeq\mpX
		+
		\roleFmt{a}\mathord{?}\stLabi\mpSeq\mpNil
	}
	\\
	\mpP_{\text{par},\roleFmt{c}}=\mpRec{\mpX}{
		\mpIf{\mpTrue}
			{\roleFmt{d}\mathord{!}\stLab\mpSeq\mpX}
			{\roleFmt{d}\mathord{!}\stLabi\mpSeq\mpNil}
	}
	&
	\mpP_{\text{par},\roleFmt{d}}=\mpRec{\mpX}{
		\roleFmt{c}\mathord{?}\stLab\mpSeq\mpX
		+
		\roleFmt{c}\mathord{?}\stLabi\mpSeq\mpNil
	}
\\
	\mpP_{\text{co},\roleFmt{a}}=\mpRec{\mpX}{
		\mpIf{\mpTrue}
			{\roleFmt{b}\mathord{!}\stLab\mpSeq
				\roleFmt{c}\mathord{!}\stLab\mpSeq\mpX}
			{\roleFmt{b}\mathord{!}\stLabi\mpSeq
				\mpRec{\mpX'}{\roleFmt{c}\mathord{!}\stLab\mpSeq\mpX'}
			}
	}
	&
	\mpP_{\text{co},\roleFmt{b}}=\mpRec{\mpX}{
		\roleFmt{a}\mathord{?}\stLab\mpSeq\mpX
		+
		\roleFmt{a}\mathord{?}\stLabi\mpSeq\mpNil
	}
	\\
	\mpP_{\text{co},\roleFmt{c}}=\mpRec{\mpX}{
		\mpIf{\mpTrue}
			{\roleFmt{d}\mathord{!}\stLab\mpSeq
				\roleFmt{a}\mathord{?}\stLab\mpSeq\mpX}
			{\roleFmt{d}\mathord{!}\stLabi\mpSeq
				\mpRec{\mpX'}{\roleFmt{c}\mathord{!}\stLab\mpSeq\mpX'}
			}
	}
	&
	\mpP_{\text{co},\roleFmt{d}}=\mpRec{\mpX}{
		\roleFmt{c}\mathord{?}\stLab\mpSeq\mpX
		+
		\roleFmt{c}\mathord{?}\stLabi\mpSeq\mpNil	
	}
\end{array}
\end{equation*}
\begin{equation*}
	\M_\text{par} = \playsrole{\roleFmt{a}}{\mpP_{\text{par},\roleFmt{a}}}
	\mpPar \playsrole{\roleFmt{b}}{\mpP_{\text{par},\roleFmt{b}}}
	\mpPar \playsrole{\roleFmt{c}}{\mpP_{\text{par},\roleFmt{c}}}
	\mpPar \playsrole{\roleFmt{d}}{\mpP_{\text{par},\roleFmt{d}}}
	\qquad
	\M_\text{co} = \playsrole{\roleFmt{a}}{\mpP_{\text{co},\roleFmt{a}}}
	\mpPar \playsrole{\roleFmt{b}}{\mpP_{\text{co},\roleFmt{b}}}
	\mpPar \playsrole{\roleFmt{c}}{\mpP_{\text{co},\roleFmt{c}}}
	\mpPar \playsrole{\roleFmt{d}}{\mpP_{\text{co},\roleFmt{d}}}
\end{equation*}
We have that, for all $\gtG$:
\[
\begin{array}{llcll}
\stFmt{\vdash}_\text{plain ind.+par}&\M_\text{par}&:&\stEnv[\text{\tiny par}]&;\\
\stFmt{\vdash}_\text{plain ind.+par}&\M_\text{par}&:&\gtG[\text{\tiny par}]&;\\
\stFmt{\vdash}_\text{plain coind.}&\M_\text{co}&:&\stEnv[\text{\tiny co}]&;\\
\stFmt{\vdash}_\text{plain coind.}&\M_\text{co}&:&\gtG[\text{\tiny co}]&;\\
\not\stFmt{\vdash}_\text{full ind.}&\M_\text{par}&:&\gtG&;\text{ and}\\
\not\stFmt{\vdash}_\text{full ind.+par}&\M_\text{co}&:&\gtG.
\end{array}
\]
\end{proof}

\begin{definition}[Interlacing]
We define a function from (a pair of closed global types without parallel composition and a partial injection from 
pairs of closed global types to recursive variables) to
(a pair containing a global type without parallel composition and
a set of recursive variables),
written $\gtSch{\gtG}{\gtGi}{f}$, by:
\begin{itemize}
    \item if $\left(\gtG,\gtGi\right)\in dom(f)$, then
	$\gtSch{\gtG}{\gtGi}{f}=\left(f\left(\gtG,\gtGi\right), \left\{ f\left(\gtG,\gtGi\right) \right\}\right)$
    \item if $\unfoldOne{\gtG}=\gtEnd$ and $\unfoldOne{\gtGi}=\gtEnd$, then $\gtSch{\gtG}{\gtGi}{f}=\left(\gtEnd,\emptyset\right)$
    \item if $\unfoldOne{\gtG}=\gtEnd$ and $\unfoldOne{\gtGi}\neq\gtEnd$, then $\gtSch{\gtG}{\gtGi}{f}=\gtSch{\gtGi}{\gtG}{f}$
    \item if $\unfoldOne{\gtG}=\gtComm{\roleP}{\roleQ}{i \in I}{\gtLab[i]}{\tyGround[i]}{\gtG[i]}$, then let:
    \begin{enumerate}
    \item
    $\gtRecVar[{\left(\gtG,\gtGi\right)}]$ be a recursion variable;
    \item
$f'=f\cup\left\{\left(\gtG,\gtGi\right)\mapsto \gtRecVar[{\left(\gtG,\gtGi\right)}] \right\}$; and
	\item
$\left(\gtGi[i],\mathcal{V}_i\right)=\gtSch{\gtGi}{\gtG[i]}{f'}$ for $i\in I$.
\end{enumerate}
\[
    \gtSch{\gtG}{\gtGi}{f}=\left\{
	\begin{array}{lr}
		\left(\gtComm{\roleP}{\roleQ}{i \in I}{\gtLab[i]}{\tyGround[i]}{\gtGi[i]},\bigcup_{i\in I}\mathcal{V}_i\right) &
		\text{for all }i\in I\text{ }\gtRecVar[{\left(\gtG,\gtGi\right)}]\not\in\mathcal{V}_i\\
		\left(\gtRec{\gtRecVar[{\left(\gtG,\gtGi\right)}]}{\gtComm{\roleP}{\roleQ}{i \in I}{\gtLab[i]}{\tyGround[i]}{\gtGi[i]}},\bigcup_{i\in I}\mathcal{V}_i\setminus\left\{\gtRecVar[{\left(\gtG,\gtGi\right)}]\right\}\right)
		 &\text{there is }i\in I\text{ }\gtRecVar[{\left(\gtG,\gtGi\right)}]\in\mathcal{V}_i
	\end{array}\right.
	\]
\end{itemize}
\end{definition}

\begin{restatable}[Interlacing is Well-Defined]{lemma}{lemInterlaceExists}
\label{lem:inter-well}
	$\gtSch{\gtG}{\gtGi}{f}$ exists \ie the algorithm above always halts.
\end{restatable}

\begin{proof}
The set $\mathcal{G}=\left\{\gtGii\suchthat \gtG\,\gtMoveStar\,\gtGii\text{ or }\gtGi\,\gtMoveStar\,\gtGii\right\}$ is finite,
say $n={\left|\mathcal{G}\right|}$.
Assume that $\dom{f}\subseteq\mathcal{G}\times\mathcal{G}$;
we can do this as if $\gtSch{\gtG}{\gtGi}{f}$ exists
then $\gtSch{\gtG}{\gtGi}{f\cup f'}$ exists and halts in fewer or the same number of steps.

We induct on $n^2-|f|$, that if $\dom{f}\subseteq\mathcal{G}\times\mathcal{G}$ and $\gtG[L],\gtG[R]\in\mathcal{G}$,
then $\gtSch{\gtG[L]}{\gtG[R]}{f}$ is defined.

If $n^2-|f|=0$, then $\dom{f}=\mathcal{G}\times\mathcal{G}$ so that $\left(\gtG[L],\gtG[R]\right)\in \dom{f}$ and
$\gtSch{\gtG[L]}{\gtG[R]}{f}=\left(f\left(\gtG[L],\gtG[R]\right), \left\{ f\left(\gtG[L],\gtG[R]\right) \right\}\right)$.

If $\unfoldOne{\gtG[L]}=\unfoldOne{\gtG[R]}=\gtEnd$, then $\gtSch{\gtG[L]}{\gtG[R]}{f}=\left(\gtEnd,\emptyset\right)$.

If $\unfoldOne{\gtG[L]}=\gtEnd$, $\unfoldOne{\gtG[R]}=\gtComm{\roleP}{\roleQ}{i \in I}{\gtLab[i]}{\tyGround[i]}{\gtG[i]}$ and
$\left(\gtG[R],\gtG[L]\right)\in \dom{f}$, then
$\gtSch{\gtG[L]}{\gtG[R]}{f}=\left(f\left(\gtG[R],\gtG[L]\right), \left\{ f\left(\gtG[R],\gtG[L]\right) \right\}\right)$.

If $\unfoldOne{\gtG[L]}=\gtEnd$, $\unfoldOne{\gtG[R]}=\gtComm{\roleP}{\roleQ}{i \in I}{\gtLab[i]}{\tyGround[i]}{\gtG[i]}$ and
$\left(\gtG[R],\gtG[L]\right)\not\in \dom{f}$, then
let $f'=f\cup\left\{\left(\gtG[R],\gtG[L]\right)\mapsto \gtRecVar[{\left(\gtG[R],\gtG[L]\right)}] \right\}$.
$n^2-|f'|<n^2-|f|$ and $\gtG[i]\in\mathcal{G}$ for all $i\in I$, so by I.H.
$\left(\gtGi[i],\mathcal{V}_i\right)=\gtSch{\gtG[L]}{\gtG[i]}{f'}$ exists for $i\in I$.\\
So,
   \[\gtSch{\gtG[L]}{\gtG[R]}{f}=\left\{
	\begin{array}{lr}
		\left(\gtComm{\roleP}{\roleQ}{i \in I}{\gtLab[i]}{\tyGround[i]}{\gtGi[i]},\bigcup_{i\in I}\mathcal{V}_i\right) &
		\text{for all }i\in I\text{ }\gtRecVar[{\left(\gtG[R],\gtG[L]\right)}]\not\in\mathcal{V}_i\\
		\left(\gtRec{\gtRecVar[{\left(\gtG[R],\gtG[L]\right)}]}{\gtComm{\roleP}{\roleQ}{i \in I}{\gtLab[i]}{\tyGround[i]}{\gtGi[i]}},\bigcup_{i\in I}\mathcal{V}_i\setminus\left\{\gtRecVar[{\left(\gtG[R],\gtG[L]\right)}]\right\}\right)
		 &\text{there is }i\in I\text{ }\gtRecVar[{\left(\gtG[R],\gtG[L]\right)}]\in\mathcal{V}_i
	\end{array}\right.\]
	
If
\[
\unfoldOne{\gtG[L]}=\gtComm{\roleP}{\roleQ}{i \in I}{\gtLab[i]}{\tyGround[i]}{\gtG[i]}\]
then
let $f'=f\cup\left\{\left(\gtG[L],\gtG[R]\right)\mapsto \gtRecVar[{\left(\gtG[L],\gtG[R]\right)}] \right\}$.
$n^2-|f'|<n^2-|f|$ and $\gtG[i]\in\mathcal{G}$ for all $i\in I$, so by I.H.
$\left(\gtGi[i],\mathcal{V}_i\right)=\gtSch{\gtG[R]}{\gtG[i]}{f'}$ exists for $i\in I$.\\
So,
    \[\gtSch{\gtG[L]}{\gtG[R]}{f}=\left\{
	\begin{array}{lr}
		\left(\gtComm{\roleP}{\roleQ}{i \in I}{\gtLab[i]}{\tyGround[i]}{\gtGi[i]},\bigcup_{i\in I}\mathcal{V}_i\right) &
		\text{for all }i\in I\text{ }\gtRecVar[{\left(\gtG[L],\gtG[R]\right)}]\not\in\mathcal{V}_i\\
		\left(\gtRec{\gtRecVar[{\left(\gtG[L],\gtG[R]\right)}]}{\gtComm{\roleP}{\roleQ}{i \in I}{\gtLab[i]}{\tyGround[i]}{\gtGi[i]}},\bigcup_{i\in I}\mathcal{V}_i\setminus\left\{\gtRecVar[{\left(\gtG[L],\gtG[R]\right)}]\right\}\right)
		 &\text{there is }i\in I\text{ }\gtRecVar[{\left(\gtG[L],\gtG[R]\right)}]\in\mathcal{V}_i
	\end{array}\right.\]
\end{proof}

\begin{lemma}
	If $\gtSch{\gtG}{\gtGi}{f}=\left(\gtGii,\mathcal{V}\right)$, then
	$\mathcal{V}=\gtFv{\gtGii}\subseteq \ran{f}$.
\end{lemma}
\begin{proof}
	We induct on the definition of $\gtSch{\gtG}{\gtGi}{f}$.
	
	If $\left(\gtG,\gtGi\right)\in \dom{f}$, then
	\[\gtSch{\gtG}{\gtGi}{f}=\left(f\left(\gtG,\gtGi\right), \left\{ f\left(\gtG,\gtGi\right) \right\}\right)\]
	and
	\[\left\{ f\left(\gtG,\gtGi\right) \right\}=\gtFv{f\left(\gtG,\gtGi\right)}\subseteq \ran{f}\]
	
	If $\unfoldOne{\gtG}=\gtEnd$ and $\unfoldOne{\gtGi}=\gtEnd$,
	then \[\gtSch{\gtG}{\gtGi}{f}=\left(\gtEnd,\emptyset\right)\]
	and \[\emptyset=\gtFv{\gtEnd}\subseteq \ran{f}\]
	
	If $\unfoldOne{\gtG}=\gtEnd$ and $\unfoldOne{\gtGi}\neq\gtEnd$,
	then \[\gtSch{\gtG}{\gtGi}{f}=\gtSch{\gtGi}{\gtG}{f}\]
	and, by I.H,
	\[\gtSch{\gtGi}{\gtG}{f}=\left(\gtGii,\mathcal{V}\right)\] with
	\[\mathcal{V}=\gtFv{\gtGii}\subseteq \ran{f}\]
	
	If $\unfoldOne{\gtG}=\gtComm{\roleP}{\roleQ}{i \in I}{\gtLab[i]}{\tyGround[i]}{\gtG[i]}$, then let
\[f'=f\cup\left\{\left(\gtG,\gtGi\right)\mapsto \gtRecVar[{\left(\gtG,\gtGi\right)}] \right\}\] and
\[\forall i\in I.\;\left(\gtGi[i],\mathcal{V}_i\right)=\gtSch{\gtGi}{\gtG[i]}{f'}\]
	By the I.H., $\forall i\in I.\;\mathcal{V}_i=\gtFv{\gtGi[i]}\subseteq \ran{f'}$.
	
	If $\gtRecVar[{\left(\gtG,\gtGi\right)}]\in\mathcal{V}_i$ for some $i\in I$, then
	\[\gtSch{\gtG}{\gtGi}{f}=\left(\gtRec{\gtRecVar[{\left(\gtG,\gtGi\right)}]}{\gtComm{\roleP}{\roleQ}{i \in I}{\gtLab[i]}{\tyGround[i]}{\gtGi[i]}},\bigcup_{i\in I}\mathcal{V}_i\setminus\left\{\gtRecVar[{\left(\gtG,\gtGi\right)}]\right\}\right)\]
	\[\begin{array}{cl}
	&\bigcup_{i\in I}\mathcal{V}_i\setminus\left\{\gtRecVar[{\left(\gtG,\gtGi\right)}]\right\}
	\\=&
	\bigcup_{i\in I}\gtFv{\gtGi[i]}\setminus\left\{\gtRecVar[{\left(\gtG,\gtGi\right)}]\right\}
	\\=&
	\gtFv{\gtRec{\gtRecVar[{\left(\gtG,\gtGi\right)}]}{\gtComm{\roleP}{\roleQ}{i \in I}{\gtLab[i]}{\tyGround[i]}{\gtGi[i]}}}
	\subseteq \bigcup_{i\in I}\ran{f'}\setminus\left\{\gtRecVar[{\left(\gtG,\gtGi\right)}]\right\}
	\\=& \ran{f}
	\end{array}
	\]
	
	If $\gtRecVar[{\left(\gtG,\gtGi\right)}]\not\in\mathcal{V}_i$ for all $i\in I$, then
	\[
	\left(\gtComm{\roleP}{\roleQ}{i \in I}{\gtLab[i]}{\tyGround[i]}{\gtGi[i]},\bigcup_{i\in I}\mathcal{V}_i\right)\]
	\[
	\gtFv{\gtComm{\roleP}{\roleQ}{i \in I}{\gtLab[i]}{\tyGround[i]}{\gtGi[i]}}=
	\bigcup_{i\in I}\gtGi[i]= \bigcup_{i\in I}\mathcal{V}_i\subseteq \bigcup_{i\in I}\ran{f'}\setminus\left\{\gtRecVar[{\left(\gtG,\gtGi\right)}]\right\}
	= \ran{f}
	\]
\end{proof}

\begin{definition}
    We define $\gtBRoles{\gtG}$ by recursion on global type syntax:
    \begin{itemize}
        \item $\gtBRoles{\gtEnd}=\emptyset$
        \item $\gtBRoles{\gtRecVar}=\emptyset$
        \item $\gtBRoles{\gtRec{\gtRecVar}{\gtG}}=\gtBRoles{\gtG}$
        \item $\gtBRoles{\gtPar{\gtG}{\gtGi}}=\gtBRoles{\gtG}\cup\gtBRoles{\gtGi}$
        \item $\gtBRoles{\gtComm{\roleP}{\roleQ}{i \in I}{\gtLab[i]}{\tyGround[i]}{\gtG[i]}}=
        \{\roleP,\roleQ\}\cup
        \bigcap_{i\in I}\gtBRoles{\gtG[i]}$
    \end{itemize}
\end{definition}

\begin{remark}
$\gtBRoles{\gtG}=\left\{\roleP\suchthat
\gtDepth{\gtG}{\roleP}\;\text{exists}\right\}$
\end{remark}

\begin{lemma}
\label{lem:bal-roles}
    Suppose that $\gtG$ and $\gtGi$ are balanced.
    Write
    \[f=\left\{(\gtG[{L,1}],\gtG[{R,1}])\mapsto\gtRecVar[1],
    \dots,(\gtG[{L,n}],\gtG[{R,n}])\mapsto\gtRecVar[n]\right\}\]
    
    Let \[\gtSch{\gtG[{L,i}]}{\gtG[{R,i}]}
    {
        \left\{(\gtG[{L,j}],\gtG[{R,j}])\mapsto\gtRecVar[j]
        \suchthat 1\leq j<i\right\}
    }=\left(\gtG[i],\mathcal{V}_i\right)\] for $1\leq i\leq n$ and
    \[\gtGii=\gtSch{\gtG}{\gtGi}{f}[\gtG[n]/\gtRecVar[n]]\dots[\gtG[1]/\gtRecVar[1]]\]
    
    Then $\gtRoles{\gtGii}
    =\gtBRoles{\gtGii}$
\end{lemma}
\begin{proof}
Induct on the derivation of $\gtSch{\gtG}{\gtGi}{f}$.
\end{proof}

\begin{restatable}{lemma}{lemInterlaceBalanced}
	If $\gtG$ and $\gtGi$ are closed global types without parallel composition and $\gtPar{\gtG}{\gtGi}$ is balanced,
	then so is $\gtSch{\gtG}{\gtGi}{\emptyset}$.
\end{restatable}

\begin{proof}
    The set $\mathcal{G}=\left\{\gtGii\suchthat
    \gtGii\text{ is of the form in \cref{lem:bal-roles}}
    \right\}$
    is closed under head reductions.
    For $\gtG\in\mathcal{G}$, $\gtRoles{\gtG}=\gtBRoles{\gtG}$.
    Therefore for $\gtG\in\mathcal{G}$, $\gtG$ is balanced.
\end{proof}

\begin{restatable}[Interlacing Works]{theorem}{thmInterlacingWorks}
\label{thm:inter-work}
    If $\gtG$ and $\gtGi$ are closed global types without parallel composition,
    $\gtProj{\roleR}{\gtPar{\gtG}{\gtGi}}{\stT}$ (by a coinductive projection with a good merge),
    
    then $\gtProj{\roleR}{\gtGii}{\stT}$, where $\gtSch{\gtG}{\gtGi}{\emptyset}=\left(\gtGii,\emptyset\right)$
\end{restatable}

\begin{proof}
	We show that if:
	\begin{enumerate}
	\item $\gtProj{\roleR}{\gtPar{\gtG}{\gtGi}}{\stT}$,
	\item $\gtSch{\gtG}{\gtGi}{f}=\left(\gtGii,\mathcal{V}\right)$,
	\item
    $f=\left\{(\gtG[{L,1}],\gtG[{R,1}])\mapsto\gtRecVar[1],\dots,(\gtG[{L,n}],\gtG[{R,n}])\mapsto\gtRecVar[n]\right\}$, and
    \item
    $\gtSch{\gtG[{L,i}]}{\gtG[{R,i}]}
    {
        \left\{(\gtG[{L,j}],\gtG[{R,j}])\mapsto\gtRecVar[j]
        \suchthat 1\leq j<i\right\}
    }=\left(\gtG[i],\mathcal{V}_i\right)$ for $1\leq i\leq n$
    \end{enumerate}
    then
    $\gtProj{\roleR}{
    \gtGii{}[\gtG[n]/\gtRecVar[n]]\dots[\gtG[1]/\gtRecVar[1]]
    }
    {\stT}$.
    
	Indeed, the relation
	\begin{equation*}
		\left\{
		\left(
		\roleR,
		\gtG{}
        [\gtG[n]/\gtRecVar[n]]\dots[\gtG[1]/\gtRecVar[1]]
        ,\stT\right)
		\suchthat
		\begin{array}{l}
		\text{there exist }\gtG[L],\gtG[R], f
        \text{ such that }\\
		\gtSch{\gtG[L]}{\gtG[R]}{f}=\left(\gtG,\gtFv{\gtG}\right)\\
		\gtProj{\roleR}{\gtPar{\gtG[L]}{\gtG[R]}}{\stT}\\
        f=\left\{(\gtG[{L,1}],\gtG[{R,1}])\mapsto\gtRecVar[1],\dots,(\gtG[{L,n}],\gtG[{R,n}])\mapsto\gtRecVar[n]\right\}\\
        \forall1\leq i\leq n.\;\\\;
        \gtSch{\gtG[{L,i}]}{\gtG[{R,i}]}
    {
        \left\{(\gtG[{L,j}],\gtG[{R,j}])\mapsto\gtRecVar[j]\suchthat 1\leq j<i\right\}
    }=\left(\gtG[i],\mathcal{V}_i\right)
		\end{array}
		\right\}
	\end{equation*}
	satisfies the rules defining a coinductive projection.
\end{proof}

\begin{remark}
	By making more careful observations than in our proofs, we can provide a bound on the size of the resultant type.
	Let
	\[
	\mathcal{G}_L=\{\gtGii\suchthat \gtG\,\gtMoveStar\,\gtGii\text{ by head transition}\}
	\]
	and
	\[
	\mathcal{G}_R=\{\gtGii\suchthat \gtGi\,\gtMoveStar\,\gtGii\text{ by head transition}\}
	\]
	Rather than using ${|\mathcal{G}|}^2-|f|$ as in~\cref{lem:inter-well}
	as our variant,
	we can instead use $2|\gtG|+2|\gtGi|-|f|$,
	as $|\gtG|\geq|\mathcal{G}_L|$,
	$|\gtGi|\geq|\mathcal{G}_R|$, and $\dom{f}\subseteq \mathcal{G}_L\times\mathcal{G}_R\cup\mathcal{G}_R\times\mathcal{G}_L$.
	Let the maximum number of choices in a communication
	in $\gtG$ and $\gtGi$ be $b$.
	We can bound the size of $\gtSch{\gtG}{\gtGi}{\emptyset}$ by $2\cdot\frac{b^{2|\gtG||\gtGi|}-1}{b-1}$.
	However, we do not know if this bound is strict, and we have not found any examples
	that are exponential in the size of the parameters.
	This is only for removing a single parallel composition;
	removing iterated parallel compositions easily causes exponential blow-up.
\end{remark}

\begin{example}[Interlacing]
Recall the global type:
\begin{equation*}
	\gtG = \gtPar
		{
			\gtRec{\gtRecVar}
			{
				\gtCommRaw{\roleFmt{a}}{\roleFmt{b}}
				{
					\begin{array}{l}%
						\gtCommChoice{first}{}{\gtRecVar}\\
						\gtCommChoice{second}{}{}
					\end{array}
				}
			}
		}
		{
			\gtRec{\gtRecVar}
			{
				\gtCommRaw{\roleFmt{c}}{\roleFmt{d}}
				{
					\begin{array}{l}%
						\gtCommChoice{first}{}{\gtRecVar}\\
						\gtCommChoice{second}{}{}
					\end{array}
				}
			}
		}
\end{equation*}
Let:
\begin{equation*}
\gtG[L]=\gtRec{\gtRecVar}
			{
				\gtCommRaw{\roleFmt{a}}{\roleFmt{b}}
				{
					\begin{array}{l}%
						\gtCommChoice{first}{}{\gtRecVar}\\
						\gtCommChoice{second}{}{}
					\end{array}
				}
			}
\qquad
\gtG[R]=			\gtRec{\gtRecVar}
			{
				\gtCommRaw{\roleFmt{c}}{\roleFmt{d}}
				{
					\begin{array}{l}%
						\gtCommChoice{first}{}{\gtRecVar}\\
						\gtCommChoice{second}{}{}
					\end{array}
				}
			}
\end{equation*}
We shall compute
\begin{equation*}
\gtSch{\gtG[L]}{\gtG[R]}{\emptyset}
\end{equation*}
for which we
shall need to compute
\begin{equation*}
\gtSch{\gtG[R]}{\gtG[L]}{\{(\gtG[L],\gtG[R])\mapsto\gtRecVar\}}\qquad
\gtSch{\gtG[R]}{\gtEnd}{\{(\gtG[L],\gtG[R])\mapsto\gtRecVar\}}
\end{equation*}
For the latter, we must compute
\begin{equation*}
\begin{split}
\gtSch{\gtEnd}{\gtEnd}{\{(\gtG[L],\gtG[R])\mapsto\gtRecVar, (\gtG[R],\gtEnd)\mapsto\gtRecVari\}}
&=(\gtEnd,\emptyset)\\
\gtSch{\gtEnd}{\gtG[R]}{\{(\gtG[L],\gtG[R])\mapsto\gtRecVar, (\gtG[R],\gtEnd)\mapsto\gtRecVari\}}
&=\gtSch{\gtG[R]}{\gtEnd}{\{(\gtG[L],\gtG[R])\mapsto\gtRecVar, (\gtG[R],\gtEnd)\mapsto\gtRecVari\}}
\\&=(\gtRecVari,\{\gtRecVari\})
\end{split}
\end{equation*}
So:
\begin{equation*}
\gtSch{\gtG[R]}{\gtEnd}{\{(\gtG[L],\gtG[R])\mapsto\gtRecVar\}}
=\gtRec{\gtRecVari}{
					\gtCommRaw{\roleFmt{c}}{\roleFmt{d}}
					{
						\begin{array}{l}
						\gtCommChoice{first}{}{\gtRecVari}\\
						\gtCommChoice{second}{}{}
						\end{array}
					}
}
\end{equation*}
For the former, we must compute
\begin{equation*}
\begin{split}
\gtSch{\gtG[L]}{\gtG[R]}{\{(\gtG[L],\gtG[R])\mapsto\gtRecVar, (\gtG[R],\gtG[L])\mapsto\gtRecVar[{(\gtG[R],\gtG[L])}]\}}
&=(\gtRecVar,\{\gtRecVar\})\\
\gtSch{\gtG[L]}{\gtEnd}{\{(\gtG[L],\gtG[R])\mapsto\gtRecVar, (\gtG[R],\gtG[L])\mapsto\gtRecVar[{(\gtG[R],\gtG[L])}]\}}
\end{split}
\end{equation*}
We now compute
\begin{equation*}
\gtSch{\gtEnd}{\gtEnd}{\{(\gtG[L],\gtG[R])\mapsto\gtRecVar, (\gtG[R],\gtG[L])\mapsto\gtRecVar[{(\gtG[R],\gtG[L])}],(\gtG[L],\gtEnd)\mapsto\gtRecVarii\}}
= (\gtEnd,\emptyset)
\end{equation*}
\begin{equation*}
\begin{array}{l}
\gtSch{\gtEnd}{\gtG[L]}{\{(\gtG[L],\gtG[R])\mapsto\gtRecVar, (\gtG[R],\gtG[L])\mapsto\gtRecVar[{(\gtG[R],\gtG[L])}],(\gtG[L],\gtEnd)\mapsto\gtRecVarii\}}\\
\qquad=\gtSch{\gtG[L]}{\gtEnd}{\{(\gtG[L],\gtG[R])\mapsto\gtRecVar, (\gtG[R],\gtG[L])\mapsto\gtRecVar[{(\gtG[R],\gtG[L])}],(\gtG[L],\gtEnd)\mapsto\gtRecVarii\}}\\
\qquad=(\gtRecVarii,\{\gtRecVarii\})
\end{array}
\end{equation*}
So
\begin{equation*}
\gtSch{\gtG[L]}{\gtEnd}{\{(\gtG[L],\gtG[R])\mapsto\gtRecVar, (\gtG[R],\gtG[L])\mapsto\gtRecVar[{(\gtG[R],\gtG[L])}]\}}
=(\gtRec{\gtRecVarii}
						{
							\gtCommRaw{\roleFmt{a}}{\roleFmt{b}}
							{
								\begin{array}{l}
									\gtCommChoice{first}{}{\gtRecVarii}\\
									\gtCommChoice{second}{}{}
								\end{array}
							}
						},\emptyset)
						\end{equation*}
We now deduce that
\begin{equation*}
\gtSch{\gtG[R]}{\gtG[L]}{\{(\gtG[L],\gtG[R])\mapsto\gtRecVar\}}
=(\gtCommRaw{\roleFmt{c}}{\roleFmt{d}}
				{
					\begin{array}{l}
					\gtCommChoice{first}{}{\gtRecVar}\\
					\gtCommChoice{second}{}
					{
						\gtRec{\gtRecVarii}
						{
							\gtCommRaw{\roleFmt{a}}{\roleFmt{b}}
							{
								\begin{array}{l}
									\gtCommChoice{first}{}{\gtRecVarii}\\
									\gtCommChoice{second}{}{}
								\end{array}
							}
						}
					}
					\end{array}
				},\{\gtRecVar\})
\end{equation*}
Hence,
\begin{equation*}
\gtSch{\gtG[L]}{\gtG[R]}{\emptyset}=(\gtRec{\gtRecVar}{
		\gtCommRaw{\roleFmt{a}}{\roleFmt{b}}
		{
			\begin{array}{l}
			\gtCommChoice{first}{}
			{
				\gtCommRaw{\roleFmt{c}}{\roleFmt{d}}
				{
					\begin{array}{l}
					\gtCommChoice{first}{}{\gtRecVar}\\
					\gtCommChoice{second}{}
					{
						\gtRec{\gtRecVarii}
						{
							\gtCommRaw{\roleFmt{a}}{\roleFmt{b}}
							{
								\begin{array}{l}
									\gtCommChoice{first}{}{\gtRecVarii}\\
									\gtCommChoice{second}{}{}
								\end{array}
							}
						}
					}
					\end{array}
				}
			}\\
			\gtCommChoice{second}{}
			{
				\gtRec{\gtRecVari}
				{
					\gtCommRaw{\roleFmt{c}}{\roleFmt{d}}
					{
						\begin{array}{l}
						\gtCommChoice{first}{}{\gtRecVari}\\
						\gtCommChoice{second}{}{}
						\end{array}
					}
				}
			}
			\end{array}
		}
	},\emptyset)
	\end{equation*}
\end{example}

\begin{example}[Interlacing is not optimal]
Let $b>1$ and let:
\begin{equation*}
	\gtG[{b,0}] = \gtRecVar\qquad
	\gtG[{b,n+1}] = \gtComm{\roleFmt{a}}{\roleFmt{b}}{1\leq i \leq b}{\gtMsgFmt{m_{i,n+1}}}{}{\gtG[{b,n}]}
\end{equation*}
For two distinct primes $p,q$, let $\gtG[p]=\gtRec{\gtRecVar}{\gtG[{b,p}]}$ and
$\gtG[q]=\gtRec{\gtRecVar}{\gtG[{b,q}]}$ with the roles $\roleFmt{a}$ and $\roleFmt{b}$ replaced by
$\roleFmt{c}$ and $\roleFmt{d}$.
We have that
\[\left|\gtSch{\gtG[p]}{\gtG[q]}{\emptyset}\right|=
\Theta(b^{2pq})
\left|\gtG[p]\right|=\frac{b^{p+1}-1}{b-1}+1
\qquad\left|\gtG[q]\right|=\frac{b^{q+1}-1}{b-1}+1
\]
Therefore,
\[\left|\gtSch{\gtG[p]}{\gtG[q]}{\emptyset}\right|=\Theta({(\left|\gtG[p]\right|\left|\gtG[q]\right|)}^2)
\]
This is suboptimal.
Consider
\begin{equation*}
	\gtG = \gtRec{\gtRecVar}{\gtG[{b,p}][\gtG[{b,q}][\roleFmt{c}/\roleFmt{a},\roleFmt{d}/\roleFmt{b}]/\gtRecVar]}
\end{equation*}
In $\gtG$, we first describe the behaviour of $\gtG[p]$ up-to a recursion variable, then
describe the behaviour of $\gtG[q]$ up-to a recursion variable, then we loop back to the start.

In general, this may not be smaller than interlacing, but
it is here, since there is only one recursive binding.
We have that $\gtPar{\gtG[p]}{\gtG[q]}\bisim \gtG$, coinductive projections are preserved, and
$|\gtG|=\frac{b^{p+q+1}-1}{b-1}+1=\Theta(\left|\gtG[p]\right|\left|\gtG[q]\right|)$.
\end{example}

Now that we can interlace two global types, without parallelism,
we can recursively remove parallel composition from global types.

\begin{definition}[Serialisation]
    We define the function $\gtSerial{\gtG}$ from a global type
    to a global type without parallelism, by recursion:
    \begin{itemize}
        \item $\gtSerial{\gtEnd}=\gtEnd$
        \item $\gtSerial{\gtRecVar}=\gtRecVar$
        \item $\gtSerial{\gtRec{\gtRecVar}{\gtG}}=\gtRec{\gtRecVar}{\gtSerial{\gtG}}$
        \item $\gtSerial{\gtComm{\roleP}{\roleQ}{i \in I}{\gtLab[i]}{\tyGround[i]}{\gtG[i]}}=
        \gtComm{\roleP}{\roleQ}{i \in I}{\gtLab[i]}{\tyGround[i]}{\gtSerial{\gtG[i]}}$
        \item $\gtSerial{\gtPar{\gtG[L]}{\gtG[R]}}=\text{fst}(\gtSch{\gtSerial{\gtG[L]}}{\gtSerial{\gtG[R]}}{\emptyset})$
    \end{itemize}
    Where $\text{fst}(l,r)=l$.
\end{definition}

\begin{remark}[Worst Case Serialisation]
	Let $f(n)$
	be the largest size of $\gtSerial{\gtG}$
	for $\gtG$ with $|\gtG|\leq n$ and
	branching number $\leq b$
	(the maximum number of choices appearing in a transmission).
	$f$ is clearly increasing at a rate
	that is faster than linear,
	so $kf(\frac{n}{k})\leq f(n)$
	for sufficiently large $n$.
	
	Also,
	$\frac{b^{2a(f(n)-a)}-1}{b-1}\leq\frac{b^{f(n)}-1}{b-1}$
	for $1\leq a \leq f(n)-1$.
	
	For $n\geq 2$,
	we have that $f(n)\leq max\left(1+f(n-1), 2\cdot\frac{b^{f(n-1)}-1}{b-1}\right)
	\leq 1+f(n-1)+2\cdot\frac{b^{f(n-1)}-1}{b-1}
	\leq \frac{b^{f(n-1)+\epsilon}}{b-1}\leq b^{f(n-1)-log_b(\frac{b-1}{2})+\epsilon}$,
	where $\epsilon\to0$ as $n\to\inf$.
	Fix $\epsilon>0$ and take $N$ large enough.
	For $n>N$,
	{\LARGE
	\begin{equation*}
	f(n)\leq 
	\underbrace{b^{\epsilon-log_b(\frac{b-1}{2})+b^{\epsilon-log_b(\frac{b-1}{2})+b^{\dots^{\epsilon-log_b(\frac{b-1}{2})+f(N)}}}}}_{n-N}
	\leq \mathcal{O}{(^{n}b)}
	\end{equation*}}
	Therefore, $f$ is worst case bounded by a hyperexponential.
	This is likely a great overestimation.
\end{remark}

\begin{remark}
    If $\gtG$ and $\gtGi$ are balanced, then so is $\gtSch{\gtG}{\gtGi}{\emptyset}$.
    Therefore, if $\gtG$ is balanced, then so is $\gtSerial{\gtG}$, by induction.
\end{remark}

\begin{restatable}[Correctness of Serialisation]{theorem}{thmSerialCorrect}
	Let $\gtProj{}{}{}$ be a coinductive projection.
    If $\gtG$ is closed, balanced, and projectable,
    say $\gtProj{}{\gtG}{\stEnvNew}$,
    then $\gtSerial{\gtG}$ is a closed and balanced global type, without parallelism,
    such that $\gtProj{}{\gtSerial{\gtG}}{\stEnvNew}$ and $\gtG\bisim\gtSerial{\gtG}$.
\end{restatable}

\begin{proof}
	We induct on $\gtSerial{}$,
	with the base case being
	$\gtG=\gtPar{\gtG[L]}{\gtG[R]}$,
	where we apply \cref{thm:inter-work}
	to $\gtG[L]$ and $\gtG[R]$,
	and note that coinductive projections preserve traces
	to observe that $\gtG\bisim\gtSch{\gtG[L]}{\gtG[R]}{\emptyset}$,
	$\gtSch{\gtG[L]}{\gtG[R]}{\emptyset}$ is closed,
	and $\gtSch{\gtG[L]}{\gtG[R]}{\emptyset}$ is balanced.
\end{proof}

\begin{example}
We can apply the serialisation
function to the same example we used for interlacing.
\begin{equation*}
	\begin{array}{ccl}
	\gtGi &:=&
	\gtSerial{
		\gtPar
		{
			\gtRec{\gtRecVar}
			{
				\gtCommRaw{\roleFmt{a}}{\roleFmt{b}}
				{
					\begin{array}{l}%
						\gtCommChoice{first}{}{\gtRecVar}\\
						\gtCommChoice{second}{}{}
					\end{array}
				}
			}
		}
		{
			\gtRec{\gtRecVar}
			{
				\gtCommRaw{\roleFmt{c}}{\roleFmt{d}}
				{
					\begin{array}{l}%
						\gtCommChoice{first}{}{\gtRecVar}\\
						\gtCommChoice{second}{}{}
					\end{array}
				}
			}
		}
	}\\&=&
	\gtSch{
		\gtRec{\gtRecVar}
			{
				\gtCommRaw{\roleFmt{a}}{\roleFmt{b}}
				{
					\begin{array}{l}%
						\gtCommChoice{first}{}{\gtRecVar}\\
						\gtCommChoice{second}{}{}
					\end{array}
				}
			}
	}{
		\gtRec{\gtRecVar}
			{
				\gtCommRaw{\roleFmt{c}}{\roleFmt{d}}
				{
					\begin{array}{l}%
						\gtCommChoice{first}{}{\gtRecVar}\\
						\gtCommChoice{second}{}{}
					\end{array}
				}
			}
	}{\emptyset}\\
	&=&
	\gtRec{\gtRecVar}{
		\gtCommRaw{\roleFmt{a}}{\roleFmt{b}}
		{
			\begin{array}{l}
			\gtCommChoice{first}{}
			{
				\gtCommRaw{\roleFmt{c}}{\roleFmt{d}}
				{
					\begin{array}{l}
					\gtCommChoice{first}{}{\gtRecVar}\\
					\gtCommChoice{second}{}
					{
						\gtRec{\gtRecVarii}
						{
							\gtCommRaw{\roleFmt{a}}{\roleFmt{b}}
							{
								\begin{array}{l}
									\gtCommChoice{first}{}{\gtRecVarii}\\
									\gtCommChoice{second}{}{}
								\end{array}
							}
						}
					}
					\end{array}
				}
			}\\
			\gtCommChoice{second}{}
			{
				\gtRec{\gtRecVari}
				{
					\gtCommRaw{\roleFmt{c}}{\roleFmt{d}}
					{
						\begin{array}{l}
						\gtCommChoice{first}{}{\gtRecVari}\\
						\gtCommChoice{second}{}{}
						\end{array}
					}
				}
			}
			\end{array}
		}
	}
	\end{array}
	\end{equation*}
	\begin{equation*}
		\gtProj{\roleFmt{a}}{\gtGi}{\stT[\roleFmt{a}]}
		\qquad
		\gtProj{\roleFmt{b}}{\gtGi}{\stT[\roleFmt{b}]}
		\qquad
		\gtProj{\roleFmt{c}}{\gtGi}{\stT[\roleFmt{c}]}
		\qquad
		\gtProj{\roleFmt{d}}{\gtGi}{\stT[\roleFmt{d}]}
\end{equation*}
\end{example}

We conclude the following:
\thmRemovePar*

In practice, using parallel composition can drastically decrease the size of
global types so avoiding it can be impractical.

A Haskell implementation of the serialisation algorithm
and some examples
are provided in this paper's artifact~\cite{anonymous_2026_21242753}.%
\section{Appendix for~\cref{sec:extended} (Mixed Choice Global Types)}
\label{app:extended}

\begin{definition}[Balanced Global Types]
\label{def:balanced-extend}
	For each participant, $\roleR$, we define the partial function $\gtDepth{}{\roleR}$ on global types recursively by
	$\gtDepth{\gtG}{\roleR}=1$ if $\roleR\notin\gtRoles{\gtG}$
	and $\gtFv{\gtG}=\emptyset$,
	and otherwise:
	\[
	\begin{array}{c}
\gtDepth{\gtCommSD{\roleP[i]}{\roleQ[i]}{i \in I}{\gtLab[i]}{\tyGround[i]}{\gtG[i]}}{\roleR}=
			\left\{\!\!\!\begin{array}{ll}
				1+max\{\gtDepth{\gtG[i]}{\roleR}\}_{i \in I} & \!\!\roleR\notin\bigcup_{i\in I}\{\roleP[i],\roleQ[i]\}\\
				1 & \!\!\roleR\in\bigcup_{i\in I}\{\roleP[i],\roleQ[i]\}
			\end{array}\right.
			\\
		\gtDepth{\gtPar{\gtG[0]}{\gtG[1]}}{\roleR}=
			\left\{\begin{array}{ll}
		\gtDepth{\gtG[0]}{\roleR}\quad\roleR\in\gtRoles{\gtG[0]}\\
                \gtDepth{\gtG[1]}{\roleR}\quad\roleR\in\gtRoles{\gtG[1]}\\
		\end{array}\right.
		\quad
		\gtDepth{\gtRec{\gtRecVar}{\gtG}}{\roleR} = \gtDepth{\gtG}{\roleR}
	\end{array}
	\]
A global type, $\gtG$, is \emph{balanced} iff for all $\gtG\,\gtMoveStar\,\gtGi$ and $\roleR\in\gtRoles{\gtGi}$,
$\gtDepth{\gtGi}{\roleR}$ exists.
\end{definition}

\begin{definition}[Weak Balanced Global Types]
\label{def:weak-bal-extend}
We define the predicate $\weakComm{\gtG}{\roleP}{\roleQ}$
inductively by the rules:
\[\begin{array}{c}
	\inference[WComm]{j\in I}{\weakComm{\gtCommSD{\roleP[i]}{\roleQ[i]}{i\in I}{\stLab[i]}{\tyGround[i]}{\gtG[i]}}{\roleP[j]}{\roleQ[j]}}
	\quad
	\inference[WPar]{
	i\in\{0,1\}
	&
	\weakComm{\gtG[i]}{\roleP}{\roleQ}
	}{\weakComm{\gtPar{\gtG[0]}{\gtG[1]}}{\roleP}{\roleQ}}
	\\[1ex]
	\inference[W$\mu$]{\weakComm{\gtG}{\roleP}{\roleQ}}{\weakComm{\gtRec{\gtRecVar}{\gtG}}{\roleP}{\roleQ}}
	\quad
	\inference[WCtx]{
	\exists j\in I&\weakComm{\gtG[j]}{\roleP}{\roleQ}
	&\{\roleP,\roleQ\}\cap\{\rolePi[j],\roleQi[j]\}=\emptyset
	}{\weakComm{\gtCommSD{\rolePi[i]}{\roleQi[i]}{i\in I}{\stLab[i]}{\tyGround[i]}{\gtG[i]}}{\roleP}{\roleQ}}
\end{array}\]
$\gtG$ is \emph{weak balanced} iff
$\forall\gtG\,\gtMoveStar\,\gtGi.\;
\weakComm{\gtGi}{\roleP}{\roleQ}
\implies$
$\gtDepth{\gtGi}{\roleP}$ and $\gtDepth{\gtGi}{\roleQ}$ exist.
\end{definition}

\begin{definition}[Send/Receive Balanced Global Types]
\label{def:extend-send-recv-bal}
	Define $\gtSendRoles{}$ and $\gtRecvRoles{}$ recursively on global types
	similarly to $\gtRoles{}$ except
	on communication constructors:
	\[
	\gtSendRoles{\gtCommSD{\roleP[i]}{\roleQ[i]}{i\in I}{\stLab[i]}{\tyGround[i]}{\gtG[i]}}
	=
	\bigcup_{i\in I}\left(\{\roleP[i]\}\cup
	\gtSendRoles{\gtG[i]}\setminus\{\roleQ[i]\}\right)
	\]
	\[
	\gtRecvRoles{\gtComm{\roleP[i]}{\roleQ[i]}{i\in I}{\stLab[i]}{\tyGround[i]}{\gtG[i]}}
	=
	\bigcup_{i\in I}\left(\{\roleQ[i]\}\cup
	\gtRecvRoles{\gtG[i]}\setminus\{\roleP[i]\}\right)
	\]
	
	$\gtG$ is:
	\begin{enumerate*}
	\item $\roleP$-send balanced iff
	for all $\gtG\,\gtMoveStar\,\gtGi$,
	if $\roleP\in\gtSendRoles{\gtGi}$ then
	$\gtDepth{\gtGi}{\roleP}$ exists;
	\item $\roleP$-receive balanced iff
	for all $\gtG\,\gtMoveStar\,\gtGi$,
	if $\roleP\in\gtRecvRoles{\gtGi}$ then
	$\gtDepth{\gtGi}{\roleP}$ exists; and
	send/receive balanced iff
	for all $\roleP$,
	$\gtG$ is $\roleP$-send/receive balanced.
	\end{enumerate*}
\end{definition}

\begin{definition}[Mixed Choice Liveness]%
  \label{def:mixed-live}
  A \emph{path} is a
  possibly infinite sequence of typing contexts
  $(\stEnvNew[n])_{n \in N}$,
  where $N = \{0,1,2,\ldots\}$ is a (finite or infinite) set of consecutive natural numbers
  and $\forall n, n+1 \!\in\! N$, $\stEnvNew[n] \stEnvMove \stEnvNew[n+1]$.
  
  We say that a path $(\stEnvNew[n])_{n \in N}$ is \emph{fair} iff,
  $\forall n \!\in\! N$:
    $\stEnvMoveGenAnnotP{\stEnvNew[n]}$
    implies there exists $\stEnvAnnotGenericSymi$ and $k$
    such that $N \!\ni\! k \ge n$ and
    $\ltsSubject{\stEnvAnnotGenericSym}\cap\ltsSubject{\stEnvAnnotGenericSymi}\neq\emptyset$
    and $\stEnvNew[k] \stEnvMoveAnnot{\stEnvAnnotGenericSymi} \stEnvNew[k+1]$.
    
  We say that a path $(\stEnvNew[n])_{n \in N}$ is:
 \begin{enumerate}[leftmargin=*, nosep]
  \item
  \emph{$\roleP$-send live} iff $\forall n \!\in\! N.\left(
  \stEnvApp{\stEnvNew[n]}{\roleP}\gtMove[{\roleQ\stEnvAnnotOutSym\stChoice{\stLab}{\tyGround}}]
  \implies
  \exists k, \stEnvAnnotGenericSym.
    \;N \ni k \ge n
    \;\wedge\;
    \roleP\in\ltsSubject{\stEnvAnnotGenericSym}
    \;\wedge\;
    \stEnvNew[k] \stEnvMoveGenAnnot
      \right)$;
  \item
  \emph{$\roleP$-receive live} iff $\forall n \!\in\! N.\left(
  \stEnvApp{\stEnvNew[n]}{\roleP}\gtMove[{\roleQ\stEnvAnnotInSym\stChoice{\stLab}{\tyGround}}]
  \implies
    \exists k, \stEnvAnnotGenericSym.\;
    N \ni k \ge n
    \;\wedge\;
    \roleP\in\ltsSubject{\stEnvAnnotGenericSym}
    \;\wedge\;
    \stEnvNew[k] \!\stEnvMoveGenAnnot\! \right)$; and
    \item \emph{live} iff it is $\roleP$-send live and $\roleP$-receive live for all $\roleP$.
  \end{enumerate}
\noindent A typing context $\stEnv$ is
\emph{$\roleP$-send live}/\emph{$\roleP$-receive live}/\emph{live},  
written
$\stEnvSendLiveRoleP{\roleP}{\stEnvNew}$/$\stEnvRecvLiveRoleP{\roleP}{\stEnvNew}$/$\stEnvLiveP{\stEnvNew}$, iff
all paths starting with $\stEnv$ that are fair are also
$\roleP$-send live/$\roleP$-receive live/live.
\end{definition}

Use the
mixed choice processes and the typing rules from \cite[Fig. 4]{PY2024}
with \inferrule{Sess}.

\begin{definition}[Mixed Choice Processes \cite{PY2024}]
\label{def:full-mixed-processes}
Define prefixes by: $\pi\bnfdef \roleQ\mathord{!}\ell\langle e \rangle \bnfsep
\roleQ\mathord{?}\ell( x )$.
Extend processes with mixed choice syntax:
$\textstyle\sum_{i\in I} \pi_i \mpSeq \mpP_i$.
Define $\stJudge{\mpEnv}{\mpP}{\stT}$ by the rules in \cref{def:type-rules}
and:
\[
	\inference[T-$\textstyle\sum$]
	{\left(
		\forall i\in I.\;
		\stJudge{\mpEnv}{e_i}{\tyGround[i]}
		\;\wedge\;
		\stJudge{\mpEnv}{\mpP[i]}{\stT[i]}\right)
		\;\wedge\;
		J'\subseteq J
		\;\wedge\;\left(
		\forall i\in J'.\;
		\stJudge{\mpEnv\mpEnvComp\mpEnvMap{x_i}{\tyGroundi[i]}}{\mpPi[i]}{\stTi[i]}
	\right)}
	{
		\stJudge{\mpEnv}{
		\textstyle\sum_{i\in I} \roleQ[i]\mathord{!}\stLab[i]\langle e_i \rangle \mpSeq \mpP_i
		+
		\textstyle\sum_{i\in J} \roleQi[i]\mathord{?}\stLabi( x_i ) \mpSeq \mpPi_i
		}{
		\stSum{\roleQ[i]}{i\in I}{\stChoice{\stLab[i]}{\tyGround[i]}\stSeq\stT[i]}{\oplus}
		+
		\stSum{\roleQi[i]}{i\in J'}{\stChoice{\stLabi[i]}{\tyGroundi[i]}\stSeq\stTi[i]}{\&}
		}
	}
\]
\end{definition}

\corTypingSessionsMixed*

\begin{proof}
	The proofs for \cref{cor:typingsessions}
	and \cref{thm:type-bal}
	still hold,
	using inversion of the system in \cite{PY2024}.
\end{proof}

\begin{definition}[Mixed Choice Subtyping]
\label{def:subtyping-mixed}
The \emph{subtyping relation $\tySub$} on mixed choice local types is coinductively defined by 
the rules:
\[
\begin{array}{c}
\cinference[\iruleStSubIn]{
  \forall i \in I
  &
  \stT[i] \tySub \tyTi[i]
}{
  \stSum{\roleP}{i \in I\cup J}{\stChoice{\stLab[i]}{\tyGround[i]} \stSeq \stT[i]}{\&}
  \tySub
  \stSum{\roleP}{i \in I}{\stChoice{\stLab[i]}{\tyGround[i]} \stSeq \stTi[i]}{\&}
}
\\[2ex]
\cinference[\iruleStSubOut]{
  \forall i \in I
  &
  \stT[i] \tySub \stTi[i]
}{
  \stSum{\roleP}{i \in I}{\stChoice{\stLab[i]}{\tyGround[i]} \stSeq \stT[i]}{\oplus}
  \tySub
  \stSum{\roleP}{i \in I\cup J}{\stChoice{\stLab[i]}{\tyGround[i]} \stSeq \stTi[i]}{\oplus}
}
\\[2ex]
\cinference[\iruleStSubEnd]{}{
  \stEnd \tySub \stEnd
}
\quad
\cinference[\iruleStSubRecL]{
  \stT{}[\stRec{\stRecVar}{\stT}/\stRecVar] \tySub \stTi
}{
  \stRec{\stRecVar}{\stT} \tySub \stTi
}
\quad
\cinference[\iruleStSubRecR]{
  \stT \tySub \stTi{}[\stRec{\stRecVar}{\stTi}/\stRecVar]
}{
  \stT \tySub \stRec{\stRecVar}{\stTi}
}
\\[2ex]
\cinference[\iruleStSubSum]{
	|K|>1
	&
	\stSum{\roleP[k]}{i \in I_k}{\stChoice{\stLab[k,i]}{\tyGround[k,i]} \stSeq \stT[k,i]}{\dagger_k}
	\tySub
	\stSum{\roleP[k]}{i \in I'_k}{\stChoice{\stLab[k,i]}{\tyGround[k,i]} \stSeq \stTi[k,i]}{\dagger_k}
}{
	\stSum{\roleP[k]}{k\in K,\,i \in I_k}{\stChoice{\stLab[k,i]}{\tyGround[k,i]} \stSeq \stT[k,i]}{\dagger_k}
	\tySub
	\stSum{\roleP[k]}{k\in K,\,i \in I'_k}{\stChoice{\stLab[k,i]}{\tyGround[k,i]} \stSeq \stTi[k,i]}{\dagger_k}
}
\end{array}
\]
\end{definition}

The following lemma is key to type checking.
Any derivation of
$\stJudge{\mpEnv}{\mpRec{\mpX}{\mpP}}{\stT}$
requires a derivation of
$\stJudge{\mpEnv\mpEnvComp\mpEnvMap{\mpX}{\stTi}}{\mpP}{\stTi}$
for some $\stTi\tySub\stT$.
Also,
$\exists \stTi\tySub\stT.\;\stJudge{\mpEnv\mpEnvComp\mpEnvMap{\mpX}{\stTi}}{\mpP}{\stTi}$
implies
$\stJudge{\mpEnv}{\mpRec{\mpX}{\mpP}}{\stT}$.
Thus type checking
$\stJudge{\mpEnv}{\mpRec{\mpX}{\mpP}}{\stT}$
requires finding $\stTi$.
This could be done via type inference (\cite[\S~5]{DBLP:journals/pacmpl/UdomsrirungruangY25}).
We propose an alternative option.

\begin{definition}[Paths to Recursion Variables]
\label{def:rec-paths}
We define $\labelpath{\mpP}{\mpX}$
to be the set of transition label vectors
observed down $\mpP$ ending in $\mpX$,
by the inductive definition:
\[
\begin{array}{rcl}
	\labelpath{\mpNil}{\mpX}&=&\emptyset
	\\
	\labelpath{\mpX}{\mpX}&=&\set{\epsilon}
	\\
	\labelpath{\mpXi}{\mpX}&=&\emptyset
	\\
	\labelpath{\mpIf{e}{\mpP}{\mpPi}}{\mpX}&=&\labelpath{\mpP}{\mpX}\cup\labelpath{\mpPi}{\mpX}
	\\
	\labelpath{\roleQ\mathord{!}\ell\langle e \rangle \mpSeq \mpP}{\mpX}&=&
	\set{
		\stFmt{\roleQ{\oplus}\ell}\vec{\stEnvAnnotGenericSym}
		\suchthat
		\vec{\stEnvAnnotGenericSym}\in\labelpath{\mpP}{\mpX}
	}
	\\
	\labelpath{\roleQ\mathord{?}\ell( x ) \mpSeq \mpP}{\mpX}&=&
	\set{
		\stFmt{\roleQ{\&}\ell}\vec{\stEnvAnnotGenericSym}
		\suchthat
		\vec{\stEnvAnnotGenericSym}\in\labelpath{\mpP}{\mpX}
	}
	\\
	\labelpath{
		\textstyle\sum_{i\in I} \pi_i \mpSeq \mpP[i]
	}{\mpX}
	&=&
	\bigcup_{i\in I}\labelpath{\pi_i \mpSeq \mpP[i]}{\mpX}
\end{array}
\]
\end{definition}

\begin{lemma}[Decidable Inversion of Recursion]
\label{lem:decide-inv-rec}
	Let $\mpEnv$, $\stT$ and $\mpP$ be given and let:
	\[
	\stT[\text{\tiny max}] =
	inf_{\tySub}\{
		\stTii
		\suchthat
		\exists
		\vec{\stEnvAnnotGenericSym[1]},\dots,
		\vec{\stEnvAnnotGenericSym[n]}\in\labelpath{\mpP}{\mpX}.\;
		\stT\,\gtMove[\vec{\stEnvAnnotGenericSym[1]}\dots\vec{\stEnvAnnotGenericSym[n]}]\,\stTii
	\}
	\]
	If
	$\stJudge{\mpEnv\mpEnvComp\mpEnvMap{\mpX}{\stTi}}{\mpP}{\stTi}$
	and $\stTi\tySub\stT$, then
	$\stT[\text{\tiny max}]$ exists,
	$\stTi\tySub\stT[\text{\tiny max}]\tySub\stT$ and
	$\stJudge{\mpEnv\mpEnvComp\mpEnvMap{\mpX}{\stT[\text{\tiny max}]}}{\mpP}{\stT[\text{\tiny max}]}$.
\end{lemma}

\begin{proof}
	By inversion of typing,
	$\forall \stTii.
		\left(\exists
		\vec{\stEnvAnnotGenericSym[1]},\dots,
		\vec{\stEnvAnnotGenericSym[n]}\in\labelpath{\mpP}{\mpX}.\;
		\stT\,\gtMove[\vec{\stEnvAnnotGenericSym[1]}\dots\vec{\stEnvAnnotGenericSym[n]}]\,\stTii\right)
		\implies
		\stTi\tySub\stTii$.
	Thus, $\stT[\text{\tiny max}]$ exists and
	$\stTi\tySub\stT[\text{\tiny max}]\tySub\stT$
	as $\stT\,\gtMove[\epsilon]\,\stT$.
	
	By induction on
	$\stJudge{\mpEnv\mpEnvComp\mpEnvMap{\mpX}{\stTi}}{\mpP}{\stTi}$,
	$\stJudge{\mpEnv\mpEnvComp\mpEnvMap{\mpX}{\stT[\text{\tiny max}]}}{\mpP}{\stT[\text{\tiny max}]}$
	since $\stT[\text{\tiny max}]\,\gtMove[\vec{\stEnvAnnotGenericSym}]\,\stT[\text{\tiny max}]$
	for all $\vec{\stEnvAnnotGenericSym}\in\labelpath{\mpP}{\mpX}$
	that must occur in the derivation.
\end{proof}

\begin{definition}[Plain Inductive Merge]
	\label{def:plainindmerge-extend}
	The plain inductive merge is defined by $\stMerge{}{\{\stT\}}=\{\stT\}$.
\end{definition}

\begin{definition}[Full Inductive Merge]
	\label{def:fullindmerge-extend}
	We define $\funcMerge{}$
	to be a partial function from non-empty and finite sets of
	local types to a local type recursively by:
	\begin{itemize}
		\item $\funcMerge{\left\{
			\stEnd
		\right\}}=\stEnd$
		\item  $\funcMerge{\left\{
			\stRecVar
		\right\}}=\stRecVar$
		\item $\funcMerge{\left\{
			\stRec{\stRecVar}{\stT[i]}\suchthat i\in I
		\right\}}=\stRec{\stRecVar}{\funcMerge{\left\{\stT[i]\suchthat i\in I\right\}}}$
		\item $\funcMerge{\left\{
			\stSum{\roleP}{j\in J}{\stChoice{\stLab[j]}{\tyGround[j]}\stSeq\stT[i,j]}{\oplus}
			+
			\stSum{\roleP}{j\in J'_i}{\stChoice{\stLabi[j]}{\tyGroundi[j]}\stSeq\stTi[i,j]}{\&}
		\right\}
		\suchthat
		i\in I
		}
		=$\\\qquad
		$\stSum{\roleP}{j\in J}{
		\stChoice{\stLab[j]}{\tyGround[j]}\stSeq
		\funcMerge{\{\stT[i,j]\suchthat i \in I\}}
		}{\oplus}
			+
		\stSum{\roleP}{j\in \bigcup_{i\in I}J'_i}{
		\stChoice{\stLabi[j]}{\tyGroundi[j]}\stSeq
		\funcMerge{\{\stT[i,j]\suchthat j\in J'_i\}}
		}{\&}
		$
	\end{itemize}
	We define \emph{full inductive merge}
	by $\stMerge{}{\mathcal{T}}=\left\{\begin{array}{lr}
		\{\funcMerge{\mathcal{T}}\}
		&\text{if it exists}
		\\
		\emptyset&\text{otherwise}
	\end{array}\right\}$
\end{definition}

\begin{definition}[Plain Coinductive Merge]
\label{def:plain-merge-extend}
We define
$\stMerge{}{\{\stT[i]\}_{i\in I}}=\{\stT\suchthat \forall i\in I,\stT\bisim\stT[i]\}$.
\end{definition}

\begin{theorem}[Standard Merges]
	\label{thm:merge-good-mixed}
	The full/plain inductive/coinductive
	merges are good.
\end{theorem}

The proofs are the same as \cref{thm:merge-good} but
using the mixed choice subtyping instead of the standard subtyping.

\thmProjGoodMixed*

\begin{proof}
The proofs are the same as \cref{thm:good-inductive-proj},
differing slightly when handling \inferrule{\iruleGtMoveCtx'}
and \inferrule{\iruleGtMoveComm'} and
proving rule \inferrule{$\projGoodType$}.
We repeat the parts that must be slightly changed,
with their modifications.

If $\gtG \,\gtMove[\ltsSendRecv{\roleP}{\roleQ}{\stChoice{\stLab}{\tyGround}}]\, \gtGi$ by
    \inferrule{\iruleGtMoveCtx'} and
    $\gtProj{\roleR}{\gtG}{\stT}$,
	then $\gtG=\gtCommSD{\rolePi[i]}{\roleQi[i]}{i \in I}{\gtLab[i]}{\tyGround[i]}{\gtGi[i]}$ and\\
    $\gtGi=\gtCommSD{\rolePi[i]}{\roleQi[i]}{i \in I}{\gtLab[i]}{\tyGround[i]}{\gtGii[i]}$ and
    $\gtGi[i] \,\gtMove[\ltsSendRecv{\roleP}{\roleQ}{\stChoice{\stLab}{\tyGround}}]\, \gtGii[i]$ for $i\in I$.\\
	$\gtProj{\roleR}{\gtG}{\stT}$, so
    there are $\stT[i]$ for $i\in I$ with
    $\gtProj{\roleR}{\gtGi[i]}{\stT[i]}$ and
    $\stMerge{}{\{\stT[i]\suchthat i\in I\}}\ni\stT$.\\
	By the I.H., there are $\stTi[i]$ with
    $\gtProj{\roleR}{\gtGii[i]}{\stTi[i]}$.\\
	If $\roleR=\roleP$, then
    $\stT[i]\,\gtMove[{\roleQ\stEnvAnnotOutSym\stChoice{\stLab}{\tyGround}}]\,\stTi[i]$ so 
    there is $\stTi$ with
    $\stMerge{}{\{\stTi[i]\suchthat i\in I\}}\ni\stTi$ and
    $\stT\,\gtMove[{\roleQ\stEnvAnnotOutSym\stChoice{\stLab}{\tyGround}}]\, \stTi$.\\
	If $\roleR=\roleQ$, then
    $\stT[i]\,\gtMove[{\roleP\stEnvAnnotInSym\stChoice{\stLab}{\tyGround}}]\,\stTi[i]$ so
    there is $\stTi$ with
    $\stMerge{}{\{\stTi[i]\suchthat i\in I\}}\ni\stTi$ and
    $\stT\,\gtMove[{\roleP\stEnvAnnotInSym\stChoice{\stLab}{\tyGround}}]\,\stTi$.\\
	If $\roleR\notin\bigcup_{i\in I}\{\roleP,\rolePi[i],\roleQ,\roleQi[i]\}$,
    $\stMerge{}{\{\stT[i],\stTi[i]\}}\ni\stT[i]$ for $i\in I$, so
    $\stMerge{}{\left(\{\stT[i]\suchthat i\in I\} \cup \{\stTi[i]\suchthat i\in I\}\right)}\ni \stT$, so
    let $\stMerge{}{\{\stTi[i]\suchthat i\in I\}}\ni \stTi$, so
    $\gtProj{\roleR}{\gtGi}{\stTi}$ and
    $\stMerge{}{\{\stTi,\stT\}} \ni \stT$.\\
	If $\roleR\in\bigcup_{i\in I}\{\rolePi[i],\roleQi[i]\}$,
    then we remove the head of $\stT$ and proceed similar to the above case
    before adding the head onto a merge of $\{\stT[i]\suchthat i\in I\}$
    by \inferrule{\mergeGoodCtx'}
    and
    reaching the same conclusion.\\
    If $\gtG \,\gtMove[\ltsSendRecv{\roleP}{\roleQ}{\stChoice{\stLab}{\tyGround}}]\, \gtGi$ by
    \inferrule{\iruleGtMoveComm'} and
    $\gtProj{\roleR}{\gtG}{\stT}$, then
    $\gtG=\gtCommSD{\roleP}{\roleQ}{i \in I}{\gtLab[i]}{\tyGround[i]}{\gtGi[i]}
    +
    \gtCommSD{\roleQ}{\roleP}{i \in I'}{\gtLabi[i]}{\tyGroundi[i]}{\gtGii[i]}
    $ and
    there is some $j\in I$ such that
    $\gtGi=\gtGi[j]$, $\stLab=\stLab[j]$, and
    $\tyGround=\tyGround[j]$.\\
	If $\roleR=\roleP$, then
    $\unfoldOne{\stT}=
\stSum{\roleQ}{i \in I}{\stChoice{\stLab[i]}{\tyGround[i]} \stSeq
	        \stTi[i]}{\oplus}
+
\stSum{\roleQ}{i \in I'}{\stChoice{\stLabi[i]}{\tyGroundi[i]} \stSeq
	        \stTii[i]}{\&}
$ and
            $\gtProj{\roleP}{\gtGi[i]}{\stTi[i]}$ for $i\in I$, so
            $\stT\,\gtMove[{\roleQ\stEnvAnnotOutSym\stChoice{\stLab[j]}{\tyGround[j]}}]\,\stTi[j]$.\\
	If $\roleR=\roleQ$, then
    $\unfoldOne{\stT}=\stSum{\roleQ}{i \in I}{\stChoice{\stLab[i]}{\tyGround[i]} \stSeq
	        \stTi[i]}{\&}
+
\stSum{\roleQ}{i \in I'}{\stChoice{\stLabi[i]}{\tyGroundi[i]} \stSeq
	        \stTii[i]}{\oplus}
$ and
            $\gtProj{\roleQ}{\gtGi[i]}{\stTi[i]}$ for $i\in I$, so
            $\stT\,\gtMove[{\roleP\stEnvAnnotInSym\stChoice{\stLab[j]}{\tyGround[j]}}]\,\stTi[j]$.
            
            Suppose that $\gtProj{\roleP}{\gtG}{\stT}$,
    $\gtProj{\roleQ}{\gtG}{\stTi}$,
    $\stT\,\gtMove[{\roleQ\stEnvAnnotOutSym\stChoice{\stLab}{\tyGround}}]$, and
    $\stTi\,\gtMove[{\roleP\stEnvAnnotInSym\stChoice{\stLabi}{\tyGroundi}}]$.\\
    So, $\roleP\in\gtRoles{\gtG}$ and
	$\unfoldOne{\stT}$ has a send to $\roleQ$ as its head.
	By induction on the derivation of
	$\roleP\in\gtRoles{\gtG}$,
	$\weakComm{\gtG}{\roleP}{\roleQ}$.
	Thus, $\gtDepth{\gtG}{\roleP}$ exists.\\
	We induct on $\gtDepth{\gtG}{\roleP}$.\\
	If $\gtDepth{\gtG}{\roleP}=1$, then
    $\unfoldOne{\gtG}$ is a parallel composition of 1 or more components, one of which is
of the form $
\gtCommSD{\roleP}{\roleQi}{i \in I}{\gtLab[i]}{\tyGround[i]}{\gtG[i]}
+
\gtCommSD{\roleQi}{\roleP}{i \in I'}{\gtLabi[i]}{\tyGroundi[i]}{\gtGi[i]}
$.\\
	So, $\unfoldOne{\stT}=
	\stSum{\roleQi}{i \in I}{\stChoice{\stLab[i]}{\tyGround[i]} \stSeq \stT[i]}{\oplus}
	+
	\stSum{\roleQi}{i \in I'}{\stChoice{\stLabi[i]}{\tyGroundi[i]} \stSeq \stTi[i]}{\&}
	$ and
    $\stT\,\gtMove[{\roleQ\stEnvAnnotOutSym\stChoice{\stLab}{\tyGround}}]$,
    hence $\roleQi=\roleQ$ and there is $j\in I$ such that
    $\stLab=\stLab[i]$ and
    $\tyGround=\tyGround[i]$.\\
	Thus, $\gtG \,\gtMove[\ltsSendRecv{\roleP}{\roleQ}{\stChoice{\stLab}{\tyGround}}]$.\\
	If $\gtDepth{\gtG}{\roleP}>1$,
    then $\unfoldOne{\gtG}$ is a parallel composition of 1 or more components, one of which contains
$\roleP$ as a participant.\\
Say this component is
$\gtCommSD{\rolePi[i]}{\roleQi[i]}{i \in I}{\gtLab[i]}{\tyGround[i]}{\gtG[i]}$ where
    $\roleP\notin\bigcup_{i\in I}\{\rolePi[i],\roleQi[i]\}$ (as $\gtDepth{\gtG}{\roleP}>1$).\\
	If $\roleQ\in\bigcup_{i\in I}\{\rolePi[i],\roleQi[i]\}$,
    then $\stTi$ will make a reduction involving the other participant, which is not $\roleP$,
    contradicting $\stTi\,\gtMove[{\roleP\stEnvAnnotInSym\stChoice{\stLabi}{\tyGroundi}}]$.\\
	So, there are $\stT[i]$, $\stTi[i]$ for $i\in I$ such that
    $\gtProj{\roleP}{\gtG[i]}{\stT[i]}$,
    $\gtProj{\roleQ}{\gtG[i]}{\stTi[i]}$,
    $\stMerge{}{\{\stT[i]\suchthat i\in I\}}\ni\stT$,
    and $\stMerge{}{\{\stTi[i]\suchthat i\in I\}}\ni\stTi$.\\
	Fix $i\in I$.\\
    We have $\gtDepth{\gtG[i]}{\roleP}\leq \gtDepth{\gtG}{\roleP}-1$,
    $\stT[i]\,\gtMove[{\roleQ\stEnvAnnotOutSym\stChoice{\stLab}{\tyGround}}]$, and
    $\stTi[i]\,\gtMove[{\roleP\stEnvAnnotInSym\stChoice{\stLabii}{\tyGroundii}}]$ for some $\stLabii$ and $\tyGroundii$.\\
	By the I.H., $\gtG[i] \,\gtMove[\ltsSendRecv{\roleP}{\roleQ}{\stChoice{\stLab}{\tyGround}}]$.\\
	This is true for all $i\in I$ so $\gtG \,\gtMove[\ltsSendRecv{\roleP}{\roleQ}{\stChoice{\stLab}{\tyGround}}]$.
\end{proof}

\begin{example}[Detail for Timeout (\cref{{ex:directed-mixed-timeout}})]
\label{ex:directed-mixed-full}
We derive the inductive/coinductive
projection of $\gtG[\text{\tiny TO}]$
using full inductive/coinductive merging.
\[\begin{array}{l}
	\gtProj{\roleFmt{A}}{
	\gtEnd
	}{
	\stEnd
	}
	\\
	\gtProj{\roleFmt{A}}{
	\gtFmt{
	\roleFmt{C}\to\roleFmt{A}:\gtMsgFmt{a_5}
	}
	}{
	\stFmt{\roleFmt{C}{\&}\stLabFmt{a_5}}
	}
	\\
	\gtProj{\roleFmt{A}}{
	\gtFmt{
	\roleFmt{B}\to\roleFmt{A}:\gtMsgFmt{a_4}{\gtSeq}
			\roleFmt{C}\to\roleFmt{A}:\gtMsgFmt{a_5}
	}
	}{
	\stFmt{\roleFmt{B}{\&}\stLabFmt{a_4}\stSeq
			\roleFmt{C}{\&}\stLabFmt{a_5}}
	}
	\\
	\gtProj{\roleFmt{A}}{
	\gtFmt{
	\roleFmt{B}\to\roleFmt{C}:\gtMsgFmt{a_3}{\gtSeq}
			\roleFmt{B}\to\roleFmt{A}:\gtMsgFmt{a_4}{\gtSeq}
			\roleFmt{C}\to\roleFmt{A}:\gtMsgFmt{a_5}
	}
	}{
	\stFmt{\roleFmt{B}{\&}\stLabFmt{a_4}\stSeq
			\roleFmt{C}{\&}\stLabFmt{a_5}}
	}
	\\
	\gtProj{\roleFmt{A}}{
	\gtFmt{
	\roleFmt{A}\to\roleFmt{C}:\gtMsgFmt{a_2}{\gtSeq}
			\roleFmt{B}\to\roleFmt{C}:\gtMsgFmt{a_3}{\gtSeq}
			\roleFmt{B}\to\roleFmt{A}:\gtMsgFmt{a_4}{\gtSeq}
			\roleFmt{C}\to\roleFmt{A}:\gtMsgFmt{a_5}
	}
	}{
	\stFmt{\roleFmt{C}{\oplus}\stLabFmt{a_2}\stSeq
			\roleFmt{B}{\&}\stLabFmt{a_4}\stSeq
			\roleFmt{C}{\&}\stLabFmt{a_5}}
	}
	\\
	\gtProj{\roleFmt{A}}{
	\gtFmt{
	\roleFmt{A}\to\roleFmt{C}:\gtMsgFmt{TOc}
	}
	}{
	\stFmt{\roleFmt{C}{\oplus}\stLabFmt{TOc}}
	}
	\\
	\gtProj{\roleFmt{A}}{
	\gtG[\text{\tiny TO}]
	}{
	\stT[\roleFmt{A}]
	}
	\\
	\gtProj{\roleFmt{B}}{
	\gtEnd
	}{
	\stEnd
	}
	\\
	\gtProj{\roleFmt{B}}{
	\gtFmt{
	\roleFmt{C}\to\roleFmt{A}:\gtMsgFmt{a_5}
	}
	}{
	\stEnd
	}
	\\
	\gtProj{\roleFmt{B}}{
	\gtFmt{
	\roleFmt{B}\to\roleFmt{A}:\gtMsgFmt{a_4}{\gtSeq}
			\roleFmt{C}\to\roleFmt{A}:\gtMsgFmt{a_5}
	}
	}{
	\stFmt{
	\roleFmt{A}{\oplus}\stLabFmt{a_4}
	}
	}
	\\
	\gtProj{\roleFmt{B}}{
	\gtFmt{
	\roleFmt{B}\to\roleFmt{C}:\gtMsgFmt{a_3}{\gtSeq}
			\roleFmt{B}\to\roleFmt{A}:\gtMsgFmt{a_4}{\gtSeq}
			\roleFmt{C}\to\roleFmt{A}:\gtMsgFmt{a_5}
	}
	}{
	\stFmt{\roleFmt{C}{\oplus}\stLabFmt{a_3}\stSeq
			\roleFmt{A}{\oplus}\stLabFmt{a_4}}
	}
	\\
	\gtProj{\roleFmt{B}}{
	\gtFmt{
	\roleFmt{A}\to\roleFmt{C}:\gtMsgFmt{a_2}{\gtSeq}
			\roleFmt{B}\to\roleFmt{C}:\gtMsgFmt{a_3}{\gtSeq}
			\roleFmt{B}\to\roleFmt{A}:\gtMsgFmt{a_4}{\gtSeq}
			\roleFmt{C}\to\roleFmt{A}:\gtMsgFmt{a_5}
	}
	}{
	\stFmt{\roleFmt{C}{\oplus}\stLabFmt{a_3}\stSeq
			\roleFmt{A}{\oplus}\stLabFmt{a_4}}
	}
	\\
	\gtProj{\roleFmt{B}}{
	\gtFmt{
	\roleFmt{A}\to\roleFmt{C}:\gtMsgFmt{TOc}
	}
	}{
	\stEnd}
	\\
	\gtProj{\roleFmt{B}}{
	\gtG[\text{\tiny TO}]
	}{
	\stT[\roleFmt{B}]
	}
	\\
	\gtProj{\roleFmt{C}}{
	\gtEnd
	}{
	\stEnd
	}
	\\
	\gtProj{\roleFmt{C}}{
	\gtFmt{
	\roleFmt{C}\to\roleFmt{A}:\gtMsgFmt{a_5}
	}
	}{
	\stFmt{\roleFmt{A}{\oplus}\stLabFmt{a_5}}
	}
	\\
	\gtProj{\roleFmt{C}}{
	\gtFmt{
	\roleFmt{B}\to\roleFmt{A}:\gtMsgFmt{a_4}{\gtSeq}
			\roleFmt{C}\to\roleFmt{A}:\gtMsgFmt{a_5}
	}
	}{
	\stFmt{
	\roleFmt{A}{\oplus}\stLabFmt{a_5}
	}
	}
	\\
	\gtProj{\roleFmt{C}}{
	\gtFmt{
	\roleFmt{B}\to\roleFmt{C}:\gtMsgFmt{a_3}{\gtSeq}
			\roleFmt{B}\to\roleFmt{A}:\gtMsgFmt{a_4}{\gtSeq}
			\roleFmt{C}\to\roleFmt{A}:\gtMsgFmt{a_5}
	}
	}{
	\stFmt{\roleFmt{B}{\&}\stLabFmt{a_3}\stSeq
			\roleFmt{A}{\oplus}\stLabFmt{a_5}}
	}
	\\
	\gtProj{\roleFmt{C}}{
	\gtFmt{
	\roleFmt{A}\to\roleFmt{C}:\gtMsgFmt{a_2}{\gtSeq}
			\roleFmt{B}\to\roleFmt{C}:\gtMsgFmt{a_3}{\gtSeq}
			\roleFmt{B}\to\roleFmt{A}:\gtMsgFmt{a_4}{\gtSeq}
			\roleFmt{C}\to\roleFmt{A}:\gtMsgFmt{a_5}
	}
	}{
	\stFmt{\roleFmt{A}{\&}\stLabFmt{a_2}\stSeq
			\roleFmt{B}{\&}\stLabFmt{a_3}\stSeq
			\roleFmt{A}{\oplus}\stLabFmt{a_5}}
	}
	\\
	\gtProj{\roleFmt{C}}{
	\gtFmt{
	\roleFmt{A}\to\roleFmt{C}:\gtMsgFmt{TOc}
	}
	}{
	\stFmt{\roleFmt{A}{\&}\stLabFmt{TOc}}}
	\\
	\funcMerge{\{
	\stFmt{\roleFmt{A}{\&}\stLabFmt{TOc}},\,
	\stFmt{\roleFmt{A}{\&}\stLabFmt{a_2}\stSeq
			\roleFmt{B}{\&}\stLabFmt{a_3}\stSeq
			\roleFmt{A}{\oplus}\stLabFmt{a_5}}
	\}}=\stT[\roleFmt{C}]
	\\
	\relMerge{\{
	\stFmt{\roleFmt{A}{\&}\stLabFmt{TOc}},\,
	\stFmt{\roleFmt{A}{\&}\stLabFmt{a_2}\stSeq
			\roleFmt{B}{\&}\stLabFmt{a_3}\stSeq
			\roleFmt{A}{\oplus}\stLabFmt{a_5}}
	\}}{\stT[\roleFmt{C}]}
	\\
	\gtProj{\roleFmt{C}}{
	\gtG[\text{\tiny TO}]
	}{
	\stT[\roleFmt{C}]
	}
	\end{array}
\]
\end{example}

\begin{proposition}[Balanced Checking]
\label{prop:directed-balanced}
	Enumerate the participants.
	
	Define $F$ from directed mixed choice global types to standard global types
	by
	\[
	F\!\left(
	\gtCommSD{\roleP}{\roleQ}{i\in I}{\gtLab[i]}{\tyGround[i]}{\gtG[i]}
	+
	\gtCommSD{\roleQ}{\roleP}{i\in J}{\gtLab[i]}{\tyGround[i]}{\gtG[i]}
	\right)
	=\gtComm{\roleP}{\roleQ}{(i,\roleR)\in I\times\{\roleP\}\cup J\times\{\roleQ\}}{\stLab[i,\roleR]}{\tyGround[i]}{\gtG[i]}
	\]
	and $\roleP<\roleQ$
	and homomorphic elsewhere.
	
	Directed mixed choice global type $\gtG$
	is:
	\begin{enumerate*}
		\item weak balanced iff $F\!\left(\gtG\right)$
		is weak balanced and
		\item balanced iff $F\!\left(\gtG\right)$
		is balanced.
	\end{enumerate*}
\end{proposition}

\begin{proof}
	By induction on $|\gtG|$,
	$\gtDepth{\gtG}{\roleP}$
	exists iff $\gtDepth{F\!\left(\gtG\right)}{\roleP}$ exists.
	
	By induction on $|\gtG|$,
	$\gtRoles{\gtG}=\gtRoles{F\!\left(\gtG\right)}$.
	
	By induction on $|\gtG|$ if $\roleP<\roleQ$, then
	$\weakComm{\gtG}{\roleP}{\roleQ}\;\vee\;\weakComm{\gtG}{\roleQ}{\roleP}
	\iff
	\weakComm{F\!\left(\gtG\right)}{\roleP}{\roleQ}$.
	
	If $\gtG\,\gtMove\,\gtGi$
	by head reduction, then
	$F\!\left(\gtG\right)\,\gtMove\,F\!\left(\gtGi\right)$.
	
	If $F\!\left(\gtG\right)\,\gtMove\,\gtGii$
	by head reduction, then
	$\gtG\,\gtMove\,\gtGi$ and $\gtGii=F\!\left(\gtGi\right)$.
	
	To check balanced and weak balanced,
	we only need to check head reductions.
	Thus,
	$\gtG$ is weak balanced iff
	$F\!\left(\gtG\right)$ is weak balanced
	and
	$\gtG$ is balanced iff
	$F\!\left(\gtG\right)$ is balanced.
\end{proof}

\section{Appendix for~\cref{sec:async} (Asynchronous Semantics)}
\label{app:async}
\subsection{Appendix for~\cref{sec:asyntax} (Syntax and Semantics)}

\begin{definition}[Roles]
	\label{def:aroles}
	We define
	the function $\gtARoles{}$,
	to return the set of participants in a global type
	that may be the subject
	of a transition,
	by the following recursive definition:
	\begin{itemize}
		\item $\gtARoles{\gtEnd}=\emptyset$
		\item $\gtARoles{\gtRecVar}=\emptyset$
		\item $\gtARoles{\gtRec{\gtRecVar}{\gtG}}=\gtARoles{\gtG}$
		\item $\gtARoles{\gtPar{\gtG[L]}{\gtG[R]}}=\gtARoles{\gtG[L]}\cup\gtARoles{\gtG[R]}$
		\item $\gtARoles{\gtComm{\roleP}{\roleQ}{i \in I}{\gtLab[i]}{\tyGround[i]}{\gtG[i]}}=\{\roleP,\roleQ\}\cup\bigcup_{i\in I}\gtARoles{\gtG[i]}$
		\item $\gtARoles{\gtCommSquig{\roleP}{\roleQ}{\gtLab}{}{\gtLab}{\tyGround}{\gtG}}=\{\roleQ\}\cup\gtARoles{\gtG}$
	\end{itemize}
		We define
	the function $\gtQRoles{}$,
	to return the set of participants in a global type
	that have sent
	an unreceived message,
	by the following recursive definition:
	\begin{itemize}
		\item $\gtQRoles{\gtEnd}=\emptyset$
		\item $\gtQRoles{\gtRecVar}=\emptyset$
		\item $\gtQRoles{\gtRec{\gtRecVar}{\gtG}}=\gtQRoles{\gtG}$
		\item $\gtQRoles{\gtPar{\gtG[L]}{\gtG[R]}}=\gtQRoles{\gtG[L]}\cup\gtQRoles{\gtG[R]}$
		\item $\gtQRoles{\gtComm{\roleP}{\roleQ}{i \in I}{\gtLab[i]}{\tyGround[i]}{\gtG[i]}}=\bigcup_{i\in I}\gtQRoles{\gtG[i]}$
		\item $\gtQRoles{\gtCommSquig{\roleP}{\roleQ}{\gtLab}{}{\gtLab}{\tyGround}{\gtG}}=\{\roleP\}\cup\gtQRoles{\gtG}$
	\end{itemize}
	We may now define
	$\gtRoles{\gtG}=\gtARoles{\gtG}\cup\gtQRoles{\gtG}$.

We define $\gtMRoles{\gtG}$ on global types
to be the set of pairs of participants that have an en-route transmission
from the first to the second,
by the recursive definition:
\begin{itemize}
\item$\gtMRoles{\gtEnd}= \emptyset$
\item$\gtMRoles{\gtRecVar}  =  \emptyset$
\item$\gtMRoles{\gtRec{\gtRecVar}{\gtG}}  =  \gtMRoles{\gtG}$
\item $\gtMRoles{\gtComm{\roleP}{\roleQ}{i \in
I}{\gtLab[i]}{\tyGround[i]}{\gtG[i]}} = \bigcup_{i\in
I}\gtMRoles{\gtG[i]}$
\item$\gtMRoles{\gtCommSquig{\roleP}{\roleQ}{\gtLab}{}{\gtLab}{\tyGround}{\gtG}} = \{(\roleP,\roleQ)\}\cup\gtMRoles{\gtG}$
\item $\gtMRoles{\gtPar{\gtG}{\gtGi}}=\gtMRoles{\gtG}\cup\gtMRoles{\gtGi}$
\end{itemize}
\end{definition}

\propQueueReductions*

\begin{restatable}[Precise Queue Reductions]{proposition}{propPreciseQueueReductions}
\label{prop:precise-queue-reductions}
	Suppose that $\gtQu{\gtG}{\roleR}$ exists and $\gtG\,\gtMoveA[\stEnvAnnotGenericSym]\,\gtGi$.
	If:
	$\stEnvAnnotGenericSym=\ltsBra{\roleP}{\roleR}{\stChoice{\stLab}{\tyGround}}$
	then
	$\gtQu{\gtG}{\roleR}=\quCons{\quMsg{\roleP}{\stLab}{\tyGround}}{\gtQu{\gtGi}{\roleR}}$;
	$\stEnvAnnotGenericSym=\ltsSel{\roleR}{\roleP}{\stChoice{\stLab}{\tyGround}}$
	then
	$\gtQu{\gtGi}{\roleR}=\quCons{\quMsg{\roleP}{\stLab}{\tyGround}}{\gtQu{\gtG}{\roleR}}$; otherwise
	$\gtQu{\gtG}{\roleR}=\gtQu{\gtGi}{\roleR}$
\end{restatable}

\begin{proof}
	Induction on the derivation of
	$\gtG\,\gtMoveA[\stEnvAnnotGenericSym]\,\gtGi$.
	
	Note that
	$\gtG\,\gtMove[\stEnvAnnotGenericSym]\,\gtGi$
	implies
	$\gtG\,\gtMoveA[\stEnvAnnotGenericSym]\,\gtGi$.
\end{proof}

Pischke et al. \cite{pischke2026asynchronousglobalprotocolsprecisely}
used the condition $\text{balanced}^+$
and combined projection of types with projection of queues.

Suppose that $\gtG$
has no en-route transmissions
and $\gtG\,\gtMoveStar\,\gtGi$.
An important observation is that for all $\roleP$ and $\roleQ$, every path down $\gtGi$ encounters the same
number of en-route transmissions from $\roleP$ to $\roleQ$.
We introduce an inductive partial function to formalise this.

\begin{definition}[Counting En-Route Messages]
\label{def:en-route-count}
We define $\gtCount{\gtG}{\roleP}{\roleQ}$ by the following recursive definition:
\begin{equation*}
	\begin{array}{rcl}
		\gtCount{\gtEnd}{\roleP}{\roleQ} & = & 0\\[4ex]
		\gtCount{\gtRecVar}{\roleP}{\roleQ} & = & 0\\[4ex]
		\gtCount{\gtRec{\gtRecVar}{\gtG}}{\roleP}{\roleQ} & = & \left\{
			\begin{array}{ll}
			\gtCount{\gtG}{\roleP}{\roleQ} & \gtRecVar\not\in\gtFv{\gtG}\text{ or }\gtCount{\gtG}{\roleP}{\roleQ}=0\\
			\text{undefined} & \text{otherwise}
			\end{array}
		\right.\\[4ex]
		\gtCount{\gtPar{\gtG[0]}{\gtG[1]}}{\roleP}{\roleQ} & = &
		\left\{
		\begin{array}{ll}
		\gtCount{\gtG[i]}{\roleP}{\roleQ}&\{\roleP,\roleQ\}\not\subseteq\gtRoles{\gtG[1-i]}\\
		\end{array}
		\right.\\[4ex]
		\gtCount{\gtComm{\rolePi}{\roleQi}{i \in I}{\gtLab[i]}{\tyGround[i]}{\gtG[i]}}{\roleP}{\roleQ} & = & \left\{
			\begin{array}{ll}
			\gtCount{\gtG[i]}{\roleP}{\roleQ} & \begin{array}{l}
				\text{for all }i,j\in I\\
                \gtCount{\gtG[i]}{\roleP}{\roleQ}=\gtCount{\gtG[j]}{\roleP}{\roleQ}\\
                (\roleP,\roleQ)\ne(\rolePi,\roleQi)
			\end{array}\\[4ex]
            0&
            \begin{array}{l}
            (\roleP,\roleQ)=(\rolePi,\roleQi)
            \\
            \forall i\in I.\; \gtCount{\gtG[i]}{\roleP}{\roleQ}=0
            \end{array}
            \\[4ex]
			\text{undefined} & \text{otherwise}
			\end{array}
		\right.\\[4ex]
		\gtCount{\gtCommSquig{\rolePi}{\roleQi}{\gtLab}{}{\gtLab}{\tyGround}{\gtG}}{\roleP}{\roleQ} & = & \left\{
			\begin{array}{ll}
			\gtCount{\gtG}{\roleP}{\roleQ}+1 & \roleP=\rolePi\text{ and }\roleQ=\roleQi\\
			\gtCount{\gtG}{\roleP}{\roleQ} & otherwise
			\end{array}
		\right.
	\end{array}
\end{equation*}
\end{definition}

\begin{definition}[Balanced Global Types]
\label{def:balanced-async}
	For each participant, $\roleR$, we define the partial function $\gtDepth{}{\roleR}$ on global types recursively by
	$\gtDepth{\gtG}{\roleR}=1$ if $\roleR\notin\gtARoles{\gtG}$
	and $\gtFv{\gtG}=\emptyset$,
	and otherwise:
	\[
	\begin{array}{c}
\gtDepth{\gtComm{\roleP}{\roleQ}{i \in I}{\gtLab[i]}{\tyGround[i]}{\gtG[i]}}{\roleR}=
			\left\{\!\!\!\begin{array}{ll}
				1+max\{\gtDepth{\gtG[i]}{\roleR}\}_{i \in I} & \!\!\roleR\notin\{\roleP,\roleQ\}\\
				1 & \!\!\roleR\in\{\roleP,\roleQ\}
			\end{array}\right.
			\\
			\gtDepth{\gtCommSquig{\roleP}{\roleQ}{\gtLab}{}{\gtLab}{\tyGround}{\gtG}}{\roleR}=\left\{\!\!\!\begin{array}{ll}
				1+\gtDepth{\gtG}{\roleR} & \!\!\roleR\neq\roleQ\\
				1 & \!\!\roleR=\roleQ
			\end{array}\right.
			\\
		\gtDepth{\gtPar{\gtG[0]}{\gtG[1]}}{\roleR}=
			\left\{\begin{array}{ll}
		\gtDepth{\gtG[0]}{\roleR}\quad\roleR\in\gtRoles{\gtG[0]}\\
                \gtDepth{\gtG[1]}{\roleR}\quad\roleR\in\gtRoles{\gtG[1]}\\
		\end{array}\right.
		\quad
		\gtDepth{\gtRec{\gtRecVar}{\gtG}}{\roleR} = \gtDepth{\gtG}{\roleR}
	\end{array}
	\]
A global type, $\gtG$, is \emph{balanced} iff for all $\gtG\,\gtMoveStar\,\gtGi$ and $\roleR\in\gtRoles{\gtGi}$,
$\gtDepth{\gtGi}{\roleR}$ exists.
\end{definition}

\begin{definition}[$\text{Balanced}^+$~{\cite[Definition~15]{pischke2026asynchronousglobalprotocolsprecisely}}]
	\label{def:bal+}
	Let $\gtCount{\gtG}{\roleP}{\roleQ}$ (\cref{def:en-route-count})
	be the number of en-route transmissions from $\roleP$
	to $\roleQ$ down any path in its syntax,
	if it exists.
	
	$\gtG$ is $\text{balanced}^+$ iff
	$\gtG$ is balanced and
	for all $\roleP$ and $\roleQ$,
	$\gtCount{\gtG}{\roleP}{\roleQ}$ exists.
\end{definition}

\begin{lemma}[Count and Queue Projection]
\label{lem:count-and-proj}
If $\gtQu{\gtG}{\roleP}$ exists,
then $\forall \roleQ.\;\gtCount{\gtG}{\roleP}{\roleQ}$ exists.
\end{lemma}

\begin{proof}
	Induction on $|\gtG|$.
\end{proof}

\begin{definition}[Asynchronous Typing Context Properties~\cite{GPPSY2023}]
\label{def:a-typing-properties}
\ \\
{\bf Safety}:
We say that
  $\predP$ is a \emph{safety property} of typing contexts
  iff, whenever $\predPApp{\stEnvNew}$, we have:
\[\begin{array}{cl}
    \inferrule{\iruleSafeComm}&
    \stEnvApp{\stEnv}{\roleP}=(\quCons{\quMsg{\roleQ}{\stLab}{\tyGround}}{\quHi},\stTi)
    \wedge
    \stEnvApp{\stEnv}{\roleQ}=(\quH,\stT)
    \wedge
    \stT\,\gtMove[{\roleP\stEnvAnnotInSym\stChoice{\stLabi}{\tyGroundi}}]
    \implies
    \stT\,\gtMove[{\roleP\stEnvAnnotInSym\stChoice{\stLab}{\tyGround}}];
    \\
    \inferrule{\iruleMoveSession}&
    \stEnvNew \stEnvMove \stEnvNewi
    \implies
    \predPApp{\stEnvNewi}\end{array}
\]
  \upshape
  We say \emph{$\stEnvNew$ is safe},
  written $\stEnvSafeP{\stEnvNew}$,
  if $\predPApp{\stEnvNew}$  holds
  for some safety property $\predP$.
\\[2ex]
{\bf Deadlock-Free}:
We say that a context $\stEnv$ is deadlock-free iff
for all $\stEnv\stEnvMoveStar\stEnvi$ if
$\stEnvNotMoveP{\stEnvi}$
then for all $\roleP\in\dom{\stEnvi}$
$\stEnvApp{\stEnvi}{\roleP}=(\quEmpty, \stT)$ s.t. $\unfoldOne{\stT}=\stEnd$.
\\[2ex]
{\bf Live}:
Let $I=\{0,1,2,\dots,n\}$ for some $n$ or $I=\{0,1,2,\dots\}$.
Let $\{\stEnv[i]\}_{i\in I}$ be a sequence of typing contexts
such that $\stEnv[i]\,\stEnvMove\,\stEnv[i+1]$ for $i,i+1\in I$.
We say that the path $\{\stEnv[i]\}_{i\in I}$ is fair iff
for all $i\in I$:
\begin{itemize}
	\item[F1] if $\stEnv[i]\stEnvMoveAnnot{\ltsSel{\roleP}{\roleQ}{\stLab(\tyGround)}}$,
	then there are $k$, $\stLabi$, and $\tyGroundi$ such that
	$I\ni k+1> i$ and $\stEnv[k]\stEnvMoveAnnot{\ltsSel{\roleP}{\roleQ}{\stLabi(\tyGroundi)}}\stEnv[k+1]$
	\item[F2] if $\stEnv[i]\stEnvMoveAnnot{\ltsBra{\roleP}{\roleQ}{\stLab(\tyGround)}}$,
	then there is $k$ such that
	$I\ni k+1> i$ and $\stEnv[k]\stEnvMoveAnnot{\ltsBra{\roleP}{\roleQ}{\stLab(\tyGround)}}\stEnv[k+1]$
\end{itemize}
We say that the path $\{\stEnv[i]\}_{i\in I}$ is live iff,
for all $i\in I$:
\begin{itemize}
	\item[L1] If $\stEnvApp{\stEnv[i]}{\roleP}
	=(\quCons{\quMsg{\roleQ}{\stLab}{\tyGround}}{\quH},\stT)$,
	then there exist $k$ and $\tyGroundi$ such that
	$I\ni k+1> i$ and
	$\stEnv[k]\stEnvMoveAnnot{\ltsBra{\roleQ}{\roleP}{\stLab(\tyGroundi)}}\stEnv[k+1]$
	\item[L2] If $\stEnvApp{\stEnv[i]}{\roleP}
	=(\quH,\stT)$ with $\unfoldOne{\stT}=\stExtSum{\roleQ}{j \in J}{\stChoice{\stLab[j]}{\tyGround[j]} \stSeq \stT[j]}$,
	then there exist $k$, $\stLabi$, and $\tyGroundi$ such that
	$I\ni k+1 > i$ and
	$\stEnv[k]\stEnvMoveAnnot{\ltsBra{\roleP}{\roleQ}{\stLabi(\tyGroundi)}}\stEnv[k+1]$
\end{itemize}
We say that a context $\stEnv$ is live iff
all fair contexts starting at $\stEnv$
are live.
\end{definition}

\subsection{Appendix for~\cref{sec:agood} (Good Projections)}

\begin{theorem}
	If a projection is a-good, then
	its restriction to global types with no en-route transmissions
	is good.
\end{theorem}

\begin{proof}
	\begin{itemize}[leftmargin=0.4in,labelindent=-\leftmargin]
		\item[\inferrule{$\projGoodEnd$}] Immediate by \inferrule{$\projGoodEnd$}.
		\item[\inferrule{$\projGoodOut$}] Apply \inferrule{$\projGoodOut$} and \inferrule{$\projGoodIn$},
		as if $\gtG\,\gtMove[\ltsSendRecv{\roleP}{\roleQ}{\stChoice{\stLab}{\tyGround}}]\,\gtGi$ then
		$\gtG\,\gtMove[\ltsSel{\roleP}{\roleQ}{\stChoice{\stLab}{\tyGround}}]\,\cdot\,\gtMove[\ltsBra{\roleQ}{\roleP}{\stChoice{\stLab}{\tyGround}}]\,\gtGi$.
		\item[\inferrule{$\projGoodIn$}] Same as above.
	 	\item[\inferrule{$\projGoodAlt$}] Apply \inferrule{$\projGoodAlt$} twice by above.
		\item[\inferrule{$\projGoodType$}] Apply \inferrule{$\projGoodType$}
		twice alongside \inferrule{$\projGoodAlt$}, \inferrule{$\projGoodOut$}, and \inferrule{$\projGoodIn$}
	\end{itemize}
\end{proof}

\thmASProjSoundness*

\begin{proof}
	Suppose that $\gtProj{\roleR}{\gtG}{(\quH[\roleR],\stT[\roleR])}$
	and $\stEnvApp{\stEnv}{\roleR}=(\quH[\roleR],\stTi[\roleR])$
	with $\stTi[\roleR]\tySub\stT[\roleR]$
	for $\roleR\in\dom{\stEnv}$.
	\\[2ex]
	Suppose that
	$\gtG\,\gtMove[\ltsSel{\roleP}{\roleQ}{\stChoice{\stLab}{\tyGround}}]$.
	By \inferrule{$\projGoodOut$},
	$\stT[\roleP]\,\gtMove[\roleQ\stEnvAnnotOutSym\stChoice{\stLab}{\tyGround}]$.
	Thus,
	$\stTi[\roleP]\,\gtMove[\roleQ\stEnvAnnotOutSym\stChoice{\stLabi}{\tyGroundi}]\,
	\stTi[1,\roleP]$
	and
	$\stT[\roleP]\,\gtMove[\roleQ\stEnvAnnotOutSym\stChoice{\stLabi}{\tyGroundi}]\,
	\stT[1,\roleP]$
	and $\stTi[1,\roleP]\tySub\stT[1,\roleP]$.
	By \inferrule{$\projGoodType$},
	$\gtG\,\gtMove[\ltsSel{\roleP}{\roleQ}{\stChoice{\stLabi}{\tyGroundi}}]\,\gtGi$.
	By \inferrule{$\projGoodOut$},
	there exists $\stTii[1,\roleP]$ such that
	$\gtProj{\roleP}{\gtGi}{(\quCons{\quMsg{\roleQ}{\stLab}{\tyGround}}{\quH[\roleP]},\stTii[1,\roleP])}$
	and $\stT[1,\roleP]\tySub\stTii[1,\roleP]$.
	By \inferrule{$\projGoodAlt$},
	there exist $\stTii[1,\roleR]$ for $\roleR\in\dom{\stEnv}\setminus\{\roleP\}$ s.t.
	$\stT[\roleR]\tySub\stTii[1,\roleR]$
	and $\gtProj{\roleR}{\gtGi}{(\quH[\roleR],\stTii[1,\roleR])}$.
	Write $\stEnv=\stEnvii\stEnvComp\stEnvMap{\roleP}{(\quH[\roleP],\stTi[\roleP])}$,
	so $\stEnv\stEnvMoveAnnot{\ltsSel{\roleP}{\roleQ}{\stChoice{\stLabi}{\tyGroundi}}}
	\stEnvi=\stEnvii\stEnvComp\stEnvMap{\roleP}{(\quCons{\quMsg{\roleQ}{\stLab}{\tyGround}}{\quH[\roleP]},\stTi[1,\roleP])}$.
	Hence, $\stEnvAssoc{\gtGi}{\stEnvi}{}$.
	\\[2ex]
	Suppose that
	$\gtG\,\gtMove[\ltsBra{\roleP}{\roleQ}{\stChoice{\stLab}{\tyGround}}]\,\gtGi$.
	By \inferrule{$\projGoodIn$},
	there exists $\quHi[\roleQ]$, $\stT[1,\roleP]$, and $\stTii[1,\roleP]$ s.t.
	$\stT[1,\roleP]\tySub\stTii[1,\roleP]$,
	$\gtProj{\roleP}{\gtGi}{(\quH[\roleP],\stTii[1,\roleP])}$,
	$\stT[\roleP]\,\gtMove[\roleQ\stEnvAnnotInSym\stChoice{\stLab}{\tyGround}]\,\stT[1,\roleP]$, and $\quH[\roleQ]=\quCons{\quMsg{\roleP}{\stLab}{\tyGround}}{\quHi[\roleQ]}$.
	$\stTi[\roleP]\tySub\stT[\roleP]$ so there exists $\stTi[1,\roleP]$ s.t.
	$\stTi[\roleP]\,\gtMove[\roleQ\stEnvAnnotInSym\stChoice{\stLab}{\tyGround}]\,\stTi[1,\roleP]$ and $\stTi[1,\roleP]\tySub\stT[1,\roleP]$.
	So $\stTi[1,\roleP]\tySub\stTii[1,\roleP]$.
	By \inferrule{$\projGoodAlt$},
	there exist $\stTii[1,\roleR]$ for $\roleR\in\dom{\stEnv}\setminus\{\roleP\}$ s.t.
	$\stT[\roleR]\tySub\stTii[1,\roleR]$,
	if $\roleR\neq\roleQ$ then $\gtProj{\roleR}{\gtGi}{(\quH[\roleR],\stTii[1,\roleR])}$,
	and $\gtProj{\roleR}{\gtGi}{(\quHi[\roleR],\stTii[1,\roleR])}$.
	Write $\stEnv=\stEnvii
	\stEnvComp\stEnvMap{\roleP}{(\quH[\roleP],\stTi[\roleP])}
	\stEnvComp\stEnvMap{\roleQ}{(\quH[\roleQ],\stTi[\roleQ])}$,
	so $\stEnv\stEnvMoveAnnot{\ltsBra{\roleP}{\roleQ}{\stChoice{\stLab}{\tyGround}}}
	\stEnvi=\stEnvii
	\stEnvComp\stEnvMap{\roleP}{(\quH[\roleP],\stTi[1,\roleP])}
	\stEnvComp\stEnvMap{\roleQ}{(\quHi[\roleQ],\stTi[\roleQ])}$.
	Hence, $\stEnvAssoc{\gtGi}{\stEnvi}{}$.
\end{proof}

\thmASProjCompleteness*

\begin{proof}
	Suppose that $\gtProj{\roleR}{\gtG}{(\quH[\roleR],\stT[\roleR])}$
	and $\stEnvApp{\stEnv}{\roleR}=(\quH[\roleR],\stTi[\roleR])$
	with $\stTi[\roleR]\tySub\stT[\roleR]$
	for $\roleR\in\dom{\stEnv}$.
	\\[2ex]
	Suppose that $\stEnv\stEnvMoveAnnot{\ltsSel{\roleP}{\roleQ}{\stChoice{\stLab}{\tyGround}}}\stEnvi$.
	So $\stTi[\roleP]\,\gtMove[\roleQ\stEnvAnnotOutSym\stChoice{\stLab}{\tyGround}]\,\stEnvApp{\stEnvi}{\roleP}$.
	Therefore, there is $\gtGi$ s.t. $\gtG\,\gtMove[{\ltsSel{\roleP}{\roleQ}{\stChoice{\stLab}{\tyGround}}}]\,\gtGi$.
	As the context transition relation is deterministic and by~\cref{thm:as_soundness_association},
	$\stEnvAssoc{\gtGi}{\stEnvi}{}$.
	\\[2ex]
	Suppose that $\stEnv\stEnvMoveAnnot{\ltsBra{\roleP}{\roleQ}{\stChoice{\stLab}{\tyGround}}}\stEnvi$.
	So
	$\stTi[\roleP]\,\gtMove[\roleQ\stEnvAnnotInSym\stChoice{\stLab}{\tyGround}]\,\stEnvApp{\stEnvi}{\roleP}$
	and $\quH[\roleQ]=\quCons{\quMsg{\roleP}{\stLab}{\tyGround}}{\quHi[\roleQ]}$.
	Therefore, there is $\gtGi$ s.t. $\gtG\,\gtMove[{\ltsBra{\roleP}{\roleQ}{\stChoice{\stLab}{\tyGround}}}]\,\gtGi$.
	As the context transition relation is deterministic and by~\cref{thm:as_soundness_association},
	$\stEnvAssoc{\gtGi}{\stEnvi}{}$.
\end{proof}

\thmAAssocCorrect*

\begin{theorem}
	If a projection is a-good, then $\predP=\{\stEnvNew\suchthat\exists\gtG\text{ with }\stEnvAssoc{\gtG}{\stEnvNew}{}\}$ is a safety property.
\end{theorem}
\begin{proof}
	\begin{itemize}
		\item[\inferrule{\iruleSafeComm}]
		
		\text{if }
    $\stEnvApp{\stEnv}{\roleP}=(\quCons{\quMsg{\roleQ}{\stLab}{\tyGround}}{\quHi},\stTi)$,\,
    $\stEnvApp{\stEnv}{\roleQ}=(\quH,\stT)$
    \,\text{and}\,
    $\stT\,\gtMove[{\roleP\stEnvAnnotInSym\stChoice{\stLabi}{\tyGroundi}}]$
		
		Suppose that $\stEnvAssoc{\gtG}{\stEnvNew}{}$,
        $\stEnvApp{\stEnv}{\roleP}=(\quCons{\quMsg{\roleQ}{\stLab}{\tyGround}}{\quHi},\stTi)$, $\stEnvApp{\stEnv}{\roleQ}=(\quH,\stT)$, and
        $\stT\,\gtMove[{\roleP\stEnvAnnotInSym\stChoice{\stLabi}{\tyGroundi}}]$.\\
	Let $\stT[\roleP]$ and $\stT[\roleQ]$ be such that
    $\gtProj{\roleP}{\gtG}{(\quCons{\quMsg{\roleQ}{\stLab}{\tyGround}}{\quHi},\stT[\roleP])}$,
    $\gtProj{\roleQ}{\gtG}{(\quH,\stT[\roleQ])}$,
    $\stTi \tySub \stT[\roleP]$, and
    $\stT \tySub \stT[\roleQ]$.\\
	So,
    $\stT[\roleQ]\,\gtMove[{\roleP\stEnvAnnotInSym\stChoice{\stLabii}{\tyGroundii}}]$
    for some $\stLabii$ and $\tyGroundii$.\\
	Now, $\gtG\,\gtMove[ \ltsBra{\roleQ}{\roleP}{\stChoice{\stLab}{\tyGround}} ]$,
    so $\stT[\roleQ]\,\gtMove[{\roleP\stEnvAnnotInSym\stChoice{\stLab}{\tyGround}}]$.\\
	Therefore,
    $\stT\,\gtMove[{\roleP\stEnvAnnotInSym\stChoice{\stLab}{\tyGround}}]$.
		\item[\inferrule{\iruleMoveSession}] Suppose that $\stEnvAssoc{\gtG}{\stEnvNew}{}$ and
        $\stEnv\stEnvMoveGenAnnot\stEnvi$.\\
        By~\Cref{thm:as_completeness_association},
        there is $\gtGi$ with $\stEnvAssoc{\gtGi}{\stEnvi}{}$.
	\end{itemize}
\end{proof}

\begin{theorem}
	If a projection is a-good, then associated contexts are deadlock-free.
\end{theorem}
\begin{proof}
	Suppose that $\stEnvAssoc{\gtG}{\stEnv}{}$ and
    $\stEnv \!\stEnvMoveStar\! \stEnvNotMoveP{\stEnvi}$.\\
	By completeness,
    there is $\gtGi$ with $\stEnvAssoc{\gtGi}{\stEnvi}{}$.
    Let $\gtProj{}{\gtGi}{\stEnvii}$ and $\stEnvi\tySub\stEnvii$.\\
	By soundness,
    $\gtNotMove{\gtGi}$ so
    $\gtGi\bisim\gtEnd$.\\
	So for all $\roleP \!\in\! \dom{\stEnvi}$,
    $\roleP\notin\gtRoles{\gtGi}$,
    so $\stEnvApp{\stEnvii}{\roleP}=(\quEmpty,\stT)$ with $\unfoldOne{\stT}=\stEnd$,
    so $\stEnvApp{\stEnvi}{\roleP}=(\quEmpty,\stTi)$ with $\unfoldOne{\stTi}=\stEnd$.
\end{proof}

\begin{theorem}[A-Good Projections ensure Liveness {\cite[Lemma~27]{pischke2026asynchronousglobalprotocolsprecisely}}]
	\label{thm:a-good-live}
	If a projection is a-good, then contexts that are associated to balanced
	global types are live.
\end{theorem}

We repeat the proof from~\cite{pischke2026asynchronousglobalprotocolsprecisely}.

\begin{definition}[Message Depth Function]
\label{def:message-depth}
For participants $\roleP$ and $\roleQ$, we define $\gtMDepth{\gtG}{\roleP}{\roleQ}$,
read as `en-route message depth of $(\roleP,\roleQ)$,
to be a partial function on global types to be
$\gtMDepth{\gtG}{\roleP}{\roleQ}=0$
if $\roleQ\notin\gtRoles{\gtG}$ and $\gtFv{\gtG}=\emptyset$,
and otherwise:
\\[2ex]
\centerline{\(
	\begin{array}{lcl}
		\gtMDepth{\gtRec{\gtRecVar}{\gtG}}{\roleP}{\roleQ} & = & \gtMDepth{\gtG}{\roleP}{\roleQ}\\[2ex]
		\gtMDepth{\gtPar{\gtG}{\gtGi}}{\roleP}{\roleQ}&=&\left\{
			\begin{array}{ll}
				\gtMDepth{\gtG}{\roleP}{\roleQ}&\roleP,\roleQ\in\gtRoles{\gtG}\\
				\gtMDepth{\gtGi}{\roleP}{\roleQ}&\text{otherwise}
			\end{array}
		\right.\\[2ex]
		\gtMDepth{\gtComm{\rolePi}{\roleQi}{i \in I}{\gtLab[i]}{\tyGround[i]}{\gtG[i]}}{\roleP}{\roleQ} & = & 
				1+max\left\{\gtMDepth{\gtG[i]}{\roleP}{\roleQ}\suchthat i\in I\right\}\\[2ex]
		\gtMDepth{\gtCommSquig{\rolePi}{\roleQi}{\gtLab}{}{\gtLab}{\tyGround}{\gtG}}{\roleP}{\roleQ} & = & 
            \left\{\begin{array}{ll}
                1 & \roleP=\rolePi\text{ and }\roleQ=\roleQi\\
				1+\gtMDepth{\gtG}{\roleP}{\roleQ} & otherwise
			\end{array}\right.
	\end{array}
\)}
\end{definition}

\begin{theorem}[Transitions preserve En-Route Counting~{\cite[Theorem~3]{pischke2026asynchronousglobalprotocolsprecisely}}]
If $\gtCount{\gtG}{\roleP}{\roleQ}$ exists
and $\gtG\,\gtMoveA[]\,\gtGi$,
then $\gtCount{\gtGi}{\roleP}{\roleQ}$ exists.
\end{theorem}

Note that $\gtG\,\gtMove[\stEnvAnnotGenericSym]\,\gtGi$
implies that $\gtG\,\gtMoveA[\stEnvAnnotGenericSym]\,\gtGi$.

\begin{lemma}[Depth is decreasing~{\cite[Lemma~4]{pischke2026asynchronousglobalprotocolsprecisely}}]
\label{lem:depth-decr}
Suppose that $\gtG\,\gtMoveA[\stEnvAnnotGenericSym]\,\gtGi$, then:
\begin{itemize}
    \item If $\roleP\not\in\ltsSubject{\stEnvAnnotGenericSym}$ and $\gtDepth{\gtG}{\roleP}$ exists, then
    $\gtDepth{\gtGi}{\roleP}\leq \gtDepth{\gtG}{\roleP}$.
    \item If $\stEnvAnnotGenericSym$ is not of the form $\ltsBra{\roleQ}{\roleP}{\stChoice{\stLab}{\tyGround}}$ and $\gtMDepth{\gtG}{\roleP}{\roleQ}$ exists, then
    $\gtMDepth{\gtGi}{\roleP}{\roleQ}\leq \gtMDepth{\gtG}{\roleP}{\roleQ}$.
\end{itemize}
\end{lemma}

\begin{lemma}[En Route Transmissions are Bounded~{\cite[Lemma~5]{pischke2026asynchronousglobalprotocolsprecisely}}]
\label{lem:en-route-bounded}
If $\gtQu{\gtG}{\roleP}$ exists, then
for $(\roleP,\roleQ)\in\gtMRoles{\gtG}$, $\gtMDepth{\gtG}{\roleP}{\roleQ}$ exists.
\end{lemma}

\begin{proof}[Proof of~\cref{thm:a-good-live}]
	This is~\cite[Lemma~27]{pischke2026asynchronousglobalprotocolsprecisely}.
	\\[2ex]
	Let $\left(\stEnv[n]\right)_{n\in N}$ with $\left(\stEnvAnnotGenericSym_{n}\right)_{n,n+1\in N}$ be a fair sequence starting from $\stEnv$.
By~\cref{thm:as_completeness_association}, find
$\left(\gtG[n]\right)_{n\in N}$ (\gtG[0]=\gtG) and 
$\left(\stEnvi[n]\right)_{n\in N}$ (\stEnvi[0]=\stEnv) such that
$\gtProj{}{\gtG[n]}{\stEnvi[n]}$,
$\stEnv[n]\tySub\stEnvi[n]$, and
$\gtG[n]\,\gtMove[{\stEnvAnnotGenericSym_n}]\,\gtG[n+1]$.
If $\stEnvApp{\stEnv[i]}{\roleP}$ has a message for $\roleQ$ in its queue, then
$(\roleP,\roleQ)\in\gtMRoles{\gtG[i]}$ so
$\gtMDepth{\gtG[i]}{\roleP}{\roleQ}$ exists.
Induct on $\gtMDepth{\gtG[i]}{\roleP}{\roleQ}$.
Suppose that this message is never dequeued.
Write $\gtG[i]=\mathbb{C}[\gtGi]$,
where $\mathbb{C}$ is a parallel composition context,
$\gtGi$ is not a parallel composition, and
$\roleP\in\gtRoles{\gtGi}$.
Now apply~\cref{thm:as_soundness_association} via fairness
at the action enabled by the head of $\gtGi$ to get to $j>i$.
By the assumption that the message is not removed from the queue,
\cref{lem:depth-decr},
and our removal of a communication above the en-route transmissions from $\roleP$ to $\roleQ$,
$\gtMDepth{\gtG[j]}{\roleP}{\roleQ}<\gtMDepth{\gtG[i]}{\roleP}{\roleQ}$.
By the I.H., the message will be dequeued.

The case for $\stEnvApp{\stEnv[i]}{\roleP}$ having a receive at the head is similar,
but we induct on $\gtDepth{\gtG[i]}{\roleP}$ instead of $\gtMDepth{\gtG[i]}{\roleP}{\roleQ}$.
\end{proof}

\begin{definition}[Asynchronous Session Deadlock-freedom]
\label{def:async-session-df}
Let $\M$ be an asynchronous session.
$\M$ is deadlock-free iff for all $\M\mpMoveStar\M'$,
either $\M'\prestruct\Pi_{\roleR\in\roleSet}(\playsrole{\roleR}{\mpNil}\mpPar\playsrole{\roleR}{\pquEmpty})$
or $\M'\mpMove$.
\end{definition}

\begin{definition}[Asynchronous Session Liveness]
\label{def:async-session-live}
Let $\M$ be an asynchronous session.
$\M$ is live iff for all $\M\mpMoveStar\M'$:
\begin{enumerate}
	\item
	if $\M'=\playsrole{\roleP}{
	\pquCons{\pquMsg{\roleQ}{\ell}{v}}{\pquH}
	}\mpPar\M''$
	then $\M''\mpMoveStar\playsrole{\roleQ}{
	\textstyle\sum_{i\in I} \roleP\mathord{?}\ell_i(x_i)\mpSeq \mpP_i 
	}\mpPar\M'''$
	where $\ell\in\{\ell_i\}_{i\in I}$; and
	\item
	if $\M'=\playsrole{\roleP}{
	\textstyle\sum_{i\in I} \roleQ\mathord{?}\ell_i(x_i)\mpSeq \mpP_i
	}\mpPar\M''$
	then $\M''\mpMoveStar\playsrole{\roleQ}{
	\pquCons{\pquMsg{\roleQ}{\ell_k}{v}}{\pquH}
	}\mpPar\M'''$
	where $k\in I$.
\end{enumerate}
\end{definition}

\begin{restatable}[Inductive and Coinductive Projections]{theorem}{thmAsyncProjGood}
\label{thm:a-good-inductive-proj}
\label{thm:a-good-coinductive-proj}
Asynchronous inductive projections and
asynchronous coinductive projections,
restricted to balanced global types, are a-good.
\end{restatable}

\begin{proof}
	The proof is similar to the proof of~\cref{thm:good-inductive-proj}.
	The only differences are our use of $\gtCount{\gtG}{\roleP}{\roleQ}$
	and $\gtDepth{\gtG}{\roleQ}$
	being defined to locate en-route transmissions.
	More detail can be found in the proof of~\cite[Lemma~19 and Lemma~21]{pischke2026asynchronousglobalprotocolsprecisely}.
\end{proof}

\thmSeparatingProjectionsAsync*

\begin{proof}
	Let $\gtProj{}{}{}$ be the asynchronous plain/full inductive/coinductive projection
	with/without parallel composition
	in \inferrule{SessAsync} in \cref{def:async-session}.
	
	Let $\gtProji{}{}{}$ be $\gtProj{}{}{}$'s synchronous counterpart
	in \inferrule{Sess} in \cref{def:type-rules}.

	If $\stJudge{}{\pquEmpty}{\quH}$
	then $\quH=\quEmpty$.
	Therefore, if
	$\tyJudge{}{\Pi_{\roleR\in\dom{\stEnv}}(\playsrole{\roleR}{\mpP[\roleR]}\mpPar\playsrole{\roleR}{\pquH})}{\stEnv}$,
	then $\forall \roleR\in\dom{\stEnv}\exists\stT.\;\stEnvApp{\stEnv}{\roleP}=(\quEmpty,\stT)$.
	
	Let
	$\gtG$ be balanced and $\gtProj{\roleR}{\gtG}{\stEnv}$.
	
	If $\forall \roleR\in\dom{\stEnv}\exists\stT.\;\stEnvApp{\stEnv}{\roleP}=(\quEmpty,\stT)$,
	then $\gtMRoles{\gtG}=\emptyset$
	by induction.
	Thus, $\gtG$ has no en-route transmissions.
	
	If $\gtG$ has no en-route transmissions \ie
	$\gtMRoles{\gtG}=\emptyset$,
	then $\forall \roleR\in\dom{\stEnv}\exists\stT.\;\stEnvApp{\stEnv}{\roleP}=(\quEmpty,\stT)$.
	
	Therefore,
	$\tyJudge{}{\Pi_{\roleR\in\dom{\stEnv}}(\playsrole{\roleR}{\mpP[\roleR]}\mpPar\playsrole{\roleR}{\pquEmpty})}{\gtG}$
	iff
	$\tyJudge{}{\Pi_{\roleR\in\dom{\stEnv}}\playsrole{\roleR}{\mpP[\roleR]}}{\gtG}$.
	
	Recall $\N_{\text{\tiny par}}$,
	$\gtG[\text{\tiny par}]$,
	$\N_{\text{\tiny co}}$, and
	$\gtG[\text{\tiny co}]$
	from \cref{thm:separating-projections}.
	
	Let $\gtProj{}{}{}$ be the plain inductive projection with parallel composition.
	We have $\tyJudge{}{\N_{\text{\tiny par}}}{\gtG[\text{\tiny par}]}$
	so
	$\tyJudge{}{\N_{\text{\tiny par}}\mpPar
	\playsrole{\roleFmt{a}}{\pquEmpty}\mpPar
	\playsrole{\roleFmt{b}}{\pquEmpty}\mpPar
	\playsrole{\roleFmt{c}}{\pquEmpty}\mpPar
	\playsrole{\roleFmt{d}}{\pquEmpty}
	}{\gtG[\text{\tiny par}]}$.
	
	Let $\gtProj{}{}{}$ be the full inductive projection without parallel composition.
	We have $\forall\gtG.\;\neg(\tyJudge{}{\N_{\text{\tiny par}}}{\gtG})$
	so
	$\forall\gtG.\;\neg(\tyJudge{}{\N_{\text{\tiny par}}\mpPar
	\playsrole{\roleFmt{a}}{\pquEmpty}\mpPar
	\playsrole{\roleFmt{b}}{\pquEmpty}\mpPar
	\playsrole{\roleFmt{c}}{\pquEmpty}\mpPar
	\playsrole{\roleFmt{d}}{\pquEmpty}
	}{\gtG})$.
	
	Let $\gtProj{}{}{}$ be the plain coinductive projection.
	We have $\tyJudge{}{\N_{\text{\tiny co}}}{\gtG[\text{\tiny co}]}$
	so
	$\tyJudge{}{\N_{\text{\tiny co}}\mpPar
	\playsrole{\roleFmt{a}}{\pquEmpty}\mpPar
	\playsrole{\roleFmt{b}}{\pquEmpty}\mpPar
	\playsrole{\roleFmt{c}}{\pquEmpty}\mpPar
	\playsrole{\roleFmt{d}}{\pquEmpty}
	}{\gtG[\text{\tiny co}]}$.
	
	Let $\gtProj{}{}{}$ be the full inductive projection with parallel composition.
	We have $\forall\gtG.\;\neg(\tyJudge{}{\N_{\text{\tiny co}}}{\gtG})$
	so
	$\forall\gtG.\;\neg(\tyJudge{}{\N_{\text{\tiny co}}\mpPar
	\playsrole{\roleFmt{a}}{\pquEmpty}\mpPar
	\playsrole{\roleFmt{b}}{\pquEmpty}\mpPar
	\playsrole{\roleFmt{c}}{\pquEmpty}\mpPar
	\playsrole{\roleFmt{d}}{\pquEmpty}
	}{\gtG})$.
\end{proof}

\begin{figure}
	\small}
\centering{
    \begin{tikzpicture}[baseline={(0,2cm)}]

\fill[yellow!50] (1cm,1.5cm) rectangle (4cm,3cm);
\fill[rounded corners, white] (1cm, 2cm) rectangle (4cm, 4cm);

\fill[rounded corners, blue!50] (1cm, 2cm) rectangle (4cm, 4cm);
\fill[white] (1cm, 3cm) rectangle (4cm, 4cm);

\draw[rounded corners, name path = safe, thick] (0, 1cm) rectangle (4cm, 5cm);

\draw[rounded corners, green,name path = df, thick] (1cm, 0cm) rectangle (5cm,4cm);

\draw[rounded corners,teal,name path = live, thick] (1cm, 2cm) rectangle (4cm, 4cm);

\draw[dashed, name path = wbal] (1cm,1.5cm) rectangle (4cm,3cm);

\node at (0.5cm,4.5cm) {$\stEnvSafePred$};
\node at (4.5cm,0.5cm) {$\stEnvDFPred$};
\node at (2.5cm,3.5cm) {$\stEnvLivePred$};
\node at (2.5cm,2.5cm) {balanced};
\node at (2.5cm,1.75cm) {projectable};

\end{tikzpicture}
\hfill%
    \begin{minipage}{\dimexpr 0.5\linewidth-\columnsep}%
        \captionsetup{singlelinecheck=off, skip=0pt, justification=raggedright}%
        \caption{Typing context properties and
corresponding global type properties.
Boxes correspond to typing context properties:
{\color{black} black for safety},
{\color{green} green for deadlock-freedom}, and
{\color{teal} teal for live}.
The shaded area corresponds to projectable global types and
{\color{blue} blue for balanced}.
}
\label{fig:async-liveness}
    \end{minipage}
\end{figure}
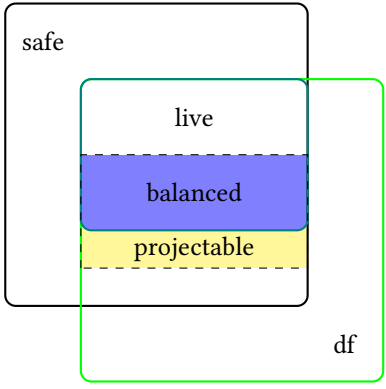

We repeat \cref{fig:liveness}
for asynchronous semantics in \cref{fig:async-liveness}.
We have not introduced the fine-grained liveness and balancedness conditions for asynchronous semantics,
which is why this figure is simpler and relegated to the appendix.

Unlike synchronous semantics,
where we know that all live contexts are associated to balanced global types
using the full coinductive projection,
there are live contexts that are not associated to balanced global types
as they require unbounded queues.

Under asynchronous semantics,
all live contexts are safe.
This is because
if a sent message from $\roleP$ to $\roleQ$ is unexpected
then $\roleQ$ will never be able to accept any message,
thus will become stuck.
\\
\centerline{
\begin{tabular}{c||cccc}
$\stEnv$&$\stEnvSafePred$&$\stEnvDFPred$&$\stEnvLivePred$&$\gtG$
\\
\hline
$\set{
\stEnvMap{\roleP}{(\quEmpty,\stFmt{\roleQ{\oplus}\stLab})}
\stEnvComp
\stEnvMap{\roleQ}{(\quEmpty,\stFmt{\roleP{\&}\stLabi})}
}$
&$\times$&$\times$&$\times$&N/A
\\
\hline
$\set{
\stEnvMap{\roleQ}{(\quEmpty,\stFmt{\roleP{\&}\stLabi})}
}$
&\checkmark&$\times$&$\times$&N/A
\\
\hline
$\set{\begin{array}{l}
\stEnvMap{\roleP}{(\quEmpty,\stFmt{\roleQ{\oplus}\stLab})}
\stEnvComp
\stEnvMap{\roleQ}{(\quEmpty,\stFmt{\roleP{\&}\stLabi})}
\stEnvComp\\
\stEnvMap{\rolePi}{(\quEmpty,\stRec{\stRecVar}{\stFmt{\roleQi{\oplus}\stLab\stSeq\stRecVar}})}
\stEnvComp
\stEnvMap{\roleQi}{(\quEmpty,\stRec{\stRecVar}{\stFmt{\rolePi{\&}\stLab\stSeq\stRecVar}})}
\end{array}
}$
&$\times$&\checkmark&$\times$&N/A
\\
\hline
N/A
&$\times$&$\times$&\checkmark&N/A
\\
\hline
$\set{
\stEnvMap{\roleQ}{(\quEmpty,\stFmt{\roleP{\&}\stLab})}
\stEnvComp
\stEnvMap{\rolePi}{(\quEmpty,\stRec{\stRecVar}{\stFmt{\roleQi{\oplus}\stLab\stSeq\stRecVar}})}
\stEnvComp
\stEnvMap{\roleQi}{(\quEmpty,\stRec{\stRecVar}{\stFmt{\rolePi{\&}\stLab\stSeq\stRecVar}})}
}$
&\checkmark&\checkmark&$\times$&N/A
\\
\hline
N/A
&\checkmark&$\times$&\checkmark&N/A
\\
\hline
N/A
&$\times$&\checkmark&\checkmark&N/A
\\
\hline
$\set{
\stEnvMap{\roleP}{(\quEmpty,\stRec{\stRecVar}{\stFmt{\roleQ{\oplus}\stLab\stSeq\roleQ{\oplus}\stLab\stSeq\stRecVar}})}
\stEnvComp
\stEnvMap{\roleQ}{(\quEmpty,\stRec{\stRecVar}{\stFmt{\roleP{\&}\stLab\stSeq\stRecVar}})}
}$
&\checkmark&\checkmark&\checkmark&N/A
\\
\hline
$\set{\begin{array}{l}
\stEnvMap{\roleP}{(\quEmpty,\stRec{\stRecVar}{\stIntSum{\roleQ}{}{
\stLab\stSeq\stRecVar,\,\stLabi\stSeq\roleR{\oplus}\stLab
}})}
\stEnvComp\\
\stEnvMap{\roleQ}{(\quEmpty,\stRec{\stRecVar}{
\stExtSum{\roleP}{}{
\stLab\stSeq\stRecVar,\,
\stLabi
}
})}
\stEnvComp
\stEnvMap{\roleR}{(\quEmpty,\stFmt{\roleP{\&}\stLab})}
\end{array}}$
&\checkmark&\checkmark&$\times$&$\gtRec{\gtRecVar}{\gtCommRaw{\roleP}{\roleQ}{\begin{array}{l}\gtLab{\gtSeq}\gtRecVar\\\gtLabi{\gtSeq}\gtCommSingle{\roleP}{\roleR}{\gtLab}{}{}\end{array}}}$
\\
\hline
$\set{
\stEnvMap{\roleP}{(\quEmpty,\stFmt{\roleQ{\oplus}\stLab})}
\stEnvComp
\stEnvMap{\roleQ}{(\quEmpty,\stFmt{\roleP{\&}\stLab})}
}$
&\checkmark&\checkmark&\checkmark&$\gtCommSingle{\roleP}{\roleQ}{\gtLab}{}{}$
\end{tabular}
}

\subsection{Appendix for~\cref{sec:aprecise} (Precise Asynchronous Subtyping)}

\begin{definition}[Forgetting Holes~{\cite[Definition~22]{pischke2026asynchronousglobalprotocolsprecisely}}]
	The forgetful holes are of the form $\forgetHole{\cdot}{i}$,
	representing the parts of the local type that
	can be `forgotten' by the asynchronous subtyping
	before reordering.
	Consequently, we must allow contexts to forget these holes.
	Given a local type context $\mathbb{L}$, we define $\overline{\mathbb{L}}$
	to be the local type context where we have removed the forgetful holes,
	by recursion:
	\begin{itemize}
		\item $\overline{[\cdot]_i}=[\cdot]_i$
		\item $\overline{\stExtSum{\roleQ}{i \in I}{\stChoice{\stLab[i]}{\tyGround[i]} \stSeq \mathbb{L}_i}}=\stExtSum{\roleQ}{i \in I}{\stChoice{\stLab[i]}{\tyGround[i]} \stSeq \overline{\mathbb{L}_i}}$
		\item $\overline{\stIntSum{\roleQ}{i \in I_L}{\stChoice{\stLab[i]}{\tyGround[i]} \stSeq \forgetHole{\cdot}{l_i}} + 
	\stIntSum{\roleQ}{i \in I_R}{\stChoice{\stLab[i]}{\tyGround[i]} \stSeq \mathbb{L}_i}}= \stIntSum{\roleQ}{i \in I_R}{\stChoice{\stLab[i]}{\tyGround[i]} \stSeq \overline{\mathbb{L}_i}}$
	\end{itemize}
\end{definition}

\lemInvSub*
This is~\cite[Lemma~16]{pischke2026asynchronousglobalprotocolsprecisely}.

\thmAAProjCompleteness*

This is, in principle,~\cite[Theorem~11]{pischke2026asynchronousglobalprotocolsprecisely}
with all references to earlier lemmas replaced by references to
\cref{def:a-good-proj} and \cref{lem:inv-subtyping}.

\thmAAProjSoundness*

This is, in principle,~\cite[Theorem~12]{pischke2026asynchronousglobalprotocolsprecisely}.
The proof is short, so we repeat it here.

\begin{proof}
Suppose that $\stEnvAssoc{\gtG}{\stEnv}{a}$ and $\gtG\,\gtMove$.
There exists $\stEnvi$ such that $\gtProj{}{\gtG}{\stEnvi}$ and $\stEnv\asubt\stEnvi$.
$\stEnvLivePred$ is downward closed under $\asubt$
and, by~\cref{thm:a-good-correct}, $\stEnvi$ is live,
so $\stEnv$ is live.
By~\cref{thm:as_soundness_association}, $\stEnvi\stEnvMove$,
so there is $\roleP$ s.t. $\stEnvApp{\stEnvi}{\roleP}=(\quH,\stT)$
with $\unfoldOne{\stT}\neq\stEnd$.
So, $\stEnvApp{\stEnv}{\roleP}=(\quH,\stTi)$ with $\stTi\asubt\stT$,
so $\unfoldOne{\stTi}\neq\stEnd$.
$\stEnv$ is live so $\stEnv\stEnvMove$.
Now apply~\cref{thm:aa_completeness_association} to conclude.
\end{proof}

\thmPreciseAssocCorrect*
\begin{proof}
	$\stEnvLivePred$ is downward closed under $\asubt$.
	Suppose that $\stEnvAssoc{\gtG}{\stEnv}{a}$
	by $\stEnvi$ with $\gtProj{}{\gtG}{\stEnvi}$ and $\stEnv\asubt\stEnvi$.
	By~\cref{thm:a-good-correct}, $\stEnvi$ is live, so
	$\stEnv$ is live.
	\\[2ex]
	Suppose that $\predPApp{\stEnv}$,
	so there exists $\gtG$ such that
	$\stEnvAssoc{\gtG}{\stEnv}{a}$.
	
	If $\stEnv\stEnvMove\stEnvi$,
	then, by \cref{thm:aa_completeness_association},
	there is $\gtGi$ such that
	$\gtG\,\gtMove\,\gtGi$ and
	$\stEnvAssoc{\gtGi}{\stEnvi}{a}$.
	Thus, $\predPApp{\stEnvi}$.
	
	If $\stEnv=\stEnvi
	\stEnvComp
	\stEnvMap{\roleP}{(\quCons{\quMsg{\roleQ}{\stLab}{\tyGround}}{\quHi},\stTi)}
	\stEnvComp
	\stEnvMap{\roleQ}{(\quH,\stT)}$
	and
	$\stT\,\gtMove[\stFmt{\roleP\stEnvAnnotInSym\stChoice{\stLabi}{\tyGroundi}}]$,
	then
	$\stT\,\gtMove[\stFmt{\roleP\stEnvAnnotInSym\stChoice{\stLab}{\tyGround}}]$
	since
	$\stEnvSafeP{\stEnv}$
	as
	$\stEnvSafePred\supseteq\stEnvLivePred$.
\end{proof}%
\section{Precise Asynchronous Subtyping}
\label{sec:precise-subtyping}
We introduce the \emph{precise asynchronous subtyping} relation $\asubt$ on local types, as defined in~\cite[Definition~3.7]{GPPSY2023}.
We exclude subtyping on basic types, as we have done elsewhere in this paper.

We first consider the tree representation 
of local type $\stT$ (denoted by $\ttree{\stT}$). 

We write $\trT$ for generic trees and additionally define three specific types of tree. 
\emph{Single-input} trees (denoted by $\stV$) are those which have only a singleton choice in all branchings, 
while \emph{single-output} trees (denoted by $\stU$) are those which have only a singleton choice in all selections. 
Trees which are both single-input and single-output are called \emph{single-input-single-output (SISO) trees} (denoted by $\stW$). 
These can all be defined coinductively by the following equations. 
\[
  \begin{array}{ll}
    \trT
    &=~
    \stExtSum{\roleP}{i \in I}{\stChoice{\stLab[i]}{\tyGround[i]} \stSeq \trT[i]} ~\mid~ \stIntSum{\roleP}{i \in I}{\stChoice{\stLab[i]}{\tyGround[i]} \stSeq \trT[i]}  ~\mid~ \stEnd
    \\[1mm]%
    \stU
    &=~
    \stExtSum{\roleP}{i \in I}{\stChoice{\stLab[i]}{\tyGround[i]} \stSeq \stU[i]} ~\mid~ \stSend{\roleP}{\stLab}{\tyGround}{\stU}  ~\mid~ \stEnd
    \\[1mm]%
    \stV
    &=~
    \stRecv{\roleP}{\stLab}{\tyGround}{\stV} ~\mid~ \stIntSum{\roleP}{i \in I}{\stChoice{\stLab[i]}{\tyGround[i]} \stSeq \stV[i]}  ~\mid~ \stEnd  
    \\[1mm]%
    \stW
    &=~
    \stRecv{\roleP}{\stLab}{\tyGround}{\stW} ~\mid~ \stSend{\roleP}{\stLab}{\tyGround}{\stW}  ~\mid~ \stEnd\\[1mm]
  \end{array}
\]

We will define reorderings of SISO trees, and to do so, we consider non-empty sequences $\Ap$ of receives not including $\roleP$ 
and $\Bp$ of sends not including $\roleP$ together with receives from any participant. 
These sequences are inductively defined (where $\roleP \ne \roleQ$) by:

\smallskip
  \centerline{\(%
  \begin{array}{ll}
    \Ap
    &=~
    \stRecvOne{\roleQ}{\stLab}{\tyGround} ~\mid~ \stRecv{\roleQ}{\stLab}{\tyGround}{\Ap}\quad 
    \Bp
~=~
    \stRecvOne{\roleR}{\stLab}{\tyGround} ~\mid~ \stSendOne{\roleQ}{\stLab}{\tyGround}  ~\mid~ \stRecv{\roleR}{\stLab}{\tyGround}{\Bp} ~\mid~ \stSend{\roleQ}{\stLab}{\tyGround}{\Bp}
  \end{array}
  \)}%
  
  \smallskip%

\noindent We define the set $\act{\stW}$ of actions of a SISO tree: 
$\act{\stEnd} = \emptyset$; 
$\act{\stRecv{\roleP}{\stLab}{\tyGround}{\stW}} = \{\roleP?\} ~\cup~ \act{\stW}$; and 
$\act{\stSend{\roleP}{\stLab}{\tyGround}{\stW}} = \{\roleP!\} ~\cup~ \act{\stW}$. 

Using these definitions, we introduce a \emph{refinement relation} ($\subttt$) defined coinductively by the following rules:

\smallskip
\centerline{\(
\begin{array}{c}

\cinference[Ref-$\mathcal{A}$]{
  \stW \subttt \Ap ; \stW'
  & 
  \act{\stW} = \act{\Ap ; \stW'}
}{
  \stRecv{\roleP}{\stLab}{\tyGround}{\stW} \subttt \Ap ; \stRecv{\roleP}{\stLab}{\tyGround}{\stWi}
}

\\[2ex]

\cinference[Ref-$\mathcal{B}$]{
  \stW \subttt \Bp ; \stW'
  & 
  \act{\stW} = \act{\Bp ; \stW'}
}{
  \stSend{\roleP}{\stLab}{\tyGround}{\stW} \subttt \Bp ; \stSend{\roleP}{\stLab}{\tyGround}{\stWi}
}

\\[2ex]

\cinference[Ref-In]{
  \stW \subttt \stWi
}{
  \stRecv{\roleP}{\stLab}{\tyGround}{\stW} \subttt \stRecv{\roleP}{\stLab}{\tyGround}{\stW'}
}

\qquad 

\cinference[Ref-Out]{
  \stW \subttt \stWi
}{
  \stSend{\roleP}{\stLab}{\tyGround}{\stW} \subttt \stSend{\roleP}{\stLab}{\tyGround}{\stW'}
}

\qquad 

\\[2ex]
\cinference[Ref-End]{}{
  \stEnd \subttt \stEnd
}
\end{array}
\)}

We can extract sets of single-input and single-output trees from a given tree using the functions $\singleIn{\cdot}$ and $\singleOut{\cdot}$.

\medskip
\centerline{\(
\begin{array}{c}
\singleIn{\stExtSum{\roleP}{i \in I}{\stChoice{\stLab[i]}{\tyGround[i]} \stSeq \trT[i]}} = \bigcup_{i \in I}\{ {\stRecv{\roleP}{\stLab[i]}{\tyGround[i]}{\stV[i]}} \mid \stV[i] \in \singleIn{\trT[i]} \} 
\\[2ex]
\singleIn{\stIntSum{\roleP}{i \in I}{\stChoice{\stLab[i]}{\tyGround[i]} \stSeq \trT[i]}} = \{ {\stIntSum{\roleP}{i \in I}{\stChoice{\stLab[i]}{\tyGround[i]} \stSeq \stV[i]}} \mid \forall i \in I : \stV[i] \in \singleIn{\trT[i]} \} 
\quad
\singleIn{\stEnd} = \{ \stEnd \} 
\\[2ex]
\singleOut{\stIntSum{\roleP}{i \in I}{\stChoice{\stLab[i]}{\tyGround[i]} \stSeq \trT[i]}} = \bigcup_{i \in I}\{ {\stSend{\roleP}{\stLab[i]}{\tyGround[i]}{\stU[i]}} \mid \stU[i] \in \singleOut{\trT[i]} \} 
\\[2ex]
\singleOut{\stExtSum{\roleP}{i \in I}{\stChoice{\stLab[i]}{\tyGround[i]} \stSeq \trT[i]}} = \{ {\stExtSum{\roleP}{i \in I}{\stChoice{\stLab[i]}{\tyGround[i]} \stSeq \stU[i]}} \mid \forall i \in I : \stU[i] \in \singleOut{\trT[i]} \} 
\quad
\singleOut{\stEnd} = \{ \stEnd \} 
\\[2ex]
\end{array}
\)}

\begin{definition}[Subtyping]
\label{def:precise_subtyping}
We consider trees that have
only singleton choices in branchings
(called \emph{single-input (SI) trees}),
or in selections (\emph{single-output (SO) trees}),
and we define the session subtyping $\asubt$ over all session types
by considering their decomposition into SI, SO, and SISO trees.\\[1mm]
\centerline{\(
\inference[Sub]{
    \forall \stU \in \singleOut{\ttree{\stT}} & \forall \stVi \in \singleIn{\ttree{\stTi}} & \exists \stW \in \singleIn{\stU} & \exists \stWi \in \singleOut{\stVi} & \stW \subttt \stWi
}
{
    \stT \asubt \stTi
}
\)}%
\smallskip

The refinement $\subttt$ captures \emph{safe permutations} of input/output messages,
that never cause deadlocks or communication errors under asynchrony;
and the subtyping relation $\asubt$ focuses on
reconciling refinement $\subttt$ with the branching
structures in session types. 
\end{definition}

\section{Appendix for \cref{sec:related} (Related Work)}
\label{app:conclusion}

\subsection{Corecursive Projection in \cite{10.1145/3678232.3678245}}
\label{sec:app:corecursive}
In this subsection, we prove that the
corecursive
projection in 
\cite{10.1145/3678232.3678245} is equivalent to 
the coinductive projection with full merging. 

\begin{lemma}[Full Coinductive Merge is Full Corecursive Merge]
\label{lem:full-coind-merge-is-corec}
	Let $\stMerge{}{}$ be the full coinductive merge.
	Let $\stMergei{}{}$ be the full corecursive merge \cite[Def. 6.1]{10.1145/3678232.3678245}.
	We have that $\stMerge{}{}=\stMergei{}{}$.
\end{lemma}

\begin{proof}
	Corecursion is smaller than coinduction, so
	$\forall \mathcal{T}.\;\stMergei{}{\mathcal{T}}\subseteq\stMerge{}{\mathcal{T}}$.
	
	Consider the derivation of $\relMerge{\mathcal{T}}{\stT}$.
	Consider $\stT$ as a coinductive type.
	A constructor in $\stT$ of depth $n$
	is determined at depth $n$ in the derivation.
	Thus, the derivation is corecursive
	so $\stT\in\stMergei{}{\mathcal{T}}$.
	
	Hence,
	$\forall \mathcal{T}.\;\stMergei{}{\mathcal{T}}=\stMerge{}{\mathcal{T}}$.
\end{proof}

\begin{proposition}[Full Coinductive is Full Corecursive]
\label{prop:full-coind-is-corec}
Let $\gtProj{}{}{}$
be the full coinductive projection.
Let $\gtProji{}{}{}$
be the full corecursive projection \cite[Def. 6.1]{10.1145/3678232.3678245}.

\begin{enumerate}
	\item If $\gtDepth{\gtGi}{\roleP}$
	exists for all $\gtG\,\gtMoveStar\,\gtGi$ and $\gtProj{\roleP}{\gtG}{\stT}$,
	then $\gtProji{\roleP}{\gtG}{\stT}$.
	\item
	If $\gtProji{\roleP}{\gtG}{\stT}$, then
	for all $\gtG\,\gtMoveStar\,\gtGi$,
	$\gtDepth{\gtGi}{\roleP}$
	exists and $\gtProj{\roleP}{\gtG}{\stT}$.
\end{enumerate}
\end{proposition}

\begin{proof}[Proof. (2)]
	Suppose that $\gtProji{\roleP}{\gtG}{\stT}$.
	Consider the corecursive data stream deriving this,
	$\stT[0],\stT[1],\dots$
	\ie
	a increasing
	(in the sense that if $i \leq j$ then
	$\stT[j]$ is more defined than $\stT[i]$)
	sequence of
	partially defined coinductive types produced by
	finite derivations
	with limit $\stT$.
	Let $i$ be such that the head of $\stT[i]$ is defined.
	By induction on the length of the derivation,
	$\gtDepth{\gtG}{\roleP}$ exists.
	By definition of the projection,
	for all $\gtG\,\gtMoveStar\gtGi$ by head reduction,
	there is $\stTi$ such that $\gtProji{\roleP}{\gtGi}{\stTi}$,
	so $\gtDepth{\gtGi}{\roleP}$ exists.
	Thus for all $\gtG\,\gtMoveStar\,\gtGi$, $\gtDepth{\gtGi}{\roleP}$ exists.
	
	Also, $\gtProj{\roleP}{\gtG}{\stT}$
	since corecursion is smaller than coinduction.
\end{proof}

\begin{proof}[Proof. (1)]
	Suppose that $\gtDepth{\gtGi}{\roleP}$
	exists for all $\gtG\,\gtMoveStar\,\gtGi$ and $\gtProj{\roleP}{\gtG}{\stT}$.
	
	Consider a derivation of $\gtProj{\roleP}{\gtG}{\stT}$
	with $\stT$ as a coinductive type.
	For every subderivation concluding $\gtProj{\roleP}{\gtGi}{\stTi}$,
	$\gtDepth{\gtGi}{\roleP}$ exists.
	By induction on $\gtDepth{\gtGi}{\roleP}$,
	the head of $\stTi$ is determined in depth
	$\gtDepth{\gtGi}{\roleP}$ in the subderivation.
	Thus, this is a corecursive derivation
	and $\gtProji{\roleP}{\gtG}{\stT}$.
\end{proof}

\subsection{Related Work with the Synthetic System
  \cite{10.1145/3776692}}
\label{sec:app:synthetic}
In this subsection, we discuss several issues with theorems and claims in 
the synthetic system,
as presented in
the paper of Castro-Perez et al.
\cite{10.1145/3776692}. 

\subsubsection{Theorem 5.9 in \cite{10.1145/3776692}}
We first show a counterexample,
disproving Theorem 5.9 in \cite{10.1145/3776692}.  

\begin{proposition}[Global types may not be well-behaved]
\label{prop:global-not-behaved}
The global type
\[\gtG=\gtRec{\gtRecVar}{\gtCommRaw{\roleP}{\roleQ}{
\begin{array}{l}
\gtLab[0]{\gtSeq}\gtCommSingle{\roleR}{\roleS}{\gtLab[2]}{}{\gtRecVar}
\\
\gtLab[1]{\gtSeq}\gtCommRaw{\roleR}{\roleS}{
\gtLab[2]{\gtSeq}\gtRecVar,\,
\gtLab[3]{\gtSeq}\gtRecVar
}
\end{array}
}}\]
does not meet the well-behavedness conditions
\cite[Def. 5.3]{10.1145/3776692}
and types a session
using the rules in \cite[Fig. 11]{10.1145/3776692}.
\end{proposition}

\begin{proof}
	Conditional commutativity \cite[Def. 5.3 (3)]{10.1145/3776692}
	fails.
	
	We have that
	\[
	\gtG\,\gtMove[\ltsSendRecv{\roleP}{\roleQ}{\gtLab[1]}{}]\,
	\gtCommRaw{\roleR}{\roleS}{
\gtLab[2]{\gtSeq}\gtRecVar,\,
\gtLab[3]{\gtSeq}\gtRecVar
}
	\,\gtMove[\ltsSendRecv{\roleR}{\roleS}{\stLab[3]}{}]
	\;
	\text{and}
	\;
		\gtG\,\gtMove[\ltsSendRecv{\roleR}{\roleS}{\stLab[2]}{}]
	\;\text{but}\;
		\gtG\,\;\;\not\!\!\!\gtMove[\ltsSendRecv{\roleR}{\roleS}{\stLab[3]}{}]
	\]
	Contradicting conditional commutativity.
	
	Consider the session:
	\[
		\M =
		\playsrole{\roleP}{\mpRec{\mpX}{
			\roleQ\mathord{!}\stLab[0] \mpSeq \mpX
		}}
		\mpPar
		\playsrole{\roleQ}{\mpRec{\mpX}{
			\textstyle\sum_{i\in \{0,1\}} \roleP\mathord{?}\stLab[i]\mpSeq \mpX
		}}
		\mpPar
		\playsrole{\roleR}{\mpRec{\mpX}{
			\roleS\mathord{!}\stLab[2] \mpSeq \mpX
		}}
		\mpPar
		\playsrole{\roleS}{\mpRec{\mpX}{
			\textstyle\sum_{i\in \{2,3\}} \roleR\mathord{?}\stLab[i]\mpSeq \mpX
		}}
	\]
	From the rules in \cite[Fig. 11]{10.1145/3776692},
	we derive that $\tyJudge{}{\M}{\gtG}$.
\end{proof}

This is due to the global type semantics in their Agda formalisation
differing from the global type semantics given in their paper.
The transition labels in the Agda formalisation contains the set
of all alternative labels,
thus requiring that each redex contains the same set of
labels if they are part of the same transition derivation.
This is violated by the transitions in \cref{prop:global-not-behaved}.

\subsection{Finite State Global Types in \cite{10.1145/3776692}}

In \cite{10.1145/3776692}, they make use of the fact that
they are using finite state MLTSs (multiparty labelled transitions systems):
\begin{quote}
``Finally, we note that as long as the MLTSs have \emph{finitely}
many states, the type system is decidable. 
First, all the typing rules in
\cite[Figure~11]{10.1145/3776692}
are structural except
\inferrule{$\vdash$-Skip} 
and, as long as the MLTS
has \emph{finitely} many states, all of their premises are decidable, and the domain of any universal
quantification is finite. In rule
\inferrule{$\vdash$-Skip},
the size of the process in the premises does not grow;
\inferrule{$\vdash$-Skip}’s premises (1), (2), and (4) are also decidable for any \emph{finite-state} MLTS; and the universal
quantification of (3) is also finite.''
\cite[\S~5.1 Page 23]{10.1145/3776692}
\end{quote}

They claim that global types without parallel composition have finite MLTSs
with size linear in the global type
\cite[A.2]{10.1145/3776692}.
However, we have shown this to be false.
Each component in
the type in \cref{thm:global-sem}
has no parallel composition
but is non-regular and does not have finite MLTSs.
\[
	\gtG=\gtRec{\gtRecVar}{
	\gtCommRaw{\roleP}{\roleQ}{
	\begin{array}{l}
		\gtLab[0]{\gtSeq}
		\gtCommRaw{\roleR}{\roleS}{
			\gtLab[2]{\gtSeq}\gtRecVar
		}
		\\
		\gtLab[1]{\gtSeq}
		\gtCommRaw{\roleR}{\roleS}{
			\gtLab[2]{\gtSeq}\gtRecVar,\,
			\gtLab[3]{\gtSeq}\gtRecVar
		}
	\end{array}
	}}
	\]
	In fact, the typing system in \cite[Figure 11]{10.1145/3776692}
	has $\gtG$ typing a session:
	\[
		\tyJudge{}{
		\playsrole{\roleP}{
		\mpRec{\mpX}{
		\roleQ\mathord{!}\stLab[0]\mpSeq\mpX
		}
		}
		\mpPar
		\playsrole{\roleQ}{\mpRec{\mpX}{
		\textstyle\sum_{i\in \{0,1\}} \roleP\mathord{?}\stLab[i]\mpSeq \mpX
		}}
		\mpPar
		\playsrole{\roleR}{
		\mpRec{\mpX}{
		\roleS\mathord{!}\stLab[2]\mpSeq\mpX
		}
		}
		\mpPar
		\playsrole{\roleS}{\mpRec{\mpX}{
		\textstyle\sum_{i\in \{2,3\}} \roleR\mathord{?}\stLab[i]\mpSeq \mpX
		}}
		}{\gtG}
	\]

\subsection{Complete Projections in \cite{Li2023-sx} are not Program Complete}
\label{sec:app:li} 
In this section, we prove that
the notion of `complete projections' in \cite{Li2023-sx}
is \emph{incorrect} from the perspective of
program realisability.

In \cite{Li2023-sx}, a global type is projectable
iff there is some correct projection that projects it.
A complete projection in \cite{Li2023-sx}
is one that can project all
projectable global types.
This is not the same as being able to type all
programs that can be typed by some correct projection.
In the setting of \cite{Li2023-sx}
\ie asynchronous sender-driven global types,
there are distinct, computable and complete projections
that type different programs.
Thus, they are incomplete from the perspective of
program realisability.

\begin{proposition}[Complete Projections can be Typability Separated]
\label{prop:complete-proj-sep}
	There are two complete and computable projection algorithms
	(in the sense of \cite{Li2023-sx})
	on asynchronous sender-driven global types,
	which have different sets of typable sessions.
\end{proposition}

\begin{proof}
	Let $\gtProj{}{}{}$
	be the projection algorithm from \cite{Li2023-sx},
	which is a function from sender-driven global types
	to typing contexts with local type syntax in
        \cite{majumdar_et_al:LIPIcs.CONCUR.2021.35}. 
	Observe that if $\gtProj{}{\gtG}{}=\stEnv$
	then
	$\forall \roleP\in \dom{\stEnv}
	\exists \stT.\;
	\stEnvApp{\stEnv}{\roleP}=(\quEmpty,\stT)
	\;\wedge\;
	\unfoldOne{\stT}\neq\stEnd
	\;\wedge\;
	\gtRoles{\stT}\subseteq\gtRoles{\gtG}
	\;\wedge\;
	\roleP\in\gtRoles{\gtG}
	$.
	
	Consider:
	$
		\gtGi = \gtCommSingle{\roleP}{\roleQ}{\gtLab}{}{}
	$.
	Let:
	\[
		\gtProji{}{\gtG}{} =
		\left\{
		\begin{array}{ll}
			\gtProj{}{\gtG}{} & \gtG\neq \gtGi
			\\
			\set{
			\stEnvMap{\roleP}{(\quEmpty,\stFmt{\roleQ{\oplus}\stLab})}
			\stEnvComp
			\stEnvMap{\roleQ}{(\quEmpty,\stFmt{\Sigma\!\left\{\roleP{\&}\stLab,\,\roleR{\&}\stLab\right\}})}
			\stEnvComp
			\stEnvMap{\roleR}{(\quEmpty,\stEnd)}
			}
			& \gtG = \gtGi
		\end{array}
		\right.
	\]
	So that $\gtProji{}{}{}$
	is also sound and complete with respect to asynchronous sender-driven global types.
	
	Consider the session:
	\[
		\M =
		\playsrole{\roleP}{
			\roleQ\mathord{!}\stLab 
		}
		\mpPar
		\playsrole{\roleQ}{
			\roleP\mathord{?}\stLab\mpSeq \mpNil
			+
			\roleR\mathord{?}\stLab\mpSeq \mpNil
		}
		\mpPar
		\playsrole{\roleR}{\mpNil}
	\]
	We type $\tyJudge{}{\M}{\gtGi}$
	using $\gtProji{}{}{}$.
	
	Suppose that $\tyJudge{}{\M}{\gtG}$
	using $\gtProj{}{}{}$.
	If $\stJudge{}{\mpNil}{\stT}$, then $\unfoldOne{\stT}=\stEnd$
	so $\roleR\notin\gtRoles{\gtG}$.
	If $\stJudge{}{\roleP\mathord{?}\stLab\mpSeq \mpNil
			+
			\roleR\mathord{?}\stLab\mpSeq \mpNil}{\stT}$,
	then $\roleR\in\gtRoles{\stT}$
	so $\roleR\in\gtRoles{\gtG}$.
	This is a contradiction.
	Thus, 
	$\forall \gtG.\;
	\neg(\tyJudge{}{\M}{\gtG})$
	using $\gtProj{}{}{}$.
\end{proof}

\subsection{History Dependent Global Types}
\label{sec:history-type}
Our projections (\cref{def:good-proj}) are ``participant-wise'', \ie
we define $\gtProj{\roleP[i]}{\gtG}{\stT[i]}$ for each participant,
and combine them to obtain
$
\stEnv=
\set{
\stEnvMap{\roleP[i]}{\stT[i]}
}_{i\in I}
$.
Alternatively,
projections could be directly defined as relations between
global types $\gtG$ and typing contexts $\stEnv$, without defining
a projection for each participant.
We call such a projection \emph{direct}.
With sender-driven choice,
participant-wise good projections
are weaker than direct.

This was remarked upon by Castagna et al. \cite{Castagna2012GlobalTypes}:

\begin{quote}
``The effective generation
of all possible projections is impossible.
The reason is that the projectability of a global type
may rely on some global knowledge that is no longer
available when working at the level of
single session types: while in a global approach we can, say,
add to some participant new synchronization offers that,
thanks to our global knowledge, we know will never be used,
this cannot be done when working at the level of single participant.''
\end{quote}

This appendix defines ``direct projections''
between global types and typing contexts,
proves that there is a global type that
is projectable by a direct projection
but not by any good participant-wise projection, and
introduces \emph{history-dependent global types}
to allow good projections to subsume
direct projection \emph{partial functions}.

\begin{definition}[Direct Projection]
\label{def:direct-projection}
The relation $\gtProj{}{}{}$
from global types to typing contexts
is a direct projection iff
\[
\forall \gtProj{}{\gtG}{\stEnv}.
\left(
\gtTrace{\gtG}=\gtTrace{\stEnv}
\wedge
\stEnvSafeP{\stEnv}
\wedge
\forall \stEnv\gtMove[\vec{\stEnvAnnotGenericSym}]\stEnvi.
\left(
\stEnvApp{\stEnvi}{\roleP}\tyNotSub\stEnd\implies
\gtG\,\gtMove[\vec{\stEnvAnnotGenericSym}]\,\gtGi\;\wedge\;
\roleP\in\gtRoles{\gtGi}
\right)
\right)
\]
\end{definition}

Consider the type $\gtG$
containing
$\roleP[0]\to\roleP[2]:\gtLab{\gtSeq}
					\roleP[1]\to\roleP[3]:\gtLab$
in two distinct positions:
\\\centerline{\(
		\gtG =
		\gtFmt{
		\Sigma\!\left\{
		\begin{array}{l}
				\roleS\to\roleP[0]\!:\!\gtLab[i]{\gtSeq}
				\roleS\to\roleP[1]\!:\!\gtLab[i]{\gtSeq}\\\quad
				\roleS\to\roleP[2]\!:\!\gtLab[i]{\gtSeq}
				\roleS\to\roleP[3]\!:\!\gtLab[i]{\gtSeq}
		\end{array}
				\Sigma\!\left\{
				\begin{array}{l}
					{\roleP[2\cdot i]}\to{\roleP[2\cdot i + 1]}\!:\!\gtLab[0]{\gtSeq}
					\colorbox{yellow}{$
					\roleP[0]\to\roleP[2]\!:\!\gtLab{\gtSeq}
					\roleP[1]\to\roleP[3]\!:\!\gtLab
					$}
					\\
					{\roleP[2\cdot i]}\to{\roleP[2\cdot i + 1]}\!:\!\gtLab[1]{\gtSeq}
					\roleP[0]\to\roleP[3]\!:\!\gtLab{\gtSeq}
					\roleP[1]\to\roleP[2]\!:\!\gtLab
				\end{array}
				\right\}
		\right\}_{i\in \{0,1\}}
		}
	\)}
	In the branch ($i=0$), $\roleP[1]$
	thinks that it can send $\stLab$ to $\roleP[2]$.
	In the branch ($i=1$), $\roleP[2]$ 
	thinks that it can receive $\stLab$ from $\roleP[1]$.
	A participant-wise projection, 
        treating these two occurrences the same,
        fails subject reduction or session fidelity.
There is a direct projection making $\gtG$ projectable.

\begin{restatable}[Sender-Driven Projection]{theorem}{thmSenderDriven}
\label{thm:sender-driven-fails}
	There is a sender-driven global type
	that is projectable by a direct projection onto
	a separated choice typing context
	without recursion,
	thus is projectable in the sense of
	\cite{majumdar_et_al:LIPIcs.CONCUR.2021.35},
	but is not projectable
	by any good projection.
	Also, any
	``participant-wise'' projection,
	that projects this type,
	causes subject reduction or session fidelity to fail.
\end{restatable}

\begin{proof}
	Projections of sender-driven global types
	can contain additional local behaviours,
	which may disappear globally after transitions
	but cannot disappear locally.
	\[
		\gtG =
		\gtFmt{
		\Sigma\!\left\{
			\begin{array}{l}
				\roleS\to\roleP[0]:\gtLab\gtSeq
				\roleS\to\roleP[1]:\gtLab\gtSeq
				\roleS\to\roleP[2]:\gtLab\gtSeq
				\roleS\to\roleP[3]:\gtLab\gtSeq\\\quad
				\Sigma\!\left\{
				\begin{array}{l}
					{\roleP[0]}\to{\roleP[1]}:\gtLab[0]\gtSeq
					\roleP[0]\to\roleP[2]:\gtLab\gtSeq
					\roleP[1]\to\roleP[3]:\gtLab
					\\
					{\roleP[0]}\to{\roleP[1]}:\gtLab[1]\gtSeq
					\roleP[0]\to\roleP[3]:\gtLab\gtSeq
					\roleP[1]\to\roleP[2]:\gtLab
				\end{array}
				\right\}
				\\
				\roleS\to\roleP[0]:\gtLabi\gtSeq
				\roleS\to\roleP[1]:\gtLabi\gtSeq
				\roleS\to\roleP[2]:\gtLabi\gtSeq
				\roleS\to\roleP[3]:\gtLabi\gtSeq\\\quad
				\Sigma\!\left\{
				\begin{array}{l}
					{\roleP[2]}\to{\roleP[3]}:\gtLab[0]\gtSeq
					\roleP[0]\to\roleP[2]:\gtLab\gtSeq
					\roleP[1]\to\roleP[3]:\gtLab
					\\
					{\roleP[2]}\to{\roleP[3]}:\gtLab[1]\gtSeq
					\roleP[0]\to\roleP[3]:\gtLab\gtSeq
					\roleP[1]\to\roleP[2]:\gtLab
				\end{array}
				\right\}
			\end{array}
		\right\}
		}
	\]
	In both branches,
	$\roleP[0]\to\roleP[2]:\gtLab\gtSeq
					\roleP[1]\to\roleP[3]:\gtLab$
	must be projected.
	In the first branch $\roleP[2]$ must be able to
	receive from $\roleP[1]$.
	In the second branch $\roleP[1]$ must be able to
	send to $\roleP[2]$.
	This contradicts \inferrule{$\projGoodType$}.
	
	\[
		\begin{array}{r@{\,}c@{\,}lr@{\,}c@{\,}l}
			\stT[0]&=&
			\stFmt{\Sigma\!\left\{
				\begin{array}{l}
					\roleS\&\stLab\stSeq
						\Sigma\!\left\{
							\begin{array}{l}
								\roleP[1]{\oplus}\stLab[0]\stSeq
								\roleP[2]{\oplus}\stLab
								\\
								\roleP[1]{\oplus}\stLab[1]\stSeq
								\roleP[3]{\oplus}\stLab
							\end{array}
						\right\}
					\\
					\roleS\&\stLabi\stSeq
						\Sigma\!\left\{
							\begin{array}{l}
								\roleP[2]{\oplus}\stLab
								\\
								\roleP[3]{\oplus}\stLab
							\end{array}
						\right\}	
				\end{array}
			\right\}}
			&
			\stT[1]&=&
			\stFmt{\Sigma\!\left\{
				\begin{array}{l}
					\roleS\&\stLab\stSeq
						\Sigma\!\left\{
							\begin{array}{l}
								\roleP[0]\&\stLab[0]\stSeq
								\roleP[3]{\oplus}\stLab
								\\
								\roleP[0]\&\stLab[1]\stSeq
								\roleP[2]{\oplus}\stLab
							\end{array}
						\right\}
					\\
					\roleS\&\stLabi\stSeq
						\Sigma\!\left\{
							\begin{array}{l}
								\roleP[2]{\oplus}\stLab
								\\
								\roleP[3]{\oplus}\stLab
							\end{array}
						\right\}	
				\end{array}
			\right\}}
			\\
			\stT[2]&=&
			\stFmt{\Sigma\!\left\{
				\begin{array}{l}
					\roleS\&\stLabi\stSeq
						\Sigma\!\left\{
							\begin{array}{l}
								\roleP[3]{\oplus}\stLab[0]\stSeq
								\roleP[0]\&\stLab
								\\
								\roleP[3]{\oplus}\stLab[1]\stSeq
								\roleP[1]\&\stLab
							\end{array}
						\right\}
					\\
					\roleS\&\stLab\stSeq
						\Sigma\!\left\{
							\begin{array}{l}
								\roleP[0]\&\stLab
								\\
								\roleP[1]\&\stLab
							\end{array}
						\right\}	
				\end{array}
			\right\}}
			&
			\stT[3]&=&
			\stFmt{\Sigma\!\left\{
				\begin{array}{l}
					\roleS\&\stLabi\stSeq
						\Sigma\!\left\{
							\begin{array}{l}
								\roleP[2]\&\stLab[0]\stSeq
								\roleP[1]\&\stLab
								\\
								\roleP[2]\&\stLab[1]\stSeq
								\roleP[0]\&\stLab
							\end{array}
						\right\}
					\\
					\roleS\&\stLab\stSeq
						\Sigma\!\left\{
							\begin{array}{l}
								\roleP[0]\&\stLab
								\\
								\roleP[1]\&\stLab
							\end{array}
						\right\}	
				\end{array}
			\right\}}
			\\
			\stT[\roleS]&=&
			\stFmt{\Sigma\!\left\{
				\begin{array}{l}
					\roleP[0]{\oplus}\stLab\stSeq
					\roleP[1]{\oplus}\stLab\stSeq
					\roleP[2]{\oplus}\stLab\stSeq
					\roleP[3]{\oplus}\stLab
					\\
					\roleP[0]{\oplus}\stLabi\stSeq
					\roleP[1]{\oplus}\stLabi\stSeq
					\roleP[2]{\oplus}\stLabi\stSeq
					\roleP[3]{\oplus}\stLabi
				\end{array}
			\right\}}
			&
			\stEnv&=&\set{
			\stEnvMap{\roleS}{\stT[\roleS]}
			\stEnvComp
			\stEnvMap{\roleP[0]}{\stT[0]}
			\stEnvComp
			\stEnvMap{\roleP[1]}{\stT[1]}
			\stEnvComp
			\stEnvMap{\roleP[2]}{\stT[2]}
			\stEnvComp
			\stEnvMap{\roleP[3]}{\stT[3]}
			}
		\end{array}
	\]
	Observe that $\gtTrace{\stEnv}=\gtTrace{\gtG}$,
	$\stEnv$ is safe and deadlock-free, and
	$\dom{\stEnv}=\gtRoles{\gtG}$.
	
	Suppose that there is some good projection $\gtProj{}{}{}$
	with $\gtProj{}{\gtG}{\stEnvi}$.
	So there is $\stEnvii$ such that
	\[\gtProj{}{\Sigma\!\left\{
				\begin{array}{l}
					{\roleP[0]}\to{\roleP[1]}:\gtLab[0]\gtSeq
					\roleP[0]\to\roleP[2]:\gtLab\gtSeq
					\roleP[1]\to\roleP[3]:\gtLab
					\\
					{\roleP[0]}\to{\roleP[1]}:\gtLab[1]\gtSeq
					\roleP[0]\to\roleP[3]:\gtLab\gtSeq
					\roleP[1]\to\roleP[2]:\gtLab
				\end{array}
				\right\}}{\stEnvii}\]
	By the criteria,
	there are $\stTi[2]$ and
	$\stTii[2]$
	such that:
	\[
	\gtProj{\roleP[2]}{\roleP[0]\to\roleP[2]:\gtLab\gtSeq
					\roleP[1]\to\roleP[3]:\gtLab}{\stTi[2]},\;
	\gtProj{\roleP[2]}{\roleP[0]\to\roleP[3]:\gtLab\gtSeq
					\roleP[1]\to\roleP[2]:\gtLab}{\stTii[2]},\;\text{and}\;
	\stEnvApp{\stEnvii}{\roleP[2]}\tySub\stTi[2],\stTii[2]\]
	We now get that
	$\stEnvApp{\stEnvii}{\roleP[2]}\gtMove[{\roleP[1]\stFmt{\&}\stLab}]$,
	so there is $\stChoice{\stLabii}{\tyGround}$
	such that
	$\stTi[2]\gtMove[{\roleP[1]\stFmt{\&}\stChoice{\stLabii}{\tyGround}}]$
	and
	$\stTi[2]\gtMove[{\roleP[0]\stFmt{\&}\stLab}]$.
	Similarly down the other branch,
	there is $\stTi[1]$ such that
	$\gtProj{\roleP[1]}{\roleP[0]\to\roleP[3]:\gtLab\gtSeq
					\roleP[1]\to\roleP[2]:\gtLab}{\stTi[1]}$,
	$\stTi[1]\gtMove[{\roleP[3]\stFmt{{\oplus}}\stChoice{\stLabiii}{\tyGroundi}}]$,
	and
	$\stTi[1]\gtMove[{\roleP[2]\stFmt{\&}\stLab}]$.
	These imply that
	$\roleP[0]\to\roleP[3]:\gtLab\gtSeq
					\roleP[1]\to\roleP[2]:\gtLab
	\gtMove[\ltsSendRecv{\roleP[1]}{\roleP[2]}{\stChoice{\stLabiii}{\tyGroundi}}]$,
	which is false.
	We also observe that $\stEnvii$ moves to
	a context where this occurs.
	
	Using characteristic processes (\cref{def:char-proc}),
	we replace the typing contexts by sessions
	and replace appeals to the good projection criteria
	with appeals to subject reduction and session fidelity.
	Thus instead of the good projection criteria failing,
	either subject reduction or session fidelity fails.
\end{proof}

We solve this issue by introducing 
\emph{history dependent global types},
while retaining all of the correctness guarantees of good projections.

\begin{definition}[History Dependent Global Types]
\emph{History dependent global types} are
3-tuples $\ENCan{\gtG,\vec{\stEnvAnnotGenericSym},\gtGi}$ 
such that $\gtG\,\gtMove[\vec{\stEnvAnnotGenericSym}]\,\gtGi$.
We define
$\ENCan{\gtG,\vec{\stEnvAnnotGenericSym},\gtGi}
\gtMove[\stEnvAnnotGenericSymi]
\ENCan{\gtG,\vec{\stEnvAnnotGenericSym}\stEnvAnnotGenericSymi,\gtGii}$
iff
$\gtGi\,\gtMove[\stEnvAnnotGenericSymi]\,\gtGii$.
\end{definition}

The type $\ENCan{\gtG,\vec{\stEnvAnnotGenericSym},\gtGi}$
represents $\gtGi$, obtained by a transition sequence
with labels $\vec{\stEnvAnnotGenericSym}$ from $\gtG$.
$\ENCan{\gtG,\epsilon,\gtG}$
is
$\gtG$ with no history;
we write $\gtG$ for $\ENCan{\gtG,\epsilon,\gtG}$
when it is clear that we are using history dependent global types.

Good projections on history dependent global types subsume
good projections on global types without history.

\begin{proposition}[Good implies History Dependent Good]
	\label{prop:good-implies-history-good}
	If $\gtProj{}{}{}$
	is a good projection on global types without history,
	then $\gtProji{}{}{}=\{(\roleP,\ENCan{\gtGi,\vec{\stEnvAnnotGenericSym},\gtG},\stT)\suchthat\gtProj{\roleP}{\gtG}{\stT}\}$
	is a good projection.
\end{proposition}

Good projections on history dependent global types subsume
direct projection \emph{partial functions}.

\begin{definition}[Lifting Direct Projections]
\label{def:lifted-proj}
Let $\gtProj{}{}{}$ be a direct projection.
Define projection $\gtProjAnnot{}{}{}{l}$
on history dependent global types by
$\gtProjAnnot{}{}{}{l}=\{
\left(
\roleP,
\ENCan{\gtG,\vec{\stEnvAnnotGenericSym},\gtGi},
\stEnvApp{\stEnvi}{\roleP}
\right)
\suchthat
\exists \stEnv.\;
\gtProj{}{\gtG}{\stEnv}
\wedge
\stEnv\stEnvMoveAnnot{\vec{\stEnvAnnotGenericSym}}\stEnvi
\}$.
\end{definition}

\begin{restatable}[Functions]{theorem}{thmTraceProjGood}
\label{thm:trace-proj}
If
$\gtProj{}{}{}$
is a direct projection and partial function,
then
$\gtProjAnnot{}{}{}{l}$ is good.
\end{restatable}

\begin{proof}
	Let $F$ denote the partial function
	defined by the relation $\gtProj{}{}{}$.

	Suppose that $\gtProjAnnot{}{\ENCan{\gtG,\vec{\stEnvAnnotGenericSym},\gtGi}}{\stEnv}{l}$,
	so $F\!\left(\gtG\right)\stEnvMoveAnnot{\vec{\stEnvAnnotGenericSym}}\stEnv$.
	
	\inferrule{$\projGoodEnd$}.
	If $\roleP\in \dom{\stEnv}$
	and $\stEnvApp{\stEnv}{\roleP}\gtMove$,
	then there is
	$\vec{\stEnvAnnotGenericSymi}$ such that
	$\roleP\in\ltsSubject{\vec{\stEnvAnnotGenericSymi}}$
	and
	$\stEnv\stEnvMoveAnnot{\vec{\stEnvAnnotGenericSymi}}$.
	Thus,
	$\gtG\,\gtMove[\vec{\stEnvAnnotGenericSym}\vec{\stEnvAnnotGenericSymi}]$.
	Thus,
	$\gtGi\,\gtMove[\vec{\stEnvAnnotGenericSymi}]$.
	Thus, $\roleP\in\gtRoles{\gtGi}$.
	
	\inferrule{$\projGoodOut$}, \inferrule{$\projGoodIn$} and
	\inferrule{$\projGoodAlt$}.
	If $\gtGi\,\gtMove[ \ltsSendRecv{\roleP}{\roleQ}{\stChoice{\stLab}{\tyGround}} ]
     \,\gtGii$, then
     $F\!\left(\gtG\right)\stEnvMoveAnnot{\vec{\stEnvAnnotGenericSym}
     \ltsSendRecv{\roleP}{\roleQ}{\stChoice{\stLab}{\tyGround}}
     }\stEnvi$
     so
     $\stEnv\stEnvMoveAnnot{
     \ltsSendRecv{\roleP}{\roleQ}{\stChoice{\stLab}{\tyGround}}
     }\stEnvi$
     and\\
     $\gtProjAnnot{}{\ENCan{\gtG,
     \vec{\stEnvAnnotGenericSym}
     \ltsSendRecv{\roleP}{\roleQ}{\stChoice{\stLab}{\tyGround}}
     ,\gtGii}}{\stEnvi}{l}$.
     
     \inferrule{$\projGoodType$}.
     If $\stEnvApp{\stEnv}{\roleP}\,\gtMove[{\roleQ\stEnvAnnotOutSym\stChoice{\stLab}{\tyGround}}]$
     and
     $\stEnvApp{\stEnv}{\roleQ}\,\gtMove[{\roleP\stEnvAnnotInSym\stChoice{\stLabi}{\tyGroundi}}]$,
     then by safety,
     $\stEnv\stEnvMoveAnnot{\ltsSendRecv{\roleP}{\roleQ}{\stChoice{\stLab}{\tyGround}}}$
     so
     $\gtGi\,\gtMove[\ltsSendRecv{\roleP}{\roleQ}{\stChoice{\stLab}{\tyGround}}]$,\\
     thus
     $\ENCan{\gtG,\vec{\stEnvAnnotGenericSym},\gtGi}
     \,\gtMove[\ltsSendRecv{\roleP}{\roleQ}{\stChoice{\stLab}{\tyGround}}]$.
\end{proof}

Note that safety of $F\!\left(\gtG\right)$
is necessary, since this is \emph{not}
implied by trace equivalence and `weak liveness' \eg
$\gtG=\gtCommSingle{\roleP}{\roleQ}{\stLab}{}{}$
and
$\stEnv=\set{
\stEnvMap{\roleP}{\stFmt{\roleQ{\oplus}\{\stLab,\,\stLabi\}}}\stEnvComp
\stEnvMap{\roleQ}{\stFmt{\roleP{\&}\stLab}}
}$
have the same traces
but $\stEnv$ is not safe.

\begin{example}[Direct Relations]
	\label{ex:ctx-proj}
	Consider the global type and contexts:
	\\\centerline{\(
		\begin{array}{c}
			\gtG = \gtCommSingle{\roleFmt{a}}{\roleFmt{b}}{\gtLab}{}{
				\gtCommSingle{\roleFmt{c}}{\roleFmt{d}}{\gtLab}{}{}
			}
			\quad
			\stEnv =
			\stEnvMap{\roleFmt{a}}{\stFmt{ \Sigma\!\left\{
				\roleFmt{b}{\oplus}\stLab,\,\roleFmt{d}{\oplus}\stLabi
			\right\}
			}}
			\stEnvComp
			\stEnvMap{\roleFmt{b}}{\stFmt{
				\roleFmt{a}{\&}\stLab
			}}
			\stEnvComp
			\stEnvMap{\roleFmt{c}}{\stFmt{
				\roleFmt{d}{\oplus}\stLab
			}}
			\stEnvComp
			\stEnvMap{\roleFmt{d}}{\stFmt{\Sigma\!\left\{
				\roleFmt{c}{\&}\stLab,\,\roleFmt{a}{\&}\stLab
			\right\}}}
			\\
			\stEnvi =
			\stEnvMap{\roleFmt{a}}{\stFmt{ \Sigma\!\left\{
				\roleFmt{b}{\oplus}\stLab,\,\roleFmt{d}{\oplus}\stLabi
			\right\}
			}}
			\stEnvComp
			\stEnvMap{\roleFmt{b}}{\stFmt{
				\roleFmt{a}{\&}\stLab
			}}
			\stEnvComp
			\stEnvMap{\roleFmt{c}}{\stFmt{
				\roleFmt{d}{\oplus}\stLab
			}}
			\stEnvComp
			\stEnvMap{\roleFmt{d}}{\stFmt{
				\roleFmt{c}{\&}\stLab
			}}
			\quad
			\stEnvii =
			\stEnvMap{\roleFmt{a}}{\stFmt{
				\roleFmt{b}{\oplus}\stLab
			}}
			\stEnvComp
			\stEnvMap{\roleFmt{b}}{\stFmt{
				\roleFmt{a}{\&}\stLab
			}}
			\stEnvComp
			\stEnvMap{\roleFmt{c}}{\stFmt{
				\roleFmt{d}{\oplus}\stLab
			}}
			\stEnvComp
			\stEnvMap{\roleFmt{d}}{\stFmt{\Sigma\!\left\{
				\roleFmt{c}{\&}\stLab,\,\roleFmt{a}{\&}\stLab
			\right\}}}
		\end{array}
	\)}
	$\gtProj{}{}{}=\{(\gtG,\stEnvi),(\gtG,\stEnvii)\}$
	is a direct projection.
	But $\gtProjAnnot{}{\gtG}{\stEnv}{l}$
	and $\stEnv$ is unsafe.
	Thus,
	typing with
	$\gtProjAnnot{}{}{}{l}$
	does not satisfy the subject reduction theorem.
	\cref{thm:trace-proj} requires direct projection
	\emph{functions}.
\end{example}

With this extension, 
algorithms for \emph{deadlock-free realisability} \cite[Section 3.2]{majumdar_et_al:LIPIcs.CONCUR.2021.35}
are good.
Note that the good projection criteria in \cref{def:good-proj}
do not require the trace condition (\cref{prop:good-proj-not-trace}).

\begin{proposition}[Good Projection Without Trace Agreement]
\label{prop:good-proj-not-trace}
There is a good projection (on global types without history), $\gtProj{}{}{}$,
and $\gtG$ and $\stEnv$
such that
$\gtProj{}{\gtG}{\stEnv}$
and
$\gtTrace{\stEnv}\neq\gtTrace{\gtG}$.
\end{proposition}

\begin{proof}
	Let $\gtProj{}{}{}$
	be the plain inductive projection.
	
	Let $f$ be a choice function on labels.
	Define $F$ recursively on local types by:
	\[
	\begin{array}{rcll}
		F\!\left(\stEnd\right)
		&=&
		\stEnd
		\\
		F\!\left(\stRecVar\right)&=&\stRecVar
		\\
		F\!\left(\stRec{\stRecVar}{\stT}\right)&=&\stRec{\stRecVar}{F\!\left({\stT}\right)}
		\\
		F\!\left(
		\stExtSum{\roleP}{i \in I}{\stChoice{\stLab[i]}{\tyGround[i]}\stSeq\stT[i]}
		\right)
		&=
		&
		\stExtSum{\roleP}{i \in I}{\stChoice{\stLab[i]}{\tyGround[i]}\stSeq
		F\!\left(\stT[i]\right)}
		\\
		F\!\left(
		\stIntSum{\roleP}{i \in I}{\stChoice{\stLab[i]}{\tyGround[i]}\stSeq\stT[i]}
		\right)
		&=
		&
		\stFmt{
		\roleP{\oplus}
		\stChoice{\stLab[j]}{\tyGround[j]}\stSeq
		F\!\left(\stT[j]\right)
		}
		&
		j=
		f(\{\stLab[i]\}_{i\in I})
	\end{array}
	\]
	$F$ makes all the internal choices.
	
	Define $F'$ by:
	\[
	\begin{array}{rcl}
		F'\!\left(\stEnd\right)
		&=&
		\stEnd
		\\
		F'\!\left(\stRecVar\right)&=&\stRecVar
		\\
		F'\!\left(\stRec{\stRecVar}{\stT}\right)&=&
		F'\!\left(\stT\subst{\stRecVar}{\stRec{\stRecVar}{\stT}}\right)
		\\
		F'\!\left(
		\stExtSum{\roleP}{i \in I}{\stChoice{\stLab[i]}{\tyGround[i]}\stSeq\stT[i]}
		\right)
		&=
		&
		\stExtSum{\roleP}{i \in I}{\stChoice{\stLab[i]}{\tyGround[i]}\stSeq
		F\!\left(\stT[i]\right)}
		\\
		F'\!\left(
		\stIntSum{\roleP}{i \in I}{\stChoice{\stLab[i]}{\tyGround[i]}\stSeq\stT[i]}
		\right)
		&=
		&
		\stIntSum{\roleP}{i \in I}{\stChoice{\stLab[i]}{\tyGround[i]}\stSeq
		F\!\left(\stT[i]\right)}
	\end{array}
	\]
	$F'$ makes all internal choices except
	the top most choice.
	
	Define $\gtProji{}{}{}$ by
	\[
		\gtProji{\roleP}{\gtG}{F\!\left(\stT\right)}
		\iff
		\gtProj{\roleP}{\gtG}{\stT}
	\]
	This is a good projection
	since
	$\gtProj{}{}{}$ is good and
	if $\gtProji{}{\gtG}{\stEnv}$
	then there is $\stEnvi$ such that
	$\stEnv\tySub\stEnvi$,
	$\gtProj{}{\gtG}{\stEnvi}$ and
	$\forall\stEnvAnnotGenericSym.\;
	\stEnv\stEnvMoveGenAnnot
	\iff
	\stEnvi\stEnvMoveGenAnnot
	$.
	
	Consider:
	\[
		\gtG =
		\gtCommSingle{\roleP}{\roleQ}{\gtLab}{}{
		\gtCommRaw{\roleP}{\roleQ}{
			\gtLab[1],\,\gtLab[2]
		}
		}
		\quad
		\stEnv
		=
		\set{
		\stEnvMap{\roleP}{
		\roleQ{\oplus}\stLab\stSeq\roleQ{\oplus}\stLab[1]
		}
		\stEnvComp
		\stEnvMap{\roleQ}{
		\roleP{\&}\stLab\stSeq\roleP{\&}\{\stLab[1],\,\stLab[2]\}
		}
		}
	\]
	We have $\gtProji{}{\gtG}{\stEnv}$
	and $\ltsSendRecv{\roleP}{\roleQ}{\stLab}\ltsSendRecv{\roleP}{\roleQ}{\stLab[2]}
	\in
	\gtTrace{\gtG}
	\setminus
	\gtTrace{\stEnv}
	$.
\end{proof}

\begin{example}[Global types from {\cite[Ex. 2.2]{Li2023-sx}}]
	\label{ex:examples-complete-mp}
	As in \cite{Li2023-sx},
	not all (history dependent) global types are
	projectable by a good projection.
	We repeat the examples from \cite[Ex. 2.2]{Li2023-sx}:
	\[
		\begin{array}{r@{\,}c@{\,}lr@{\,}c@{\,}l}
			\gtG[r]&=&
			\gtFmt{
			\Sigma\!\left\{
			\begin{array}{l}
				\roleP\to\roleQ:\gtMsgFmt{o}{\gtSeq}
				\roleQ\to\roleR:\gtMsgFmt{o}{\gtSeq}
				\roleP\to\roleR:\gtMsgFmt{o}
				\\
				\roleP\to\roleQ:\gtMsgFmt{m}{\gtSeq}
				\roleP\to\roleR:\gtMsgFmt{o}{\gtSeq}
				\roleQ\to\roleR:\gtMsgFmt{o}
			\end{array}
			\right\}
			}
			&
			\gtG[s]&=&
			\gtFmt{
			\Sigma\!\left\{
			\begin{array}{l}
				\roleP\to\roleQ:\gtMsgFmt{o}{\gtSeq}
				\roleR\to\roleQ:\gtMsgFmt{o}
				\\
				\roleP\to\roleQ:\gtMsgFmt{m}{\gtSeq}
				\roleR\to\roleQ:\gtMsgFmt{m}
			\end{array}
			\right\}
			}
			\\
			\gtGi[r]&=&
			\gtFmt{
			\Sigma\!\left\{
			\begin{array}{l}
				\roleP\to\roleQ:\gtMsgFmt{o}{\gtSeq}
				\roleQ\to\roleR:\gtMsgFmt{o}{\gtSeq}
				\roleR\to\roleP:\gtMsgFmt{o}{\gtSeq}
				\roleP\to\roleR:\gtMsgFmt{o}
				\\
				\roleP\to\roleQ:\gtMsgFmt{m}{\gtSeq}
				\roleP\to\roleR:\gtMsgFmt{o}{\gtSeq}
				\roleR\to\roleQ:\gtMsgFmt{o}{\gtSeq}
				\roleQ\to\roleR:\gtMsgFmt{o}
			\end{array}
			\right\}
			}
			&
			\gtGi[s]&=&
			\gtFmt{
			\Sigma\!\left\{
			\begin{array}{l}
				\roleP\to\roleQ:\gtMsgFmt{o}{\gtSeq}
				\roleR\to\roleQ:\gtMsgFmt{b}
				\\
				\roleP\to\roleQ:\gtMsgFmt{m}{\gtSeq}
				\roleR\to\roleQ:\gtMsgFmt{b}
			\end{array}
			\right\}
			}
		\end{array}
	\]
	There is no good projection
	that projects $\ENCan{\gtG[r],\epsilon,\gtG[r]}$
	or $\ENCan{\gtG[s],\epsilon,\gtG[s]}$.
	
	Suppose that $\gtProj{}{}{}$ is good and
	\[\gtProj{}{\ENCan{\gtG[r],\epsilon,\gtG[r]}}{\stEnv}\]
	Thus,
	\[\gtProj{}{\ENCan{\gtG[r],\ltsSendRecv{\roleP}{\roleQ}{\stLabFmt{o}},
	\roleQ\to\roleR:\gtMsgFmt{o}{\gtSeq}
	\roleP\to\roleR:\gtMsgFmt{o}}}{\stEnvi}\]
	\[
	\gtProj{}{\ENCan{\gtG[r],
	\ltsSendRecv{\roleP}{\roleQ}{\stLabFmt{m}},
	\roleP\to\roleR:\gtMsgFmt{o}{\gtSeq}
	\roleQ\to\roleR:\gtMsgFmt{o}
	}}{\stEnvii}
	\]
	\[
		\stEnvApp{\stEnv}{\roleR}\tySub\stEnvApp{\stEnvi}{\roleR},\stEnvApp{\stEnvii}{\roleR}
	\]
	Thus,
	$\stEnvApp{\stEnvi}{\roleR}\gtMove[\stFmt{\roleQ{\&}\stLabFmt{o}}]$
	and
	$\stEnvApp{\stEnvi}{\roleR}\gtMove[\stFmt{\roleP{\&}\stChoice{\stLabi}{\tyGroundi}}]$.
	\[\gtProj{}{\ENCan{\gtG[r],\ltsSendRecv{\roleP}{\roleQ}{\stLabFmt{o}}
	\ltsSendRecv{\roleQ}{\roleR}{\stLabFmt{o}},
	\roleP\to\roleR:\gtMsgFmt{o}}}{\stEnviii}\]
	Thus,
	$\stEnvApp{\stEnvi}{\roleP}\tySub\stEnvApp{\stEnviii}{\roleP}\gtMove[\stFmt{\roleR{\oplus}\stLabFmt{o}}]$.
	Thus, $\stEnvApp{\stEnvi}{\roleP}\gtMove[\stFmt{\roleR{\oplus}\stChoice{\stLabii}{\tyGroundii}}]$.
	However,
	\[
	\ENCan{\gtG[r],\ltsSendRecv{\roleP}{\roleQ}{\stLabFmt{o}},
	\roleQ\to\roleR:\gtMsgFmt{o}{\gtSeq}
	\roleP\to\roleR:\gtMsgFmt{o}}
	\quad\not\!\!\!\!\!\!\gtMove[\ltsSendRecv{\roleP}{\roleR}{\stChoice{\stLabi}{\tyGroundi}}]
	\]
	Hence, \inferrule{$\projGoodType$} fails.
	
	Suppose that $\gtProj{}{}{}$ is good and
	\[\gtProj{}{\ENCan{\gtG[s],\epsilon,\gtG[s]}}{\stEnv}\]
	Thus,
	\[
		\gtProj{}{
		\ENCan{
			\gtG[s],
			\ltsSendRecv{\roleP}{\roleQ}{\gtMsgFmt{o}},
			\gtFmt{
			\roleR\to\roleQ:\gtMsgFmt{o}
			}
		}
		}{\stEnvi}
	\]
	\[
		\gtProj{}{
		\ENCan{
			\gtG[s],
			\ltsSendRecv{\roleP}{\roleQ}{\gtMsgFmt{m}},
			\gtFmt{
			\roleR\to\roleQ:\gtMsgFmt{m}
			}
		}
		}{\stEnvii}
	\]
	\[
		\stEnvApp{\stEnv}{\roleR}\tySub\stEnvApp{\stEnvi}{\roleR},\stEnvApp{\stEnvii}{\roleR}
	\]
	We have:
	$\stEnvApp{\stEnvi}{\roleR}\gtMove[\stFmt{\roleQ}{\oplus}\stLabFmt{o}]$
	by \inferrule{$\projGoodOut$};
	$\stEnvApp{\stEnvii}{\roleR}\gtMove[\stFmt{\roleQ}{\oplus}\stLabFmt{m}]$
	by \inferrule{$\projGoodOut$};
	$\stEnvApp{\stEnvi}{\roleQ}\gtMove[\stFmt{\roleR}{\&}\stLabFmt{o}]$
	by \inferrule{$\projGoodIn$};
	$\stEnvApp{\stEnvii}{\roleQ}\gtMove[\stFmt{\roleR}{\&}\stLabFmt{m}]$
	by \inferrule{$\projGoodIn$}.
	By \inferrule{$\projGoodType$},
	if $\stEnvApp{\stEnvi}{\roleR}\gtMove[\stFmt{\roleQ}{\oplus}\stChoice{\stLabi}{\tyGroundi}]$
	then
	$\stChoice{\stLabi}{\tyGroundi}=\stLabFmt{o}$.
	By \inferrule{$\projGoodType$},
	if $\stEnvApp{\stEnvii}{\roleR}\gtMove[\stFmt{\roleQ}{\oplus}\stChoice{\stLabi}{\tyGroundi}]$
	then
	$\stChoice{\stLabi}{\tyGroundi}=\stLabFmt{m}$.
	Hence,
	$\stEnvApp{\stEnv}{\roleR}\gtMove[\stFmt{\roleQ}{\oplus}\stLabFmt{o}]$
	and
	$\stEnvApp{\stEnv}{\roleR}\gtMove[\stFmt{\roleQ}{\oplus}\stLabFmt{m}]$.
	Hence,
	$\stEnvApp{\stEnvi}{\roleR}\gtMove[\stFmt{\roleQ}{\oplus}\stLabFmt{m}]$;
	this is a contradiction.
	
	Note that this also means that they cannot be projected by
	any direct projection.
	If $\gtProj{}{}{}$ is a direct projection relation and
	$\gtProj{}{\gtG}{\stEnv}$,
	then $\set{ (\gtG,\stEnv) }\subseteq \gtProj{}{}{}$
	is a direct projection partial function.
	Thus, $\set{ (\gtG,\stEnv) }_{l}$
	is a good projection on history dependent
	global types, making
	$\gtG$ projectable.
	\\[2ex]
	$\gtGi[r]$ and $\gtGi[s]$
	can be projected without using history dependent types.
	
	Let $\gtProj{}{}{}$
	be the plain inductive projection.
	\[
		\gtProj{\roleP}{\gtGi[s]}{
			\stFmt{\Sigma\!\left\{\roleQ{\oplus}\stLabFmt{o},\,\roleQ{\oplus}\stLabFmt{m}\right\}}
		}
		\quad
		\gtProj{\roleQ}{\gtGi[s]}{
			\stFmt{\Sigma\!\left\{\roleP{\&}\stLabFmt{o}\stSeq\roleR{\&}\stLabFmt{b},
			\,\roleP{\&}\stLabFmt{m}\stSeq\roleR{\&}\stLabFmt{b}\right\}}
		}
		\quad
		\gtProj{\roleR}{\gtGi[s]}{\stFmt{\roleQ{\oplus}\stLabFmt{b}}}
	\]
	\[
		\stEnvi[s] =
		\set{
		\stEnvMap{\roleP}{\stFmt{\Sigma\!\left\{\roleQ{\oplus}\stLabFmt{o},\,\roleQ{\oplus}\stLabFmt{m}\right\}}}
		\stEnvComp
		\stEnvMap{\roleQ}{\stFmt{\Sigma\!\left\{\roleP{\&}\stLabFmt{o}\stSeq\roleR{\&}\stLabFmt{b},
			\,\roleP{\&}\stLabFmt{m}\stSeq\roleR{\&}\stLabFmt{b}\right\}}}
		\stEnvComp
		\stEnvMap{\roleR}{\stFmt{\roleQ{\oplus}\stLabFmt{b}}}
		}
	\]
	Thus, $\gtProj{}{\gtGi[s]}{\stEnvi[s]}$
	
	Define $\gtProj{}{}{}$ by:
	\[
		\begin{array}{l}
			\gtProj{\roleP}{}{} =\\
			\quad\left\{
			
			(
			\gtGi[r],
			\stFmt{
			\Sigma\!\left\{
			\roleQ{\oplus}\stLabFmt{o}\stSeq
			\roleR{\&}\stLabFmt{o}\stSeq
			\roleR{\oplus}\stLabFmt{o},\,
			\roleQ{\oplus}\stLabFmt{m}\stSeq
			\roleR{\oplus}\stLabFmt{o}
			\right\}
			}
			)
			,
			(
				\gtFmt{\roleQ\to\roleR:\gtMsgFmt{o}{\gtSeq}
				\roleR\to\roleP:\gtMsgFmt{o}{\gtSeq}
				\roleP\to\roleR:\gtMsgFmt{o}},
			\stFmt{
			\roleR{\&}\stLabFmt{o}\stSeq
			\roleR{\oplus}\stLabFmt{o}
			}
			)
			,\right.\\\quad
			(
				\gtFmt{
				\roleR\to\roleP:\gtMsgFmt{o}{\gtSeq}
				\roleP\to\roleR:\gtMsgFmt{o}},
			\stFmt{
			\roleR{\&}\stLabFmt{o}\stSeq
			\roleR{\oplus}\stLabFmt{o}
			}
			)
			,
			(
				\gtFmt{
				\roleP\to\roleR:\gtMsgFmt{o}},
			\stFmt{
			\roleR{\oplus}\stLabFmt{o}
			}
			),\\\quad
			(
			\gtFmt{
				\roleP\to\roleR:\gtMsgFmt{o}{\gtSeq}
				\roleR\to\roleQ:\gtMsgFmt{o}{\gtSeq}
				\roleQ\to\roleR:\gtMsgFmt{o}
			},
			\stFmt{
			\roleR{\oplus}\stLabFmt{o}
			}
			),
			(
			\gtFmt{
				\roleR\to\roleQ:\gtMsgFmt{o}{\gtSeq}
				\roleQ\to\roleR:\gtMsgFmt{o}
			},
			\stEnd
			),\\\quad
			(
			\gtFmt{
				\roleQ\to\roleR:\gtMsgFmt{o}
			},
			\stEnd
			),
			(\gtEnd,\stEnd)
			\}
			\\[1ex]
			\gtProj{\roleQ}{}{} =\\
			\quad\left\{
			
			(
			\gtGi[r],
			\stFmt{
			\Sigma\!\left\{
			\roleP{\&}\stLabFmt{m}\stSeq
			\roleR{\&}\stLabFmt{o}\stSeq
			\roleR{\oplus}\stLabFmt{o},\,
			\roleP{\&}\stLabFmt{o}\stSeq
			\roleR{\oplus}\stLabFmt{o}
			\right\}
			}
			)
			,
			(
				\gtFmt{\roleP\to\roleR:\gtMsgFmt{o}{\gtSeq}
				\roleR\to\roleQ:\gtMsgFmt{o}{\gtSeq}
				\roleQ\to\roleR:\gtMsgFmt{o}},
			\stFmt{
			\roleR{\&}\stLabFmt{o}\stSeq
			\roleR{\oplus}\stLabFmt{o}
			}
			)
			,\right.\\\quad
			(
				\gtFmt{
				\roleR\to\roleQ:\gtMsgFmt{o}{\gtSeq}
				\roleQ\to\roleR:\gtMsgFmt{o}},
			\stFmt{
			\roleR{\&}\stLabFmt{o}\stSeq
			\roleR{\oplus}\stLabFmt{o}
			}
			)
			,
			(
				\gtFmt{
				\roleQ\to\roleR:\gtMsgFmt{o}},
			\stFmt{
			\roleR{\oplus}\stLabFmt{o}
			}
			),\\\quad
			(
			\gtFmt{
				\roleQ\to\roleR:\gtMsgFmt{o}{\gtSeq}
				\roleR\to\roleP:\gtMsgFmt{o}{\gtSeq}
				\roleP\to\roleR:\gtMsgFmt{o}
			},
			\stFmt{
			\roleR{\oplus}\stLabFmt{o}
			}
			),
			(
			\gtFmt{
				\roleR\to\roleP:\gtMsgFmt{o}{\gtSeq}
				\roleP\to\roleR:\gtMsgFmt{o}
			},
			\stEnd
			),\\\quad
			(
			\gtFmt{
				\roleP\to\roleR:\gtMsgFmt{o}
			},
			\stEnd
			),
			(\gtEnd,\stEnd)
			\}
			\\[1ex]
			\gtProj{\roleR}{}{} =\\
			\quad\{
			
			(
			\gtGi[r],
			\stFmt{
			\Sigma\!\left\{
			\roleP{\&}\stLabFmt{o}\stSeq
			\roleQ{\oplus}\stLabFmt{o}\stSeq
			\roleQ{\&}\stLabFmt{o},\,
			\roleQ{\&}\stLabFmt{o}\stSeq
			\roleP{\oplus}\stLabFmt{o}\stSeq
			\roleP{\&}\stLabFmt{o}
			\right\}
			}
			)
			,\\\quad
			(
				\gtFmt{\roleP\to\roleR:\gtMsgFmt{o}{\gtSeq}
				\roleR\to\roleQ:\gtMsgFmt{o}{\gtSeq}
				\roleQ\to\roleR:\gtMsgFmt{o}},
			\stFmt{
			\Sigma\!\left\{
			\roleP{\&}\stLabFmt{o}\stSeq
			\roleQ{\oplus}\stLabFmt{o}\stSeq
			\roleQ{\&}\stLabFmt{o},\,
			\roleQ{\&}\stLabFmt{o}\stSeq
			\roleP{\oplus}\stLabFmt{o}\stSeq
			\roleP{\&}\stLabFmt{o}
			\right\}
			}
			)
			,\\\quad
			(
				\gtFmt{
				\roleR\to\roleQ:\gtMsgFmt{o}{\gtSeq}
				\roleQ\to\roleR:\gtMsgFmt{o}},
			\stFmt{
			\roleQ{\oplus}\stLabFmt{o}\stSeq
			\roleQ{\&}\stLabFmt{o}
			}
			)
			,
			(
				\gtFmt{
				\roleQ\to\roleR:\gtMsgFmt{o}},
			\stFmt{
			\roleQ{\&}\stLabFmt{o}
			}
			),\\\quad
			(
			\gtFmt{
				\roleQ\to\roleR:\gtMsgFmt{o}{\gtSeq}
				\roleR\to\roleP:\gtMsgFmt{o}{\gtSeq}
				\roleP\to\roleR:\gtMsgFmt{o}
			},
			\stFmt{
			\Sigma\!\left\{
			\roleP{\&}\stLabFmt{o}\stSeq
			\roleQ{\oplus}\stLabFmt{o}\stSeq
			\roleQ{\&}\stLabFmt{o},\,
			\roleQ{\&}\stLabFmt{o}\stSeq
			\roleP{\oplus}\stLabFmt{o}\stSeq
			\roleP{\&}\stLabFmt{o}
			\right\}
			}
			),\\\quad
			(
			\gtFmt{
				\roleR\to\roleP:\gtMsgFmt{o}{\gtSeq}
				\roleP\to\roleR:\gtMsgFmt{o}
			},
			\stFmt{\roleP{\oplus}\stLabFmt{o}\stSeq
			\roleP{\&}\stLabFmt{o}}
			),
			(
			\gtFmt{
				\roleP\to\roleR:\gtMsgFmt{o}
			},
			\stFmt{\roleP{\&}\stLabFmt{o}}
			),
			(\gtEnd,\stEnd)
			\}
		\end{array}
	\]
	\[
		\stEnvi[r]
		=
		\left\{\begin{array}{l}
		\stEnvMap{\roleP}{
		\stFmt{
			\Sigma\!\left\{
			\roleQ{\oplus}\stLabFmt{o}\stSeq
			\roleR{\&}\stLabFmt{o}\stSeq
			\roleR{\oplus}\stLabFmt{o},\,
			\roleQ{\oplus}\stLabFmt{m}\stSeq
			\roleR{\oplus}\stLabFmt{o}
			\right\}
			}
		}
		\stEnvComp\\
		\stEnvMap{\roleQ}{
		\stFmt{
			\Sigma\!\left\{
			\roleP{\&}\stLabFmt{m}\stSeq
			\roleR{\&}\stLabFmt{o}\stSeq
			\roleR{\oplus}\stLabFmt{o},\,
			\roleP{\&}\stLabFmt{o}\stSeq
			\roleR{\oplus}\stLabFmt{o}
			\right\}
		}
		}
		\stEnvComp\\
		\stEnvMap{\roleR}{
		\stFmt{
			\Sigma\!\left\{
			\roleP{\&}\stLabFmt{o}\stSeq
			\roleQ{\oplus}\stLabFmt{o}\stSeq
			\roleQ{\&}\stLabFmt{o},\,
			\roleQ{\&}\stLabFmt{o}\stSeq
			\roleP{\oplus}\stLabFmt{o}\stSeq
			\roleP{\&}\stLabFmt{o}
			\right\}
		}
		}
		\end{array}
		\right\}
	\]
	The projection $\gtProj{}{}{}$
	is good and $\gtProj{}{\gtGi[r]}{\stEnvi[r]}$.
\end{example}

Observe that
for global types with non-standard choice constructors
\eg full mixed choice, sender-driven choice, directed mixed choice, \etc,
projectability under asynchronous semantics
does not imply projectability under synchronous semantics.
The projection algorithm in \cite{Li2023-sx}
is complete for asynchronous semantics
but \emph{incomplete} for synchronous semantics.  

\begin{restatable}[\cite{Li2023-sx} is Incomplete in Synchronous Semantics]{theorem}{thmProjIncomplete}
\label{thm:proj-incomplete}
Let\\
$\gtG =
		\gtFmt{
		\Sigma\!\left\{
			\roleP[0]\to\roleQ[i]:\gtLab{\gtSeq}
			\roleP[0]\to\roleP[1]:\gtLab{\gtSeq}
			\roleP[1]\to\roleQ[1-i]:\gtLab
		\right\}_{i\in\{0,1\}}
		}$.
There is a good projection $\gtProj{}{}{}$ and
a typing context $\stEnv$
such that $\gtProj{}{\gtG}{\stEnv}$
and $\gtTrace{\stEnv}=\gtTrace{\gtG}$
but the projection algorithm in \cite{Li2023-sx}
cannot project $\gtG$.
\end{restatable}

\begin{proof}
	Consider the global type:
	\[
		\gtG =
		\gtFmt{
		\Sigma\!\left\{
			\begin{array}{l}
			\roleP[0]\to\roleQ[0]:\gtLab{\gtSeq}
			\roleP[0]\to\roleP[1]:\gtLab{\gtSeq}
			\roleP[1]\to\roleQ[1]:\gtLab
			\\
			\roleP[0]\to\roleQ[1]:\gtLab{\gtSeq}
			\roleP[0]\to\roleP[1]:\gtLab{\gtSeq}
			\roleP[1]\to\roleQ[0]:\gtLab
			\end{array}
		\right\}
		}
	\]
	Consider the projection:
	\[
		\gtProj{}{}{}
		=
		\left\{\begin{array}{l}
			(\roleP[0],\gtG,
			\stSum{\roleQ[i]}{i\in \{0,1\}}{
			\stLab\stSeq\roleP[1]{\&}\stLab
			}{\oplus}
			),
			(\roleP[0],\gtFmt{\roleP[0]\to\roleP[1]:\gtLab{\gtSeq}\roleP[1]\to\roleQ[1]:\gtLab},\stFmt{\roleP[1]{\oplus}\stLab}),\\
			(\roleP[0],\gtFmt{\roleP[0]\to\roleP[1]:\gtLab{\gtSeq}\roleP[1]\to\roleQ[0]:\gtLab},\stFmt{\roleP[1]{\oplus}\stLab}),
			(\roleP[0],\gtFmt{\roleP[1]\to\roleQ[1]:\gtLab},\stEnd),\\
			(\roleP[0],\gtFmt{\roleP[1]\to\roleQ[0]:\gtLab},\stEnd),
			(\roleP[0],\gtEnd,\stEnd),\\
			(\roleP[1],\gtG,
			\stFmt{\roleP[0]{\&}\stLab\stSeq
			\stSum{\roleQ[i]}{i\in \{0,1\}}{
			\stLab
			}{\oplus}}
			),
			(\roleP[1],\gtFmt{\roleP[0]\to\roleP[1]:\gtLab{\gtSeq}\roleP[1]\to\roleQ[0]:\gtLab},
			\stFmt{\roleP[0]{\&}\stLab\stSeq
			\stSum{\roleQ[i]}{i\in \{0,1\}}{
			\stLab
			}{\oplus}}
			),\\
			(\roleP[1],\gtFmt{\roleP[0]\to\roleP[1]:\gtLab{\gtSeq}\roleP[1]\to\roleQ[1]:\gtLab},
			\stFmt{\roleP[0]{\&}\stLab\stSeq
			\stSum{\roleQ[i]}{i\in \{0,1\}}{
			\stLab
			}{\oplus}}
			),
			(\roleP[1],\gtFmt{\roleP[1]\to\roleQ[1]:\gtLab},\stSum{\roleQ[i]}{i\in \{0,1\}}{
			\stLab
			}{\oplus}),\\
			(\roleP[1],\gtFmt{\roleP[1]\to\roleQ[0]:\gtLab},\stSum{\roleQ[i]}{i\in \{0,1\}}{
			\stLab
			}{\oplus}),
			(\roleP[1],\gtEnd,\stEnd),\\
			(\roleQ[0],\gtG,
			\stSum{\roleP[i]}{i\in \{0,1\}}{
			\stLab
			}{\&}
			),
			(\roleQ[0],\gtFmt{\roleP[0]\to\roleP[1]:\gtLab{\gtSeq}\roleP[1]\to\roleQ[0]:\gtLab},
			\stSum{\roleP[i]}{i\in \{0,1\}}{
			\stLab
			}{\&}
			),\\
			(\roleQ[0],\gtFmt{\roleP[1]\to\roleQ[0]:\gtLab},
			\stSum{\roleP[i]}{i\in \{0,1\}}{
			\stLab
			}{\&}
			),
			(\roleQ[0],\gtFmt{\roleP[0]\to\roleP[1]:\gtLab{\gtSeq}\roleP[1]\to\roleQ[1]:\gtLab},
			\stEnd
			),\\
			(\roleQ[0],\gtFmt{\roleP[1]\to\roleQ[1]:\gtLab},
			\stEnd
			),
			(\roleQ[0],\gtEnd,
			\stEnd
			),\\
			(\roleQ[1],\gtG,
			\stSum{\roleP[i]}{i\in \{0,1\}}{
			\stLab
			}{\&}
			),
			(\roleQ[1],\gtFmt{\roleP[0]\to\roleP[1]:\gtLab{\gtSeq}\roleP[1]\to\roleQ[1]:\gtLab},
			\stSum{\roleP[i]}{i\in \{0,1\}}{
			\stLab
			}{\&}
			),\\
			(\roleQ[1],\gtFmt{\roleP[1]\to\roleQ[1]:\gtLab},
			\stSum{\roleP[i]}{i\in \{0,1\}}{
			\stLab
			}{\&}
			),
			(\roleQ[1],\gtFmt{\roleP[0]\to\roleP[1]:\gtLab{\gtSeq}\roleP[1]\to\roleQ[0]:\gtLab},
			\stEnd
			),\\
			(\roleQ[1],\gtFmt{\roleP[1]\to\roleQ[0]:\gtLab},
			\stEnd
			),
			(\roleQ[1],\gtEnd,
			\stEnd
			)\end{array}
		\right\}
	\]
	\[
		\stEnv=\left\{
		\begin{array}{l}
			\stEnvMap{\roleP[0]}{
			\stSum{\roleQ[i]}{i\in \{0,1\}}{
			\stLab\stSeq\roleP[1]{\oplus}\stLab
			}{\oplus}
			}
			\stEnvComp
			\stEnvMap{\roleP[1]}{
			\stFmt{
			\roleP[0]{\&}\stLab\stSeq
			\stSum{\roleQ[i]}{i\in \{0,1\}}{
			\stLab
			}{\oplus}
			}}
			\stEnvComp
			\stEnvMap{\roleQ[0]}{
			\stSum{\roleP[i]}{i\in \{0,1\}}{
			\stLab
			}{\&}
			}
			\stEnvComp
			\stEnvMap{\roleQ[1]}{
			\stSum{\roleP[i]}{i\in \{0,1\}}{
			\stLab
			}{\&}
			}
		\end{array}
		\right\}
	\]
	The projection $\gtProj{}{}{}$
	is good and $\gtProj{}{\gtG}{\stEnv}$
	and $\gtTrace{\gtG}=\gtTrace{\stEnv}$.
	
	$\gtG$ is not projectable in asynchronous semantics,
	since: $\roleP[0]$
	could send $\stLab$ to $\roleQ[0]$,
	then $\roleP[0]$ could send $\stLab$ to $\roleP[1]$,
	then $\roleP[1]$ could receive $\stLab$ from $\roleP[0]$,
	then $\roleP[1]$ could send $\stLab$ to $\roleQ[0]$.
	So any projection is deadlocked in asynchronous semantics.
	Therefore, the algorithm fails to
	output a projection for $\gtG$,
	since it is sound with respect to asynchronous semantics.
\end{proof}%

\fi
\end{document}